%% file: main.tex
\documentclass[a4paper,openright,twoside,11 pt]{report}
\pdfoutput=1 
\usepackage[toc,page]{appendix} 

\usepackage{graphicx}
\usepackage{amsmath}
\usepackage{amsfonts}
\usepackage{amssymb}
\usepackage{slashed}
\usepackage{url}
\usepackage{bm}
\usepackage{color}

\usepackage{float} %floats placement

\usepackage{ulem}

\usepackage{bm}
\usepackage{caption}

\input{macros.tex}

\author{Simone Biondini}
\title{PhD Thesis}

\begin{document}

%%%%%%%%%%%%%%%%%%%%%%%%%%%%%%%%%%%%%%%%%%%%%%%%%%%%%%%%%%%%%%%%%%%%%%%%%%%%%%%
%   Title Page                                                                %
%%%%%%%%%%%%%%%%%%%%%%%%%%%%%%%%%%%%%%%%%%%%%%%%%%%%%%%%%%%%%%%%%%%%%%%%%%%%%%%
\begin{titlepage}
\begin{center}

%\vspace{1.\titleheight}

\parbox{.38\textwidth}{\begin{flushleft}\quad\phantom{a}\includegraphics[height=1.5cm]{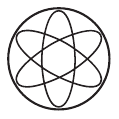}\end{flushleft}}\hfill
\parbox{.58\textwidth}{\begin{flushright} \includegraphics[height=1.5cm]{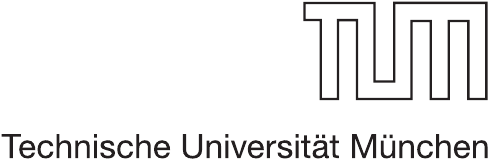}\quad\phantom{a}\end{flushright}}

\vspace{1cm}
TECHNISCHE UNIVERSIT\"AT M\"UNCHEN\\
Institut f{\"u}r Theoretische Physik T30f\\

\vspace{2.5cm}

{\Huge \bf Effective field theories for heavy Majorana neutrinos in a thermal bath}\\
\vspace{2mm}
%{\Huge \bf for Heavy Quarkonia}\\
\vspace{2mm}
%{\Huge \bf at Finite Temperature}
% \rule{.55\textwidth}{1pt}\\[0pt]
% 

\vspace{1.5cm}
 
{Simone Biondini
}
\end{center}
\vfill

{\noindent Vollst{\"a}ndiger Abdruck der von der Fakult{\"a}t f{\"u}r Physik der Technischen Universit{\"a}t M{\"u}nchen zur Erlangung des akademischen Grades eines
\vspace{.4cm}
\begin{center} 
\emph{Doktors der Naturwissenschaften (Dr.~rer.~nat.)}
\end{center}
\vspace{.4cm}
genehmigten Dissertation.

\vspace{1cm}
\begin{tabular}{lll}
Vorsitzender:&&Univ.-Prof. Dr. Lothar Oberauer \\\\
Pr{\"u}fer der Dissertation:&\qquad1.&Univ.-Prof. Dr. Nora Brambilla\\
&\qquad 2.&Univ.-Prof. Dr. Alejandro Ibarra 
\end{tabular}
\vspace{1cm}

\noindent Die Dissertation wurde am 16.03.2016 bei der Technischen Universit{\"a}t M{\"u}nchen eingereicht und durch die Fakult{\"a}t f{\"u}r Physik am 06.05.2016 angenommen.
}

\end{titlepage}
	
\cleardoublepage
	%\newpage
	 %\vspace*{2in}
	 %{\it }
	 %\end{flushright}
	 %\cleardoublepage
\section*{Zusammenfassung}
\input{zusammenfassung.tex}
\section*{Abstract}
\input{abstract.tex}

\tableofcontents

\chapter*{Introduction}
\input{introduction.tex}

		\addcontentsline{toc}{chapter}{{}Introduction}

\chapter{Baryon Asymmetry in the Early Universe}
\label{chap:baryo_lept}
\input{baryo_lept.tex}

\chapter{Baryogenesis via Leptogenesis}
\label{chap:lepto}
\input{lepto.tex}

\chapter{Effective field theories}
\label{chap:eff_the}
\input{eff_the.tex}

\chapter{Thermal field theory in a nutshell}
\label{chap:therm_tex}
\input{therm_the.tex}

\chapter{EFT approach for right-handed neutrinos in a thermal bath}
\label{chap:part_prod}
\input{part_prod.tex}

\chapter{CP asymmetries at finite temperature: the nearly degenerate case}
\label{chap:CPdege}
\input{CP_dege.tex}

\chapter{CP asymmetries at finite temperature: the hierarchical case}
\label{chap:CPhiera}
\input{CP_hiera.tex}

\chapter{Flavoured CP asymmetries}
\label{chap:CPfla}
\input{fla_CP.tex}

\chapter*{Conclusions and Outlook}
\input{conclusions.tex}

\begin{appendices}
%\chapter{Thermal field theory}
%\label{appA:thermapp}
%\input{thermapp.tex}

\chapter{Matching coefficients for the thermal width}
\label{appB:matchwidth}
\input{matchwidth.tex}

\chapter{Matching coefficients for the CP asymmetry: nearly degenerate case}
\label{appC:CPdegematch}
\input{CPdegematch.tex}

\chapter{Matching coefficients for the CP asymmetry: hierarchical case}
\label{appD:CPhieramatch}
\input{CPhieramatch.tex}

\end{appendices}

\bibliographystyle{ieeetr}  
\bibliography{biblio}

\end{document}

%% file: macros.tex
\def\siml{{\ \lower-1.2pt\vbox{\hbox{\rlap{$<$}\lower6pt\vbox{\hbox{$\sim$}}}}\ }} 
\def\simg{{\ \lower-1.2pt\vbox{\hbox{\rlap{$>$}\lower6pt\vbox{\hbox{$\sim$}}}}\ }}

\def\vbfD{{\ \lower-8pt\vbox{\hbox{\rlap{$\!\leftrightarrow$}\lower8pt\vbox{\hbox{$\!\bf D$}}}}\ }} 
\def\dsl{\,\raise.15ex\hbox{/}\mkern-13.5mu D}

\newcommand{\nn}{\nonumber}
\newcommand{\Tint}{\hbox{$\sum$}\!\!\!\!\!\!\!\int\,}
\newcommand{\be}{\begin{equation}} 
\newcommand{\ee}{\end{equation}}
\newcommand{\bea}{\begin{eqnarray}} 
\newcommand{\eea}{\end{eqnarray}}
\newcommand{\beq}{\begin{equation}}
\newcommand{\eeq}{\end{equation}}
\newcommand{\bqa}{\begin{eqnarray}}
\newcommand{\eqa}{\end{eqnarray}}

\newcommand{\pa}{\partial}
\def \bra0 {\langle 0}
\def \ket0 {0 \rangle}
\def \tr {{\rm{tr}}}

\newcommand{\Appendix}[1]%
    {%
     \section{#1}%
      }

%% file: zusammenfassung.tex
Schwere Majorana-Neutrinos treten in vielen Szenarien der Physik
jenseits des Standardmodells auf: Im urspr\"unglichen See-Saw-Mechanismus
liefern sie eine nat\"urliche Erkl\"arung f\"ur die kleinen Massen der
Neutrinos im Standardmodell, w\"ahrend sie im Rahmen der einfachsten
Leptogenesis-Modelle f\"ur die Baryonasymmetrie im Universum
verantwortlich sind. In dieser Doktorarbeit entwickeln wir eine
effektive Feldtheorie f\"ur nichtrelativistische Majorana-Teilchen, die
analog ist zur effektiven Theorie f\"ur schwere Quarks. Wie wenden die auf
diese Weise erhaltene effektive Feldtheorie an, um die Rechnungen in
einem hei\ss en Medium durchzuf\"uhren, welche die fr\"uheren Stufen der
Evolution des Universums modellieren sollen. Insbesondere wenden wir
dies auf den Fall an, in dem schwere Majorana-Neutrinos in einem hei\ss en
und dichten Plasma der Standardmodellteilchen zerfallen, dessen
Temperatur viel kleiner ist, als die Masse der Majorana-Neutrinos, aber
immerhin viel gro\ss er ist, als die elektroschwache Skala. Die thermischen
Korrekturen zu der Zerfallsbreite, die in der effektiven Feldtheorie
berechnet wurden, stimmen mit den aktuellen Ergebnissen \"uberein, welche
mit Hilfe von anderen Methoden gewonnen wurden, wobei die hier
vorgestellte Herleitung einfacher zu sein scheint. Indem wir dieselbe
Hierarchie zwischen den Massen der schweren Neutrinos und der Temperatur
annehmen, berechnen wir systematisch die thermischen Korrekturen  zu den
direkten und indirekten CP-Asymmetrien in Zerf\"allen der
Majorana-Neutrinos. Diese gehen als Schl\"usselelemente in die Gleichungen
ein, welche die thermodynamische Evolution der induzierten
Leptonenasymmetrie beschreiben, welche eventuell zu der
Baryonenasymmetrie im Universum f\"uhrt. Wir betrachten den Fall von zwei
Majorana-Neutrinos mit nahezu entarteten Massen, was eine resonante
Verst\"arkung der CP-Asymmetrie zul\"asst, sowie ein hierarchisches Spektrum
bei dem ein schweres Neutrino deutlich leichter ist, als die anderen
Spezies. Flavoureffekte werden ebenfalls bei der Herleitung der
CP-Asymmetrien bei endlicher Temperatur ber\"ucksichtigt. Die hier
vorgestellte effektive Feldtheorie eignet sich auch f\"ur eine Vielzahl
von unterschiedlichen Modellen, welche nichtrelativistische
Majorana-Fermionen beinhalten.

%% file: abstract.tex
Heavy Majorana neutrinos enter in many scenarios of physics
beyond the Standard Model: in the original seesaw mechanism they provide a
natural explanation for the small masses of the Standard Model
neutrinos and in the simplest leptogenesis framework they are at the origin of
the baryon asymmetry in the universe. In this thesis, we develop an
effective field theory for non-relativistic Majorana particles,
which is analogous to the heavy-quark effective theory. We apply the 
effective field theory so obtained to address calculations in a hot medium which models the early stages 
of the universe evolution. In particular, we apply it to the case of a heavy Majorana neutrino decaying in a hot  plasma of Standard Model particles, whose temperature is
much smaller than the mass of the Majorana neutrino but still much larger 
than the electroweak scale. The thermal corrections to the decay width computed in the 
effective field theory agree with recent results obtained using 
different methods, whereas the derivation appears to be simpler.
Assuming the same hierarchy between heavy neutrino masses and the temperature, we compute systematically thermal 
corrections to the direct and indirect CP asymmetries in the Majorana neutrino decays. 
These are key ingredients entering the equations that describe the thermodynamic evolution 
of the induced lepton-number asymmetry eventually leading to the baryon asymmetry in the universe.
We consider the case of two Majorana neutrinos with nearly degenerate masses, that allows for a 
resonant enhancement of the CP asymmetry, and a hierarchical 
spectrum with one heavy neutrino much lighter than the other neutrino species. Flavour effects
are also taken into account in the derivation of the CP asymmetries at finite temperature.
The effective field theory presented here is suitable to be used 
for a variety of different models involving non-relativistic Majorana fermions.

%% file: introduction.tex
Neutrino flavour oscillations, the large matter-antimatter asymmetry
of the universe and dark matter are commonly interpreted as major
experimental observations that require going beyond the Standard Model
(SM) of particle physics.  Among the many possible extensions of the
SM that have been proposed, a minimal extension would consist in the
inclusion of some generations of right-handed neutrinos.  Right-handed
neutrinos are singlet under the SM gauge groups, therefore they are
often called \textit{sterile} neutrinos.  Models have been considered with
different sterile neutrino generations and with neutrino masses
spanning from the eV to $10^{15}$ GeV scale. We refer to~\cite{Drewes:2013gca,Adhikari:2016bei} 
for recent reviews and a large body of references therein.

The experimental observation of neutrino mixing~\cite{Fukuda:1998mi,Ahmed:2003kj} 
implies that neutrinos carry a finite mass. 
A simple model capable of giving mass to the observed SM neutrinos and
at the same time providing a natural explanation for its smallness is
the seesaw mechanism originally proposed in~\cite{Minkowski:1977sc,GellMann:1980vs,Mohapatra:1979ia}.
In this model, right-handed neutrinos, whose mass, $M$, is much larger
than the electroweak scale, $M_W$, are coupled to lepton doublets like
right-handed leptons in the SM are. The small ratio $M_W/M$ ensures
the existence of very light mass eigenstates that may be identified
with the observed light neutrinos.  Concerning the baryon asymmetry of
the universe, although the SM contains all the requirements necessary
to dynamically generate the asymmetry, it fails to explain an asymmetry as large
as the one observed~\cite{Dolgov:1991fr}, and now accurately
determined by cosmic microwave background anisotropy
measurements~\cite{Komatsu:2008hk, Larson:2010gs}.  Baryogenesis through
leptogenesis in the original formulation of~\cite{Fukugita:1986hr} is a possible mechanism to explain
the baryon asymmetry.  In this scenario, heavy right-handed neutrinos provide 
both a source of lepton number and CP violation, moreover, they can be out of
equilibrium at temperatures where the SM particles are still thermalized.
Finally, together with many other candidates~\cite{Bertone:2004pz},
light right-handed neutrinos, minimally coupled to SM
particles like in the seesaw mechanism, may provide suitable
candidates for dark-matter particles~\cite{Boyarsky:2009ix}.

Heavy right-handed neutrinos play therefore a crucial role in models
trying to explain the neutrino masses and mass hierarchy, and in
leptogenesis. What qualifies a
neutrino as heavy in this context is that its mass is much larger than
the electroweak scale, and consequently of any SM particle. 
This allows for a temperature window in the early universe, where
the temperature is larger than the electroweak scale, but much smaller
than the neutrino mass.  In this temperature range the heavy neutrino
is out of equilibrium, and therefore contributing to the lepton
asymmetry of the universe, while the SM particles may be seen as part
of an in-equilibrium plasma at a temperature $T$.  For such 
temperatures the relevant hierarchy of energy scales is $M \gg T \, \gg M_W$ and it calls for a non-relativistic 
treatment of the heavy neutrino. Because right-handed neutrinos can be embedded 
into Majorana fields, we may want to construct a non-relativistic effective field theory (EFT)
for Majorana fermions along the same line as a non-relativistic EFT for heavy quarks, 
the heavy quark effective theory (HQET), has been built for Dirac fermions \cite{Isgur:1989vq,Eichten:1989zv}.
The construction and the application to leptogenesis of an EFT for Majorana fermions is the original part of the present thesis. 

It is a fundamental aspect of leptogenesis to take place during the early stages of the universe evolution. Therefore Majorana neutrinos are part of a thermal bath made of SM relativistic degrees of freedom.
Interactions with the medium modify the neutrino dynamics (thermal production rate, mass, ...) 
and affect the thermodynamic evolution of the lepton asymmetry. Taking into account properly thermal effects can be achieved in the framework of quantum field theories at finite temperature. The derivation of observables at finite temperature poses both conceptual and technical challenges. The thermal production rate of right-handed neutrinos has been recently
studied in~\cite{Laine:2013lka} in the relativistic and ultra-relativistic regimes.
The non-relativistic regime also turns out to be interesting for leptogenesis since it is conceivable that the CP
asymmetry is effectively generated when the temperature of the plasma
drops below the heavy-neutrino mass. In this regime the thermal
production rate for heavy Majorana neutrinos has been addressed in~\cite{Salvio:2011sf,Laine:2011pq}. A two-loop thermal field theory computation is necessary to describe the processes that account for the presence of a heat bath, namely a medium made of SM particles. The neutrino production rate is then expressed as a series in the SM couplings and powers of $T/M$. 

In the non-relativistic regime, where the EFT approach may be used, we show how to simplify the derivation of the neutrino production rate in terms of the neutrino thermal width as the pole of the heavy-neutrino propagator~\cite{Biondini:2013xua}. The advantages of an EFT treatment for heavy particles over exploiting 
the hierarchy $M \gg T$ in the course of fully relativistic calculations 
in thermal field theory are manifold. First, the EFT makes manifest, already at the 
Lagrangian level, the non-relativistic nature of the Majorana particle and a natural power counting in $T/M$ for corrections to a given observable of interest.
Second, it allows to separate the computation of relativistic and thermal corrections: 
relativistic corrections are computed setting $T=0$ and contribute to 
the Wilson coefficients of the EFT, whereas thermal corrections are computed in the EFT as small 
perturbations affecting the propagation of the non-relativistic Majorana particles in the plasma. 

Another key ingredient in leptogenesis is the CP asymmetry generated in heavy neutrino decays into leptons and 
antileptons in different amounts. Due to the CP violating phases of the Yukawa couplings the decay rate into particles can differ from that into antiparticles. Then the matter-antimatter imbalance in the lepton sector is partially reprocessed into a baryon asymmetry by the sphaleron transitions in the SM~\cite{Kuzmin:1985mm}. The CP asymmetry is originated from the interference between the tree-level and the one-loop self-energy and vertex diagrams. The contribution from the interference with the self-energy diagram is often called indirect 
contribution, while the one from the interference with the vertex diagram is called direct contribution. 
The relative importance of the indirect and direct contributions depends on the heavy-neutrino mass spectrum. 
For example, the vertex contribution is half of the self-energy contribution in the hierarchical case,
when the mass of one species of neutrinos is much lighter than the others~\cite{Liu:1993tg,Covi:1996wh}. 
The situation is rather different when two heavy neutrinos are almost degenerate in mass. 
In this case, the self-energy diagram can develop a resonant enhancement that can be traced back to a mixing
phenomenon similar to the one found in kaon physics \cite{Flanz:1996fb}. 
An analysis from first principles has been carried out in~\cite{Buchmuller:1997yu,Garny:2011hg,Garbrecht:2011aw}. 
The main phenomenological outcome is that the scale of the heavy right-handed neutrino masses 
can be lowered down to energy scales of $\mathcal{O}$(TeV)~\cite{Pilaftsis:2003gt}, welcoming collider searches. 

A recent endeavour aims at treating the CP asymmetry in a finite temperature framework, as for the right-handed neutrino production rate. The lepton-number asymmetry has been considered for a generic heavy-neutrino mass spectrum, 
e.g., in~\cite{Covi:1997dr,Giudice:2003jh,Garny:2010nj,Anisimov:2010dk,Kiessig:2011fw} within different approaches. 
Thermal effects are included using thermal masses for the Higgs boson and leptons and 
taking into account thermal distributions for the Higgs boson and leptons as decay products of the heavy Majorana neutrinos. In particular, resumming thermal masses in the Higgs and lepton propagators is justified in the high temperature regime $T \gg M$ \cite{Davidson:2008bu, Fong:2013wr}. To the best of our knowledge, such results are not on the same footing of those obtained for the neutrino production rate~\cite{Salvio:2011sf,Laine:2011pq,Biondini:2013xua}, namely, the expansion in the SM couplings has not been included in the CP asymmetry. 

The main difficulty in including systematically interactions involving SM particles of the heat bath is due to the technical complexity of the required calculation. Indeed a three-loop calculation in thermal field theory would be needed. Facing the computation directly in a fully relativistic field theory seems, to date, not an affordable task. The state of the art can be found in~\cite{Laine:2013vpa}, where the most complicated two-loop topology and the corresponding master integrals at finite temperature are discussed. If we give up insisting on a fully relativistic treatment and restrict ourselves to the non-relativistic regime, the EFT developed for heavy Majorana neutrinos may be useful to address thermal corrections to the CP asymmetry. The three-loop thermal calculation of the original theory splits into the calculation of the 
imaginary parts of two-loop diagrams that match the Wilson coefficients of the effective operators of the EFT, 
a calculation that can be performed in vacuum, and the calculation of a thermal one-loop diagram in the EFT. The program is pretty much close to that carried out for the right-handed neutrino production rate apart going one loop higher in the matching. 
In its range of applicability, the EFT framework provides a significantly simpler method of calculation and most importantly it provides a way to address systematically thermal corrections to the CP asymmetry in leptogenesis. The method is applied for two heavy neutrinos with nearly degenerate masses in \cite{Biondini:2015gyw}, whereas the hierarchical case is studied in \cite{Biondinihier}.  

The outline of the thesis is as follows. In chapter~\ref{chap:baryo_lept} the origin of the observed baryon asymmetry is discussed in the contest of the early universe. The basic requirements for any particle physics model to generate a matter-antimatter imbalance are also addressed. Then baryogenesis via leptogenesis is introduced in chapter~\ref{chap:lepto}, where the simplest realization of thermal leptogenesis in its original formulation by Fukugita and Yanagida is presented. The right-handed neutrino production rate and CP asymmetry are introduced that enter the Boltzmann equations governing the time evolution of heavy neutrinos and lepton-asymmetry number densities. The results obtained in the thesis rely on EFT and thermal field theory tools. Therefore chapter~\ref{chap:eff_the} and \ref{chap:therm_tex} are respectively devoted to a brief introduction to those subjects. The construction of the EFT for non-relativistic Majorana neutrinos together with the re-derivation of the thermal right-handed neutrino production rate in the EFT is the content of chapter~\ref{chap:part_prod}. The CP asymmetries at finite temperature are studied in chapter~\ref{chap:CPdege} for two heavy neutrinos nearly degenerate in mass, whereas the results for a hierarchical mass spectrum are collected in chapter~\ref{chap:CPhiera}. The impact of lepton flavour on our approach is discussed in chapter~\ref{chap:CPfla}, together with the expressions of the CP asymmetries in the flavoured case. Finally some conclusions and outlook are drawn, whereas technical details on the calculations are collected in the appendices.

%% file: baryo_lept.tex
In this chapter the basic concepts and notation related to the physics of the early universe are introduced.  To the best of our knowledge, the universe is evolving today from a very dense and hot phase. The Big-Bang cosmology and the thermal history of the universe are discussed in section \ref{hot_sec1}.  The early universe sets the stage for many interesting phenomena, such as the dark matter production, the generation of the baryon asymmetry and the nucleosynthesis of light elements. In section \ref{hot_sec2} we address in some detail the framework for a dynamical generation of the baryon asymmetry  discussing the Sakharov conditions together with a toy model to show their implementation. Finally the baryon and lepton number violation within the SM is presented, which is induced by the sphaleron processes in the early universe. The discussion aim at showing why one has to invoke some new physics beyond the SM to quantitatively explain the observed baryon asymmetry in the universe.
\section{Big-Bang Cosmology}
\label{hot_sec1}
At least on large scale our universe appears to us as isotropic and homogeneous, and this matter  of fact is often attached to the so-called \textit{cosmological principle} stating that the universe looks the same to all observers. The expansion of the universe is a natural consequence of any isotropic and homogeneous cosmological model based on General Relativity (GR). 
The very fact that the universe expands today implies that it was denser and warmer in the past. On the basis of GR and thermodynamics, we can extrapolate that matter had higher and higher temperature and density at earlier and earlier epochs, and that at most stages the entire system was in thermal equilibrium. The Big-Bang would then be the initial point in space-time from which we can start to study and address the early universe physics. 

The formulation of the Big-bang model began in the 1940s with the idea that the abundances of light chemical elements had a cosmological origins. In their pioneering work \cite{Gamow:1946eb, Alpher:1948ve}, George Gamow and his collaborators, Alpher and Herman, supposed that the universe was hot and dense enough to allow a nucleosynthetic processing of the hydrogen, and has expanded and cooled down to the present state. Later in 1948, Alpher and Herman predicted an important consequence of a hot universe \cite{Alpher:1949,Alpher:1949b}: a transition from a plasma of baryons, electrons and photons to a gas of atoms and free electromagnetic radiation. At this stage the atomic gas gets transparent to photons, and a relic background radiation is expected to be associated with this transition. Indeed the Cosmic Microwave Background (CMB) was detected sixteen years after its prediction \cite{PenzWils} and it has been the first experimental proof that our universe had a hot past. 
\subsection{Dynamics of an expanding universe}
We address briefly the dynamics of an expanding universe by using GR. We aim at capturing the main features relevant to our discussion: in the past the universe was smaller, denser and hotter. We focus on the epoch in which the universe was filled with relativistic particles, namely with typical momenta much bigger than their mass. The present discussion  follows standard text book derivations, such as \cite{RubaHOT}.  

Starting from the observation of an isotropic and homogeneous universe, its overall geometry can be described in terms of few independent parameters entering the Einstein equations of GR. In particular we start from the well known equation
\be 
R_{\mu \nu}- \frac{1}{2} g_{\mu \nu} R = 8 \pi G T_{\mu \nu} \, ,
\label{hotbb_1}
\ee
that connects the space-time geometry with the energy content of the universe, where $R_{\mu \nu}$ is the Ricci tensor, $R$ is the Ricci scalar, $T_{\mu \nu}$ is the energy-momentum tensor and $G$ is the gravitational constant. Natural units $c=\hbar =1$ are adopted throughout the thesis. One can find the explicit form of (\ref{hotbb_1}) for an isotropic and homogeneous metric, known as Friedmann-Lemaitre-Robertson-Walker (FLRW) metric:
\be 
ds^2=dt^2-a(t) \left[ \frac{dr^2}{1-\kappa r^2 } +r^2 (d \theta^2 +\sin^2 \theta \, d\varphi^2)\right]  \, ,
\label{hotbb_2}
\ee      
which has a maximally symmetric 3-D subspace of a 4-D space-time. In eq.~(\ref{hotbb_2}) $t$ is the time variable, $(r,\theta,\varphi)$ are the polar coordinates, $\kappa$ is a constant related to the spacial curvature. Its possible values are $-1$, $0$ and $+1$ accommodating a 3-hyperboloid, a 3-plane and  a 3-sphere respectively and describing an open, flat or close universe. The quantity $a(t)$ is called scale factor and it measures how rapidly the universe expands through the definition of the Hubble parameter
\be 
H(t)= \frac{\dot{a}(t)}{a(t)} \, ,
\label{hotbb_3}
\ee
where the dot stands for the time derivative. Assuming a FLRW geometry the left-hand side of eq.~(\ref{hotbb_1}) becomes (the $00$ component)
\be 
R_{00}-\frac{1}{2}R= 3 \left( \frac{\dot{a}^2}{a^2}  + \frac{\kappa}{a^2}\right) \, . 
\label{hotbb_4}
\ee

Let us now consider the energy momentum tensor on right-hand side in eq.~(\ref{hotbb_1}). We notice that, for cosmological epochs relevant to us, the content of the universe can be described as a homogeneous fluid with energy density $\varepsilon(t)$ and pressure $p(t)$.  If we consider this fluid as a whole at rest with respect to a comoving reference frame, then the only non-zero component of the fluid velocity, $u_\mu$, is $u_0=1$. Hence, the $00$ component of the energy momentum tensor gives
\be 
T_{00}= (\varepsilon + p) u_0 u_0 - g_{00} p=\varepsilon \, .
\label{hotbb_5}
\ee
Combining (\ref{hotbb_4}) and (\ref{hotbb_5}) we obtain the Friedmann equation:
\be 
\left( \frac{\dot{a}}{a} \right)^2 = \frac{8\pi G}{3} \varepsilon - \frac{\kappa}{a^2} \, ,
\label{hotbb_6}
\ee 
that relates the rate of the cosmological expansion with the total energy density, $\varepsilon$, and space curvature, $\kappa$. 
The Friedmann equation has to be supplemented with an additional equation since two unknown functions of time appear: $a(t)$ and $\varepsilon(t)$. That  equation can be obtain from the covariant conservation of the energy momentum tensor $T_{\mu \nu}$, that  brings to 
\be 
\dot{\varepsilon} + 3 \frac{\dot{a}}{a} (\varepsilon + p)=0 \, .
\label{hotbb_7}
\ee
Last but not the least, we add the equation of state of matter. This is necessary to close the system of equations that governs the universe expansion, and it can be written as follows
\be 
p=p(\varepsilon) \, ,
\label{hotbb_8}
\ee
enforcing the pressure to be some function of the energy density.  The equation of state  (\ref{hotbb_8}) is not a consequence of GR.  

Since we are going to deal with a heat bath of SM particles at high temperatures, it is instructive to inspect more closely the Friedmann equation in the case the universe consists, almost entirely, of relativistic degrees of freedom. Indeed we want to study the dynamics of very heavy particles inducing a baryon asymmetry in a background of either massless particles or with a mass much smaller than the typical three-momentum scale, provided by the temperature of the plasma, $T$. This epoch in the early universe is often denoted as \textit{radiation dominated era}. 
In the case of a plasma made almost entirely of relativistic particles, the equation of state in (\ref{hotbb_8}) reads: 
\be 
p = \frac{\varepsilon}{3}.
\label{hotbb_9}
\ee 
We further assume a flat geometry, $\kappa=0$, which is indeed very close to the real universe, so that the Friedmann equation (\ref{hotbb_6}) becomes
\be 
\left( \frac{\dot{a}}{a}\right)^2 = \frac{8 \pi G}{3}  \varepsilon \, . 
\label{hotbb_10}
\ee
Inserting the equation of state (\ref{hotbb_9}) into (\ref{hotbb_7}) we obtain for the energy density and Friedmann equation in (\ref{hotbb_10}) respectively
\bea
&&\varepsilon= \frac{K}{a^4} \, ,
\label{hotbb_11}
\\
&&\left( \frac{\dot{a}}{a}\right)^2 = \frac{8 \pi}{3} G  \,   \frac{K}{a^4} \, ,
\label{hotbb_11bis}
\eea
where $K$ is a constant that embeds the energy density and scale factor at some initial time $t_0$. %If we assume the initial time $t_0=0$, then $t$ becomes really the age of the universe. 
One can easily find from (\ref{hotbb_11bis}) that $a(t) \propto \sqrt{t}$ and hence the Hubble rate is $H=1/(2t)$. The energy density as a function of time can be obtained from the Friedmann equation (\ref{hotbb_10}), once the scale factor $a(t)$ has been eliminated in favour of $t$:
\be  
\varepsilon=\frac{3}{8 \pi G }H^2 = \frac{3}{32 \pi G} \frac{1}{t^2} \, .
\label{hotbb_12}
\ee
Already from this last simple relation we see that the smaller the age of the universe the bigger the energy density. 

It is useful to relate the Hubble parameter with the temperature of the universe. This will help to clarify that earlier times correspond to higher temperatures.  Considering a relativistic massless particle specie, labelled with the subscript $i$, as part of a heat bath in thermal equilibrium and neglecting chemical potentials,
the corresponding energy density reads
\be 
\varepsilon_i(T)= g_i \int \frac{d^3\bm{p}}{(2 \pi)^3} \, p \,f_{i}(p)  = g_i \begin{cases}
 \frac{\pi^2}{30} \, T^4  \, , \; \text{(boson)} \, ,
\\
 \frac{7}{8} \, \frac{\pi^2}{30} \, T^4  \, ,  \text{(fermion)} \, .
\end{cases} 
\label{hotbb_13}
\ee
In thermal equilibrium the distribution $f_i(p)$ in (\ref{hotbb_13}) is either the Bose-Einstein or the Fermi-Dirac distribution, namely
\be 
n_{B}= \frac{1}{e^{\beta E(p)}-1} \, , \quad n_{F}= \frac{1}{e^{\beta E(p)}+1} \, ,
\label{thermeq_4}
\ee
where $E(p)$ is the energy of the particle, $\beta = 1/T$ and written in a reference frame at rest with respect to the thermal bath. For highly relativistic particles the energy is $E(p)=p$ and $p \equiv |\bm{p}|$ stands for the modulo of the three-momentum of the particle with internal degree of freedom $g_i$ (for example spin polarizations). Hence for a thermal bath made of different relativistic particle species, the total energy density is
\be 
\varepsilon = \left( \sum_{i} g_{b,i} + \frac{7}{8}\sum_{i} g_{f,i} \right)  \frac{\pi^2}{30} T^4 = g_{*} \frac{\pi^2}{30} T^4 \, ,
\label{hotbb_14}
\ee
where we define the effective number of degrees of freedom, $g_*$, as the sum over bosonic, $g_{b,i}$, and fermionic, $g_{f,i}$, degrees of freedom (the latter weighted for the statistical factor $7/8$ coming from the integration of the Fermi-Dirac distribution). In general $g_*$ is temperature dependent because the number of relativistic particle species may change during the universe evolution. Now we rewrite eq.~(\ref{hotbb_12}) substituting the expression for the energy density in (\ref{hotbb_14}) as follows
\be 
H=  \frac{T^2}{M^{*}_{Pl}} \, ,
\label{hotbb_15}
\ee
where we used $G=M_{Pl}^{-2}$, where $M_{Pl}$ is the Planck mass, and the definition of the effective Planck mass, which depends on the number of effective degrees of freedom:
\be 
M^{*}_{Pl}= \sqrt{\frac{90}{8 \pi^3 g_*}} M_{Pl} \simeq \frac{1}{1.66 \sqrt{g_*}} M_{Pl}\, .
\label{hotbb_16}
\ee
We notice that   $M^{*}_{Pl}$ is temperature dependent because it is a function of $g_*$. This dependence is rather weak and it is a good approximation to take $M^{*}_{Pl}$ as a constant discussing the early universe at some stage of its evolution. Finally by comparing eq.~(\ref{hotbb_11}) and (\ref{hotbb_14}) we obtain
\be 
T(t) \propto \frac{1}{a(t)} \, ,
\label{hotbb_17}
\ee
where the relation holds exactly when the number of relativistic degrees of freedom does not change over the considered period of time. Due to the weak dependence on $g_*$ with the temperature, the relation  (\ref{hotbb_17}) provides an important observation: at a smaller scale factor corresponds a higher temperature. In summary we say that going back in time the universe was smaller, denser and warmer.

Let us conclude this section with a brief discussion about thermal equilibrium. We are going to consider processes that occur in an expanding universe filled with particles. The rates of interactions between these particles are often much higher than the expansion rate of the universe, so that the cosmic medium is in thermal equilibrium at any moment of time. However, we note that as a rule of thumb
the most interesting periods in the cosmological evolution are those when one or
another reaction goes out of equilibrium. In this case the abundance of some particle species freezes out and decouples from the heat bath. Nevertheless the laws of equilibrium thermodynamics are still useful since they enable us to estimate the time of departure from equilibrium and determine
the direction of non-equilibrium processes. Moreover most of the constituents of the heat bath, understood as a background for a given process of interest deviating the equilibrium conditions, are in thermal equilibrium. 

The thermodynamical description of a system with various particle species is usually
made in terms of a chemical potential $\mu$ for each type of particle. Given the reaction involving different particles labelled with $A_i$ and $B_j$ as follows
\be 
A_{1} + A_2 + \, ... \,  + A_n = B_{1} + B_2 + \, ... \,  + B_m \, ,
\label{thermeq_1}
\ee
the corresponding chemical potentials in thermal equilibrium, or better in chemical equilibrium, obey to the following relation
\be 
\mu_{A_{1}} + \mu_{A_2} + \, ... \,  + \mu_{A_n} = \mu_{B_{1}} + \mu_{B_2} + \, ... \,  + \mu_{B_m} \, .
\label{thermeq_2}
\ee
For example the chemical potential of the photon is zero and for a particle and its corresponding  antiparticle the chemical potentials are the same but opposite in sign. Let us consider the process $e^+ e^- \to 2\gamma$. We say that it is in equilibrium if it is equally likely as the back reaction $2 \gamma \to e^+ e^-$. 

Being the particle interactions in the thermal plasma fairly weak, we can take the equilibrium distributions to be the Bose--Einstein and Fermi--Dirac ones, as anticipated when writing (\ref{thermeq_4}). 
%\be 
%n_{B}= \frac{1}{e^{\beta P^{\mu}u_{\mu}}-1} \, , \quad n_{F}= \frac{1}{e^{\beta P^{\mu}u_{\mu}}+1} \, ,
%\label{thermeq_3}
%\ee
%when they are given in a covariant Lorentz frame where $u^{\mu}$ is the velocity with respect to the heat bath frame and $P^{\mu}$ the four-momentum of the particle.  we obtain 
Upon integrating the distribution function over the three-momentum one obtains the corresponding
number density of the particle species $i$
\be 
n_i = g_i \int \frac{d^3\bm{p}}{(2 \pi)^3} \,  f_i(E(p)) \, ,
\label{thermeq_5}
\ee
where $f_i$ can be either $n_B$ or $n_F$ in (\ref{thermeq_4}) and $g_i$ are the internal degrees of freedom of the particle. For example for the photons one finds ($m_\gamma=0$ and $\mu_\gamma=0$) 
\be
n_\gamma = \frac{2 T^3}{\pi^2} \zeta(3) \, ,
\label{photon_de}
\ee
where $\zeta(3)=1.202$, being $\zeta(x)$ the Riemann zeta function. More details on the thermodynamics of the early universe can be found e.~g.~in~\cite{RubaHOT} or in the appendix of~\cite{Davidson:2008bu}. 

\subsection{Brief thermal history of the universe}
\begin{figure}
\centering
\includegraphics[scale=0.355]{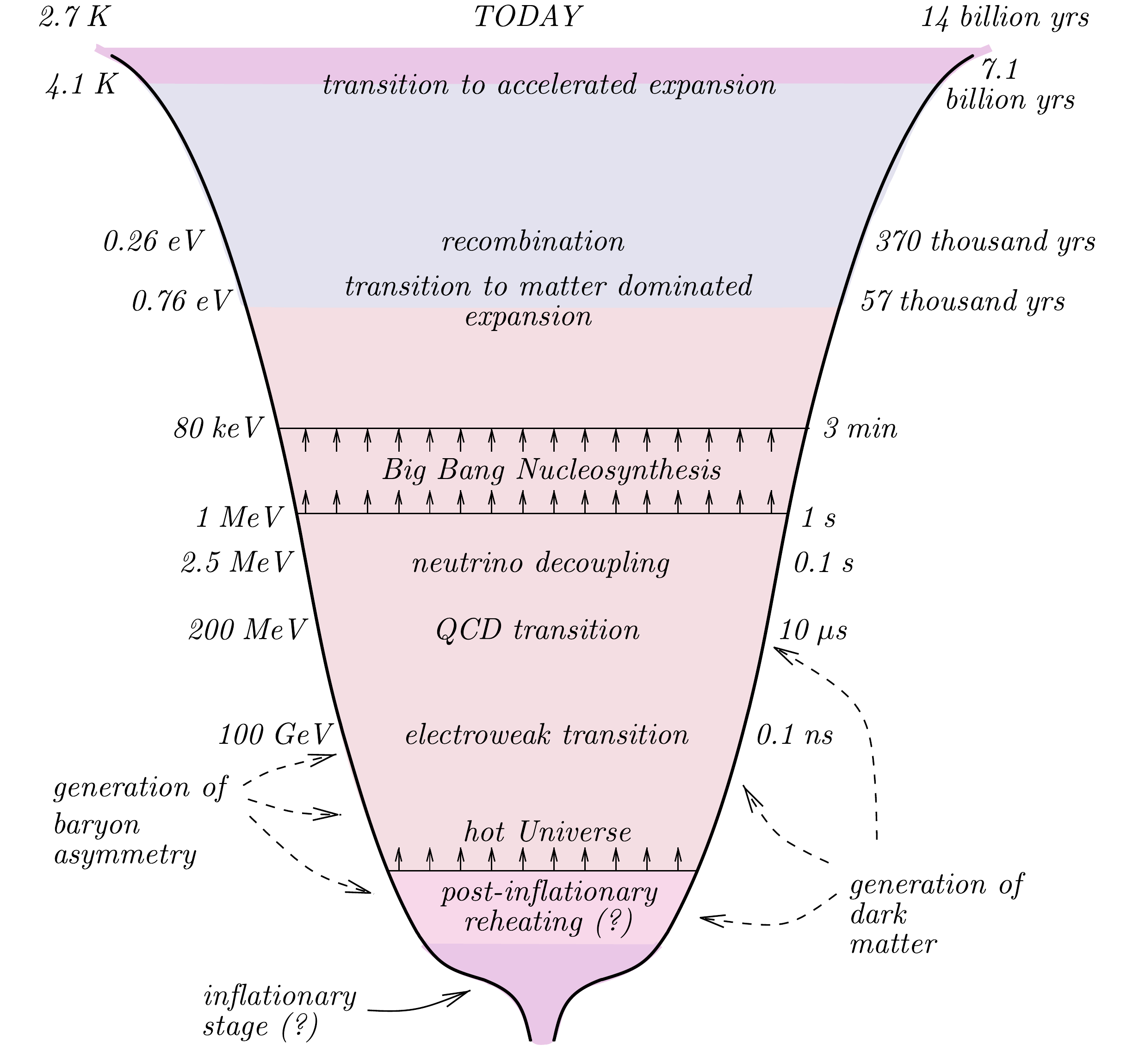}
\caption{\label{fig:hotbb_1} Stages of the universe evolution from inflation (bottom) to the present era (top). Typical temperatures, on the left, and age of the universe, on the right, are shown. Figure from \cite{RubaHOT}.}
\end{figure}

We discussed how the cosmological principle leads to an expanding universe with a hot past. Going back in time means looking at a smaller and smaller universe filled with particles at higher and higher temperatures. We can pin point some relevant periods in the universe evolution, shown in figure \ref{fig:hotbb_1}, and we aim at discussing them briefly in order to arrive at the topic of interest: the generation of the baryon asymmetry in the universe. %Indeed it is important to have in mind that a given epoch in the universe is connected with the ones coming first and has consequences on the following one. 

We start with the \textit{recombination} period, also called photon decoupling or last scattering. The plasma of hadrons, mainly hydrogen, electrons and photons turns into a gas of atoms. Before recombination the temperature was too high to allow for bound states of nuclei and electrons, so that the photons were continuously scattered off the charged particles and trapped in the hot plasma. The transition temperature from  the plasma to the gas of atoms can be naively estimated to be of order of $T \sim$10 eV, even though more accurate analysis give fraction of the eV scale, $T \sim 0.3$ \cite{RubaHOT}. From this moment onwards, the cross section with neutral atoms is so small that the average photon has not interacted with matter ever since: the medium became transparent to photons. The CMB carries information about this very moment, giving access to the universe when its temperature was about 3000 $K$ ($T \sim 0.3$ eV) and 370 000 years old. We have already mentioned that the high degree of CMB isotropy shows that
the Universe was pretty much homogeneous at recombination: the density perturbations
were comparable with temperature fluctuations and were roughly of order $\delta T/T \sim 10^{-5}$.
Nevertheless, these perturbations have grown and have given rise to structures: first stars,
then galaxies, then clusters of galaxies. The CMB provides the earliest direct probe of
universe structure that we can study in great detail. 

Proceeding back in time we find the Big-Bang Nucleosynthesis (BBN) \cite{Olive:1999ij, Yao, Steigman:2005uz, Cyburt:2004yc}. The temperature is set by the nuclei biding energy, namely  $T \sim 1$ MeV. Accurate analysis provides somewhat smaller temperatures though, namely fractions of MeV. From an earlier phase where protons and neutrons were free in the hot plasma, as the temperature dropped during the universe expansion, neutron capture and thermonuclear reactions became possible. At this stage light elements were formed: mainly Deuterium, D, Helium isotopes, $^{3}$He and $^{4}$He, and small amount of Lithium, $^{7}$Li. Quantitative calculations based on GR and kinetic equations provides the primordial abundances of the element species. These predictions depend on essentially a single parameter, called the baryon-to-photon ratio and defined as follows
\be 
\eta_B = \frac{n_B-n_{\bar{B}}}{n_{\gamma}} \, ,
\label{hotbb_18}
\ee
where $n_B$, $n_{\bar{B}}$ and $n_{\gamma}$ are the baryon, antibaryon and photon number densities. The final light-element abundances are highly sensitive to this parameter, which characterizes the baryon-photon plasma during the nucleosynthesis process. The population of D and $^{3}$He depends on $\eta_B$, and the cross sections of the processes leading to the formation of the heavier elements, like the $^{4}$He, inherits the dependence on the baryon-to-photon ratio. The larger $\eta_B$ the later the process generating the $^{4}$He will stop, and consequently the smaller the freeze-out abundances of the reacting elements D and  $^{3}$He.
Today the direct measurement of primordial abundances is pretty accurate, and this is a cornerstone of the early universe physics and the standard hot big bang cosmology. Indeed there is a range of $\eta_B$ which is consistent with all four abundances (D,  $^{3}$He,  $^{4}$He and $^7$Li), which at (95\% CL) reads \cite{Yao}   
\be 
4.7 \times 10^{-10} \leq \eta_B \leq 6.5 \times 10^{-10} \, .
\label{hotbb_19}
\ee

From now on, going back in time requires educated extrapolations. We cannot infer solid statements on our universe when it was hotter than $T\sim\text{MeV}$. 
%We do not have direct access to the universe when it was warmer than $T \sim$MeV. 
However, it is possible and desirable  that higher temperatures occurred in our universe. From the theoretical point of view this offers a very interesting scenario to test the laws of particle physics to extreme conditions. As we shall see the explanation of a baryon asymmetry naturally asks for some higher temperature regimes.  By assuming that temperatures of order of the GeV scale and higher were possible, we can list additional epoch comprising \textit{phase transitions}. Briefly we can summarize them as follows
\begin{itemize}
\item[1)] A transition (better a crossover) from a hadron gas to a quark-gluon plasma where the chiral symmetry is possibly restored. The transition temperature can be estimated from the QCD non-perturbative scale, $\Lambda_{{\rm{QCD}}} \sim 250$ MeV , even though more accurate simulations from lattice QCD provide the crossover to occur at $T_c = 154 \pm 9$ MeV \cite{Bazavov:2014pvz}. For $T \geq T_c$ quarks and gluons are not bounded any more in colourless hadrons, rather they interact as individual particles. %Today the study of this QCD matter is accessible at the heavy ion-collisions programs at CERN. 

\item[2)] Electroweak phase transition. Above the electroweak scale, $T_W \sim 100$ GeV, the Higgs condensate is absent and the $W$'s and $Z$ boson are then massless. The gauge group would be an unbroken SU(2)$_{L}\times$U(1)$_Y$, and all the SM fermion are massless as well. Further elaborations on the subject will be provided in the next sections. 

\item[3)] A more speculative transition is that involving the grand unification scale. This is related to the hypothesis that at higher energies, $T_{\rm{\small GUT}} \sim 10^{16}$, the fundamental strong, weak and electromagnetic forces are unified into a single force. The supersymmetric extension of the SM provides some motivation for such speculation. %However, such high temperatures may be not be reached in the early universe as many inflation models seem to suggest.
\end{itemize}

The next cosmological period we can see in figure \ref{fig:hotbb_1} is the reheating phase after inflation. Here two relevant processes might have occurred that represent a contemporary challenge in particle physics and cosmology: the generation of the baryon asymmetry in the universe and the production of dark matter. Since we are going to discuss the former in the upcoming section we spent some words here on the latter. %We refer to extensive reviews on the subject for a detailed discussion.

There are many experimental observations that suggest the presence of an additional component in the matter content of the universe. At galactic and sub-galactic scales, this evidence includes galactic rotation curves \cite{Borriello:2000rv}, the weak gravitational lensing of distant galaxies by foreground structure \cite{Hoekstra:2002nf}, and the weak modulation of strong lensing
around individual massive elliptical galaxies \cite{Metcalf:2003sz}. 
Furthermore, velocity dispersions of stars in some dwarf galaxies imply that they contain as much
as one thousand times more mass than can be assigned to their luminosity, and the same was observed quite some time ago at the scale of galaxy clusters in 1933 by Fritz Zwicky \cite{Zwicky:1933gu}.  On cosmological scales, observations of the anisotropies in the cosmic microwave
background have lead to a determination of the total matter density of $\Omega_{\hbox
{\scriptsize mat}} h^2=0.1326 \pm 0.0063$ \cite{Komatsu:2008hk}, where $h$ is the reduced Hubble constant. Moreover, this information combined with measurements of the light chemical element abundances leads to an accurate estimate of the baryonic
density given by $\Omega_{B} h^2 = 0.02273 \pm 0.00062$ \cite{Komatsu:2008hk}. Taken together, these observations strongly suggest that more than 80\% of the matter in the universe (by
mass) consists of non-luminous and non-baryonic particles, called dark matter. 

On the other hand, there is almost total lack of information on dark matter from the particle physics point of view leading to a  difficult assessment of the production mechanism in the early universe. Besides the fact that dark matter does not interact with photons, our knowledge of its fundamental interaction is scarce. We demand dark matter to be generated in the early stages of the universe evolution because it is an essential ingredient for the clumping of matter in the primordial gravitational potential wells that eventually formed stars, galaxies and large scale structures.  The process of the formation of large scale structures through the gravitational clustering of collisionless dark matter particles can be studied using
N-body simulations. When the observed structures in our universe are compared to the results of cold dark matter simulations good agreement has been found \cite{Tegmark:2003uf}. Here cold means the dark matter to be non-relativistic at time of structure formation. 
Many candidates has been put forward e.~g.~gravitinos and neutralinos from supersymmetry, axions and sterile neutrinos. We refer to \cite{Hooper:2009zm, Gelmini:2015zpa, Steffen:2008qp} for extensive reviews on dark matter candidates, as well as for discussions on dark matter production mechanisms in the early universe.   

Finally we comment on the epoch of reheating and how some of the issues related to the Big-Bang Cosmology are treated. This stage comes right after the \textit{inflationary stage}. Many of the problems that affect the Big-Bang theory arise from the very special initial conditions one has to require. At a qualitatively level, the Big-Bang model  cannot explain why our universe is so large, almost spatially flat, homogeneous and isotropic. Another issue refers to the primordial density perturbations detected in the CMB, which are the seeds for the generation of the matter structures we see today (stars, galaxies, clusters and so on and so forth). The hot Big-Bang theory does not contain a way to generate those perturbations and they have to be put ``by hands". 
The aforementioned problems find an elegant solution in the inflationary model, according to which the hot phase of the early universe was preceded by a phase of exponential expansion.  An initially small region of typical length of the Planck scale, $l_{Pl} \sim 1/M_{Pl}$, was inflated to very large sizes even larger than those of the visible present universe horizon. This explains eventually the dilution of any initial anisotropy, the homogeneity and the flatness. Moreover the model introduces a new field, the inflaton, which drives the exponential expansion and after the inflation epoch ends, it transfers its energy into the ordinary matter that populate the early universe. This is usually called the  \textit{reheating phase}. The primordial matter and energy perturbations are understood as quantum fluctuations of the inflaton field. The basic ideas of inflation were originally proposed by Guth \cite{Guth:1980zm} and Sato \cite{Sato:1980yn} independently, which were reviewed and brought to the modern fashion by Linde \cite{Linde:1981mu}, and Albrecht and Steinhardt \cite{Albrecht:1982wi}. The inflationary epoch plays an important role with respect to the baryon asymmetry in the universe, as we are going to discuss in the upcoming section, and more in general it provides a reasonable explanation for the existence of a thermal bath of particles in the very early stages of the universe evolution.

\section{Dynamical generation of the baryon asymmetry}
\label{hot_sec2}
Observations suggest that the number of baryons in the universe is different from the number of antibaryons. The almost total absence of antimatter on Earth, in our solar system and in cosmic rays indicates that the universe is baryonically asymmetric. A more accurate reasoning could bring us to admit that matter and antimatter galaxies could coexist in clusters. However we would expect a detectable background of photon radiation coming from nucleus-antinucleus annihilation within the clusters \cite{Riotto:2011zz}. This argument can be further generalized to large hypothetical domains of matter an antimatter in the universe, but the missing observation of any induced distortion on the CMB discards this possibility. As Cohen, de Rujula and Glashow have compellingly argued, if there were to exist large amounts of antimatter in the universe they could only be at a cosmological scale from us \cite{Cohen:1997ac}. It therefore seems that our universe is fundamentally matter-antimatter asymmetric. 

There are observables to make this statement more quantitative. In particular we refer to the baryon-to-photon ratio, already introduced in section \ref{hot_sec1}, and we recall it here with the experimental value attached
\be 
\eta_B = \frac{n_B-n_{\bar{B}}}{n_\gamma} = (6.21 \pm 0.16) \times 10^{-10} \, .
\label{bau_1}
\ee
Such precise measurement comes from the study of the CMB anisotropies \cite{Larson:2010gs}. %In (\ref{bau_1}) the numerator could be regarded effectively as if $n_B=0$, being antibaryons not detected at all. 
As regards the CMB analysis, the parameter $\eta_B$ plays a crucial role in determining the relative amplitudes of even and odd peaks of the power spectrum of the microwave background.  This is in turn related to the acoustic oscillations of the baryon-photon fluid at the time of recombination. 
It is astonishing the high level of agreement with an independent prediction: the abundances of the light elements provided by BBN. As discussed in the previous section~\ref{hot_sec1}, the generation of elements like H, $^3$He, $^4$He and $^7$Li occurred before the last scattering in a hot plasma. It is found that their abundances can be obtained by an input of a single parameter, $\eta_B$. The range for this parameter, predicted by  BBN and written in~(\ref{hotbb_19}), agrees with the value extracted from the CMB analysis in (\ref{bau_1}) establishing an extraordinary matching between two independent measurements.

The challenge both from the cosmology and particle physics side is to explain the observed value in (\ref{bau_1}). The standard cosmological model dramatically fails in reproducing even only the order of magnitude of the baryon-to-photon ratio if we start with a matter-antimatter symmetric phase at high temperatures. Let us consider the reaction $p + \bar{p} \leftrightarrow 2 \gamma$, at temperature of the order of one GeV. Protons and neutrons constitute the baryon content of the universe at this epoch. As the universe cools down the process $2 \gamma \to p + \bar{p}$ becomes ineffective due to Boltzmann suppression, and therefore the annihilation process $p + \bar{p} \to 2 \gamma$ takes over. The same reactions stand for neutrons and antineutrons. Eventually the number of baryons and antibaryons is strongly reduced with respect to the photon number density, a straightforward calculation provides  \cite{RubaHOT, Turner} 
\be 
\frac{n_B}{n_\gamma} \approx \frac{n_{\bar{B}}}{n_\gamma} \approx 10^{-18} \, , 
\label{bau_2}
\ee
which is far too smaller than the value required for a successful nucleosynthesis, see (\ref{hotbb_19}), and than the one in (\ref{bau_1}) from CMB analysis.  It is hard to figure out processes at temperatures below one GeV able to enhance the small ratio between baryon and photon number densities induced by annihilations (an exception is provided by the Affleck-Dine Baryogenesis \cite{Affleck:1984fy}). Because of the strong disagreement between (\ref{bau_1}) and (\ref{bau_2}), we come to the conclusion that a primordial matter-antimatter asymmetry had to exist already before BBN, and more specifically at temperatures of the GeV scale. 
%The nucleons-antinucleons number densities decreases due to the annihilation processes as long as the annihilation rate is larger then the Universe expansion rate as given in (\ref{hotbb_15}). Estimating the annihilation rate with $\Gamma_{B-\bar{B}} \approx n_B \langle \sigma \, v \approx m_{\pi}^2$, we obtain a freeze-out temperature $T \approx 20$ MeV. Nucleons and antinucleons are so rare that annihilations are not possible anymore and from (\ref{bau_2}) one obtains
%\be 
%\frac{n_B}{n_\gamma} \approx \frac{n_{\bar{B}}}{n_\gamma} \approx 10^{-18} \, ,
%\label{bau_3}
%\ee
%

The observed baryon asymmetry  could be set as an initial condition for the universe evolution. However, it would require a high fine tuning and the ad hoc baryon asymmetry would have not survived the huge dilution induced by the inflationary epoch. This is why the scenario of a dynamically generated baryon asymmetry is more appealing.
The dynamical generation of a baryon asymmetry in the context of quantum field theory is called \textit{baryogenesis} \cite{Sakharov:1967dj}. Indeed, quoting A. Riotto, ``the guiding principle of modern cosmology aims at explaining the initial conditions required by standard cosmology on the basis of quantum field theories of elementary particle physics in a thermal bath"~\cite{Riotto:2011zz}. 

\subsection{The Sakharov conditions}
Assuming a vanishing initial matter-antimatter asymmetry, it can be dynamically generated in an expanding universe if the particle interactions and the cosmological evolution satisfy the three Sakharov conditions \cite{Sakharov:1967dj}:
\begin{itemize}
\item[1)] baryon number violation,
\item[2)] C and CP violation,
\item[3)] departure from thermal equilibrium.
\end{itemize}    
In the following $B$ stands for the baryon number understood as the total baryonic charge of a given process. 
Any particle physics model that aims at generating an imbalance between matter and antimatter has to account for the aforementioned necessary conditions. 
Originally introduced in the framework of GUTs, we briefly discuss the Sakharov conditions also in relation with the SM of particle of physics in order to show that all the three requirements are fulfilled. However all attempts to reproduce quantitatively the observed baryon asymmetry have failed within the SM.

Since we start from a baryon symmetric universe, we need processes that violate the baryon (antibaryon) number to somehow evolve into a situation in which it holds $\eta_B \neq 0$. Processes are required that change the number density of baryons and antibaryons entering the definition of $\eta_B$ . If C and CP are exact symmetries, then one can show that the rate for any process which produces a baryon excess is equal to the rate of the complementary process generating antibaryons. Hence no net imbalance can be produced. CP violation can be implemented in a model either introducing complex phases in the Lagrangian which cannot be reabsorbed by field redefinition (explicit breaking), or if some Higgs field generates a complex vacuum expectation value (spontaneous breaking). 

Of the three Sakharov conditions, the first two can be investigated only after a particle physics model is specified. The third one, the departure from thermal equilibrium, can be discussed in a more general way. The baryon number $B$ is odd under C and CP discrete transformations. Using this property of $B$ together with the requirement that the Hamiltonian of the system commutes with the combination CPT, where T here is the time-reversal discrete symmetry, the thermal average of $B$ reads
\bea 
\langle B \rangle_T & \equiv & \frac{1}{Z}\tr\left[ e^{-\frac{H}{T}} B\right] =  \frac{1}{Z} \tr \left[ (CPT)(CPT)^{-1} e^{-\frac{H}{T}} B \right]  
\nn
\\
&=& \frac{1}{Z} \tr \left[ e^{-\frac{H}{T}}(CPT)^{-1}  B (CPT) \right] =-  \frac{1}{Z}\tr \left[ e^{-\frac{H}{T}} B  \right]   \nonumber \\
&=& -\langle B \rangle_T \, .
\label{bau_4}
\eea
In (\ref{bau_4}) $H$ and $Z$ are the Hamiltonian and the partition function of the system respectively (see also (\ref{therm_02}) in chapter~\ref{chap:therm_tex} for more details on the partition function). 
Therefore we see that in thermal equilibrium $\langle B \rangle_T=0$, and the same stands for the antibaryon number. The outcome is the following: if we start with a baryonic symmetric phase, processes in thermal equilibrium cannot alter the initial value for the baryon and antibaryon number and hence $\eta_B$ remains zero. Put in other words, processes generating a net baryon number are equally likely as those destroying it.   

\begin{figure}
\centering
\includegraphics[scale=0.575]{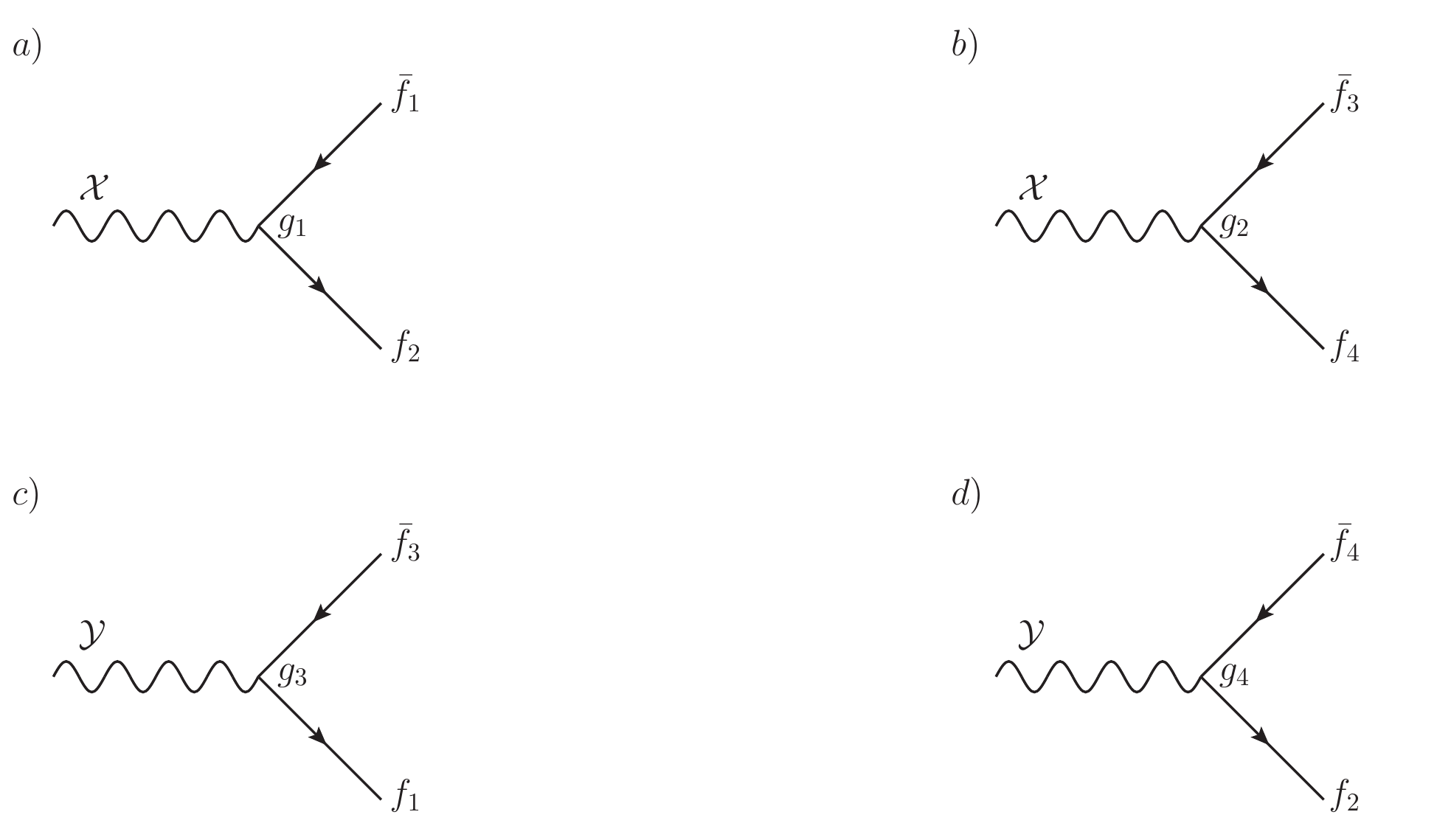}
\caption{\label{fig:bau_2} Tree level diagrams for the decay processes in (\ref{bau_6}) and (\ref{bau_6bis}). The heavy gauge bosons $\mathcal{X} $ and $\mathcal{Y}$ are the wiggled lines, solid lines stand for fermions. Similar diagrams for the charge conjugate processes are not shown.}
\end{figure} 
In order to illustrate the Sakharov conditions we choose a toy model, similar to the one in \cite{TurnerAndKolb}, that is inspired to GUTs. Baryon number violation occurs naturally in this class of models because quarks and leptons are embedded in the same irreducible representations. Then heavy gauge bosons and scalars are introduced that can mediate interactions leptons and quarks at the same vertex. The toy model consists of some exotic particles, the gauge bosons $\mathcal{X}$ and $\mathcal{Y}$, and four massless fermions, $f_1,...f_4$, each of the latter carrying a baryon number $B_1,...,B_4$ respectively. The interaction Lagrangian of the toy model reads
\be 
\mathcal{L}^{\hbox{\tiny toy}}_{\hbox{\tiny int}} = g_1 \mathcal{X} \bar{f}_2 f_1 + g_2 \mathcal{X} \bar{f}_4 f_3 + g_3 \mathcal{Y} \bar{f}_1 f_3 + g_4 \mathcal{Y}  \bar{f}_2 f_4 + h.c. \, ,
\label{bau_5}
\ee
where $g_1,...,g_4$ are dimensionless complex coupling constants. The induced decay processes are
\bea 
&& \mathcal{X} \to \bar{f}_1 + f_2 \, , \quad \mathcal{X} \to \bar{f}_3 + f_4  \, ,
\label{bau_6}
\\
&& \mathcal{Y} \to \bar{f}_3 + f_1 \, , \quad \mathcal{Y} \to \bar{f}_4 + f_2  \, .
\label{bau_6bis}
\eea
The tree level diagrams for the decay processes are shown in figure \ref{fig:bau_2}. %The possible baryon number reaction are listed for the toy model in (\ref{bau_5}) for the particle $\mathcal{X}$ in table. 
Let us discuss the decay rates. At tree level we can parametrize the decay rate for the process $\mathcal{X} \to \bar{f}_1 + f_2$ as follows
\be 
\Gamma^{(0)}(\mathcal{X} \to \bar{f}_1 + f_2 )= |g_1|^2 A_{\mathcal{X}} \, ,
\label{bau_7}
\ee
where $A_{\mathcal{X}}$ contains the two-body phase space factor (the subscript stands for a decaying $\mathcal{X}$). For the charge conjugate process, that involves the particles $f_1$ and $\bar{f}_2$ in the final state, we have
\be 
\Gamma^{(0)}(\bar{\mathcal{X}} \to f_1 + \bar{f}_2 )= |g_1^*|^2 A_{\mathcal{\bar{X}}} = |g_1|^2 A_{\mathcal{X}}  \, ,
\label{bau_8}
\ee
and we conclude that no asymmetry can be generated at tree level as the kinematic factors $A_{\mathcal{X}}$ and $A_{\mathcal{\bar{X}}}$ are equal. However the first Sakharov condition is already met: we start from a gauge boson with zero baryon number and we end up with a final state with a net baryon number $B_2+\overline{B}_1=B_2-B_1$ for the first process in (\ref{bau_6}). Of course one has to require $B_1 \neq B_2$.

\begin{figure}
\centering
\includegraphics[scale=0.61]{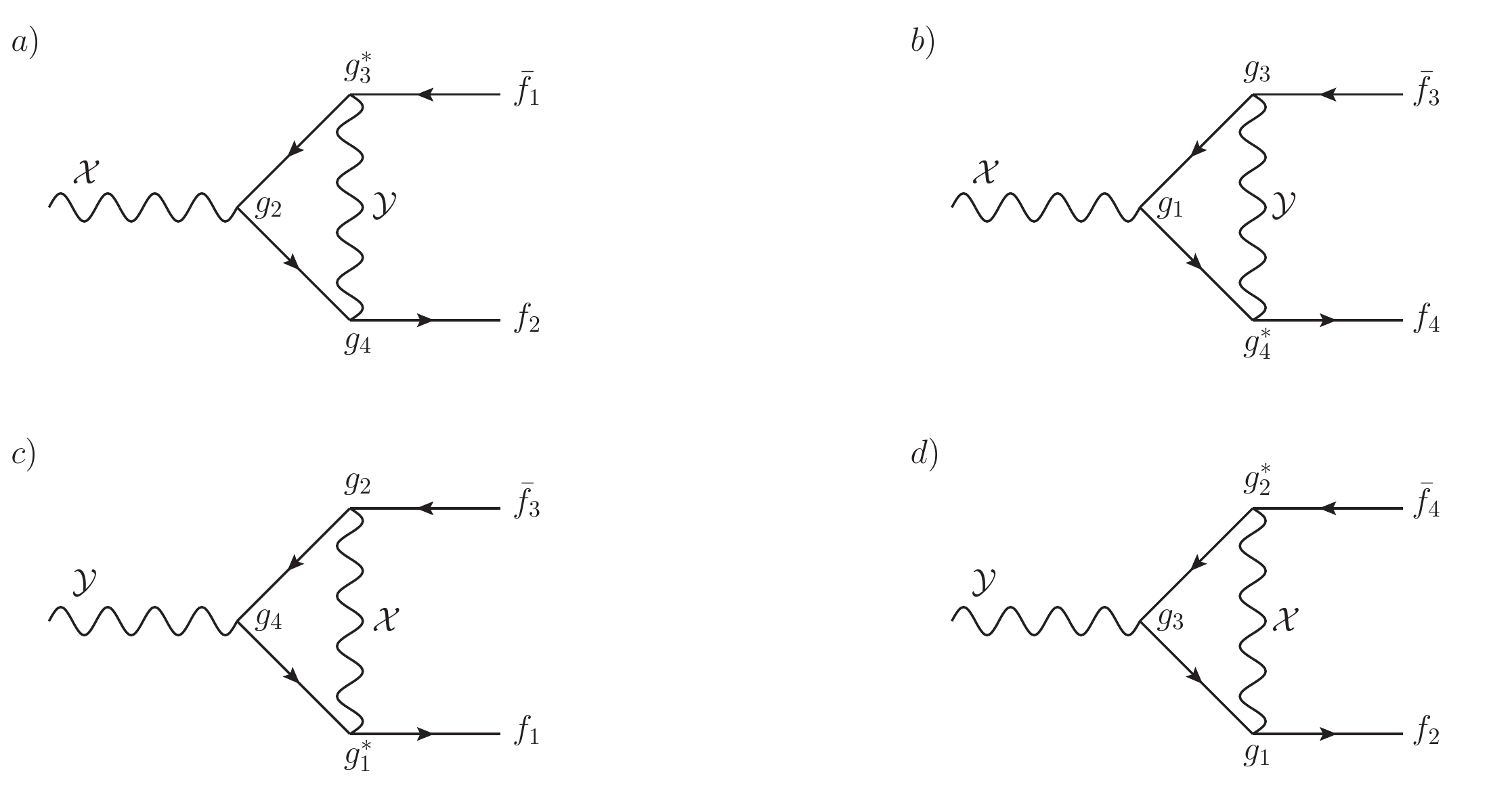}
\caption{\label{fig:bau_3} One-loop diagrams for the decay processes in (\ref{bau_6}) and (\ref{bau_6bis}).  Similar diagrams for the charge conjugate processes can be drawn.}
\end{figure}
Clearly we have to go beyond tree level to obtain different rates for the  decay of $\mathcal{X}$. The one-loop diagrams describing the decay processes (\ref{bau_6}) are shown in figure \ref{fig:bau_3}, upper raw. They are built by allowing for the exchange of a virtual heavy scalar $\mathcal{Y}$. This time the decay width also comes from the interference between tree level and a one-loop amplitudes that give (we pick only the $\mathcal{O}(g^4)$ terms)
\bea 
&&\Gamma^{\hbox{\tiny(1)}}(\mathcal{X} \to \bar{f}_1 + f_2 )=g_1g_2^* g_3g_4^* \, B^*_{\mathcal{X}} + g_1^* g_2 g^*_3g_4 \, B_{\mathcal{X}} \, ,
\label{bau_9}
\\
&& \Gamma^{\hbox{\tiny(1)}}(\bar{\mathcal{X}} \to f_1 + \bar{f}_2 )= 
g_1 g^*_2 g_3 g^*_4 \, C_{\bar{\mathcal{X}}} + g_1^* g_2 g^*_3g_4 \, C^*_{\bar{\mathcal{X}}} \, ,
\label{bau_10}
\eea 
where $B_{\mathcal{X}}$ and $C_{\bar{\mathcal{X}}}$ comprise both the two-body phase space and the one-loop amplitude corresponding to the triangle topology in figure {\ref{fig:bau_3}}. In general the loop amplitude is a complex quantity, the imaginary part corresponding to the sum of the cuts that put different particles simultaneously on shell. The explicit calculation gives $B_{\mathcal{X}}=C_{\bar{\mathcal{X}}}$. We further elaborate the details of a very similar derivation in the case of leptogenesis in chapter~\ref{chap:CPdege}. Then we do find a non-vanishing difference in the decay rates
\be 
\Gamma(\mathcal{X} \to \bar{f}_1 + f_2 ) - \Gamma(\bar{\mathcal{X}} \to f_1 + \bar{f}_2 ) = 4 \, {\rm{Im}}(g_1 g^*_2 g_3 g^*_4 ) \,  {\rm{Im}}(B_{\mathcal{X}}) \, ,
\label{bau_11}
\ee
where the decay rates $\Gamma(\mathcal{X} \to \bar{f}_1 + f_2 )$ and $\Gamma(\bar{\mathcal{X}} \to f_1 + \bar{f}_2 )$ are understood as the sum of the tree-level and one-loop contributions as given in eqs.~(\ref{bau_7}) and (\ref{bau_8}), and in eqs.~(\ref{bau_9}) and (\ref{bau_10}) respectively. Similarly we have for the other decay mode the result
\be 
\Gamma(\mathcal{X} \to \bar{f}_3 + f_4 ) - \Gamma(\bar{\mathcal{X}} \to f_3 + \bar{f}_4 ) = -4 \, {\rm{Im}}(g_1 g^*_2 g_3 g^*_4 ) \,  {\rm{Im}}(B_{\mathcal{X}}) \, ,
\label{bau_12}
\ee
the derivation follows closely the one outlined. The loop amplitude in (\ref{bau_12}) is the same as in (\ref{bau_11}) because the very same particle content (the massless fermions and the intermediate gauge boson $\mathcal{Y}$) propagates in the triangle topologies $a)$ and $b)$ of figure \ref{fig:bau_3}.
Besides the loop diagrams in the first raw in figure  \ref{fig:bau_3}, one could also consider those with the $\mathcal{X}$ as internal propagating gauge boson. However, this would lead to vanishing coupling combinations, such as ${\rm{Im}}(g_1 g^*_1 g_2 g^*_2 ) =0$, and eventually provide a vanishing difference in (\ref{bau_11}) and (\ref{bau_12}).  It is now clear how the second Sakharov condition enters: the decay rates for the process $\mathcal{X} \to \bar{f}_1 + f_2 $ and $\bar{\mathcal{X}} \to f_1 + \bar{f}_2 $ can be different due to the interference between tree-level and one-loop diagrams that involve C and CP violating processes.  
Moreover, there have to be two distinct heavy gauge bosons, coupling differently to the fermions and being heavier than the sum of the decaying products. The latter condition ensures the loop amplitude to have a non vanishing imaginary part, ${\rm{Im}}(B_\mathcal{X})$.

The baryon asymmetry generated in the decays of the heavy gauge boson $\mathcal{X}$ can be expressed as follows
\be 
\epsilon_{\mathcal{X}}= \frac{(B_2-B_1)\Delta \Gamma(\mathcal{X} \to \bar{f}_1 + f_2)+ (B_4-B_3)\Gamma(\mathcal{X} \to \bar{f}_3 + f_4) }{\Gamma_{\mathcal{X}}} \, ,
\label{bau_13}
\ee 
where we define
\bea 
&&\Delta \Gamma(\mathcal{X} \to \bar{f}_1 + f_2) = \Gamma(\mathcal{X} \to \bar{f}_1 + f_2 ) - \Gamma(\bar{\mathcal{X}} \to f_1 + \bar{f}_2 ) \, ,
\label{bau_14}
\\
&&\Delta \Gamma(\mathcal{X} \to \bar{f}_3 + f_4) = \Gamma(\mathcal{X} \to \bar{f}_3 + f_4 ) - \Gamma(\bar{\mathcal{X}} \to f_3 + \bar{f}_4 ) \, ,
\label{bau_15}
\eea
and the total width reads
\bea 
\Gamma_{\mathcal{X}} &=& \Gamma(\mathcal{X} \to \bar{f}_1 + f_2 ) + \Gamma(\bar{\mathcal{X}} \to f_1 + \bar{f}_2 ) 
\nn
\\
&+& \Gamma(\mathcal{X} \to \bar{f}_3 + f_4 ) + \Gamma(\bar{\mathcal{X}} \to f_3 + \bar{f}_4 )  \, .
\label{bau_16}
\eea
Finally by using the results in (\ref{bau_11}) and (\ref{bau_12}) we obtain for the baryon asymmetry generated in the $\mathcal{X}$ decays
\be 
\epsilon_{\mathcal{X}} = \frac{4}{\Gamma_{\mathcal{X}} } \left[ (B_2-B_1)-(B_4-B_3) \right] {\rm{Im}}(g_1 g^*_2 g_3 g^*_4 ) \,  {\rm{Im}}(B_{\mathcal{X}}) \, ,
\label{bau_17}
\ee
where we remind that $B_{\mathcal{X}}$ is not the baryon number of the heavy gauge boson, but the factor containing the loop amplitude. In order to have a non vanishing baryon asymmetry (\ref{bau_17}), both the couplings combination and the loop amplitude have to be complex quantities. 

It is interesting to note that the two heavy gauge bosons cannot be degenerate in mass. Indeed the baryon asymmetry for the $\mathcal{Y}$ heavy gauge boson reads
\be 
\epsilon_{\mathcal{Y}} = \frac{4}{\Gamma_{\mathcal{Y}} } \left[ (B_1-B_3)-(B_2-B_4) \right] {\rm{Im}}(g_1 g^*_2 g_3 g^*_4 ) \,  {\rm{Im}}(B'_{\mathcal{Y}}) \, ,
\label{bau_18}
\ee
where $B'_{\mathcal{Y}}$ is the loop amplitude for a decaying $\mathcal{Y}$ boson. 
If the gauge bosons are mass degenerate then the condition $B_{\mathcal{X}}=B'_{\mathcal{Y}}$ holds, and then $\epsilon_{\mathcal{X}} + \epsilon_{\mathcal{Y}}=0$ holds as well. This is because the two-particle phase space is the same in the decay processes for $\mathcal{X}$ and $\mathcal{Y}$, and the only difference in the corresponding loop amplitudes is the mass of the heavy intermediate boson (see figure \ref{fig:bau_3}), whereas all the fermions are massless. 

Now we come to the third Sakharov condition: the departure from thermal equilibrium. In this toy model such condition is achieved as follows. Let us consider the heavy boson $\mathcal{X}$. If the decay rate $\Gamma_{\mathcal{X}}$ is smaller than the expansion rate of the universe, the particles $\mathcal{X}$ cannot decay on the time scale of the universe expansion. Then the interactions governing the $\mathcal{X}$ dynamics are so weak that they cannot catch up with the expanding system and the heavy gauge bosons $\mathcal{X}$ decouple from the thermal plasma. If the decoupling occurs when the particles are still relativistic, namely for $M_{\mathcal{X}} < T$, the heavy bosons remain as abundant as photons, $n_{\mathcal{X}} \approx n_{\bar{\mathcal{X}}} \propto T^3$ (see eq.~(\ref{photon_de})), also at later times. Therefore, at time such that $M_{\mathcal{X}} \simeq T$, they populate the universe with an abundance much larger than the equilibrium one:
\bea 
%&& n_{\mathcal{X}} \approx n_{\bar{\mathcal{X}}} \approx n_{\gamma} \, ,\: \text{for}  \: T \leq M_{\mathcal{X}} \, ,
%\\
&& n_{\mathcal{X}} \approx n_{\bar{\mathcal{X}}} \approx (M_{\mathcal{X}} T)^{\frac{3}{2}} e^{-\frac{M_{\mathcal{X}}}{T}} \ll n_{\gamma}   \, ,
\label{bau_19}
\eea
which holds for $T \leq M_{\mathcal{X}}$ and it is Boltzmann suppressed when $M_{\mathcal{X}} <T$. 
The heavy particles are more abundant than their corresponding equilibrium population at temperature below $M_{\mathcal{X}}$: this is exactly what out-of-equilibrium dynamics means in this class of models. In other words, the heavy gauge bosons  generate the baryon asymmetry through their CP violating decays and the back reactions, the inverse decays, are exponentially suppressed because the massless fermions populate the thermal plasma with mean energies much smaller than the heavy states mass, $M_{\mathcal{X}}$.  

In general the out-of-equilibrium condition requires the typical interaction rate for the gauge boson $\mathcal{X}$ to be
\be 
\left.  \Gamma_{\mathcal{X}} < H \right|_{T=M_{\mathcal{X}}} \, ,
\label{bau_20}
\ee
where $H$ stands for the Hubble rate as given in (\ref{hotbb_15}). Evaluating $H$ at $T=M_{\mathcal{X}}$, one can obtain from (\ref{bau_20}) a condition on the model parameters. The decay rate goes like $\Gamma_{\mathcal{X}} \sim |g_i|^2 M_{\mathcal{X}}$ and if the couplings are taken as spanning from $10^{-2}$ to $10^{-3}$, and $g_*$ is taken at about $10^2$, we obtain \cite{Riotto:2011zz}
\be 
M_{\mathcal{X}} > \left[ 10^{-4},   10^{-3}\right]  M_{Pl} \approx \left[ 10^{15},  10^{16}\right]  \, \text{GeV} \, .
\label{bau_21}
\ee  
Such energy scale window sets the typical mass of the heavy states in GUT models, within which the first convincing realization for baryogenesis has been proposed \cite{Sakharov:1967dj}. Quite recently it has been suggested that the reheating temperature after the inflation cannot be higher than 10$^{15}$ GeV as accounted for the CMB analysis \cite{Giudice:2000ex}. The thermal production of these heavy particles predicted by GUT models seems then seriously affected, undermining the very basis of such scenario for a successful baryogenesis.  
%In the SM the baryon ($B$) and lepton number ($L$) are called \textit{accidental symmetries}. They are individually conserved at tree level but are violated at quantum level via Adler-Bell-Jackiw triangular anomalies REF. More specifically in 1976 t'Hooft realized that non-perturbative effects, called instantons, 

\subsection{Baryogenesis: a call for New Physics}
\label{sec:baryonew}
Baryogenesis can already be implemented in SM framework, however, there are severe limitations in providing a quantitative solution for the baryon asymmetry generation. Indeed in order to reproduce the experimental value in (\ref{bau_1}) some new physics is needed together with an interesting and challenging overlap between cosmology and particle physics. 
As anticipated before, the SM contains all the ingredients required by the Sakharov conditions. The following discussion will also help to set some important and relevant aspects for the topic in the next chapter: baryogenesis via leptogenesis. 

Let us start with the baryon number violation in the electroweak theory. 
In the SM the baryon and lepton number, $B$ and $L$, are called \textit{accidental symmetries}. They are individually conserved at tree level but are violated at quantum level via Adler-Bell-Jackiw triangular anomalies \cite{Adler:1969gk, Dimopoulos:1978kv}. More specifically in 1976 t'Hooft realized that non-perturbative effects \cite{'tHooft:1976up}, called instantons, may induce processes which violate the combination $(B+L)$ but conserve $(B-L)$. The probability for these processes to occur today in our universe is pretty much low, being exponentially suppressed. However, in the early stages of the universe evolution, namely at much higher temperatures, baryon and lepton number violation processes could occur more likely enough to play a role in baryogenesis.   Let us express the baryon and lepton numbers as follows 
\be 
B=\int d^3x \, J_{0}^B (x) \, , \quad L=\int d^3x \, J_{0}^L (x) \, ,
\label{eleW_1}
\ee 
where the currents read
\bea
&& J_{\mu}^B= \frac{1}{3} \sum_i  \left( \bar{Q}_{i} \gamma_{\mu} P_L Q_{i} + \bar{U}_{i} \gamma_{\mu} P_R U_i +\bar{D}_{i} \gamma_{\mu} P_R  D_i \right) \, ,
\label{eleW_2}
\\
&&J_{\mu}^L= \sum_i \left( \bar{L}_i  \gamma_\mu P_L  L_i + \bar{E}_i \gamma_\mu P_R  E_i \right) \, . 
\label{eleW_3}
\eea
\begin{table}[t]
\centering
\begin{tabular}{c c c c c c c}
\hline
 & & $Q_{1}=\left( \begin{array}{c}
u \\
d \end{array} \right)_L$ &  $u_R$ & $d_R $ &  $L_{1}=\left( \begin{array}{c}
\nu_e \\
e \end{array} \right)_L$   & $e_R$ 
\vspace{0.15 cm}
\\
\hline 
& $ \begin{array}{c}
B \\
L \end{array}$  & $ \begin{array}{c}
1/3 \\
0 \end{array} $ &  $ \begin{array}{c}
1/3 \\
0 \end{array} $ & $ \begin{array}{c}
1/3 \\
0 \end{array} $ &  $ \begin{array}{c}
0 \\
1 \end{array} $  & $ \begin{array}{c}
0 \\
1 \end{array} $
\\
\hline
\end{tabular}
\caption{\label{Tab:tab1} SM fermions and their baryon and lepton numbers for the first generation. The particles are given as SU(2)$_L$ doublets and singlets.}
\end{table} 
The fields $Q_i$ stand for the SU(2)$_L$ doublet quarks, $U_i$ and $D_i$ for the SU(2)$_L$ singlets quarks, then $L_i$ refers to the SU(2)$_L$ lepton doublets and $E_i$ for the SU(2)$_L$ lepton singlets. The left- and right-handed chiral projectors are $P_L=(1-\gamma^5)/2$ and $P_R=(1+\gamma^5)/2$, the index $i$ refers to the fermion generation. For example we have $E_1=e$, $E_2=\mu$ and $L_3= (\nu_\tau, \tau)^T$ for leptons, $U_1$, $U_2$ and $U_3$ are the  SU(2)$_L$ singlet up, charm and top quarks respectively. We summarize the $B$ and $L$ numbers in table \ref{Tab:tab1} for the first generation (they read the same for the second and third generation). The baryon and lepton number are classically conserved but the divergences of the currents in (\ref{eleW_2}) and (\ref{eleW_3}) do not vanish at quantum level 
\be 
\pa^{\mu}   J_{\mu}^B = \pa^{\mu}   J_{\mu}^L = \frac{N_f}{32 \pi^2} \left( g^2 W^{a}_{\mu \nu} \widetilde{W}^{a, \mu \nu} - g'^2 F_{\mu \nu} \widetilde{F}^{ \mu \nu} \right) \, , 
\label{eleW_4}
\ee
where $W^{a}_{\mu \nu}$ and $F_{\mu \nu}$ are the SU(2)$_L$ and U(1)$_Y$ field strength tensors respectively, with corresponding gauge couplings $g$ and $g'$, and $N_f$ is the number of the fermion generations, $\widetilde{W}^{a, \mu \nu}$ and $\widetilde{F}^{ \mu \nu}$ the dual field strength tensors. From (\ref{eleW_4}) it is clear that 
\be 
\pa^{\mu} (J_{\mu}^B - J_{\mu}^L)=0 \, ,
\label{eleW_5}
\ee
so that $(B-L)$ is conserved. On the other hand, the combination $(B+L)$ is violated and we have
\be 
\pa^{\mu} (J_{\mu}^B + J_{\mu}^L) = 2 N_f \pa_\mu \mathcal{K}^{\mu} \, ,
\label{eleW_6}
\ee
with 
\bea 
\mathcal{K}^{\mu}= &-&\frac{g^2}{32 \pi^2} \, 2 \epsilon^{\mu \nu \rho \sigma} W^{a}_{\nu} (\pa_\rho W^a_\sigma + \frac{g}{3} \epsilon^{abc} W^b_\rho W^c_\sigma ) 
\nn \\
&+&\frac{g'^2}{32 \pi^2} \epsilon^{\mu \nu \rho \sigma} B_\nu B_{\rho \sigma} \, .
\eea
\begin{figure}
\centering
\includegraphics[scale=0.65]{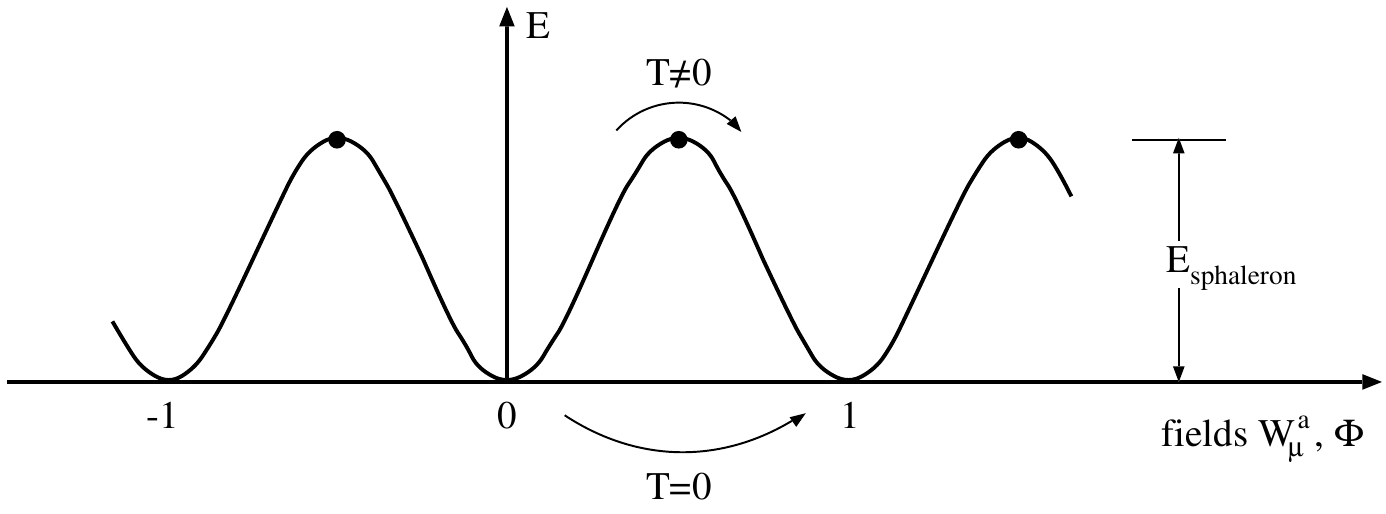}
\caption{\label{fig:bau_4} Vacuum structure of the electroweak theory, three different degenerate state are shown. Chern-Simon numbers, $N_{cs}=0,\pm 1$, are along the $x$ axis and are related to the field configurations. The sphaleron energy corresponds to the height of the barrier. At $T=0$ transitions are possible only via  tunnelling, whereas at finite temperature they can be induced by thermal fluctuations. Figure adapted from \cite{Chen:2007fv}.}
\end{figure} 

It is important to notice that the violation of the current combination (\ref{eleW_6}) is related to the vacuum structure of the electroweak theory. There are infinite degenerate ground states separated by a potential barrier as shown in figure \ref{fig:bau_4}, and a topological charge called Chern-Simons number, $N_{cs}$, is attached to each of the vacua. The change of the  baryon (lepton) number with time can be then associated with the change in the Chern-Simons number, that is in turn due to a change from a vacuum state to another:
\be 
\Delta B \equiv B(t_f) -B(t_i) = N_f \left[ N_{cs}(t_f) -  N_{cs}(t_i) \right]= N_f \Delta N_{cs} \, ,  
\ee  
where $t_i$ and $t_f$ are the initial and final time respectively and $N_f$ the number of fermion generations.
Going from one ground state to another implies having $\Delta N_{cs} = \pm 1, \, \pm 2 \, \, ...$ as also shown in figure \ref{fig:bau_4}. In the SM there are three fermion generations, so that $\Delta B = \Delta L = N_f \Delta N_{CS} = \pm 3 n$, with $n$ a positive integer. That is to say that a vacuum to vacuum transition changes $\Delta B$ and $\Delta L$ by multiples of three units, and each transition generates 9 left-handed quarks (3 colors for each generation) and 3 left-handed leptons (one per generation).

In a semi-classical view, the probability of going to one vacuum state to another is determined by an \textit{instanton} configuration. %More in detail, the solution of the equation of motion at the saddle point, namely at the maximal for the potential describing the ridge of different vacuum configurations, is referred as the \textit{sphaleron}.   
The transition rate has a very different form whether it is calculated at zero temperature or at finite temperature. In the former case, the probability of baryon and lepton non-conserving processes has been computed by t'Hooft and it is highly suppressed by a factor $e^{-4\pi/ \alpha_W} \approx \mathcal{O}(10^{-165})$ \cite{'tHooft:1976up}, where $\alpha_W= g^2/(4\pi)$. The instantons do not threaten
the stability of the proton \cite{'tHooft:1976up}. In a thermal bath the situation may be quite different. It was suggested by Kuzmin, Rubakov and Shaposhnikov that transitions between vacua can be induced by thermal fluctuations of the electroweak field configurations \cite{Kuzmin:1985mm}.  So instead of tunnelling from one vacuum to another we may have a transition induced by thermal fluctuations over the barrier (see figure \ref{fig:bau_4}). In the case temperatures are larger than the typical barrier height the exponential $T=0$  suppression is weakened and the $(B+L)$ violating processes may profuse and be in equilibrium in the expanding universe. 

Finite temperature transitions among different ground states of the electroweak theory are governed by the sphaleron configurations which are static configurations corresponding to unstable solutions for the equation of motion of the theory \cite{Klinkhamer:1984di}. The transition rate is quite different according to the corresponding temperature to be higher or lower than $T_W$,  the temperature of the electroweak phase transition. In particular for $T < T_{EW}$ one finds the transition rate per unit volume \cite{Arnold:1987mh}
\be 
\frac{\Gamma_{B+L}}{V} = \mu M_W^4 \left( \frac{M_W}{\alpha_W T} \right)^3  e^{-\frac{E_{\hbox{\tiny sph}}}{\alpha_W  T}} \, ,
\ee
where $M_W$ is the W boson mass, $\mu$ a constant of order one and $E_{\hbox{\tiny sph}} \equiv M_W(T)/a_W$ is the sphaleron energy. The latter is temperature dependent through the finite temperature expectation value of the Higgs boson. The rate is still pretty much suppressed at temperatures below the electroweak scale. However the exponential suppression is expected to vanish when the electroweak symmetry is restored. In the symmetric phase, $T > T_{EW}$ the same rate has been found to be \cite{Arnold:1996dy}
\be 
\frac{\Gamma_{B+L}}{V} \sim \alpha^5 T^4 \ln \frac{1}{\alpha_W} \, . 
\ee 
Hence at temperature of order $T \sim 10^2$ GeV, the baryon number violating processes are not suppressed and are in equilibrium up to temperature of order $\mathcal{O}(10^{12})$ GeV \cite{Burnier:2005hp}. The first Sakharov condition is satisfied in the early universe already within the SM. 

Let us come to the C and CP violation in the SM. It is known that C is maximally violated since only left-handed fermions couple to the SU(2) gauge fields. The CP violation was observed in the quark sector, more specifically in strange and beauty mesons decays \cite{Christenson:1964fg,Burkhardt:1988yh, Barr:1993rx}. Then the second Sakharov condition is also fulfilled.  However the CP phases provided within the quark sector are far too small to account for $\eta_B \sim \mathcal{O}(10^{-10})$. In short, the only CP phase in the SM originates in the CKM matrix, connecting the mass and interaction (electroweak) eigenstates of the left-handed quarks \cite{Kobayashi:1973fv}. There is a more quantitative way to express the amount of CP violation by means of the Jarlskog invariant that comes out to be  $J \sim \mathcal{O}(10^{-20})$ \cite{Jarlskog:1985ht}. Being not present any significant enhancement of the baryon asymmetry due to processes within the SM in the early universe \cite{Gavela:1994ds,Gavela:1994dt}, it seems impossible to fill the many orders of magnitude gap to reproduce the baryon-to-photon ratio in eq.~(\ref{bau_1}). 

Let us come to the third Sakharov condition. The departure from thermal equilibrium in the SM is provided  by the electroweak phase transition. This mechanism gives the name to a class of models, which the SM belongs to, that provides the generation of the baryon asymmetry: \textit{electroweak baryogenesis}.  However in order to provide a sufficient deviation from equilibrium, the electroweak transition is required to be strongly first order and this sets a severe bound on the Higgs mass, $m_\phi \leq 72$ GeV \cite{Kajantie:1995kf}. Thus, viable models of electroweak baryogenesis need a modification of the scalar potential such that the nature of the electroweak phase transition is modified, together with new sources of CP violation (for example see \cite{Losada:1996ju,Carena:2008vj}).

In summary, despite the Sakharov conditions are comprised in the SM, we cannot achieve a successful baryogenesis. Additional sources of CP violation are invoked, together with some alternative mechanism for a strong enough departure from thermal equilibrium: the generation of the observed baryon asymmetry requires some new physics. Besides GUT baryogenesis, briefly discussed in the toy model in section~\ref{hot_sec2}, alternatives comprise Affleck--Dine mechanism \cite{Affleck:1984fy} and spontaneous baryogenesis \cite{Cohen:1988kt}. Another interesting and appealing framework is baryogenesis via leptogenesis \cite{Fukugita:1986hr} (see Chapter~\ref{chap:lepto}). In this class of models an asymmetry is generated in the leptonic sector. Then due to the connection between baryon and lepton number provided by the sphaleron transitions, the lepton asymmetry is partially reprocessed into a baryon one.  We already set the basis for leptogenesis discussing the toy model for GUT baryogenesis. Indeed new heavy states are added to the SM particle content: in its original formulation, heavy neutrinos with a large Majorana mass. In the following we discuss how baryon and lepton asymmetries can be related to each other. 

\subsection{Relating baryon and lepton asymmetries} 
In this section we deal with the relation between baryon and lepton number at high temperatures. Beside being an interesting application of sphaleron transitions and equilibrium dynamics, such discussion introduces a fundamental ingredient for leptogenesis. Our aim is to show that a matter-antimatter imbalance stored in the baryon sector implies a lepton asymmetry and viceversa. In the present discussion we stick to the SM particle content and the derivation follows the one given in~\cite{Buchmuller:2005eh,Chen:2007fv}.  

Let us consider a weakly coupled plasma at temperature $T$. We can assign a chemical potential $\mu_i$ to each of the quarks, leptons and Higgs field in the heat bath. Since there are left-handed lepton and quark SU(2) doublets, right-handed  quarks and lepton SU(2) singlets (see table~\ref{Tab:tab1}) and one Higgs doublet, we can assign $5 N_f +1$ chemical potentials, where $N_f$ stands for the number of fermion generations. If we consider  the degrees of freedom in the thermal bath as massless, the asymmetries in the number densities of particle and antiparticles  read
\be 
n_i - \bar{n}_i = \frac{g_iT^3}{6} \begin{cases}
\beta \mu_i + \mathcal{O}((\beta \mu_i)^3) \, ,
\\
2\beta \mu_i + \mathcal{O}((\beta \mu_i)^3) \, ,
\end{cases}
\label{barlep_1}
\ee
where the first line holds for fermions, whereas the latter for bosons and $g_i$ stands for the internal degrees of freedom of the particle (antiparticle). The key observation is that one can deduce the particle-antiparticle asymmetries from the chemical potentials. We can find some relations among the chemical potentials of the different particles participating the  interactions in the early universe \cite{Harvey:1990qw}. Quarks, leptons and Higgs bosons interact via Yukawa and gauge couplings and, in addition, via the non-perturbative sphaleron processes. In thermal equilibrium all these processes yield constraints between the various chemical potentials. The effective 12-fermion interactions induced by the sphalerons lead to 
\be 
\sum_i ( 3 \mu_{Q_{i}} +  \mu_{L_{i}} )=0 \, .
\label{barlep_2}
\ee
where the sum runs over the quark and lepton generations (the meaning of the index is the same as given in \ref{sec:baryonew}).
The SU(3) QCD instanton processes \cite{Mohapatra:1991bz}, which generate an effective interaction between left- and right-handed quarks,  provide the following relation 
\be 
\sum_i (2 \mu_{Q_{i}} - \mu_{U_{i}} - \mu_{D_{i}})=0  \, .
\label{barlep_3}
\ee
A third condition, valid at all temperatures, is obtained by requiring that the total
hypercharge of the plasma vanishes. From eq.~(\ref{barlep_1}) and the known hypercharges one
derives
\be 
\sum_i \left(  \mu_{Q_{i}} +2 \mu_{U_{i}} - \mu_{D_{i}} -  \mu_{L_{i}} - \mu_{E_{i}}  + \frac{2}{N_f} \mu_{\phi} \right)  =0 \, ,
\label{barlep_4}
\ee
where $\mu_\phi$ is the chemical potential of the Higgs doublet (all the components have the same chemical potential). The Yukawa interactions yield relations between the chemical potentials of left-handed and right-handed fermions (with different flavours)
\be 
\mu_{Q_{i}} -\mu_{D_{j}}- \mu_{\phi} =0 \, , \phantom{xx} \mu_{Q_{i}} -\mu_{U_{j}} + \mu_{\phi} =0 \, , \phantom{xx}  \mu_{L_{i}}- \mu_{E_{j}} - \mu_{\phi} =0 \, .
\label{barlep_5}
\ee
The relations (\ref{barlep_2})-(\ref{barlep_5}) hold if the corresponding interactions are in thermal equilibrium. In the temperature range 10$^2$ GeV $< T <$ 10$^{12}$ GeV,  gauge interactions are in equilibrium.  On the other hand, Yukawa interactions are in
equilibrium in a more restricted temperature range that depends on the strength of
the Yukawa couplings \cite{Harvey:1990qw}. We ignore this slight complication in the present discussion. 

We define the baryon- and lepton-asymmetries number density as follows according to (\ref{barlep_1})
\be 
n_{\Delta B}= \frac{g_B}{6} \Delta B \, T^2 \, , \quad n_{\Delta L}= \frac{g_L}{6} \Delta L \, T^2 \, ,
\label{barlep_6}
\ee
with 
\bea
&& \Delta B= \sum_i  (2 \mu_{Q_{i}} + \mu_{U_{i}} + \mu_{D_{i}}) \, ,
\label{barlep_7}
\\
&& \Delta L = \sum_i (2 \mu_{L_{i}} + \mu_{E_{i}}) \, .
\label{barlep_8}
\eea
and we assume that the asymmetry in each generation is the same, e.~g.~$\mu_{L_e}=\mu_{L_\mu} = \mu_{L_\tau} \equiv \mu_L$. Then $g_B$ and $g_L$ are the degrees of freedom of the baryons and leptons. The relations (\ref{barlep_2})-(\ref{barlep_5}) can be solved them in terms of a single chemical potential. If one takes $\mu_{L}$ the baryon and lepton asymmetries are found to be~\cite{Buchmuller:2005eh}
\bea
&&\Delta B =  -\frac{4}{3} N_f \mu_{L} \, ,
\label{barlep_9}
\\
&& \Delta L =  \frac{14 N_f^2 + 9 N_f}{6 N_f +3} \mu_{L} \, .
\label{barlep_10}
\eea
This implies the important connection between the $\Delta B$, $\Delta (B - L)$ and $\Delta L$ asymmetries, that reads \cite{Khlebnikov:1988sr}
\bea 
&&\Delta B=c_s \Delta (B - L) \, , 
\label{barlep_11}
\\
&&\Delta L = (c_s -1) \Delta (B - L) \, ,  
\label{barlep_12}
\eea
with 
\be 
c_s = \frac{8N_f + 4}{22 N_f +13} \, .
\label{barlep_13}
\ee
Looking at ({\ref{barlep_11}) one finds that, in order to have a baryon asymmetry, $B-L$ violating interactions have to occur in the early universe. Moreover, since the $B-L$ combination is conserved by sphaleron interactions, the baryon asymmetry today is the same as the one present at the freeze-out of the sphaleron processes. There is another way to look at the relations (\ref{barlep_11}) and (\ref{barlep_12}). An asymmetry generated in the lepton sector induces automatically a baryon asymmetry when sphalerons are in equilibrium:
\be 
\Delta B = \frac{c_s}{c_s -1} \Delta L. 
\label{barlep_14}
\ee
A baryon asymmetry can be achieved also in those models where only lepton number is violated. This welcome the possibility to explain the generation of a matter-antimatter imbalance via lepton violating processes, namely baryogenesis via leptogenesis. 

%% file: lepto.tex
The Sakharov conditions were first implemented in the contest of GUTs where heavy scalar or gauge boson decays generate the imbalance between baryons and antibaryons. However the necessary conditions for the generation of a matter-antimatter asymmetry can be embedded in different scenarios besides GUT models. Indeed, one of the most promising framework for explaining the baryon asymmetry in the universe is via leptogenesis \cite{Fukugita:1986hr}. In its original formulation, the new heavy states are Majorana neutrinos with large Majorana masses that decays into leptons and antileptons in different amounts. The net asymmetry in the lepton sector is then partially reprocessed into a baryon one through the sphaleron transitions in the SM \cite{Kuzmin:1985mm}, that connect the baryon and lepton number. 
%\cite{Fukuda:1998mi}
  
The increasing popularity of leptogenesis is also due to its deep connection with neutrino physics. %For the following discussion we turn into a particle physics point of view, and we come back to the baryogenesis prospective in a while. 
The recent amount of literature on leptogenesis has been triggered by the discovery of neutrino oscillations \cite{Fukuda:1998mi}. Such experimental evidence has shown that the strict prediction of the SM, namely that neutrino are massless, is wrong and a mechanism to account for neutrino masses is necessary. The absolute neutrino mass scale cannot be inferred by means of oscillation data: only two mass squared differences are available. Complementary experimental searches provide upper bounds on the absolute neutrino mass scale. It comes out that neutrino masses lies in the eV scale and then the question why these particles are much lighter than other SM fermions arises quite naturally. In section~\ref{sec_lepto1}  this topic is briefly introduced. Then in section~\ref{sec_lepto2} we discuss the simplest realization of leptogenesis. An interesting development, especially from the phenomenological point of view, is addressed in section~\ref{sec_lepto3} with a brief discussion on resonant leptogenesis. Finally the recent advancements as regards the thermal aspects of leptogenesis together with open challenges are presented in section~\ref{sec_lepto4}.    

\section{Neutrino oscillations and seesaw type I} 
\label{sec_lepto1}
Neutrino oscillation experiments have shaped and fixed an important feature for the most elusive SM particles: neutrinos mix and therefore different neutrino mass eigenstates exist. The weak and mass eigenstates are not the same and they are connected with a unitary transformation:
\be 
\nu_{L,f} = \sum_{i=1}^n U_{f i} \nu_{L,i} \, .
\label{lepto_1_def}
\ee     
In (\ref{lepto_1_def}) $\nu_{L,f}$ stands for the left-handed neutrino of flavour $f=e,\mu,\tau$, $\nu_{L,i}$ is the left-handed neutrino  with a definite mass $m_{i}$. The left-handed neutrino fields (right-handed antineutrinos) are the chiral field component participating the weak interactions in the SM. All compelling neutrino oscillation data can be described assuming 3-neutrino mixing in vacuum, so that $n=3$ in (\ref{lepto_1_def}). According to this choice $U_{fi}$ is a 3 $\times$ 3 matrix, often called leptonic-mixing matrix or Pontecorvo--Maki--Nakagawa--Sakata (PMNS) matrix \cite{Pontecorvo:1967fh,Pontecorvo:1957cp,Maki:1962mu}. Similarly to the mixing matrix in the quark sector, the leptonic-mixing matrix is expressed in terms of some physical parameters: in this case 3 mixing angles and three complex phases, two Majorana phases and one Dirac phase. The matrix reads~\cite{Blanchet:2012bk,DiBari:2012fz}
\be 
U=\left( \begin{array}{ccc}
c_{12}\,c_{13} & s_{12}\,c_{13} & s_{13}\,e^{-{\rm i}\,\delta} \\
-s_{12}\,c_{23}-c_{12}\,s_{23}\,s_{13}\,e^{{\rm i}\,\delta} &
c_{12}\,c_{23}-s_{12}\,s_{23}\,s_{13}\,e^{{\rm i}\,\delta} & s_{23}\,c_{13} \\
s_{12}\,s_{23}-c_{12}\,c_{23}\,s_{13}\,e^{{\rm i}\,\delta}
& -c_{12}\,s_{23}-s_{12}\,c_{23}\,s_{13}\,e^{{\rm i}\,\delta}  &
c_{23}\,c_{13}
\end{array} \right)
\, {\rm diag}\left(e^{i\,\rho}, 1, e^{i\,\sigma}
\right)\, ,
\label{lepto_2_bis}
\ee
where $s_{ij} \equiv \sin\theta_{ij}$ and $c_{ij}\equiv\cos\theta_{ij}$ and $\theta_{ij}$ stand for the mixing angles, $\delta$ is the Dirac phase and $\sigma$ and $\rho$ are the Majorana phases. On the basis of the existing neutrino data it is impossible to establish weather the massive neutrinos are Dirac or Majorana fermions. Recent and updated values for the mixing angles can be found in \cite{Fogli:2012ua}. 

\begin{figure}[t]
\centering
\includegraphics[scale=0.42]{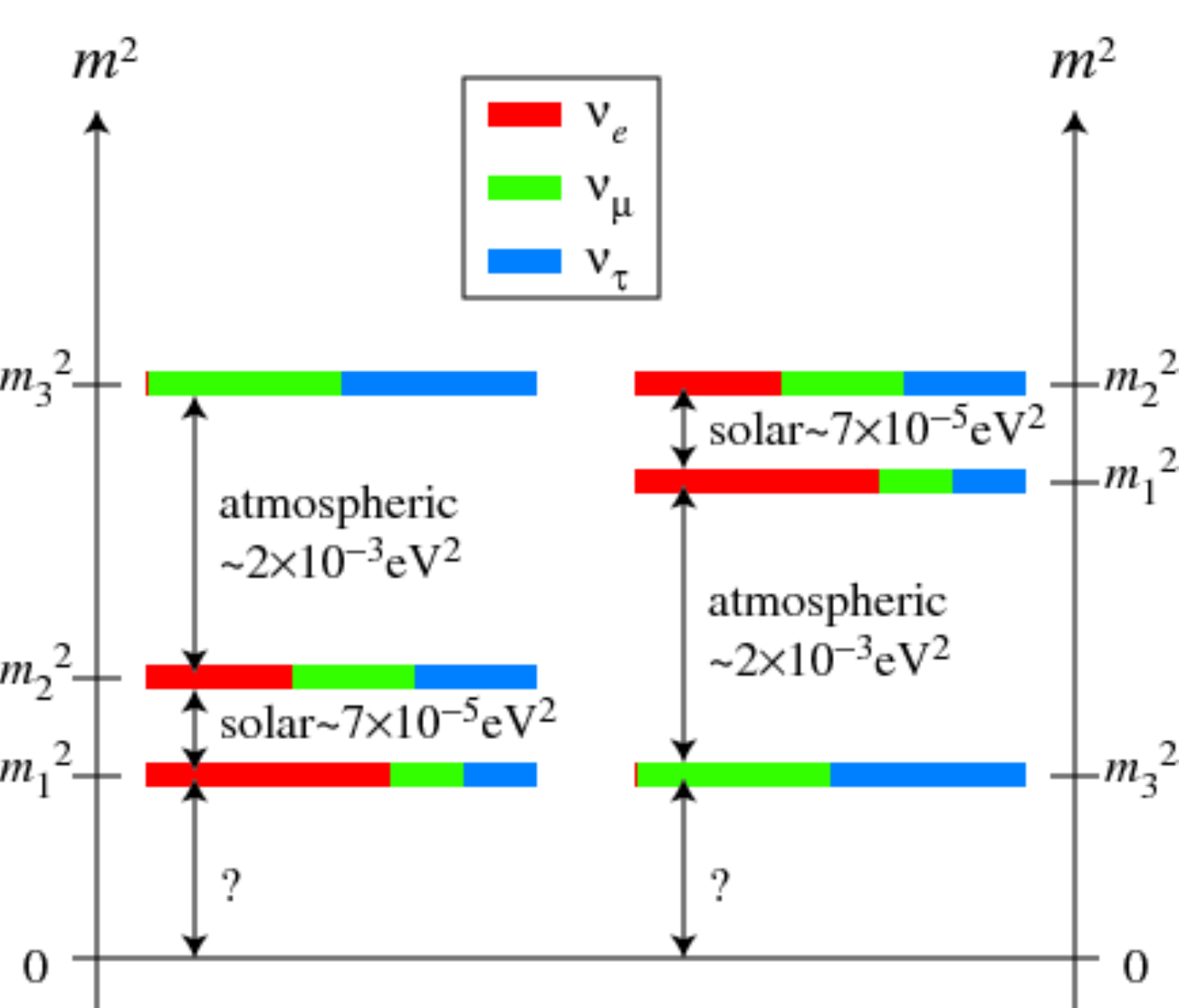}
\caption{\label{fig:lepto_1} Normal (left) and inverted ordering (right) for the neutrino mass squares. The color code assigns the corresponding flavour content to the squared mass. Figure from \cite{King:2013eh}.}
\end{figure}
Moreover oscillation experiments show that at least two neutrinos have to be massive.  The oscillation data are sensitive to two independent mass squared differences \cite{Fogli:2012ua}: 
\be
\Delta m^2_{21}=7.54 \times 10^{-5} \; \text{eV}^2 \, , \quad |\Delta m^2_{31(32)}|=2.47 \times 10^{-3}  \text{eV}^2 \, .
\label{masssquared}
\ee
The numbering of the massive neutrinos $\nu_{L,i}$ is arbitrary. We adopt here the convention which allows to associate $\theta_{13}$ with the smallest mixing angle in the PMNS matrix $U$, and $\Delta m^2_{21}>0$ and $|\Delta m^2_{31(32)}|$ with the parameters which drive respectively the solar and the atmospheric oscillations. Hence the mass squared differences in (\ref{masssquared}) are sometimes denoted as $\Delta m^2_{{\rm{sol}}}$ and  $\Delta m^2_{{\rm{atm}}}$ respectively. The subscripts of the latter notation are inherited from the type of neutrinos used in the experiments, namely solar or atmospheric. Due to the nature of the available observables, there is some degree of uncertainty in the hierarchy of the neutrino mass eigenstates. We find then two viable options, as shown in figure~\ref{fig:lepto_1}: a first one is called \textit{normal ordering} (NO) and corresponds to $m_1 < m_2 < m_3$ with 
\be
m_3^2 - m^2_2 = \Delta m^2_{32}  \hspace{5mm} \mbox{\rm and}
\hspace{5mm}
m_2^2 - m^2_1 = \Delta m^2_{21}\, .
\label{lepto_3}
\ee
On the other hand a second option is represented by the \textit{inverted ordering} (IO), namely $m_3 < m_1 < m_2$ and in this case we write
\be
m_3^2 - m^2_1 = \Delta m^2_{31}  \hspace{5mm} \mbox{\rm and} \hspace{5mm}
m_2^2 - m^2_1 = \Delta m^2_{12} \,  .
\label{lepto_4}
\ee
It may be convenient to introduce the atmospheric neutrino mass scale \cite{Schwetz:2011zk}
\be 
m_{\rm atm} \equiv \sqrt{\Delta m^2_{\rm atm}+\Delta m^2_{\rm  sol}} = (0.049\pm 0.001)\,{\rm eV} \, ,
\label{lepto_5}
\ee
and the solar neutrino mass scale
\be 
m_{\rm sol} \equiv \sqrt{\Delta m^2_{\rm  sol}}=(0.0087 \pm 0.0001)\,{\rm eV} \, ,
\label{lepto_6}
\ee 
in order to have a rough idea of what scale for neutrino masses one can reasonably expect. However,  
the lightest neutrino mass can be arbitrarily small, down to the limit of being massless. 

Upper bounds on the lightest neutrino mass, or in general on the absolute neutrino mass scale $m_i$, are provided by complementary experiments to those studying oscillations. We mention experimental techniques based on tritium beta decay \cite{Kraus:2004zw}, the neutrinoless double beta decay ($00\nu \beta$) \cite{Andreotti:2010vj,Ackermann:2012xja,Henning:2009tt} and cosmological observations from the WMAP collaboration \cite{Abazajian:2011dt,Hannestad:2006as}. The last one provides a stringent bound on the sum of neutrino masses
\be 
\sum_i m_i \leq 0.58 \text{eV} \, (\text{95} \% \; \text{C. L.}) ,
\label{lepto_7}
\ee
which translates in an upper bound on the lightest neutrino mass: $m_1 \siml 0.19$ eV. 

Within the SM, neutrinos are massless and come as left-handed fields that couple to electroweak gauge bosons. Right-handed neutrinos and left-handed antineutrinos are not introduced in the SM particle content.  A way to naturally implement a mass for neutrinos is to make a carbon copy of all the other Dirac fermions: allow for helicity transitions from left-handed to right-handed fields, so that right-handed neutrino fields are necessary to build a Dirac mass term. Clearly these states do not participate the weak interactions, and the right-handed neutrinos (left-handed antineutrinos) are inert or \textit{sterile}, i.e. neutral under the  SU(2)$_L \times$U(1)$_Y$ gauge group. A neutrino mass term is then generated via the coupling with the Higgs field. The price to pay is that one  makes drastically more broadened the Higgs-fermion Yukawa couplings range in order to account for such small fermion masses (recall that $m_\nu \sim \mathcal{O}(\text{eV}) \ll m_e \approx 10^5 \text{eV}$). However a unique feature of neutrinos has to be taken into account: the neutrino is the sole elementary fermion in the SM which may be its own antiparticle, more precisely a Majorana fermion.

A minimal extension of the SM, able to explain not only why neutrinos
are massive but also why they are much lighter than all the other massive fermions,
is represented by the seesaw mechanism \cite{Minkowski:1977sc,GellMann:1980vs}. There exist different realizations of such mechanism, however  
in the minimal seesaw type-I, one adds
right-handed neutrinos, $\nu_{R,I}$, to the SM Lagrangian with a Majorana mass term that violates lepton number.
In the case that right-handed neutrinos are represented by Majorana fermion fields, 
the Lagrangian may be written as follows~\cite{Fukugita:1986hr} (we adopt some of the notation of~\cite{Asaka:2006rw}):
\begin{equation}
\mathcal{L}=\mathcal{L}_{\rm{SM}} + \frac{1}{2} \bar{\psi}_{I} i \slashed{\partial}  \psi_{I}  
- \frac{M_{I}}{2} \bar{\psi}_{I}\psi_{I} - F_{f I}\bar{L}_{f} \tilde{\phi} P_{R}\psi_{I}  - F^{*}_{f I}\bar{\psi}_{I} P_{L} \tilde{\phi}^{\dagger}  L_{f} \, ,
\label{lepto_8}
\end{equation} 
where $\psi_{I}=\nu_{R,I}+\nu^{c}_{R,I}$ is the Majorana field
comprising the right-handed neutrino $\nu_{R,I}$ of type $I=1,2,3$
and mass $M_{I}$; $\mathcal{L}_{\rm{SM}}$ is the SM Lagrangian with unbroken SU(2)$_L\times$U(1)$_Y$ gauge symmetry (see eq.~\ref{SMlag} in appendix~\ref{appB:matchwidth}), 
$\tilde{\phi}=i \sigma^{2} \, \phi^*$ embeds the SM Higgs doublet, 
$L_{f}$ is the SM lepton doublet of flavour $f$, $F_{fI}$ is a complex Yukawa coupling, 
and the right-handed and left-handed projectors are denoted by $P_R = (1 + \gamma^5)/2$ 
and $P_L = (1 - \gamma^5)/2$ respectively. Without loss of generality, we have chosen the basis where the Majorana mass term is diagonal.

The physical mass states for the right-handed neutrinos can naturally be much larger than the electroweak scale, being the $\psi_I$ field a singlet under the SM gauge group. The Lagrangian in (\ref{lepto_8}) is valid at high energies and makes right-handed neutrinos participate in particle interactions in the early universe. However, at temperatures below $T_W$, we can replace the Higgs filed with its vacuum expectation value, $v$, and we define a  Dirac mass matrix as $(m_D)_{f I} \equiv F_{f I} v$.  The Lagrangian in (\ref{lepto_8}) then reads
\be 
\mathcal{L}=\mathcal{L}_{\rm{SM}} + \frac{1}{2} \bar{\psi}_{I} i \slashed{\partial}  \psi_{I}  
- \frac{M_{I}}{2} \bar{\psi}_{I}\psi_{I} - (m_D)_{f I}  \bar{\nu}_f P_{R} \psi_{I}  - (m_D^*)_{f I}  \bar{\psi}_{I} P_{L} \nu_f \, ,
\label{lepto_8EW}
\ee
where $\nu_f$ stands for the active (left-handed) SM neutrino with flavour $f$. The neutrino mass matrix takes the form
\be 
\left( \begin{array}{c c}
0 & m_D \\
m_D^T & M
\end{array} \right)  \, ,
\label{seesawMat}
\ee 
which can be block diagonalized in the seesaw limit $m_D \ll M$ leading to two different sets of eigenvalues: a light and a heavy one. Three light eigenvalues are suppressed by a factor $m_D M^{-1}$ and correspond to the small active neutrinos masses that are found by diagonalizing the mass matrix obtained by the seesaw formula \cite{Minkowski:1977sc,GellMann:1980vs,Mohapatra:1979ia}
\be 
 m_\nu = -m_{D} \frac{1}{M} m_D^T  \, ,
\label{lepto_9}
\ee  
where $m_\nu$ is a 3$\times$3 matrix of active neutrino masses, mixing angles, and (possible) CP-violating phases. An  analysis of eq.~(\ref{lepto_9}) shows  that the number of right-handed neutrinos must be at least two
to fit neutrino oscillation data. If there were only one sterile neutrino, then
the two active neutrinos would be massless.  The matrix $m_\nu$ in (\ref{lepto_9}) can be diagonalized by a unitary matrix $U_\nu$ \cite{Blanchet:2012bk,DiBari:2012fz} 
\be 
D_{\nu} \equiv {\rm{diag}} \left( m_1, m_2, m_3 \right) = -U_\nu^{\dagger} m_\nu U^*_\nu \, .
\label{lepto_10}
\ee 
In a basis where the charged lepton mass matrix is diagonal (terms not displayed in (\ref{lepto_8}), see (\ref{eq_fla_1})), the unitary matrix $U_\nu$ coincides with the leptonic mixing matrix in (\ref{lepto_2_bis}). The masses $M_1$, $M_2$ and $M_3$ correspond, in a good approximation,  to the eigenstates of the Majorana mass matrix, already diagonal in the Lagrangian (\ref{lepto_8}) and they are the set of large eigenvalues of the neutrino matrix in (\ref{seesawMat}). In this way the lightness of ordinary neutrinos is explained just as an algebraic by-product. If the largest eigenvalue in the Dirac neutrino mass
matrix, $m_D$, is assumed to be of the order of the electroweak scale, as for the other massive
fermions, then for example the atmospheric neutrino mass scale $m_{\rm atm} $
can be naturally reproduced for $M_3 \sim 10^{14}-10^{15}$ GeV, close to the grand-unified scale~\cite{Blanchet:2012bk,DiBari:2012fz}. This is the minimal version of the seesaw mechanism. Other options are viable \cite{Magg:1980ut,Schechter:1980gr,Foot:1988aq,Ma:1998dn} which are not addressed here. 

The seesaw formula (\ref{lepto_9}) allows the mass of singlet neutrinos to be a free parameter. Indeed multiplying $m_D$ by any number $x$, namely changing the Yukawa couplings, and $M_I$ by
$x^2$ does not alter the right-hand side of the formula. Therefore, the choice of $M_I$ is a matter of theoretical prejudice that cannot be fixed by active-neutrino experiments alone. %Different models provide some constraints on the Majorana masses $M_I$ (for reviews see). 
In the following we mention three benchmark examples \cite{Drewes:2013gca}:
\begin{itemize}
\item $M_I \simg 10^9$ GeV:  this mass scale is motivated by embedding the Lagrangian (\ref{lepto_8}) in GUT scenarios \cite{Georgi:1974my}, such as SO(10) unification \cite{Fritzsch:1974nn,Maltoni:2000iq}. For $m_D$ of order of the electroweak scale, hence $F$ of order one, right-handed neutrino masses $M_I \sim 10^9 - 10^{14}$ GeV allow for the explanation of neutrino oscillation data via (\ref{lepto_9}). A baryon asymmetry can be attained within such a framework via standard thermal leptogenesis (see section~\ref{sec_lepto2}). 

\item If one assumes the Majorana matrix $M_I$ to have two eigenvalues of the order of the electroweak scale, $\mathcal{O}(10^2)$ GeV, and one in the keV range, we reduce to the so-called neutrino minimal standard model ($\nu$MSM) \cite{Asaka:2005pn}. This choice does not demand any new scale between the Planck and the electroweak scale, but it does require small Yukawa couplings $F$. Besides accommodating successfully neutrino oscillations data, the model can be adjusted to account for both the baryon asymmetry generation via leptogenesis and a viable dark matter candidate.   

\item Right-handed neutrino masses at the eV scale may explain the anomalies seen in some short baseline and reactor neutrino experiments \cite{Palazzo:2013me} and/or account for the fits on cosmological data that require additional radiation (relativistic particles) \cite{Dolgov:2003sg,Cirelli:2004cz}.
\end{itemize}
For more details we refer to extensive reviews on right-handed neutrino phenomenology and implications in cosmology \cite{Drewes:2013gca,Adhikari:2016bei}. 
As far as we are concerned with leptogenesis in the thesis, we focus on right-handed neutrino mass ranges that allow for the generation of a matter-antimatter asymmetry in the early universe. The Lagrangian in (\ref{lepto_8}), besides accommodating the neutrino oscillation data, provides new heavy fields suitable for a successful implementation of baryogenesis via leptogenesis, which is the subject of the next two sections.

\section{Vanilla leptogenesis}
\label{sec_lepto2}
In this section we come back to the matter-antimatter generation and we discuss how leptogenesis works. In order to introduce all the basic concepts on the subject we start with the simplest and original realization of leptogenesis, often called \textit{vanilla leptogenesis} \cite{Blanchet:2012bk}. Despite the various assumptions and simplifications, this scenario comprises all the main ideas behind leptogenesis and enables us to highlight important connections with the active (low mass) neutrino parameters. 

In this scenario three right-handed neutrinos with large and hierarchically ordered Majorana masses, far above the electroweak scale, are introduced and participate the dynamics in the early inverse.  Yukawa interactions among right-handed neutrinos, SM lepton and Higgs doublets in the thermal bath allow for an equilibrium abundance of these heavy states in the very early stages of the universe after inflation. This requires the reheating temperature to be at least of order of the lightest heavy neutrino mass, $M_1$. Since
in most models of neutrino masses embedding the type-I seesaw the lightest RH
neutrino mass is $M_1 \ll 10^{15}$ GeV, the condition of thermal leptogenesis can be satisfied
compatibly with the upper bound on the reheating temperature, $T_{RH} \siml 10^{15}$ GeV, from CMB observations \cite{Giudice:2000ex}. As mentioned a hierarchically ordered spectrum for the heavy Majorana neutrino mass pattern is assumed, in particular one usually requires that $M_1 \ll M_{2,3}$. This last condition has an important consequence: the CP asymmetries are effectively generated by the decays of the lightest heavy neutrino. Indeed, any previous asymmetry due to the heavier states is erased by the fast interactions mediated by the lightest heavy neutrino. Therefore it suffices to consider only the decays of $\nu_{R,1}$ into leptons and antileptons as being relevant for the generation of a  matter-antimatter imbalance. %In the following we consider only $\nu_{R,1}$ as a dynamical degree of freedom, whereas $\nu_{R,2}$ and $\nu_{R,3}$ participate the interactions as virtual states.

The following discussion is similar to that carried out in the contest of GUT baryogenesis in section~\ref{hot_sec2}. We recall that leptogenesis belongs to models where the matter-antimatter asymmetry is generated in decays of very heavy particles. We shall introduce two key ingredients: the heavy neutrino decay widths (and production rate) and the CP asymmetry. The heavy neutrino decay processes
\bea 
&&\nu_{R,1} \to \ell_{f} + \phi \, ,
\label{lepto_11a}
\\
&&\nu_{R,1} \to \bar{\ell}_{f} + \phi^{\dagger} \, , 
\label{lepto_11b}
\eea
violate the lepton number, $L$. We denote with $\ell_f$ a lepton, either charged or neutral, belonging to the SU(2)$_L$ lepton doublet. For a Majorana neutrino, the condition $\psi^c = C\bar{\psi}^T = \psi$, is invariant with respect to global $U(1)$ gauge transformations of the field $\psi$ carrying a $U(1)$ charge, $Q$, only if $Q=0$. As a result, $\psi$ cannot carry non-zero additive quantum numbers, such as a lepton number $L$. Since the Higgs boson do not carry any lepton number, the processes (\ref{lepto_11a}) and (\ref{lepto_11b}) violate lepton number by one unit $|\Delta L|=1$, as well as the inverse decay processes do.
The first Sakharov condition is met.
\begin{figure}[t]
\centering
\includegraphics[scale=0.58]{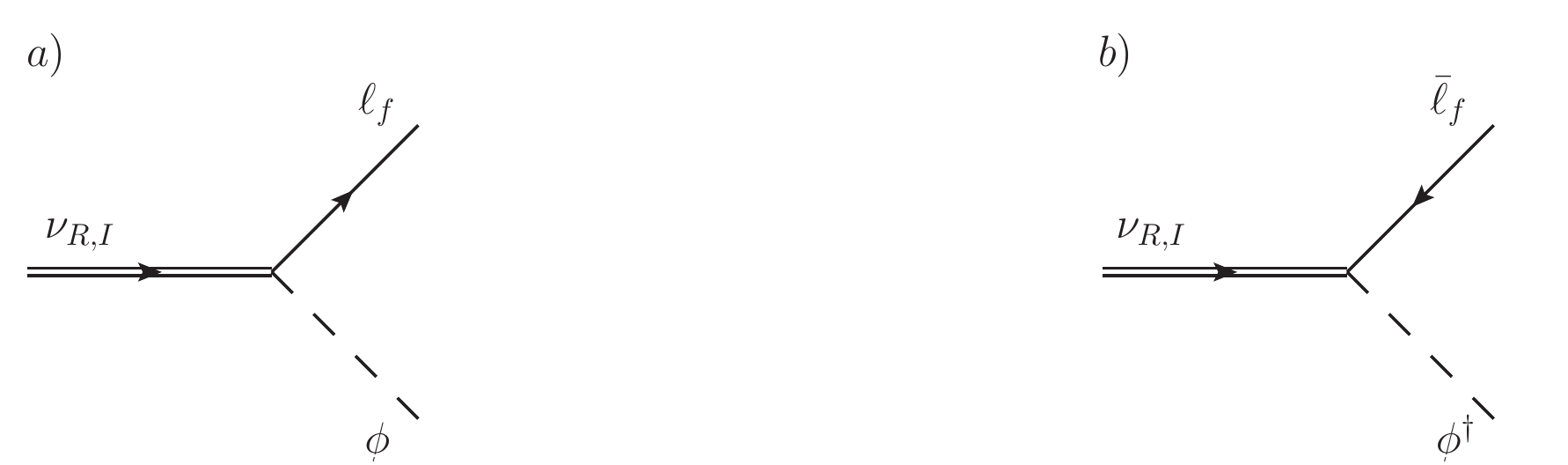}
\caption{\label{fig:lepto_2} Tree-level diagrams for the decay processes $\nu_{R,1} \to \ell_{f} + \phi$ and $\nu_{R,1} \to \bar{\ell}_{f} + \phi^{\dagger}$. Double solid lines stand for heavy right-handed neutrino propagators (forward arrow corresponds to $\langle 0| T(\psi \bar{\psi}) |0\rangle$),  
solid lines for lepton propagators and dashed lines for Higgs boson propagators.}
\end{figure}

The decay widths into leptons and antileptons can be calculated straightforwardly from (\ref{lepto_8}). At tree level, namely at order $|F|^2$ in the Yukawa couplings, they read as follows
\bea 
&&\Gamma(\nu_{R,1} \to \ell_{f} + \phi)  = \frac{|F_{f1}|^2}{16 \pi} M_1 \, , 
\label{lepto_12a}
\\
&&\Gamma(\nu_{R,1} \to \bar{\ell}_{f} + \phi^\dagger)  = \frac{|F_{f1}|^2}{16 \pi} M_1 \, ,
\label{lepto_12b} 
\eea
and the corresponding diagrams are given in figure \ref{fig:lepto_2}. We see that the leptonic and antileptonic widths in (\ref{lepto_12a}) and (\ref{lepto_12b}) are the same and therefore, at this order, no lepton asymmetry can be generated in the heavy Majorana neutrino decays. Indeed we can define a CP asymmetry, which is the analogue of that written in (\ref{bau_13}), as follows
\be 
\epsilon_{1f}= \frac{\Gamma(\nu_{R,1} \to \ell_{f} + \phi) - \Gamma(\nu_{R,1} \to \bar{\ell}_{f} + \phi^\dagger) }{ \sum_f \, \Gamma(\nu_{R,1} \to \ell_{f} + \phi) + \Gamma(\nu_{R,1} \to \bar{\ell}_{f} + \phi^{\dagger})}   ,
\label{lepto_13} 
\ee   
where the difference in the numerator is due to the corresponding lepton number of the final state: $L=+1$ and $L=-1$ for a lepton and an antilepton respectively. The asymmetry is then normalized to the total width, summed over the lepton flavour $f$. The quantity $\epsilon_{1f}$ is a measure of the CP asymmetry generated by the decay of the lightest heavy neutrino,  and we will refer to it in this way. We notice that the CP asymmetry carries two indices, one related to the heavy neutrino species and one to the flavour of the lepton (antilepton) produced in the decays. 

An assumption of vanilla leptogenesis is the \textit{single-flavour} approximation, or  \textit{unflavoured} regime. In short, this amounts  at  assuming  that  the  leptons  and  antileptons  which  couple to the right handed neutrinos maintain their coherence as  flavour superpositions throughout the leptogenesis era. Therefore the interactions occurring in the thermal bath do not 
distinguish different lepton flavours. The unflavoured regime is found to be an appropriate choice at high temperatures, namely $T > 10^{12}$ GeV,  
while the different lepton flavours are resolved at lower temperatures~\cite{Nardi:2005hs, Nardi:2006fx}. According to the unflavoured regime the expression in (\ref{lepto_13}) reads
\be 
\epsilon_{1}=
\frac{\sum_{f} \Gamma(\nu_{R,1} \to \ell_{f} + \phi) - \Gamma(\nu_{R,1} \to \bar{\ell}_{f} + \phi^\dagger)  }
{\sum_{f}  \Gamma(\nu_{R,1} \to \ell_{f} + \phi) + \Gamma(\nu_{R,1} \to \bar{\ell}_{f} + \phi^{\dagger})} \, .
\label{eq:adef}
\ee 
where the sum runs over the SM lepton flavours. We refer to the unflavoured version of the CP asymmetry in the rest of the discussion and we define the total width as follows
\be 
\Gamma_1 = \sum_f \Gamma(\nu_{R,1} \to \ell_{f} + \phi) + \Gamma(\nu_{R,1} \to \bar{\ell}_{f} + \phi^\dagger) = \frac{|F_{1}|^2}{8 \pi} M_1 \, .
\label{totwidth}
\ee
\begin{figure}[t]
\centering
\includegraphics[scale=0.52]{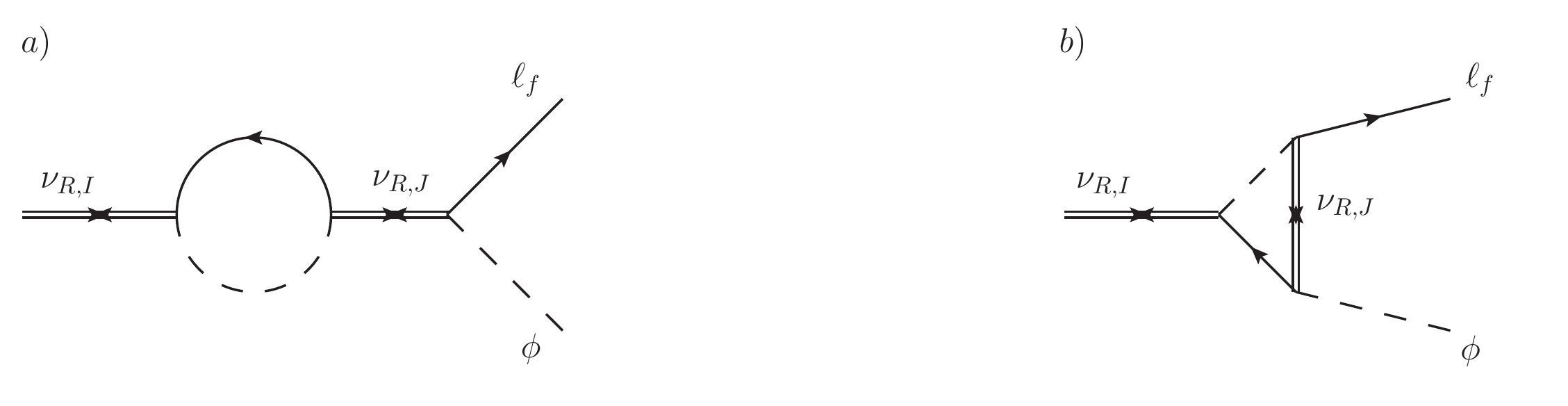}
\caption{\label{fig:lepto_3}  One-loop self-energy and vertex diagrams that interfere with the tree level diagrams for the decay process $\nu_{R,1} \to \ell_{f} + \phi$. The notation of the particles is the same as given in figure \ref{fig:lepto_2}. The neutrino propagators with forward-backward arrows correspond to $\langle 0| T(\psi \psi) |0\rangle$ 
or $\langle 0| T(\bar{\psi} \bar{\psi})|0 \rangle$. Similar diagrams exist for the decay process $\nu_{R,1} \to \bar{\ell}_{f} + \phi^\dagger$.}
\end{figure}

Similarly to the GUT toy model in section~\ref{hot_sec2}, the CP asymmetry in leptogenesis is originated from the interference between the tree-level diagrams, shown in figure~\ref{fig:lepto_2}, and the one-loop self-energy and vertex diagrams in figure~\ref{fig:lepto_3}. The lepton appears as the final state and the corresponding diagrams with an antilepton in the final state are not shown.  
The contribution from the interference of the tree-level diagram with the self-energy diagram is often called \textit{indirect} 
contribution, while the one arising from the interference with the vertex diagram is called \textit{direct} contribution. 
The relative importance of the indirect and direct contributions to the CP asymmetry depends on the heavy neutrino mass spectrum. For not too degenerate neutrino masses, the direct contribution reads (second line shows the hierarchical limit) \cite{Covi:1996wh,Fong:2013wr}
\begin{eqnarray}
\epsilon_{1,{\rm{direct}}}&=& \frac{M_i}{M_1} \left[ 1-\left( 1+\frac{M^2_i}{M^2_1} \right)  \ln \left( 1+ \frac{M_1^2}{M_i^2} \right) \right] \frac{ {\rm{Im}}\left[ \left( F_1^* F_i\right)^2\right] }{8 \pi|F_1|^2} \nonumber \\
&\underset{M_1 \ll M_i}{=}& -\frac{1}{16 \pi } \frac{M_1}{M_i}\frac{ {\rm{Im}}\left[ \left( F_1^* F_i\right)^2\right] }{|F_1|^2} +\mathcal{O}\left( \frac{M_1}{M_i}\right)^3 \, ,
\label{CPvert}
\end{eqnarray}
whereas the contribution originated by the interference of the tree level process with the self-energy diagram reads (second line shows the hierarchical limit) \cite{Covi:1996wh,Fong:2013wr}
\begin{eqnarray}
\epsilon_{1,{\rm{indirect}}}&=& \frac{M_1 M_i}{M_1^2-M_i^2} \frac{ {\rm{Im}}\left[ \left( F_1^* F_i\right)^2\right] }{8 \pi|F_1|^2} \nonumber \\
&\underset{M_1 \ll M_i}{=}&-\frac{1}{8 \pi } \frac{M_1}{M_i}\frac{ {\rm{Im}}\left[ \left( F_1^{*} F_i\right)^2\right] }{|F_1|^2}  +\mathcal{O}\left( \frac{M_1}{M_i}\right)^3  .
\label{CPself}
\end{eqnarray}
where $M_i$ are the heavier states, with $i=2,3$, and a sum over the repeated index $i$ is understood. Due to the assumption $M_1 \ll M_i$, one selects automatically the situation where the heavy neutrino mass difference, $|M_1-M_i|$ is much bigger than the heavy neutrino widths or the mixing terms. This is a relevant aspect we are going to discuss in section~\ref{sec_lepto3}. 
The second Sakharov condition is also met: given that the Yukawa couplings are complex, the C and CP violation arise from the interference between the decay process at tree level and one-loop, the latter generating a non-zero absorbative term. Details on the calculation of the CP asymmetry are provided in chapters~\ref{chap:CPdege} and \ref{chap:CPhiera}, where we study the problem within an EFT approach. 

We address the third Sakharov condition, namely the out-of-equilibrium dynamics. The required deviation from thermal equilibrium is provided by the expansion of the universe. When the temperature has cooled down to values of order of the heavy neutrino mass, their equilibrium number density should become exponentially suppressed. However, if the heavy neutrinos are sufficiently weakly coupled with the heat bath, they cannot follow the rapid change of the equilibrium particle distributions, remaining as abundant as earlier times. This is made manifest when the temperature drops below $M_1$ and we can say that the deviation from thermal equilibrium consists in a too large number density of heavy neutrinos with respect to their equilibrium density \cite{Riotto:2011zz}. In particular this requires the total decay width given in (\ref{totwidth}) to be smaller than the Hubble rate, $H$, at the time defined by $T \sim M_1$. Strictly speaking we have to impose    
\be 
\Gamma^{T=0}_{1} <  H (T=M_1) \, ,
\label{condout}
\ee
where the superscript in the total width signals that it is taken in the zero temperature limit \cite{Buchmuller:2004nz,Buchmuller:2005eh}, hence it corresponds to  the quantity in (\ref{totwidth}).
One can rephrase both the total width and Hubble rate in term of low mass neutrino parameters \cite{Buchmuller:1996pa} by defining the \textit{decay parameter} \cite{Buchmuller:2004nz} as follows
\be 
K_1 \equiv \frac{\Gamma^{T=0}_{1}}{ H (T=M_1)} = \frac{\tilde{m}_1}{m_*} \, ,
\label{decaypar}
\ee
where the \textit{effective neutrino mass} and the \textit{equilibrium neutrino mass} read respectively \cite{Buchmuller:1996pa}
\bea 	
&&\tilde{m}_1 = \frac{|F_1|^2 v^2}{M_1} \, ,
\label{effnu}
\\
&&m_* \simeq 8 \pi \sqrt{g_*} \, 1.66 \, \frac{v^2}{M_{Pl}} \simeq 1.1 \times 10^{-3}\; \text{eV} \, . 
\label{equnu}
\eea
We notice that the decay parameter might be already improved at the level of its definition inserting a finite temperature version of the decay width for $T \sim M_1$, which does exist \cite{Laine:2013lka}. The effective neutrino mass can be also understood as a measure of the strength of the coupling between $\nu_{R,1}$ and the thermal bath.   The deviation from thermal equilibrium is naively established requiring $\tilde{m}_1 < m_*$ according to (\ref{condout}) and (\ref{decaypar}).

\subsection{Boltzmann equations, weak and strong washout}
In order to see if the leptogenesis may explain the observed baryon asymmetry a detail and careful numerical analysis is needed. The quantitative description of this non-equilibrium dynamics is achieved in terms of kinetic rate equations: the Boltzmann equations \cite{Kolb:1979qa} or their quantum mechanical generalization known as Kadanoff-Baym equations \cite{Anisimov:2010dk,Frossard:2012pc,Garny:2009rv,Garny:2009qn}. It is shown that successful leptogenesis is possible for $\tilde{m}_1 < m_*$ as well as $\tilde{m}_1 > m_*$ \cite{Buchmuller:2004nz}. These two situations are called \textit{weak} and \textit{strong washout} respectively. In the present discussion, a washout process is what works against the generation of a the lepton asymmetry. For example, in a simplified version of the processes relevant to leptogenesis, one can could think of heavy neutrino decays in (\ref{lepto_11a}) and (\ref{lepto_11b}), that generate a lepton asymmetry, whereas the corresponding inverse decays $\ell_f  + \phi  \to \nu_{R,1}$ and $\bar{\ell}_f + \phi^\dagger \to \nu_{R,1}$ erase the matter-antimatter imbalance.  
\begin{figure}[t]
\centering
\includegraphics[scale=0.4]{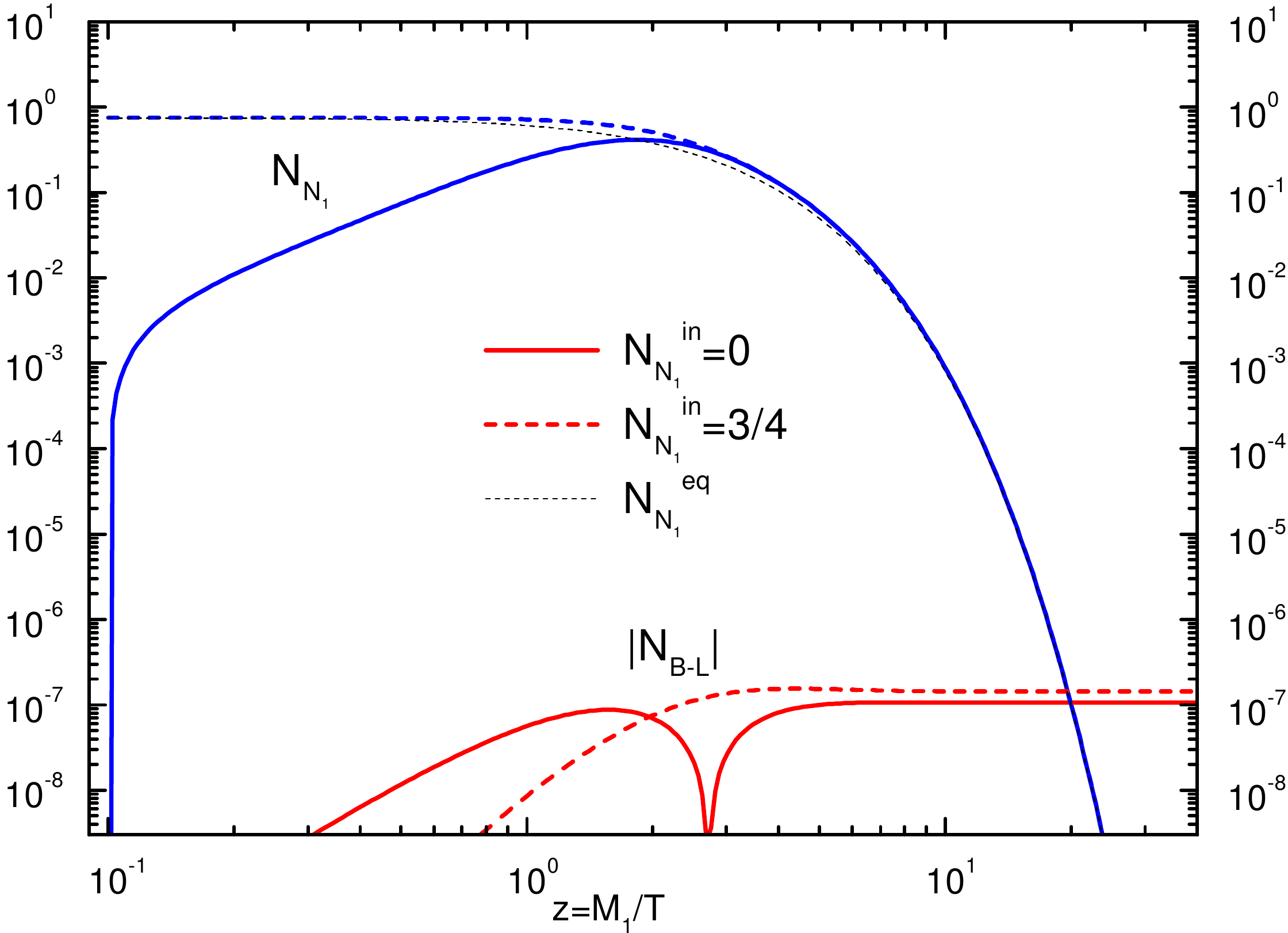}
\caption{\label{fig:lepto_4} Evolution of the heavy neutrino and the $B-L$ abundance for typical leptogenesis parameters: $M_1= 10^{10}$ GeV, $\tilde{m}_1=10^{-3}$ eV and $\epsilon_1=10^{-6}$. The normalization for the heavy neutrino and $B-L$ number density are calculated in a portion of comoving volume that contains one photon at the onset of leptogenesis \cite{Buchmuller:2004nz} (Figure from~\cite{Anisimov:2010dk}).}
\end{figure}

From eqs.~(\ref{effnu}) and (\ref{equnu}) we see that $m_*$ is a fixed parameter whereas $\tilde{m}_1$ can be related to the experimental data providing the light neutrino mass scale. One can show that $\tilde{m}_1 \geq m_{\hbox{\scriptsize min}}$ \cite{Fujii:2002jw}, where $m_{\hbox{\scriptsize min}}$ is the smallest light neutrino mass. Then if one takes  $\tilde{m}_1 \geq m_{\hbox{\scriptsize sol}}$ \cite{Buchmuller:2003gz} the strong wash out condition is satisfied, namely it holds $\tilde{m}_1 > m_*$. In this regime decays and inverse decays rapidly thermalize at $T \sim M_1$ so that any initial asymmetry possibly present before the onset of leptogenesis is erased. This may be understood in terms of the decay parameter in (\ref{decaypar}). Being $K_1 \approx \mathcal{O}(10)$ for $\tilde{m}_1 \geq m_{\hbox{\scriptsize sol}}$, or $K_1 \approx \mathcal{O}(50)$ if we take $\tilde{m}_1 \geq m_{\hbox{\scriptsize atm}}$, the heavy neutrinos remain coupled with the thermal bath even at temperatures $T < M_1$ and their distribution tracks closely the equilibrium one. Moreover the dependence on the initial conditions is absent in the strong washout regime and leptogenesis may be highly predictive also in its simplest formulation. The weak washout regime shows opposite features. The asymmetry generated before the onset of leptogenesis, for $T > M_1$, is not erased by fast processes at $T \sim M_1$. Hence the assumptions on the initial abundance of right-handed neutrinos, either a vanishing or an equilibrium one, as well as the asymmetry generated for temperatures bigger than $M_1$ enter the quantitative description \cite{Buchmuller:2004nz}. 

It is conceivable that the heavy neutrino decays leading to the generation of the matter-antimatter asymmetry occur in the non-relativistic regime, i.~e.~$T < M_1$. This is true both for the strong and weak washout. The difference is that the lepton asymmetry is generated right after the time $T \sim M_1$ in the strong washout, being neutrinos very close to the equilibrium distribution, whereas in the latter case the matter-antimatter imbalance is typically generated at later times. In figure \ref{fig:lepto_4} we show the evolution of the heavy neutrino and $B-L$ abundances, in turn related to baryon and lepton asymmetries, in the case of $m_1 \sim m_*$. Such results are obtained from the rate equations that we discuss right now. 

We introduce the Boltzmann equations for a simplified situation: we take into account only the inverse decays of the heavy neutrinos as washout process that may erase the matter-antimatter asymmetry \cite{Buchmuller:2004nz}. A detailed study of the Boltzmann equations for leptogenesis involving the complete class of relevant processes is found e.~g.~in \cite{Nardi:2007jp}. Indeed we are interested in introducing the role of the total width, the right-handed neutrino production rate and CP asymmetry in the rate equations governing the evolution of the lepton asymmetry. However for consistency one has to include the $|\Delta L|=2$ scattering processes like $\ell \phi \leftrightarrow \bar{\ell} \phi^\dagger $ and $\ell \ell \leftrightarrow \phi \phi$, to take care of the real intermediate state subtraction \cite{Kolb:1979qa,Frossard:2012pc}.  In this simplified scenario the Boltzmann equations read \cite{Buchmuller:2004nz, Giudice:2003jh}
\bea
&& szH\frac{dY_{\nu_{R,1}}}{dz}= - \left( \frac{Y_{\nu_{R,1}}}{Y_{\nu_{R,1}}^{\rm{eq}}} - 1 \right) \gamma_{\nu_{R,1}} \, ,
\label{Boltz1}
\\
&& szH\frac{dY_{\Delta L }}{dz} = \epsilon_1 \left( \frac{Y_{\nu_{R,1}}}{Y_{\nu_{R,1}}^{\rm{eq}}} - 1 \right) \gamma_{\nu_{R,1}} -  \frac{Y_{\Delta L}}{Y_\ell} \gamma_{|\Delta L|=2} \, ,
\label{Boltz2}
\eea
where $Y_i=n_i/s$ are the number densities normalized to the entropy density $s= (2 \pi^2)g_* T^3 /45$, so that the reduction of the particle number density due to the universe expansion is already accounted for (recall the relation between the scale factor and the temperature, $a \propto T^{-1}$, in eq~\eqref{hotbb_17}). Then $H$ is the Hubble parameter and $z=M_1/T$, the latter being a variable suitable to study the particle evolution, $Y_{\Delta L} = 2( Y_{\ell} - Y_{\bar{\ell}})$, where the factor of two comes from the SU(2)$_{L}$ lepton doublet. For a detailed derivation of the Boltzmann equations for leptogenesis we refer to \cite{Davidson:2008bu,Garayoa:2009my}.

The first relevant quantity for the rate equations in (\ref{Boltz1}) and (\ref{Boltz2}) is the space-time density of the rate at
which the thermal plasma at temperature $T$ creates quanta of the lightest right-handed neutrino.
In thermal equilibrium, the creation rate equals the destruction rate, such that both quantities
are usually named \textit{equilibrium interaction rate}. Even though we are interested in out-of-equilibrium dynamics to address the lepton asymmetry generation, the right-handed neutrino production rate can be extracted in terms of an equilibrium distribution function by assuming kinetic equilibrium (in this case the non-equilibrium distributions functions are proportional to the equilibrium ones). 
At leading order  $\gamma_{\nu_{R,1}}$ is given by the thermal average of the total decay rate  given in (\ref{totwidth}), it reads \cite{Salvio:2011sf}
\be 
\gamma_{\nu_{R,1}}^{{\hbox{\scriptsize LO}}}= 2 \int \frac{d^3 \bm{k}}{(2\pi)^3}  \, \frac{\Gamma_1^{T=0}}{E_{\nu_{R,1}}} \,  n_F(E_{\nu_{R,1}}) \, ,
\label{prod_right}
\ee 
where $E_{\nu_{R,1}}=\sqrt{k^2+M_1^2}$ and $n_F$ is the Fermi-Dirac distribution, the factor of two is due to the spin polarization of the heavy neutrino. It is possible to simplify the expression given in (\ref{prod_right}) as follows
\be 
\gamma_{\nu_{R,1}}^{{\hbox{\scriptsize LO}}} =  \frac{T^3}{\pi^2} \, \frac{|F_1|^2 M_1}{8 \pi}\,  z^2 \mathcal{K}_1(z) \, ,
\label{prod_right_bis}
\ee
where $\mathcal{K}_1(z)$ is the modified Bessel function of the first kind (see for example \cite{TableInt}). The expression in eq.~(\ref{prod_right_bis}) enters the standard numerical analysis for leptogenesis. 

The second key ingredient is the CP asymmetry defined in (\ref{eq:adef}). The
lepton asymmetry arises because the decay rate of right-handed neutrinos into matter differs from the one into antimatter. The CP asymmetry quantifies how efficiently a matter-antimatter symmetry is generated in the heavy neutrino decays. In vanilla leptogenesis typical values are $\epsilon_1 \sim 10^{-6}$ \cite{Buchmuller:2004nz,Buchmuller:2005eh,Davidson:2008bu} that eventually provide $\eta_B \sim 10^{-10}$ (the CP asymmetry is reduced by other efficiency factors, such as heavy neutrino over entropy number density for $T \gg M_1$ and washout processes \cite{Davidson:2008bu}). The lepton asymmetry induced by the out-of-equilibrium decays of the heavy neutrinos is then reprocessed into a baryon asymmetry thanks to the sphaleron transitions. The rate of change is established by the relations (\ref{barlep_13}) and (\ref{barlep_14}), that for three fermion generations gives $\Delta B = -(28/51) \Delta L$.  In the case of a hierarchically ordered neutrino mass the CP asymmetry can be read off (\ref{CPvert}) and (\ref{CPself}). 

Let us conclude this section by shortly discussing the Davidson-Ibarra bound for vanilla leptogenesis~\cite{Davidson:2002qv}. Especially in this framework, the information attained from the experiments looking at neutrino oscillations and mixing parameters can shed light and provide  constraints on leptogenesis . The Davidson--Ibarra bound sets a lower bound on the lightest heavy neutrino mass~\cite{Davidson:2002qv}, $M_1 \gtrsim  10^{9}$ GeV, which is  obtained combining the observed baryon asymmetry and the light neutrino masses. This bound gives a clear hint on the energy scale of leptogenesis, at least in its simplest realization, together with the typical temperatures needed for a thermal production of the heavy neutrinos in the early universe. Indeed it implies that the
right-handed neutrinos must be produced at temperatures $T \simg  10^9$ GeV which in turn implies the reheating
temperature after inflation to be of the same order to ensure thermal production of right-handed neutrinos in the early universe. Such bound holds if and only if the following conditions apply: $\nu_{R,1}$ dominates the contribution to leptogenesis, the mass spectrum of the heavy neutrinos is hierarchical, $M_1 \ll M_2,M_3$, and leptogenesis occurs in the unflavoured regime. Violation of one or more of these conditions allows for lowering down the leptogenesis scale and searching for heavy neutrino at present day colliders.

\section{Resonant leptogenesis}
\label{sec_lepto3}
In the previous section we have introduced the CP asymmetry as a key ingredient for the generation of a lepton asymmetry, eventually leading to the observed baryon asymmetry. Moreover we have presented the case where the lightest neutrino dominates the contribution to leptogenesis. We consider now the case where two neutrinos are on the same footing with respect the CP asymmetry generation. We  label them as neutrino of type~1 and type~2 with masses $M_1$ and $M_2$ respectively. 

The  one-loop processes necessary to obtain a different decay rate into leptons and antileptons are shown in figure \ref{fig:lepto_3}. The same topology stands for the neutrino of type~2. Let us consider the contribution arising from the self-energy diagram, namely the indirect asymmetry, given in eq.~(\ref{CPself}). It is straightforward to grasp what happens in the case of nearly degenerate masses for the heavy neutrinos. By assuming two heavy neutrinos with masses $M_1$ and $M_2 = M_1 +\Delta$, with $\Delta \ll M_1$, the indirect contribution goes like $\sim 1/ \Delta$. A similar behaviour is not shown in the direct CP asymmetry in eq.~(\ref{CPvert}). For very small values of the mass splitting, $\Delta$, the indirect CP asymmetry may become several order of magnitudes larger than the direct CP. This raises concerns about the validity of perturbation theory that breaks down in the degenerate limit $\Delta \to 0$. 

A more accurate analysis shows that the physics behind this apparent ill-defined situation is well known: the indirect CP asymmetry can be regarded to be the analogue of  mesonic states mixing, for example in the kaon system \cite{Lee:1957qq}, as it has been originally proposed in \cite{Flanz:1996fb}. Later on different approaches have been considered in order to regularize the  $\Delta \to 0$ limit and obtain a meaningful result for the indirect CP asymmetry. Among them, we mention briefly the one based on an effective LSZ-type formalism that aims at comprising both the mixing and the decays of heavy neutrinos \cite{Pilaftsis:1997jf,Pilaftsis:2003gt}. Indeed the approach is rather close to what we are going to present in chapter~\ref{chap:CPdege} in an effective field theory fashion.  The idea is that the heavy neutrino may undergo many interactions before decaying effectively into a lepton and Higgs boson pair. The neutrino with mass $M_1$ can turn into a neutrino with mass $M_2$ and back many times before decaying. One obtains, in the case of two heavy neutrinos, a 2$\times$2 propagator matrix: finite widths and mixing vertices are resummed in the expression of the neutrino propagators that can be safely used in constructing meaningful amplitudes. For example, the CP asymmetry induced in the decays of the neutrino of type~1 is found to be \cite{Pilaftsis:1997jf}
\be 
\epsilon_{1,{\rm{indirect}}} = \frac{ {\rm{Im}}\left[ \left( F_1^* F_2\right)^2\right] }{8 \pi|F_1|^2} \frac{M_1 M_2 (M_1^2-M_2^2)}{(M_1^2 -M_2^2)^2+M_1^2 \Gamma_2^2} \, ,
\label{reso_1}
\ee 
where the decay width for the neutrino of type~2 is defined as follows
\be 
\Gamma_2 = \frac{|F_2|^2M_2}{8 \pi} \, ,
\label{reso_2}
\ee
in complete analogy with the total width in (\ref{totwidth}) for the neutrino of type~1. In the limit $M_1 \to M_2$, namely $\Delta \to 0$, the quantity in (\ref{reso_1}) is regularized by the neutrino type~2 width. A similar expression holds for the indirect CP asymmetry in the neutrino type~2 decays, where the width regularizing the observable is $\Gamma_1$. 

The expression of the CP asymmetry given in (\ref{reso_1}) provides an interesting speculation: requiring the following condition 
\be 
M_1-M_2 \approx \frac{\Gamma_2}{2} \, ,
\label{reso_3}
\ee
the indirect CP asymmetry get resonantly enhanced and its expression reads
\be 
\epsilon_{1,{\rm{indirect}}} \approx \frac{1}{2} \frac{ {\rm{Im}}[ \left( F_1^* F_2\right)^2]}{|F_1|^2 |F_2^2|} \, .
\label{reso_4}
\ee
Thus, in the resonant case, the asymmetry is suppressed by neither the smallness of the light neutrino
masses, nor the smallness of their mass splitting, nor small ratios between the  heavy neutrino masses.
Actually, the CP asymmetry could be of order one, more precisely $\epsilon_{1,{\rm{indirect}}} \siml  1/2$, if we further require \cite{Pilaftsis:1997jf}
\be 
\frac{ {\rm{Im}}\left[ \left( F_1^* F_2\right)^2\right] }{|F_1|^2 |F_2|^2} \approx 1 \, .
\label{reso_5}
\ee
The fact that the asymmetry could be large, independently of the sterile neutrino masses, allows
for the possibility of low scale leptogenesis, down to the TeV scale (one at least requires sphalerons to be in equilibrium to reprocess the lepton asymmetry into a baryon one). Searches for the heavy neutrino states have been undertaken at the LHC \cite{ATLAS:2012ak,Khachatryan:2014dka,Khachatryan:2015gha} without positive result so far. Indeed the Yukawa couplings are pretty small due to the seesaw type-I scheme, see (\ref{lepto_9}). The parameter space can be also explored with indirect searches like the effect of low heavy neutrino mass state in rare decays \cite{Drewes:2015iva,Ibarra:2011xn,Abada:2014kba}, when the $\nu$MSM is considered. 

With resonant leptogenesis, the Boltzmann equations are different \cite{Pilaftsis:1997jf}. The densities of neutrino type~1 and type~2 are followed, since both contribute to the asymmetry. Moreover the relevant time scales are different with respect vanilla leptogenesis. For instance, the typical time scale to build up coherently the CP asymmetry is particularly long, of order $1/\Delta$ , and it can be larger than the time scale for the change of the abundance of the sterile neutrinos. This situation implies that for resonant leptogenesis quantum effects in the Boltzmann equations can be significant \cite{DeSimone:2007gkc,Cirigliano:2007hb,Garny:2011hg,Garbrecht:2011aw}. 

Finally different approaches lead to different results for the regulator in (\ref{reso_1}). These are reviewed and scrutinised in \cite{Garny:2011hg}, where the authors provide a first principle analysis of the CP asymmetries obtained form the Kadanoff-Beym equations. Similar derivations from first principle are found in \cite{Buchmuller:1997yu, Garbrecht:2011aw}. 

\section{Open challenges in thermal leptogenesis}
\label{sec_lepto4}
Leptogenesis takes place in the early universe: a hot and dense plasma made of thermalized SM particles sets the stage for the heavy neutrinos dynamics. On general grounds thermal effects are expected to play a role and a first quantitative study on the subject can be found in \cite{Giudice:2003jh}. The authors show how thermal  corrections affect  several ingredients in the analysis: coupling constants, particle propagators (of the SM particles) and CP asymmetries. 

Renormalization of gauge and Yukawa couplings in a thermal plasma is studied in \cite{Kajantie:1995dw}. In
practice, it is a good approximation to use the zero-temperature renormalization group equations for
the couplings, with a renormalization scale $\mu \sim 2 \pi T$ \cite{Giudice:2003jh}. The value $\mu > T$ is related to the fact that
the average energy of the colliding particles in the plasma is larger than the temperature (this can be related to the expressions of the thermal condensates of the Higgs, fermions and gauge bosons, that show powers of $\pi$).
In the thermal plasma, any particle with sizeable couplings to the background acquires a thermal mass
which is proportional to the plasma temperature \cite{Comelli:1996vm,Weldon:1982bn}. Consequently, decay and scattering rates get modified. Explicit expressions for the thermal masses that enter the relevant leptogenesis processes are collected in \cite{Giudice:2003jh}. The relevance of thermal masses depends on the temperature regime though. 

Let us consider the decays and inverse decays of the heavy neutrino with mass $M_1$ into a lepton and a Higgs boson. Since thermal corrections to the Higgs mass are particularly large,
$m_\phi(T ) \approx 0.4 \, T$,  decays and inverse decays become kinematically forbidden in the temperature range $m_\phi(T) - m_\ell (T) < M_1 <  m_\phi(T) + m_\ell (T)$. Rough estimates give the  kinematically forbidden range $2 \siml T/M_1 \siml 5$ \cite{Davidson:2008bu}. However we notice that if the heavy neutrino number density and
its $L$-violating reactions reach thermal equilibrium at $T \sim M_1$, any memory
of the specific conditions at higher temperatures is erased quite efficiently. Consequently, in the strong washout regime, these
corrections have practically no effect on the final value of the baryon asymmetry.

In temperature range $T < M_1$, which we call non-relativistic regime, thermal corrections to the heavy neutrino production rate have been addressed in \cite{Salvio:2011sf}. This is one of the key ingredients entering the Boltzmann equations (\ref{Boltz1}) and (\ref{Boltz2}). The authors provided a first NLO evaluation of the neutrino production rate. In particular they take into account radiative and thermal corrections to $\gamma_{\nu_{R,1}}$. The latter are shown to be of the form $g_{{\rm SM}} (T/M_1)^n$ for dimensional reasons. The leading thermal correction, proportional to the Higgs four-coupling, $\lambda$, has been evaluated in \cite{Salvio:2011sf}. Then the calculation has been further extended in \cite{Laine:2011pq} where the contributions proportional to the top-Yukawa coupling, $\lambda_t$, and the SU(2)$_L \times$U(1)$_Y$ gauge couplings, $g$ and $g'$, have been included. A two-loop calculation in a relativistic thermal field theory has been performed in both the derivations. In \cite{Biondini:2013xua} we used an effective field theory approach to describe the interaction between non-relativistic Majorana neutrino and SM particles at finite temperature, assuming $M_1 \gg T$. We simplify the derivation of NLO thermal corrections to the neutrino production rate over exploiting EFT techniques, namely implementing from the beginning the non-relativistic nature of the heavy neutrinos. This will be the subject of chapter~\ref{chap:part_prod}. 

We notice that without embedding the SM  couplings in the expression of the heavy neutrino production rate, the observable is taken as if it were in vacuum (a part from the thermal average performed on the heavy neutrino distribution density, see (\ref{prod_right})). Indeed without the NLO corrections the heavy neutrinos do not see the surrounding thermal plasma and they just decay into Higgs-lepton pair due to the Yukawa interaction. Providing thermal corrections as an expansion in the SM couplings allows an actual description of the right-handed neutrino in a heat bath. The right-handed neutrino production rate has been recently embedded in the rate equations for leptogenesis in the non-relativistic regime \cite{Bodeker:2013qaa}, where it is highlighted as well as in \cite{Salvio:2011sf}, that the analogue NLO expression for the CP asymmetry is still missing. The thermal production rate of right-handed neutrinos has been addressed in the relativistic and ultra-relativistic regime in \cite{Laine:2013lka}.  

CP asymmetries are the second key ingredient in the rate equations describing the evolution of the lepton asymmetry. They are expected to be affected by thermal corrections as well. Indeed in \cite{Giudice:2003jh,Covi:1997dr} a first attempt to generalize the CP asymmetries at finite temperature has been carried out in the framework of thermal field theory: the zero-temperature propagators were replaced  with their finite temperature
versions in the matrix elements of the Boltzmann equations. In so-doing the cuts on the one-loop amplitude generating the absorbative part get a temperature dependence because of the distribution function of the internal particles put on-shell. In \cite{Giudice:2003jh} the effect of thermal masses, for the Higgs and lepton doublets, has been also included in the evaluation of the CP asymmetry. A hierarchically ordered spectrum for the neutrino mass is considered in these works.

A novel approach to thermal leptogenesis has been recently developed in the contest of non-equilibrium quantum field theory \cite{Schwinger:1960qe,Keldysh:1964ud,Calzetta:1986cq}.  The formalism provides a first principle derivation of the classical Boltzmann equations that can be recovered from their quantum version known as Kadanoff--Baym equations \cite{Danielewicz:1982kk,Knoll:2001jx,Ivanov:1998nv,Weinstock:2005jw}. The latter are evolution equations for two-point functions in which a loop expansion can be performed in the close-time-path (CPT) formalism. In this contest thermal corrections to the CP asymmetries induced by the vertex and self-energy diagrams are studied in \cite{Garny:2009qn,Garny:2010nj,Garny:2009rv}. 
The corresponding results differ from the findings in previous works \cite{Giudice:2003jh,Covi:1997dr}. Thermal corrections obtained by substituting naively thermal propagators in the one-loop self-energy and vertex diagram in figure~\ref{fig:lepto_3} provides \cite{Giudice:2003jh,Covi:1997dr}
\be
\epsilon_1(T) = \epsilon_1^{T=0}(1-n_F(E_\ell)+n_B(E_\phi) -2n_F(E_\ell) n_B(E_\phi)) \, ,
\label{chall_1}
\ee 
whereas the derivation within CTP gives \cite{Garny:2009qn,Garny:2010nj,Garny:2009rv}
\be 
\epsilon_1(T) = \epsilon_1^{T=0}(1-n_F(E_\ell)+n_B(E_\phi) ) \, ,
\label{chall_2}
\ee
where $n_F(E_\ell)$ and $n_B(E_\phi)$  are the lepton and Higgs distribution functions and $\epsilon_1^{T=0}$ stands for the zero temperature CP asymmetry given in (\ref{CPself}) and (\ref{CPvert}). Some comments are in order. The authors of \cite{Covi:1997dr} noticed that there is a cancellation of the thermal contributions in their result. Indeed  if the same argument $E=E_\phi=E_\ell$ is kept in the distribution functions, it holds $n_B(E)-n_F(E)=2 \, n_B(E)n_F(E)$. Physically this cancellation can be understood as a compensation between stimulated emission and Pauli blocking. Only if the the Higgs boson and the lepton enter with the same energy, an exact cancellation holds. In \cite{Giudice:2003jh} this issue was solved by inserting the (different) thermal masses for the lepton and the Higgs boson. However the result in (\ref{chall_1}) is missing a contribution that is not included in the naive substitution of the thermal propagators in the $T=0$ topology of the one-loop diagrams shown in figure~\ref{fig:lepto_3}. The CTP formalism provides an additional term, exponentially suppressed, that cancels exactly the term quadratic in the distribution functions in (\ref{chall_1}). We discuss a similar issue in some detail in chapter \ref{chap:part_prod} for the heavy neutrino thermal width. Clearly the SM couplings are not comprised in the results (\ref{chall_1}) and (\ref{chall_2}), a part from their inclusion in the thermal masses, which well justified   in the regime $T \gg M_1$. 

A thermal treatment of the lepton-number asymmetry in the resonant case, 
i.e. when the mass difference of the heavy neutrinos is of the order of magnitude of their decay widths, 
can be found for instance in~\cite{Garny:2011hg}, where the Boltzmann equations 
are superseded by the Kadanoff--Baym equations.  
The lepton-number asymmetry has been also considered for a generic heavy neutrino mass spectrum, 
such as in~\cite{Garny:2010nj,Anisimov:2010dk,Kiessig:2011fw,Garbrecht:2011aw} within different approaches. 
The thermal effects considered include using thermal masses for the Higgs boson and leptons and 
taking into account thermal distributions for the Higgs boson and leptons as decay products of the heavy Majorana neutrinos.

In the following chapters, we aim at treating systematically thermal effects to the CP asymmetry in the non-relativistic regime, namely when the temperature of the plasma is smaller than the heavy neutrino mass scale. 
These effects lead to corrections in terms of series in the SM couplings and in $T/M$ 
in the same way as they do for the heavy Majorana neutrino production rate~\cite{Salvio:2011sf,Laine:2011pq}. The calculation is based on the EFT approach developed for and tested on the right-handed neutrino thermal production rate. To our knowledge a NLO treatment of the thermal correction to the CP asymmetry has been not presented.
We will derive such thermal corrections for the case of two Majorana neutrinos with nearly degenerate masses in chapter~\ref{chap:CPdege} , whereas we address the hierarchical case in chapter~\ref{chap:CPhiera}. Finally the results for flavoured CP asymmetries at finite temperature are shown in chapter~\ref{chap:CPfla}.

%% file: eff_the.tex
In this chapter the main concepts about effective field theories are introduced. In particular we are going to provide and discuss the topics and techniques necessary to understand the results given in chapters~\ref{chap:part_prod}-\ref{chap:CPfla}. In section  \ref{eff_sec_1} the main idea behind EFTs is presented together with the example of the Fermi effective interaction. A general strategy to construct an EFT starting from a given fundamental theory is provided in section \ref{eff_sec_2}, where we also show explicitly how to obtain a low-energy effective Lagrangian. Finally in section \ref{eff_sec_3} the heavy quark effective theory (HQEFT) is introduced, being the prototype for the development of the EFT describing heavy Majorana neutrinos interacting with light SM fields. 

\section{What is an EFT?}
\label{eff_sec_1}
Nature comes to us in many scales. We can think of galaxies, our Earth, molecules and nuclei that are very different sizes and held together with rather disparate binding energies. However it is true that we do not have to understand what happens at all scales at once in order to figure out how a physical system works at a particular scale of interest. 

For example, the derivation of the chemistry laws can be  traced back to the electromagnetic interactions. However it does not help much starting a quantitative analysis from the fundamental Quantum Electro Dynamics (QED) among quark and leptons. In order to understand the most relevant physics at the atomic scale, it will suffice a simpler description in terms of non-relativistic electrons orbiting around and bounded to the nucleus through a Coulomb potential. Thus, at good approximation, the behaviour of chemical elements can be understood in terms of the electron mass and the fine structure constant $\alpha \approx 1/137$, whereas the proton mass is a higher energy scale necessary to assess the possible relevant corrections. 

In order to study a particular system it is necessary to single out the most relevant degrees of freedom which are the building blocks to attain a simple description of the problem at hand.  It is crucial to make the appropriate choice of these variables capable to capture the most important effects at a given scale. The degrees of freedom that become relevant at any higher energy scale are not taken into account and do not appear explicitly in the formulation of the theory. At this point we mention another useful example for the discussion. A heavy particle cannot be created at an energy scale smaller
than its mass, $M$. Therefore a theory, and its corresponding Lagrangian, valid at such small energies does
not contain this degree of freedom. This is rigorously  ensured by
the decoupling theorem proved by Appelquist and Carazzone \cite{Appelquist:1974tg}, who showed
that heavy degrees of freedom actually decouple
at energy scales much lower than their mass. In this respect, decoupling means that any
effect of the heavy degrees of freedom is, up to logarithmic contributions, suppressed by inverse powers of the heavy scale $M$. In the strict heavy mass limit, $M \to \infty$, the heavy state does not provide any correction.
\begin{figure}[t]
\centering
\includegraphics[scale=0.60]{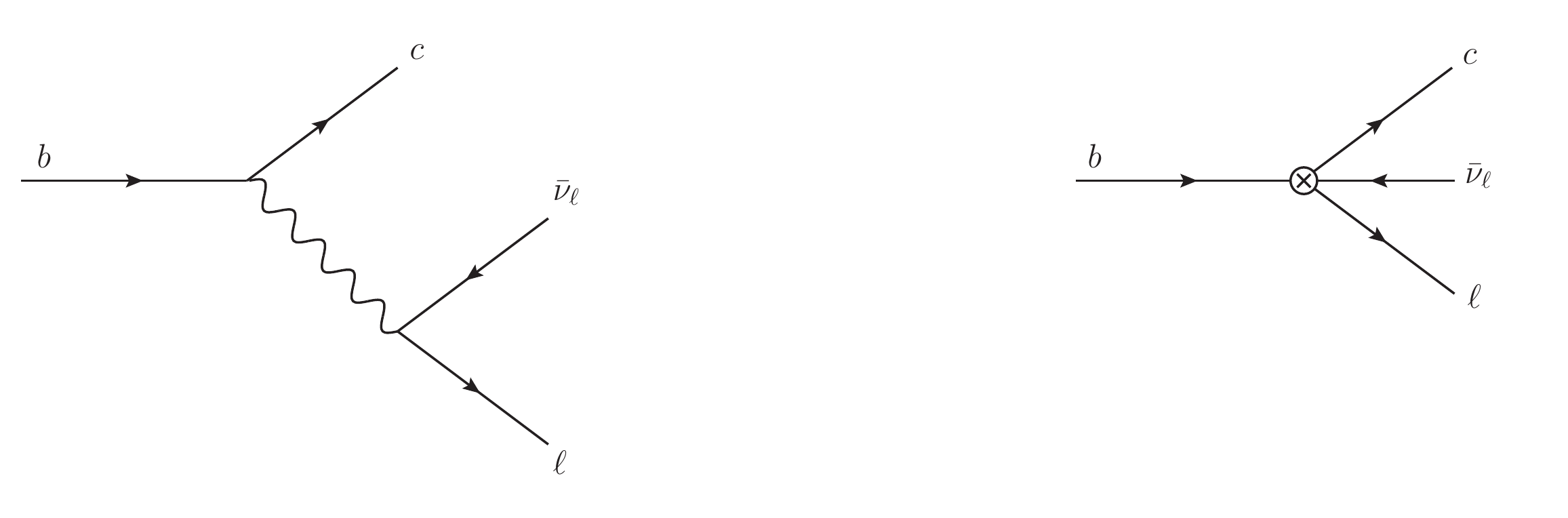}
\caption{\label{fig:eft_1} The diagrams describing the process $ b \to c \ell \bar{\nu}_{\ell}$ are shown both in the fundamental theory  (left) and in the effective theory (right). Wiggled line stands for the $W$ boson, which is shrunk into a point, the crossed vertex, in the four-particle effective interaction.}
\end{figure}

Before going to the principle of construction of an EFT, let us briefly discuss the effective four-fermion interaction mediated by the $W$ boson. This will serve to illustrate some of the main points and it is also a relevant example in the realm of particle physics. Let us consider the decay $b \to c \ell^- \bar{\nu}_{\ell} $, shown in figure \ref{fig:eft_1} (on the left). We can write the process amplitude exploiting the full SM Lagrangian that comprises, at tree level, two left-handed fermion currents and the $W$ boson propagator. Let us look at the energy scales appearing in the problem. They are the $b$ quark and $c$ quark masses, $m_b$ and $m_c$, the lepton masses which are already negligible with respect to the heavy quark masses, the $W$ boson mass, $M_W$, and the maximal momentum transfer $q_{\rm{max}}^2=(m_b-m_c)^2$. Given $m_b \approx 5$ GeV, $m_c \approx 1.3$ GeV and $M_W \approx 80$ GeV , we can perform an expansion for the $W$ boson propagator in momentum space (Feynman gauge) 
\be 
\frac{1}{q^2-M_W^2} = -\frac{1}{M^2_W} \left( 1 + \frac{q^2}{M_W^2} + \cdots \right) \, , 
\label{eft_1} 
\ee 
being the momentum transfer $q \leq q_{\rm{max}} \ll M_W$, and the dots stand for higher order terms in $q/M_W$. This corresponds to an expansion of the $W$ propagator into local terms as follows
\bea 
\bra 0 | T \left( W^{+}_{\mu}(x)W^{-}_{\nu}(0) \right) | \ket0  &=& -\int \frac{d^4q}{(2\pi)^4} \frac{i \, g_{\mu \nu}}{q^2 -M^2_W} \, e^{-i q \cdot x} \nn \\
&=& \frac{i \, g_{\mu \nu}}{M_W^2} \left( 1 - \frac{\pa^2}{M_W^2} + \cdots \right) \delta^4(x) \, . 
\label{eft_2}
\eea
The above expansion has a simple physical interpretation in terms of distances and interaction ranges: the distance scale of the $W$ boson propagation is of order $1/M_W$ which is seen as local at typical distances of order $1/q$ (by assumption $1/q \gg 1/M_W$). In summary, instead of using the full SM Lagrangian, one can exploit the leading term of the expansion in inverse powers of the $W$ boson mass to effectively describe the transition $b \to c \ell \bar{\nu}_{\ell} $. The corresponding Lagrangian reads  
\be 
\mathcal{L}_{{\hbox{\tiny Fermi}}}= -\frac{g^2}{2 M_W^2} \left( \bar{b} \gamma_{\mu} P_L c \right) \left( \bar{\nu}_{\ell} \gamma^{\mu} P_L \ell \right) + h.c. \, ,
\label{eft_3}
\ee
where $g$ is the coupling of the SU(2)$_L$ gauge group and $P_L=(1-\gamma^5)/2$ is the left-handed projector. In this way the well-known Fermi interaction is recovered, which is the leading term in the systematic expansion shown in (\ref{eft_2}).

Clearly the typical energies involved in the transition are of the order of the $m_b$ (if we further take $m_c \ll m_b$), so that there is not enough energy to produce a real $W$ boson which is not included as a dynamical field in the low-energy theory (\ref{eft_3}). The weak decay and the corresponding transition amplitude are well described by an effective interaction shown in figure \ref{fig:eft_1} (on the right) and induced by the dimension-six operators in (\ref{eft_3}). We notice the appearance of an effective coupling with mass dimension $-2$, which can be determined by requiring that the low-energy Lagrangian, $\mathcal{L}_{{\rm{Fermi}}}$, provides the same physical result of the full SM theory in the low-momentum region, $q \ll M_W$. A systematic improvement of the effective Lagrangian (\ref{eft_3}) is possible.  According to the propagator expansion in (\ref{eft_2}), one would obtain higher order fermionic operators that improve the accuracy at relative order $(q/M_W)^2$ and so on and so forth. 

Having introduced some relevant aspects when dealing with low-energy effective theories, we move now to a more detailed discussion on how to refine the ideas discussed so far.

\section{Principles of construction}
\label{eff_sec_2}
The starting point for the construction of any EFT is the presence of separated energy scales, at least two. The main point is that the physics at a given scale does not sensibly depend on the details of the physics at the other higher scales. Then one has to identify the parameters of the system that are very smaller or bigger than the relevant scale of interest, and put them to zero and to infinity respectively. Despite this sounds a sensible approximation to treat the problem, a systematic improvement is possible in terms of corrections induced by the higher energy scales neglected at first. The following discussion is based on \cite{Pich:1998xt}. 

Let us take the Lagrangian that describes a given physical system and we label it simply as $\mathcal{L}$. We assume that in the theory described by $\mathcal{L}$ there is a separation of the energy and momentum scales, $m$ and $M$, such that $M \gg m$ (the large and small scales do not have to be necessarily parameters of the fundamental Lagrangian, $\mathcal{L}$). Say that we are interested in the physics at the scale $m$, which is the small scale in the problem at hand. From the fundamental theory, we want to extract a second theory  valid only at low energies, namely that describes degrees of freedom with typical energy and momentum of order $m$.  Of course the low-energy theory, with Lagrangian $\mathcal{L}_{\hbox{\tiny EFT}}$, has the same infrared (IR) behaviour of the fundamental theory (but a different ultraviolet (UV) one). It is useful to introduce a cutoff scale, $\Lambda$, that enables us to separate the high energy modes of order $M$ from those of order $m$: $M \gg \Lambda \gg m$. The EFT comprises and describes the dynamics of degrees of freedom with typical energies smaller than $\Lambda$ and the higher energy modes are said to be \textit{integrated out} from the theory. Indeed they do not appear explicitly in the low-energy theory. 

We write the EFT Lagrangian and explain the procedure to specify its field content and parameters. It reads
\be 
\mathcal{L}_{\hbox{\tiny EFT}}=\sum_{i} \,c_{i}\left( \frac{\Lambda}{M}\right) \frac{\mathcal{O}^{(d_i)}_i(\Lambda,m)}{M^{d_i-4}}  \, .
\label{eft_4}
\ee 
The Lagrangian is organized in terms of operators, $\mathcal{O}_i$, of arbitrary dimension, $d_i$, that develop a dependence on the low-energy scale $m$ and on the cutoff scale $\Lambda$. The latter cancels against the dependence of the matching coefficients, $c_i$, on the very same scale. Indeed the cutoff scale cannot appear in the observables obtained from (\ref{eft_4}), being $\Lambda$ an auxiliary scale introduced for the construction of the EFT. The matching coefficients are also called Wilson coefficients and they include the contributions from the high energy and momentum modes of order $M$. 

The effective operators are constructed with the fields that are still dynamical at the scale $m$ and the operator dimension fixes the corresponding relative importance at low energies. We can distinguish three types of operators: \textit{relevant} ($d_i < 4$ ), \textit{marginal} ($d_i = 4$) and \textit{irrelevant} ($d_i > 4$). The definition refers to their behaviour at small energies. In our example of the Fermi four-fermion interaction in (\ref{eft_3}), dimension-six operators appear and their effects scale as powers of $q/M_W$.  Therefore they are small at small energies but this does not mean they are not important: the dimension-six operators in (\ref{eft_3}) give the leading contribution to describe weak decays at energies smaller than the electroweak scale. 

We now list the main steps to build the EFT in eq.~(\ref{eft_4}):
\begin{itemize}
\item[1)] identify the hierarchy of scales in the physical system ($M \gg m$) and the corresponding high- and low-energy degrees of freedom. The latter will be the field content of the low-energy Lagrangian;
\item[2)] constrain the from of the effective Lagrangian by symmetry arguments. The symmetry set of the fundamental Lagrangian, $\mathcal{L}$, can be entirely or only in part present in the $\mathcal{L}_{\hbox{\tiny EFT}}$. One has to consider only the symmetry sub-set that corresponds to the low-energy theory;
\item[3)] write down the most general Lagrangian as in eq.~(\ref{eft_4}) made of the light degrees of freedom and satisfying the symmetries specified in the previous two points. If the large scale $M$ corresponds to the mass of a heavy particle, then the string of operators in eq.~(\ref{eft_4}) stands for local interactions between the low-energy degrees of freedom (as the Fermi interaction in (\ref{eft_3}));
\item[4)] the EFT comes with a power counting, namely the ratio $m/M$ induced by the form of the Lagrangian (\ref{eft_4}). This clearly defines the number of operators with increasing dimension to be included in the low-energy Lagrangian to achieve the desired accuracy of a given physical observable, say at order $(m/M)^n$. One has to include a finite number of operators and parameters, the Wilson coefficients $c_i$, and hence only a finite number of divergent amplitudes can appear due to the series truncation. The EFT is therefore \textit{renormalizable order by order};
\item[5)] determine the parameters of the EFT Lagrangian through the \textit{matching}. This procedure provides the explicit form of the coefficients in (\ref{eft_4}) and then the EFT is ready to use.  Basically one has to evaluate Green's functions both in the fundamental theory and in the EFT, and match the two sets so obtained at energies smaller than the cutoff scale $\Lambda$, where $M \gg \Lambda \gg m$. Indeed the EFT is expected to reproduce exactly the fundamental theory in the low-energy domain, in which the two quantum field theories have to provide the same physical results.   
\end{itemize}
The points listed above provide a general scheme to obtain an EFT starting from a given fundamental theory, the latter being valid for a wider range of energy scales. On the other hand the EFT is valid in the low-energy domain and it is a simpler theory suited to address observables in its range of applicability. We notice that what we called fundamental theory can in turn be itself an EFT and the given procedure can be iterated. In this case a tower of EFTs can be obtained (we discuss such case in chapter \ref{chap:CPhiera}). 

Some more comments are in order regarding the matching procedure. This is the most technical and involved step from the computational point of view. As already mentioned Green's functions are calculated in the fundamental theory and matched to those obtained in the EFT. The matching is organized order by order in the expansion parameter, namely $m/M$ in our notation. This expansion is performed in the fundamental theory side and eventually matches the operator expansion on the EFT side. Because the matching occurs at a scale much smaller than $M$, any external momenta $q_j$ of order $m$ allow  for an expansion in powers of $q_j/M$ in the Green's functions of the fundamental theory. 

A regularization scheme has to be adopted for dealing with divergent amplitudes, since loop diagrams may enter the Green's functions exploited in the matching. A useful and common choice is Dimensional Regularization (DR) \cite{'tHooft:1972fi}. In DR scaleless integrals vanish by construction and this turns out to greatly simplify the matching procedure. Indeed any lower scale appearing in the physical system can be set to zero in the matching, that is realized at energy smaller than the cutoff $\Lambda$. Being $m$ the natural scale of the low-energy theory, all loop diagrams on the EFT side vanish because $m \to 0$ and they effectively become scaleless.  
Due to the presence of loop amplitudes, UV divergences appear in the fundamental theory and they are accounted for standard renormalization procedure. The main consequence is a possible induced dependence on the renormalization scale $\mu$ for the Wilson coefficients, typically through logarithms of $\mu / M$. This way the matching coefficients contain the effects of the high energy scale and degrees of freedom. Eventually the scale $\mu$ does not appear in the observables. 

We conclude this section by mentioning that EFTs have been widely used in different contexts, from very small systems in particle physics, nuclear and atomic physics up to the description of the largest structures in our universe. The literature on the topic is really vast and we quote here few examples about particle physics~\cite{Pich:1998xt,Manohar:1996cq, Kaplan:1995uv, Meissner:1993ah, Caswell:1985ui}.  In this thesis we are going to show a novel application of the EFT approach to treat the dynamics of heavy Majorana neutrinos in a thermal bath (see chapters \ref{chap:part_prod}-\ref{chap:CPfla}).  

\subsection{An example for a matching calculation: a heavy scalar particle}
\begin{figure}[t]
\centering
\includegraphics[scale=0.6]{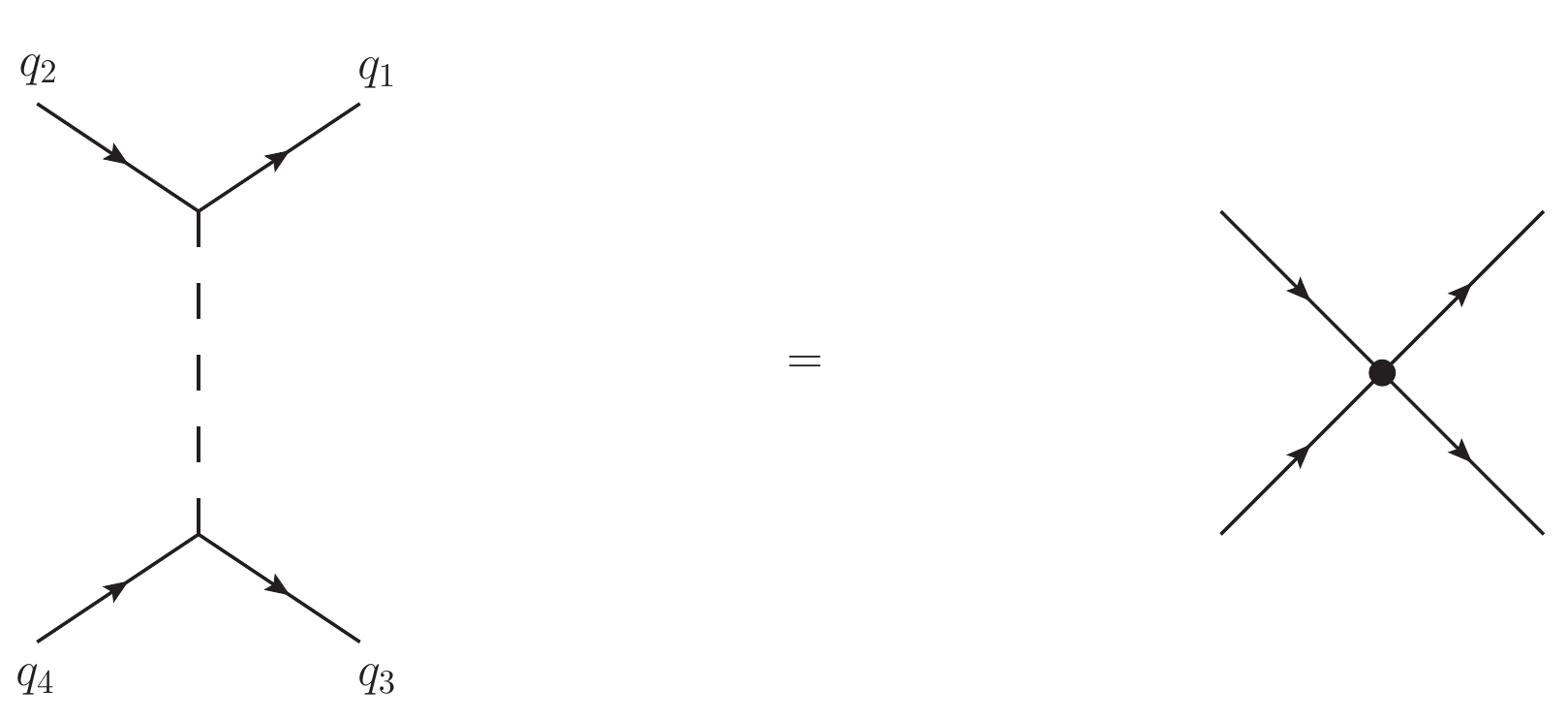}
\caption{\label{fig:eft_2} Tree level diagram in the full theory (left-hand side of the equality) contributing to the matching of the dimension-six operator in the EFT (right-hand side of the equality). Solid lines stand for fermions, whereas the dashed line for the heavy scalar particle. The effective vertex is shown in a black dot.}
\end{figure}
In this section we shall consider a simple example in order to illustrate the general procedure to obtain an EFT. Moreover some notation adopted throughout the thesis is introduced. Let us consider a fermion and its antifermion with mass $m$ described by the field $\psi$, and a scalar particle with mass $M$ described by the real field $\varphi$. We allow them to be interacting fields and the corresponding Lagrangian reads
\be 
\mathcal{L}= \frac{1}{2} (\pa \varphi)^2 -\frac{1}{2}M^2 \varphi^2 + \bar{\psi} i \slashed{\pa} \psi -m\bar{\psi}\psi + \tilde{g}   \varphi \bar{\psi} \psi \, .
\label{eft_5}
\ee
Despite the very simple field and interaction content in eq.~(\ref{eft_5}), we can already practice the EFT language. The two kinetic terms and the interaction term are marginal operators, indeed of dimension four. The dimensionless coupling constant $\tilde{g}$ is responsible for the Yukawa interaction between fermions and the scalar. On the other hand the mass terms are relevant operators. Their coefficients are of dimension two, $\left[ M^2 \right] =2$, and one respectively, $\left[m\right] = 1$.

Let us now consider the process of a fermion-fermion scattering in this theory: $f \, f \to f \, f$, shown in the left-hand side of the equality in figure \ref{fig:eft_2}. How does the same process appear at typical energies of order $m \ll M$? We can guess by analogy with the weak decays discussed in section~\ref{eff_sec_1} that we are going to obtain a local four-fermion interaction as well. In this low-energy theory the heavy scalar does not appear as a physical degree of freedom and we can only have light fermions with energies smaller than $M$. Our aim is to construct explicitly the EFT for this toy model and show the techniques used throughout the thesis. The EFT Lagrangian reads (see \ref{eft_4})
\be 
\mathcal{L}_{\hbox{\tiny EFT}}= \bar{\psi} i \slashed{\pa} \psi -m\bar{\psi}\psi + \frac{\tilde{c}}{M^2} (\bar{\psi}\psi) (\bar{\psi}\psi) + \cdots
\label{eft_6}
\ee
The dots stand for higher order operators with higher dimensions, $\tilde{c}$ is the matching coefficient and $\psi$ describes fermions with energy smaller than $M$. The heavy scalar particle is not explicitly present in the theory, since  energy and momenta of order $M$ are integrated out. The effects of the heavy scalar are embedded into a local interaction, namely the dimension-six operator in (\ref{eft_6}). It corresponds to the leading term of the operator string we can consider in the EFT and it is an irrelevant operator because its dimension is $d=6$. 

We want to describe the $ff \to ff$ scattering at tree level, therefore the asymptotic states are four fermions and the corresponding Green's function we need for the matching reads
\be 
-i  \int d^{4}x\,e^{i q_1 \cdot x} \int d^{4}y \int d^{4}z\,e^{i q_3 \cdot y} e^{-iq_4 \cdot z}\, 
\langle \Omega | T(\psi^{\mu}(x) \bar{\psi}^{\nu }(0) \psi^{\alpha}(y) \bar{\psi}^{\beta}(z) )| \Omega \rangle ,
\label{eft_7}
\ee
where $| \Omega \rangle$ is the ground state of the fundamental theory. Then $\mu$, $\nu$, $\alpha$ and $\beta$ are Lorentz indices and $q_1$, $q_2$, $q_3$ and  $q_4$ are the external momenta carried by the scattering fermions. Here and in the rest of the thesis we consider for the diagrammatic counterpart of the Green's functions the quantity $-i\mathcal{D}$, where $\mathcal{D}$ is a generic Feynman diagram amputated of the external legs. In order to match the Green's function in eq.~(\ref{eft_7}), one has to evaluate it in the fundamental theory (\ref{eft_5}) and then expand the result in powers of $q_j/M$. Indeed we want to device the low-energy theory where particles carry energies and momenta $q_j \sim m \ll M$. We focus here on a tree level matching and discard loop amplitudes. The tree-level diagram in the fundamental theory for the  $f \,f \to f \, f $ scattering is shown in figure \ref{fig:eft_2}, and one obtains the following result 
\bea
&&-i  \int d^{4}x\,e^{i q_1 \cdot x} \int d^{4}y \int d^{4}z\,e^{i q_3 \cdot y} e^{-iq_4 \cdot z}\, 
\langle \Omega | T(\psi^{\mu}(x) \bar{\psi}^{\nu }(0) \psi^{\alpha}(y) \bar{\psi}^{\beta}(z) )| \Omega \rangle \nn \\
&&= -\frac{\tilde{g}^2}{(q_1-q_2)^2-M^2} \delta^{\mu \nu} \delta^{\alpha \beta} = \frac{\tilde{g}^2}{M^2}  \delta^{\mu \nu} \delta^{\alpha \beta} + \cdots \, ,
\label{eft_8}
\eea
where we retain only first term in the expansion $(q_1-q_2)/M \ll 1$. To keep the notation simple we drop, here and in the rest of the thesis, propagators on external legs and we label to so-obtained amputated Green's function with the same indices as the unamputated ones. 

As regards the EFT we have to evaluate the diagram on the right-hand side of the equality shown in figure \ref{fig:eft_2} by using the Lagrangian (\ref{eft_6}). The result reads
\bea
&&-i  \int d^{4}x\,e^{i q_1 \cdot x} \int d^{4}y \int d^{4}z\,e^{i q_3 \cdot y} e^{-iq_4 \cdot z}\, 
\langle \Omega | T(\psi^{\mu}(x) \bar{\psi}^{\nu }(0) \psi^{\alpha}(y) \bar{\psi}^{\beta}(z) )| \Omega \rangle \nn \\
&&= \frac{\tilde{c}}{M^2} \delta^{\mu \nu} \delta^{\alpha \beta}  \, .
\label{eft_9}
\eea
Comparing eqs.~(\ref{eft_8}) and (\ref{eft_9}), namely matching the Green's functions, we obtain the Wilson coefficient of the dimension-six operator: $\tilde{c}=\tilde{g}^2$. This enables to use the low-energy theory instead of the fundamental theory, the former being exactly equivalent to the latter at order $(q_j/M)^0$ (this is the order at which we worked here).  

Of course we can include higher order operators in the low-energy Lagrangian, that contain derivatives acting on the fermion fields. Accordingly one has to go further in the expansion in powers of $q_j/M$ shown in (\ref{eft_8}), in order to match the new terms induced by the additional higher order operators on the EFT side. The accuracy of the low-energy Lagrangian is systematically improved this way.
  
\section{An EFT prototype for heavy particles: the HQEFT} 
\label{eff_sec_3}
\begin{figure}[t]
\centering
\includegraphics[scale=0.7]{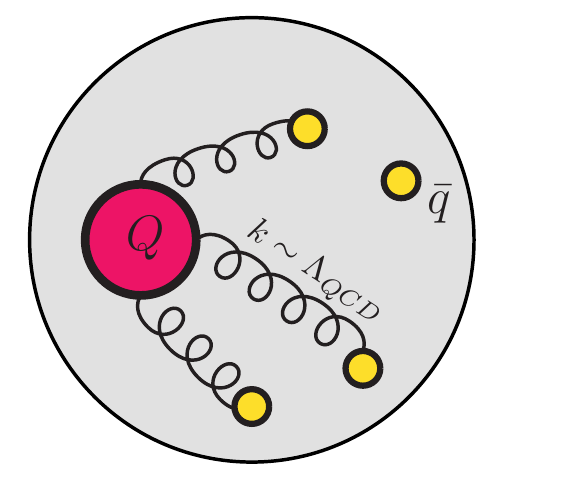}
\caption{\label{fig:eft_3} Schematic representation of a heavy-light meson, with valence quark $Q$ and antiquark $\bar{q}$. The typical momentum transfer between the heavy quark and light partons (yellow dots) in the bound state is of order $\Lambda_{{\rm{QCD}}}$.}
\end{figure}
In the following chapters we are going to discuss the dynamics of heavy Majorana neutrinos in a thermal bath of SM particles. In particular we want to study the medium induced modifications on the neutrino width and on the CP asymmetries generated in heavy-neutrino decays. It is conceivable that the generation of the CP asymmetry takes place when the Majorana neutrinos are non-relativistic, namely when their typical three-momentum is much smaller than their mass. In this case, at variance with the examples discussed so far, the heavy particle is a fundamental ingredient of the low-energy theory: its non-relativistic (low-energy) excitations participate the dynamics by interacting with the light fields. Here we discuss the original formulation for such EFT: the heavy quark effective theory (HQEFT) \cite{Isgur:1989vq,Isgur:1989ed, Shifman:1987rj,Falk:1990yz,Grinstein:1990mj}, that provides the prototype for developing the EFT for non-relativistic Majorana fermions.  

The physical system described by the HQEFT is a heavy-light meson, a color singlet state made of a quark and an antiquark bound by the non-perturbative gluon dynamics. In particular the up, down and strange quarks are understood as light quarks, whereas the charm and bottom quarks are taken as heavy (the top quark does not provide bound states). This distinction is also based on the comparison between quark masses and the dynamically generated scale in QCD, that is $\Lambda_{{\rm{QCD}}}$. 
Let us label such heavy-light meson with $Q\bar{q}$, assuming the bound state made of a valence heavy quark $Q$ and light antiquark $\bar{q}$ with mass $m_{Q} \gg \Lambda_{{\rm{QCD}}}$ and $m_q \ll \Lambda_{{\rm{QCD}}}$ respectively. We give a simplified representation of such system in figure \ref{fig:eft_3}. The typical size of such bound state is of order $1/ \Lambda_{{\rm{QCD}}}$, which is also the natural momentum transfer between the heavy and light partons in the meson.

An important consequence of the comparison among typical momentum transfer, sizes and mass scales in the heavy-light meson is the following: the velocity of the heavy quark, $v$, is almost unchanged by the strong interaction dynamics in the bound state. Assuming the momentum of the heavy quark to be $m_{Q}v$, the effect of the interactions with light partons mediated by the binding gluons can amount at changing the heavy quark momentum by $\Delta p \approx \Lambda_{{\rm{QCD}}}$, and hence $\Delta v \approx \Lambda_{{\rm{QCD}}}/m_{Q}$. Moreover the heavy quark is a non-relativistic object in the heavy-light meson because its mass is bigger than the three-momentum of order $\Lambda_{{\rm{QCD}}}$. It holds $v < 1$ and benchmark values are $v \sim 0.3 (0.5)$  for  beauty (charm) mesons.

Our aim is to show the field content of the EFT that describes the physics at energies much smaller than the heavy-quark mass. The procedure to obtain such low-energy theory is the same discussed in the previous sections. However, in this case, the fundamental theory is a more involved one, namely QCD.  It contains positive powers of the heavy-quark mass, and it reads
\be 
\mathcal{L}_{{\rm{QCD}}}=-\frac{1}{4} F^{a \, \mu \nu} F^a_{\mu \nu} + \bar{Q} (i \slashed{D}-m_{Q}) Q +  \bar{q}_i (i \slashed{D}-m_{q_i}) q_i \, ,
\label{eft_hq_1}
\ee 
where counterterms are understood and as well as the sum over the light quarks index $i$. The first term is the Yang-Mills sector describing the gluons, $a$ is the color index, $\mu$ and $\nu$ are Lorentz indices.  Regards the fermion sector we distinguish the heavy quark term from the light quarks one. Colour, Lorentz  and flavour indices are understood in the quark terms.  The covariant derivative is $D_{\mu}=\pa_{\mu}+ i g_s A_{\mu}^{a} T^{a}$, where $T^{a}$ are the SU(3) generators in the adjoint representation, $g_s$ is the strong coupling constant and $A_{\mu}^{a}$ are the gluon fields. The effective theory is constructed by making sure that Green's functions in the effective theory are equal to those in QCD at a given order in $1/m_Q$ and  $\alpha_s=g_s^2/(4 \pi)$.

\subsection{The HQEFT Lagrangian}
In order to construct the HQEFT we have to integrate out energy modes of order $m_{Q}$. We know from (\ref{eft_4}) that positive powers of $m_{Q}$ are not present anymore in the EFT Lagrangian, at variance with the QCD Lagrangian (\ref{eft_hq_1}). We proceed as follows to derive the effective Lagrangian in the heavy mass limit for a non-relativistic heavy quark in the heavy-light meson. We consider an off-shell heavy quark interacting with the surrounding light fields. Its momentum can be written as $p^{\mu}=m_Qv^{\mu}+k^{\mu}$, with $v^2=1$. In particular the four-momentum $k$ determines the amounts of which the heavy quark is off-shell due to the interactions. We call it residual momentum and it is of order $\Lambda_{{\rm{QCD}}}$ by construction. This condition allows for a simplification of the Dirac heavy quark propagator
\be 
\bra0 | T (Q(x)\bar{Q}(0)) | \ket 0 = \int \frac{d^4p}{(2 \pi)^4} \frac{i(\slashed{p}+m_Q)}{p^2-m_Q^2+i \eta}  e^{-ip \cdot x} \, .
\label{eft_hq_2}
\ee   
By taking the heavy quark propagator in momentum space and substituting the parametrization $p^{\mu}=Mv^{\mu}+k^{\mu}$, we obtain for $k \sim \Lambda_{{\rm{QCD}}} \ll m_Q$
\be 
\frac{i(\slashed{p}+m_Q)}{p^2-m_Q^2+i \eta}  = \frac{i(m_Q \slashed{v} + m_Q +\slashed{k})}{2 m_Q \, v \cdot k + k^2 +i \eta } =  \frac{1+\slashed{v}}{2} \frac{i}{v \cdot k + i \eta} + \mathcal{O} \left( \frac{1}{m_Q} \right) \, ,
\label{eft_hq_3}
\ee
where higher order corrections besides the leading term are not shown. They vanish in the strict heavy mass limit, $m_Q \to \infty$. We notice that the heavy quark propagator contains a velocity-dependent projector
\be 
\hat{P} \equiv  \frac{1+\slashed{v}}{2} \, ,
\label{eft_hq_4}
\ee  
that is also called non-relativistic projector and in the heavy-quark rest frame $v^{\mu}=(1,\bm{0})$ it becomes $\hat{P}=(1+\gamma^0)/2$. 

The original heavy quark field, $Q(x)$, can be decomposed into a large component $H(x)$, whose energy is of order $m_Q$ and a small component $h(x)$, whose energy is much smaller than $m_Q$, by using the non-relativistic projectors 
\be 
Q(x)= \frac{1+\slashed{v}}{2} Q(x) +  \frac{1-\slashed{v}}{2} Q(x) = h(x) + H(x) \, ,
\label{eft_hq_5}
\ee
where $\check{P}=(1-\slashed{v})/2$ complete the basis of the velocity projectors. 
The small component field, $h(x)$, is the degree of freedom that remains dynamical in the low-energy theory, whereas the field $H(x)$ is integrated out and it will not appear in the EFT. Then $h(x)$ is the field, made of two independent components, that describes in the HQEFT Lagrangian the low-energy modes of the heavy quark. Moreover it annihilates a heavy quark but \textit{does not create an antiquark}. The field $h(x)$ satisfies 
\be 
\frac{1+\slashed{v}}{2} h(x) = h(x) \, ,
\label{eft_hq_6}
\ee
and the equal time anti-commutation relations \cite{Dugan:1991ak} 
\begin{eqnarray}
\left\lbrace h^{\alpha}(t, \bm{x}) ,h^{\beta}(t, \bm{y}) \right\rbrace &=& 
\left\lbrace \bar{h}^{\alpha }(t, \bm{x}) , \bar{h}^{\beta }(t, \bm{y}) \right\rbrace = 0 \,,
\label{eft_hq_7}
\\
\left\lbrace h^{\alpha}(t, \bm{x}), \bar{h}^{\beta }(t, \bm{y}) \right\rbrace &=&  
\frac{1}{v^0} \left( \frac{1+\slashed{v}}{2} \right)^{\alpha \beta} \delta^{3}(\bm{x}-\bm{y}) \,.
\label{eft_hq_8}
\end{eqnarray}
From the expansion of the full propagator in (\ref{eft_hq_3}) and the discussion on the low-energy degree of freedom for the heavy quark, we write the HQEFT Lagrangian at leading order in the $1/m_Q$ expansion, namely in the static limit, as follows
\be 
\mathcal{L}^{(0)}_{\hbox{\tiny HQEFT}}= \bar{h} \,  (i v \cdot D)  \, h \, .
\label{eft_hq_9}
\ee
Due to the presence of a covariant derivative in (\ref{eft_hq_9}), we can also specify the heavy quark-gluon vertex at tree level to be $-ig_s T^a v^\mu$ instead of $-ig_s T^a \gamma^\mu$. This change is due to the sandwich of the gamma matrix between two propagators of the non-relativistic heavy quarks, giving $\hat{P} \gamma^\mu \hat{P}=v^\mu \hat{P}$.  

An analogous derivation can be made by defining the original heavy quark field as follows
\be 
Q(x)= e^{-i m_Q \, v \cdot x} \left[ h(x) + H(x) \right] \, , 
\label{eft_hq_10}
\ee 
where the exponential prefactor has the effect of subtracting the quantity $m_Q v^{\mu}$ from the heavy quark momentum. Substituting the decomposition (\ref{eft_hq_10}) for $Q(x)$ into the QCD Lagrangian (\ref{eft_hq_1}) and neglecting any effect of the large component $H(x)$ one gets back the Lagrangian (\ref{eft_hq_9}). However this last procedure allows for considering the higher order corrections in the heavy quark mass expansion. It is beyond the present discussion to derive those corrections and we refer to \cite{textMan} for details. Only the strategy is outlined here. If one keeps the large component field $H(x)$ when plugging the decomposition (\ref{eft_hq_10}) into the Lagrangian (\ref{eft_hq_1}), the HQEFT Lagrangian is found to be
\be 
\mathcal{L}_{\hbox{\tiny HQEFT}}= \bar{h} \,  (i v \cdot D)  \, h - \bar{H}( i v \cdot D + 2 m_Q )H + \bar{h} \, i \slashed{D} \, H + \bar{H}  \, i \slashed{D} \,  h \, .
\label{eft_hq_11}
\ee 
One may see that the large component field, $H(x)$, describes the heavy quark field excitations of order $m_Q$, whereas the small component, $h(x)$, does not. It is useful to define the perpendicular component of the covariant derivative with respect to the four velocity, 
$D^{\mu}_{\perp} = D^{\mu} - (v \cdot D) v^{\mu}$. This provides the substitutions  $\slashed{D} \to \slashed{D}_{\perp}$ in (\ref{eft_hq_11}). Since the field $H(x)$ corresponds to quantum excitations of order $m_Q$, it can be integrated out when the assumptions of the HQEFT are valid. This is done by solving the equations of motion for $H(x)$ derived form the Lagrangian (\ref{eft_hq_11}) so that the field $H(x)$ can be eliminated in favour of $h(x)$. Finally the correction at order $1/m_Q$ to the free HQEFT Lagrangian (\ref{eft_hq_9}) reads
\be 
\mathcal{L}^{(1)}_{\hbox{\tiny HQEFT}}=-\bar{h} \, \frac{D^2_{\perp}}{2 m_Q} \, h -g_s \, \bar{h} \, \frac{\sigma_{\mu \nu} F^{\mu \nu}}{4 m_Q} \, h \, ,
\label{eft_hq_12}
\ee
where $\sigma_{\mu \nu}=i[\gamma_{\mu}, \gamma_\nu]/2$ and $F^{\mu \nu}=[D^{\mu},D^{\nu}]$ is the field strength tensor. The first term in eq.~(\ref{eft_hq_12}) is the heavy-quark kinetic energy and the second term is the magnetic moment interaction, the latter describes the interaction between the heavy quark and the gluons carrying energy and momenta of order $\Lambda_{{\rm{QCD}}} \ll m_Q$. The operators in eq.~(\ref{eft_hq_12})  are of dimension five and hence suppressed in one power of the high energy scale $m_Q$ whose corresponding energy modes have been integrated out. 
\subsection{Concluding remarks}

The formalism shown with the example of the HQEFT is quite general. It applies every time we want to keep non-relativistic excitations of a heavy particle in the low-energy theory.  What do we mean exactly for a heavy particle in the EFT framework?  We call a particle heavy if its mass, $M$,
is much larger than any other scale, $E$, characterizing the system. The scale $E$ may include the spatial momentum of the heavy particle, 
scales that appear from dimensional transmutation like $\Lambda_{\rm{QCD}}$, symmetry breaking scales, masses of other particles, the temperature of a medium and any other energy or momentum scale that describes the heavy particle and its environment \cite{Vairo:2014xea}.  Under this condition the heavy
particle turns out the be also non-relativistic. 

\begin{figure}[t]
\centering
\includegraphics[scale=0.7]{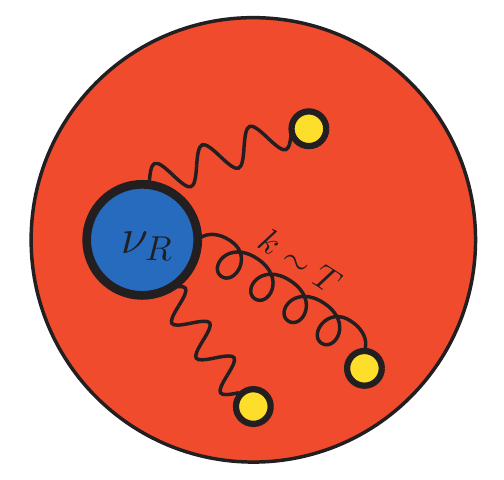}
\caption{\label{fig:eft_4} Schematic representation of a heavy Majorana neutrino in a heat bath of SM particles (yellow dots). The typical momentum transfer between the heavy right-handed neutrino and the light SM fields is of order of the temperature $T$.}
\end{figure}

Our aim is to describe the dynamics of heavy Majorana neutrinos in a thermal bath of SM particles. Then the temperature enters to describe the hot plasma in which the heavy neutrinos find themselves. Moreover we are interested in the regime $M \gg T$, relevant for leptogenesis, and this enables us to adopt an EFT approach. The situation is sketched graphically in figure \ref{fig:eft_4}, where the heavy neutrino is in a thermal bath of SM particles modelling the early universe. The heavy Majorana neutrino is kicked continuously by light particles in the heat bath with momentum transfer of order $T$. From the EFT prospective it does not make any difference if the small scale is the temperature instead of $\Lambda_{\rm{QCD}}$ like it was for the HQEFT. In matching the fundamental theory onto the low-energy one, the small scales can be put to zero. According to our assumption on the hierarchy between $M$ and $T$, we can set $T \to 0$. \textit{Hence the matching can be done at zero temperature}. 

The EFT we are going to obtain comprises non-relativistic excitation of the heavy Majorana neutrinos and the light SM particles.
The dynamical scale of the EFT is the temperature of the heat bath and therefore the observables calculated in this theory may depend on the temperature. This is why we need to introduce some notation and topics about the thermal aspect of the problem at hand. We do this in the next chapter discussing thermal field theory.  
Finally we stated that the field $h$ of HQEFT annihilates a heavy quark but it does not create an heavy antiquark. To account for an heavy anti-quark one has to consider the charge conjugate field of $Q(x)$ and redo the decomposition in a small and large component. This reflects the Dirac nature of the heavy-quark field. In this thesis we deal with heavy Majorana neutrinos and the main difference with the heavy quark is that we take the neutrinos to be Majorana particles: the field is equal to its charge conjugate. This is an element to keep in mind when constructing the low-energy theory as we are going to show in chapter \ref{chap:part_prod}. 

%% file: therm_the.tex
In this chapter we deal with the formalism of thermal field theory, namely we discuss the quantum field theory techniques that allow for perturbative calculations of observables involving particles in a medium. 
In section~\ref{sec_therm_1} we introduce the physics context for which thermal field theory is needed and we show how to obtain the thermal version of a scalar field propagator in section~\ref{sec_therm_2}. The imaginary-time formalism is briefly discussed, whereas some more details are provided for the real-time formalism adopted in the following chapters. We provide a comparison between the two formalizations of thermal field theory in section~\ref{sec_therm_3} by explicitly calculating tadpole diagrams in the $\lambda \phi^4$ theory. In section~\ref{sec_therm_4} the physical correlators are introduced and we discuss the analytic continuation from imaginary to real times. Finally we address the particle production rate for the case of right-handed neutrinos as a relevant application of the thermal field theory formalism in section~\ref{sec_therm_5}.

\section{Why thermal field theory?}
\label{sec_therm_1}
Thermal field theory is used to describe a large ensemble of interacting particles in a thermodynamical environment. This might seem the same as the classical statistical mechanics. However there are some important differences with the older and more familiar kinetic or many-body theory \cite{PhysRev.168.233,ManyTheBook}:  the usage of the path integral approach, the possibility to account for non-abelian interactions like QCD, a Lorentz-covariant formulation. A quantitative study and understanding of phase transitions in quantum field theory has been the first success of thermal field theory. 

Before coming closer to the formalism, let us consider some relevant applications of thermal field theory on the particle physics and cosmology side. If one is interested in studying hot and dense plasmas, then the early universe is a good example. Indeed at any time before recombination the mean free path between subsequent particle interactions was much smaller than the entire system size, so that one can speak of a thermalized medium. This is also strongly supported by  CMB analysis which shows that the universe at the time of the last scattering exhibits an almost uniform black-body radiation spectrum, up to small fluctuations $\delta T / T \ll 1$. Going back in time and higher in temperatures, it is expected the universe to behave as a thermodynamical system in which very interesting processes happened. Contemporary challenges in the field include dark matter production, the generation of the baryon asymmetry in the universe, the reheating dynamics after the inflation. It is important to notice that weak (or very weak) interactions play the major role in driving the production, decay rates and relic abundances of particles in the early universe. 

Another important example is the hot QCD medium which is established in the transition, better a crossover, from the hadron phase to a quark-gluon plasma (QGP) at high temperatures. Being the original idea rather old \cite{Cabibbo:1975ig}, accurate lattice simulations provides nowadays the crossover temperature $T_c=154 \pm 9$ MeV \cite{Bazavov:2014pvz}. The way to achieve and study the QGP is by making heavy ions collide at relativistic energies. In this case we have a system dominated by strong interactions, and the maximal temperature is of order of few hundreds of MeV. The advantage with respect to the early universe is that we can have direct access to the hot QCD medium in present day experiments at the Relativistic Heavy Ion Collider (RHIC) at Brookhaven National Laboratory and at the Large Hadron Collider (LHC) at CERN.  Typical interesting observables are the yields of different particles, jet quenching, transport coefficients and the hydrodynamics parameters of the plasma, like the temperature. 

In both the cases, interactions among particles occur in a medium that may be characterized by some thermodynamical parameters. Despite the different questions one addresses in studying the early stages of the universe and the QGP, there are some similarities especially regards the calculation techniques. Even though strong interactions drive the dynamics of the QGP formation and evolution, at high temperatures (energies), they become weaker due to asymptotic freedom. Resummation techniques developed and adopted in one field can then be exploited also for the other. In the end, either in the case of QGP or the early universe, we are interested in calculating observables in a rigorous way by means of a quantum field theory at finite temperature. We highlight here the main difference between the early universe and the medium produced in the heavy ion collisions: the latter expands much faster. This makes more difficult to attain thermal equilibration in the QGP than in the early universe.  

\begin{figure}
\centering
\includegraphics[scale=0.57]{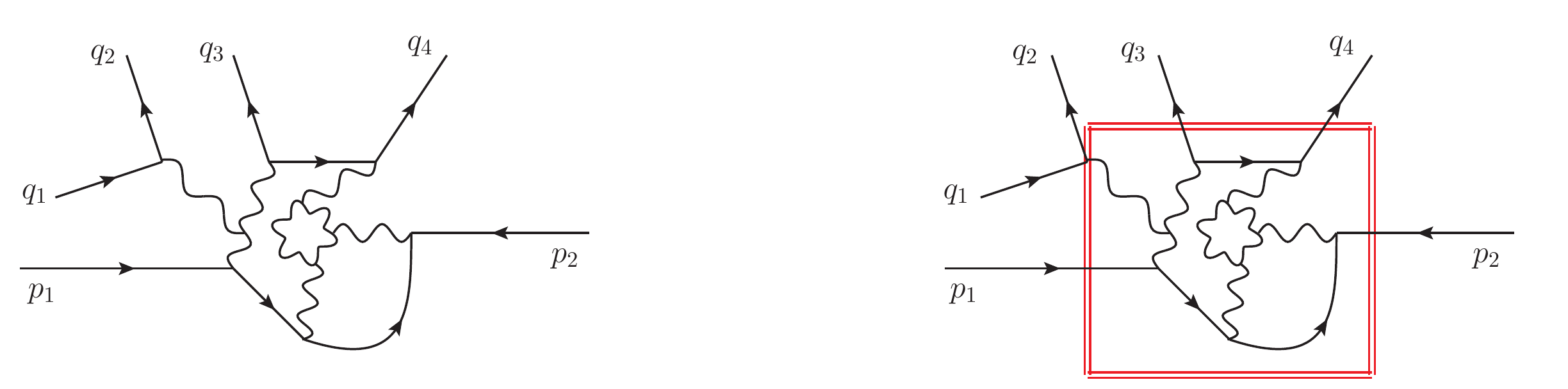}
\caption{\label{fig:therm_0} A many-particle scattering reaction is shown. The momenta of the two incoming particles are $p_1$ and $p_2$, whereas $q_1,...\,,q_4$ are those of the four out-coming particles. The red box on the right panel stands for a thermal average of the same process on the left. Figure adapted from \cite{LaineICTP}.}
\end{figure}
Let us now come closer to the formal foundation of thermal field theory. We want to compare the two situations shown in the left and right panel of figure \ref{fig:therm_0}. On the left, a complicated process at zero temperature is shown where there are two incoming particles, with momenta $p_i$, and four out-coming particles, with momenta $q_j$, the latter created by the interactions. The calculation of a multi-particle scattering is, in general, technically very complicated. We look now at the same process from a different prospective, as shown on the right of figure \ref{fig:therm_0}. With the solid red box we mean that we only care of some average properties of the system governed by the same interactions  as for the $T=0$ case. If we observed the system for a long enough period of time, the statistical description could be invoked and the problem be treated in terms of few quantities characterizing the entire system. Examples are the temperature, the chemical potentials and some conserved charges. Hence a multi-scattering process like the one to the left of figure~\ref{fig:therm_0} become more tractable if understood as a thermal system. 
Of course it is important to understand under which conditions one can adopt a statistical description starting from the microscopic and quantum mechanical view. Let us put this way: consider a system defined by an initial state and a Lagrangian (Hamiltonian) describing the microscopic interactions. Then the system has some dynamical thermalisation scale, and only if we observe the system over time periods longer than that scale, we are able to appreciate a thermodynamical behaviour.  

The starting point of thermal field theory calculations is the definition of the expectation value of a given observable in a thermal bath:
\be 
\langle A \rangle_{\beta}= \frac{1}{Z} \tr \left\lbrace  e^{-\beta H} \, A \right\rbrace \, ,
\label{therm_01}
\ee
where $H$ is the Hamiltonian of the system and  $Z$ is the partition function of the system in turn defined by 
\be 
Z=  \tr \left\lbrace  e^{-\beta H} \right\rbrace \, \, .
\label{therm_02}
\ee
Then $\beta = 1/T$, $T$ being the temperature and we recall that we take the Boltzmann constant equal to one. The trace is performed over all the accessible states of the system, either discrete or continuous states. A common choice is to consider the eigenstates of the Hamiltonian operator, namely $H | n \rangle = E_n | n \rangle$, as adopted throughout the thesis. For simplicity the chemical potential in the Boltzmann factor is not considered, $\mu=0$. We notice that we already used the definition (\ref{therm_01}) in chapter~\ref{chap:baryo_lept} when discussing the baryon number in thermal equilibrium (see eq.~(\ref{bau_4})). %It is possible to formulate the Boltzmann factor in a Lorentz-covariant way, by writing it as $exp\left( -\beta u^{\mu} P_{\mu}\right) $ with $u^{\mu}$ and $P^{\mu}$ the four-velocity of the heat-bath and the four-momentum associated to a given state respectively. We choose the reference frame of the heat bath $u^{\mu}=(1,\bm{0})$, so that only $P_0=E$ is relevant. 

There exists two different and equivalent formulations of thermal field theory: the imaginary-time formalism (ITF) and real-time formalism (RTF). Matsubara was the pioneer in the former \cite{Matsubara:1955ws}, where a purely imaginary time was included in the evolution operator. On the other hand, Schwinger, Mills and Keldysh developed an alternative formalism by choosing a particular contour in the complex plane to allow for real times \cite{Schwinger:1960qe, Keldysh:1964ud}. In the following both the realizations of thermal field theory are discussed and we highlight differences between them. We remind the reader to text books like \cite{LeBellac, DasBook} for an extensive treatment of the subject.

\section{Green's functions at finite temperature}
\label{sec_therm_2}
Thermal field theory can be also seen as a combination of quantum field theory and statistical mechanics. In order to address perturbative calculations in a hot and dense medium we need to understand how the $T=0$ formulation of the Feynman rules changes in the finite temperature case. It is usual to start the study of quantum field theory by looking at the free propagation of a scalar particle with mass $m$, described by the field $\phi$, and obtain the corresponding two-point Green's function. In the following we want to address the same quantity  at finite temperature.  Throughout the chapter we discuss mostly a scalar field theory to introduce the fundamental techniques of thermal field theory. 

In order to set the notation we recall the free scalar in-vacuum propagator, that corresponds to a quantum field theory at zero temperature. The scalar propagator is defined by 
\be 
i \Delta (x-y) \equiv \bra0 | T \left( \phi(x) \phi(y) \right) | \ket0 \, , 
\label{therm_1}
\ee
where the time ordered product of the scalar fields is defined as follows
\be 
T\left( \phi(x) \phi(y) \right) = \begin{cases}
\phi(x) \phi(y) \, , x^0 > y^0 \\ 
\phi(y) \phi(x) \, , x^0 < y^0 \, 
\end{cases}  \, ,
\label{therm_2}
\ee
and it has not to be confused with the same symbol used also for the temperature. The field $\phi$ can be expressed in terms of a Fourier decomposition with creation and annihilation operators, it reads
\be 
\phi(x) = \int \frac{d^3\bm{k}}{(2 \pi)^3} \frac{1}{2 E_k} \left[ a(\bm{k}) \, e^{-ik \cdot x} + a^{\dagger}(\bm{k}) \, e^{+i k \cdot x} \right]   \, .
\label{therm_3}
\ee
In (\ref{therm_3}) $k=(k_0,\bm{k})$ is the four momentum of the scalar field and the energy is given by $E_k=k_0=\sqrt{\bm{k}^2+m^2}$, $a(\bm{k})$ is the annihilation operator and it acts on the vacuum state as follows $a(\bm{k})| \ket0 =0$, whereas $a^{\dagger}(\bm{k})$ is the creation operator. 
Then using the expression for the scalar field (\ref{therm_3}) in the definition of the propagator  (\ref{therm_1}) one obtains 
\be 
i \Delta(x-y) = \int \frac{d^4 k}{(2 \pi)^4} \frac{i}{k^2+m^2+i \eta} \, e^{-i k \cdot (x-y)} \, ,
\label{therm_4}
\ee
that provides the corresponding Feynman rule for the propagation of a scalar particle, from the space-time point $x$ to $y$, if $x^0 < y^0$, and from $y$ to $x$ in the other case. 

Let us now come to the finite temperature case. According to the definitions in (\ref{therm_01}) and (\ref{therm_02}) , we have to change the expectation value for the two-point Green's function as follows
\be 
i \Delta^{T} (x-y) \equiv \langle T \left( \phi(x) \phi(y) \right) \rangle_\beta  = \frac{1}{Z} \sum_n \langle n |  T \left( \phi(x) \phi(y) \right) | n \rangle \, e^{-\beta E_n} \, ,
\label{therm_5}
\ee
where the superscript, $\beta$, stands for \textit{the thermal version of the scalar propagator}. In order to obtain a more explicit form for the two-point function, one has to insert the field decomposition (\ref{therm_3}) in into the definition (\ref{therm_5}). The multi-boson states $| n \rangle$ are obtained by the creation operator acting repeatedly on the vacuum state of the theory, namely
\be 
| n \rangle = | n_1(\bm{k}_1) , \, n_2(\bm{k}_2) , \, ... \rangle = \prod_i \frac{\left[  a^{\dagger}(\bm{k}_i)\right]^{n(\bm{k}_i)} }{\sqrt{n_i(\bm{k}_i)!}} | 0 \rangle \, . 
\label{therm_6}
\ee
Because the eigenstates are orthonormal  one derives for (\ref{therm_5}) the following expression
\be 
i \Delta^{T} (x-y) = \frac{1}{Z}  \int \frac{d^3\bm{k}}{(2 \pi)^3} \frac{1}{2E_k} \sum_n e^{-\beta E_n } \left[  (n(\bm{k}) +1) e^{-i k \cdot (x-y)} + n(\bm{k}) e^{i k \cdot (x-y)} \right]  \, ,
\label{therm_7}
\ee
where $n(\bm{k})$ is the occupation number of a bosonic state with three-momentum $\bm{k}$. We can simplify the sum over the eigenstates $| n \rangle$ weighted by the Boltzmann factor. We use the definition of the Bose-Einstein distribution
\be 
n_{B}(E_k) \equiv \frac{1}{Z } \sum_n n(\bm{k}) \, e^{-\beta E_n }=\frac{1}{e^{\beta E_k}-1} \, ,
\label{therm_8}
\ee
and we finally obtain 
\be 
i \Delta^{T} (x-y) =   \int \frac{d^3\bm{k}}{(2 \pi)^3} \frac{1}{2E_k}  \left[  (n_{B}(E_k) +1) e^{-i k \cdot (x-y)} + n_{B}(E_k) e^{i k \cdot (x-y)} \right]  \, .
\label{therm_9}
\ee
There is a physical interpretation of the thermal scalar propagator in (\ref{therm_9}). As in the zero temperature case we have, for $y_0 < x_0$, the creation of a scalar particle at the space-time point $y$ and the corresponding annihilation at $x$. However, in addition, there is a medium induced creation and annihilation of scalar particles at different energies, governed by the Bose-Einstein factor $n_{B}(E_k)$ that acts as a statistical weight. For $T \to 0$ ($\beta \to \infty$) we recover the in-vacuum result (one can see this integrating on $k^0$ the expression given in (\ref{therm_4})).
\subsection{Imaginary-time formalism}
We now look at the derivation of the two-point Green's function from a different prospective. Moreover we allow also for the interaction term between scalar fields with a four-particle interaction, the Lagrangian reads
\be 
\mathcal{L}_{\lambda \phi^4}= \frac{1}{2} (\pa \phi)^2 - \frac{1}{2}m^2 \phi^2 -\frac{\lambda}{4!} \phi^4 \, .
\label{imf_0}
\ee
We are going to use this simple field theory to provide the complete set of Feynman rules at finite temperature: the scalar propagator (it has been already derived, we will just show an equivalent expression of (\ref{therm_9})), the interaction vertex and the way to express loop integrals.   

In the imaginary-time formalism, a purely imaginary time is incorporated in the evolution operator. Indeed, one can regard the Boltzmann factor in (\ref{therm_01}), and hence that in (\ref{therm_5}), as an evolution operator once the following assignment is made
\be 
\beta \equiv \tau = i t \, .
\label{imf_1}
\ee
Therefore we can think of an operator evolving in time according to 
\be 
A(\tau)=e^{\tau H} A(0) e^{-\tau H} \, .
\ee
Since $\tau$ is complex the transformation is unitary. It is possible to obtain a partition function for the theory in (\ref{imf_0}), a generating functional with a source, and hence a diagrammatic approach as in the zero temperature case. There is however a first important difference. Having defined an imaginary time $\tau=\beta$ we enforce the evolution operator to be restricted only to a finite time interval. Indeed we have given up the time variable in favour of the temperature. One can see this looking at the two-point function for the scalar field. % An interesting property can be obtain for the two-point Green's function: the thermal propagator become periodic with $\beta$ (actually this is true for any $n$-pint Green's function). 
By taking the space-time arguments of the field operator as $x=(\tau_x,\bm{x})$ and $y=(\tau_y,\bm{y})$, with $\tau_x=\tau$ and $\tau_y=0$, one obtains ($0 \leq \tau  \leq \beta$)
\bea
&& \Delta^T(x-y)= \Delta^T(\tau,\bm{x}-\bm{y}) \nn
\\
&&=\frac{1}{Z} \tr \left[ e^{-\beta H} T \left( \phi(\tau,\bm{x})\phi(0,\bm{y}) \right) \right]   =\frac{1}{Z} \tr \left[ e^{-\beta H}  \phi(\tau,\bm{x}) \phi(0,\bm{y}) \right] \nn 
\\
&&=\frac{1}{Z} \tr \left[ e^{-\beta H} e^{ \beta H}   \phi(0,\bm{y}) e^{- \beta H}  \phi(\tau,\bm{x}) \right]  = \frac{1}{Z} \tr \left[  e^{-\beta H} \phi(\beta,\bm{y}) \phi(\tau,\bm{x})   \right] \nn \\
&&=  \frac{1}{Z} \tr \left[  e^{-\beta H} T \left(  \phi(\tau,\bm{x})  \phi(\beta,\bm{y})   \right)   \right] = \Delta^T(\tau-\beta, \bm{x}-\bm{y}) \, .
\label{imf_2}
\eea  
This kind of property for the propagator is named a Kubo-Martin-Schwinger (KMS) relation \cite{Kubo:1957mj,Martin:1959jp}. More in general it can be written as
\be 
\Delta^T(\tau,\bm{x}-\bm{y}) = \Delta^T(\tau+ n \beta, \bm{x}-\bm{y}) \, ,  n \in \mathbb{Z} \, .
\label{imf_3}
\ee
This has an important consequence: the time argument is restricted to the interval $\left[ 0,\beta \right] $.  It is rather clear that one loses contact with real-time quantities within the ITF, and that one is restricted to the evaluation of static thermodynamical quantities. This is the context in which such formalism was originally derived. In order to calculate time-dependent quantities from the ITF, one has to perform a non trivial analytical continuation to real times after all the diagrams of interest are calculated. 

As in the zero temperature case, the evaluation of Feynman diagrams is easier in momentum space. Then  we  address a second aspect of the imaginary-time formalism: the Matsubara sum. Going to imaginary times $0 \leq it \equiv \tau \leq \beta$, and summing over discrete energies $k_0 \equiv \omega_n =2 \pi i T n$ instead of integrating 
\be 
\int \frac{d k_0}{(2 \pi)} \to i T \, \sum_{n=-\infty}^{+\infty}  \, ,
\label{imf_4}
\ee
the propagator in (\ref{therm_9}) can be rewritten as 
\be 
i \Delta^T(x-y)= i T \sum_{n=-\infty}^{+\infty}   \int \frac{d^3\bm{k}}{(2 \pi)^3}  \frac{i}{k^2-m^2} \, e^{-ik \cdot (x-y)} \, ,
\label{imf_5}
\ee
where $k$ in the denominator and the exponent is the four-momentum. An explicit derivation of~(\ref{imf_5}) can be found, e.~g., in~\cite{Thoma:2000dc}. We see that the scalar propagator acquires a non-trivial dependence on the temperature via the so-called Matsubara frequencies, $\omega_n$. The interaction vertex is left untouched. In summary, we can list the Feynman rules  for the model Lagrangian in (\ref{imf_0}) in momentum space in ITF as follows
\begin{itemize}
\item[1)] the scalar propagator is 
\be 
\begin{minipage}[c]{0.05\linewidth}
\includegraphics[scale=0.75]{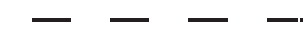} 
\end{minipage}
\hspace{2.2 cm}
= \, \frac{i}{k^2-m^2} \, ,
\label{imf_6}
\ee
with the $k_0=2\pi i  T n$, namely restricted to the Matsubara frequencies;
\item[2)] the vertex is (as in the $T=0$ case)
\be 
\begin{minipage}[c]{0.05\linewidth}
\includegraphics[scale=0.75]{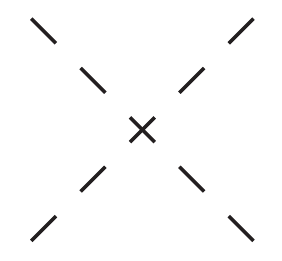} 
\end{minipage}
\hspace{2.0 cm}
= \, -i \lambda \, ;
\label{imf_7}
\ee
\item[3)] in loop integrals we have to make the replacement for the loop momentum $\ell$ as follows
\be 
\int \frac{d^4 \ell}{(2 \pi)^4} \, \to \,  \Tint \equiv i T \sum_{n=-\infty}^{+\infty} \int \frac{d^3\bm{\ell}}{(2 \pi)^3} \, ,
\label{imf_8}
\ee
where $\ell_0$ takes discrete values over the Matsubara frequencies;
\item[4)] topologies and symmetry factors are the same as in the $T=0$ case. 
\end{itemize}  

\subsection{Real-time formalism}
We already mentioned that the ITF was originally derived to calculate static quantities, for instance the free energy or the pressure. If one is interested in real-time quantities  that evolve with time, one can still stick on the ITF and perform an analytic continuation to real times after the Matsubara sums have been carried out. This procedure may be cumbersome and  one could better start with the real time variables from the very beginning. The RTF of thermal field theory is suited for addressing time evolving observables, such as phase transitions. Non-equilibrium dynamics is more naturally accounted for in the RTF. The price to pay is the so-called \textit{doubling of the degrees of freedom} that make the calculations rather involved. RTF provides a more transparent organization of the thermal content in actual calculations since the in-vacuum and finite temperature terms in particle propagators are disentangled from the beginning.   

\begin{figure}
\centering
\includegraphics[scale=0.71]{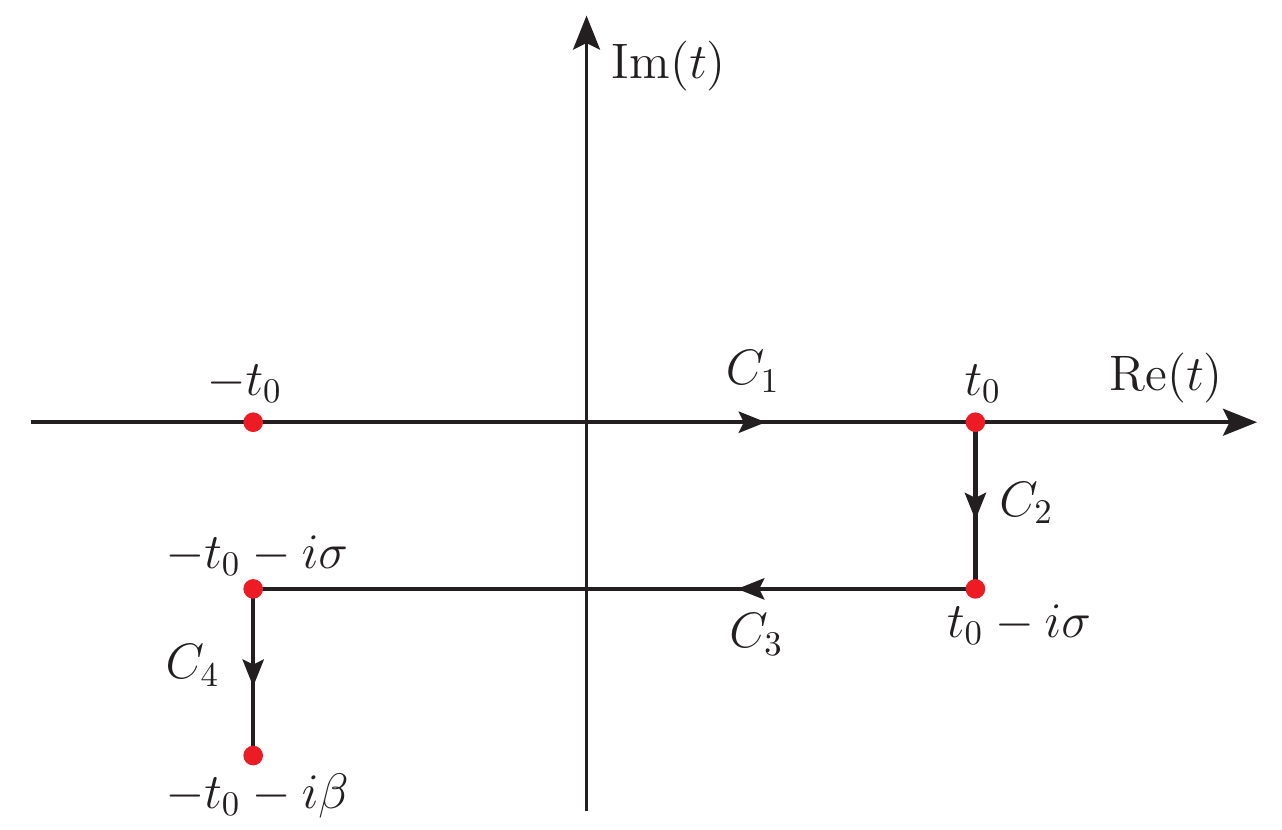}
\caption{\label{fig:therm_4} Schwinger--Keldysh contour \cite{Schwinger:1960qe, Keldysh:1964ud} in the complex time plane for a given choice of $0<\sigma < \beta$. The red dots stand for the boundaries along the paths $C_1$, ..., $C_4$. }
\end{figure}
The aim of this discussion is to provide the same set of Feynman rules (\ref{imf_6})-(\ref{imf_8}) for the $\lambda \phi^4$ theory in the RTF. The way to allow for real time arguments in the Green's functions is to consider a particular contour in the complex $t$ plane, as shown in figure \ref{fig:therm_4}. The Schwinger--Keldysh contour  is a deformation of the straight line in the complex plane from $t$ to $t-i\beta$ one has to consider in the ITF. Let us describe the path in figure \ref{fig:therm_4}. We start with $C_1$, standing for the path from the initial time $t_i \equiv -t_0$ on the negative real axis, with $t_0 >0$, to a positive valued real time $t_0$. One moves down from there according to the path $C_2$, where the time gets an imaginary part $t_0-i\sigma$, with $0<\sigma <\beta$. Then one goes back to a negative real part moving horizontally along $C_3$ arriving at $-t_0 - i \sigma$. Finally along $C_4$ we end up with $-t_0-i\beta$. This is the  choice for the contour if one aims at working with real times. Taking now $t_0 \to \infty$ the entire real time axis is spanned.

The generating functional of the theory has to be evaluated on the Schwinger--Keldysh contour. By analytic reasons it can be shown that the contributions along the path $C_2$ and $C_4$ can be neglected, indeed factorizing in the generating functional as constants irrelevant to the field dynamics \cite{Niemi:1983nf}. %Accordingly we adopt the choice $\sigma=0$ for the Schwinger--Keldysh contour in the following (in general one can keep $\sigma \neq 0$ that enters the propagator expressions). 
Therefore we are left with the possibilities for the two-point function time arguments, $x_0$ and $y_0$, to lie either on $C_1$ or $C_3$. This suggest that the propagators at finite temperature have a richer structure than the in-vacuum counterparts. In terms of ensemble averages they have the following expressions
\bea 
&&i \Delta^T_{11}(x_0-y_0,\bm{x}-\bm{y}) =  \langle T (\phi(x) \phi(y)) \rangle_\beta\, , \quad x_0  \in C_1 \; \text{and} \;  y_0 \in C_1
\label{rtf_1}
\\
&&i \Delta^T_{12}(x_0-y_0,\bm{x}-\bm{y}) =  \langle \phi(y) \phi(x) \rangle_\beta \, , \quad \phantom{xxx} x_0 \in C_1 \; \text{and} \;  y_0 \in C_3
\label{rtf_2}
\\
&&i \Delta^T_{21}(x_0-y_0,\bm{x}-\bm{y}) = \langle \phi(x) \phi(y) \rangle_\beta \, , \quad \phantom{xxx}  x_0 \in C_3 \; \text{and} \;  y_0 \in C_1
\label{rtf_3}
\\
&&i \Delta^T_{22}(x_0-y_0,\bm{x}-\bm{y}) = \langle \overline{T} (\phi(x) \phi(y)) \rangle_\beta \, , \quad x_0 \in C_3 \; \text{and} \;  y_0 \in C_3 \, ,
\label{rtf_4}
\eea 
where $\overline{T}$ stand for the anti-time ordering along the $C_3$ branch. We call a field of ``type~1" that having the time coordinate on the upper brunch, $C_1$, whereas for fields of ``type~2" the time variable lies on the lower branch, $C_3$. Because of the orientation along the contour, times on the lower brunch come always after those on the upper one and times on the lower brunch are conversely ordered (a later time comes first).    % They are labelled as ``11" propagator for $x_0 \in C_1$ and $y_0 \in C_1$ and so on and so forth. A field of type 1 is associated to time variables belonging to $C_1$, whereas field of type 2 are those having a time variable on the $C_2$ path. The time ordering is given by the contour arrows in figure \ref{fig:therm_4}, so that field of type 2 comes always after fields of type 1. This is the doubling of the degrees of freedom in the RTF. More comments are in order.
The necessary appearance of the field of type~2 in the construction of the RTF stands for the doubling of the degrees of freedom mentioned before. 

Let us briefly comment on the quantum field theory structure when adding fields of type~2.
First, they never enter matrix elements as asymptotic states but they appear only in internal lines. They are indeed not physical and they act as ghost fields (not to be confused with the Fadeev--Popov ghost though).  Second, the real-time propagator can be recast in a 2$\times$2 matrix, according to the different combinations of the field of type 1~and 2:
\bea
\Delta^T(x-y) &=& \left( 
\begin{array}{c c}
\Delta_{11}(x-y) & \Delta_{12}(x-y) \\
\Delta_{21}(x-y) & \Delta_{22}(x-y)
\end{array} \right) \, ,
\label{rtf_5}
\eea  
The off-diagonal elements are often denoted as the Wightman propagators, $\Delta^{<}$ and $\Delta^{>}$ instead of $\Delta_{12}$ and $\Delta_{21}$ respectively. In order to write explicitly the propagator components we have to choose the value of the parameter $\sigma$ and we stick to the popular choice $\sigma=0$. This is also the original one adopted by Schwinger--Keldysh. Not setting $\sigma=0$, the parameter will enter explicitly the propagator expression \cite{DasBook}. We give the 2$\times$2 propagator matrix in momentum space, for a detailed derivation we refer to \cite{LeBellac}.
It reads:
\bea 
&&i \Delta^T(k)=\left( 
\begin{array}{c c}
\frac{i}{k^2-m^2+i\eta} & \theta(-k_0) \, 2 \pi \delta(k^2-m^2)  \\
\theta(k_0) \, 2 \pi \delta(k^2-m^2)  & -\frac{i}{k^2-m^2-i\eta}
\end{array} 
\right) 
\nn \\
&&\phantom{xxxxxxxxxxxxxxxxxxxxxx}+ 2 \pi \delta(k^2-m^2)  \, n_{B}(|k_0|)  \left( 
\begin{array}{c c}
1 &  1 \\  
1 & 1
\end{array} 
\right) \, . \nn \\
\phantom{x}
\label{rtf_6}
\eea 
As one may see from (\ref{rtf_6}), there is a more transparent separation between the vacuum and thermal part than the propagator in ITF. Indeed for the physical propagator $\Delta^T_{11}(k)$ we simply have the sum of the in-vacuum scalar propagator and a thermal piece, the latter manifestly disentangled from the former. This was not the case for the Matsubara propagator in (\ref{imf_5}) where the temperature dependence is somehow encrypted in the Matsubara frequencies.  Moreover the thermal part is made of a Dirac delta function, that enforces the thermal particles to be on-shell, and weighted by the Bose-Einstein distribution. This clearly shows that the thermal propagator comprises thermalized on-shell particle contributions even in the free case. It is also clear that in the in-vacuum limit, $T \to 0$, one reduces to the zero temperature scalar propagator. 

We are now ready to write the set of Feynman rules for the $\lambda \phi^4$ theory:
\begin{itemize}
\item[1)] the physical propagator, $\Delta_{11}(k)$, reads
\be
\begin{minipage}[c]{0.05\linewidth}
\includegraphics[scale=0.75]{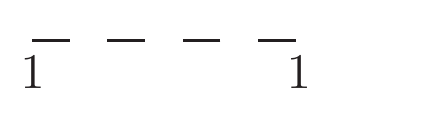} 
\end{minipage}
\hspace{1.8 cm} =\,  i\Delta_{11}(k)= \, \frac{i}{k^2-m^2+i\eta} + 2 \pi n_{B}(|k_0|) \delta(k^2-m^2) \, .
\label{rtf_7}
\ee
The other propagators in eq.~(\ref{rtf_6}), namely $\Delta_{12}(k)$, $\Delta_{21}(k)$ and $\Delta_{22}(k)$, are drawn as (\ref{rtf_7}) accounting for different labels of the dashed line, 12, 21 and 22 respectively;
\item[2)] regarding the the four-particle vertex, fields of type~1 are not mixed with fields of type~2. The vertex involving fields of type 2 has a relative minus sing coming from the anti-time ordering:
\be 
\begin{minipage}[c]{0.05\linewidth}
\includegraphics[scale=0.75]{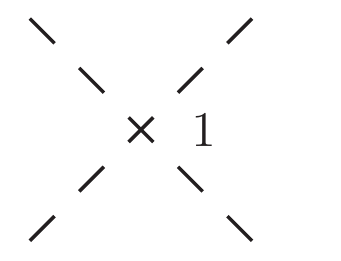} 
\end{minipage}
\hspace{2.0 cm}
= \, -i \lambda \, , \hspace{1.5 cm} 
\begin{minipage}[c]{0.05\linewidth}
\includegraphics[scale=0.75]{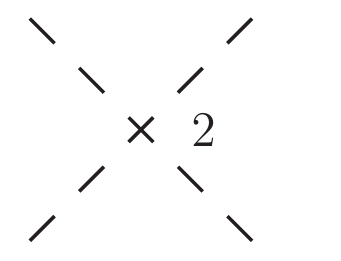} 
\end{minipage}
\hspace{2.0 cm}
= \, i \lambda \, ;
\label{rtf_8}
\ee
\item[3)] for any loop one has to integrate on the loop momentum $\ell$ as in the $T=0$ case
\be 
\int \frac{d^4 \ell}{(2 \pi)^4} \,,
\label{rtf_9}
\ee
and one has to include, when necessary, internal lines with fields of type~2;
\item[4)] topologies and symmetry factors are the same as in the $T=0$ case.
\end{itemize}
The Feynman rules in the two formalisms of thermal field theory are quite different. In the RTF there is a closer contact with the corresponding $T=0$ theory despite the addition of the field of type~2. 

Before addressing an actual calculation, we further comment on the real time 2$\times$2 propagator in eq.~(\ref{rtf_5}). The four components of the propagator are not independent. This can be traced back to the periodicity condition on the propagators in the ITF that we have discussed (see \ref{imf_2}). So we can think of Kubo-Schwinger relations also in the RFT, that reads as follows
\be 
\Delta_{11}-\Delta_{12}-\Delta_{21}+\Delta_{22} =0 \, .
\label{imf_10}
\ee
It is rather straightforward to verify it, one has just plug in eq.~(\ref{imf_10}) the components explicitly given in (\ref{rtf_6}). 
%\bea 
%\Delta_{11}-\Delta_{12}-\Delta_{21}+\Delta_{22} =\frac{i}{k^2-m^2+i\eta} - \frac{i}{k^2-m^2-i\eta} + 2 \pi  \delta(k^2-m^2) \, .
%\label{imf_11}
%\eea
%Then using the following identity for the Dirac delta
%\be 
%\frac{1}{x+i\eta} - \frac{1}{x-i\eta}= -2 i \pi \delta(x) \, ,
%\label{imf_12}
%\ee
%the relation in (\ref{imf_10}) is verified. 
We introduce a useful representation of the RTF, called\textit{ Keldysh representation}. It is constructed by linear combinations of the two-point Green's functions in eq.~(\ref{rtf_5}). The propagator components are called retarded, advanced and symmetric and they read respectively
\bea 
\Delta_{R} \equiv \Delta_{11}- \Delta_{12} \, , \quad \Delta_{A} \equiv \Delta_{11}- \Delta_{21} \, , \quad \Delta_S \equiv \Delta_{11}+ \Delta_{22} \, . 
\label{imf_13}
\eea
The three propagator components are sufficient because of the relation (\ref{imf_10}) and explicitly they read
\bea
&&i\Delta_{R}(k)  = \frac{i}{k^2-m^2+i{\rm{sgn}}(k_0)} \, ,
\label{imf_14} \\
&&i\Delta_{A}(k)  = \frac{i}{k^2-m^2-i{\rm{sgn}}(k_0)}  \, ,
\label{imf_15}
\\
&&i\Delta_S(k)= 2  \pi \delta(k^2-m^2) \left[ 1+ n_B(|k_0|)\right]  \, .
\label{imf_16}
\eea
The poles are both above (below) the real axis for the retarded (advanced) propagator; only the symmetric propagator contain a thermal distribution. The latter usually helps in pin pointing the thermal contribution in  actual calculations.

An analogous derivation holds for the thermal propagator of a massive particle with spin. As for the scalar propagator one finds a 2$\times$2 matrix in the real-time formalism, whereas the Bose--Einstein distribution is replaced by the Fermi--Dirac one. The fermion propagator reads  \cite{LeBellac, DasBook}
\bea 
&&i S^T(k)=(\slashed{k}+m) \left[ \left( 
\begin{array}{c c}
\frac{i}{k^2-m^2+i\eta} & \theta(-k_0) \, 2 \pi \delta(k^2-m^2)  \\
\theta(k_0) \, 2 \pi \delta(k^2-m^2)  & -\frac{i}{k^2-m^2-i\eta}
\end{array} 
\right)  \right. 
\nn \\
&& \left. \phantom{xxxxxxxxxxxxxxxxxxxxxx}- 2 \pi \delta(k^2-m^2)  \, n_{F}(|k_0|)  \left( 
\begin{array}{c c}
1 &  1 \\  
1 & 1
\end{array} 
\right) \right]  \, . \nn \\
\phantom{x}
\label{imf_17}
\eea 
The four components are not independent and the corresponding condition of eq.~(\ref{imf_10}) may be obtained for the fermion propagator. Moreover we can define the advanced, retarded and symmetric propagators in complete analogy with the bosonic case:
\bea
&&iS_{R}(k)  = \frac{i(\slashed{k}+m)}{k^2-m^2+i{\rm{sgn}}(k_0)} \, ,
\label{imf_14}
\\
&&iS_{A}(k)  = \frac{i(\slashed{k}+m)}{k^2-m^2-i{\rm{sgn}}(k_0)}  \, ,
\label{imf_15}
\\
&&iS_S(k)= 2 \pi \, (\slashed{k}+m) \,  \delta(k^2-m^2) \left[ 1- n_F(|k_0|)\right]  \, .
\label{imf_16}
\eea     

Regards the fermion propagator in the ITF we observe that the periodicity condition on the two-point point Green's function is 
\be 
S^T(\tau, \bm{x} - \bm{y})= (-1)^n S^T (\tau + n \beta, \bm{x} - \bm{y} ) \, , \quad n \in \mathbb{Z},
\ee
where the minus sign comes from the anti-commutating fermion fields, and from that the Matsubara sum goes on $k_0=2 \pi i T (n+1)$, with $n \in \mathbb{Z}$. 
\section{Comparison between the ITF and RTF: a tadpole computation}
\label{sec_therm_3}
\begin{figure}[t]
\centering
\includegraphics[scale=0.69]{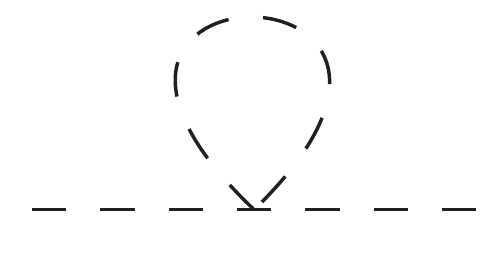}
\caption{\label{fig:therm8} The tadpole diagram in $\lambda \phi^4$ theory.}
\end{figure}

In order to show an application of both the ITF and RTF, we consider the calculation of the tadpole diagram in the $\lambda \phi^4$ theory, as shown in figure \ref{fig:therm8}. Besides the purpose to discuss a comparison between a loop computation in the two formalisms, this example is also of particular relevance to the thesis. Indeed thermal contributions to the heavy neutrino width are encoded in tadpole diagrams in the EFT (see chapter~\ref{chap:part_prod}). 

Let us start with the ITF. Using the Feynman rules given in section~\ref{sec_therm_2} we write the self-energy induced by the tadpole diagram, $\Pi_{\hbox{\tiny tad}}$, as follows
\bea
-i\Pi_{\hbox{\scriptsize tad}}&=& -i \lambda \, \frac{1}{2} \, i T \, \Tint \frac{i}{(\ell_0)^2-\bm{\ell}^2-m^2}
\nn
\\ &=&  i\frac{\lambda}{2} \, T \,\sum_{n=-\infty}^{+\infty} \int \frac{d^3\bm{\ell}}{(2\pi)^3} \frac{1}{(2 \pi i T n)^2-E^2_\ell} 
\label{comp_1}
\eea
where the diagram has to be understood as amputated of the external legs and  $1/2$  is the symmetry factor.  The energy has been substituted with the Matsubara modes. To simplify the expression in (\ref{comp_1}), we make the substitution $\mathcal{E}=E_\ell /(2 \pi T)$, and we evaluate the series on the integer $n$ 
\bea 
&&\sum_{n=-\infty}^{+\infty} \frac{1}{n^2 + \mathcal{E}^2} = 2 \sum_{n=1}^{+\infty} \frac{1}{n^2 + \mathcal{E}^2} + \frac{1}{\mathcal{E}^2} 
\\
&&= \frac{\pi}{\mathcal{E}} \coth (\pi \mathcal{E}) = \frac{\pi}{\mathcal{E}} \left( 1+ \frac{2}{e^{2 \pi \mathcal{E}}-1} \right)  \, ,
\label{comp_2}
\eea
where we used the representation for the $\coth(\pi x)$ that reads \cite{TableInt}
\be 
\coth(\pi x)= \frac{1}{\pi x } + \frac{2x}{\pi}\sum_{n=1}^{+\infty} \frac{1}{n^2 + x^2}   \, .
\label{comp_3}
\ee

At this point the Matsubara sum has been performed, whereas the integration in the three-momentum is left. We substitute back into (\ref{comp_1}) the result of the Matsubara sum in (\ref{comp_2}) and we obtain 
\be 
\Pi_{\hbox{\scriptsize tad}} =  \frac{\lambda}{4}  \int \frac{d^3\bm{\ell}}{(2\pi)^3} \frac{1}{E_\ell}  \left( 1+2 n_{B}(E_\ell)\right) \, .
\label{comp_4}
\ee
The in-vacuum and thermal contribution are now disentangled. The in-vacuum term is UV divergent and one can take care of it by standard renormalization. Moreover the renormalization of the $T=0$ suffices to make the theory finite at $T \neq 0$. The reason may be understood as follows: the temperature scale does not modify the theory at distances much smaller than $1/T$, and therefore the short-distance singularities are the same as in the $T=0$ case.      
We focus on the thermal part that can be evaluated analytically for a massless scalar, $m=0$, and it gives 
\be 
\Pi_{\hbox{\scriptsize tad}}^T \equiv \delta m_T^2= \frac{\lambda}{24 } T^2 \, .
\label{comp_5}
\ee
The result in eq.~(\ref{comp_5}) can be understood as a thermal correction to the mass of the scalar field at order $\lambda$. If one considers also the $T=0$ correction to the mass coming from the first term in (\ref{comp_4}), after having subtracted the divergent part, the overall correction to the mass reads $\delta m^2= \delta m^2_{T=0} + \delta m_T^2 $.

\begin{figure}[t]
\centering
\includegraphics[scale=0.65]{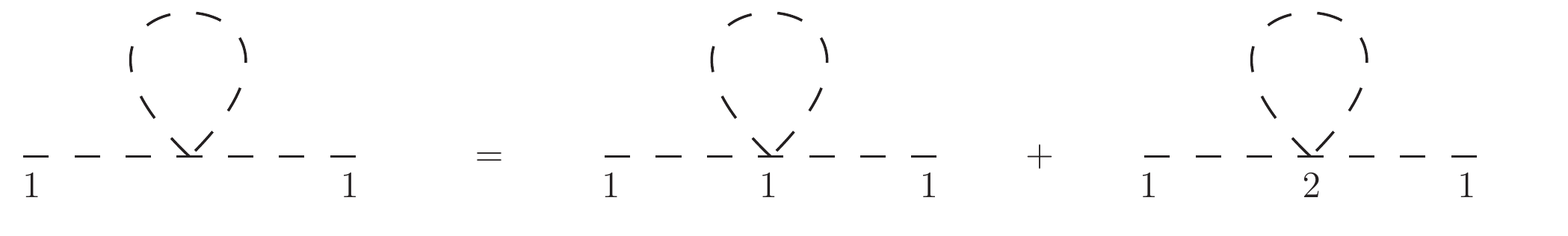}
\caption{\label{fig:therm9} The tadpole diagram in $\lambda \phi^4$ theory for the physical type~1 field. The tadpole diagram comprises two contributions: $\Pi_{\hbox{\scriptsize tad},11}$ and $\Pi_{\hbox{\scriptsize tad},22}$. In the latter the non-physical type~2 field enter.}
\end{figure}

Let us now move to the calculation of the very same diagram in the RTF.  The self energies are assigned the same index labelling as the propagators. Therefore we can have $\Pi_{11}$, $\Pi_{12}$,  $\Pi_{21}$ and $\Pi_{22}$. The relevant self-energies are $\Pi_{11}$ and $\Pi_{22}$, as shown in figure \ref{fig:therm9}. The external fields have to be the physical ones, namely of type~1. The self-energies $\Pi_{12}$ and $\Pi_{21}$ do not exists in this theory since all the legs of a vertex must have the same index. Then we write, following the Feynman rules in RTF given in eqs.~(\ref{rtf_7})-(\ref{rtf_9})
\be 
-i\Pi_{\hbox{\scriptsize tad,11}} = -i \lambda \, \frac{1}{2} \, \int \frac{d^4 \ell}{(2 \pi)^4} \left(  \frac{i}{\ell^2-m^2+i\eta} + 2 \pi n_{B}(|\ell_0|) \delta(\ell^2-m^2 ) \right) \, ,
\label{comp_6}
\ee
where the propagator $\Delta_{11}(\ell)$ enters and the $1/2$ is the symmetry factor. Here the zero temperature and thermal contributions are already disentangled from  the very beginning. Again one can notice the divergent part arising from the $T=0$ momentum integration.  By considering only the thermal part in (\ref{comp_6}) we can write
\bea 
\Pi_{\hbox{\scriptsize tad},11}^{T} &=& \frac{\lambda}{2} \int \frac{d^4 \ell}{(2 \pi)^4} \,  2 \pi n_{B}(|\ell_0|) \delta(\ell^2-m^2 ) \nn
\\
&=& \frac{\lambda}{2}   \int \frac{d^3\bm{\ell}}{(2\pi)^3} \frac{n_{B}(E_\ell) }{E_\ell}   \, ,
\label{comp_7}
\eea
which is the same as we found in eq.~(\ref{comp_4}) for the in-medium contribution. It is clear that simplifying further the last result in (\ref{comp_7}), we find, in the $m=0$ case, the same expression for the self-energy 
\be 
\Pi_{\hbox{\tiny tad},11}^{T}\equiv \delta m^2_{11,T}= \frac{\lambda}{24}T^2 \, .
\label{comp_8}
\ee
Calculating the self-energy with the field of type~2 we find exactly the same result as in (\ref{comp_8}) 
\be
\Pi_{\hbox{\scriptsize tad},22}^{T}\equiv \delta m^2_{22,T}= \frac{\lambda}{24}T^2 \, ,
\label{comp_9}
\ee
and also for the $T=0$ parts one finds the same contribution coming from $\Pi_{\hbox{\scriptsize tad,11}}$ and $\Pi_{\hbox{\scriptsize tad,22}}$. Hence $\delta m^2_{11} = \delta m^2_{22}$,  where we include the in-vacuum contributions in the mass corrections. We can express the one-loop resummation for the $\Delta_{11}$ propagator as follows
\bea
i \Delta_{11}^{(1)} &=& i \Delta_{11} + (-i \delta m^2_{11} ) (i\Delta_{11})^2 + (i \delta m^2_{11} ) (i\Delta_{12}) (i\Delta_{21})  \nonumber \\
&=& i \Delta_{11}  - i \delta m^2_{11} \left[ (i\Delta_{11})^2 - (i\Delta_{12}) (i\Delta_{21}) \right] \, ,
\label{comp_10}
\eea
where with $i\Delta_{11}^{(1)}$ and $i \Delta_{11}$ we mean the one-loop resummed and leading order scalar propagator respectively. 
We conclude this section by highlighting that the perturbative expansion in~(\ref{comp_10}) is well defined. The pinch singularities, that would appear within a naive perturbation theory involving only type~1 fields \cite{Dolan:1973qd}, cancel out in the combination of the propagators constructed with the type~2 fields. 

\section{Self-energies and discontinuities}
\label{sec_therm_4}
In the previous section we introduced the RTF of thermal field theory for the $\lambda \phi^4$ theory. We aim at providing a more general treatment in the present section. We have already mentioned that the RTF is a suitable choice if one is interested in calculating time dependent observables. These are usually expressed in terms of a Minkowskian time, $t$, and the temperature, $T$. The production rate of weakly interacting particles, oscillations in a hot and dense plasma, transport coefficients such as thermal conductivities are examples of such observables. 
From the practical point of view, most of the observables can be obtained calculating two-point functions of elementary or composite operators.

Since we are going to use coordinates both in Minkowsky and Euclidean space, we recall some notation in short. We denote Euclidean space-time coordinates by $X=(\tau,\bm{x})$ and momenta $K=(k_n,\bm{k})$, whereas the Minkowskian counterparts with $x=(t,\bm{x})$ and $k=(k_0,\bm{k})$. The Matsubara energy modes are understood in the Euclidean momentum and the Wick rotation is $\tau \to it$ ($k_n \to -ik^0$). The scalar products are defined as usual, namely $X \cdot K=\tau k_n - \bm{x} \cdot \bm{k}$ and $x \cdot k=t k_0 - \bm{x} \cdot \bm{k}$. %The thermal ensamble is defined by the density matrix
%\be 
%\hat{\rho}=\frac{1}{Z} e^{-\beta \hat{H}} \, ,
%\label{spec_1} 
%\ee   
%and the expectation value of an operator or products of operator is 
%\be 
%\langle A \rangle \equiv \tr \left\lbrace \hat{\rho} A \right\rbrace \, ,
%\label{spec_2}
%\ee
%the last equation being equivalent to (\ref{fig:therm_0}).
 
\subsection{Bosonic case}
We consider complex field operators that describe bosonic degrees of freedom. We denote with $\phi_i$ and $\phi^{\dagger}_j$ such operators, where the subscript may be understood as labelling a generic quantum number. If these operators are in momentum or coordinate space is made explicit by their arguments. We can define the following class of correlation functions
\bea 
&&\Pi^{>}_{i j}(k)= \int \frac{d^4x}{(2 \pi)^4} \langle \phi_i(x) \phi^{\dagger}_j(0) \rangle_\beta \, e^{i x \cdot k} \, ,
\label{spec_3}
\\
&&\Pi^{<}_{ij}(k)= \int \frac{d^4x}{(2 \pi)^4} \langle\phi^{\dagger}_j(0)   \phi_i(x)  \rangle_\beta \, e^{i x \cdot k} \, ,
\label{spec_4}
\\
&&\rho_{ij}(k) = \frac{1}{2} \int \frac{d^4x}{(2 \pi)^4} \langle \left[ \phi_i(x), \, \phi^{\dagger}_j(0)  \right] \rangle_\beta  \, e^{i x \cdot k}\, ,
\label{spec_5}
%\\
%&&\Delta_{ij}(k) = \frac{1}{2} \frac{d^4x}{(2 \pi)^4} \langle \left\lbrace \phi_i(x), \, \phi^{\dagger}_j(0) \right\rbrace \rangle \, e^{i x \cdot k} \, ,
%\label{spec_6}
\eea 
where $\Pi^{>}_{ij}$  and $\Pi^{<}_{ij}$ are called Wightman functions and $\rho_{ij}$ is the \textit{spectral function}. We do not introduce a fourth correlation function, the statistical correlator, being not relevant for our discussion (its definition can be found e.~g.~in \cite{LaineBasics,LeBellac}). Our goal is to derive the relation between the different correlators and to show that they can be expressed in terms of the spectral function. Moreover we want to establish the equation capturing the analytic continuation from the ITF to the RTF. 
The retarded and advanced correlators can be defined as follows
\bea 
&&\Pi^{R}_{i j}(k)= i \int \frac{d^4x}{(2 \pi)^4} \langle \left[ \langle \phi_i(x) \, , \phi^{\dagger}_j(0) \right] \theta(t) \rangle_\beta \, e^{i x \cdot k} \, ,
\label{spec_7}
\\
&&\Pi^{A}_{ij}(k)=i \int \frac{d^4x}{(2 \pi)^4} \langle -\left[ \langle \phi_i(x) \, , \phi^{\dagger}_j(0) \right] \theta(-t) \rangle_\beta \, e^{i x \cdot k} \, \, .
\label{spec_8}
\eea
It is worth noticing that the retarded propagator involves only positive times, and then due to
\be 
e^{i k_0 t} = e^{i t \,  {\rm{Re}}k_0 } e^{-t\, {\rm{Im}}k_0 } \, , 
\label{spec_9}
\ee
there is an exponentially suppressed factor for $k_0>0$. Therefore $\Pi_{ij}^R$ is an analytic function in the upper-half $k_0$-plane. The same holds for the advanced propagator in the lower-half $k_0$-plane. This turns out to be particularly useful and does not apply in general for the other correlation functions. 

We complete the list with two more correlators one may encounter in practical computations. They are the time-ordered two-point correlator:
\be 
\Pi_{ij}^T(k)=\int \frac{d^4x}{(2 \pi)^4} \langle \phi_i(x) \phi^{\dagger}_j(0) \theta(t)  + \phi^{\dagger}_j(0) \phi_i(x)\theta(-t) \rangle \, e^{i x \cdot k} \, , 
\label{spec_10}
\ee
which is the thermal counterpart of the zero-temperature version used in perturbation theory, and the Euclidean correlator 
\be 
\Pi_{ij}^E(K)= \int \frac{d^4X}{(2 \pi)^4}  \langle \phi_i(X) \phi^{\dagger}_j(0) \rangle \, e^{i X \cdot K} \, ,
\label{spec_11}
\ee 
the latter appears typically in non-perturbative calculations. Being restricted to $0 \leq \tau \leq \beta$, the Euclidean correlator is also time-ordered and can be evaluated by standard imaginary-time functional integrals. 

It turns out that all correlators can be related to each other (as we have briefly discussed for the free propagators in the previous section) and expressed in terms of the spectral function. The latter can be in turn related to an analytic continuation of the Euclidean correlator in (\ref{spec_11}). To show these statements we start with the Wightman correlators in (\ref{spec_3}) and (\ref{spec_4}). A complete set of the Hamiltonian eigenstates can be inserted in their definition and one finds (details can be found in~\cite{LaineBasics, LeBellac}) 
\be 
\Pi^<_{ij}(k)=  e^{-\beta k_0} \Pi^>_{ij}(k) \, .
\label{spec_12}
\ee
This is a KMS relation for the correlators in RTF. From eqs.~(\ref{spec_3})-(\ref{spec_5}) and (\ref{spec_12}) we find that 
\be 
\rho_{ij}(k)= \frac{1}{2} \left[ \Pi^>_{ij}(k) - \Pi^<_{ij}(k)\right] = \frac{1}{2} \left(e^{\beta k_0} -1 \right)\Pi^<_{ij}(k) \, , 
\label{spec_13}
\ee
and inverting the relation either for $\Pi^>_{ij}(k)$ or $\Pi^<_{ij}(k)$ one obtains
\bea
&&\Pi^<_{ij}(k) =2 n_B (k_0) \rho_{ij}(k) \, ,
\label{spec_14}
\\
&&\Pi^>_{ij}(k)= 2 \left[ 1+  n_B (k_0) \right] \rho_{ij}(k) \, .
\label{spec_15}
\eea 
%For the statistical correlator we have 
%\be 
%\Delta_{ij}(k) = \frac{1}{2} \left[\Pi^>_{ij}(k) + \Pi^<_{ij}(k) \right] =  \left[ 1+ 2 n_B (k_0) \right] \rho_{ij}(k) \, .
%\label{spec_15}
%\ee

Let us come to $\Pi^R(k)$ and $\Pi^A(k)$. One has to interpret the commutator in (\ref{spec_7}) as an inverse transformation of the spectral function definition in (\ref{spec_5}). Inserting the representation for the $\theta(t)$
\be 
\theta(t)= i \int_{-\infty}^{\infty} \frac{d \omega}{2 \pi}  \frac{e^{-i \omega t}}{\omega + i\eta } \, ,
\label{spec_16}
\ee 
we obtain the following result for the retarded correlator
\bea 
\Pi^R(k)&=& i  \int \frac{d^4x}{(2 \pi)^4} 2 \theta(t) e^{i k \cdot x} \, \int \frac{d^4p}{(2 \pi)^4} \rho_{ij}(p) \, e^{-i p \cdot x} 
\nn
\\
&=& \int_{-\infty}^{\infty} \frac{dp_0}{\pi} \frac{\rho_{ij}(p_0,\bm{k})}{p_0-k_0-i\eta} \, .
\label{spec_17}
\eea
A similar expression for the advanced correlator holds
\bea
\Pi^A(k)=% -i  \int \frac{d^4x}{(2 \pi)^4} 2 \theta(-t) e^{i k \cdot x} \, \int \frac{d^4p}{(2 \pi)^4} \rho_{ij}(p) \, e^{-i p \cdot x} 
%\nn
%\\
 \int_{-\infty}^{\infty} \frac{dp_0}{\pi} \frac{\rho_{ij}(p_0,\bm{k})}{p_0-k_0+i\eta} \, .
\label{spec_18}
\eea
Making use of 
\be 
\frac{1}{x \pm i\eta}=P \left( \frac{1}{x}\right) \mp i\pi \delta(x) \, , 
\label{spec_19}
\ee
we obtain, assuming $\rho_{ij}$ real, the relations
\be 
{\rm{Im}} \, \Pi^R(k)=\rho_{ij}(k) \, , \quad{\rm{Im}} \, \Pi^A(k)=-\rho_{ij}(k) \, .
\label{spec_20}
\ee
We move to the Euclidean correlator. Its expression in terms of the spectral function reads~\cite{LaineBasics, LeBellac}
%They can be expressed in terms of the spectral function as well. Details are given in the appendix , whereas we provide here the final expressions that read
%\bea
%\Pi^T_{ij}(k)&=& \int_{-\infty}^{\infty} \frac{dP_0}{\pi} \frac{i \, \rho_{ij}(p_0,\bm{k})}{p_0-k_0-i\eta} + 2 \rho_{ij}(k_0,\bm{k}) n_B(k_0)  
%\\
%&-=& i \Pi^R(k)+  \Pi^<_{ij}(k)  \, ,
%\label{spec_21}
%\eea
%and 
\be 
\Pi^E_{ij}(K)= \int_{-\infty}^{\infty} \frac{dp_0}{\pi}\frac{\rho_{ij}(p_0,\bm{k})}{p_0-ik_n} \, ,
\label{spec_22}
\ee
and comparing (\ref{spec_17}) and (\ref{spec_22}), we can write
\be 
\Pi^R_{ij}(k) = \Pi^E_{ij}(k_n \to -i k_0, \bm{k}) \, .
\label{spec_23}
\ee
The relation in (\ref{spec_23}) captures the meaning of the analytic continuation from the ITF, related to the Matsubara sums, and the physical Minkowskian space-time, related to the RTF \cite{Cuniberti:2001hm, Burnier:2011jq}. The two-point correlators defined in this section may be seen as propagators, at zeroth order in perturbation theory, or as self-energies so that a loop expansion can be implemented. 

We notice that the spectral representation of the Euclidean correlator in (\ref{spec_22}) can be inverted, once performed the limit $p_0 \to k_0$, by using (\ref{spec_19}). One finds 
\bea 
\rho_{ij}(k)&=&\frac{1}{2i} {\rm{Disc}} \, \Pi^E_{ij}(k_n \to -i k_0, \bm{k})
\nn
\\
&=&\frac{1}{2i} \left[ \Pi^E_{ij}(-i (k_0+i\eta), \bm{k}) - \Pi^E_{ij}(-i (k_0-i\eta), \bm{k}) \right]
\nn
\\
&=&\left. {\rm{Im}} \Pi^E_{ij}(K) \right|_{k_n \to -i [k_0+i\eta]} \, .
\label{spec_24}
\eea

The set of correlators in eqs.~(\ref{spec_3})-(\ref{spec_5}) can be defined for fermion field operators as well. Since there is no conceptual difference but the fact that one deals with anti-commutating field, we do not show them here. A detailed discussion is found in~\cite{LaineBasics}.

\section{Particle production rates: right-handed neutrinos in a heat bath}
\label{sec_therm_5}
In this final section of the chapter we derive an observable within the thermal field theory formalism: the particle production rate in a thermal bath. This observable is relevant to the topics and content of the present thesis. We are going to show the framework adopted in one of the first quantitative derivation of the right-handed neutrino production rate at finite temperature \cite{Asaka:2006rw, Laine:2011pq}.

Our starting point is a physical system where some particles interact strongly enough to keep thermal equilibrium, whereas some others interact weakly and cannot maintain thermal equilibration (their distribution is not the equilibrium one). The latter particles decouple from the thermal bath, either escaping the system if it is of finite size or staying within the system without interacting anymore. We can quote typical examples: the decoupling of weakly interacting particles in the early universe like dark matter or heavy particle responsible for baryogenesis, the production of electromagnetic hard probes like muons or photons in the QGP generated in heavy-ion collisions experiments.

Let us start by considering a concrete model: the addition of right-handed (sterile) neutrinos to the SM Lagrangian. This model may account for a successful leptogenesis and also it provides a dark matter candidate (within the $\nu$MSM). The Lagrangian has been already discussed in chapter~\ref{chap:lepto} when we introduced leptogenesis in eq.~(\ref{lepto_8}). Here we consider only one right-handed neutrino species, $\nu_{R}$, embedded in the Majorana field $\psi=\nu_{R}+\nu_{R}^c$ and the Lagrangian reads
\begin{equation}
\mathcal{L}=\mathcal{L}_{\rm{SM}} + \frac{1}{2} \bar{\psi} i \slashed{\partial}  \psi  
- \frac{M}{2} \bar{\psi}\psi - F_{f }\bar{L}_{f} \tilde{\phi} P_{R}\psi  - F^{*}_{f}\bar{\psi} P_{L} \tilde{\phi}^{\dagger}  L_{f} \, ,
\label{prodrate_1}
\end{equation}  
where we suppress the index generation for the right-handed neutrino and $\mathcal{L}_{\rm{SM}}$ contains the thermalized degrees of freedom. \textit{The goal is to derive an equation that relates the right-handed neutrino production rate to Green's functions at finite temperature}.

We consider the density matrix $\hat{\rho}$ describing all the degrees of freedom in the thermal bath, the thermalized SM particles and the right-handed neutrinos. We denote the density matrix with a ``hat" in order not to confuse it with the spectral function, $\rho$ (accordingly we denote the Hamiltonian as $\hat{H}$ in this section). The time evolution for the density matrix can be written as follows
\be 
i \frac{d \hat{\rho}(t)}{d t} = \left[ \hat{H}, \hat{\rho} (t)\right] \, , 
\label{prodrate_2}
\ee  
where $\hat{H}$ is the full Hamiltonian that can be split as follows
\be 
\hat{H}= \hat{H}_{\rm{SM}} + \hat{H}_{\psi} + \hat{H}_{{\rm{int}}} \, .
\label{prodrate_3}
\ee
In (\ref{prodrate_3}) $\hat{H}_{\rm{SM}}$ refers to the SM degrees of freedom, $\hat{H}_{\psi}$ is the free Hamiltonian for the right-handed neutrinos and $\hat{H}_{{\rm{int}}}$ describes the interactions between the right-handed neutrinos and SM particles and reads
\be 
\hat{H}_{{\rm{int}}} = \int d^3 \bm{x} \left[ F_{f }\bar{L}_{f} \tilde{\phi} P_{R}\psi +  F^{*}_{f}\bar{\psi} P_{L} \tilde{\phi}^{\dagger}  L_{f} \right]   \, .
\label{prodrate_4}
\ee
All the fields have to be interpreted as field operators. 

In order to solve the equation for the density matrix (\ref{prodrate_2}), we need to set an initial condition. We assume that the initial population of right-handed neutrinos is zero and then we can write
\be 
\hat{\rho}(0) =  \hat{\rho}_{\rm{SM}} \times | \ket0 \bra0 | \, , 
\label{prodrate_5}
\ee
where $| \ket0 $ is the vacuum state for right-handed neutrinos and 
\be 
\hat{\rho}_{\rm{SM}}= \frac{1}{Z_{\rm{SM}}} e^{-\hat{H}_{\rm{SM}} \beta} \, ,
\label{prodrate_6}
\ee
where $Z_{\rm{SM}}$ stands for the SM partition function and $\beta=1/T$. 
We have to formulate the evolution of the density matrix in terms of quantities in the interaction picture, hence we split the Hamiltonian in the free $\hat{H}_{0}=\hat{H}_{\rm{SM}} + \hat{H}_{\psi} $ and interacting part, $\hat{H}_{\rm{int}}$.
We put in the free part the whole $\hat{H}_{\rm{SM}}$ because the interactions among SM degrees of freedom cannot change the number of right-handed neutrinos. Then one obtains 
\be 
i \frac{d \hat{\rho}_{I}(t)}{d t} = \left[ \hat{H}_{I}, \hat{\rho}_{I} (t)\right] \, ,
\label{prodrate_7}
\ee
where 
\be 
\hat{\rho}_{I} \equiv e^{i\hat{H}_0 t} \hat{\rho}  e^{-i\hat{H}_0t} \, , \quad \hat{H}_{I} \equiv e^{i\hat{H}_0t} \hat{H}_{\rm{int}}  e^{-i\hat{H}_0t} \, , 
\label{prodrate_8}
\ee
are the density matrix and interaction Hamiltonian in the interaction picture respectively.  
Now we can use perturbation theory with respect to $\hat{H}_I$ and obtain for the density matrix 
\be 
\hat{\rho}_I(t)= \hat{\rho}(0) -i \int_0^t dt' [\hat{H}_I(t'),\hat{\rho}_0 ] + (-i)^2  \int_0^t dt' \int_0^{t'} dt''  [\hat{H}_I(t'),[\hat{H}_I(t''),\hat{\rho}_0 ]] + \cdots \, ,
\label{prodrate_9}
\ee 
where $\hat{\rho}(0) \equiv \hat{\rho}_I(0)$ and the dots stand for higher order terms in the perturbative expansion. We stress that the perturbative series breaks down if the abundance of right-handed neutrinos is too close to the equilibrium one, and we cannot rely anymore on small changes of the initial vanishing abundance. We have to assume $t < t_{\rm{eq}}$, where $t_{\rm{eq}}$ is the time for the right-handed neutrino equilibration. 

Now we make the connection between the  density matrix and the number operator of the right-handed neutrinos. Let us write down the Fourier decomposition of the Majorana neutrino fields as follows
\be 
\psi(x)= \int \frac{d^3\bm{k}}{(2\pi)^3 } \frac{1}{2 E_k}  \sum_s \left[ a_s(\bm{k}) u(\bm{k},s) \, e^{-ik \cdot x} + a_s^{\dagger}(\bm{k}) v (\bm{k},s) \, e^{+i k \cdot x} \right] \, ,
\label{prodrate_10}
\ee
\be 
\bar{\psi}(x)= \int \frac{d^3\bm{k}}{(2\pi)^3 }  \frac{1}{2 E_k} \sum_s \left[ a^{\dagger}_s(\bm{k}) \bar{u}(\bm{k},s) \, e^{+ik \cdot x} + a_s(\bm{k}) \bar{v} (\bm{k},s) \, e^{-i k \cdot x} \right] \, ,
\label{prodrate_11}
\ee 
where the sum is understood over the spin polarizations of the Majroana fermion, then $k_0=\sqrt{\bm{k}^2+M^2}$ and the spinor sums read
\be 
\sum_s  \, u(\bm{k},s) \bar{u}(\bm{k},s) = \slashed{k} +M \, , \quad \sum_s  \, v(\bm{k},s) \bar{v}(\bm{k},s) = \slashed{k} -M \, . 
\label{prodrate_12}
\ee
The creation and annihilation operators satisfy 
\be 
\left\lbrace a_s(\bm{k}), a^{\dagger}_r(\bm{q})\right\rbrace = (2\pi)^3 \delta^3(\bm{k-q})  \delta_{sr} \, ,
\label{prodrate_13}
\ee
and we notice that only one type of creation operator is needed because of the Majorana nature of the fermion.
Then the right-handed neutrino number operator can be defined as follows \cite{Asaka:2006rw}
\be 
\frac{d n_{\nu_R}}{d^3\bm{x} \, d^3\bm{k}} \equiv \frac{1}{V} \sum_s a^{\dagger}_s(\bm{k}) a_s(\bm{k})  \, ,
\label{prodrate_14}
\ee
and the distribution function that really gives the number of right-handed neutrinos in the thermal bath per $d^3\bm{x}$ and $d^3\bm{k}$ reads 
\be 
\frac{d n_{\nu_R}(\bm{x},\bm{k})}{d^3\bm{x} \, d^3\bm{k}}= \tr \left[\frac{d n_{\nu_R}}{d^3\bm{x} \, d^3\bm{k}} \hat{\rho}_I(t) \right] \, .
\label{prodrate_15}
\ee
Indeed an equivalent form for the thermal expectation value of a given operator in (\ref{therm_01}) can also be written as $\langle A \rangle_\beta=\tr \left\lbrace \hat{\rho} A \right\rbrace $, due to the definition $\hat{\rho}=1/Z e^{-\beta \hat{H}}$. We plug in (\ref{prodrate_15}) the time derivative of the perturbative expansion for $\hat{\rho}_I(t)$ in (\ref{prodrate_9}). The first term leads to a time-independent result, the second term is linear in the creation and annihilation operator and then we still obtain a vanishing quantity. The first non-trivial term is the one comprising the interaction Hamiltonian twice and gives 
\be 
\frac{d n_{\nu_R}(x,\bm{k})}{d^4 x \, d^3\bm{k}}= -\frac{1}{V} \tr \left\lbrace \sum_s a^{\dagger}_s(\bm{k}) a_s(\bm{k}) \int_0^t dt  [\hat{H}_I(t),[\hat{H}_I(t'),\hat{\rho}_0 ]]  \right\rbrace \equiv R(T,\bm{k}) \, ,
\label{prodrate_16}
\ee
where $R(T,\bm{k}) $ is called \textit{particle production rate} and it is understood as a function of the three-momentum of the right-handed neutrinos and the temperature of the plasma. The expression in (\ref{prodrate_16}) is obtained at order $F^2$ in the Yukawa couplings. Inserting the field operators (\ref{prodrate_10}) and  (\ref{prodrate_11}) in the interaction Hamiltonian in (\ref{prodrate_4}) we can rewrite it as follows
\be 
\hat{H}_{\rm{int}}= \int d^3 \bm{x}  \int \frac{d^3\bm{k}}{(2\pi)^3 } \frac{1}{2 E_k} \sum_s \left[ \bm{J}^{\dagger}_{\bm{k},s} (x) a_s(\bm{k}) e^{-i k \cdot x} + a^{\dagger}_s(\bm{k}) \bm{J}_{\bm{k},s} (x)  e^{i k \cdot x} \right]  \, ,
\label{prodrate_17}
\ee
where the following definitions hold
\be 
\bm{J}_{\bm{k},s} (x)=-F_{f'} \bar{j}_{f'}(x) P_R v(\bm{k},s) + F_f^{*}  \bar{u} (\bm{k},s) P_L j_{\alpha}(x) \, ,
\label{prodrate_18}
\ee
\be 
\bm{J}^{\dagger}_{\bm{k},s} (x) = - F_f^{*}  \bar{v} (\bm{k},s) P_L j_{f}(x) + F_{f'} \bar{j}_{f'}(x) P_R u(\bm{k},s)   \, ,
\label{prodrate_19}
\ee
with 
\be 
j_f(x)= \tilde{\phi}^{\dagger}(x) L_{f}(x) \, , \quad  \bar{j}_{f'}(x)=  \bar{L}_{f'}(x) \tilde{\phi}(x) \, .
\label{prodrate_20}
\ee
Now the setting is complete and we list the steps to be performed with the intermediate results:
\begin{itemize}
\item[1)] we insert the interaction Hamiltonian (\ref{prodrate_17}) in the particle production rate in (\ref{prodrate_16}), and we get rid of the right-handed neutrino creation and annihilation operators by using (\ref{prodrate_13}). In the trace evaluation only terms of the type $\bra 0 | a a^{\dagger} a a^{\dagger} | \ket0 $ survive and one finds 
\bea
&&R(T,\bm{k})=\frac{1}{V} \frac{1}{(2 \pi)^3 2 E_{k}} \sum_{s} \int_0^t dt' \int d^3 \bm{x}  \int d^3 \bm{x}' 
\nn
\\
&&\hspace{1 cm} \times \langle \bm{J}^{\dagger}_{\bm{k},s} (x')\bm{J}_{\bm{k},s} (x) e^{i k \cdot (x-x')} + \bm{J}^{\dagger}_{\bm{k},s} (x)\bm{J}_{\bm{k},s} (x') e^{-i k \cdot (x-x') } \rangle \, .
\label{prodrate_21}
\eea 
The expectation value now refers exclusively to $\hat{\rho}_{\rm{SM}}$.
\item[2)] Substituting the expression (\ref{prodrate_18}) and (\ref{prodrate_19})  in  (\ref{prodrate_21}), we notice that correlators of the type $\langle j_{f'}(x') j_f(x) \rangle$ and  $\langle \bar{j}_{f'}(x') \bar{j}_f(x) \rangle$ vanish. This is due to the conservation of lepton number in the SM.  The particle production rate becomes
\bea
&&R(T,\bm{k})=\frac{1}{V} \frac{1}{(2 \pi)^3 2 E_{k}} \sum_{s} \int_0^t dt' \int d^3 \bm{x}  \int d^3 \bm{x}'  F^*_f F_{f'}
\nn
\\
&&\hspace{0.7 cm} \times \langle  \left[ \bar{v} (\bm{k},s) P_L \, j_{f}(x') \, \bar{j}_{f'}(x) P_R v(\bm{k},s)   + \bar{j}_{f'}(x') \, P_R  u(\bm{k},s) \bar{u}(\bm{k},s) P_L \, j_{f}(x) \right]  
\nn
\\
&&\hspace{0.7 cm} \times  e^{i k \cdot (x-x')}  + (x \leftrightarrow x') \rangle  \, .
\label{prodrate_22}
\eea 
\item[3)] We can use the completeness relations for the spinors in (\ref{prodrate_12}). To this aim we have to write for example 
\be
\bar{v} (\bm{k},s) P_L j_{f}(x') \, j_{f'}(x) P_R v(\bm{k},s) = \tr \left\lbrace v(\bm{k},s) \bar{v} (\bm{k},s) P_L j_{f}(x') \, j_{f'}(x) P_R  \right\rbrace \, ,
\ee
and the sum over the spin polarization may be now used. The combination involving the particle spinor $u(\bm{k},s)$ is already in a form suitable for a direct evaluation of the spin sum. 
The terms proportional to the neutrino mass $M$ get projected out by the chiral projectors and the rate reads
\bea
&&R(T,\bm{k})=\frac{1}{V} \frac{1}{(2 \pi)^3 2 E_{k}} \sum_{s} \int_0^t dt' \int d^3 \bm{x}  \int d^3 \bm{x}'  F^*_f F_{f'}
\nn
\\
&&\hspace{0.7 cm} \times \langle   \left\lbrace  (P_R \, \slashed{k} \, P_L)^{\beta \alpha}  \left[  j^\alpha_{f}(x') \, \bar{j}^{\beta}_{f'}(x) + \bar{j}^\beta_{f'}(x') \, j^{\alpha}_{f}(x) \right] \right\rbrace   e^{i k \cdot (x-x')} \rangle 
\nn
\\
&&\hspace{0.7 cm} + (x \leftrightarrow x') \, .
\label{prodrate_22}
\eea  
where we have now made explicit the Lorentz index carried by the lepton doublet operator and contracted with the corresponding right-handed neutrino Lorentz index. 
\item[4)] We have to rewrite the two-point correlator 
\be 
\langle j^\alpha_{f}(x') \, \bar{j}^\beta_{f'}(x) + \bar{j}^\beta_{f'}(x') \, j^\alpha_{f}(x) \rangle \, .
\label{prodrate_23}
\ee
This can be understood as self energies of the right-handed neutrino and using the Wightman correlation functions for fermion fields, we write (these expression are the fermionic counterpart of (\ref{spec_3}) and (\ref{spec_4}) for bosonic operators)
\bea 
&&\Pi^{>, \, \alpha \beta}_{f f'}(k)\equiv \int \frac{d^4 x}{(2 \pi)^4} e^{ik \cdot (x-x')} \langle j^\alpha_{f}(x) \, \bar{j}^\beta_{f'}(x') \rangle
\label{prodrate_24}
\\ 
&&\Pi^{<, \, \alpha \beta}_{f f'}(k)\equiv \int \frac{d^4 x}{(2 \pi)^4} e^{ik \cdot (x-x')} \langle - \bar{j}^{\beta}_{f'}(x') j^\alpha_{f}(x) \,  \rangle \, .
\label{prodrate_25}
\eea
Inverting the relations (\ref{prodrate_24}) and (\ref{prodrate_25}) after exploiting translational invariance to recast the space-time arguments as written in (\ref{prodrate_23}), we obtain for the combination in (\ref{prodrate_23}) the following expression
\be 
\langle j^\alpha_{f}(x') \, \bar{j}^\beta_{f'}(x) + \bar{j}^\beta_{f'}(x') \, j^\alpha_{f}(x) \rangle = \int \frac{d^4 q}{(2\pi)^4} e^{-iq \cdot (x-x')} \left[ \Pi^{>, \, \alpha \beta}_{ff'}(-q) - \Pi^{<, \, \alpha \beta}_{ff'}(q) \right] \, .
\label{prodrate_26}
\ee
\item[5)] We still have to perform the integration over the time and space coordinates that appear in the production rate. The result reads, taking the limit for large time $t$
\be 
\lim_{t \to \infty} \int d^3 \bm{x}  \int d^3 \bm{x}'  \int_0^t dt' \left[ e^{i(k-q) \cdot (x-x')} +e^{-i(k-q) \cdot (x-x')} \right] = V (2 \pi)^4 \delta^4(k-q) \, ,
\label{prodrate_27}
\ee
that allows to cancel the factor $1/V$ in the production rate and remove the integration on the momentum $q$ in (\ref{prodrate_26}). Then we find ($E_k = k_0$)
\be 
R(T,\bm{k})= \frac{1}{(2\pi)^3 2 k_0}  F^*_f F_{f'}  \tr \left\lbrace \slashed{k} P_L \left[ \Pi^{>}_{f f'}(-k) - \Pi^{<}_{ff'}(k)\right] P_R   \right\rbrace  \, ,
\label{prodrate_28}
\ee
where we have written the trace over the Lorentz indices for the spinor part.
\item[6)] Using the definition of the self-energies in terms of the spectral function $\rho_{ff'}$ 
\bea 
&&\Pi^{>}_{f f'}(-k) = 2\left[ 1 - n_F(-k_0) \right] \rho_{ff'}(-k)=2n_F(k_0)  \rho_{ff'}(-k) \, ,
\label{prodrate_29}
\\
&&\Pi^{<}_{f f'}(k)= -2n_F(k_0)  \rho_{ff'}(k) \, ,
\label{prodrate_30}
\eea
we finally obtain 
\be 
R(T,\bm{k})= \frac{n_F(k_0)}{(2\pi)^3  k_0}  \sum_{f=1}^3 |F_f|^2   \tr \left\lbrace \slashed{k} P_L \left[ \rho_{f f}(-k)  +\rho_{ff}(k) \right] P_R   \right\rbrace \, ,
\label{prodrate_31}
\ee
because within the SM the lepton flavour conservation forces $f=f'$.  
\end{itemize}
If the thermal plasma is charge symmetric $\rho_{f f}(-k)=\rho_{ff}(k)$ and the two terms in (\ref{prodrate_31}) can be combined.  

The result in (\ref{prodrate_31}) is the master equation adopted in recent works that address  the right-handed neutrinos  production in a thermal plasma of SM particles \cite{Asaka:2006rw, Laine:2011pq}. This observable is of particular relevance for understanding quantitatively leptogenesis and/or the dark matter in the early universe.    
We make one further comment: the spectral function can be traced back to self-energies at finite temperature, in particular as we shown in (\ref{spec_23}) and (\ref{spec_24})
\be 
\rho = {\rm{Im}} \Pi_R = {\rm{Im}} \Pi^E_{k_n \to -i \left[ k_0 + i\eta  \right]  } \, .
\label{prodrate_32}
\ee
%Together with the assumption that the medium is charge symmetric, we can set $\rho_{f f}(-k)  =\rho_{ff}(k)$ and hence we can trace back the particle production rate to the retarded self-energy (or Euclidean correlator) as follows
%\be 
%R(T,\bm{k})= \frac{n_F(k_0)}{(2\pi)^3  k_0}  \sum_{f=1}^3 |F_f|^2    
%\ee
Hence, defining a differential decay rate in accordance with \cite{Salvio:2011sf} as 
\be 
\frac{d n_{\nu_R}(k)}{d^4 x \, d^3 \bm{k}} \equiv \frac{2 n_F(k^0)}{(2\pi)^3} \Gamma(k) \, ,
\label{prodrate_33}
\ee 
we can establish a correspondence between the imaginary part of the self-energies, $\Pi^E$ or $\Pi_R$, and a \textit{thermal  width} as follows
\be 
\Gamma(k)=\frac{1}{k_0} \, {\rm{Im}} \Pi^E(K)_{k_n \to -i \left[ k_0 + i\eta  \right]  } =  {\rm{Im}} \Pi_R(k) \, ,
\label{prodrate_34}
\ee
where we have used the relation given in (\ref{prodrate_32}). The thermal width for a non-relativistic Majorana neutrino at order $|F|^2$ and at leading order in the SM couplings reads \cite{Salvio:2011sf,Laine:2011pq}
\bea
\Gamma(k) = \frac{|F|^2 M_1}{8 \pi \sqrt{\bm{k}^2+M^2}}\left\lbrace 1 -\lambda \frac{T^2}{M^2} -|\lambda_t|^2 \left[ \frac{21}{2(4 \pi)^2} + \frac{7 \pi^2}{60} \left(\frac{T^4}{M^4} +\frac{4}{3} \frac{\bm{k}^2 T^4}{M^6} \right)   \right]  \right. \nn 
\\ 
\left. \phantom{x} +(3g^2+g'^2) \left[\frac{29}{8(4 \pi)^2} -\frac{\pi^2}{80} \left(\frac{T^4}{M^4} +\frac{4}{3} \frac{\bm{k}^2 T^4}{M^6} \right)   \right] \right\rbrace  \, .
\label{result_widthLaine}
\eea
The thermal width is the object we exploit to derive the neutrino production rate within the EFT approach. In particular the thermal width can be understood as the pole of the non-relativistic heavy Majorana neutrino propagator in the low-energy theory, as we are going to show in chapter~\ref{chap:part_prod}.

%% file: part_prod.tex
The subject of this chapter is the dynamics of a heavy Majorana neutrino in a thermal bath of SM particles. In particular we aim at treating the problem from an EFT prospective: figure out the different energy scales appearing in the system, pick the suitable degrees of freedom to describe the physics at a given scale and construct a low-energy Lagrangian specifying its parameters. Our assumption is that the heavy-neutrino mass is much larger than the temperature of the thermal bath. Therefore, as discussed in chapter~\ref{chap:eff_the}, the temperature scale can be set to zero in the matching as well as any other low-energy scale. The heavy neutrinos are non-relativistic particles in the EFT and we derive the corresponding Majorana propagator in section~\ref{sec_maj}. Then we study the operator content of the EFT Lagrangian describing the interactions between the non-relativistic excitations of the heavy Majorana neutrino and SM particles (Higgs boson, fermions and gauge bosons) in section~\ref{EFT_partprod}, together with the expressions of the Wilson coefficients. As a proof of concept and a non-trivial application of the EFT we derive the thermal width at order $|F|^2$ and at leading order in the SM couplings in section~\ref{thermwidth_partprod}. Finally in section \ref{expansion_partprod} the convergence of the $T/M$ expansion to the exact result is addressed, being a critical issue of the presented EFT approach. 

\section{Non-relativistic Majorana fermions}
\label{sec_maj}
In this section, we derive some general properties of a free Majorana fermion in 
the limit where its mass $M$ is much larger than the energy and momentum of any other 
particle in the system. Our aim is to identify the low-energy modes,
write the Majorana free propagator and construct the corresponding Lagrangian.
Low-energy modes are those that may be excited at energies below $M$. 
In the next sections, we will identify the Majorana fermion studied here with a Majorana neutrino, 
and the low-energy degrees of freedom with the low-energy modes 
of the neutrino and the SM particles.

If $\psi$ is a spinor describing a relativistic Majorana particle, then 
\begin{equation}
\psi=\psi^{c}=C  \bar{\psi}^{\,T} \, ,
\label{eq:Majodef}
\end{equation}
where $\psi^{c}$ denotes the charge-conjugate spinor and $C$  
the charge-conjugation matrix that satisfies $C^\dagger=C^T=C^{-1}=-C$ and $C\,\gamma^{\mu\,T}\,C = \gamma^\mu$.\footnote{
A possible choice for $C$ is $C=-i\gamma^{2}\gamma^{0}$.} 
Thus a Majorana spinor has only two independent components.
It is different from a Dirac spinor that has instead four independent components
corresponding to a distinguishable particle and antiparticle.
The relativistic propagators for a free Majorana particle are:
\begin{eqnarray}
\langle 0 | T( \psi^{\alpha} (x) \bar{\psi}^{\beta} (y)  )| 0 \rangle &=& 
i \int \frac{d^{4}p}{(2 \pi)^{4}} \, \frac{(\slashed{p}+M)^{\alpha \beta}}{p^{2}-M^{2}+i\eta}  \, e^{-ip \cdot (x-y)} \,,
\label{eq5_partprod}
\\
\langle 0 | T( \psi^{\alpha}(x) \psi^{\beta} (y) )| 0 \rangle &=& 
-i \int \frac{d^{4}p}{(2 \pi)^{4}} \, \frac{\left[  (\slashed{p}+M) C \right]^{\alpha \beta}  }{p^{2}-M^{2}+i\eta} \,  e^{-ip \cdot (x-y)} \,,
\label{eq6_partprod}
\\
\langle 0 | T( \bar{\psi}^{\alpha} (x) \bar{\psi}^{\beta} (y)  ) | 0 \rangle &=& 
-i \int \frac{d^{4}p}{(2 \pi)^{4}} \, \frac{ \left[  C (\slashed{p}+M) \right]^{\alpha \beta}}{p^{2}-M^{2}+i\eta} \, e^{-ip \cdot (x-y)} \,,
\label{eq7_partprod}
\end{eqnarray}
where $\alpha$ and $\beta$ are Lorentz indices and $T$ stands for the time-ordered product.
Note that, due to the Majorana nature of the fermions and at variance with the Dirac fermion case, 
the combinations $\langle 0 | \psi\psi| 0 \rangle $ and $\langle 0 | \bar{\psi}\bar{\psi}| 0 \rangle$ 
do not vanish. This is a feature that has to be accounted for in the 
relativistic theory when computing amplitudes, since Majorana fields 
may be contracted with vertices involving either particle or antiparticle fields.

In order to identify the low-energy modes of a heavy Majorana field, $\psi$,  
let us assume first that $\psi$, rather than a Majorana field, is a Dirac field describing a heavy quark.
Low-energy modes of a non-relativistic Dirac field have been studied in the framework of HQEFT~\cite{Neubert:1993mb} as we have briefly discussed in chapter~\ref{chap:eff_the} (see section~\ref{eff_sec_3}). We repeat part of the discussion here by rearranging slightly the notation in order to show differences and similarities with Majorana fermions. 
In a given reference frame, the momentum of a non-relativistic heavy quark of mass $M$ is $Mv^\mu$, where $v^2=1$,  
up to fluctuations whose momenta, $k^\mu$, are much smaller than $M$. These fluctuations may come 
from the interactions with other particles that, by assumption, carry energies and momenta 
much smaller than $M$. The Dirac field describing a heavy quark can be split 
into a large component, $\psi_{>}$, whose energy is of order $M$, 
and a small component, $\psi_{<}$, whose energy is much smaller than $M$:
\begin{equation}
\psi=\left(\frac{1+\slashed{v}}{2} \right) \psi + \left(\frac{1-\slashed{v}}{2} \right) \psi \equiv \psi_{<} + \psi_{>} \,.
\label{psi}
\end{equation}
According to the above definition: $({1+\slashed{v}})/{2} \times \psi_{<} = \psi_{<}$ and $({1-\slashed{v}})/{2} \times \psi_{>} = \psi_{>}$. 
The small component field, $\psi_{<}$, is eventually matched into the field $h$ of HQET.
This is the field, made of two independent components, that describes in HQET 
the low-energy modes of the heavy quark. It satisfies
\begin{equation}
\frac{1+\slashed{v}}{2} h = h  \,.
\end{equation}
The field $h$ annihilates a heavy quark but does not create an antiquark. 
It satisfies the following equal time anti-commutation relations \cite{Dugan:1991ak}:
\begin{eqnarray}
\left\lbrace h^{\alpha}(t, \bm{x}) ,h^{\beta}(t, \bm{y}) \right\rbrace &=& 
\left\lbrace \bar{h}^{\alpha }(t, \bm{x}) , \bar{h}^{\beta }(t, \bm{y}) \right\rbrace = 0 \,,
\label{7ab_partprod}
\\
\left\lbrace h^{\alpha}(t, \bm{x}), \bar{h}^{\beta }(t, \bm{y}) \right\rbrace &=&  
\frac{1}{v^0} \left( \frac{1+\slashed{v}}{2} \right)^{\alpha \beta} \delta^{3}(\bm{x}-\bm{y}) \,.
\label{7c_partprod}
\end{eqnarray}
The charge conjugated of \eqref{psi} is 
\begin{equation}
\psi^c=\left(\frac{1-\slashed{v}}{2} \right)(C\gamma^{0} \psi^{*}_{<}) + \left(\frac{1+\slashed{v}}{2} \right)(C\gamma^{0} \psi^{*}_{>}) \,,
\label{eq9_partprod}
\end{equation}
whose small component, $C\gamma^{0} \psi^{*}_{>}$, may be eventually matched into a 
HQET field, made again of two independent components, that describes the low-energy modes of a heavy antiquark. 
Clearly this field is independent from the one describing the heavy quark:
it annihilates a heavy antiquark but does not create a quark.
It satisfies similar equal time anti-commutation relations as the field $h$.

Let us now go back to consider $\psi$ a field describing a heavy Majorana particle whose momentum 
in some reference frame is $Mv^\mu$ up to fluctuations, $k^\mu$, that are much smaller than~$M$.
Like in \eqref{psi} we may decompose the four-component Majorana spinor into a large and a small component. 
From \eqref{eq:Majodef} it follows, however, that in this case \eqref{psi} and \eqref{eq9_partprod} describe the same field, hence 
\begin{equation}
\psi_{<}= C\gamma^{0}\psi^{*}_{>} \, , \quad  \psi_{>}= C\gamma^{0}\psi^{*}_{<}\,.
\label{eq9a_partprod}
\end{equation}
This implies that the small component of the Majorana particle field 
coincides with the small component of the Majorana antiparticle field.
In the EFT that describes the low-energy modes of non-relativistic Majorana fermions, 
both the particle and antiparticle modes are described by the same field $N$. 
The field $N$ matches $\psi_{<}$ in the fundamental theory and fulfils
\begin{equation}
\frac{1+\slashed{v}}{2} N = N  \,.
\end{equation}
This is consistent with the Majorana nature of the fermion: we cannot
distinguish a particle from its antiparticle. Note that, while in the fundamental theory 
a Majorana fermion and antifermion are described by the same spinor $\psi$ that is self conjugated, 
in the non-relativistic EFT a Majorana fermion and  antifermion are 
described by the same spinor $N$ that is not self conjugated but 
has by construction only two independent components.
Analogously to the field $h$ in HQET, the field $N$ annihilates a heavy Majorana fermion 
(or antifermion). It satisfies the following equal time anti-commutation relations:
\begin{eqnarray}
\left\lbrace N^{\alpha}(t, \bm{x}) , N^{\beta}(t, \bm{y}) \right\rbrace &=& 
\left\lbrace \bar{N}^{\alpha}(t, \bm{x}) , \bar{N}^{\beta}(t, \bm{y}) \right\rbrace = 0 \,,
\\
\left\lbrace N^{\alpha}(t, \bm{x}) , \bar{N}^{\beta}(t, \bm{y}) \right\rbrace &=& 
\frac{1}{v^0} \left( \frac{1+\slashed{v}}{2} \right)^{\alpha \beta} \delta^{3}(\bm{x}-\bm{y}) \,,
\end{eqnarray}
which may be also derived from the full relativistic expression of the Majorana 
spinors given in~\cite{Mannheim:1980eb}.
Finally, we provide the expression for the non-relativistic Majorana propagator. 
Starting from eqs.~(\ref{eq5_partprod})-(\ref{eq7_partprod}), projecting on the small components of the Majorana fields 
and putting $p^{\mu}=Mv^{\mu}+k^{\mu}$, where $k^2 \ll M^2$, we obtain in the large $M$ limit (keeping only the $(1/M)^0$ term)
\begin{equation}
\langle 0 | T ( N^{\alpha}(x) \bar{N}^{\beta}(y) )  | 0 \rangle 
= \left( \frac{1+\slashed{v}}{2}\right)^{\alpha \beta} \int \frac{d^{4}k}{(2 \pi)^{4}} \, e^{-ik(x-y)}\, \frac{i}{v\cdot k +i\eta}  \,  ,
\label{effpropagator}
\end{equation} 
whereas the other possible time-ordered combinations vanish as they contain only creation or annihilation operators. 
The corresponding Lagrangian for a free Majorana fermion is like the HQET Lagrangian in the static limit: 
\begin{equation}
\mathcal{L}^{(0)}_{\hbox{\tiny N}} = \bar{N}\, i v \cdot \partial \, N \,.
\label{effLag00}
\end{equation}
An analysis of heavy Majorana fermions in an EFT framework analogous to the one presented in this section 
can be also found in~\cite{Kopp:2011gg,Hill:2011be}.

\section{EFT for non-relativistic Majorana neutrinos}
\label{EFT_partprod}
Starting from this section we will assume an extension of the SM that has been 
implemented in several leptogenesis scenarios~\cite{Fukugita:1986hr,Luty:1992un,Buchmuller:2005eh,Davidson:2008bu}.
It consists of the addition to the SM of some sterile neutrinos with 
masses much larger than the electroweak scale.\footnote{
A similar model but with neutrinos not heavier than the electroweak scale is in~\cite{Asaka:2005pn,Asaka:2005an}.
}
The mechanism has been discussed in some detail in chapter~\ref{chap:lepto}. Assuming that we have well separated neutrino masses, the production of a net lepton asymmetry 
starts when the lightest of the sterile neutrinos, whose mass, $M_1 \equiv M$, is above the electroweak scale,   
decouples from the plasma reaching an out-of-equilibrium condition. 
This happens when the temperature drops to $T \sim M$. 
During the universe expansion, the sterile neutrino continues to decay in the regime $T<M$. 
For $T<M$ the recombination process is almost absent (exponentially suppressed) and a net lepton asymmetry is generated. If the temperature of the system, $T$, is such that standard thermal 
leptogenesis is efficiently active, then $T$ is also above the electroweak scale. 
Hence the hierarchy of scales for the problem at hand reads
\be 
M \gg T \gg M_W \, ,
\label{hiera_width}
\ee 
and our aim is to device an EFT to reproduce the thermal width as given in eq.~\eqref{result_widthLaine} that in turn enters the definition of the right-handed neutrino production rate in eq.~(\ref{prodrate_33}). 
\subsection{Green's functions for $M \gg T$}
In chapter~\ref{chap:therm_tex} we made a clear connection between the imaginary part of self-energies at finite temperature and the width of a right-handed neutrino. In \cite{Laine:2011pq} the thermal width for a non-relativistic Majorana neutrino in a heat bath of SM particles was derived by using the ITF. It was extracted from the imaginary part of a Euclidean correlator, $\Pi^E(K)$, at finite temperature. Even though an operator product expansion (OPE) is mentioned and partially adopted in the calculation, the derivation is carried out within a fully relativistic thermal field theory. Therefore the temperature scale enters the computation from the very beginning through Matsubara sums and exponentially suppressed terms of the type $e^{-M/T}$ are kept in intermediate steps. 

The hierarchy of scales in (\ref{hiera_width}) allows for a simplification of the computational scheme in the non-relativistic case. In particular, the relation $M \gg T$ calls for an effective field theory treatment: we can device a quantum field theory that does not comprise ab initio exponentially suppressed terms like $e^{-M/T}$, being $T$ a small scale in the problem. This is a general characteristic of the EFT approach: in any analytic expansion performed within the low-energy theory exponentially suppressed terms vanish. 
\begin{figure}
\centering
\includegraphics[scale=0.59]{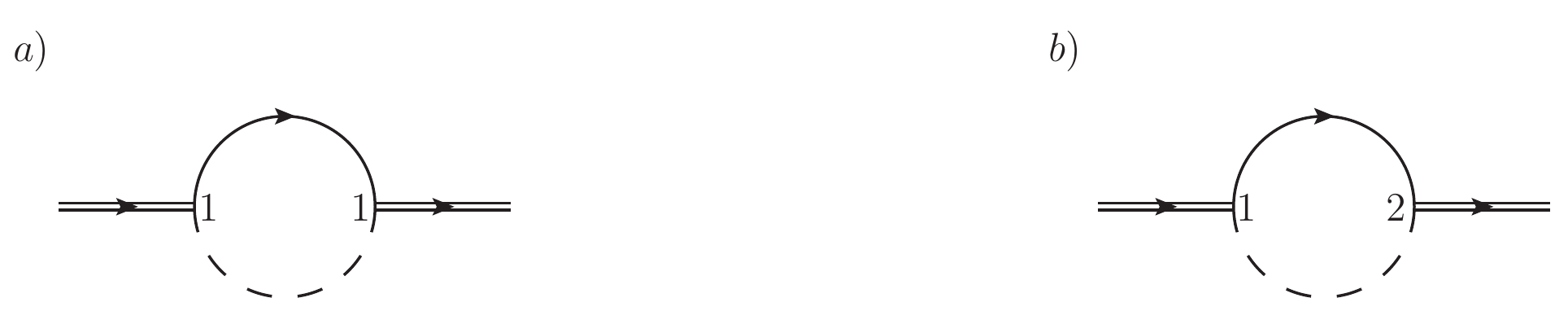}
\caption{\label{fig:selfRTF}The $\Pi_{11}$ and $\Pi_{12}$ self-energy diagrams in the RTF are shown.}
\end{figure}

Let us try to make the point by looking at the diagrams relevant for the neutrino thermal width at zeroth order in the SM couplings and shown in figure~\ref{fig:selfRTF}. We work in the RTF of thermal field theory. The imaginary part of the retarded self-energy is the relevant quantity for our scope. In turn, the retarded self energy may be written as $\Pi_R = \Pi_{11} + \Pi_{12}$, 
where $\Pi_{11}$ is the self energy when the initial and final neutrinos are on the physical branch of the Keldysh contour, 
and $\Pi_{12}$ is the self energy when the initial neutrino is on the physical branch whereas the final neutrino 
is on the complex branch of the Keldysh contour~\cite{LeBellac,DasBook}. Adopting the cutting rules at finite temperature one is able to extract the imaginary part of the diagrams, and the calculation would go ``thermal" as in \cite{Laine:2011pq}. However the separation of the energy scales, $M \gg T$, may have some impact. Let us consider first the loop diagram $a)$ in figure~\ref{fig:selfRTF}, $\Pi_{11}$. The lepton and Higgs boson propagator are the ``11" components of the 2$\times$2 scalar and fermion propagators given in (\ref{rtf_6}) and (\ref{imf_17}) respectively. In the case the incoming neutrino is taken at rest, $v^\mu=(1,\bm{0})$, they read in momentum space 
\bea
&&i \Delta_{11}(Mv-\ell)= \frac{i}{(Mv-\ell)^2 + i \eta} + 2 \pi \delta((Mv-\ell)^2)n_B(|M-\ell^0|) \, ,
\label{11higgs} 
\\
&&i S_{11}(\ell)= \slashed{\ell} \left[ \frac{i}{\ell^2 + i \eta} - 2 \pi \delta(\ell^2)n_F(|\ell^0|) \right] \, .
\label{11lepto}
\eea
The natural momentum scale in the loop is of order of the heavy neutrino mass and hence the lepton and Higgs boson momentum are of order $M$. Indeed we want to look at the heavy neutrino decay, so that the energy of order $M$ is shared between the two massless decay products. Relying on the limit $(M-\ell) \sim \ell \sim M \gg T$, the thermal parts in (\ref{11higgs}) and (\ref{11lepto}) become exponentially suppressed (for $T \to 0$ they vanish) and only the in-vacuum terms survive. This amounts at taking as vanishing the small scale in the problem according to the EFT approach, namely $T \to 0$. 

Now we consider the diagram $b)$ in figure~\ref{fig:selfRTF}, $\Pi_{12}$. In this case the ``12" Higgs and lepton propagator components enter the loop amplitude and they read
\bea
&&i \Delta_{12}(M-\ell) = 2 \pi \delta((Mv-\ell)^2) 
\nn
\\
&&\hspace{2.2 cm}\times \left[ \theta(\ell^0-M) (1+n_B(|M-\ell^0|)) + \theta(M-\ell^0) n_B(|M-\ell^0|) \right]  \, ,
\nn
\\
\phantom{x}
\label{12higgs}
\\
&&i S_{12}(\ell)= 2\pi \delta(\ell^2) \slashed{\ell} \left[ \theta(-\ell^0) (1-n_F(|\ell^0|)) - \theta(\ell^0)n_F(|\ell^0|) \right] \, .
\label{12lepto}
\eea   
There is only one kinematically allowed combination for the product of the propagator components (\ref{12higgs}) and (\ref{12lepto}), providing for the corresponding self energy 
\be 
\Pi_{12} \approx  n_B(M/2)n_F(M/2) \, ,
\ee
which contains only an exponentially suppressed term and it does vanishes in the limit $M \gg T \to 0$. The main outcome is that the second diagram in figure~\ref{fig:selfRTF} involving the fields of type~2 is not necessary in the strict limit $M \gg T$. It is sufficient to take all the fields as of type~1 and work effectively in a quantum field theory at zero temperature for the calculation of the Green's functions at energies much larger than $T$. These will be eventually matched with the corresponding Green's function of the low-energy theory at a scale $\Lambda$ such that $M \gg \Lambda \gg T$ (see section~\ref{eff_sec_2}). Put in other words, the heavy neutrino field of type~2 decouples in the non-relativistic limit as it was pointed out in the case of a heavy quark in \cite{Brambilla:2008cx}.
\begin{figure}[t]
\centering
\includegraphics[scale=0.59]{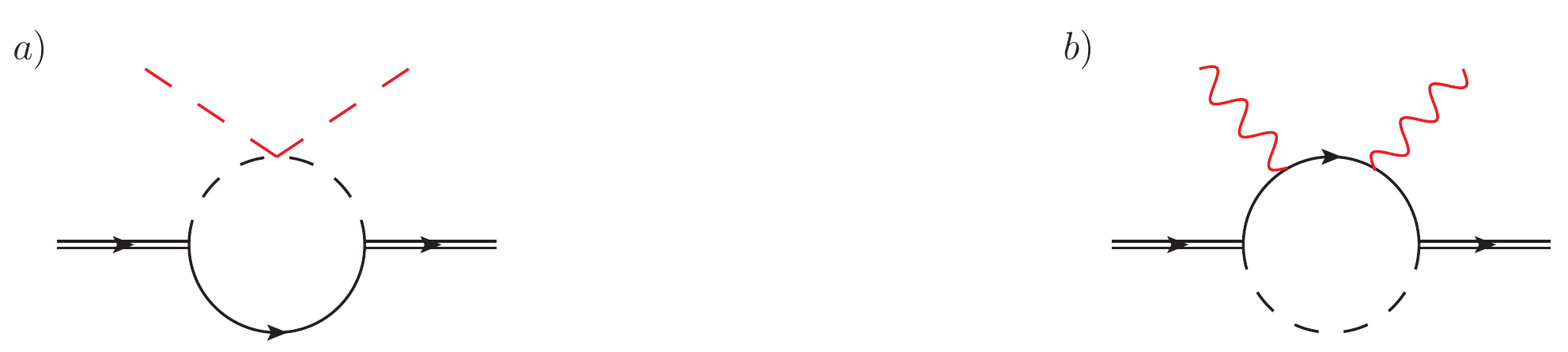}
\caption{\label{fig:selfRTF_2}Scattering between a heavy Majorana neutrino and a Higgs (gauge) boson in diagram $a)$ (diagram $b)$). Wiggled lines stand for the gauge boson. The Higgs and gauge bosons carry a momentum $q \sim T$ and their are shown in red dashed and wiggled lines respectively.}
\end{figure}

Of course, the two-point Green's function is not the only one we can consider. The argument aforementioned holds for a generic $n$-point scattering amplitude if the additional external particles carry momenta of order $T \ll M$ (in general any scale smaller than the neutrino mass). In figure~\ref{fig:selfRTF_2} we show an example where Higgs and gauge bosons are external particles carrying momenta $q \sim T$ (the soft particles are in red dashed and wiggled line respectively). They inject/carry away an energy scale that is either put to zero in the four-point Green's functions describing the diagrams in figure~\ref{fig:selfRTF_2}, or, more in general, that induces an expansion in $T/M$. Such procedure eventually provides the expanded Green's functions in the fundamental theory that match those on the EFT side (see section~\ref{eff_sec_2} in chapter~\ref{chap:eff_the}).

In summary, the EFT approach for the calculation of the thermal width can be outlined as follows:
\begin{itemize}
\item[1)] Write down a low-energy Lagrangian valid at energy scales much smaller than the heavy-neutrino mass, $M$. In this theory the heavy neutrino is non-relativistic. Any low-energy scale, included the temperature of the heat bath, is set to zero in the matching which is performed in vacuum. The doubling of the degrees of freedom does not affect the matching calculation and Feynman rules at  $T=0$ can be used.
\item[2)] In the  EFT so obtained the temperature is a dynamical scale. Observables, such as the heavy-neutrino thermal width, have to be evaluated in a thermal field theory framework, either the ITF or the RTF. Therefore the finite temperature treatment can be postponed at the level of the EFT which is, by construction, simpler than the fundamental theory in its range of applicability.
\end{itemize}

\subsection{EFT Lagrangian at order $1/M^3$}
Within a top-down approach for deriving an EFT Lagrangian, one has to start with a fundamental Lagrangian valid in a wider range of energies. We will consider in the following the simple case of a SM extension involving only one heavy right-handed neutrino. 
The Lagrangian has been already written in the previous chapter in (\ref{prodrate_1}), we recall it here~\cite{Asaka:2006rw}:
\begin{equation}
\mathcal{L}=\mathcal{L}_{\hbox{\tiny SM}} 
+ \frac{1}{2} \,\bar{\psi} \,i \slashed{\partial}  \, \psi  - \frac{M}{2} \,\bar{\psi}\psi 
- F_{f}\,\bar{L}_{f} \tilde{\phi} P_{R}\psi  - F^{*}_{f}\,\bar{\psi}P_{L} \tilde{\phi}^{\dagger}  L_{f} \, ,
\label{eq3_partprod}
\end{equation} 
where $\psi = \nu_R + \nu_R^c$ is the Majorana field embedding the right-handed neutrino field $\nu_R$, 
$\tilde{\phi}=i \sigma^{2} \, \phi^*$, with $\phi$ the Higgs doublet, and $L_{f}$ are lepton doublets with flavour $f$.
The Majorana neutrino has mass $M$, $F_f$ is a (complex) Yukawa coupling and 
$P_L = (1 - \gamma^5)/2$, $P_R = (1 + \gamma^5)/2$ are the left-handed and right-handed projectors respectively.
Lepton doublets, $L_{f}$, carry SU(2) indices, which are contracted with those of the Higgs doublet, $\phi$,
and Lorentz indices, which are contracted with those carried by the Majorana fields.
Right-handed neutrinos are sterile, hence their interaction has not been gauged.
Because we are considering the Lagrangian \eqref{eq3_partprod} for a neutrino mass $M$ and a temperature $T$
much larger than the electroweak scale, the SM Lagrangian, $\mathcal{L}_{\hbox{\tiny SM}}$, 
is symmetric under an unbroken SU(2)$\times$U(1) gauge symmetry and its particles are massless (see the Lagrangian in eq.~(\ref{SMlag}) in appendix~\ref{appB:matchwidth}).

By construction, an EFT suitable to describe non-relativistic Majorana neutrinos 
must be, under the condition \eqref{hiera_width}, equivalent to our fundamental theory \eqref{eq3_partprod} 
order by order in $\Lambda/M$. The scale $\Lambda$ is the ultraviolet cut-off of the EFT and is such that $T \ll \Lambda \ll M$. 
The relevant degrees of freedom at the temperature energy scale are the non-relativistic Majorana field, $N$, 
introduced in section~\ref{sec_maj}, which describes the Majorana neutrino, and the SM particles accounted for in $\mathcal{L}_{\hbox{\tiny SM}}$.
The EFT is written as an expansion in local operators and powers of $1/M$.
The higher the dimension of the operator, the more its contribution to physical observables 
is suppressed by powers of $T/M$. In the following, we will consider only operators 
up to dimension seven, i.e. contributing up to order $1/M^3$ to physical observables. 

The EFT Lagrangian has the general structure 
\begin{equation}
\mathcal{L}_{\hbox{\tiny EFT}}= \mathcal{L}_{\hbox{\tiny SM}}+\mathcal{L}_{\hbox{\tiny N}}+\mathcal{L}_{\hbox{\tiny N-SM}}\,,
\label{eq10}
\end{equation}
where $\mathcal{L}_{\hbox{\tiny N}}$ describes the propagation of the non-relativistic Majorana 
neutrino and $\mathcal{L}_{\hbox{\tiny N-SM}}$ its interaction with the SM particles.
The Lagrangian's parts $\mathcal{L}_{\hbox{\tiny N}}$ and $\mathcal{L}_{\hbox{\tiny N-SM}}$ are determined 
by matching at the scale $\Lambda$ matrix elements in the EFT with matrix elements computed in~\eqref{eq3_partprod}.
A crucial observation is that, in the matching, $T$ can be set to zero because $\Lambda \gg T$; 
hence $\mathcal{L}_{\hbox{\tiny EFT}}$ can be computed in the vacuum.
In the following two paragraphs, we will write $\mathcal{L}_{\hbox{\tiny N}}$ and $\mathcal{L}_{\hbox{\tiny N-SM}}$
at the accuracy needed to compute the Majorana neutrino thermal width 
at first order in the SM couplings and at order $T^4/M^3$ (see eq.~\ref{result_widthLaine}).
In a given reference frame the momentum of the Majorana neutrino is $Mv^\mu$ up to fluctuations of order $T$.

At order $1/M^0$ the Lagrangian $\mathcal{L}_{\hbox{\tiny N}}$ would coincide with \eqref{effLag00}, if the Majorana 
neutrino would be stable at zero temperature. However, the Majorana neutrino may decay into a Higgs and a lepton.
Accounting for this modifies the Lagrangian \eqref{effLag00} into 
\begin{equation}
\mathcal{L}_{\hbox{\tiny N}}= \bar{N} \left(iv \cdot \partial + \frac{i\Gamma^{T=0}}{2} \right)N + \mathcal{O}\left(\frac{1}{M}\right) \,,
\label{effLag0}
\end{equation}
where $\Gamma^{T=0}$ is the decay width at zero temperature in the heavy-mass limit at order $|F|^2$ and zeroth order in the SM couplings.
It has been computed previously in the literature~\cite{Salvio:2011sf,Laine:2011pq} and reads at leading order 
\begin{equation}
\Gamma^{T=0}=\frac{|F|^2M}{8 \pi} \,, 
\label{dt0}
\end{equation}
where $|F|^2=\sum_{f=1}^3F^*_f F_f$. We have already written the heavy neutrino width at zero temperature in chapter~\ref{chap:lepto} in eq.~(\ref{totwidth}), and here we suppress the generation index accordingly with the rest of the present chapter.  

\begin{figure}[t]
\centering
\includegraphics[scale=0.51]{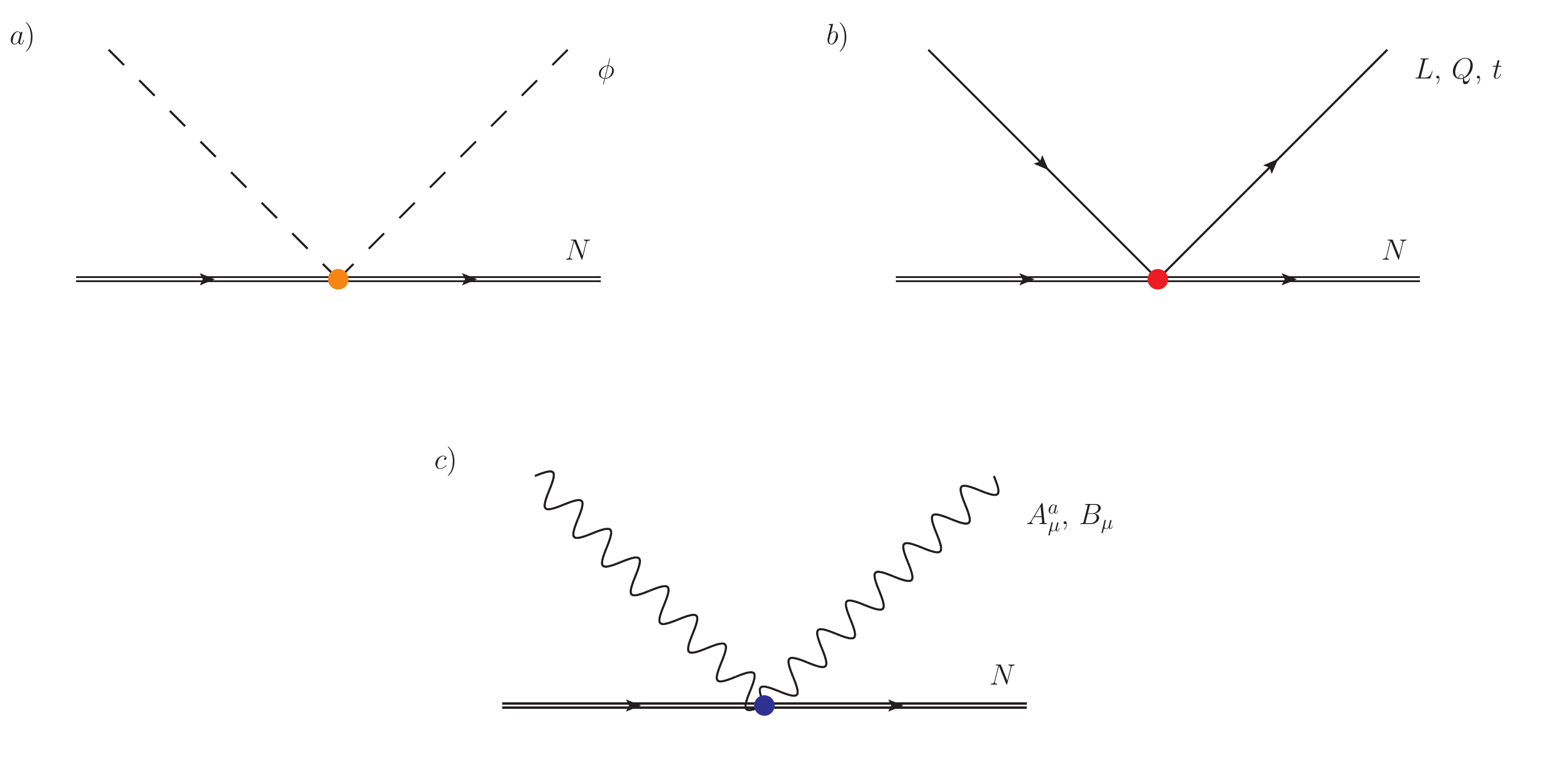}
\caption{\label{Fig1} Diagrams showing the different types of vertices 
induced by the EFT Lagrangian $\mathcal{L}_{\hbox{\tiny N-SM}}$.
These involve interactions between heavy Majorana neutrinos and Higgs fields in $a)$, 
fermions in $b)$ and the gauge bosons in $c)$.}
\end{figure}

The Lagrangian $\mathcal{L}_{\hbox{\tiny N-SM}}$, organized in an expansion in $1/M$, reads
\begin{equation}
\mathcal{L}_{\hbox{\tiny N-SM}}=\frac{1}{M}\mathcal{L}_{\hbox{\tiny N-SM}}^{(1)}
+\frac{1}{M^2}\mathcal{L}_{\hbox{\tiny N-SM}}^{(2)}
+\frac{1}{M^3}\mathcal{L}_{\hbox{\tiny N-SM}}^{(3)}
+\mathcal{O}\left(\frac{1}{M^4}\right),
\end{equation}
where $\mathcal{L}_{\hbox{\tiny N-SM}}^{(n)}$ includes all operators of dimension $4+n$.
They describe the effective interactions between the Majorana neutrino and 
the Higgs field $\phi$, the lepton doublets $L_f$ of all flavours $f$,
the heavy-quark doublets $Q^T=(t,b)$, where $t$ stands for the top field and $b$ for the bottom field, 
the right-handed top field and the SU(2)$\times$U(1) gauge bosons. 
We consider only Yukawa couplings with the top quark and neglect Yukawa couplings with other quarks and leptons, 
for the ratio of Yukawa couplings is proportional to the ratio of the corresponding fermion masses
when the gauge symmetry is spontaneously broken.
The number of operators contributing to $\mathcal{L}_{\hbox{\tiny N-SM}}$
may be further significantly reduced by assuming the Majorana neutrino at rest and 
by selecting only operators that could contribute to the Majorana neutrino thermal width 
at first order in the SM couplings and at order $T^4/M^3$.
At first order in the SM couplings, thermal corrections are encoded into tadpole diagrams. 
Hence we need to consider only operators with imaginary coefficients (tadpoles do not develop an imaginary part), 
made of two Majorana fields with no derivatives acting on them (the Majorana neutrino is at rest), 
coupled to bosonic operators with an even number of spatial and time derivatives 
(the boson propagator in the tadpole is even for space and time reflections) 
and to fermionic operators with an odd number of derivatives 
(the massless fermion propagator in the tadpole is odd for space-time reflections).
Finally, we may use field redefinitions to get rid of operators containing terms like 
$\slashed{\partial}$(fermion field) or $\partial^2$(boson field). 
The Lagrangian $\mathcal{L}_{\hbox{\tiny N-SM}}^{(1)}$ reads 
\begin{equation}
\mathcal{L}_{\hbox{\tiny N-SM}}^{(1)}=a \; \bar{N} N \, \phi^{\dagger} \phi \, .
\label{Wa}
\end{equation}
The Lagrangian $\mathcal{L}_{\hbox{\tiny N-SM}}^{(2)}$ does not contribute to our observable 
because it involves either boson fields with one derivative or fermion fields with no derivatives.
The Lagrangian $\mathcal{L}_{\hbox{\tiny N-SM}}^{(3)}$ reads
\begin{eqnarray}
\mathcal{L}_{\hbox{\tiny N-SM}}^{(3)} &=&  
~~ b \; \bar{N}N \,  \big(v\cdot D \phi^{\dagger}\big)  \,  \big(v\cdot D \phi\big) 
\nonumber  \\ 
&& + c^{ff'}_{1} \; \left[ \left(\bar{N}P_{L} \, iv\cdot D L_{f}\right) \left(\bar{L}_{f'} P_{R} N \right) \right.
\nonumber \\
&&
\left.
\hspace{1cm}  
+ \left( \bar{N} P_{R} \, i v\cdot D L^{c}_{f'} \right) \left( \bar{L}^{c}_{f} P_{L}  N \right)  \right]  
\nonumber \\
&& 
+ c^{ff'}_{2} \; \left[  \left(\bar{N} P_{L} \, \gamma_{\mu} \gamma_{\nu} \, iv\cdot D L_{f} \right) 
\left( \bar{L}_{f'} \, \gamma^{\nu} \gamma^{\mu} \, P_{R} N \right)  \right.
\nonumber \\
&&
\left.
\hspace{1cm}  
+ \left(\bar{N}  P_{R} \, \gamma_{\mu} \gamma_{\nu} \, iv\cdot D L^{c}_{f'} \right) 
\left( \bar{L}^{c}_{f} \, \gamma^{\nu} \gamma^{\mu} \, P_{L} N \right) \right]  
\nonumber \\ 
&& 
+ c_{3} \; \bar{N}N \, \left( \bar{t}P_{L} \, v^\mu v^\nu \gamma_{\mu} \, i D_{\nu} t \right) 
+ c_{4} \; \bar{N}N \, \left( \bar{Q}P_{R} \, v^\mu v^\nu \gamma_{\mu} \, i D_{\nu} Q \right)
\nonumber \\
&& 
+ c_{5} \; \bar{N}\,\gamma^5\gamma^\mu\,N \, \left( \bar{t}P_{L} \, v\cdot\gamma \, i D_{\mu} t \right) 
+ c_{6} \; \bar{N}\,\gamma^5\gamma^\mu\,N \, \left( \bar{Q}P_{R} \, v\cdot\gamma \, i D_{\mu} Q \right) 
\nonumber \\
&& 
+ c_{7} \; \bar{N}\,\gamma^5\gamma^\mu\,N \, \left( \bar{t}P_{L} \, \gamma_\mu \, i v\cdot D t \right)
+ c_{8} \; \bar{N}\,\gamma^5\gamma^\mu\,N \, \left( \bar{Q}P_{R} \, \gamma_\mu \, i v\cdot D Q \right)
\nonumber \\
&& 
- d_{1} \; \bar{N}N \,  v^\mu v_\nu W^{a}_{\alpha\mu} W^{a\,\alpha\nu}  \
- d_{2} \; \bar{N}N \,  v^\mu v_\nu F_{\alpha\mu} F^{\alpha\nu} 
\nonumber \\
&&
+ d_{3} \; \bar{N}N \, W^{a}_{\mu\nu} W^{a\,\mu\nu}  \
+ d_{4} \; \bar{N}N \, F_{\mu\nu} F^{\mu\nu} 
\, .
\label{Wb}
\end{eqnarray}
The fields $W^a_{\mu\nu}$ and $F_{\mu\nu}$ are the field strength tensors of the SU(2) 
gauge fields, $A^a_{\mu}$, and U(1) gauge fields, $B_{\mu}$, respectively.
For the operators multiplying $c^{ff'}_{1}$ and $c^{ff'}_{2}$
the SU(2) indices of $L_{f}$ and $\bar{L}_{f'}$ are contracted with each other 
while their Lorentz indices are contracted with gamma matrices and Majorana fields.

The Wilson coefficients $a$, $b$, $c^{ff'}_{i}$, $c_i$ and $d_i$ encode all contributions coming from the high-energy modes 
of order $M$ that have been integrated out when matching from the fundamental theory \eqref{eq3_partprod} to the
EFT  \eqref{eq10}. We are interested only in their imaginary parts. At first order in the SM couplings they read
\begin{eqnarray}
&& {\rm{Im}}\,a = -\frac{3}{8\pi}|F|^{2}\lambda \,, 
\label{coa}\\
&& {\rm{Im}}\,b =- \frac{5}{32 \pi} (3g^{2}+g'^{\,2})|F|^{2}\,, 
\label{cob}\\
&& {\rm{Im}}\,c^{ff'}_{1} = \frac{3}{8\pi}|\lambda_{t}|^{2}{\rm{Re}}\left(  F_{f'} F^{*}_{f} \right) -  \frac{3}{16 \pi}(3g^{2}+g'^{\,2}){\rm{Re}}\left(  F_{f'} F^{*}_{f} \right)   \,, 
\label{co1}\\
&& {\rm{Im}}\,c^{ff'}_{2} = \frac{1}{384 \pi} (3g^{2}+g'^{\,2}){\rm{Re}}\left(  F_{f'} F^{*}_{f} \right)  \,,
\label{co2}\\
&& {\rm{Im}}\,c_{3} = \frac{1}{24\pi}|\lambda_{t}|^{2}|F|^{2} \,, 
\qquad {\rm{Im}}\,c_{4} = \frac{1}{48\pi}|\lambda_{t}|^{2}|F|^{2} \,,
\label{co34}\\
&& {\rm{Im}}\,c_{5} = \frac{1}{48\pi}|\lambda_{t}|^{2}|F|^{2} \,,
\qquad {\rm{Im}}\,c_{6} = \frac{1}{96\pi}|\lambda_{t}|^{2}|F|^{2} \,,
\label{co56}\\
&& {\rm{Im}}\,c_{7} = \frac{1}{48\pi}|\lambda_{t}|^{2}|F|^{2} \,,
\qquad {\rm{Im}}\,c_{8} = \frac{1}{96\pi}|\lambda_{t}|^{2}|F|^{2} \,,
\label{co78}\\
&& {\rm{Im}}\,d_{1} = - \frac{1}{96\pi}g^{2}|F|^{2} \,, 
\qquad {\rm{Im}}\,d_{2} =- \frac{1}{96\pi}g'^{\,2}|F|^{2}\,,
\label{cod12}\\
&& {\rm{Im}}\,d_{3} = - \frac{1}{384\pi}g^{2}|F|^{2} \,, 
\qquad\!\!\! {\rm{Im}}\,d_{4} =- \frac{1}{384\pi}g'^{\,2}|F|^{2}\,,
\label{cod34}
\end{eqnarray}
where $g$ is the SU(2) coupling, $g'$ the U(1) coupling, 
$\lambda$ the four-Higgs coupling and $\lambda_t$ the top Yukawa coupling.
We refer to appendix~\ref{appB:matchwidth} for details on the calculation.

The EFT Lagrangian derived in this section follows from symmetry arguments and standard (one-loop) perturbation theory.  
Owing to the hierarchy \eqref{hiera_width}, the temperature could be set to zero when computing the Wilson coefficients.
Thermal effects factorize. This factorization may be considered as the main advantage in the use of the EFT. 
The calculation of the Majorana neutrino thermal width will turn out to be very simple. 
Indeed, already at this level, the structure and power counting of the EFT allow to make some general statements 
about the origin and size of the different contributions. The width will be the sum of contributions coming 
from the scattering with Higgs, SM fermions (either leptons or left-handed heavy quarks or right-handed tops) 
and gauge fields in the early universe plasma. We call these contributions 
$\Gamma_{\phi}$, $\Gamma_{\rm fermions}$ and $\Gamma_{\rm gauge}$ respectively.
The leading operator responsible for the interaction of the Majorana neutrino with the Higgs 
is  the dimension five operator \eqref{Wa}, hence the natural power counting of the EFT implies 
\begin{equation}
\Gamma_{\phi} \sim \frac{T^{2}}{M}  \,.
\label{eq0}
\end{equation}
This is also the leading contribution to the thermal width of the Majorana neutrino.
The interaction of the Majorana neutrino with the SM fermions and the gauge bosons is mediated 
in \eqref{Wb} by operators of dimension seven, hence 
\begin{equation}
\Gamma_{\hbox{\tiny fermions}} \sim \frac{T^{4}}{M^{3}} \,, \qquad 
\Gamma_{\hbox{\tiny gauge}} \sim \frac{T^{4}}{M^{3}} \, .
\label{eq}
\end{equation}
In section~\ref{thermwidth_partprod}, we will compute $\Gamma_{\phi}$, $\Gamma_{\hbox{\tiny fermions}}$ and $\Gamma_{\hbox{\tiny gauge}}$
at first order in the SM couplings.

\subsection{Matching the dimension-five operator}
The matching of the dimension-five operator is discussed in order to show in some detail the calculation of a four-point Green's function both in the fundamental theory and the EFT. The effective theory must reproduce the fundamental one at energies below its cut-off~$\Lambda$. 
A way to enforce this is by matching low-energy matrix elements in the two theories.
The matching fixes the Wilson coefficients of the EFT, which encode, order by order in the 
couplings, the contributions from the high-energy modes that have been integrated out.
Because in the matching we are integrating out only high-energy modes, we can set to zero 
any low-energy scale appearing in loops. A consequence is that, in the matching, loop diagrams in the EFT vanish in dimensional 
regularization because scaleless. We adopt dimensional regularization in all loop calculations of the thesis.
\begin{figure}
\centering
\includegraphics[scale=0.52]{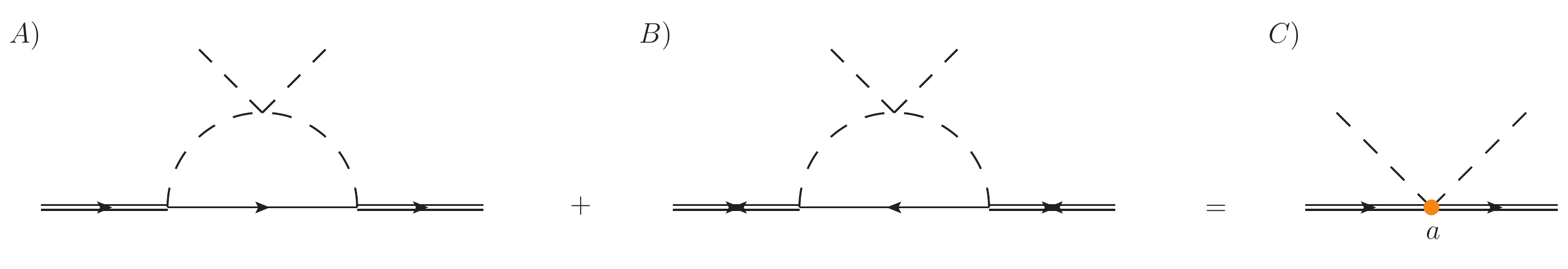}
\caption{\label{fig:amatch_part} Diagrams in the full theory (left-hand side of the equality)
contributing to the Majorana neutrino-Higgs four-field operators in the EFT (right-hand side). The neutrino propagator with forward arrow corresponds to $\langle 0| T(\psi \bar{\psi}) |0\rangle$,  whereas the neutrino propagators with forward-backward arrows correspond to $\langle 0| T(\psi \psi) |0\rangle$ 
or $\langle 0| T(\bar{\psi} \bar{\psi})|0 \rangle$.}
\end{figure}

We perform the matching in the reference frame $v^{\mu}=(1,\bm{0}\,)$, 
where we assume the plasma to be at rest.
Since we are interested in the imaginary parts of the Wilson coefficients,
we evaluate the imaginary parts of $-i {\mathcal{D}}$, 
where ${\mathcal{D}}$ are generic Feynman diagrams,  
by taking the Majorana neutrino momentum at $M+i\eta$.
We may also choose the incoming and outgoing SM particles to carry the same momentum $q^\mu$.
Because $q^{\mu}$ is much smaller than $M$, diagrams in the fundamental theory are expanded 
in powers of $q^{\mu}$. This expansion matches the operator expansion in the EFT.

There is only one diagram contributing to the matching of the dimension-five operator and we show it in figure~\ref{fig:amatch_part}. In order to determine the corresponding Wilson coefficient, $a$, we compute in the fundamental theory the matrix element
\begin{equation}
-i \left.\int d^{4}x\,e^{i p \cdot x} \int d^{4}y \int d^{4}z\,e^{i q \cdot (y-z)}\, 
\langle \Omega | T(\psi^{\mu}(x) \bar{\psi}^{\nu }(0) \phi_{m}(y) \phi_{n}^{\dagger}(z) )| \Omega \rangle
\right|_{p^\mu =(M + i\eta,\bm{0}\,)},
\label{match_partprod_1}
\end{equation} 
where $\mu$ and $\nu$ are Lorentz indices, $m$ and $n$ are SU(2) indices and $|\Omega \rangle$ is the ground state 
of the fundamental theory. The matrix element \eqref{match_partprod_1} describes a $2 \rightarrow 2$ scattering between 
a heavy Majorana neutrino at rest and a Higgs boson carrying momentum $q^\mu$.

From figure~\ref{fig:amatch_part} it is clear that a one-loop matching between the amplitudes in the fundamental theory and in the EFT is needed in order to determine the Wilson coefficient. When computing matrix elements involving Majorana fermions, one has to consider 
that the relativistic Majorana field may be contracted in more ways than if it was a Dirac field, 
this reflecting the indistinguishability of the Majorana particle and antiparticle.
The different contractions give rise to the different propagators listed in \eqref{eq5_partprod}-\eqref{eq7_partprod} and diagrams $A)$ and $B)$ in figure~\ref{fig:amatch_part}. We refer to appendix~\ref{appB:matchwidth} for more details on this aspect.
When contracting the Majorana fields in \eqref{match_partprod_1} according to \eqref{eq5_partprod}, one obtains at leading order 
\begin{eqnarray}
\left[\hat{P}\left( -i \mathcal{D}_{\hbox{\tiny A}} \right)\hat{P}\right]^{\mu\nu}
= 6 |F|^2 \lambda \, \delta_{mn} \,  \int \frac{d^{4} \ell }{(2\pi)^{4}}  
\left( \hat{P} \, P_L \slashed{\ell} \, \hat{P} \right)^{\mu \nu} \frac{i}{\ell^{2}+i \eta} 
\left( \frac{i}{(M v - \ell )^2+i \eta} \right)^{2} ,
\nn
\\
\label{match_partprod_2}
\end{eqnarray}
where we have dropped all external propagators and $\mathcal{D}$ is the amputated diagram shown in figure~\ref{fig:amatch_part}.
The external heavy neutrino propagators reduce in the non-relativistic limit and in the rest frame 
to a matrix proportional to $\hat{P}=(1+\gamma^{0})/2$ (see \eqref{effpropagator}).
We have kept the matrix $\hat{P}$ on the left- and right-hand side of \eqref{match_partprod_2}, because it helps projecting 
out the contributions relevant in the heavy-neutrino mass limit, e.g., $ \hat{P} \, P_L \, \hat{P} = \hat{P}/2$.
After projection, also the matrix $\hat{P}$ may be eventually dropped from the left- and right-hand side of the matching equation.

Since we are interested in ${\rm{Im}}\left( -i \mathcal{D}\right)$ for the matching, it is enough to extract the imaginary part of the loop amplitude. In order to calculate the loop integral in \eqref{match_partprod_2} standard in-vacuum techniques are adopted. The direct application of the Feynman parameters leads to 
\bea 
I_a&=&\int \frac{d^4 \ell}{(2\pi)^4} \frac{i\slashed{\ell}}{\ell^{2}+i \eta}  \left( \frac{i}{(M v - \ell )^2+i \eta} \right)^{2} \nn 
\\
&=& -i \frac{\slashed{v}}{16 \pi  M} - \frac{\slashed{v}}{(4 \pi)^2  M} \left[  \frac{1}{\varepsilon} + \ln \left( \frac{4 \pi \mu^2}{M^2}\right)  -\gamma_E +1 \right] 
\label{match_partprod_3}
\eea
where a real and imaginary part appear. Alternative methods may be used that are more suitable for automated loop calculations. In particular tensor reduction to scalar integrals is one of the most popular~\cite{Passarino:1978jh}. 
Hence for the matrix element in (\ref{match_partprod_2}) we obtain, dropping also the non-relativistic projector,  
\be 
{\rm{Im}}\left( -i \mathcal{D}_{\hbox{\tiny A}}\right) = - \frac{3 |F|^2 \lambda}{16 \pi M} \delta_{mn} \delta^{\mu \nu}  \, .
\label{match_partprod_4}
\ee
Diagram $B$ in figure \ref{fig:amatch_part} provides exactly the same result given in (\ref{match_partprod_4}).  Therefore from the fundamental theory side one finds 
\be 
{\rm{Im}}\left( -i \mathcal{D}_{\hbox{\tiny A}}\right) + {\rm{Im}}\left( -i \mathcal{D}_{\hbox{\tiny B}}\right) = - \frac{3 |F|^2 \lambda}{8\pi M} \delta_{mn} \delta^{\mu \nu} \, .
\label{match_partprod_5}
\ee

The symmetries of the EFT enforce that the matrix element \eqref{match_partprod_1} is reproduced by the following expression, where we simply apply the Feynman rule for a vertex in the EFT (the extra $-i$ is due to our definition $-i\mathcal{D}$)
\begin{equation}
{\rm Im} (-i \mathcal{D}_C )={\rm Im} \left( -i  \frac{ia}{M} \right)  \delta_{mn}\delta^{\mu \nu}   + \dots = \frac{\delta_{mn}\delta^{\mu \nu}}{M}    {\rm Im} \, a + \cdots     \,,
\label{match_partprod_6}
\end{equation}
where the dots stand for contributions coming from operators that are not listed in \eqref{Wa}. 
Finally comparing eqs.~(\ref{match_partprod_5}) with (\ref{match_partprod_6}) fixes the imaginary part of $a$:
\begin{equation}
{\rm Im}\,a = -\frac{3}{8\pi}|F|^{2}\lambda \, .
\label{match_final_a}
\end{equation}

\section{Thermal width in the EFT}
\label{thermwidth_partprod}
\begin{figure}[htb]
\centering
\includegraphics[scale=0.6]{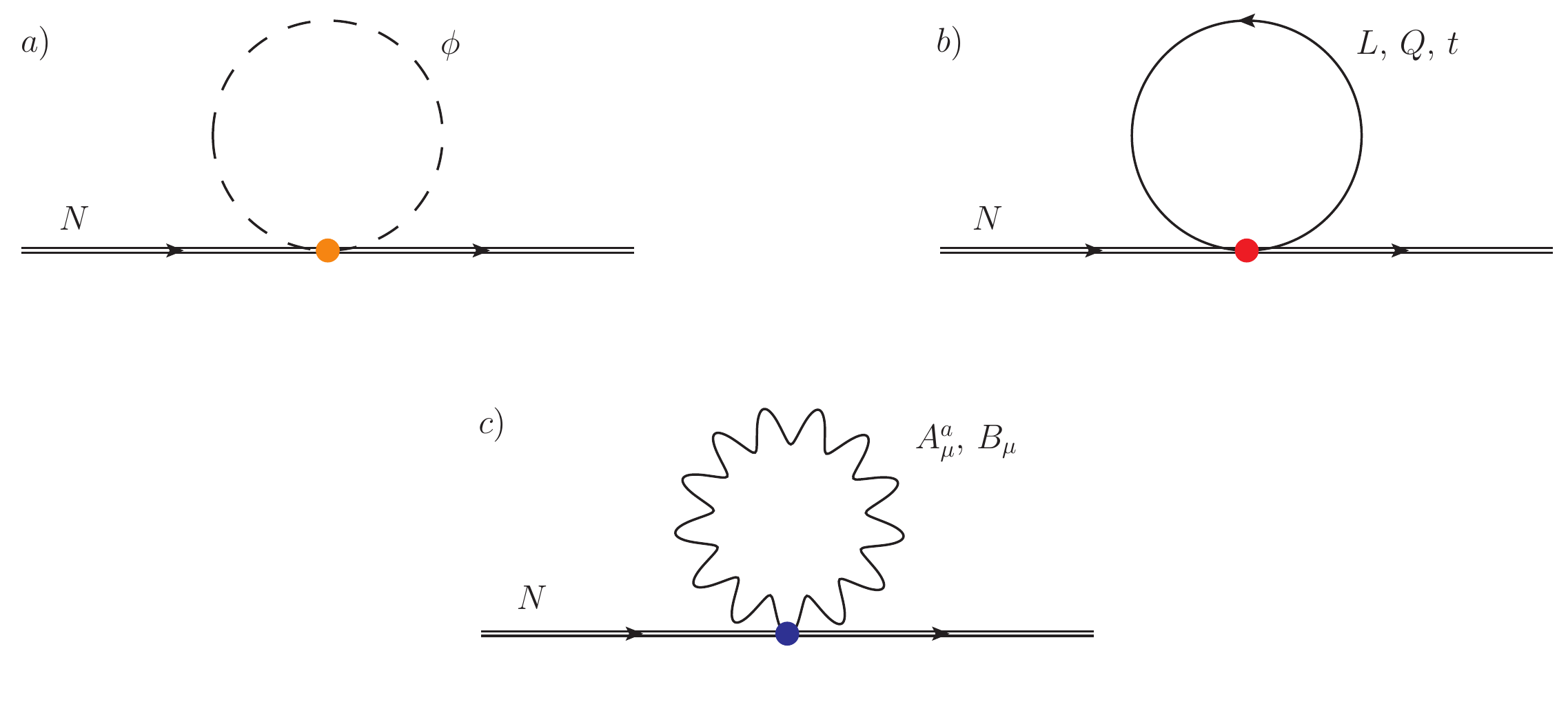}
\caption{\label{fig:fig_tadpoles_partprod} Tadpole diagrams contributing to the thermal
width of a heavy Majorana neutrino at first order in the SM couplings.
The heavy Majorana neutrino is represented by a double line, the
Higgs propagator by a dashed line, fermion propagators (leptons, heavy
quark doublets and top singlet) by a continuous line and gauge bosons by
a wiggled line.}
\end{figure}

A Majorana neutrino in a plasma of SM particles thermalized at some temperature $T$ decays with a width 
$\Gamma = \Gamma^{T=0}+\Gamma^T$, where $\Gamma^{T=0}$ is the in-vacuum width and 
$\Gamma^{T}$ encodes the thermal corrections to the width induced by the interaction with the particles in the medium.
We call $\Gamma^{T}$ the Majorana neutrino thermal width.
The decay of the Majorana neutrino happens at a distance of order $1/M$. The neutrino 
releases a large amount of energy of the order of its mass into a   
Higgs and lepton pair. The interaction vertex is described by the Lagrangian \eqref{eq3_partprod}.
At such small distances the neutrino is insensitive to the plasma and the decay 
happens as in the vacuum. The width is $\Gamma^{T=0}$, which at leading order can be read off eq. \eqref{dt0}.\footnote{
Next-to-leading order corrections in the SM couplings to $\Gamma^{T=0}$ have been calculated in~\cite{Salvio:2011sf,Laine:2011pq}.
Those corrections may be taken over as they are in the EFT to improve the expression 
of the zero-temperature Majorana neutrino width in $\mathcal{L}_{\hbox{\tiny N}}$.
}
At distances of order $1/T$, the vertices involving Majorana neutrinos in the fundamental Lagrangian \eqref{eq3_partprod}
cannot be resolved, instead the Majorana neutrino effectively interacts with Higgs, 
fermion and gauge boson pairs as shown in figure~\ref{Fig1}. 
These are the vertices in the EFT that can be read off eqs.~\eqref{Wa} and \eqref{Wb}.
The effective couplings of these vertices are the Wilson coefficients listed in \eqref{coa}-\eqref{cod34}.
They are all of first order in the SM couplings $g^2$, $g'^{\,2}$, $\lambda$ and $|\lambda_t|^2$. 
Hence, at that order, only tadpole diagrams of the type shown in 
figure~\ref{fig:fig_tadpoles_partprod} can contribute to the Majorana neutrino width.
Tadpoles do not vanish (in dimensional regularization) only if the momentum circulating in the loop is of the order 
of the plasma temperature, instead  they induce a thermal correction, $\Gamma^{T}$, to the width.
In the following, we will calculate $\Gamma^T$ assuming that the thermal bath 
of SM particles is at rest with respect to the Majorana neutrino.
Moreover, we choose our reference frame such that $v^\mu = (1,\bm{0}\,)$. 

We calculate finite temperature effects in the RTF. 
This amounts at modifying the contour of the time integration in the partition function to allow for real time.
A consequence of this is that in the real-time formalism the degrees of freedom double (see chapter~\ref{chap:therm_tex}). 
One usually refers to them as degrees of freedom 
of type $1$ and $2$. The physical degrees of freedom, those describing 
initial and final states, are of type $1$. Propagators can mix fields 
of type $1$ with fields of type $2$, while vertices do not couple 
fields of different types. 
It has been shown in~\cite{Brambilla:2008cx} that because the $12$ component of a 
heavy-field propagator vanishes in the heavy-mass limit, heavy fields 
of type $2$ decouple from the theory and can be neglected.  
This also applies to the Majorana neutrino field $N$, 
which may be considered of type $1$ only.  
In our case, we will calculate the tadpole diagrams shown in figure~\ref{fig:fig_tadpoles_partprod}. 
Because there the SM fields couple directly to the neutrino field $N$, 
also the SM fields may be considered to be of type $1$ only.
This is a significant simplification in the calculation that the non-relativistic 
EFT makes manifest from the beginning.

Tadpole diagrams like those shown in figure~\ref{fig:fig_tadpoles_partprod} involve only $11$ components 
of the real-time propagators of the SM fields. 
The $11$ component is the time-ordered propagator of the physical field; 
for a bosonic (scalar) field propagating from~$0$ to~$x$ it reads  
\begin{equation}
i\Delta_{11}(x)= \int \frac{d^{4}q}{(2 \pi)^{4}} \,e^{-i q \cdot x}\left[ \frac{i}{q^{2}+i\eta } 
+ 2\pi n_{\hbox{\tiny B}}(|q_{0}|)\delta(q^{2}) \right] \,,
\label{eq12}
\end{equation}
where  $n_{\hbox{\tiny B}}(|q_{0}|)=1/(e^{|q_{0}|/T}-1)$ is the Bose--Einstein distribution in the rest frame, 
and for a fermionic field propagating from $0$ to $x$ 
\begin{equation}
iS_{11}(x)= \int \frac{d^{4}q}{(2 \pi)^{4}}\,e^{-i q \cdot x}\,\slashed{q} \left[ \frac{i}{q^{2}+i\eta } 
- 2\pi n_{\hbox{\tiny F}}(|q_{0}|)\delta(q^{2}) \right] \,,
\label{eq13}
\end{equation}
where  $n_{\hbox{\tiny F}}(|q_{0}|)=1/(e^{|q_{0}|/T}+1)$ is the Fermi--Dirac distribution in the rest frame.
We recall that SM particles are massless in the high-temperature regime \eqref{hiera_width}.

Thermal corrections to the decay width can be computed from the Majorana neutrino propagator in momentum space:
\begin{equation}
\int d^{4}x\,\,e^{i k \cdot x}\, \langle T(N^{\alpha}(x) N^{\dagger\,\beta }(0)) \rangle_T^{\hbox{\tiny int}} \,,
\label{fullNN}
\end{equation}
where $\langle \cdots \rangle_T^{\rm int}$ stands for the thermal average evaluated on the action
$\displaystyle \int d^4x\,\mathcal{L}_{\hbox{\tiny EFT}}$.
In the $v^\mu = (1,\bm{0}\,)$ frame, the Majorana neutrino propagator has the general form (cf. with \eqref{effpropagator})
\begin{eqnarray}
&&\left( \frac{1+\gamma_0}{2}\right)^{\alpha\beta} \frac{iZ}{k^{0}- E+i{\Gamma}/{2}}=
\nn
\\
&& \left( \frac{1+\gamma_0}{2}\right)^{\alpha\beta} Z\left[ 
\frac{i}{k^{0}+i\eta} - \left(i E+\frac{\Gamma}{2}\right)\left( \frac{i}{k^{0}+i\eta} \right)^{2}  
+  \cdots \right].
\label{eq15}
\end{eqnarray}
The wave function normalization $Z$, mass shift $E$ and width $\Gamma$ are determined by self-energy diagrams.
In our case, we consider only the tadpole diagrams shown in figure~\ref{fig:fig_tadpoles_partprod}. 
Because $Z-1$ is given by the derivative of the self-energy with respect to the incoming momentum 
and because tadpole diagrams do not depend on the incoming momentum, we have that $Z=1$.
In the expansion \eqref{eq15}, the width $\Gamma$ is then twice the real part of the residue of the double pole in $k^0=0$. 

We start by considering the contribution to the decay width from the Higgs tadpole 
(diagram $a$ in figure~\ref{fig:fig_tadpoles_partprod}). A Higgs tadpole may contribute to \eqref{fullNN} either 
through the dimension five operator \eqref{Wa} or through the dimension seven operator in the 
first line of \eqref{Wb} or through higher-order operators. 
Expanding \eqref{fullNN} in $\mathcal{L}_{\hbox{\tiny N-SM}}$, we obtain
\begin{eqnarray}
&& i \frac{a}{M} \int d^{4}x\,e^{i k \cdot x}\, \langle \int d^{4}z\, T( N^{\alpha}(x) N^{\dagger \,\beta}(0) \, 
N^{\dagger \,\mu}(z) N_{\mu}(z) \phi^{\dagger}(z) \phi(z) )\rangle_T^{\hbox{\tiny free}} 
\nonumber\\
&+& i \frac{b}{M^3} \int d^{4}x\,\,e^{i k \cdot x}\, \langle \int d^{4}z\, T( N^{\alpha}(x) N^{\dagger \,\beta}(0) \, 
N^{\dagger \,\mu}(z) N_{\mu}(z) \partial_0\phi^{\dagger}(z) \partial_0\phi(z) )\rangle_T^{\hbox{\tiny free}} 
\nonumber\\
&+& \hbox{contributions of higher order in}~1/M\,, 
\label{gammaab}
\end{eqnarray}
where $\langle \cdots \rangle_T^{\hbox{\tiny free}}$ stands for the thermal average evaluated on the action
$\displaystyle \int d^4x\,(\mathcal{L}_{\hbox{\tiny SM}}+\mathcal{L}_{\hbox{\tiny N}})$.
The Wilson coefficients $a$ and $b$ can be read off eqs. \eqref{coa} and \eqref{cob} respectively.
Because the Majorana neutrinos do not thermalize, we have that 
\begin{equation}
\langle \hbox{(Majorana fields)} \times \hbox{(SM fields)}\rangle_T^{\hbox{\tiny free}} = 
\langle 0|\hbox{(Majorana fields)} |0\rangle \times \langle \hbox{(SM fields)}\rangle_T\,,
\end{equation}
where $\langle 0|\hbox{(Majorana fields)} |0\rangle$ is a free Green's function that can be  
computed by contracting the Majorana neutrino fields according to \eqref{effpropagator}, and 
$\langle \cdots \rangle_T$ is a thermal average of SM fields weighted by the SM partition function.
Comparing \eqref{gammaab} with \eqref{eq15}, we obtain 
\begin{eqnarray}
\Gamma_{\phi} &=& 2 \frac{{\rm Im} \, a}{M} \langle \phi^{\dagger}(0) \phi(0) \rangle_{T}  
+ 2 \frac{{\rm Im} \, b}{M^3} \langle \partial_0\phi^{\dagger}(0) \partial_0\phi(0) \rangle_{T}  
\nonumber\\
&=&  \frac{{\rm Im} \, a}{3} \frac{T^2}{M} +  \frac{2\pi^2}{15} {\rm Im} \, b \frac{T^4}{M^3}\,.
\label{gammaphi_part}
\end{eqnarray}
The last line follows from having computed the Higgs thermal condensates at leading order:
\begin{eqnarray}
\langle \phi^{\dagger}(0) \phi(0) \rangle_{T}  &=&  
2 \int  \frac{d^{4}q}{(2\pi)^{4}}\, 2\pi n_{\hbox{\tiny B}}(|q_{0}|)\delta(q^{2})  = \frac{T^2}{6}\,,
\label{cond1}\\
\langle  \partial_0\phi^{\dagger}(0) \partial_0\phi(0) \rangle_{T}  &=&  
2 \int  \frac{d^{4} q}{(2\pi)^{4}} \, q_0^2 \, 2\pi n_{\hbox{\tiny B}}(|q_{0}|)\delta(q^{2})  = \frac{\pi^2}{15}T^4\,.
\label{cond2}
\end{eqnarray}
We have used dimensional regularization to get rid of the vacuum contributions. 
We observe that bosonic condensates involving an odd number of spatial or time derivatives  
would give rise to vanishing momentum integrals. 

In a similar way we can compute the contribution to the decay width from the fermion tadpoles
(diagram $b$ in figure~\ref{fig:fig_tadpoles_partprod}):
\begin{eqnarray}
\Gamma_{\hbox{\tiny fermions}} &=& - \left(\frac{{\rm Im}\, c^{ff'}_{1}}{2M^3} + \frac{ 2 {\rm Im}\, c^{ff'}_{2} }{M^3} \right)
\langle \bar{L}_{f'}(0)\gamma^0 iD_0 L_f(0) \rangle_{T}  
\nonumber\\
&& + 2 \frac{{\rm Im}\, c_{3}}{M^3} \langle \bar{t}(0) P_L\gamma^0 iD_0 t(0) \rangle_{T}  
+ 2 \frac{{\rm Im}\, c_{4}}{M^3} \langle \bar{Q}(0) P_R\gamma^0 iD_0 Q(0) \rangle_{T}  
\nonumber\\
&=&  \left( -{\rm Im}\, c^{ff}_{1} -4 {\rm Im}\, c^{ff}_{2}  + 3 {\rm Im}\, c_{3} + 6 {\rm Im}\, c_{4} \right) 
\frac{7 \pi^2}{60} \frac{T^4}{M^3}\,,
\label{gammafermions}
\end{eqnarray}
where the Wilson coefficients $c^{ff}_{i}$ and $c_i$ can be read off eqs. \eqref{co1}-\eqref{co34}.
The last line of \eqref{gammafermions} follows from having computed the lepton thermal condensate at leading order, 
\begin{equation}
\langle \bar{L}_{f'}(0)\gamma^0 iD_0 L_f(0) \rangle_{T} =
-2 \delta_{ff'} \int  \frac{d^{4}q}{(2\pi)^{4}}\, q_0 \, {\rm Tr}\, \left\{ \gamma^0\slashed{q} \right\} 
\, (-2\pi) n_{\hbox{\tiny F}}(|q_{0}|)\delta(q^{2})  =\delta_{ff'}  \,  \frac{7 \pi^2}{30} T^4\,,
\label{cond3}
\end{equation}
and similarly the quark condensates, $\langle \bar{t}(0) P_L\gamma^0 iD_0 t(0) \rangle_{T}  = 7 \pi^2 T^4/40$ 
and $\langle \bar{Q}(0) P_R\gamma^0$ $\times iD_0 Q(0) \rangle_{T} = 7 \pi^2 T^4/20$.
We note that fermionic condensates involving an even number of derivatives would give rise to vanishing momentum integrals. 

Tadpole diagrams generated by operators multiplying the Wilson coefficients $c_5$, $c_6$, $c_7$ and $c_8$ 
in \eqref{Wb} provide a contribution to the width that depends on the spin coupling of the Majorana neutrino with the medium.\footnote{
The operator $N^{\dagger}\,\gamma^5\gamma^i\,N$ can be also written as $-2 N^{\dagger}\,S^i\,N$, where $\vec{S}$ is the spin 
operator.} If the medium is isotropic, this coupling is zero.

Finally, the contribution to the decay width from the gauge boson tadpoles (diagram $c$ in figure~\ref{fig:fig_tadpoles_partprod}) gives
\begin{eqnarray}
\Gamma_{\hbox{\tiny gauge}} &=&  2 \frac{{\rm Im}\, d_{1}}{M^3} \langle W^a_{0i}(0) W^a_{0i}(0) \rangle_{T}  
+ 2 \frac{{\rm Im}\, d_{2}}{M^3} \langle F_{0i}(0) F_{0i}(0) \rangle_{T}  
\nonumber\\
&=&  \left( 3 {\rm Im}\, d_{1} +  {\rm Im}\, d_{2} \right) \frac{2 \pi^2}{15} \frac{T^4}{M^3}\,,
\label{gammagauge}
\end{eqnarray}
where the Wilson coefficients $d_{i}$ can be read off eq. \eqref{cod12}.
The last line of \eqref{gammagauge} follows from having computed the gauge boson thermal electric condensates 
at leading order~\cite{Brambilla:2008cx}: 
$\langle W^a_{0i}(0) W^a_{0i}(0) \rangle_{T}   = \pi^2 T^4/5$ 
and $ \langle F_{0i}(0) F_{0i}(0) \rangle_{T} =  \pi^2 T^4/15$.
The operators $\bar{N}N \, W^{a}_{\mu\nu} W^{a\,\mu\nu}$ and $\bar{N}N \, F_{\mu\nu} F^{\mu\nu}$ in the last line
of \eqref{Wb} do not contribute to the thermal width because at leading order    
$\langle W^a_{\mu\nu}(0) W^{a\,\mu\nu}(0) \rangle_{T}   = \langle F_{\mu\nu}(0) F^{\mu\nu}(0) \rangle_{T} = 0$.

If the Majorana neutrino is not at rest, then we need to add to \eqref{Wb} 
operators that depend on the neutrino momentum. The leading operator is the dimension seven operator (embedding SM degrees of freedom)
\begin{equation}
\mathcal{L}_{{\hbox{\tiny N-k}}} = - \frac{1}{2 M^3} a \; \bar{N} \left[\partial^2-(v\cdot \partial)^2\right] N \, \phi^{\dagger} \phi \, .
\label{Wmomdep}
\end{equation}
The Wilson coefficient of this operator is fixed by the relativistic dispersion relation
\begin{equation}
\bar{N}N\, \left(\sqrt{(M+\delta m)^2+\bm{k}^{\,2}} - M \right) = 
\bar{N}N\,\left(\delta m + \frac{\bm{k}^{\,2}}{2M} - \delta m\, \frac{\bm{k}^{\,2}}{2M^2} + \dots\right) \,,
\end{equation}
with $\delta m = - a\,\phi^{\dagger} \phi/M$, or by methods similar to those developed in~\cite{Brambilla:2003nt}.
Therefore there is a thermal width induced by the operator \eqref{Wmomdep} that is going to be momentum dependent. It reads
\begin{eqnarray}
\Gamma_{\phi,\hbox{\tiny mom.\,dep.}} &=& 
2 \frac{{\rm Im} \, a}{M} \left(-\frac{\bm{k}^{\,2}}{2M^2} \right) \langle \phi^{\dagger}(0) \phi(0) \rangle_{T}  
= -  \frac{{\rm Im} \, a}{6}\frac{\bm{k}^{\,2}T^2}{M^3} \,.
\label{gammamomdep}
\end{eqnarray}

The above expressions for the thermal decay widths induced by Higgs, fermions and gauge bosons 
are consistent with the estimates \eqref{eq0} and \eqref{eq} obtained by sole power-counting arguments.
Summing up $\Gamma_{\phi}$, $\Gamma_{\phi,\hbox{\tiny mom.\,dep.}}$, $\Gamma_{\hbox{\tiny fermions}}$ and $\Gamma_{\hbox{\tiny gauge}}$ 
and using the explicit expressions of the Wilson coefficients, we get 
at first order in the SM couplings and at order $T^4/M^3$ the Majorana neutrino thermal width:
\begin{equation}
\Gamma^T = 
\frac{|F|^{2}M}{8\pi}\left[-\lambda \left( \frac{T}{M} \right)^{2} 
+ \frac{\lambda}{2} \frac{\bm{k}^{\,2}\,T^2}{M^4}
- \frac{\pi^{2}}{80}\left( \frac{T}{M} \right)^{4}(3g^{2}+g'^{\,2}) 
- \frac{7\pi^{2}}{60}\left( \frac{T}{M} \right)^{4}|\lambda_{t} |^{2} \right] .
\label{eq20}
\end{equation} 
If the neutrino is at rest, we can set $\bm{k}=\bm{0}$.
Equation \eqref{eq20} agrees with the analogous expression derived in~\cite{Salvio:2011sf} 
up to order $T^2/M$. It also agrees with the result of~\cite{Laine:2011pq}, given in eq.~(\ref{result_widthLaine}), up to order $T^4/M^3$.
In~\cite{Laine:2011pq} also corrections of order $\bm{k}^{\,2}T^4/M^5$ have been computed.
We note that we could express our results \eqref{gammaphi_part}, \eqref{gammamomdep}, \eqref{gammafermions} 
and \eqref{gammagauge} also in terms of Higgs, lepton, quark and gauge field condensates.
This appears to be a straightforward consequence of the EFT, which requires, 
at the order considered here, that thermal corrections are encoded into tadpole diagrams. 
In relation to $\Gamma^T$, condensates have been also discussed in~\cite{Laine:2011pq}.

\section{The $T/M$ expansion}
\label{expansion_partprod}
In this chapter we have computed the thermal corrections to the neutrino thermal width as an expansion in the SM couplings and in $T/M$. This quantity enters the production rate expression that has been computed in a similar fashion in~\cite{Salvio:2011sf,Laine:2011pq,Biondini:2013xua}. 
Up to the order to which it is known, the expansion in $T/M$ is well behaved, i.e., for reasonably small values of $T/M$ it converges. 

\begin{figure}[ht]
\centering
\includegraphics[scale=0.8]{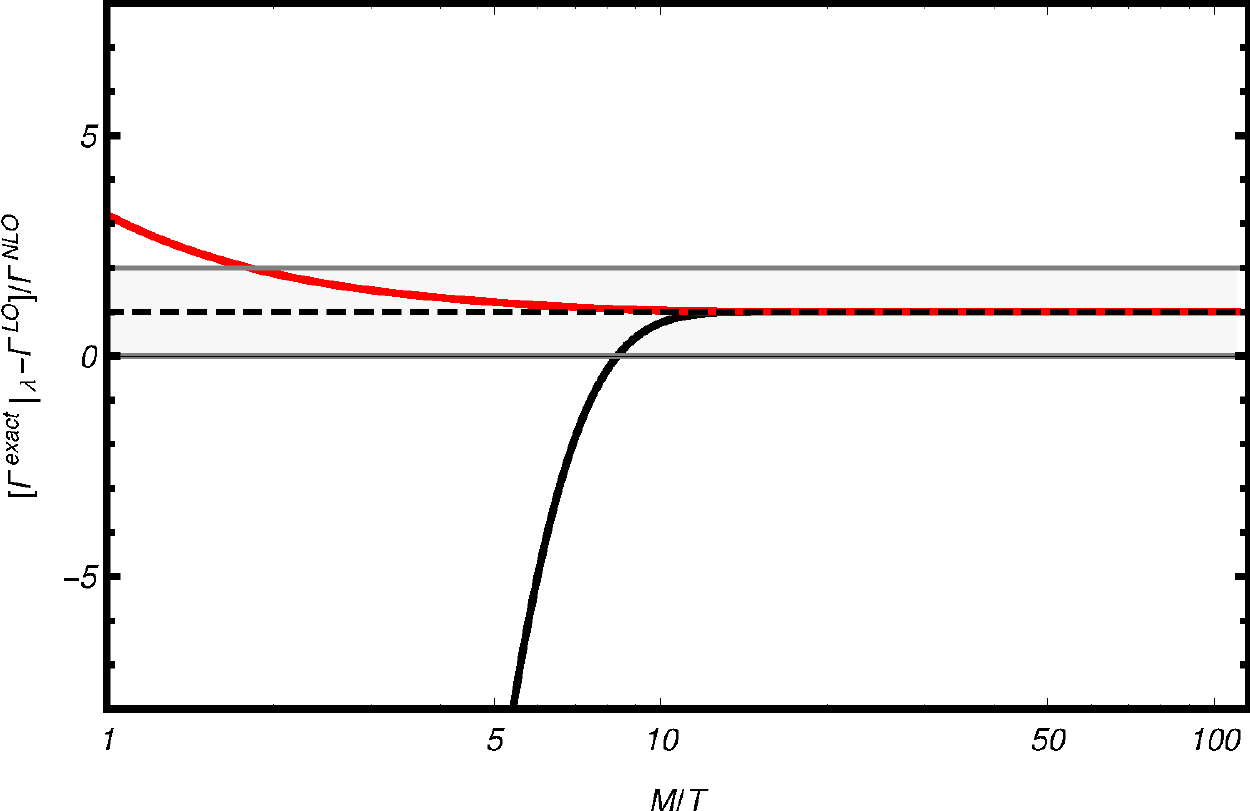}
\caption{The black line shows the difference between the exact neutrino production rate up to order $\lambda$ (top-Yukawa and gauge couplings set to zero) 
and the neutrino production rate at leading order in $T/M$ divided by the neutrino production rate at next-to-leading order in $T/M$.
The red line is as above but with the leading-order neutrino production rate multiplied by $(1 + n_B(M/2) - n_F(M/2))$. The neutrino is taken at rest.
The one-loop running four-Higgs coupling, $\lambda$, is taken $\lambda(10^7~{\rm GeV}) \approx 0.02$ ($\lambda(125~{\rm GeV}) \approx 0.126$)~\cite{Rose15}.
}
\label{Gammaexpansion} 
\end{figure}

Despite the above fact, it has been remarked in~\cite{Laine:2013lka} that, 
when comparing the production rate for heavy Majorana neutrinos in the $T/M$ expansion with the exact result, 
which is known at leading order in the SM couplings, the two results overlap only at very small values of $T/M$, i.e., values around $1/10$ or smaller.
In the same work, it has been also noticed that for values of $T/M$ larger than $1/10$ 
not only the discrepancy between the exact and the approximate result appears larger than the last known term in the expansion, but also of opposite sign.
The situation is well illustrated by the black curve in figure~\ref{Gammaexpansion}. 
It shows the difference between the exact neutrino production rate at order $\lambda$ (top-Yukawa and gauge couplings are set to zero) taken from~\cite{Laine:2013lka}
and the neutrino production rate at leading order in $T/M$ divided by the neutrino production rate at next-to-leading order in $T/M$. 
At next-to-leading order in $T/M$ the production rate depends only on the SM coupling $\lambda$.

Here we want to inspect the origin of this behaviour  
and devise a strategy to improve the expansion in $T/M$ in such a way that it overlaps with the exact result for reasonably small, 
not only very small, values of $T/M$. We will say that the expansion overlaps with the exact result if the discrepancy between the exact 
and the approximate result is not larger than the last known term in the expansion.

The problem is rather general. In the form we have it here, it happens when dealing with a double expansion 
where one of the expansion parameters is much smaller than the other one.
In our case $\lambda$ is much smaller than $T/M$ for a relatively wide range of temperatures.
Under this circumstance, exponentially suppressed terms of the type $e^{-M/T}$ 
may become numerically as large as next-to-leading order terms of the type $\lambda\,(T/M)^2$.
In fact $e^{-M/T}$ is larger than or very close to $\lambda\,(T/M)^2$ for $T/M \simg 1/8$.
One should recall that exponentially suppressed terms vanish in any analytic expansion.

The solution of the problem consists in keeping exponentially suppressed terms in the not-so-small parameter 
at leading order in the small-parameter expansion. 
In our case, this amounts at keeping terms of the type $e^{-M/T}$ in the computation of the neutrino observables at zeroth-order in the SM couplings. 
Let us illustrate how this works in the case of the neutrino production rate.
The relevant diagrams are the self-energies shown in figure~\ref{fig:selfRTF}, which, in the following, we will call $\Pi$.
The neutrino production rate is proportional to the retarded self energy, $\Pi_R$.
We already mentioned in section~\ref{EFT_partprod} that the retarded self energy may be written as $\Pi_R = \Pi_{11} + \Pi_{12}$. 
The ``12'' component of a heavy-particle propagator vanishes exponentially in the heavy-mass limit~\cite{Brambilla:2008cx}. 
For this reason we did not need to consider $\Pi_{12}$ in section~\ref{EFT_partprod}. 
But we need to consider it here if we want to keep exponentially suppressed terms. 
Cutting $\Pi_{11}$ and keeping the thermal distributions of the lepton and Higgs boson gives for a neutrino at rest 
$\Pi_{11} = \left[T=0~{\rm result}\right] \times (1+n_B(M/2)) (1-n_F(M/2))$, where $n_B$ and $n_F$ are the Bose and Fermi distributions respectively.
Cutting $\Pi_{12}$ gives $\Pi_{12} = \left[ T=0~{\rm result} \right]  \times n_B(M/2) n_F(M/2)$.
Summing the two contributions gives $\Pi_R = \left[ T=0~{\rm result} \right]  \times (1 + n_B(M/2) - n_F(M/2))$.
Hence, we can improve the neutrino production rate at leading order in the SM coupling by multiplying the $T=0$ result by 
\begin{equation}
1 + n_B(M/2) - n_F(M/2) \approx 1 + 2\,e^{-M/T}  + ...\;,
\label{expimprovement}
\end{equation}
which amounts at keeping (at least) terms of the type $e^{-M/T}$.

In figure~\ref{Gammaexpansion} the red curve shows the difference between the exact neutrino production rate at order $\lambda$ 
(top-Yukawa and gauge couplings set to zero) and the neutrino production rate at leading order in $T/M$ 
multiplied by $(1 + n_B(M/2) - n_F(M/2))$ divided by the neutrino production rate at next-to-leading order in $T/M$.
The grey band shows the region where the discrepancy between the exact production rate and the next-to-leading order one 
is not larger than the next-to-leading order one. 
We see that now the curve is in the grey band for $T/M \siml 1/2$.
Moreover, higher-order corrections in $T/M$ do not change the sign of the next-to-leading order correction.
The result is consistent with our understanding of the problem and in fact provides a simple way to solve it.
This computational scheme could be also implemented in the case of the CP asymmetry (see chapters~\ref{chap:CPdege} and \ref{chap:CPhiera}). 
%For the direct CP asymmetry, the leading-order diagrams are in this case given by the two-loop diagrams shown in figure~\ref{Fig2}.
%Because we are cutting them and taking the imaginary parts of the remaining one-loop subdiagrams, 
%exponentially suppressed contributions can be computed straightforwardly taking into account the combinatorics of all possible physical 
%and unphysical degrees of freedom contributing to $\Pi_{11}$ and $\Pi_{12}$ at two loops.
%A computation along this line is in~\cite{Garny:2010nj}.
%For the indirect CP asymmetry, the computation may be done in the EFT, whose parameters are the thermal decay widths and masses.
%The exponential improvement of the widths has been discussed in the previous paragraphs. 

Finally, we comment about the neutrino three momentum $\bm{k}$, rather of its absolute value $|\bm{k}|$. Strictly speaking the non-relativistic expansion is an expansion in $T/M$ and $|\bm{k}|/M$ 
and is as good as these two parameters are small. If $|\bm{k}|$ is chosen to be equal to $T$ or smaller, 
as we did in figure~\ref{Gammaexpansion}, then $T/M$ is the relevant expansion parameter.
But if $|\bm{k}| = 2T$, $|\bm{k}|= 3T$, ... then this is $|\bm{k}|/M$.
In particular, one has to expect (naively) the exact result to overlap with the result of the perturbative series 
at temperature $2$, $3$, ... times smaller than one would have for $|\bm{k}| \le T$.

%% file: CP_dege.tex
In this chapter the formalism developed for the right-handed neutrino production rate is applied to 
the evaluation of CP asymmetries in heavy Majorana neutrino decays. Indeed the EFT approach has provided a simpler 
derivation of the neutrino thermal width and we shall exploit such tools in order to inspect processes at higher order. 
We consider an extension of the SM that includes two 
generations of heavy Majorana neutrinos with nearly degenerate masses, $M$ and $M+\Delta$, and
we compute the leading thermal corrections to the direct and indirect CP asymmetries 
in heavy neutrino decays into leptons and antileptons.
In section~\ref{sec:rev} we review the basic set-up of the EFT for non-relativistic Majorana neutrinos, this time applied to two nearly mass degenerate states.
In section~\ref{sec:zeroT} we re-derive the zero temperature direct CP asymmetry from the vertex diagram and relate it to the EFT. In section~\ref{sec:finiteT} we explain how to match the relevant dimension-five operators of the EFT at two loops. The leading thermal corrections to the direct CP asymmetry are computed in section~\ref{sec:direct}
and  the leading thermal corrections to the indirect CP asymmetry in section~\ref{sec:indirect}. The result is organized in a $T/M$ expansion and as series in the SM couplings.
\section{Non-relativistic Majorana neutrinos with nearly degenerate masses}
\label{sec:rev}
We want to study the dynamical generation of the CP asymmetries in heavy-neutrino decays occurring in the early universe. The CP asymmetry is defined as fallows 
\begin{equation}
\epsilon_{I}=
\frac{\sum_{f} \Gamma(\nu_{R,I} \to \ell_{f} + X)-\Gamma(\nu_{R,I} \to \bar{\ell}_{f}+ X )  }
{\sum_{f} \Gamma(\nu_{R,I} \to \ell_{f} + X ) + \Gamma(\nu_{R,I} \to \bar{\ell}_{f}+ X)} \, .
\label{CP_CPdege}
\end{equation}
The sum runs over the SM lepton flavours, $\nu_{R,I}$ stands for the
$I$-th heavy right-handed neutrino species, $\ell_{f}$ is a SM lepton with flavour $f$ 
and $X$ stands for any other SM particle not carrying a lepton number. In this chapter we do not address flavour effects on the CP asymmetry which are treated in chapter~\ref{chap:CPfla}. The definition in eq.~(\ref{CP_CPdege}) replaces that given in eq.~(\ref{eq:adef}) when considering more general combinations of the decay products and more than one neutrino species contributing to the CP asymmetry.

Interactions with the medium modify the neutrino dynamics (thermal production rate, mass, ...)  and affect the thermodynamic evolution of the lepton asymmetry. Since we are interested in the temperature regime
\begin{equation}
M \gg T \gg M_{W} \, ,
\label{hiera_CPdege}
\end{equation}  
the EFT constructed in the previous chapter for a non-relativistic Majorana neutrino can be used as a starting point. One can read its general structure up to dimension-seven operators in eqs.~\eqref{eq10}, \eqref{Wa} and \eqref{Wb}. In the temperature window \eqref{hiera_CPdege} and in an expanding universe the heavy
neutrino is likely out of equilibrium, which is one of the Sakharov 
conditions necessary for generating a lepton asymmetry~\cite{Sakharov:1967dj}. 

However there is a slight modification one has to take into account about the degrees of freedom in the EFT, in order to address a successful generation of the CP asymmetry: at least two different heavy Majorana neutrino species interacting with different Yukawa couplings are needed. This fact is closely related to the generation of a non-vanishing phase in the Yukawa couplings combination entering the processes responsible for the CP asymmetry. We have encountered a similar situation for the toy model described in section~\ref{hot_sec2}. We comment on this point later in the next sections once we have worked out explicitly the expressions for the CP asymmetry.  

In the following, we will consider only two heavy neutrinos and assume that they have masses above the electroweak scale.  
In the case right-handed neutrinos are represented by Majorana fermion fields, 
the Lagrangian read off \eqref{lepto_8}, where the neutrino generation index is then $I=1,2$. The corresponding mass eigenstates are $M_1$ and $M_2$ and in order to consider two heavy neutrinos with nearly degenerate masses we ask $M_1 \equiv M$ and $M_2=M+\Delta$, where $\Delta$ is the mass splitting such that $\Delta \ll M$. The system with two nearly degenerate neutrinos is still characterized by one large scale, $M$. Therefore we integrate out momentum and energy modes of order $M$ from the 
fundamental Lagrangian~\eqref{lepto_8} and replace it by a suitable EFT aimed 
at describing the non-relativistic dynamics of the Majorana neutrinos.
The EFT is organized as an expansion in operators of increasing dimension suppressed by powers of $1/M$.
The Wilson coefficients of the operators encode the high-energy modes of the fundamental theory and can be 
evaluated by setting $T=0$.  Then we compute thermal corrections to the Majorana neutrino leptonic (antileptonic) widths as thermal averages weighted by the partition function of the EFT. This procedure is the same adopted in chapter~\ref{chap:part_prod} for the neutrino thermal width at order $F^2$. However, we shall see that we have to work out the decay widths at order $F^4$ in the Yukawa couplings to calculate the CP asymmetry (see also section~\ref{sec_lepto2}). 

The EFT Lagrangian up to operators of dimension five is
\begin{equation}
\mathcal{L}_{\rm{EFT}}=\mathcal{L}_{\rm{SM}} 
+ \bar{N}_I \left( i v \cdot \partial -\delta M_I \right) N_I +
\frac{i\Gamma^{T=0}_{IJ}}{2}\bar{N}_I N_J+\frac{a_{IJ}}{M_I}\bar{N}_IN_J \phi^{\dagger}\phi + \dots,
\label{eq:efflag_CPdege}
\end{equation}
where $N_{I}$ is the field describing the low-energy modes of the $I$-th non-relativistic Majorana neutrino, 
$\delta M_1=0$, $\delta M_2=\Delta$, $\Gamma^{T=0}_{IJ}$ is the decay matrix at $T=0$ 
and $a_{IJ}$ are the Wilson coefficients of the dimension-five operators $\bar{N}_IN_J \phi^{\dagger}\phi$
describing the interaction of the Majorana neutrinos with the Higgs doublet of the SM.  These are the only operators of dimension five that give thermal corrections to the neutrino widths and masses. 
The dots in \eqref{eq:efflag_CPdege} stand for higher-order operators that contribute with subleading 
corrections. Being the CP asymmetry a dimensionless quantity, we are able to size the thermal correction induced by the dimension-five operators to be of order $(T/M)^2$. This is due to the temperature dependence developed by the Higgs condensate at leading order, $\langle \phi^\dagger(0)\phi(0)\rangle_T  \propto T^2$ (see eq.~\eqref{cond1}), hence two inverse powers of the neutrino mass have to appear. Higher order operators in \eqref{eq:efflag_CPdege} induce parametrically $T/M$ suppressed corrections. 
The natural dynamical scale of the EFT Lagrangian is the temperature, $T$. Since $T$ is taken larger than the electroweak 
scale, $\mathcal{L}_{\rm{SM}}$ is still the SM Lagrangian with unbroken SU(2)$_L\times$U(1)$_Y$ gauge symmetry.

The Lagrangian \eqref{eq:efflag_CPdege} has been obtained by integrating out the mass $M=M_1$ from the Lagrangian~\eqref{lepto_8}; 
$\delta M_2=\Delta \ll M$ is the residual mass of the neutrino of type 2. 
In~\eqref{eq:efflag_CPdege} masses are understood as on-shell masses,
as it is typical of non-relativistic EFTs, which implies that off-diagonal elements 
of the mass matrix vanish. Moreover, in the diagonal terms we will neglect terms that would contribute 
to the CP asymmetry at order $F^6$ or smaller~\cite{Kniehl:1996bd,Anisimov:2005hr}. 
Off-diagonal elements do not vanish for the absorptive parts $i\Gamma^{T=0}_{IJ}/2$. The specification $T=0$ recalls that they are computed at $T=0$.
Finally, the Lagrangian \eqref{eq:efflag_CPdege} has been written in a reference frame where the Majorana neutrinos 
have momentum $M v^\mu$ ($v^2=1$) up to a residual momentum that is much smaller than $M$.
In the following, we will assume that the thermal bath of SM particles is comoving with the 
Majorana neutrinos. For the matching calculation, a convenient choice of the reference frame is the rest frame $v^{\mu}=(1,\bm{0})$.

The expression for the non-relativistic Majorana
propagator in the EFT (\ref{eq:efflag_CPdege}), both for the neutrino of type~1 and type~2, can be obtained by projecting (\ref{eq5_partprod})-(\ref{eq7_partprod}) on the small components of the Majorana fields. 
Putting $p^{\mu}=Mv^{\mu}+k^{\mu}$, where $k^2 \ll M^2$, we obtain in the large $M$ limit
\bea
&&\langle 0 | T ( N_1^{\alpha}(x) \bar{N}_1^{\beta}(y) )  | 0 \rangle 
= \left( \frac{1+\slashed{v}}{2}\right)^{\alpha \beta} \int \frac{d^{4}k}{(2 \pi)^{4}} \, e^{-ik\cdot(x-y)}\, \frac{i}{v\cdot k +i\epsilon}  \,  ,
\label{effpropagator_bis}
\\
&&\langle 0 | T ( N_2^{\alpha}(x) \bar{N}_2^{\beta}(y) )  | 0 \rangle 
= \left( \frac{1+\slashed{v}}{2}\right)^{\alpha \beta} \int \frac{d^{4}k}{(2 \pi)^{4}} \, e^{-ik\cdot(x-y)}\, \frac{i}{v\cdot k -\Delta +i\epsilon}  \,  ,
\label{effpropagator2}
\eea 
where $M_1=M$ and $\Delta = M_2-M_1$. The other possible time-ordered combinations vanish. We notice the presence of a residual mass, $\Delta$, in the neutrino type~2 propagator in (\ref{effpropagator2}). We stress that the expressions ``neutrino of type~1" and ``neutrino of type~2" are referred to the heavy neutrino species. They do not have to be confused with the field of type~1 and type~2 on the Keldysh contour in the RTF of thermal field theory (see chapter~\ref{chap:therm_tex}).

\begin{figure}[t]
\centering
\includegraphics[scale=0.565]{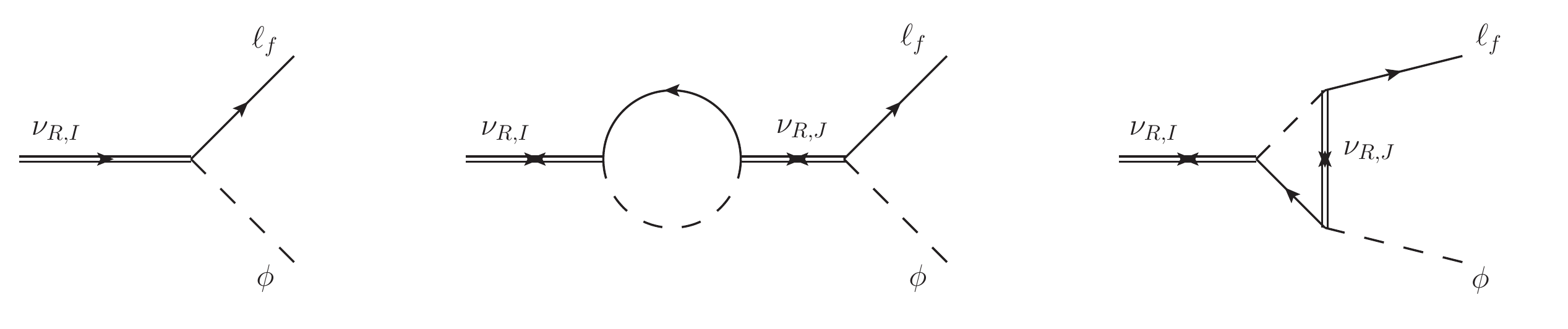}
\caption{From left to right: tree-level, and one-loop self-energy and vertex diagrams.
Double solid lines stand for heavy right-handed neutrino propagators,
solid lines for lepton propagators and dashed lines for Higgs boson propagators.
The neutrino propagator with forward arrow corresponds to $\langle 0| T(\psi \bar{\psi}) |0\rangle$,  
whereas the neutrino propagators with forward-backward arrows correspond to $\langle 0| T(\psi \psi) |0\rangle$ 
or $\langle 0| T(\bar{\psi} \bar{\psi})|0 \rangle$. Decay into antileptons are not shown.}
\label{fig:dirind_degeCP} 
\end{figure} 
\subsection{Set up of the CP asymmetries in the EFT} 
\label{sec:dirindi}
In chapter~\ref{chap:lepto} we distinguished between indirect and direct CP asymmetry, 
the distinction being based on the leading-order processes shown in figure~\ref{fig:dirind_degeCP}. 
In this chapter, we extend that distinction beyond leading order by calling contributions to the indirect 
CP asymmetry, $\Delta\Gamma_{I,{\rm indirect}}$, those that allow for  the phenomenon of resonant enhancement. The resonant leptogenesis was discussed briefly in chapter~\ref{chap:lepto} and is characterized by
a large enhancement of the asymmetry when $\Delta$ is of the order of the largest 
between the neutrino width difference and the mixing vertices.
In the framework of a strict perturbative expansion in the Yukawa couplings, such a behaviour is induced 
by Feynman diagrams (like the second of figure~\ref{fig:dirind_degeCP}) becoming singular in the limit $\Delta\to 0$, 
which signals a break down of the expansion in that limit. 
The singularity is eventually removed by resumming certain classes of diagrams, 
like those responsible for the width and/or the mixing of the different neutrinos.
Viceversa, we call contributions to the direct CP asymmetry, $\Delta\Gamma_{I,{\rm direct}}$, those that 
do not exhibit this phenomenon. Order by order in an expansion in the Yukawa couplings, 
Feynman diagrams that contribute to the direct CP asymmetry are not singular in the limit $\Delta\to 0$.
The CP asymmetry is related to the sum of these two kind of contributions (see definition \eqref{CP_CPdege}): 
\begin{eqnarray}
\sum_{f} \Gamma(\nu_{R,I} \to \ell_{f} + X)-\Gamma(\nu_{R,I} \to \bar{\ell}_{f}+ X ) &=& \Delta\Gamma_{I,{\rm direct}} + \Delta\Gamma_{I,{\rm indirect}} \,.
\label{asysec2}
\end{eqnarray}

The term $\Delta\Gamma_{I,{\rm direct}}$ includes all contributions to the CP asymmetry that originate from single operators in the EFT 
and all contributions that come from mixing of operators in the EFT that do not show the phenomenon of resonant enhancement. 
Concerning the first class of contributions, at the accuracy of the Lagrangian~\eqref{eq:efflag_CPdege} 
there are only dimension-three and dimension-five operators that may have imaginary Wilson coefficients.
Concerning the second class of contributions, we will denote them $\Delta\Gamma_{I,{\rm direct}}^{\rm mixing}$.
At the order we are working, the only relevant contribution of this kind affects the heavier Majorana neutrino of type 2 
and will be computed in section~\ref{sec:direct2}.
Hence, $\Delta\Gamma_{I,{\rm direct}}$ reads 
\begin{equation}
\Delta\Gamma_{I,{\rm direct}} = \left(\Gamma^{\ell,T=0}_{II}-\Gamma^{\bar{\ell},T=0}_{II}\right) 
+ \left(\Gamma^{\ell,T}_{II,{\rm direct}}-\Gamma^{\bar{\ell},T}_{II,{\rm direct}}\right) 
+ \Delta\Gamma_{I,{\rm direct}}^{\rm mixing}\,,
\label{asysec2bis}
\end{equation}
with
\begin{equation}
\Gamma^{\ell,T}_{II,{\rm direct}} = \frac{2}{M}{\rm Im}\,a_{II}^\ell \,\langle \phi^\dagger(0)\phi(0)\rangle_T,
\qquad 
\Gamma^{\bar{\ell},T}_{II,{\rm direct}} = \frac{2}{M}{\rm Im}\,a_{II}^{\bar\ell} \,\langle \phi^\dagger(0)\phi(0)\rangle_T,
\label{asysec3}
\end{equation}
where the subscripts $\ell$ and $\bar{\ell}$ isolate the leptonic and antileptonic contributions.
The first term in the right-hand side of~\eqref{asysec2bis}, $\Gamma^{\ell,T=0}_{II}-\Gamma^{\bar{\ell},T=0}_{II}$,  
is the zero temperature contribution to the direct CP asymmetry, which we will compute in section~\ref{sec:zeroT}.
The second term, $\Gamma^{\ell,T}_{II,{\rm direct}}-\Gamma^{\bar{\ell},T}_{II,{\rm direct}}$,
isolates the dominant thermal correction to the direct CP asymmetry, therefore  $a_{II}^\ell$ and  $a_{II}^{\bar{\ell}}$ have to be computed to derive its explicit expression. 

In equation \eqref{asysec3} the thermal dependence is encoded in the Higgs thermal condensate $\langle \phi^\dagger(0)\phi(0)\rangle_T$, 
which at leading order is written in~\eqref{cond1}.
%\begin{equation}
%\langle \phi^\dagger(0)\phi(0)\rangle_T  = \frac{T^2}{6}.
%\label{higgscondensate}
%\end{equation}
The relative size of the thermal correction to the direct CP asymmetry is therefore $T^2/M^2$.
High-energy contributions induced by loops with momenta of the order of the neutrino mass 
are encoded in the Wilson coefficients $a_{II}^\ell$ and $a_{II}^{\bar{\ell}}$. 
Since the condensate is real, to compute the widths we need the imaginary parts of $a_{II}^\ell$ and $a_{II}^{\bar{\ell}}$. 
Their expressions, at order $F^2$ in the Yukawa couplings, can be easily inferred from eq.~(\ref{match_partprod_4}) taking into account that such expression refers to the leptonic contribution. A detailed derivation disentangling the lepton and antilepton contributions is given in appendix~\ref{appC:CPdegematch} and the result reads
\begin{equation}
{\rm Im}\, a_{II}^\ell = {\rm Im}\, a_{II}^{\bar\ell} = -\frac{3}{16\pi}|F_{I}|^2\lambda.
\label{agen}
\end{equation}
The coupling $\lambda$ is the four-Higgs coupling. 
We have defined $|F_{I}|^2 \equiv \sum_{f} F_{f I} F^{*}_{f I}$ and, for further use, $F_{J}F_{I}^{*} \equiv \sum_f F_{fJ}F_{fI}^{*}$.

A necessary condition to produce a CP asymmetry, i.e., to get a non-vanishing difference from a final state 
with a lepton and one with an antilepton, is for ${\rm Im}\,a_{II}^\ell$ and ${\rm Im}\,a_{II}^{\bar{\ell}}$
to be sensitive to the phases of the Yukawa couplings $F_{fI}$.
At order $F^2$, ${\rm Im}\, a_{II}^\ell$ and ${\rm Im}\, a_{II}^{\bar\ell}$ are not.
Hence, to produce a non-vanishing direct CP asymmetry, one needs to compute at least corrections of order $F^4$. 
In fact, corrections proportional to $(F_{1}F^{*}_{2})^2$ are sensitive to the phases of the Yukawa couplings.
From the optical theorem the imaginary part of a two-loop diagram proportional to $(F_{1}F^{*}_{2})^2$ may be 
understood as the interference between a tree-level and a one-loop amplitude developing an imaginary part. We are going to clarify this aspect in the following section where we derive the direct CP asymmetry at $T=0$ at zeroth order in the SM couplings.

\section{Matching $\Gamma^{T=0}_{II}$: direct asymmetry at zero temperature}
\label{sec:zeroT}
The width difference \eqref{asysec2bis}, and hence the direct CP asymmetry, depends on the Wilson coefficients $\Gamma^{T=0}_{II}$ and $a_{II}$ of~\eqref{eq:efflag_CPdege}. 
In this section we compute the leptonic, $\Gamma^{\ell,T=0}_{II}$, and antileptonic, $\Gamma^{\bar{\ell},T=0}_{II}$, components of $\Gamma^{T=0}_{II}$.
In so doing we re-derive the expression for the direct CP asymmetry at zero temperature~\cite{Fukugita:1986hr}. 
Considerations made here will be used in the next section to select the parts of the Wilson coefficients 
${\rm Im}\,a_{II}^\ell$ and ${\rm Im}\,a_{II}^{\bar{\ell}}$ relevant for the thermal corrections to the direct CP asymmetry.

\begin{figure}[ht]
\centering
\includegraphics[scale=0.55]{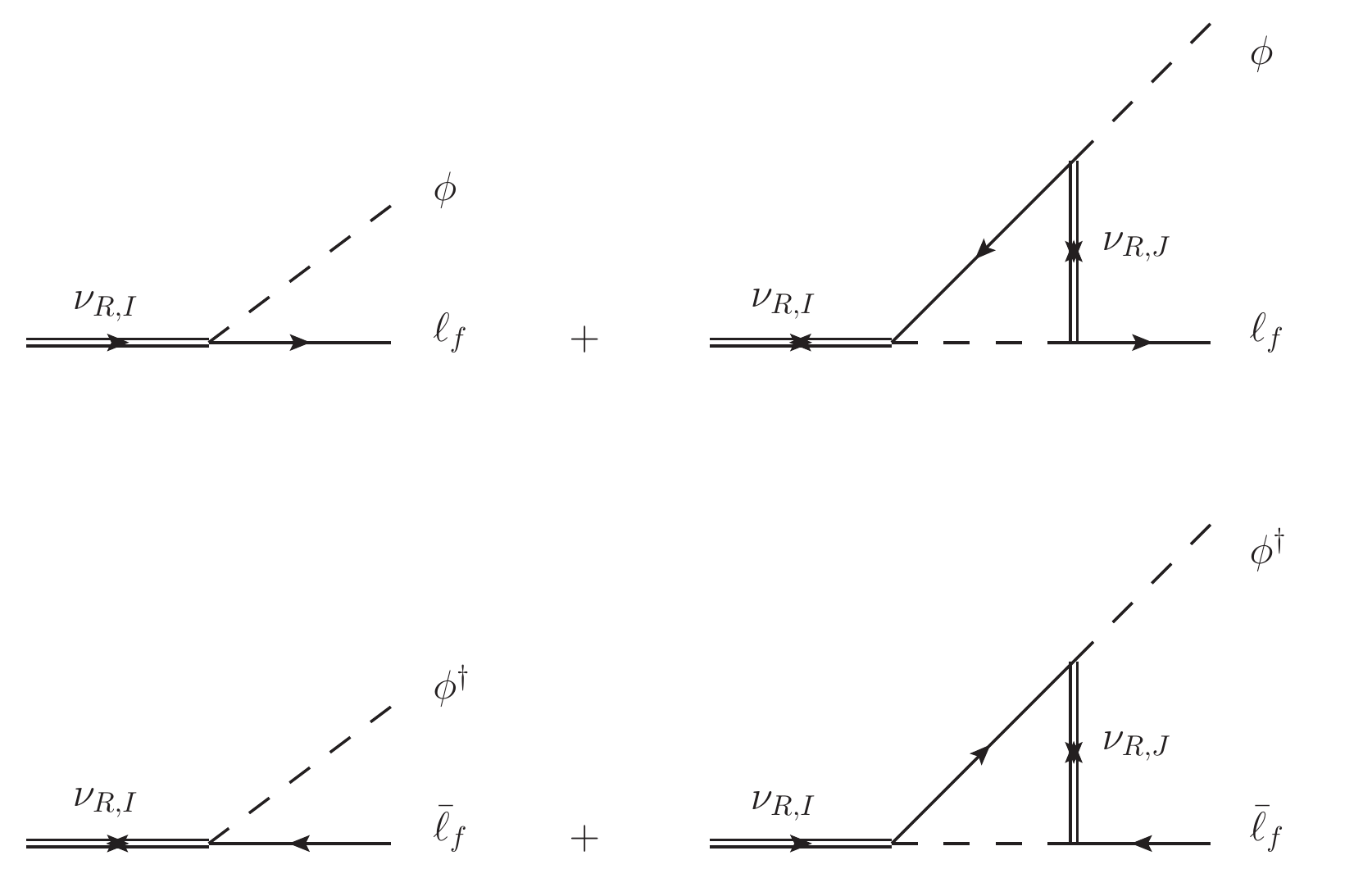}
\caption{Tree-level and one-loop diagrams contributing to the direct CP asymmetry.
The subscript $I$ stands either for 1 or 2. 
The first and second raw show decays into leptons and antileptons respectively.}
\label{fig:fig_2_CPdege} 
\end{figure}

We start considering the decay of a heavy right-handed neutrino of type 1, $\nu_{R,1}$, into leptons. 
Up to one loop the amplitude has the following form (see the two upper diagrams in figure~\ref{fig:fig_2_CPdege} that display only direct 
contributions): 
\begin{equation}
\mathcal{M}(\nu_{R,1} \rightarrow \ell_{f}+X) =  A \left[ F_{f  1} + \sum_{J} (F^{*}_{ f' 1} F_{f'  J})F_{f  J} \, B \right] ,
\label{CPdege_b2}
\end{equation} 
where $A$ and $B$ are functions that parametrize the amplitude at tree-level and one-loop respectively. 
We obtain the total decay width into leptons by squaring the amplitude and summing over the lepton flavours. 
Up to $\mathcal{O}(F^4)$ it reads
\begin{eqnarray}
&& \sum_{f} \Gamma(\nu_{R,1} \rightarrow \ell_f + X) =  |A|^2\left[ |F_{1}|^2 
+ \sum_{J} \left( (F^{*}_{1}F_{J})^2 \, B + ( F_{1} F^{*}_{J})^2 \, B^{*} \right) \right]  
\nonumber \\
&& \hspace{1.2cm}
=  |A|^2 \left\lbrace  |F_{1}|^2    + \sum_{J} \left(   2 \, {\rm{Re}}(B){\rm{Re}} \left[ (F_{1}^{*}F_{J})^2\right] 
-2 \, {\rm{Im}}(B) {\rm{Im}} \left[ (F_{1}^{*}F_{J})^2\right] \right)  \right\rbrace .
\nn
\\
\phantom{x}
\label{CPdege_b3}
\end{eqnarray} 
We may write similar relations for the decay into antileptons:
\begin{equation}
\mathcal{M}(\nu_{R,1} \rightarrow \bar{\ell_{f}} + X) =  A \left[ F^{*}_{f 1} + \sum_{J} (F_{f'1} F^{*}_{f' J})F^{*}_{fJ} \, C \right] ,
\label{CPdege_b5}
\end{equation}
and 
\begin{eqnarray}
&& \sum_{f} \Gamma(\nu_{R,1} \rightarrow \bar{\ell}_f+X) = |A|^2\left[ |F_{1}|^2 
+ \sum_{J} \left( (F^{*}_{1}F_{J})^2 \, C^{*} + (F_{1}F^{*}_{J})^2 \, C \right) \right]  
\nonumber \\
&& \hspace{1.2cm}
=  |A|^2 \left\lbrace  |F_{1}|^2    + \sum_{J} \left( 2 \, {\rm{Re}}(C){\rm{Re}} \left[ (F_{1}^{*}F_{J})^2\right] +2 
\, {\rm{Im}}(C) {\rm{Im}} \left[ (F_{1}^{*}F_{J})^2\right] \right) \right\rbrace ,
\nn
\\
\phantom{x}
\label{CPdege_b6}
\end{eqnarray} 
where $C$ is the analogous of $B$ in~(\ref{CPdege_b2}). 
The CP asymmetry~(\ref{CP_CPdege}), due to the decay of $\nu_{R,1}$, is then 
\begin{equation}
\epsilon_{1}= \sum_{J}  \, \frac{\left( {\rm{Re}}(B)-{\rm{Re}}(C) \right) {\rm{Re}} \left[ (F_{1}^{*}F_{J})^2\right] 
- \left( {\rm{Im}}(B)+{\rm{Im}}(C) \right) {\rm{Im}} \left[ (F_{1}^{*}F_{J})^2 \right] }{|F_{1}|^2} \, .
\label{CPdege_cpzero}
\end{equation}
\begin{figure}[t!]
\centering
\includegraphics[scale=0.45]{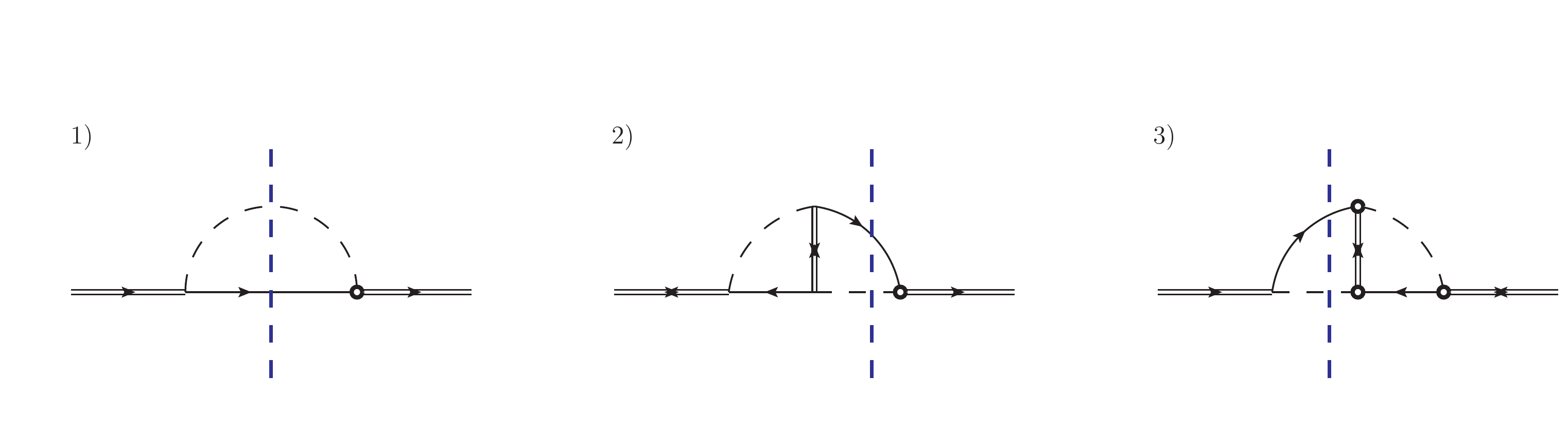}
\caption{One-loop and two-loops self-energy diagrams in the fundamental theory \eqref{lepto_8} 
contributing to the decay of a heavy Majorana neutrino into leptons.
Vertical blue dashed lines are the cuts selecting a final state made of a Higgs boson and a lepton.
Circled vertices and propagators are defined in appendix~\ref{appC:CPdegematch}.}
\label{fig:fig3_CPdege} 
\end{figure}
The functions $A$, $B$ and $C$ can be computed by cutting one and two-loop diagrams contributing to the propagator of a neutrino of type~1:
\begin{equation}
-i \left. \int d^{4}x \, e^{ip\cdot x} \, 
\langle \Omega | T \left( \psi_{1}^{\mu}(x) \bar{\psi}_{1}^{\nu}(0) \right) | \Omega \rangle \right|_{p^\alpha =(M + i\eta,\bm{0}\,)} \, ,
\label{CPdege_matrixFund}
\end{equation}
where $|\Omega\rangle$ is the ground state of the fundamental theory 
and where we have chosen the rest frame $v^{\alpha}=(1,\bm{0})$, so that the incoming momentum is $p^\alpha =(M,\bm{0}\,)$.
Diagrams with cuts through lepton propagators contribute to $A$ and $B$ (see figure~\ref{fig:fig_2_CPdege}), while diagrams with 
cuts through antilepton propagators contribute to $A$ and $C$.
An analogous equation to \eqref{CPdege_cpzero} holds for $\epsilon_{2}$.

We consider the in-vacuum diagrams in figure~\ref{fig:fig3_CPdege} for incoming and outgoing neutrinos of type~1. 
The cuts select the contribution to the width into leptons (for details on cutting rules see appendix~\ref{appC:CPdegematch}).
We call $\mathcal{D}^{\ell}_{1}$, $\mathcal{D}^{\ell}_{2}$ and $\mathcal{D}^{\ell}_{3}$ respectively
the diagrams shown in figure~\ref{fig:fig3_CPdege} with amputated external legs.
The quantity ${\rm Im}\left[-i(\mathcal{D}^{\ell}_{1}+\mathcal{D}^{\ell}_{2}+\mathcal{D}^{\ell}_{3})\right]$ provides 
$\delta^{\mu\nu}\,\sum_f \Gamma(\nu_{R,1} \rightarrow \ell_f+X)/2$ at $T=0$ in the fundamental theory \eqref{lepto_8}, 
which matches $\delta^{\mu\nu}\,\Gamma^{\ell,T=0}_{11}/2$ in the EFT~\eqref{eq:efflag_CPdege}.
The quantities $\Gamma^{\ell,T=0}_{II}$ and $\Gamma^{\bar{\ell},T=0}_{II}$ are the leptonic and antileptonic components of $\Gamma^{T=0}_{II}$ respectively. 
At leading order $\Gamma_{II}^{T=0}=\Gamma_{II}^{\ell,T=0}+\Gamma_{II}^{\bar{\ell},T=0}$. 
An explicit calculation up to order $\Delta/M$ gives
\begin{eqnarray}
&&
\delta^{\mu\nu}\,\frac{\Gamma^{\ell,T=0}_{11}}{2} = 
{\rm{Im}}\left[ -i (\mathcal{D}^{\ell}_{1} +\mathcal{D}^{\ell}_{2} +\mathcal{D}^{\ell}_{3})  \right]  = 
\nonumber \\
&& \hspace{1cm}
 \delta^{\mu \nu} \frac{M}{16 \pi}\left\lbrace  \frac{|F_1|^2}{2}  - \frac{\sum_{J=1}^2 {\rm{Re}}\left[ (F_1^* F_J)^2\right]}{(4 \pi)^2}  
\left[ \left( 1-\frac{\pi^2}{6}\right) + \left( 1-\frac{\pi^2}{12} -4 \ln 2\right) \frac{\Delta}{M} \right]  \right.  
\nonumber \\
&& \hspace{1cm}
\left. -\frac{\sum_{J=1}^2 {\rm{Im}}\left[ (F_1^* F_J)^2\right] }{16 \pi} \left[   (-1 +2 \ln 2) + (-3 + 4 \ln 2)\frac{\Delta}{M}   \right]  \right\rbrace .
\label{CPdege_eq15}
\end{eqnarray}
The sum over $J$ comes from the generation of the intermediate Majorana neutrino exchanged in the two-loop diagrams in figure~\ref{fig:fig3_CPdege}, 
clearly $\sum_J {\rm{Im}} (F_1^* F_J)^2 =  {\rm{Im}} (F_1^* F_2)^2$.
We have not considered cuts through the intermediate neutrino, which would correspond to neutrino transitions 
involving the emission of a lepton and an antilepton, because they do not contribute to the CP asymmetry.

The analogous calculation for $\sum_f \Gamma(\nu_{R,1} \rightarrow \bar{\ell}_f+X)$ at $T=0$ in the fundamental theory, 
which matches $\Gamma^{\bar{\ell},T=0}_{11}$ in the EFT, requires the calculation of the one-loop diagram with a virtual antilepton and 
the two-loop diagrams shown in figure~\ref{fig:fig3_CPdege} but with cuts through antilepton propagators. Up to order $\Delta/M$, we obtain 
\begin{eqnarray}
&&
\delta^{\mu\nu}\,\frac{\Gamma^{\bar{\ell},T=0}_{11}}{2} = 
{\rm{Im}}\left[ -i (\mathcal{D}^{{\bar\ell}}_{1} +\mathcal{D}^{{\bar\ell}}_{2} +\mathcal{D}^{{\bar\ell}}_{3})  \right]  = 
\nonumber \\
&& \hspace{1cm}
\delta^{\mu \nu} \frac{M}{16 \pi}\left\lbrace  \frac{|F_1|^2}{2}  - \frac{\sum_{J=1}^2 {\rm{Re}}\left[ (F_1^* F_J)^2\right] }{(4 \pi)^2}  
\left[ \left( 1-\frac{\pi^2}{6}\right) + \left( 1-\frac{\pi^2}{12} -4 \ln 2\right) \frac{\Delta}{M} \right]  \right.  
\nonumber \\
&& \hspace{1cm}
\left. +\frac{\sum_{J=1}^2 {\rm{Im}}\left[ (F_1^* F_J)^2 \right] }{16 \pi} \left[   (-1 +2 \ln 2) + (-3 + 4 \ln 2)\frac{\Delta}{M}   \right] \right\rbrace .
\label{CPdege_eq16}
\end{eqnarray}
The right-hand side of \eqref{CPdege_eq16} differs from the right-hand side of \eqref{CPdege_eq15} only for the sign of the term proportional to ${\rm{Im}}\left[ (F_1^* F_J)^2 \right]$. 
It is precisely this term that originates the CP asymmetry.

From \eqref{CPdege_eq15} and \eqref{CPdege_eq16} it follows:
\begin{eqnarray}
&&\Gamma^{\ell,T=0}_{11}-\Gamma^{\bar{\ell},T=0}_{11} =
- \frac{M}{64 \pi^2} \left[   (-1 +2 \ln 2) + (-3 + 4 \ln 2)\frac{\Delta}{M}   \right] {\rm{Im}}\left[ (F_1^* F_2)^2 \right] ,  
\nn
\\
\label{CPdege_EQ17}\\
&&\Gamma^{T=0}_{11} =\Gamma^{\ell,T=0}_{11} + \Gamma^{\bar{\ell},T=0}_{11} = \frac{M}{8\pi}|F_1|^2,
\label{CPdege_Gamma1T0}
\end{eqnarray}
where in the last line we have neglected terms of order $F^4$.
The direct CP asymmetry at $T=0$ for the leptonic decay of a neutrino of type 1 follows from the definition~\eqref{CP_CPdege}.
In the EFT, equation \eqref{CP_CPdege} translates into the ratio of the above two quantities and reads (including corrections of order $\Delta/M$)
\begin{equation}
\epsilon_{1,{\rm direct}}^{T=0}= \frac{\Gamma^{\ell,T=0}_{11}-\Gamma^{\bar{\ell},T=0}_{11}}{\Gamma^{T=0}_{11}} = 
\left[   (1 -2 \ln 2) + (3 - 4 \ln 2)\frac{\Delta}{M}   \right] \frac{{\rm{Im}}\left[ (F_{1}^{*}F_{2})^2\right]}{8 \pi |F_{1}|^2} .
\label{CPdege_b14}
\end{equation}
Similarly we may obtain the direct CP asymmetry for the leptonic decay of a neutrino of type 2 just by 
changing $F_1 \leftrightarrow F_2$ and $\Delta \to -\Delta$ in the above formula:
\begin{equation}
\epsilon_{2,{\rm direct}}^{T=0}= -\left[   (1 -2 \ln 2) -(3 - 4 \ln 2)\frac{\Delta}{M}   \right] \frac{{\rm{Im}}\left[ (F_{1}^{*}F_{2})^2\right]}{8 \pi |F_{2}|^2}, 
\label{CPdege_b15}
\end{equation}
where we have used ${\rm Im}\left[ (F_{2}^{*}F_{1})^2\right] = - {\rm Im}\left[ (F_{1}^{*}F_{2})^2\right]$.
The result agrees with the original result~\cite{Covi:1996wh} and following confirmations, 
like the more recent~\cite{Fong:2013wr}, after accounting for the different definition of the Yukawa couplings\footnote{
Our couplings are the complex conjugate of the couplings in~\cite{Covi:1996wh} and~\cite{Fong:2013wr}.}. 

It is useful to compare equations \eqref{CPdege_eq15} and \eqref{CPdege_eq16} with \eqref{CPdege_b3} and \eqref{CPdege_b6} respectively. 
It follows that 
\begin{eqnarray}
&& |A|^2 = \frac{M}{16 \pi},\\
&& {\rm{Re}}(B)={\rm{Re}}(C),\\
&& {\rm{Im}}(B)={\rm{Im}}(C)=\frac{1}{16 \pi} \left[   (-1 +2 \ln 2) + (-3 + 4 \ln 2)\frac{\Delta}{M}   \right] \, .
\end{eqnarray}
Replacing the above expressions in \eqref{CPdege_cpzero} one gets back \eqref{CPdege_b14}.
The condition ${\rm{Re}}(B)={\rm{Re}}(C)$ requires both ${\rm{Im}}(B)$ and ${\rm Im}\left[ (F_{1}^{*}F_{J})^2\right]$ 
to be different from zero to produce a non-vanishing CP asymmetry.
The first request is at the origin of the following statement: 
the relevant two-loop diagrams for the CP asymmetry are those that can be cut with two cuts into three tree-level
diagrams. This guarantees that after a first cut through the lepton (or antilepton) propagator 
the remaining one-loop diagram (what is called $B$ above) develops a complex phase.
The second request is fulfilled if there are at least two Majorana neutrino generations with different complex Yukawa couplings. 
In fact only $J = 2$ contributes to the asymmetry in \eqref{CPdege_eq15} and \eqref{CPdege_eq16}. This is why one needs at least two different neutrino species.

Regarding the latter condition we can add a comment. In the exact degenerate case the CP phases can be rotated away leading to purely
real Yukawa couplings, and, therefore, to a vanishing CP asymmetry~\cite{Buchmuller:1997yu}. We can understand it as follows. The heavy neutrino mass matrix, $M_{I}$, has been chosen to be diagonal (see section~\ref{sec_lepto1}). If we furthermore set $M_1=M_2$ a unitary transformation on the sterile neutrino fields, $\psi_{I} \to (U \psi)_{I}$, leaves unchanged the free sterile neutrino Lagrangian whereas in the interaction part we have to redefine accordingly the Yukawa couplings as $F_{fI} \to (F U)_{fI}$. We then notice that the combination $(F^* F)_{IJ}$, entering the CP asymmetry in \eqref{CPdege_b14}, is an hermitian matrix and the unitary transformation on the Yukawa coupling leads to $(F^* F)_{IJ} \to (U^\dagger \, F^* F \, U)_{IJ}$. Therefore the hermitian matrix $F^* F$ is diagonalized by the unitary transformation in a matrix with real eigenvalues and no physical phases can appear. 

\section{Matching $a_{II}$}
\label{sec:finiteT}
In order to evaluate the leading thermal correction to the direct CP asymmetry, i.e., $\Gamma^{\ell,T}_{II,{\rm direct}}-\Gamma^{\bar{\ell},T}_{II,{\rm direct}}$, 
we need to compute the Wilson coefficients $a_{II}$ of the dimension-five operators in~\eqref{eq:efflag_CPdege}.
We have seen that at order $F^2$ in the Yukawa couplings the coefficients $a_{II}$ do not contribute to the 
asymmetry, hence, in this section, we will give them at order~$F^4$. 
They also depend linearly on some SM couplings, in particular the four-Higgs and gauge couplings.
The coefficients $a_{II}$ are determined by matching four-point Green's functions with two external Majorana neutrinos and 
two external Higgs bosons computed in the fundamental theory with the corresponding vertices in the EFT. 
In particular, we may consider a Higgs boson with momentum  $q^\alpha \sim T \ll M$ scattering 
off a Majorana neutrino at rest in the reference frame  $v^{\alpha}=(1,\bm{0})$.
In the matching, we integrate out loop momenta of order $M$, hence 
the momentum of the Higgs boson can eventually be set to zero and the matching done in the vacuum.
Thermal corrections do not affect the matching but the CP asymmetry through the Higgs thermal condensate.
Because the Higgs thermal condensate is real, we just need to compute the imaginary parts of $a_{II}$. 
This can be done by using standard cutting rules at $T=0$. 
\begin{figure}[t]
\centering
\includegraphics[scale=0.55]{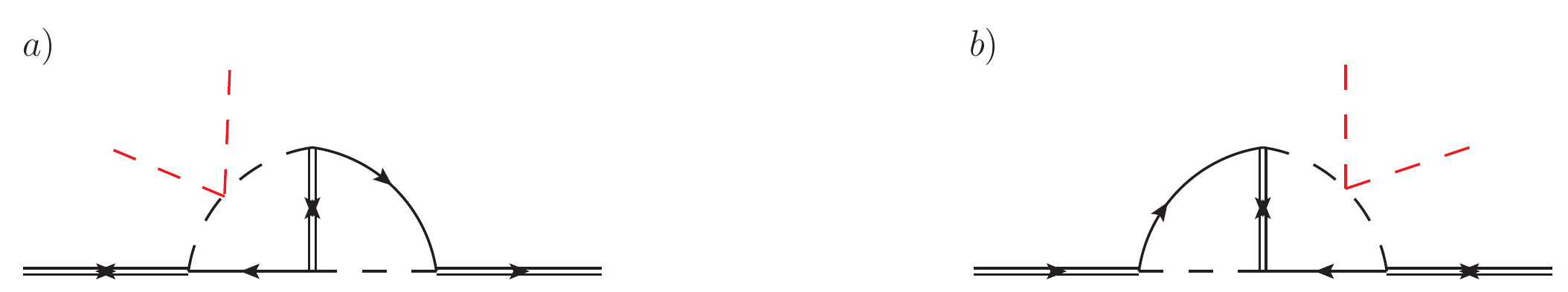}
\caption{\label{fig:Higgs1_CPdege} Diagrams at order $\lambda F^4$ contributing to the matching coefficients of the dimension-five operator. They preserve the topology of the diagrams at order $F^4$ once the Higgs-four coupling is removed.}
\end{figure}

Diagrams with cuts through lepton propagators contribute to the leptonic component of $a_{II}$, 
$a_{II}^\ell$, while diagrams with cuts through antilepton propagators contribute to the antileptonic component 
of $a_{II}$, $a_{II}^{\bar{\ell}}$. Not the entire cut diagram contributes to the asymmetry.
The part of the cut diagram that contributes to the asymmetry can be isolated using the same arguments 
developed in section~\ref{sec:zeroT} and is proportional to~${\rm Im}\left[ (F_{1}^{*}F_{2})^2\right]$.
\subsection{Diagrams with Higgs-four interaction}
\label{sec:CPdege_lambda}
\begin{figure}[t]
\centering
\includegraphics[scale=0.55]{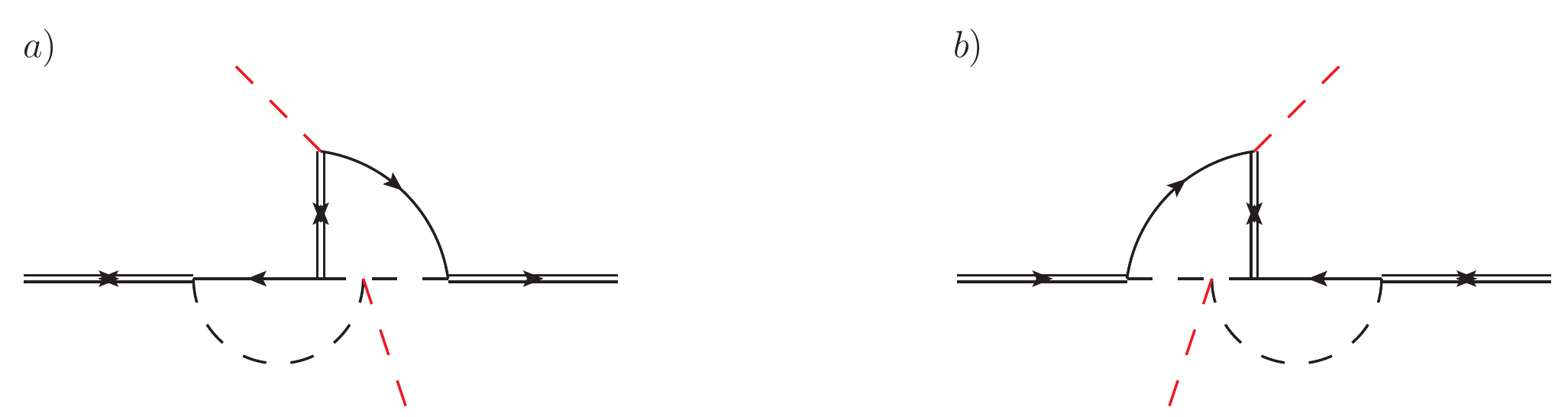}
\caption{\label{fig:Higgs2_CPdege}Diagrams at order $\lambda F^4$ contributing to the matching coefficients of the dimension-five operator. They do not preserve the topology of the diagrams at order $F^4$.}
\end{figure}

The complete set of diagrams for the matching of ${\rm Im}\,a_{II}^{\ell}$ and ${\rm Im}\,a_{II}^{\bar{\ell}}$
at order $F^4$ and at first order in the SM couplings, together with details of the calculation, can be found in appendix~\ref{appC:CPdegematch}. Here we discuss the systematics for diagrams involving the four-Higgs coupling, $\lambda$. We need to match four-point Green's functions with two external Higgs bosons and two heavy Majorana fields,
\begin{equation}
-i \left.\int d^{4}x\,e^{i p \cdot x} \int d^{4}y \int d^{4}z\,e^{i q \cdot (y-z)}\, 
\langle \Omega | T(\psi^{\mu}(x) \bar{\psi}^{\nu }(0) \phi_{m}(y) \phi_{n}^{\dagger}(z) )| \Omega \rangle
\right|_{p^\alpha =(M + i\eta,\bm{0}\,)},
\label{matrix_CPexa}
\end{equation} 
so that the external legs appearing in the diagrams for the matching are fixed. Then the diagrams have to admit cuts through lepton (antilepton) lines such that a one-loop subdiagram is still left. A first set of diagrams is obtained from the two-loop self-energies in figure \ref{fig:fig3_CPdege} by adding a four-Higgs interaction, as shown in figure \ref{fig:Higgs1_CPdege}. Once the four-Higgs vertex is removed, the diagrams of figure~\ref{fig:Higgs1_CPdege} preserve the topology of the $T=0$ two-loop diagrams of figure~\ref{fig:fig3_CPdege}. A second second set is found by opening up one of the Higgs line in the diagrams of figure \ref{fig:fig3_CPdege} and adding a four-Higgs interaction to the internal Higgs line. Then one is left with four Higgs external lines and two of them have to be linked together to provide a two-loop diagram. The result of such procedure brings, for example, to the diagrams shown in figure \ref{fig:Higgs2_CPdege}.

Let us focus on diagram $a)$ of figure~\ref{fig:Higgs1_CPdege}. We outline the strategy for the matching where we consider a Majorana an incoming and outgoing neutrino with mass $M$, namely the lightest of the two. The heavy neutrino is taken in its rest frame according to \eqref{matrix_CPexa} and we organize the computation in the following steps: 
\begin{itemize}
\item[1)] write down the corresponding matrix element obtained from the fundamental Lagrangian \eqref{lepto_8}. The  amplitude so obtained is at two-loop and it reads
\bea
&&\left[\hat{P}\left( -i \mathcal{D}_{\hbox{\tiny a}} \right)\hat{P}\right]^{\mu\nu}
= -6 \lambda (F_1^* F_2)^2 \, \delta_{mn} \,  \int \frac{d^{4} \ell }{(2\pi)^{4}}  \int \frac{d^{4} Q }{(2\pi)^{4}}  
\left( \hat{P} \, P_L \slashed{\ell} (M\slashed{v}+\slashed{Q}) \, \hat{P} \right)^{\mu \nu} 
\nn \\
&&\frac{i}{\ell^{2}+i \eta} \frac{i }{(M v + Q )^2+i \eta}
 \frac{i }{(M v - \ell )^2+i \eta}  \frac{iM}{(\ell + Q)^2-M_2^2+i\eta} \left( \frac{i}{Q^2+i\eta}\right)^2    ,
 \nn 
 \\
 \phantom{x}
\label{match1_CPdege}
\eea
where $M_2=M+\Delta$. We keep the non-relativistic projector, $\hat{P}$, as for the computation in section~\ref{EFT_partprod}. Eventually they are also dropped from the amputated matrix element.  
\item[2)] Cut the diagram through the lepton and Higgs line, whose momentum is $\ell^{\mu}$ and $(Mv -\ell)^\mu$ respectively, as follows
\begin{equation}
\begin{minipage}[c]{0.05\linewidth}
\includegraphics[scale=0.525]{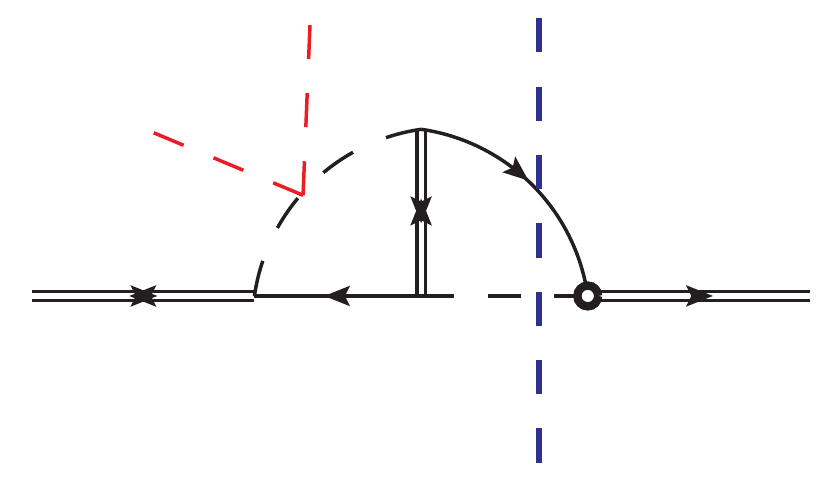}
\end{minipage} \phantom{xxxxxxxxxxxxxxxxxxx} \begin{cases}
\frac{i\slashed{\ell}}{\ell^2+i\eta} \, \to 2\pi \, \slashed{\ell} \, \theta(\ell^0) \delta(\ell^2) 
\\
\frac{i }{(M v - \ell )^2+i \eta} \to 2\pi \, \theta(M-\ell^0) \delta((Mv-\ell)^2)  \, .
\end{cases}
\label{match2_CPdege}
\end{equation}
This way one selects a process with a lepton  in the final state (eventually leading to a contribution to ${\rm{Im}} \, a_{11}^{\ell}$).  A one-loop diagram is still left after the cut. For details on the cutting rules at $T=0$ see appendix~\ref{appC:CPdegematch}.

\item[3)] The loop diagram has to be evaluated and we only need its real part that provides eventually the term relevant for the CP asymmetry, i.e.~the one proportional to the combination ${\rm{Im}}\left[ (F_{1}^{*}F_{2})^2\right]$.  The loop diagram reads
\bea
J(M,\Delta)=\int \frac{d^4 Q}{(2 \pi)^4} \frac{i^4 \, (M \slashed{v}+\slashed{Q})}{\left[ (Mv + Q)^2+i \eta \right] \left[(Q+\ell)^2-M_2^2+i \eta\right] \left[ Q^2+i \eta \right]^2} \,
\label{match3_CPdege}
\eea
where the kinematics after the cut on the lepton and the Higgs boson gives $\ell^2=0$ and $ Mv \cdot \ell =M^2/2$. The integral can be computed with standard $T=0$ techniques and the result is 
\be 
J(M,\Delta)= -\frac{\slashed{v}}{16 \pi M^3} \left( \ln 2 - \frac{\Delta}{M}  \right) \, + \cdots \, ,
\label{match4_CPdege}
\ee 
where the dots stand for higher order terms in the $\Delta/M$ as well as $i$ times the imaginary part of the integral irrelevant for the CP asymmetry. The latter combines with the Yukawa coupling combination ${\rm{Re}}\left[ (F_{1}^{*}F_{2})^2\right]$ similarly to what we have written in~\eqref{CPdege_eq15}. In (\ref{match4_CPdege}) we do not display  terms proportional to $\slashed{\ell}$ because they vanish. This is due to the presence of an additional $\slashed{\ell}$ in the amplitude (\ref{match1_CPdege}), giving then $\slashed{\ell} \slashed{\ell} =\ell^2$. Then together with a $\delta(\ell^2)$ imposed by the cut on the lepton line, see (\ref{match2_CPdege}), such terms are zero. 
\item[4)] The amputated Green's function reads
\be 
{\rm{Im}}\left( -i \mathcal{D}^{\ell}_{\hbox{\tiny a}} \right)= +\frac{3\lambda{\rm{Im}}[(F_1^*F_2)^2]}{2(16 \pi)^2} \left[ \ln 2 -(1-\ln 2) \frac{\Delta}{M}\right] \delta^{\mu \nu} \delta_{mn} + \cdots
\label{match5_CPdege}
\ee
where the dots stand for terms irrelevant for the CP asymmetry, higher order terms in the $\Delta/M$ expansion and the superscript on $\mathcal{D}$ signals that we cut on a lepton line. The same result is obtained from diagram $b)$ in figure~\ref{fig:Higgs1_CPdege}. This can be understood in the following way. First, the diagram is proportional to $(F_1F_2^*)^2$ instead of $(F_1^*F_2)^2$ and hence there is an overall minus in the imaginary part of the Yukawa couplings combination. Second the two diagrams differ for an odd number of circled vertices and an even number of complex propagators when applying the cutting rules. This gives another relative sign that cancels the former one due to the Yukawa couplings. The complete set of diagrams involving the Higgs-four coupling is discussed in appendix~\ref{appC:CPdegematch}. The result in \eqref{match5_CPdege} has to be matched with the matrix element (\ref{matrix_CPexa}) on the EFT side, namely ${\rm{Im}} \, a_{11}^\ell\delta^{\mu \nu} \delta_{mn}$. 
\end{itemize}
In order to obtain the corresponding contribution to $a_{11}^{\bar{\ell}}$ one has to consider the cuts on antileptons in diagram $a)$ and $b)$ in figure \ref{fig:Higgs1_CPdege} and follow the outlined procedure. (Actually the results for antileptons are simply obtained with the substitution $F_1 \rightarrow F_1^*$ and $F_2 \rightarrow F_2^*$).

\begin{figure}[h!]
\centering
\includegraphics[scale=0.6]{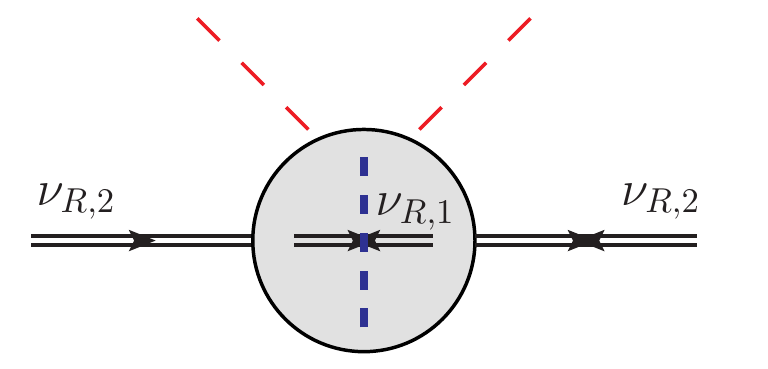}
\caption{\label{fig:NuCut} A schematic representation of a two-loop amplitude (grey blob) with two external heavy neutrinos of type~2 and two soft Higgs bosons. The cut, represented with a blue dash line, put the intermediate neutrino of type~1 on-shell, leaving energy and momenta of order $\Delta \ll M$ running in the remaining loop amplitude.}
\end{figure}

We add here a last remark. 
In the previous example we considered the lightest neutrino as the incoming one. Of course the diagrams in figure \ref{fig:Higgs1_CPdege} and \ref{fig:Higgs2_CPdege} stand also for an incoming neutrino of type~2, slightly heavier in mass. The result for the matrix element in~\eqref{match5_CPdege} can be used to infer the corresponding expression for the neutrino of type~2 after the substitutions $F_1 \leftrightarrow F_2$, $M \to M_2$ and $\Delta \to -\Delta$ are made. In general,  that the above substitutions work follows from the fact that the real transition from a heavier neutrino of type 2 to a lighter neutrino of type 1, 
which is a decay channel absent in the case of neutrinos of type 1, is a process accounted for by the EFT (see section~\ref{sec:direct2}), 
and, therefore, it does not contribute to the matching. Figure~\ref{fig:NuCut} shows the cut through the intermediate neutrino of type~1 in a generic two-loop amplitude (grey blob) with an incoming neutrino of type~2 and an external Higgs. When cutting the heavy neutrino of type~1 together with a lepton (antilepton) only a residual small energy, $\Delta$, is available in the remaining loop amplitude. Energy and momenta turn out to be of order $\Delta$ and hence not typical of the matching (we integrate out energies modes of order $M \gg \Delta$). For example, the diagrams in figure~\ref{fig:Higgs2_CPdege} admit such cuts through the intermediate neutrino and lepton line that leave a one-loop sub-diagram. However these contributions are not included in the matching because of the argument just given. 
\subsection{Diagrams with gauge bosons}
\label{sec:CPdege_gauge}
At order $F^4$, other SM couplings besides the Higgs-four coupling may enter the matching of the leading dimension-five operators in (\ref{eq:efflag_CPdege}). This does not happen in the case of the neutrino production rate calculated at order $F^2$ (see chapter~\ref{chap:part_prod}). In particular gauge interactions can be accounted for systematically in the EFT approach and we show how they enter the dimension-five operators in \eqref{eq:efflag_CPdege}. In this section we discuss some of the diagrams involving gauge bosons and  how to organize  the corresponding matching calculation. Further details are given in appendix~\ref{appC:CPdegematch}.

\begin{figure}[t]
\centering 
\includegraphics[scale=0.46]{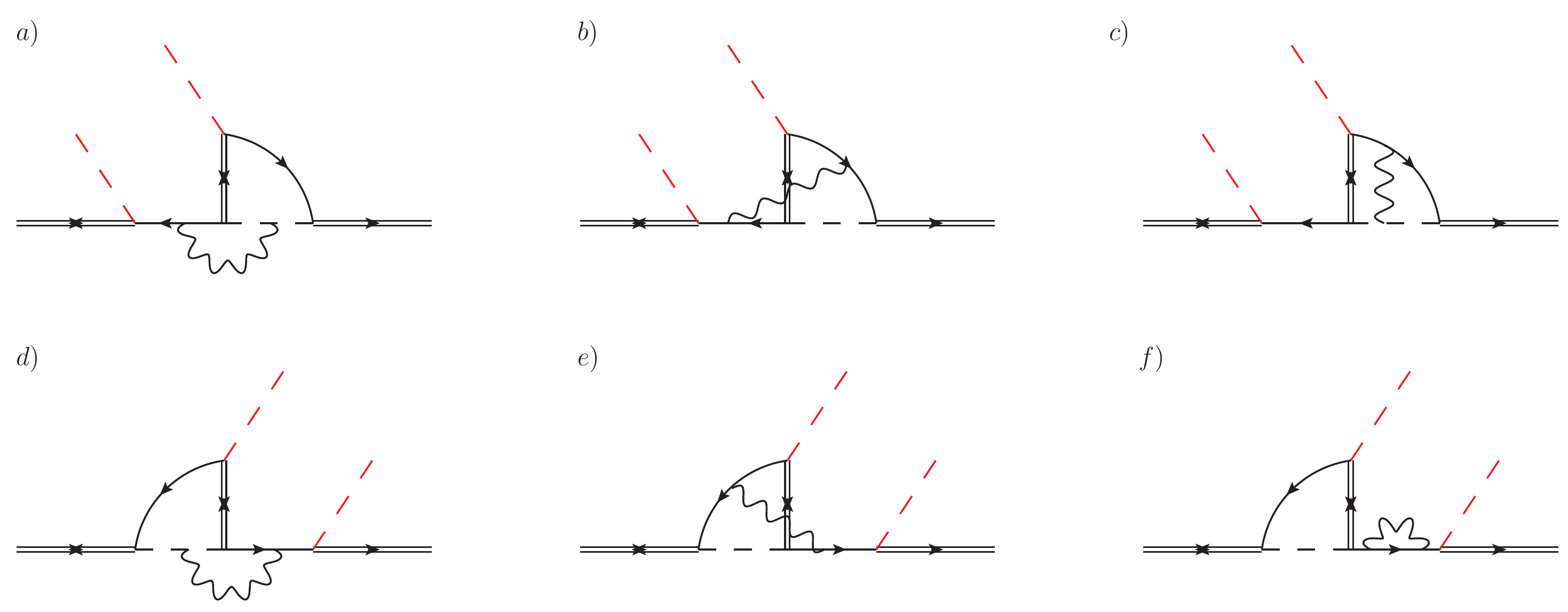}
\caption{\label{fig:example_gauge_CPdege}If the incoming and outgoing Majorana neutrinos are conventionally chosen to be of type 1, 
then the displayed diagrams contribute to $a^\ell_{11}$ at order $F^4$ and at first order in the gauge couplings. 
The diagrams contribute also to $a^{\bar{\ell}}_{11}$ if cut through the antilepton.
Only diagrams proportional to $(F_{1}^{*} F_{2})^2$ are displayed.}
\end{figure}

\begin{figure}[t]
\centering 
\includegraphics[scale=0.51]{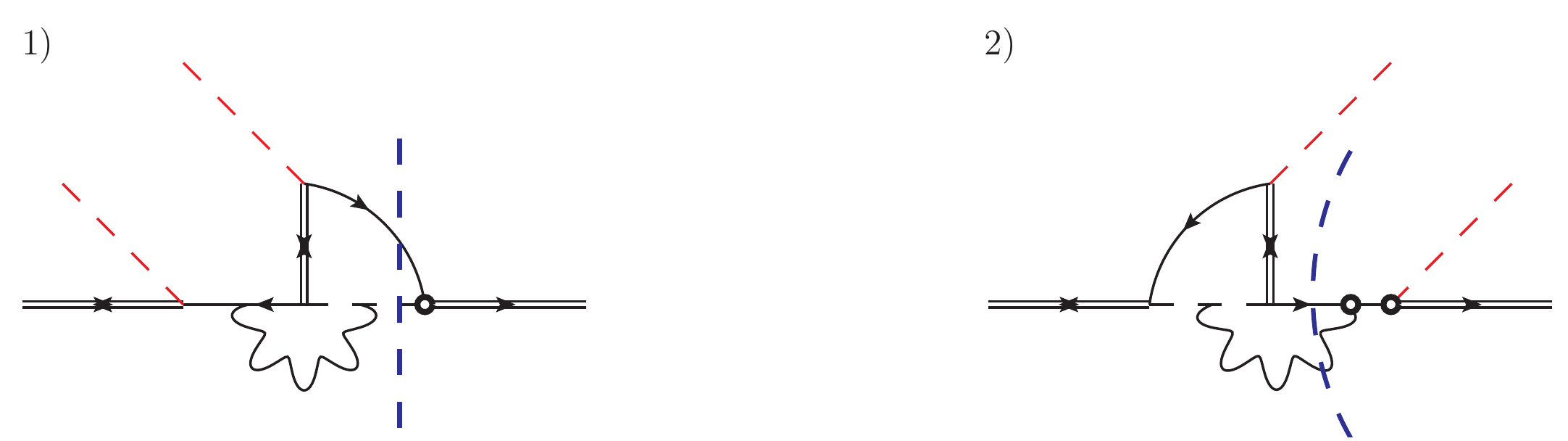}
\caption{\label{fig:gauge_examplecut} Diagrams $a$ and $d$ of figure~\ref{fig:example_gauge_CPdege} with a cut on the lepton line. In the diagram on the left, the lepton is cut together with a Higgs boson (the gauge boson enters as a virtual particle in the remaining loop), whereas in the diagram on the right the gauge boson is cut together with the lepton.}
\end{figure}

Diagrams with gauge interactions contribute to the matching coefficient of the dimension-five operators, provided that one considers a four-point Green's function with two heavy neutrinos and two Higgs bosons as external legs as given in \eqref{matrix_CPexa}. In order to build the relevant diagrams we start again by looking at those shown in figure~\ref{fig:fig3_CPdege}. Opening up a Higgs line and adding one gauge boson, one finds a first set of diagrams shown in figure~\ref{fig:example_gauge_CPdege}. The calculation strategy will follow the one outlined in section~\ref{sec:CPdege_lambda} for diagrams involving the Higgs self interaction. By cutting lepton lines one obtains a contribution to $a^{\ell}_{II}$ whereas cutting on antilepton lines to $a^{\bar{\ell}}_{II}$. However there is a main difference that is worth highlighting. By cutting the diagrams in figure~\ref{fig:example_gauge_CPdege} we distinguish two different type of processes:
\begin{itemize}
\item[1)] processes without a gauge boson in the final state, e.g.~see diagram $1)$ in figure~\ref{fig:gauge_examplecut};
\item[2)] processes with a gauge boson in the final state, e.g.~see diagram $2)$ in figure~\ref{fig:gauge_examplecut}.
\end{itemize}  
These being two distinct physical processes, we can compute them in different gauges. We found convenient to adopt the Landau gauge in the first type of processes and the Coulomb gauge in the second one. The advantage of such a choice is twofold and discussed in detail in appendix~\ref{appC:CPdegematch}. Here we just mention that all the diagrams with a gauge boson attached to an external Higgs line can be discarded for the matching of the dimension-five operators in \ref{eq:efflag_CPdege}, indeed contributing to operators with a higher dimension. Moreover one avoids spurious singularities by using the Coulomb gauge for those diagrams where the gauge boson is cut.  %First we can neglect all the diagrams with a gauge boson attached to an external Higgs leg in the matching of dimension-five operators in (). Indeed both in the Coulomb and Landau gauge the interaction between a Higgs and a gauge boson can be written with a derivative acting on the Higgs field. The corresponding diagrams will therefore develop 

\begin{figure}[t!]
\centering 
\includegraphics[scale=0.51]{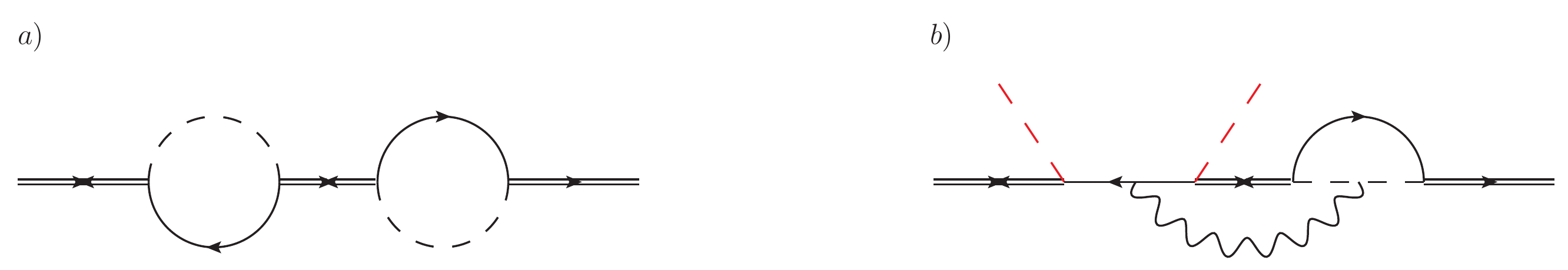}
\caption{\label{fig:fig_gauge_ind_dir} Diagram $a$, two-loop heavy neutrino self energy relevant for the indirect CP asymmetry at $T=0$. Diagram $B$ is obtained from opening up one Higgs leg and adding one gauge boson. }
\end{figure}
In section~\ref{sec:dirindi} we provided an alternative definition for direct and indirect asymmetry. Going beyond leading order we find an interesting result: diagrams that can possibly show a resonant enhancement at $T=0$, namely at order $(F_1F^*_2)^2$, may loose this property when going at order $g^2(F_1F^*_2)^2$ or $g'^2(F_1F^*_2)^2$, namely when adding a gauge boson. Respectively $g$ and $g'$ are the SU(2)$_L$ and U(1)$_Y$ gauge couplings. The argument can be understood as follows. In chapter~\ref{chap:lepto} the indirect contribution to the asymmetry was introduced as the one originated from the interference between the tree-level and one loop self-energy diagram for the heavy neutrino decay (diagrams shown in figure~\ref{fig:dirind_degeCP}). The corresponding two-loop self-energy diagram in the fundamental theory is shown in figure~\ref{fig:fig_gauge_ind_dir}, diagram $a)$. The indirect asymmetry arising from diagrams with such topology are studied in section~\ref{sec:indirect} in the EFT. Following the outlined strategy for seeking diagrams suitable for the matching of the four-pint Green's function \eqref{matrix_CPexa}, we open up one Higgs line and add a gauge boson to obtain two-loop diagrams potentially contributing to the matching of the dimension-five operators. We show an example in figure~\ref{fig:fig_gauge_ind_dir}, diagram $b)$, which cannot become resonant when the gauge boson carries away an energy of order $M$ and, according to the definition adopted in section~\ref{sec:dirindi}, it contributes to the direct CP asymmetry. 
Clearly it does contribute to the Wilson coefficients ${\rm{Im}} \,a_{II}^\ell$ and ${\rm{Im}} \,a_{II}^{\bar{\ell}}$.

By considering the complete set of diagrams relevant for the matching calculation we obtain up to order $\Delta/M$ (only terms contributing to the asymmetry are displayed):
\begin{eqnarray}
{\rm{Im}} \, a^{\ell}_{11}=-{\rm{Im}} \, a^{\bar{\ell}}_{11} &=&
\frac{{\rm{Im}}\left[ (F_1^{*}F_2)^2\right] }{(16\pi)^2} 
\left\lbrace   6 \lambda \left[ 1+\ln 2-\left( 2-\ln 2 \right)\frac{\Delta}{M}\right]  
\right.
\nonumber\\
&& \hspace{1.1cm}
- \frac{3}{8}g^2 \left[ 2 - \ln2 +\left(3 - 5 \ln 2\right) \frac{\Delta}{M}  \right]  
\nonumber\\
&& \hspace{1.1cm}
\left.
- \frac{g'^2}{8}\left[ 4 - \ln2 +\left(1 - 5 \ln 2\right) \frac{\Delta}{M}  \right]  \right\rbrace ,
\label{match1} 
\end{eqnarray}
\begin{eqnarray}
{\rm{Im}} \, a^{\ell}_{22}=-{\rm{Im}} \, a^{\bar{\ell}}_{22} &=& 
-\frac{{\rm{Im}}\left[ (F_1^{*}F_2)^2\right] }{(16\pi)^2} 
\left\lbrace   6 \lambda \left[ 1+\ln 2+\left( 2-\ln 2 \right)\frac{\Delta}{M}\right]  
\right.
\nonumber\\
&& \hspace{1.2cm}
- \frac{3}{8}g^2\left[ 2 - \ln2 -\left(3 - 5 \ln 2\right) \frac{\Delta}{M}  \right]  
\nonumber\\
&& \hspace{1.1cm}
\left.
- \frac{g'^2}{8}\left[ 4 - \ln2 -\left(1 - 5 \ln 2\right) \frac{\Delta}{M}  \right]  \right\rbrace ,
\label{match2}
\end{eqnarray}
where $\lambda$ is the four-Higgs coupling, and $g$ and $g'$ are the SU(2)$_L$ and U(1)$_Y$ gauge couplings respectively.
Note the sign difference between ${\rm{Im}} \, a^{\ell}_{II}$ and ${\rm{Im}} \, a^{\bar{\ell}}_{II}$.
We remark that at this order the result does not depend on the top-Yukawa coupling, $\lambda_t$. Further elaboration on the subject are found in appendix~\ref{appC:CPdegematch}.

\section{Thermal corrections to the direct asymmetry}
\label{sec:direct}

We may now proceed to calculate the thermal corrections to the widths and CP asymmetries
of the two Majorana neutrinos, assuming that the thermal bath of SM particles is at rest with respect to the
Majorana neutrinos and the reference frame. It is convenient to split both the neutrino width, 
$\Gamma_{II}=\Gamma_{II}^{T=0}+ \Gamma_{II}^{T}$, and the CP asymmetry, 
$\epsilon_{I}=\epsilon^{T=0}_{I} + \epsilon_{I}^{T}$, into a zero temperature and a thermal part. We find convenient to divide the discussion on the thermal correction for the two different neutrino species. 

\begin{figure}[htb]
\centering
\includegraphics[scale=0.6]{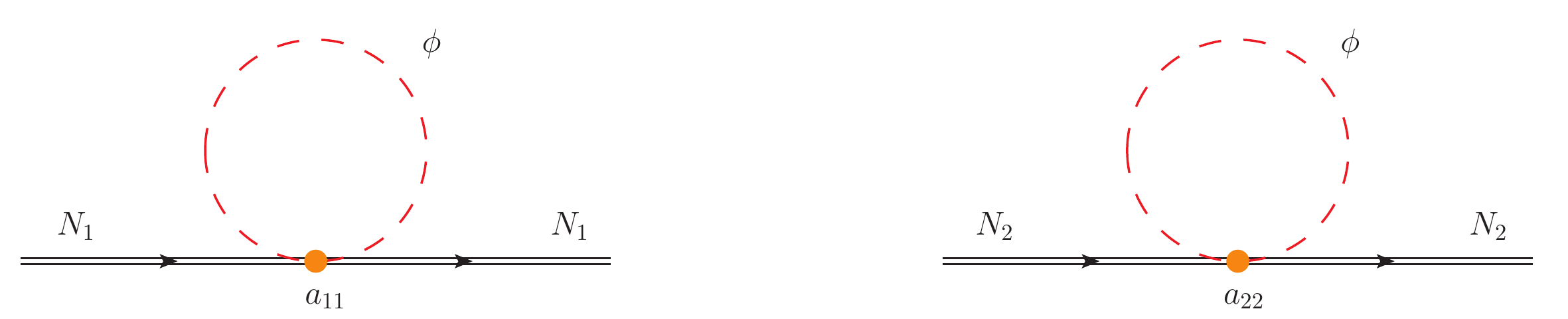}
\caption{Tadpole diagrams responsible for the leading thermal corrections to the neutrino widths 
and CP asymmetries in the EFT. We show in red particles belonging to the thermal bath whose momentum is of order $T$.}
\label{fig:tadpoles_CPdege} 
\end{figure}

\subsection{Neutrino of type 1}
\label{sec:direct1}
We consider first neutrinos of type 1, which are assumed to be lighter than those of type~2.
The zero-temperature width at leading order has been written in \eqref{CPdege_Gamma1T0}.
The leading thermal correction to the width has been calculated in~\cite{Salvio:2011sf,Laine:2011pq,Biondini:2013xua}
and can be easily re-derived from \eqref{asysec3}, \eqref{cond1} and \eqref{agen}.
The expression of the width up to order  $F^2 \lambda \times (T/M)^2$ reads  
\begin{equation}
\Gamma_{11} = \Gamma_{11}^{T=0}+\Gamma_{11}^{T} =\frac{M}{8\pi} |F_1|^2 \left[ 1-\lambda\left(\frac{T}{M}\right)^2 \right]  .
\label{deno_CPdege}
\end{equation}

The in-vacuum part of the direct CP asymmetry, $\epsilon^{T=0}_{1,{\rm direct}}$, can be read off~\eqref{CPdege_b14}. 
In order to obtain $\epsilon^{T}_{1,{\rm direct}}$, one has to evaluate $\Gamma^{\ell,T}_{11,{\rm direct}}-\Gamma^{\bar{\ell},T}_{11,{\rm direct}}$.
Thermal corrections are encoded into the Higgs thermal condensate represented by the first tadpole diagram shown in figure~\ref{fig:tadpoles_CPdege}.
From  \eqref{asysec3}, \eqref{cond1} and \eqref{match1} it follows
\begin{eqnarray}
\Gamma^{\ell,T}_{11,{\rm direct}}-\Gamma^{\bar{\ell},T}_{11,{\rm direct}} &=& 
\frac{{\rm{Im}}\left[ (F^{*}_1 F_2)^2\right] }{64 \pi^2}  
\left\lbrace   \lambda \left[ 1 + \ln 2-\left( 2-\ln 2 \right)\frac{\Delta}{M}\right]  
\right.
\nonumber\\
&& \hspace{-4cm}
\left.
- \frac{g^2}{16}\left[ 2- \ln2 +\left( 3 - 5 \ln 2\right) \frac{\Delta}{M}  \right]  
- \frac{g'^2}{48}\left[ 4- \ln2 +\left( 1 - 5 \ln 2\right) \frac{\Delta}{M}  \right]  \right\rbrace  \frac{T^2}{M}.
\label{gammaphi} 
\end{eqnarray} 
From \eqref{asysec2bis}, \eqref{CPdege_EQ17}, \eqref{deno_CPdege} and \eqref{gammaphi}, and considering that 
$\Delta\Gamma_{1,{\rm direct}}^{\rm mixing} =0$, we obtain then the thermal part of the CP asymmetry 
generated from the decay of Majorana neutrinos of type 1 at leading order in the SM couplings, at order $T^2/M^2$ and at order $\Delta/M$:
\begin{eqnarray}
\epsilon^{T}_{1,{\rm direct}} &=&\frac{{\rm{Im}}\left[ (F^{*}_1 F_2)^2\right] }{8 \pi |F_{1}|^2}  \left(  \frac{T}{M} \right)^2 
\left\lbrace   \lambda \left[ 2-\ln 2+\left( 1-3\ln 2 \right) \frac{\Delta}{M}\right]  
\right.
\nonumber \\
&& 
\left.
- \frac{g^2}{16}\left[ 2- \ln2 +\left( 3 - 5 \ln 2\right) \frac{\Delta}{M}  \right]  
- \frac{g'^2}{48}\left[ 4- \ln2 +\left( 1 - 5 \ln 2\right) \frac{\Delta}{M}  \right]  \right\rbrace.
\label{CPnu1}
\nonumber
\\
\end{eqnarray}

\subsection{Neutrino of type 2}
\label{sec:direct2}
The in-vacuum contribution to the CP asymmetry of Majorana neutrinos of type~2 can be read off~\eqref{CPdege_b15}.
Thermal contributions of the type \eqref{asysec3}, can be computed as for neutrinos of type 1,  
the relevant diagram being the second diagram of figure~\ref{fig:tadpoles_CPdege}. 
They may be read off \eqref{gammaphi} and \eqref{CPnu1} after the replacements 
$F_1 \leftrightarrow F_2$, $M \to M_2$ and $\Delta \to -\Delta$.

\begin{figure}[hbt]
\centering
\includegraphics[scale=0.52]{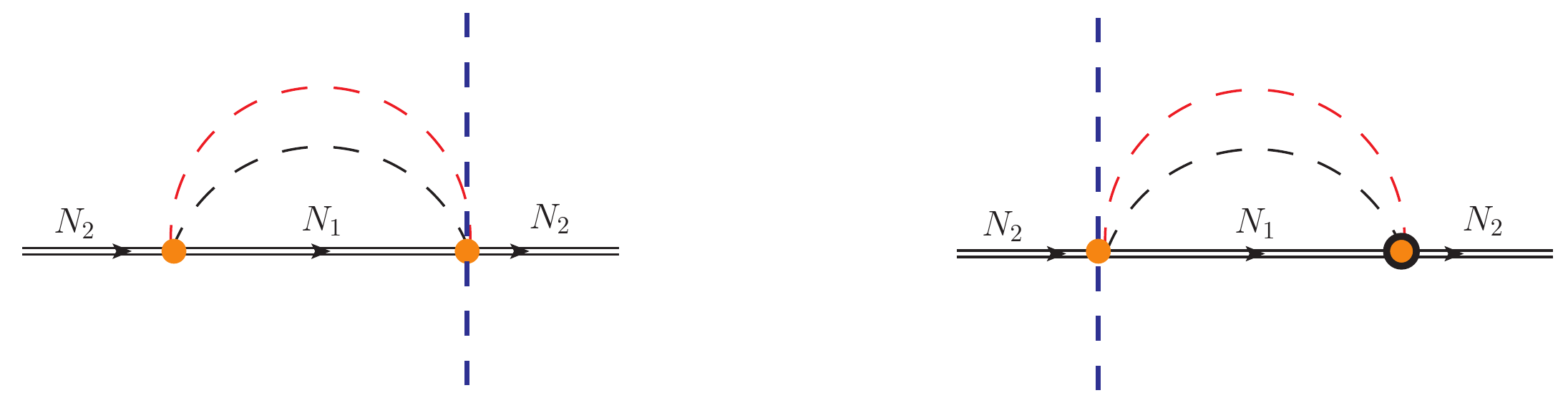}
\caption{Diagrams contributing in the EFT to the CP asymmetry of the Majorana neutrino of type 2 (see text). 
The orange dot stands for the vertex $-i {\rm Re}\,(F^{*}_1 F_2)/M$; the circled dot has opposite sign.
The dot with a cut selects the leptonic (or antileptonic) decay components: 
$-3(F_1 F_2^*)\lambda/(8\pi M)$ (or $-3(F_2 F_1^*)\lambda/(8\pi M)$) for incoming neutrino of type 1. 
Propagators on the right of the cut are complex conjugate.
Red dashed lines indicate thermal Higgs bosons, while black dashed lines indicate 
Higgs bosons carrying a momentum and energy of order~$\Delta$. 
}
\label{fig:DeltaEFT}
\end{figure}

If the neutrino of type 2 is heavier than the neutrino of type 1, there may be an additional source of CP asymmetry coming from 
diagrams where, after the cut through the lepton (or antilepton), the remaining one-loop subdiagram develops 
an imaginary part because of the kinematically allowed transition $\nu_{R,2}\to\nu_{R,1} + $ Higgs boson.
Such a transition involves a momentum transfer of order $\Delta$. Since $\Delta \ll M$, terms coming from momentum 
regions of order $\Delta$ have been excluded from the matching and do not contribute to $a_{IJ}$.
However, they do contribute in the EFT. 

\begin{figure}[t]
\centering
\includegraphics[scale=0.52]{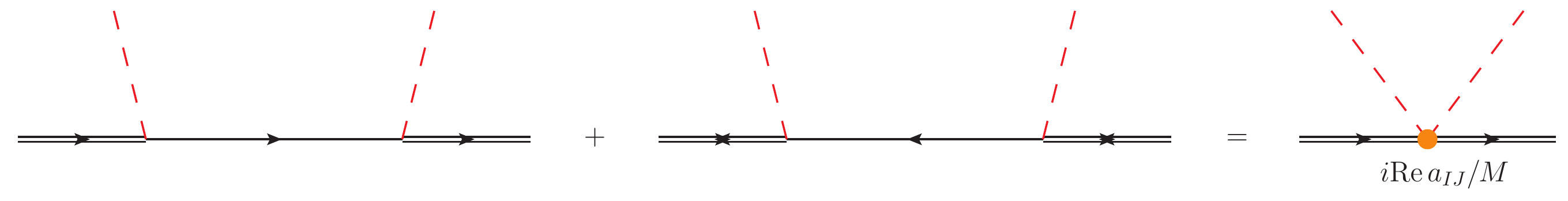}
\caption{On the left-hand side the diagrams in the fundamental theory that match the real part of $a_{IJ}$ at order $F^2$ (right-hand side).
Red dashed lines indicate external Higgs bosons with a soft momentum much smaller than the mass of the Majorana neutrinos.}
\label{fig:treeMatch} 
\end{figure}

The leading order diagrams in the EFT are shown in figure~\ref{fig:DeltaEFT}.\footnote{
The corresponding diagrams in the full theory are diagrams 1)-6) in figure~\ref{fig:new_Higgs_direct}.} 
They may be understood as the mixing of two dimension five operators in the EFT, 
hence they contribute to the direct CP asymmetry \eqref{asysec2bis} through the term $\Delta\Gamma_{2,{\rm direct}}^{\rm mixing}$. Since we are interested in corrections at leading order in the Higgs-four coupling, one has to consider only one of the two dimension-five operators with an effective coupling comprising $\lambda$. They read off \eqref{eq:efflag_CPdege} and the corresponding matching coefficients can be inferred from (\ref{agen}) by generalizing to different neutrino species as follows
\be 
{\rm Im}\, a_{IJ} = -\frac{3}{16\pi}(F_{J}F_{I}^{*}+F_{I}F_{J}^{*})\lambda.
\label{agenbis}
\ee
These vertices are shown in orange in figure~\ref{fig:DeltaEFT} and we are interested in the imaginary needed to obtain a width.
At our accuracy, for the uncut vertex, we just need to consider the real parts of the dimension five operators 
mixing neutrinos of type 1 with neutrinos of type 2. The corresponding vertex, 
shown with an orange dot in figure~\ref{fig:DeltaEFT}, is $i {\rm Re}\,a_{12}/M$. 
The real part of $a_{IJ}$ can be computed at order $F^2$ by matching the two tree-level diagrams 
shown in the left-hand side of figure~\ref{fig:treeMatch} with the corresponding vertex in the EFT. 
The result reads
\begin{equation}
{\rm{Re}}\, a_{IJ}=-\frac{F_IF_J^*+F_I^*F_J}{2}.
\label{re}
\end{equation}
The contribution from the cut is $-2 \times 1/M \times (3\,F_{I}^*F_{J}\lambda/(16\pi))$ for the leptonic cut 
and $-2 \times 1/M \times (3\,F_{J}^*F_{I}\lambda/(16\pi))$ for the antileptonic one, where $I$ is the outgoing neutrino and $J$ the ingoing one (the factor $-2$ comes from the cutting rules, see appendix~\ref{appC:CPdegematch}).

The momentum flowing in the diagrams of figure~\ref{fig:DeltaEFT} can be of order $T$ or of order $\Delta$. 
If the momentum flowing in both loops is of order $T$ this contributes to the 
asymmetry $\Gamma^{\ell,T}_{22,{\rm direct}}-\Gamma^{\bar{\ell},T}_{22,{\rm direct}}$ at order $T^3/M^2$; 
if the momentum flowing in both loops is of order $\Delta$ this contributes to the 
asymmetry at order $\Delta^3/M^2$. Both contributions are beyond our accuracy.
If instead one Higgs boson carries a momentum and energy of order $T$ and the other a momentum and energy of order $\Delta$, 
then this momentum region contributes to the asymmetry at order $T^2\Delta /M^2$, which is inside our accuracy.
The color code used for the Higgs bosons in figure~\ref{fig:DeltaEFT} identifies this specific momentum region. 
Its contribution to the direct asymmetry of Majorana neutrinos of type 2 is
\begin{equation}
\Delta\Gamma_{2,{\rm direct}}^{\rm mixing} = \frac{{\rm{Im}}\left[ (F^{*}_1 F_2)^2\right] }{16 \pi^2} \lambda \frac{T^2\Delta}{M^2}.
\label{gammaDelta}
\end{equation}
Summing this to the CP asymmetry of the Majorana neutrino of type 2 obtained from the tadpole diagram 
of figure~\ref{fig:tadpoles_CPdege}, and discussed at the beginning of this section, we obtain that 
the thermal correction to the direct CP asymmetry of the Majorana neutrino of type 2 
at leading order in the SM couplings, at order $T^2/M^2$ and at order $\Delta/M$ is
\begin{eqnarray}
\epsilon^{T}_{2,{\rm direct}} &=& -\frac{{\rm{Im}}\left[ (F^{*}_1 F_2)^2\right] }{8 \pi |F_{2}|^2}  \left(  \frac{T}{M} \right)^2 
\left\lbrace   \lambda \left[ 2-\ln 2-\left( 9 - 5\ln 2 \right) \frac{\Delta}{M}\right]  \right.
\nonumber \\
&&
\left.
- \frac{g^2}{16}\left[ 2- \ln2 - 7 \left( 1 - \ln 2\right) \frac{\Delta}{M}  \right]  
- \frac{g'^2}{48}\left[ 4- \ln2 -\left( 9 - 7 \ln 2\right) \frac{\Delta}{M}  \right]  \right\rbrace.
\nonumber 
\\
\label{CPnu2}
\end{eqnarray}

We observe that in the exact degenerate limit ($\Delta \to 0$), the single direct CP asymmetries $\epsilon_{1,{\rm direct}}$
and $\epsilon_{2,{\rm direct}}$ do not vanish. However, the sum of \eqref{CPdege_EQ17} with \eqref{gammaphi}, and 
with the corresponding expressions for the type 2 neutrino does vanish. 
This sum is the CP-violating parameter defined in~\cite{Pilaftsis:1998pd}.

\section{Indirect asymmetry}
\label{sec:indirect}
The indirect CP asymmetry is made of all contributions that may exhibit the phenomenon of resonant enhancement (see section~\ref{sec:rev}).
It can be understood as originating from the mixing between the different neutrino species 
that makes the mass eigenstates different from the CP eigenstates~\cite{Flanz:1996fb}.
This mixing can be described by the EFT. 
In the following we will compute the indirect CP asymmetry at leading order and its first thermal correction. 
Besides the hierarchies $M\gg T \gg M_{W}$ and $M \gg \Delta$ we will not assume any special relation between 
$\Delta$ and the neutrino decay widths. In particular we will allow for the resonant case 
$\Delta \sim \Gamma_{11},\Gamma_{22}$ and resum the widths in the neutrino propagators.
Nevertheless we will treat the mixing perturbatively, which amounts at requiring
$\Delta^2 + (\Gamma_{22}-\Gamma_{11})^2/4 \gg M^2 \, [{\rm Re}(F_1^*F_2)]^2/(16\pi)^2$ 
(this condition can be inferred from the right-hand side of the following equation~\eqref{gammalt0indirect}; 
see also~\cite{Garny:2011hg}).\footnote{
Relaxing this condition does not pose conceptual problems. A non-perturbative mixing will affect, however, 
both the direct and indirect CP asymmetries and make their analytical expressions less compact.
For the indirect asymmetry, this has been considered without resummation of the widths in~\cite{Flanz:1996fb}.}  

Mixing between the different neutrino generations in the effective Lagrangian \eqref{eq:efflag_CPdege} 
is induced by the off-diagonal elements of $\Gamma_{IJ}^{T=0}$, 
\begin{equation}
\Gamma_{IJ}^{T=0} = \frac{M}{16\pi}\left( F_I^*F_J + F_J^*F_I \right),
\label{gammamixing}
\end{equation}
which can be obtained from the absorptive part of diagram $1)$ in figure~\ref{fig:fig3_CPdege} and the corresponding 
one with an antilepton in the loop~\cite{Flanz:1996fb,Buchmuller:1997yu}
(for $I=J=1$ \eqref{gammamixing} gives back \eqref{CPdege_Gamma1T0}), and by the off-diagonal elements of $a_{IJ}$.
The imaginary part of $a_{IJ}$  read off \eqref{agenbis}, whereas
%\begin{equation}
%{\rm Im}\, a_{IJ} = -\frac{3}{16\pi}(F_{J}F_{I}^{*}+F_{I}F_{J}^{*})\lambda.
%\end{equation}
the real part of $a_{IJ}$ has been computed at order $F^2$ in the previous section and can be read off~\eqref{re}. 

\begin{figure}[t!]
\centering
\includegraphics[scale=0.52]{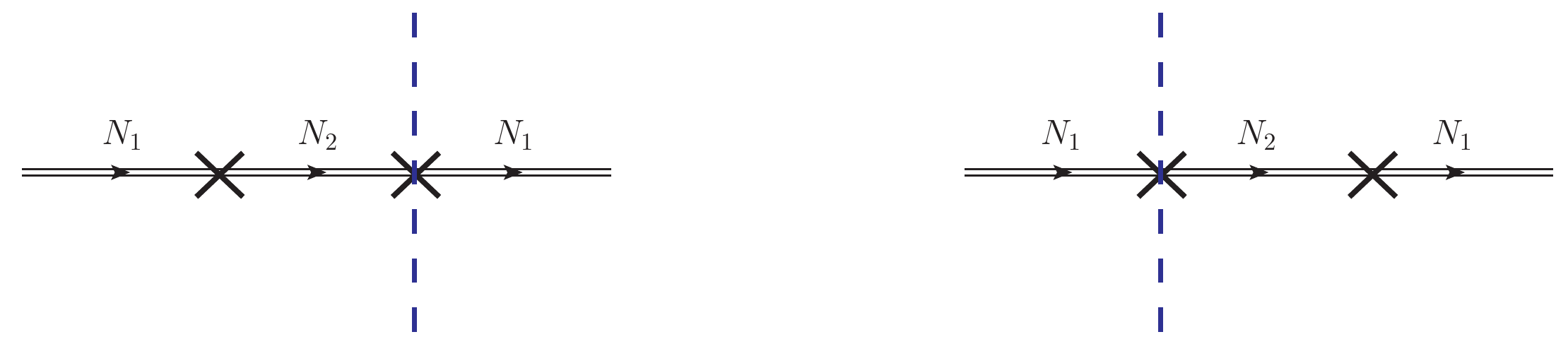}
\caption{Diagrams showing in the EFT a neutrino of type 1 decaying into a lepton after mixing with a neutrino of type 2.
The cross stands for the mixing vertex $-\Gamma_{IJ}^{T=0}/2$. 
The cross with a cut selects the leptonic (or antileptonic) decay components: 
$M(F_I^*F_J)/(16\pi)$ (or $M(F_J^*F_I)/(16\pi)$).
Propagators on the right of the cut are complex conjugate.
Because the mixing vertex is real, circled and uncircled vertices coincide~\cite{Denner:2014zga}.
}
\label{fig:indirectEFT}
\end{figure}

At zero temperature and at order $F^4$ the width of a neutrino of type 1 that decays into a lepton after mixing with a neutrino of type 2 
is given in the EFT by the sum of the cuts on the diagrams shown in figure~\ref{fig:indirectEFT} (in the fundamental theory the diagrams look like diagram $a)$ in figure~\ref{fig:fig_gauge_ind_dir}).
The diagrams are amputated of the external legs and evaluated at the pole of the propagator of the (incoming and outgoing) neutrino of type 1.
If the width is of the order of $\Delta$, then it should be resummed so that the (complex) pole of the neutrino of type 1 is at $-i\Gamma_{11}^{T=0}/2$
and the pole of the intermediate neutrino of type 2 is at  $\Delta -i\Gamma_{22}^{T=0}/2$.
The crossed vertex in figure~\ref{fig:indirectEFT} stands for the mixing vertex $-\Gamma_{IJ}^{T=0}/2$, 
where $I$ identifies the outgoing and $J$ the incoming neutrino. 
The cut through the vertex selects the decay into a lepton or an antilepton.
In the first case, the value of the cut is $M(F_I^*F_J)/(16\pi)$, in the second case it is $M(F_J^*F_I)/(16\pi)$.
For leptonic cuts the diagrams in figure~\ref{fig:indirectEFT} give
\begin{equation}
\Gamma^{\ell,T=0}_{11,{\rm indirect}} = \frac{M}{16\pi} F_1^*F_2\frac{i}{-\Delta+i(\Gamma_{22}^{T=0}-\Gamma_{11}^{T=0})/2}
\left(-\frac{M}{16\pi}\right)\frac{F_1^*F_2+F_2^*F_1}{2} + {\rm c.c.},
\label{gammalt0indirect}
\end{equation}
where c.c. stands for complex conjugate. For antileptonic cuts the diagrams in figure~\ref{fig:indirectEFT}  give 
$\Gamma^{\bar{\ell},T=0}_{11,{\rm indirect}}$, which is the same as \eqref{gammalt0indirect} but with the change $F_1^*F_2 \leftrightarrow F_2^*F_1$ in the mixing vertices.
The indirect CP asymmetry at $T=0$ for a Majorana neutrino of type 1 is then
\begin{equation}
\epsilon_{1,{\rm indirect}}^{T=0}= \frac{\Gamma^{\ell,T=0}_{11,{\rm indirect}}-\Gamma^{\bar{\ell},T=0}_{11,{\rm indirect}}}{\Gamma^{T=0}_{11}} 
= - \frac{{\rm{Im}}\left[ (F^{*}_1 F_2)^2\right] }{16 \pi |F_{1}|^2}  \frac{M \, \Delta}{\Delta^2 + (\Gamma_{22}^{T=0}-\Gamma_{11}^{T=0})^2/4}.
\label{indirect1T0}
\end{equation}
Similarly one obtains the indirect CP asymmetry at $T=0$ for a Majorana neutrino of type~2
\begin{equation}
\epsilon_{2,{\rm indirect}}^{T=0}= \frac{\Gamma^{\ell,T=0}_{22,{\rm indirect}}-\Gamma^{\bar{\ell},T=0}_{22,{\rm indirect}}}{\Gamma^{T=0}_{22}} 
= - \frac{{\rm{Im}}\left[ (F^{*}_1 F_2)^2\right] }{16 \pi |F_{2}|^2}  \frac{M \, \Delta}{\Delta^2 + (\Gamma_{22}^{T=0}-\Gamma_{11}^{T=0})^2/4}.
\label{indirect2T0}
\end{equation}
We recall that $\Gamma_{II}^{T=0} = M|F_I|^2/(8\pi)$.

The above result for the indirect asymmetry at $T=0$ agrees with~\cite{Buchmuller:1997yu} (see also~\cite{Garny:2011hg} and discussion therein). 
It agrees with~\cite{Anisimov:2005hr} by remarking that the additional term proportional to $\log(M^2_2/M^2_1)$ there 
is a contribution of relative order $F^6$ to the CP asymmetry and therefore beyond our accuracy.
Whenever we can neglect the width $\Gamma_{11}^{T=0}$, equations~\eqref{indirect1T0} and \eqref{indirect2T0}
agree with~\cite{Pilaftsis:1997dr,Pilaftsis:1997jf,Pilaftsis:1998pd,Pilaftsis:2003gt,Dev:2014laa}.
Finally, we notice that in the framework of the Kadanoff--Baym evolution equations 
(see for instance \cite{Garny:2011hg,Frossard:2012pc,Garbrecht:2014aga}) the quantity related to the CP asymmetry 
is a modification of the above one that accounts for coherent transitions between the Majorana neutrino mass eigenstates.

The computation done above shows that, although at $T=0$ there should be in general no advantage in using the EFT, there is some in computing the indirect CP asymmetry.
In fact, the EFT naturally separates the physics of the Majorana neutrino decay, which goes into the widths and the mixing vertices, 
from the quantum-mechanical physics of the neutrino oscillations. 
This separation is well depicted in the Feynman diagrams of figure~\ref{fig:indirectEFT}.
It also makes more apparent the potentially resonant behaviour of the contribution.

Thermal corrections to \eqref{gammalt0indirect} affect masses, widths and mixing vertices. 
From \eqref{asysec3} (generalized to off-diagonal elements), \eqref{cond1} and 
\eqref{agenbis} it follows that the leading thermal correction to the width matrix is of relative size $\lambda T^2/M^2$:
\begin{equation}
\Gamma^T_{IJ}=-\frac{\lambda T^2}{16\pi M}(F_IF_J^*+F_I^*F_J).
\label{gammat}
\end{equation}
The thermal correction to the mass matrix follows from \eqref{re} and \eqref{cond1}, and is of relative size $T^2/M^2$:
\begin{equation}
M^T_{IJ}=\frac{T^2}{12 M}(F_IF_J^*+F_I^*F_J).
\label{mass}
\end{equation}
The mass thermal correction \eqref{mass} differs from the one used in~\cite{Pilaftsis:1997jf} and taken from~\cite{Weldon:1982bn}. 
The reason for the difference is that the thermal correction computed in~\cite{Weldon:1982bn}
refers to a massless neutrino while the one written above refers to a neutrino in the heavy mass limit.
In the massless case the neutrino gets a thermal mass both from fermions and bosons in the medium, whereas 
in the heavy-mass case, as can be immediately read off the effective Lagrangian \eqref{eq:efflag_CPdege},
fermion contributions are suppressed in $T/M$ and only Higgs bosons contribute. 

If we restrict to the leading corrections, we may neglect the thermal correction to the 
decay matrix, which is suppressed by $\lambda$, and keep only the thermal correction to the mass matrix. 
This modifies the mixing vertex in figure~\ref{fig:indirectEFT} from 
 $-\Gamma_{IJ}^{T=0}/2$ to  $-\Gamma_{IJ}^{T=0}/2 -i M^T_{IJ}$ and the mass $\Delta$ in the 
intermediate propagator to $\Delta + M^T_{22}-M^T_{11}$. The former modification comes from the vertex induced by dimension-five operators in (\ref{eq:efflag_CPdege}), $i{\rm{Re}}\,a_{IJ} \times (1/M_I) \times (T^2/6)$, where the real part of the matching coefficients read off (\ref{re}) and the the Higgs thermal condensate has been used.
If we neglect corrections of relative order $\lambda$, cuts are not affected by thermal effects, so that 
\begin{eqnarray}
\Gamma^{\ell,T}_{11,{\rm indirect}} &=&\bigg[ \frac{M}{16\pi} F_1^*F_2\frac{i}{-\Delta
- (|F_2|^2-|F_1|^2)T^2/(6M) +i(\Gamma_{22}^{T=0}-\Gamma_{11}^{T=0})/2}
\nonumber\\
&&
\hspace{1.5cm}
\times\left(-\frac{M}{16\pi} -i \frac{T^2}{6M}\right)\frac{F_1^*F_2+F_2^*F_1}{2} + {\rm c.c.}\bigg]
- \Gamma^{\ell,T=0}_{11,{\rm indirect}} \;,
\nn
\\
\label{gammaltindirect}
\end{eqnarray}
which is valid at leading order in $T/M$.
Similarly $\Gamma^{\bar{\ell},T}_{11,{\rm indirect}}$ is given by \eqref{gammaltindirect} but with the change $F_1^*F_2 \leftrightarrow F_2^*F_1$ in the mixing vertices.
The leading thermal correction to the indirect CP asymmetry for a Majorana neutrino of type 1 is then 
\begin{equation}
\epsilon_{1,{\rm indirect}}^{T}= -\frac{\epsilon_{1,{\rm indirect}}^{T=0}}{3} \,\left(|F_2|^2-|F_1|^2\right)\,
\frac{M\Delta}{\Delta^2 + (\Gamma_{22}^{T=0}-\Gamma_{11}^{T=0})^2/4}\,\frac{T^2}{M^2},
\label{indirect1T}
\end{equation}
and analogously the thermal correction to the indirect CP asymmetry for a neutrino of type~2 is 
\begin{equation}
\epsilon_{2,{\rm indirect}}^{T}= -\frac{\epsilon_{2,{\rm indirect}}^{T=0}}{3} \,\left(|F_2|^2-|F_1|^2\right)\,
\frac{M\Delta}{\Delta^2 + (\Gamma_{22}^{T=0}-\Gamma_{11}^{T=0})^2/4}\,\frac{T^2}{M^2}.
\label{indirect2T}
\end{equation}
Note that the indirect asymmetry vanishes for each neutrino type in the exact degenerate limit 
$\Delta \to 0$~\cite{Buchmuller:1997yu,Pilaftsis:1998pd}.

%% file: CP_hiera.tex
In this chapter we come back to the simplest realization of thermal leptogenesis. This scenario was introduced already in chapter~\ref{chap:lepto} and we referred to as vanilla leptogenesis. A hierarchical spectrum for the heavy-neutrino masses is assumed together with an unflavoured regime. Despite the energy scale corresponding to the lightest heavy neutrinos is not directly accessible at present day colliders, $M_1 \simg 10^9$ GeV, vanilla leptogenesis still offers a valid framework to address many aspects of the matter-antimatter generation in the early universe. We are going to study thermal corrections to the CP asymmetry in the lightest heavy neutrino decays as a series in the SM couplings and an expansion in $M_1/M_i$ and $T/M_1$. Indeed we have a different hierarchy of scales, as explained in section~\ref{CP_hiera_sec00}, with respect to the nearly degenerate case discussed in chapter~\ref{chap:CPdege}. First of all there is a separation between the heavy neutrino masses, $M_1 \ll M_i$ with $i=2,3$. In section~\ref{CPhiera_sec1} we integrate out energy modes of order $M_i$ and we are left with an EFT where only the lightest heavy neutrino is dynamical together with the SM particles. Then, in section~\ref{CPhiera_sec2}, we device a second EFT by integrating out the scale $M_1$. In this second EFT non-relativistic excitations of the lightest heavy neutrino field take part in the dynamics and the typical scale is the temperature of the heat bath. Thermal corrections to the CP asymmetry are calculated in section~\ref{CPhiera_sec3} where some effects induced by the heavy neutrino motion are also considered.

\section{A tower of EFTs}
\label{CP_hiera_sec00}
In this chapter we are going to work within vanilla leptogenesis. In this scenario one assumes one heavy Majorana neutrino, with mass $M_1$, much lighter than the other heavy states and the one-flavour regime. Under this assumption the final CP asymmetry is produced by the lightest heavy neutrino decays, and it reads
\begin{equation}
\epsilon_1=  \frac{ \sum_f \, \Gamma(\nu_{R,1} \to \ell_{f} + X)-\Gamma(\nu_{R,1} \to \bar{\ell}_{f}+ X )  }{\sum_f \, \Gamma(\nu_{R,1} \to \ell_{f} + X )+\Gamma(\nu_{R,1} \to \bar{\ell}_{f}+ X)} \, ,
\label{CPhiera_def1}
\end{equation}
where the sum runs over the lepton flavours. In (\ref{CPhiera_def1}) $\nu_{R,1}$ stands for the lightest heavy Majorana neutrino, $\ell_f$ is a SM lepton with flavour $f$ and $X$ represents any other SM particle not carrying a lepton number. Flavour effects are studied in chapter~\ref{chap:CPfla}. 
In the unflavoured regime the leptogenesis scale can be inferred combining neutrino oscillations and mixing data with the observed baryon asymmetry. An important example is the Davidson--Ibarra bound that provides a lower bound on the lightest heavy neutrino mass \cite{Davidson:2002qv, Buchmuller:2002rq}, $M_1 \gtrsim  10^{9}$ GeV. This bound sets the energy scale of leptogenesis, at least in its simplest realization, together with the typical temperatures needed for a thermal production of the heavy neutrinos in the early stages of the universe evolution. The Davidson--Ibarra bound also rises a possible issue for vanilla leptogenesis: if one tries to embed vanilla leptogenesis in a supersymmetric framework, it is hard to reconcile the stringent lower bound on the reheating temperature imposed by the gravitino decays with the corresponding upper bound required by leptogenesis \cite{Kawasaki:2004qu}.

There is a crucial moment for the generation of the lepton asymmetry and it occurs when the temperature of the thermal plasma, $T$, equals the mass of the lightest heavy neutrino, $T \sim M_1$. This is the time at which out-of-equilibrium dynamics may take place and the heavy neutrino evolves towards a non-relativistic regime. 
One can then distinguish between the following two situations: $T>M_1$ and $T<M_1$. The CP asymmetry originated in the former regime can be efficiently erased if the so-called strong washout regime is considered. This seems to be the favoured scenario according to the present values of solar and atmospheric neutrino oscillation data. Therefore the final asymmetry  is independent of the initial abundance of the lightest heavy neutrino and is effectively generated when the temperature dropped below $M_1$ \cite{Buchmuller:2004nz,Blanchet:2012bk}. The heavy neutrino can be considered non-relativistic at this stage. 

Let us point out the relations among the energy scales relevant for the problem at hand: first, a hierarchy between the lightest right-handed neutrino mass, $M_1$, and those of the heavier states $M_i$, $i =2,3$ \footnote{We consider three species for the heavy neutrinos, though in general the model may account for a generic number of species. However at least two heavy neutrino species are necessary to have non-vanishing CP asymmetries.}. Second, a hierarchy between the temperature of the thermal plasma, $T$, and the mass of the lightest heavy neutrino. The former is due to a hierarchically ordered mass spectrum, whereas the latter is related with the universe expansion and the resulting establishment of a non-relativistic dynamics for the lightest heavy neutrino. In summary we have
\begin{equation}
M_{i} \gg M_{1} \gg T \gg M_W \, ,
\label{CPhiera_hiera}
\end{equation} 
and therefore an EFT approach can be considered. The last inequality ensures that temperatures are above the electroweak scale and then the SM sector is described by an unbroken SU(2)$_L \times$U(1)$_Y$ gauge group. We aim at modelling the decays and the generation of the CP asymmetry at finite temperature and we are going to exploit two different EFTs. A first one will serve to integrate out the energy modes of the order of the heavier neutrino masses, namely $M_i \gg M_1$. This is constructed by introducing effective vertices between SM leptons and the Higgs boson~\cite{Buchmuller:2000nd}. We call this effective field theory EFT$_{1}$ throughout the chapter. 
\begin{figure}
\centering
\includegraphics[scale=0.6]{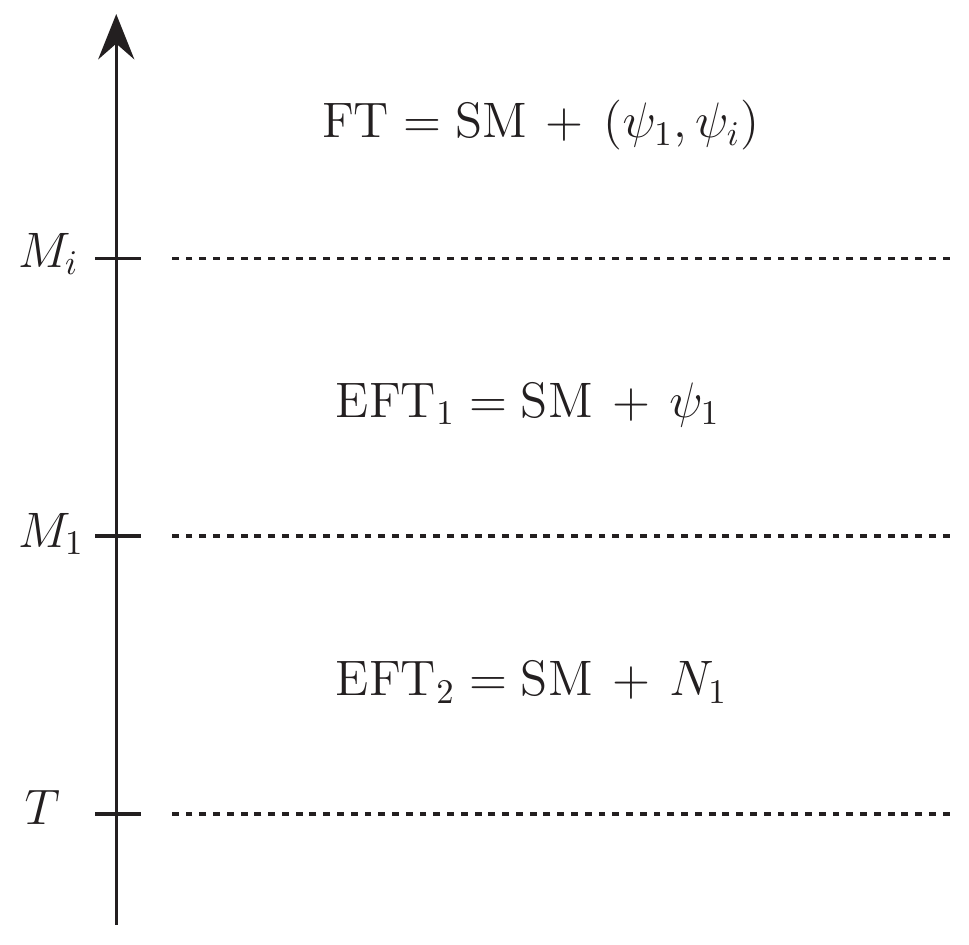}
\caption{\label{fig:EFT_diag}The hierarchy of scales is shown together with the quantum field theories and the degrees of freedom relevant at a given scale. FT stands for the fundamental theory in~\eqref{lepto_8}, where all the three heavy Majorana neutrinos are dynamical. Then integrating out the scale $M_i$ and $M_1$ we obtain two subsequent EFTs, EFT$_1$ and EFT$_2$ respectively. In the former the lightest heavy neutrino is still relativistic ($\psi_1$), whereas in the latter there are only its non-relativistic excitations ($N_1$).}
\end{figure}

In a second step we integrate out the high-energy excitations corresponding to energies and momenta of order $M_1$. Then, only non-relativistic modes for the lightest heavy Majorana neutrino are left together with the SM particles with energies of order $T \ll M_1$. The hierarchy of energy scales, the quantum field theories and the degrees of freedom relevant at a given scale of interest are shown in figure~\ref{fig:EFT_diag}.  We compute thermal corrections to the CP asymmetry within this second EFT, labelled as EFT$_{2}$, at leading order in the SM couplings and at order $(T/M_1)^2$.  We exploit the techniques developed in \cite{Biondini:2015gyw, Biondini:2013xua} and discussed in chapter~\ref{chap:part_prod} and \ref{chap:CPdege}. In particular in the previous chapter, the corrections to the CP asymmetry induced by the leading dimension five operator have been obtained. It was also shown that the top-quark Yukawa coupling, $\lambda_t$, does not enter the CP asymmetry at order $(T/M_1)^2$. However, within the EFT approach, a systematic improvement of the  calculation of the CP asymmetry is achievable in terms of effective operators that describe the interactions between the  heavy neutrino and the SM particles. Here we consider some dimension-seven operators accounting for the interaction between the heavy Majorana neutrino and the top-quark singlet, heavy-quark doublet and lepton doublet that induce a correction of order $|\lambda_t|^2(T/M_1)^4$ to the CP asymmetry. 

Similarly to what done in the case of two heavy neutrinos nearly degenerate in mass, in order to obtain the thermal corrections to the CP asymmetry, we need the expression of the Wilson coefficients of the EFT$_2$. Since $T \ll M_1$ the matching can be done by setting the temperature to zero. This amounts at evaluating two-loop cut diagrams in vacuum for both the dimension-five and dimension-seven operators. Once the Wilson coefficients are known, a tadpole computation in the EFT$_2$ is all what is needed to compute the thermal corrections to the decay widths into leptons and antileptons entering in turn the expression of the CP asymmetry in (\ref{CPhiera_def1}). Reducing the complexity of the required three-loop calculation in a fully relativistic theory  comes as the main advantage of the EFT approach.

Despite thermal effects are expected not to affect strongly the neutrino dynamics in the regime $T \ll M_1$ \cite{Giudice:2003jh}, the knowledge of the thermal corrections to the CP asymmetry may be useful to proceed towards a complete theory of leptogenesis and provide more precisely the parameters entering the rate equations for the heavy-neutrino and lepton-asymmetry number densities. Indeed in the Boltzmann equations, the heavy neutrino production rate and the CP asymmetry enter as key ingredients. Thermal corrections to the right-handed neutrinos production rate  has been derived in the non-relativistic case in \cite{Salvio:2011sf,Laine:2011pq}. In order to connect those results with leptogenesis, a treatment within Boltzmann-like equations in the non-relativistic regime has been carried out in \cite{Bodeker:2013qaa}, where the thermally corrected production rate has been embedded in the rate equations describing the out-of-equilibrium dynamics of leptogenesis. Studies in this direction may be further improved by inserting the expression for the CP asymmetry we propose here. 

\section{CP asymmetry at zero temperature and EFT$_{1}$}
\label{CPhiera_sec1}
\begin{figure}[tbp]
\centering
\includegraphics[scale=0.55]{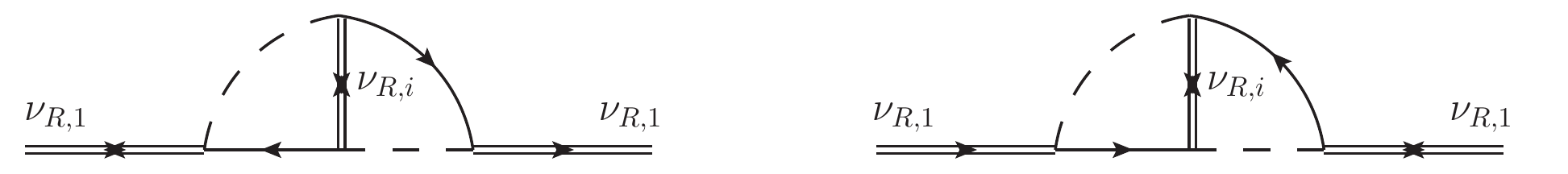}
\caption{\label{fig:self1hiera} Self-energy diagrams for the lightest heavy Majorana neutrino, labelled with $\nu_{R,1}$, corresponding to the mass eigenstate $M_1$. The diagrams are generated from the interference between the tree level and the one-loop vertex diagram in figure~\ref{fig:dirind_degeCP}. }
\end{figure}

\begin{figure}[tbp]
\centering
\includegraphics[scale=0.52]{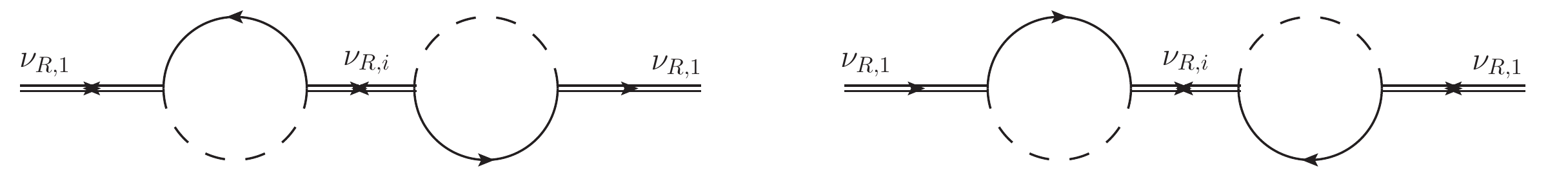}
\caption{\label{fig:self2hiera} Self-energy diagrams for the lightest heavy Majorana neutrino, labelled with $\nu_{R,1}$, corresponding to the mass eigenstate $M_1$. The diagrams are generated from the interference between the tree level and the one-loop self-energy diagram in figure~\ref{fig:dirind_degeCP}.}
\end{figure}

The CP asymmetry can be calculated considering the interference between tree-level and one-loop diagrams that we show in figure \ref{fig:dirind_degeCP}. The loop diagrams are often called self-energy and vertex diagram and their contribution to the CP asymmetry depends on the heavy-neutrino mass spectrum. It is well known that in the case of a hierarchical neutrino mass spectrum and in the unflavoured regime, the two contributions are of the same order and in particular the one originated by the self-energy diagram is twice as big as the vertex one~\cite{Liu:1993tg,Covi:1996wh}.
The calculation of the CP asymmetry can be traced back to the extraction of the imaginary parts of the heavy neutrino self-energy at one and two-loop (up to order $F^4$). We have presented in detail how this works for the vertex topology in the nearly degenerate case in  section~\ref{sec:zeroT}. We may exploit the same formalism to obtain the CP asymmetry due to the vertex diagram, $\epsilon_{1,{\rm{direct}}}$, and the one due to the self-energy diagram, $\epsilon_{1,{\rm{indirect}}}$ in the hierarchical case. The CP asymmetry in (\ref{CPhiera_def1}) may be rewritten as follows 
\begin{eqnarray}
\epsilon_{1}&=& - \sum_{I}  \, \frac{
 2 \,  {\rm{Im}}(B_{\hbox{\tiny direct}}) {\rm{Im}} \left[ (F_{1}^{*}F_{I})^2 \right] }{|F_{1}|^2}   - \sum_{I}  \, \frac{
 2 \,  {\rm{Im}}(B_{\hbox{\tiny indirect}}) {\rm{Im}} \left[ (F_{1}^{*}F_{I})^2 \right] }{|F_{1}|^2} \, ,
\label{CPhiera_def2}
\end{eqnarray}
where the functions $B_{\hbox{\tiny direct}}$ and $B_{\hbox{\tiny indirect}}$ can be calculated by cutting the two-loop diagrams shown in figure \ref{fig:self1hiera} and \ref{fig:self2hiera} and contributing to the propagator of the lightest heavy neutrino
\begin{equation}
-i \left. \int d^{4}x \, e^{ip\cdot x} \, 
\langle \Omega | T \left( \psi_{1}^{\mu}(x) \bar{\psi}_{1}^{\nu}(0) \right) | \Omega \rangle \right|_{p^\alpha =(M_1 + i\eta,\bm{0}\,)} \, ,
\label{matrixFund_hiera}
\end{equation}
where $| \Omega \rangle$ stands for the ground state of the fundamental theory. 
In particular, the function $B_{\hbox{\tiny direct}}$ and $B_{\hbox{\tiny indirect}}$ can be extracted cutting on leptons lines in figure \ref{fig:self1hiera} and \ref{fig:self2hiera} respectively, and evaluating the remaining loop. In (\ref{CPhiera_def2}) and throughout the chapter, we use the notation $(F_{J}F_{I}^{*}) \equiv \sum_f F_{fJ}F_{fI}^{*}$ and  $|F_{I}|^2 \equiv \sum_{f} F_{f I} F^{*}_{f I}$. We already simplified the expression of the CP asymmetry in (\ref{CPhiera_def2}) by imposing that the loop functions of the antileptons coincide with those of the leptons ($B$ and $C$ in chapter~\ref{chap:CPhiera}). 

In the following we give the result for an arrangement of the heavy neutrino masses away from the nearly degenerate case, where a resummation of the intermediate neutrino widths and/or mixing vertex would be needed, and then we show the limit $M_1 \ll M_i$ in the same equation. 
The result for the CP induced by the vertex diagrams in figure~\ref{fig:self1hiera} reads (we already wrote the following results in eqs.~\eqref{CPvert} and \eqref{CPself}) 
\begin{eqnarray}
\epsilon_{1,{\rm{direct}}}&=& \frac{M_i}{M_1} \left[ 1-\left( 1+\frac{M^2_i}{M^2_1} \right)  \ln \left( 1+ \frac{M_1^2}{M_i^2} \right) \right] \frac{ {\rm{Im}}\left[ \left( F_1^* F_i\right)^2\right] }{8 \pi|F_1|^2} \nonumber \\
&\underset{M_1 \ll M_i}{=}& -\frac{1}{16 \pi } \frac{M_1}{M_i}\frac{ {\rm{Im}}\left[ \left( F_1^* F_i\right)^2\right] }{|F_1|^2} +\mathcal{O}\left( \frac{M_1}{M_i}\right)^3 \, ,
\label{CPhiera_vert}
\end{eqnarray}
where the ratio $M_1/M_i$ is the expansion parameter of the EFT$_1$. We keep only the leading order term in the $M_1/M_i$ expansion and a sum over repeated indices is understood if not differently specified. On the other hand the CP asymmetry generated by the self-energy diagrams in figure \ref{fig:self2hiera} is
\begin{eqnarray}
\epsilon_{1,{\rm{indirect}}}&=& \frac{M_1 M_i}{M_1^2-M_i^2} \frac{ {\rm{Im}}\left[ \left( F_1^* F_i\right)^2\right] }{8 \pi|F_1|^2} \nonumber \\
&\underset{M_1 \ll M_i}{=}&-\frac{1}{8 \pi } \frac{M_1}{M_i}\frac{ {\rm{Im}}\left[ \left( F_1^{*} F_i\right)^2\right] }{|F_1|^2}  +\mathcal{O}\left( \frac{M_1}{M_i}\right)^3  .
\label{CPself_vert}
\end{eqnarray}
Due to the assumption $M_1 \ll M_i$, one selects automatically the situation where the heavy neutrino mass difference, $|M_1-M_i|$, is much bigger than the heavy neutrino widths or the mixing terms. This prevents a resonant behaviour of the CP asymmetry for the mass pattern considered here.  The second line in eqs.~(\ref{CPhiera_vert}) and (\ref{CPself_vert}) shows the hierarchical limit of the more general corresponding expressions and the agreement with the known results \cite{Covi:1996wh,Fong:2013wr}. We notice that $\epsilon_{1,{\rm{indirect}}}=2\epsilon_{1,{\rm{direct}}}$ in the limit $M_1 \ll M_i$. 

Our first task is to consider an EFT that is obtained by integrating out degrees of freedom with energy and momenta of order $M_i \gg M_1$, that we call EFT$_1$.  Our aim is to use the EFT$_{1}$ as a starting point for the construction of the EFT$_2$. Within the former EFT one may reproduce the expanded results in eqs.~(\ref{CPhiera_vert}) and  (\ref{CPself_vert}) order by order in the $M_1/M_i$ expansion. This has been already considered in~\cite{Buchmuller:2000nd}, where it was recognized that the full Lagrangian in (\ref{lepto_8}) can be simplified to have only the lightest neutrino as a dynamical degree of freedom. Being the temperature much smaller than the heavy neutrino mass in the non-relativistic regime, the temperature can be set to zero in the matching between the full theory in (\ref{lepto_8}) and the EFT$_1$. In the following we briefly show how the procedure works. 
\begin{figure}[t!]
\centering
\includegraphics[scale=0.58]{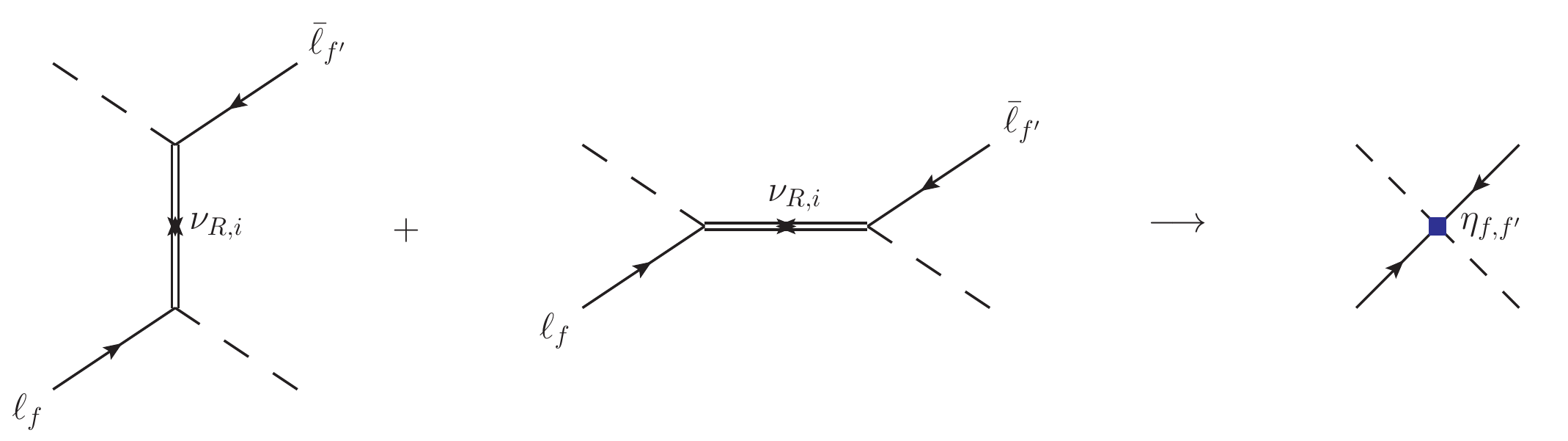}
\caption{\label{fig:eft1_hiera} The figure shows the tree-level matching between the fundamental theory and the EFT$_1$. The two diagrams on the left hand side are the $t$-channel and $s$-channel that appear as sub-diagrams in the vertex and self-energy two-loop topology in figures \ref{fig:self1hiera} and \ref{fig:self2hiera}. On the right-hand side the four-particle diagram stands for the effective interaction in the EFT$_1$.}
\end{figure}

We start by isolating the Higgs-lepton scatterings that enter the diagrams relevant for the CP asymmetry in figure~\ref{fig:self1hiera} and \ref{fig:self2hiera}. In order to have a non-vanishing CP asymmetry, a heavy Majorana neutrino with mass $M_{i}$ has to run as internal propagator (complex phases in the Yukawa couplings combination, see eq.~\eqref{CPhiera_def2}). Therefore the mass of the exchanged heavy neutrino is much bigger than the typical energies carried by the Higgs boson and the lepton, when they come from the decay of a $\nu_{R,1}$ with $M_1 \ll M_i$. The high energy modes of order $M_{i}$ can be then integrated out from the theory and we are left with a four-particle effective vertex interaction, as shown in figure~\ref{fig:eft1_hiera}. This is analogous to the situations studied in chapter~\ref{chap:eff_the}. As regards of the diagrammatic matching shown in figure~\ref{fig:eft1_hiera} two comments are in order. First, we do not show the corresponding diagrams with an outgoing (ingoing) lepton (antilepton). We do take them into account in the matching calculation and eventually in the EFT$_1$ Lagrangian.  Second, we do not consider the diagrams in which the exchanged heavy neutrino propagator comes from $\langle 0 | T(\psi \bar{\psi})|0\rangle$. This paring induces two-loop diagrams with a purely real Yukawa couplings combination, eventually leading to a vanishing CP asymmetry in the unflavoured regime. We study those diagrams in chapter~\ref{chap:CPfla}, anticipating here that their contribution to the CP asymmetry is suppressed by a relative power $M_1/M_i$ with respect to the expressions given in \eqref{CPhiera_vert} and (\ref{CPself_vert}).   
Further details on the tree level matching are given in appendix \ref{appD:CPhieramatch}. The difference between the vertex and self-energy diagrams amounts at a different kinematic channel, more specifically a $t$-channel for the vertex diagram and $s$-channel for the self-energy one. After the two processes are matched onto the four-particle vertex as shown in figure~\ref{fig:eft1_hiera}, the direct and indirect contribution to the CP asymmetry become indistinguishable. Therefore the sum of the second lines in equations (\ref{CPhiera_vert})  and (\ref{CPself_vert}) can be reproduced when calculating the CP asymmetry within the EFT$_1$.

The effective interaction between Higgs bosons and lepton doublets is comprised in the Lagrangian of the EFT$_1$ that reads at order $1/M_i$
\begin{eqnarray}
\mathcal{L}_{{\rm{EFT}_1}}=\mathcal{L}_{\hbox{\tiny SM}} 
&+& \frac{1}{2} \,\bar{\psi}_{1} \,i \slashed{\partial}  \, \psi_{1}  - \frac{M_1}{2} \,\bar{\psi}_{1}\psi_{1}  - F_{f1}\,\bar{L}_{f} \tilde{\phi} P_{R}\psi_{1}  - F^{*}_{f1}\,\bar{\psi}_{1}P_{L} \tilde{\phi}^{\dagger}  L_{f}
\nonumber 
\\
&+& \left( \frac{\eta_{ff'}^i}{M_i} \bar{L}_f \tilde{\phi} \, C P_R \,  \tilde{\phi}^T  \bar{L}^T_{f'} + h. c.\right) + \cdots
\label{Lag2_hiera}
\end{eqnarray}
where  $C$ the charge conjugation matrix, $T$ stands for the transpose of the lepton doublet field and $\eta^i_{ff'}$ is the Wilson coefficient of dimension-five operator (also called Weinberg operator). The dots stand for higher order terms in the expansion $1/M_i$ and for the four-particle operators involving Yukawa coupling combinations, such as $F_{f,i} F^{*}_{f',i}$, that do not contribute eventually to the unflavoured CP asymmetry.  
The matching coefficient reads
\begin{equation}
\eta^i_{f,f'}=\frac{1}{2} \left( F_{f,i}F_{f',i} \right) \, ,
\label{eq8}
\end{equation}
where the index $i$ is not summed on the right-hand side of (\ref{eq8}) in this case. The matching condition comprises the contribution from the $s$-channel and $t$-channel Higgs-lepton scatterings that are subdiagrams of the self-energy and vertex topology of the two-loop self energy diagrams generating the CP asymmetry. 
\begin{figure}[tbp]
\centering
\includegraphics[scale=0.6]{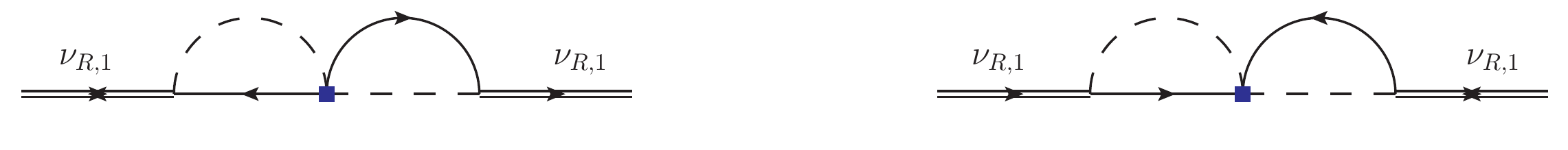}
\caption{\label{fig:eft1cp} Two-loop self-energy diagrams for the lightest neutrino $\nu_{R,1}$ in the EFT$_{1}$. The internal propagator corresponding to the heavier neutrino states is shrunk into a point, accounting for the effective vertices described in figure~\ref{fig:eft1_hiera}. A single topology now describes the two-loop self energy and vertex diagram.}
\end{figure}
The main outcome is that now the CP asymmetry can be represented by a sole topology, shown in figure \ref{fig:eft1cp}, at variance of those previously considered in figure \ref{fig:self1hiera} and \ref{fig:self2hiera}. These diagrams obtained from the Lagrangian (\ref{Lag2_hiera}) will be used in matching the dimension-three and higher order operators of the EFT$_2$.
%Details on the computation of the CP asymmetry are given in appendix~\ref{appD:CPhieramatch}, where diagrams .  
\subsection{Effective Higgs mass}
\label{sec_higgsmass}
\begin{figure}[ht]
\centering
\includegraphics[scale=0.585]{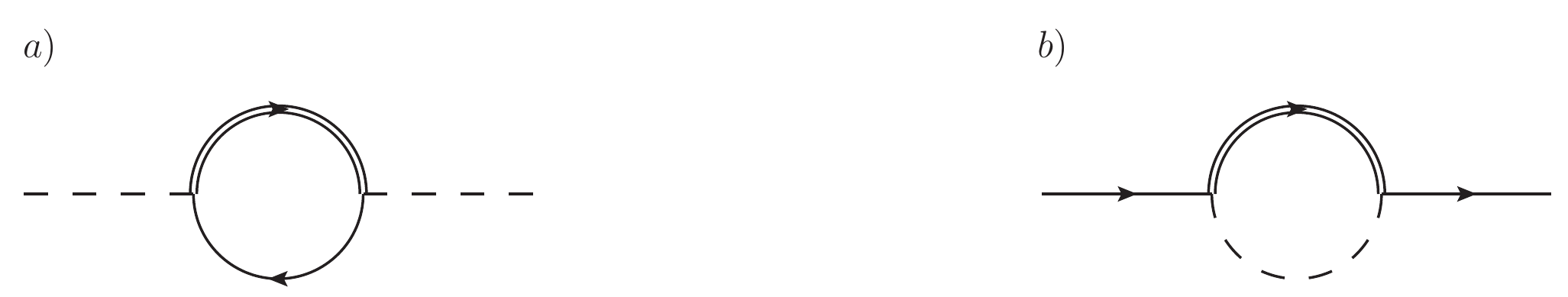}
\caption{One-loop self-energy diagrams for the Higgs, diagram $a)$, and lepton-doublet propagators, diagram $b)$, 
in the fundamental theory. The solid double line in the loop stands for the propagators 
of the heavier Majorana neutrinos with masses $M_i$ ($i=2,3$).}
\label{fig:self_mi0}
\end{figure}

At the level of the EFT$_1$ a finite Higgs mass is generated from matching loop corrections to the Higgs propagator in the fundamental theory, 
which involve heavy Majorana neutrinos with mass $M_i$, with the EFT$_1$ operator $-m_\phi^2 \phi^\dagger \phi$.
The relevant one-loop diagram is diagram~$a)$ of figure~\ref{fig:self_mi0}.
Note that, because of chiral symmetry, the one-loop correction to the lepton-doublet propagator vanishes (see diagram~$b)$ of figure~\ref{fig:self_mi0}).

From the self-energy diagram $a)$ of figure~\ref{fig:self_mi0} one obtains, after renormalizing in the $\overline{{\rm{MS}}}$ scheme, 
\begin{equation} 
m_\phi^2  = 2\frac{M_i^2|F_i|^2}{(4 \pi)^2}\left[1 +  \ln \left( \frac{\mu^2}{M_i^2}  \right) \right] \,.
\label{mass_Higgs_M}
\end{equation} 
A sum over the index $i$ is understood. Implications of the above formula for bounds on the heavy neutrinos masses 
and Yukawa couplings can be found in~\cite{Boyarsky:2009ix}.
The correction induced to the width and to the CP asymmetry by the finite Higgs mass is of relative order $m^2_\phi/M^2_1 \sim  |F_i|^2 M_i^2/M^2_1$, 
hence it is parametrically suppressed by two Yukawa couplings with respect to the other corrections considered in this work. 
Since we systematically neglect higher-order corrections in the Yukawa couplings, 
in the following we will also neglect the effects due to the finite Higgs-boson mass \eqref{mass_Higgs_M}. 
\section{Matching the decays of the lightest neutrino and EFT$_{2}$}
\label{CPhiera_sec2}
In this section we set up the calculation of the matching coefficients of the effective field theory valid at energies much lower than the lightest neutrino mass, $M_1$, that is the  next relevant scale  according to the hierarchy in (\ref{CPhiera_hiera}). We call such effective field theory EFT$_2$.  By integrating out energy modes of order $M_1$, we end up with a quantum field theory where the degrees of freedom are non-relativistic heavy neutrinos and SM particles with typical energies $T \ll M_1$. 

To specify the Wilson coefficients of EFT$_2$ is necessary to match EFT$_1$ and EFT$_2$. Once again the temperature can be set to zero and in-vacuum matrix elements are considered because the matching occurs at a scale $\Lambda$ such that $M_1 \gg \Lambda\gg T$. The Lagrangian so obtained exhibits an expansion in the lightest heavy neutrino mass, and the expression  for the EFT$_2$ reads, following the notation of chapter~\ref{chap:part_prod}
\begin{equation}
\mathcal{L}_{\text{EFT}_2}=\mathcal{L}_{\hbox{\tiny SM}} + \bar{N} \left(iv \cdot \partial +i \frac{\Gamma^{T=0}}{2} \right)N  + \frac{\mathcal{L}_{\hbox{\tiny N-SM}}^{(1)}}{M_1} +  \frac{\mathcal{L}_{\hbox{\tiny N-SM}}^{(3)}}{M_1^3} + \cdots  \, .
\label{Lag3_hiera}
\end{equation}
In (\ref{Lag3_hiera}) $N$ is the field describing the low-energy modes of the lightest heavy neutrino,  $\mathcal{L}_{\hbox{\tiny N-SM}}^{(1)}$ and $\mathcal{L}_{\hbox{\tiny N-SM}}^{(3)}$ comprise dimension-five and dimension-seven operators respectively and the dots stand for higher order operators further suppressed in the scale $M_1$. We do not consider $\mathcal{L}_{\hbox{\tiny N-SM}}^{(2)}$ because it contains operators not contributing to the thermal tadpoles (see section~\ref{EFT_partprod}). Hence no thermal widths and in turn no thermal contributions to CP asymmetry in heavy Majorana neutrino decays can be originated by those operators. 
\subsection{Matching dimension-three operators}
\begin{figure}[t]
\centering
\includegraphics[scale=0.5]{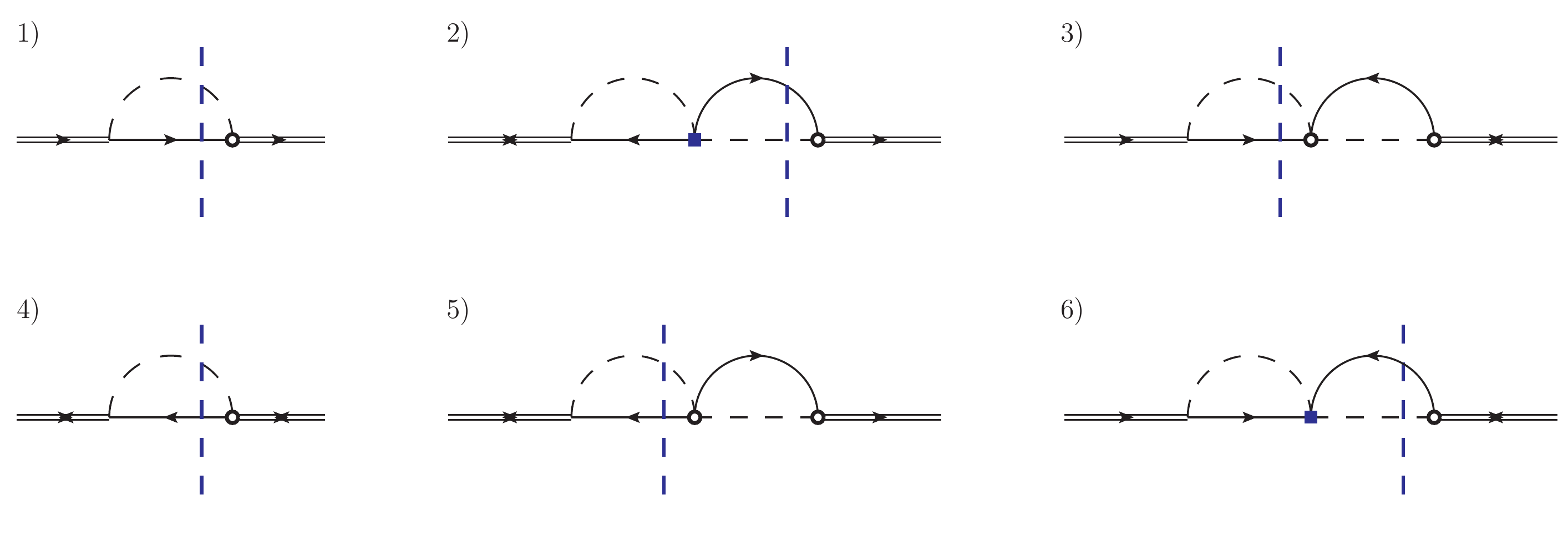}
\caption{\label{fig:CP_hiera_match} One loop and two-loop self energy diagrams in the EFT$_1$.  In the first raw the cut on leptons are shown with vertical blue dashed lines. In the lower raw the one loop diagram involves antileptons, whereas the two-loop self energy diagrams are the same but with  cuts through antileptons lines. }
\end{figure}

The width at zero temperature, $\Gamma^{T=0}$, can be obtained at order $F^2$ and $F^4$ by applying cutting rules to the one and two-loop self-energy diagrams shown in figure~\ref{fig:CP_hiera_match}. Cuts on leptons and antileptons are performed in the diagrams in the upper and lower raw respectively so that the leptonic and antileptonic decay widths are separated. The procedure is exactly the same as the one used in chapter~\ref{chap:CPdege}. 

We call $\mathcal{D}^{\ell}_{1}$, $\mathcal{D}^{\ell}_{2}$ and $\mathcal{D}^{\ell}_{3}$ respectively
the diagrams shown in figure~\ref{fig:CP_hiera_match}, upper raw, with amputated external legs. The quantity ${\rm Im}\left[-i(\mathcal{D}^{\ell}_{1}+\mathcal{D}^{\ell}_{2}+\mathcal{D}^{\ell}_{3})\right]$ matches the neutrino width into leptons $\delta^{\mu \nu} \Gamma^{\ell,T=0}/2$. The explicit calculation gives
\begin{eqnarray}
\delta^{\mu\nu}\,\frac{\Gamma^{\ell,T=0}}{2} &=&
{\rm{Im}}\left[ -i (\mathcal{D}^{\ell}_{1} +\mathcal{D}^{\ell}_{2} +\mathcal{D}^{\ell}_{3})  \right]  
\nn 
\\
&=& 
\delta^{\mu \nu} \frac{M_1}{16 \pi}\left\lbrace  \frac{|F_1|^2}{2} 
 % -  3  \frac{M_1}{M_i} \, \frac{ {\rm{Re}}\left[ (F_1^* F_i)^2\right]}{2(4 \pi)^2}  \, \left[ \frac{2}{\varepsilon} +\ln \left( \frac{4 \pi \mu^2}{M_1^2}\right) + 2 -\gamma_E   \right] 
 - 3  \frac{M_1}{M_i} \,  \frac{ {\rm{Im}}\left[ (F_1^* F_i)^2\right] }{32 \pi}   + \cdots \right\rbrace ,
\label{lepwidth}  
\end{eqnarray}
where the dots stand for the terms proportional to ${\rm{Re}}\left[ (F_1^* F_i)^2\right]$, irrelevant for the CP asymmetry, and higher order terms in the $M_1/M_i$ expansion. 
In order to extract the decay width into antileptons we have to consider the cuts on antileptons as shown in the lower raw of figure~\ref{fig:CP_hiera_match}. Notice that the one loop diagram is different from that in the upper raw, whereas the two-loop self energy diagrams are the same but with the cut through the virtual antilepton.  We call $\mathcal{D}^{\bar{\ell}}_{4}$, $\mathcal{D}^{\bar{\ell}}_{5}$ and $\mathcal{D}^{\bar{\ell}}_{6}$ respectively
the diagrams shown in figure~\ref{fig:CP_hiera_match}, lower raw, with amputated external legs. The result reads 
\begin{eqnarray}
\delta^{\mu\nu}\,\frac{\Gamma^{\bar{\ell},T=0}}{2} &=& 
{\rm{Im}}\left[ -i (\mathcal{D}^{\bar{\ell}}_{4} +\mathcal{D}^{\bar{\ell}}_{5} +\mathcal{D}^{\bar{\ell}}_{6})  \right]  
\nonumber \\
&= & 
 \delta^{\mu \nu} \frac{M_1}{16 \pi}\left\lbrace  \frac{|F_1|^2}{2}  + 3   \frac{M_1}{M_i} \,  \frac{ {\rm{Im}}\left[ (F_1^* F_i)^2\right] }{32 \pi} + \cdots  \right\rbrace ,
\label{antilepwidth}  
\end{eqnarray}
where there is only a change of sign in the coefficient of the Yukawa couplings combination ${\rm{Im}} \left[ (F_1^* F_i)^2 \right] $ between (\ref{lepwidth}) and (\ref{antilepwidth}). The CP asymmetry, as defined in (\ref{CPhiera_def1}), reads 
\begin{equation}
\epsilon^{T=0}_1=\frac{\Gamma^{\ell,T=0}-\Gamma^{\bar{\ell},T=0}}{\Gamma^{\ell,T=0}+\Gamma^{\bar{\ell},T=0}}= -\frac{3}{16 \pi} \frac{M_1}{M_i} \frac{{\rm{Im}} \left[ (F_1^* F_i)^2 \right] }{|F_1|^2} \, ,
\label{CPhiera}
\end{equation}
using the result for the leptonic and antileptonic width. The last result coincides with the sum of the direct and indirect contribution obtained in the hierarchical limit of the expressions in (\ref{CPhiera_vert}) and (\ref{CPself_vert}) (see second line of each equation).  

\subsection{Matching higher dimension operators}
We discuss now the operators of higher dimension that appear in (\ref{Lag3_hiera}). The Lagrangian $\mathcal{L}_{\hbox{\tiny N-SM}}^{(1)}$ contains just one dimension-five operator that reads 
\begin{equation}
\mathcal{L}_{\hbox{\tiny N-SM}}^{(1)}=a \; \bar{N} N \, \phi^{\dagger} \phi \, ,
\label{Ope_Higgs}
\end{equation} 
where $a$ is the corresponding matching coefficient. Diagrams contributing to the matching and to the CP asymmetry are of order $F^4$, and depend on SM couplings. In particular, each of those diagram gives a leptonic contribution to $a$, that we label $a^{\ell}$, when cutting through a lepton line. The same diagram with cuts on antileptons gives the antileptonic contribution, $a^{\bar{\ell}}$. The diagrams and corresponding cuts are given in appendix~\ref{appD:CPhieramatch}. The derivation is close to that carried out in the case of two heavy neutrinos with nearly degenerate neutrino masses in chapter~\ref{chap:CPdege}, and the diagrams necessary for the matching involve the Higgs-four coupling and the gauge couplings of the SU(2)$_L \times$U(1)$_Y$ gauge group. 
\begin{table}[t!]
\centering
\begin{tabular}{l | l  l  l}
\hline
& $\phantom{xxx}\lambda$ & $\phantom{xxx}|\lambda_t|^2$ & $\phantom{x}3g^2+g'^2$ 
\\
\hline
$T=10^9$ GeV & $\approx 0.004$ & $\approx 0.316$ & $ \approx 1.158$
\\
$T=10^7$ GeV & $\approx 0.020$ & $\approx 0.393$ & $ \approx 1.213$
\\
$T=10^3$ GeV & $\approx 0.096$ & $\approx 0.732$ & $ \approx 1.553$
\\
\hline
\end{tabular}
\caption{\label{Tab:tab1_hiera} The SM couplings $\lambda$, $|\lambda_t|^2$ and $3g^2+g'^2$ at different temperatures (energies) are shown~\cite{Rose15,Buttazzo:2013uya}, respectively $T=10^9$ GeV, $T=10^7$ GeV and $T=10^3$ GeV.}
\end{table}

In this chapter we investigate also the effect of some dimension-seven operators in $\mathcal{L}_{\hbox{\tiny N-SM}}^{(3)}$. In particular we single out thermal corrections involving the top-quark Yukawa coupling, $\lambda_t$. Despite these corrections are parametrically suppressed by $(T/M_1)^2$ with respect to those induced by the operator in (\ref{Ope_Higgs}), large differences in the value of the SM couplings and constants appearing in the fermion thermal condensates may alter the numerical relevance of the different corrections.  A similar situation is realized for the neutrino thermal width~\cite{Laine:2011pq,Biondini:2013xua}. In table~\ref{Tab:tab1_hiera} we show the values of SM couplings at high temperatures~\cite{Rose15,Buttazzo:2013uya} for $T=10^9$ GeV, $T=10^7$ GeV and $T=10^3$ GeV, and it is clear that corrections of order $|\lambda_t|^2(T/M_1)^4$ may be of the same order or larger than those of order $\lambda(T/M_1)^2$. Of course there is the same issue with corrections of order $(3g^2+g'^2)(T/M_1)^4$. However we should consider a rather large number of additional diagrams to address the complete derivation of the latter corrections. That is why we stick to thermal corrections involving the top-Yukawa coupling because it suffices to consider a quite limited number of diagrams (see appendix~\ref{appD:CPhieramatch} and section~\ref{sec_hiera_top_close}).

To this aim, the Majorana neutrino-top-quark singlet and heavy-quark doublet effective operators have to be considered together with the Majorana neutrino-lepton doublet operator. As regards the former ones we consider only those that give a non-vanishing contribution in an isotropic medium\footnote{We do not display the operators that develop an interaction between the heavy neutrino spin with the medium. They give vanishing thermal tadpoles in an isotropic medium. They are listed in eq.~(\ref{Wb}) in chapter~\ref{chap:part_prod}.}. They read
\begin{eqnarray}
&&\mathcal{L}_{\hbox{\tiny N-t}}^{(3)}=c_{3} \; \bar{N}N \, \left( \bar{t}P_{L} \, v^\mu v^\nu \gamma_{\mu} \, i D_{\nu} t \right) \,,
\label{Ope_t}
\\
&&\mathcal{L}_{\hbox{\tiny N-Q}}^{(3)}= c_{4} \; \bar{N}N \, \left( \bar{Q}P_{R} \, v^\mu v^\nu \gamma_{\mu} \, i D_{\nu} Q \right) \, ,
\label{Ope_Q}
%\\
%&&\mathcal{L}_{\hbox{\tiny N-L}}^{(3)}=c_{1}^{ff'} \, \left( \bar{N} P_{R} \, i v\cdot D L^{c}_{f'} \right) \left( \bar{L}^{c}_{f} P_{L}  N \right) \,,
%\label{Ope_L}
\end{eqnarray}
where $t$ is the top-quark singlet field and $Q$ is the heavy-quark SU(2) doublet. For the heavy neutrino-lepton doublet low-energy interaction we have
\begin{eqnarray}
\mathcal{L}_{\hbox{\tiny N-L}}^{(3)}= c^{hh'}_{1,c} \;  \left( \bar{N} P_{R} \, i v\cdot D L^{c}_{h'} \right) \left( \bar{L}^{c}_{h} P_{L}  N \right)  
+ c^{hh'}_{1} \left(\bar{N}P_{L} \, iv\cdot D L_{h}\right) \left(\bar{L}_{h'} P_{R} N \right)    \, .
\label{Ope_L}
\end{eqnarray} 

From  (\ref{Lag3_hiera}), and  (\ref{Ope_Higgs})-(\ref{Ope_L})  the thermal corrections to the difference between the leptonic and antileptonic decays of the lightest heavy Majorana neutrino can be written as 
\begin{equation}
\sum_f \, \Gamma(\nu_{R,1} \to \ell_{f} + X)-\Gamma(\nu_{R,1} \to \bar{\ell}_{f}+ X )  = \left( \Gamma^{\ell,T=0}-\Gamma^{\bar{\ell},T=0} \right)   + \left( \Gamma^{\ell,T}-\Gamma^{\bar{\ell},T} \right) \, ,
\label{def_gammaT}
\end{equation}
with 
\begin{equation}
\Gamma^{\ell,T}=\Gamma^{\ell,T}_{\phi}+\Gamma^{\ell,T}_{{\rm{fermions}}} \, ,
\end{equation}
and
\begin{eqnarray}
&&\Gamma^{\ell,T}_{\phi}=2 \frac{{\rm{Im}} \, a^{\ell}}{M_1}  \,\langle \phi^\dagger(0)\phi(0)\rangle_T \, , 
\label{Gamma_lep_higgs} \\
&&\Gamma^{\ell,T}_{{\rm{fermions}}}= 2 \frac{{\rm{Im}} \, c_3^{\ell}}{M_1^3} \, \langle \bar{t}(0) P_L\gamma^0 iD_0 t(0) \rangle_{T} 
+ 2 \frac{{\rm{Im}} \, c_4^{\ell}}{M_1^3}  \langle \bar{Q}(0) P_R\gamma^0 iD_0 Q(0) \rangle_{T} \nonumber \\
&&\hspace{1.5 cm}- \frac{{\rm{Im}} \, c_{1,c}^{hh', \ell}}{4 M_1^3} \langle \bar{L}_{h'}(0) \gamma^0 iD_0 L_h(0) \rangle_{T}  \, ,
\label{Gamma_lep_ferm}
\end{eqnarray}
for the leptonic contribution. Similar expressions hold for $\Gamma^{\bar{\ell},T}$, one has only to replace the matching coefficients in (\ref{Gamma_lep_higgs}) and  (\ref{Gamma_lep_ferm}) with their antileptonic counterparts.  
We observe that $\Gamma^{T=0}= \Gamma^{\ell,T=0}+\Gamma^{\bar{\ell},T=0}=|F_1|^2M_1/(8 \pi)$, which enters (\ref{Lag3_hiera}), is the neutrino decay width in vacuum, and it can be calculated from the heavy neutrino self-energy diagrams at order $F^2$. Moreover the in-vacuum combination in (\ref{def_gammaT}) can be obtained from (\ref{lepwidth}) and (\ref{antilepwidth}), and it reads
\begin{equation}
\Gamma^{\ell,T=0}-\Gamma^{\bar{\ell},T=0} =- \frac{6}{(16 \pi)^2} \frac{M_1^2}{M_i}  {\rm{Im}}\left[ \left( F^{*}_1 F_i\right)^2\right] \, .
\label{Diff_zero}
\end{equation} 
So we need to calculate the thermal part as indicated in (\ref{def_gammaT}), namely the imaginary parts of the matching coefficients appearing in (\ref{Gamma_lep_higgs}) and (\ref{Gamma_lep_ferm}) and the corresponding antileptonic counterparts.

We illustrated and discussed the methodology for the matching calculation in chapter~\ref{chap:CPdege}, hence we recall the main points in short. Two loop diagrams in the fundamental theory (in the present case it is the EFT$_1$) are matched onto a four-particle effective vertices between heavy neutrinos and SM particles in the EFT$_2$. In the case of the dimension-five operator, one has to consider diagrams with two Higgs bosons and two heavy Majorana neutrinos as external legs and the corresponding four-point Green's function. The external Higgs are given typical momentum $q^{\mu} \sim T$, which can be set to zero in the matching.  The complete set of diagrams is shown and discussed in appendix~\ref{appD:CPhieramatch}. Then leptons and antileptons can be put on shell by properly cutting each diagram, so to select the contribution to $a^{\ell}$ and $a^{\bar{\ell}}$ respectively. The result is
\begin{equation}
{\rm{Im}} \, a^{\ell}=-{\rm{Im}} \, a^{\bar{\ell}}=\frac{3}{(16 \pi)^2} \frac{M_1}{M_i} \left[  8 \lambda - \frac{\left( 2g^2+g'^2\right)}{4} \right]   {\rm{Im}}\left[ \left( F^{*}_1 F_i\right)^2\right] \, ,
\label{match_a}
\end{equation} 
where the corresponding result for antileptons can be obtained substituting $F_1 \leftrightarrow F_{i}$, $\lambda$ is the Higgs four-coupling and $g$ and $g'$ are the SU(2)$_L$ and U(1)$_Y$ gauge couplings respectively. 
%We notice that the ratio of the scales $M_1/M_i$ is imprinted in the matching coefficient in (\ref{match_a}) as a remnant of the first matching between the fundamental theory in (\ref{2}) and the EFT$_1$ in (\ref{Lag2}). 

We find that the top-quark Yukawa coupling does not enter the matching coefficient of the dimension-five operator, in analogy with the nearly degenerate case (see eqs.~(\ref{match1}) and (\ref{match2})). In order to have such a coupling in the expression of the CP asymmetry, we add the study of some dimension-seven operators. We pick those that induce a dependence on the top-quark Yukawa coupling, $\lambda_t$. These are the top quark and heavy quark doublet-heavy neutrino operators in (\ref{Ope_t}) and (\ref{Ope_Q}), and also the lepton doublet-heavy neutrino operator in (\ref{Ope_L}).  In the former case two top quarks (two heavy-quark doublets) are considered as external particles together with two heavy Majorana neutrinos appearing in the four-point Green's function relevant for the matching. Lepton doublets are taken as external particles together with heavy Majorana neutrinos in the latter case. The external momentum of the SM particles cannot be put to zero, since we look for contributions containing the low-energy momentum $q$ that matches with the powers of momentum in the effective operators (\ref{Ope_t})-(\ref{Ope_L}). Details on the relevant diagrams and the corresponding cuts are given in the appendix~\ref{appD:CPhieramatch} and section~\ref{sec_hiera_top_close}. The result reads 
\begin{eqnarray}
&&{\rm{Im}} \, c_{3,f}^{\ell}=-{\rm{Im}} \, c^{\bar{\ell}}_3=-\frac{5|\lambda_t|^2}{2(16 \pi)^2} \frac{M_1}{M_i}  {\rm{Im}}\left[ (F_{1}^{*}F_{i})^2\right] \, ,
\label{match_ct}
\\
&&{\rm{Im}} \, c_{4,f}^{\ell}=-{\rm{Im}} \, c^{\bar{\ell}}_4=-  \frac{5|\lambda_t|^2}{4(16 \pi)^2} \frac{M_1}{M_i} {\rm{Im}}\left[ (F_{1}^{*}F_{i})^2 \right] \, ,
\label{match_cQ}  
\\
&&{\rm{Im}} \, c_{1,c}^{hh', \ell}=  -\frac{9 |\lambda_t|^2}{(16 \pi)^2} \frac{M_1}{M_i} {\rm{Im}} \left[(F_1^* F_i)(F^*_{h1}F_{h'i})-(F_1F_i^*)(F_{h'1}F^*_{hi}) \right]  \, ,
\label{match_cL} 
\\
&&{\rm{Im}} \, c_1^{hh', \bar{\ell}}=  -\frac{9 |\lambda_t|^2}{(16 \pi)^2} \frac{M_1}{M_i} {\rm{Im}} \left[(F_1 F^*_i)(F_{h1}F^*_{h'i})-(F^*_1F_i)(F^*_{h'1}F_{hi}) \right] \, ,
\label{match_cLbar} 
\end{eqnarray} 
where $\lambda_t$ is the top Yukawa coupling.

\section{CP asymmetry at finite temperature}
\label{CPhiera_sec3}
\begin{figure}[t!]
\centering
\includegraphics[scale=0.5]{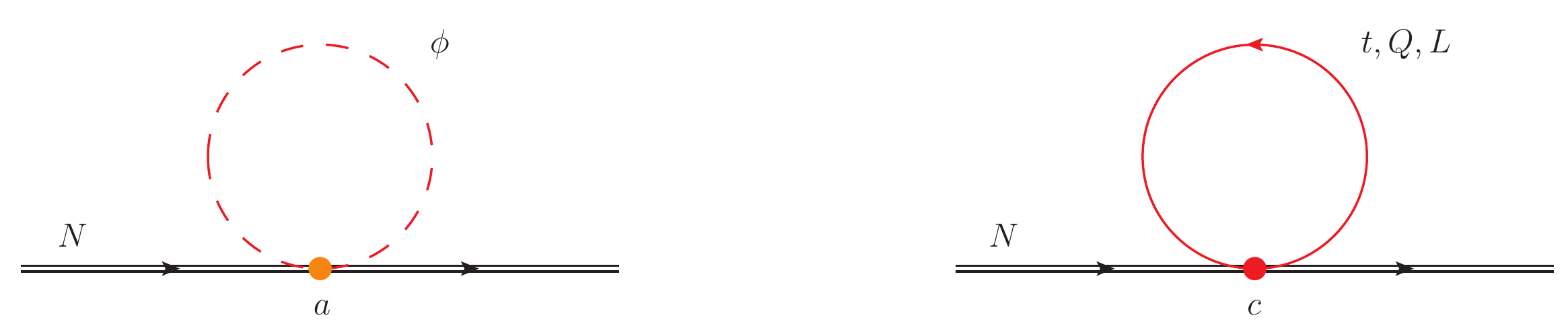}
\caption{\label{fig:tadpolesEFT2} Tadpole diagrams inducing thermal corrections to the heavy neutrino widths and CP asymmetry. We show particles belonging to the thermal plasma in red, Higgs bosons on the left and top quark (heavy quark doublet) on the right. With the vertex label $c$ we mean the different matching coefficients of the top quark, heavy quark doublet and lepton doublet-heavy neutrino operators.}
\end{figure}

In this section we show the result for the thermal corrections to the CP asymmetry. As already explained we compute these corrections in the EFT$_2$, and they  are encoded in tadpole diagrams as shown in figure \ref{fig:tadpolesEFT2}. In the following we assume that the thermal bath is at rest with respect to the lightest heavy Majorana neutrino and we choose the neutrino reference frame such that $v^{\mu}=(1,\bm{0})$. We find convenient to split both the neutrino width and the CP asymmetry into a vacuum and thermal part, namely $\Gamma=\Gamma^{T=0}+\Gamma^{T}$ and $\epsilon_1=\epsilon_1^{T=0}+\epsilon_1^{T}$. This parametrization will be useful to single out the thermal part in the CP asymmetry.  
Let us start with the heavy neutrino width. Since we aim at considering also thermal corrections involving the top-Yukawa coupling, we need to include the corresponding terms that go like $|F_1|^2 |\lambda_t|^2(T/M_1)^4$ in the neutrino width. Those terms are easily isolated following the EFT derivation in chapter~\ref{chap:part_prod}, and the total neutrino width reads
\begin{equation}
\Gamma=\Gamma^{T=0}+\Gamma^{T}=\frac{|F_1|^2M}{8 \pi } \left[ 1-\lambda \left( \frac{T}{M_1}\right)^2 - |\lambda_t|^2 \frac{7 \pi^2}{60}  \left( \frac{T}{M_1}\right)^4 \right] \, . 
\label{widthfour_hiera}
\end{equation}

The in-vacuum part for the CP asymmetry, $\epsilon_1^{T=0}$, at leading order in $M_1/M_i$ can be found in eq.~\eqref{CPhiera}. The other quantity needed for the derivation of $\epsilon_1^{T}$ is the second term on the right hand side in eq.~(\ref{def_gammaT}). Using the following expressions for the condensates at leading order
\begin{eqnarray}
&&\langle \phi^{\dagger}(0) \phi(0) \rangle_{T}  = \frac{T^2}{6} \, , \quad \langle \bar{t}(0) P_L \gamma^0 iD_0 t(0) \rangle_{T}  = \frac{7 \pi^2 T^4}{40} \, ,  \\ 
&&\langle \bar{Q}(0) P_R\gamma^0   iD_0 Q(0) \rangle_{T} = \frac{7 \pi^2 T^4}{20} \, , \; \langle \bar{L}_{h'}(0)\gamma^0 iD_0 L_h(0) \rangle_{T} = \frac{7 \pi^2 T^4}{30} \delta_{h,h'} \, ,
\label{condensate_hiera}
\end{eqnarray}
and using the matching coefficients in (\ref{match_a})-(\ref{match_cLbar}) we obtain  
\begin{equation}
\Gamma^{\ell,T}-\Gamma^{\bar{\ell},T}=\frac{1}{64 \pi^2} \frac{M^2_1}{M_i}  {\rm{Im}}\left[ \left( F^{*}_1 F_i\right)^2\right] \left[   \left( 4 \lambda - \frac{2g^2+g'^2}{8} \right)  \frac{T^2}{M^2_1} - |\lambda_t|^2 \frac{7 \pi^2}{20} \left( \frac{T}{M_1}\right)^4   \right]  \, .
\label{difftherm_hiera}
\end{equation}
Finally from eqs.~(\ref{Diff_zero}), (\ref{widthfour_hiera}) and (\ref{difftherm_hiera}) we obtain, at order $M_1/M_i$, fully at order $(T/M_1)^2$ and at order $|\lambda_t|^2(T/M_1)^4$, the following result:
\begin{equation}
\epsilon_1^{T}=-\frac{3}{16 \pi} \frac{ {\rm{Im}}\left[ \left( F^{*}_1 F_i\right)^2\right]}{|F_1|^2} \frac{M_1}{M_i} \left[   \left(  -\frac{5}{3} \lambda + \frac{2g^2+g'^2}{12} \right) \left( \frac{T}{M_1}\right)^2  +\frac{7\pi^2}{20} |\lambda_t|^2 \left( \frac{T}{M_1}\right)^4   \right] \, .
\label{finalRes}
\end{equation}
The expression (\ref{finalRes}) comprises all thermal corrections at relative order $(T/M_1)^2$, whereas as regards the thermal corrections at order $(T/M_1)^4$ only that proportional to the top-Yukawa coupling is included. Corrections going like $(3g^2+g'^2)(T/M_1)^4$ are not calculated here.
\begin{figure}[ht]
\centering
\includegraphics[scale=1.1]{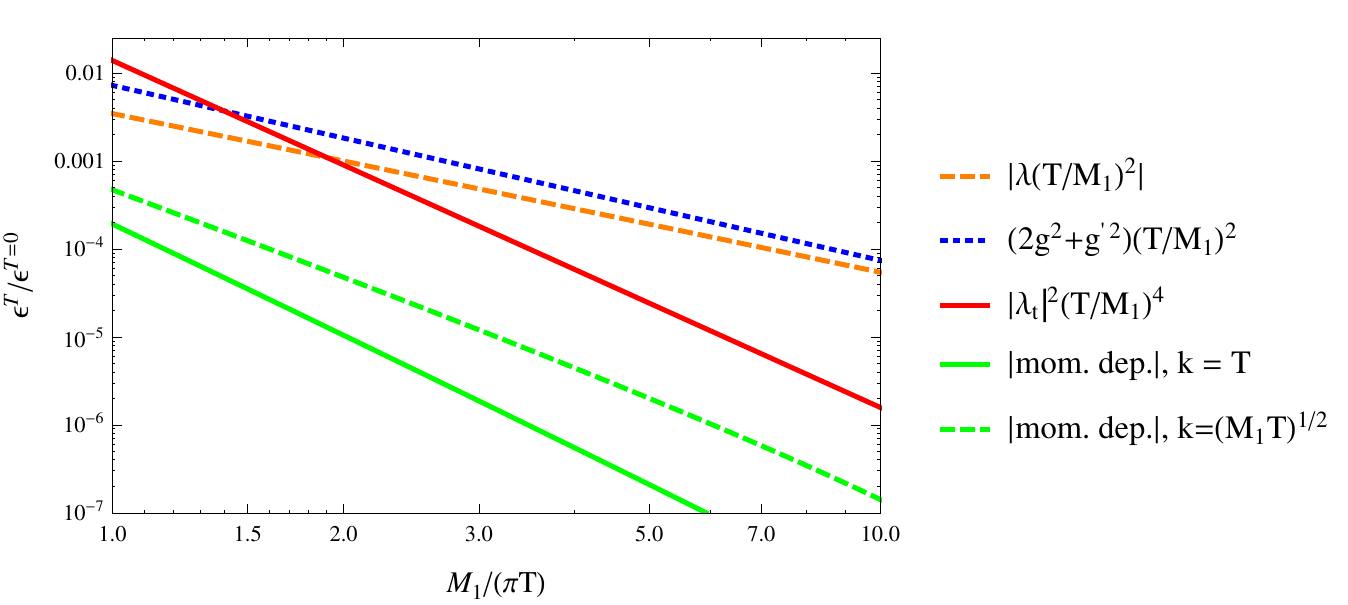}
\caption{Thermal corrections to the CP asymmetry of a Majorana neutrino decaying into leptons and anti-leptons as a function of the temperature.
The orange dashed line shows the contribution proportional to the Higgs self-coupling (the sign of the contribution 
has been changed to make it positive), the blue dotted line shows the contribution proportional to the gauge couplings 
and the red continuous line shows the contribution proportional to the top Yukawa coupling. 
These three contributions can be read from \eqref{finalRes} and refer to a neutrino at rest.
The green lines show the leading thermal contribution proportional to the neutrino momentum, which can be read from \eqref{finRes2_hiera} 
(also in this case the sign of the contribution has been changed to make it positive).
For the green continuous line we take the neutrino momentum to be $T$, whereas for the green dashed line we take it to be $\sqrt{M_1T}$.
The SM couplings have been computed at $\pi T$ with one-loop running~\cite{Rose15}. 
The different thermal contributions to the CP asymmetry have been normalized with respect to \eqref{CPhiera} at leading order in $M_1/M_i$.
The neutrino mass has been taken $M_1 = 10^7$~GeV.}
\label{fig_plotCPcontributions}
\end{figure}

\subsection{Thermal corrections and neutrino motion}
Let us conclude this section with the addition of the leading term  induced by the heavy-neutrino motion to the CP asymmetry. So far we have always considered the neutrino to be at rest. This is clear from the effective Lagrangian in (\ref{Lag3_hiera}). If the neutrino is not at rest, then one has to add operators that depend on the neutrino momentum. By noticing that such operators still have to describe the low-energy interaction between SM particle and the heavy neutrino in order to to generate thermal corrections, we find the leading one to be  \cite{Biondini:2013xua}
\begin{equation}
\mathcal{L}_{{\hbox{\tiny N-k}}}=-\frac{a}{2M_1^3} \bar{N} \left[ \partial^2-(v \cdot \partial)^2 \right] N \phi^{\dagger} \phi \, ,
\label{Ope_mom_hiera}
\end{equation}
as written already in eq.~(\ref{Wmomdep}). The Wilson coefficient, $a$, in (\ref{Ope_mom_hiera}) comes out to be exactly the same of the dimension-five operator in eq.~(\ref{Ope_Higgs}). This can be inferred from the relativistic dispersion relations or by using methods discussed in \cite{Brambilla:2003nt}. When the Wilson coefficient is calculated at order $F_1^2$ in the Yukawa coupling, one obtains  a momentum dependent thermal correction to the total neutrino width, that reads off eq.~(\ref{gammamomdep}).
In this chapter, we evaluate the same matching coefficient at order $(F_1^*F_i)^2$ and therefore the operator in (\ref{Ope_mom_hiera}) can induce different widths into leptons and antileptons as follows   
\begin{equation}
\Gamma^{\ell,T}_{\phi,{\hbox{\tiny mom.\,dep.}}} - \Gamma^{\bar{\ell},T}_{\phi,{\hbox{\tiny mom.\,dep.}}}= -\frac{1}{64 \pi^2} \frac{M^2_1}{M_i}  {\rm{Im}}\left[ \left( F^{*}_1 F_i\right)^2\right] \left[   \left( 2 \lambda - \frac{2g^2+g'^2}{16} \right)  \frac{\bm{k}^2 \,T^2 }{M^4_1}  \right] \, .
\label{mom_diff_hiera}
\end{equation}
Therefore we obtain from eqs.~(\ref{Diff_zero}), (\ref{gammamomdep}) and (\ref{mom_diff_hiera}) a thermal contribution to the CP asymmetry that depends on the heavy neutrino momentum, that reads 
\begin{equation}
\epsilon_{1,{\hbox{\tiny mom.\,dep.}}}^{T}=-\frac{3}{16 \pi} \frac{ {\rm{Im}}\left[ \left( F^{*}_1 F_i\right)^2\right]}{|F_1|^2} \frac{M_1}{M_i} \left[   \left(  \frac{5}{6} \lambda - \frac{2g^2+g'^2}{24} \right) \frac{\bm{k}^2 \,T^2 }{M^4_1} \right] \, .
\label{finRes2_hiera}
\end{equation}

We have computed them at leading order in the SM couplings and for each coupling we have provided 
the leading thermal corrections. The leading thermal corrections proportional
to the Higgs self-coupling, $\lambda$, and to the gauge couplings, $2g^2+g'^2$, are of relative order $(T/M_1)^2$, 
whereas those proportional to the top Yukawa coupling, $|\lambda_t|^2$, are of relative order $(T/M_1)^4$.
We show the different contributions in figure~\ref{fig_plotCPcontributions}. 
At low temperatures, thermal corrections proportional to the Higgs self-coupling and to the gauge couplings dominate.
However, at temperatures closer to the neutrino mass, the suppression in $T/M_1$ becomes less important 
and the numerically most relevant corrections turn out to be those proportional to the top Yukawa coupling.
In figure~\ref{fig_plotCPcontributions} we also show the thermal contribution to the CP asymmetry due to a moving Majorana neutrino, 
which has been computed in \eqref{finRes2_hiera}. We plot this contribution for the case of a neutrino with momentum $T$ 
and for the case of a neutrino in thermal equilibrium with momentum $\sqrt{M_1T}$. 
We see that for the considered momenta the effect of a moving neutrino on the thermal CP asymmetry is tiny. 
\section{A closer look at processes at order $|\lambda_t|^2(F_1^*F_i)^2$}
\label{sec_hiera_top_close}
\begin{figure}[t!]
\centering
\includegraphics[scale=0.565]{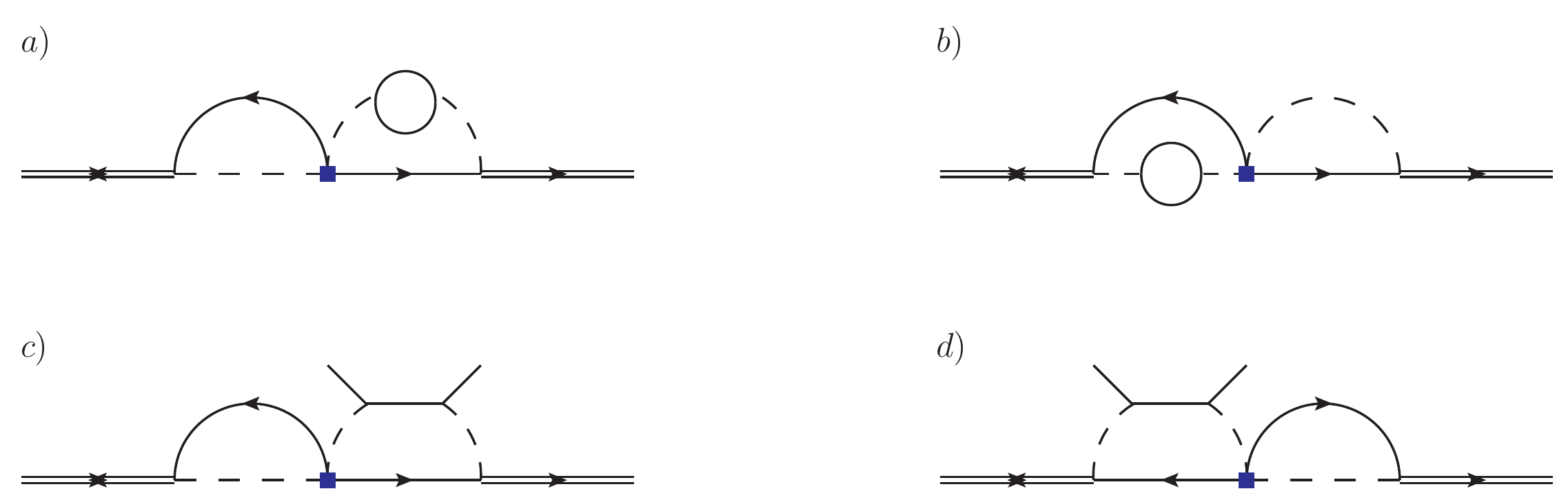}
\caption{\label{fig:three_top}Heavy neutrino self-energies at order $|\lambda_t|^2(F_1^*F_i)^2$ (at three loop), diagram $a)$ and $b)$. Heavy neutrino-top quark scattering at two loop, diagram $c)$ and $d)$, entering the matching of the dimension-seven operator in (\ref{Ope_t}). Solid lines with no arrows stand for top quarks (heavy-quark doublet), in order not to confuse with leptons.}
\end{figure}
\begin{figure}[t!]
\centering
\includegraphics[scale=0.585]{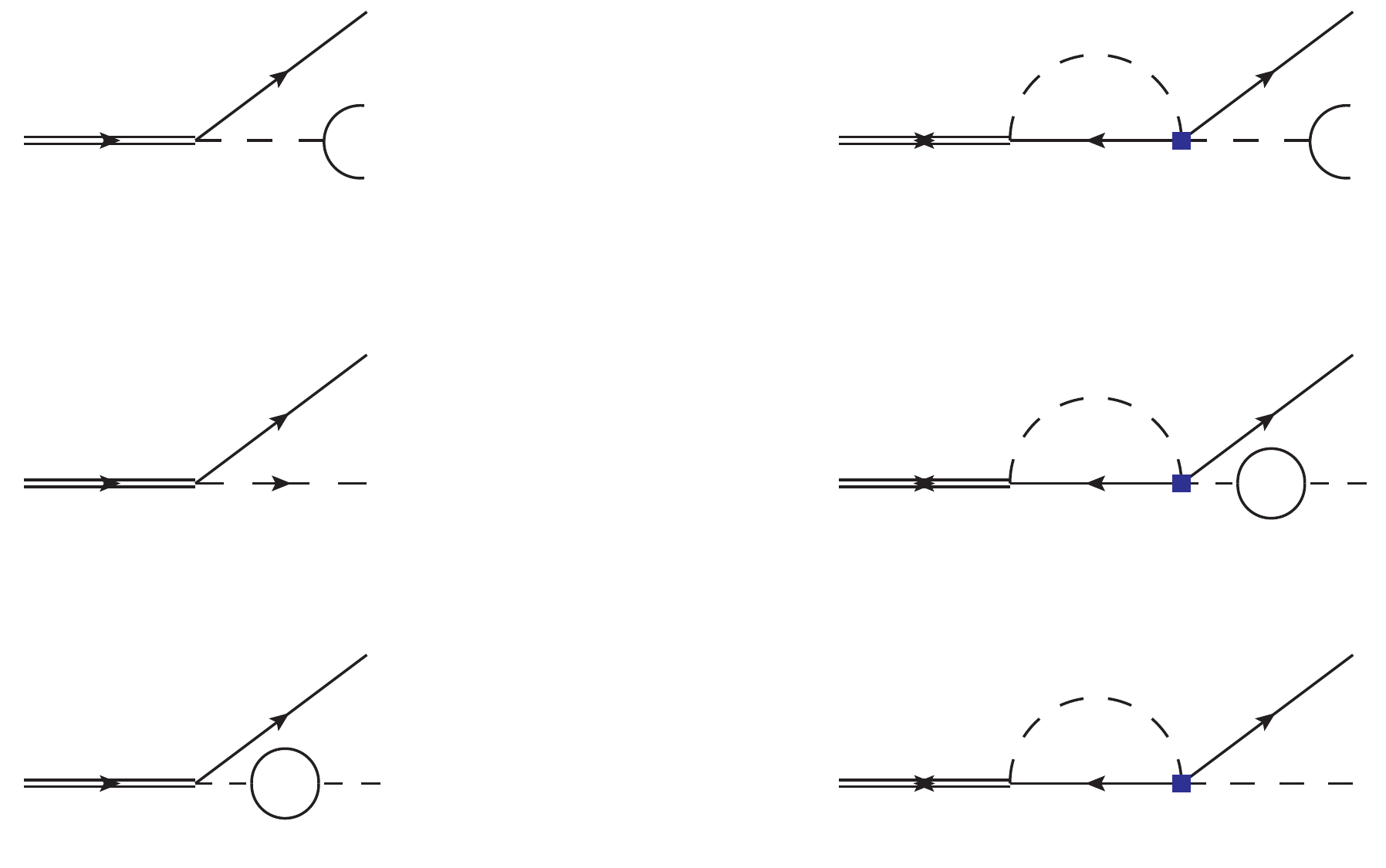}
\caption{\label{fig:three_top_decay}Tree level and loop diagrams for the heavy Majorana neutrino decay processes $\nu_{R,1} \to \ell_f + t + \bar{Q}$ and $\nu_R \to \ell_f + \phi$. The interference of each couple of diagrams in each raw gives the diagram $a)$ in figure~\ref{fig:three_top}.}
\end{figure}

We discuss in some detail the diagrams involving the top-Yukawa coupling, $\lambda_t$. In particular we aim at showing the systematics to obtain them and the connection with a known problem in the literature. Let us start with the two-loop self-energy diagrams in figure~\ref{fig:eft1cp}. Those diagrams are at order $(F_1^*F_i)^2$  in the Yukawa couplings and originate a CP asymmetry in heavy Majorana neutrino decays at $T=0$, see eq.~\eqref{CPhiera}. After the cuts through a lepton and a Higgs line shown in figure~\ref{fig:CP_hiera_match}, the diagrams are divided into a tree-level and a one-loop diagram that describe the decay process $\nu_{R,1} \to \ell_f + \phi$. To find out which diagrams we have to consider for the matching of the dimension-seven operators involving the top-Yukawa coupling, we switch on the interactions allowed by the Feynman rules in the SM and we obtain the diagrams $a)$ and $b)$ in figure~\ref{fig:three_top}. Let us focus on diagram $a)$. This diagram can be constructed also starting from the interference between the tree level and one-loop diagrams responsible for the decay process $\nu_{R,1} \to \ell_f + t + \bar{Q}$, shown in the first raw in figure~\ref{fig:three_top_decay}.  It can also be obtained by making interfere the diagrams responsible for the heavy-neutrino decay process $\nu_R \to \ell_f + \phi$, second and third raw in figure~\ref{fig:three_top_decay}, where there is a self-energy correction of order $|\lambda_t|^2$ for the Higgs boson.  Therefore the topology of diagram $a)$ in figure~\ref{fig:three_top} comprises both the processes: $\nu_{R,1} \to \ell_f + t + \bar{Q}$ and  $\nu_R \to \ell_f + \phi$, according to different cuts: either two particles are put on shell, a lepton and a Higgs boson, or three particles are put on shell, a lepton and the top-quark pair. In summary, if the diagrams $a)$ and $b)$ in figure~\ref{fig:three_top} are understood at $T=0$, they would give the zero temperature radiative corrections to the CP asymmetries in heavy Majorana neutrino decays, which are not calculated at the best of our knowledge. Conversely if the same diagrams are understood at finite temperature, one would obtain also the thermal contributions to to the CP asymmetries in heavy Majorana neutrino decays. This is an example of the calculation at three-loop in thermal field theory one has to tackle.  

\begin{figure}
\centering
\includegraphics[scale=0.585]{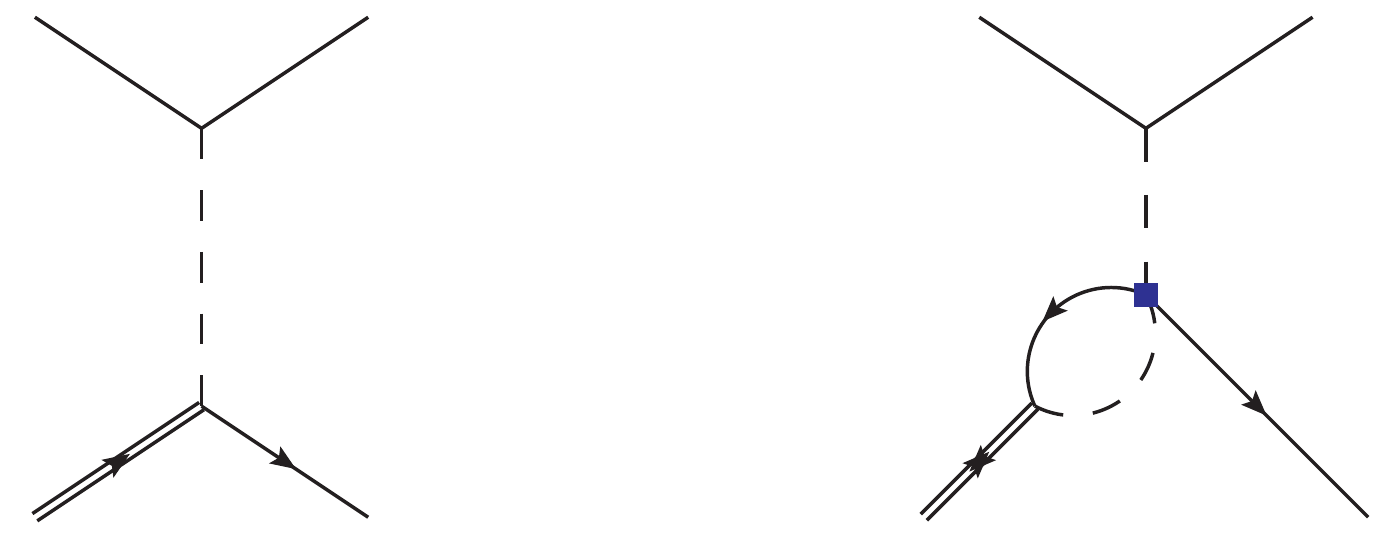}
\caption{\label{fig:top_sca}Tree-level and loop-diagrams for the heavy Majorana neutrino-top scattering  processes $\nu_{R,1} + \bar{t} \to \ell_f + \bar{Q}$ in the $t$-channel. The interference of these two diagrams gives the two-loop diagram $c)$ of figure~\ref{fig:three_top} that we have to consider for the matching of the operator in~(\ref{Ope_t}). The charge conjugate process is not shown.}
\end{figure}

Let us turn to the EFT prospective. We always considered four-particle effective vertices like those in figure~\ref{Fig1} as the first step for the derivation of the neutrino thermal width induced by SM particle in the heat bath. A matching of four-point Green's functions describing the scatterings between a heavy Majorana neutrino at rest and soft particles of the thermal bath are involved. Opening-up the three-loop topologies in figure~\ref{fig:three_top} one can obtain diagrams with either lepton doublets or top-quark singlets (heavy-quark doublets) as external legs. This provides the systematics in the EFT approach and the corresponding diagrams are listed in appendix~\ref{appD:CPhieramatch}. We show those with singlet top quarks (heavy-quark doublets) as external particles in figure~\ref{fig:three_top}, diagrams $c)$ and $d)$. If we consider the top-quark singlet as external legs, we notice that those diagrams can be also understood as the interference between the tree level and one-loop scattering process, shown in figure~\ref{fig:top_sca}, in the EFT$_1$. The  process is a $t$-channel  scattering, $\nu_{R,1} + \bar{t} \to \bar{Q} + \ell_f$, and it can produce a CP asymmetry as well, not in heavy neutrino decays but rather in heavy neutrino scattering with top quarks \cite{Luty:1992un,Plumacher:1996kc}. The analogue of the $\Delta \Gamma^{T=0}= \Gamma^{\ell,T=0}-\Gamma^{\bar{\ell},T=0}$  would be $\Delta \sigma^{T=0}= \sigma^{\ell,T=0}-\sigma^{\bar{\ell},T=0}$ , where $\sigma^{\ell,T=0} \, (\sigma^{\bar{\ell},T=0})$ is the cross section for producing a lepton (antilepton) in the final state of the scattering process. It was also pointed out that the CP asymmetry arising from such scattering process are affected by IR divergences, due to the exchange of a massless Higgs boson \cite{Nardi:2007jp,Abada:2006ea,Pilaftsis:2003gt,Pilaftsis:2005rv}. Then different solutions have been proposed and the most popular is to consider a finite mass for the Higgs boson that comes from thermal corrections and it does regularize the divergence. However a resummation of degrees of freedom from the temperature scale, those inducing a Higgs thermal mass, is rigorously justified in the regime $T \gg M_1$ (in a similar fashion of the hard thermal loop resummation~\cite{Braaten:1989mz}).

Even though we are concerned about the matching coefficients for the width (see eqs.~\eqref{match_ct} and~\eqref{match_cQ}), we elaborate a bit on the exclusive processes that contain the IR divergence due to the massless Higgs boson and the solution adopted in the literature. Indeed we make our proposal in order to obtain a finite Higgs mass at the level of the EFT$_1$, valid at $M_1 \gg T$, that would regularize the divergence in each single cut, as well as a thermal Higgs mass would do in the regime $T \gg M_1$. This observation would be relevant when studying the CP asymmetry in the scattering processes. Our observation is that a finite Higgs mass may be originated by integrating out high energy modes of order of the heavier neutrino masses, $M_i$. Let us explain our point. The matching between the EFT$_1$ and the EFT$_2$ occurs at a scale much lower than $M_i$, due to the hierarchy in \eqref{CPhiera_hiera}. In general effects of higher energy modes are encoded into operators suppressed in powers of $M_i$, see the dimension-five operators in the EFT$_1$ Lagrangian \eqref{Lag2_hiera}. However we can consider an additional effect due to energy modes of order $M_i$. All the SM particles are massless because of the unbroken symmetry SU(2)$_L\times$U(1)$_Y$ and hence the whole set of self-energy diagrams that would provide a mass shift to the Higgs boson vanishes in dimensional regularization, but one. It was shown in figure~\ref{fig:self_mi0}, diagram $a)$. 

In summary, we suggest that a finite Higgs mass arising from the one-loop $T=0$ diagram in figure~\ref{fig:self_mi0} may work as well to regularize the divergences that appear in each single cut (but not in the sum) when calculating the contribution to the CP asymmetry in diagram $c)$ of figure~\ref{fig:three_top_decay}.

%% file: fla_CP.tex
In the previous two chapters we have computed the CP asymmetry, both direct and indirect, in the so-called unflavoured approximation, 
i.e., we have computed the CP asymmetry, defined in eq.~\eqref{CP_CPdege}, as a sum over lepton flavours.
This is the relevant CP asymmetry when the flavour composition of the quantum states of the leptons (antileptons) in the thermal plasma 
has no influence on the final lepton asymmetry. However this scenario is justified only at high temperatures, typically $T > 10^{12}$ GeV, and a quantitative analysis  of leptogenesis for a wider temperature window requires flavour to be included. In section~\ref{sec_fla1} the mechanism to resolve different lepton flavours in the early universe is introduced, together with the definition of a flavoured CP asymmetry. In section~\ref{sec_fla2} we show the results for the CP asymmetries for two nearly degenerate heavy-neutrino masses whereas the case of a hierarchical mass spectrum is discussed in section~\ref{sec_fla3}, this time including the flavour of the final lepton (antilepton).  
\section{General discussion on flavour in leptogenesis}
\label{sec_fla1}
We often highlighted that leptogenesis occurs in a hot and dense plasma and, therefore, the medium effects have to be properly taken into account. The effect of a heat bath of SM particles has been included both in the heavy-neutrino production rate and the CP asymmetries in heavy-neutrino decays. In an EFT approach, such effects are organized as a series in the SM couplings and powers of $T/M$. There is another aspect of the interactions in the thermal bath that may play a role during leptogenesis  that we have neglected so far. SM lepton doublets come with different flavours, $f=e,\mu,\tau$, and we always assumed this feature not to have any influence on the leptogenesis dynamics. Indeed the derivation of the CP asymmetries in chapter~\ref{chap:CPdege} and \ref{chap:CPhiera} was carried out summing over different lepton flavours, providing the so-called \textit{one-flavour} or \textit{unflavoured} regime. 

The unflavoured regime is found to be an appropriate choice at high temperatures, namely $T > 10^{12}$ GeV,  
whereas different lepton flavours are resolved at lower temperatures~\cite{Nardi:2005hs, Nardi:2006fx}. 
In~\cite{Campbell:1992jd, Cline:1993bd} it was shown how to estimate the temperature at which lepton flavours are resolved 
considering the interactions induced by charged lepton Yukawa couplings in the most general seesaw type-I Lagrangian \cite{Nardi:2005hs,Davidson:2008bu}
\bea
&&\mathcal{L}=\mathcal{L}_{\rm{SM}} + \frac{1}{2} \bar{\psi}_{I} i \slashed{\partial}  \psi_{I}  
- \frac{M_{I}}{2} \bar{\psi}_{I}\psi_{I} - F_{f I}\bar{L}_{f} \tilde{\phi} P_{R}\psi_{I}  - F^{*}_{f I}\bar{\psi}_{I} P_{L} \tilde{\phi}^{\dagger}  L_{f} \, 
\nn
\\
&&\hspace{1.6 cm} -h^*_f \bar{e}_f \phi^\dagger P_L L_f - h_f \bar{L}_f  \phi \, P_R e_f \, ,
\label{eq_fla_1}
\eea
where $e_f$ is the SU(2) lepton singlet with flavour $f$ and $h_f$ are the charged lepton Yukawa couplings. The Lagrangian (\ref{eq_fla_1}) is written in a basis in which the right-handed neutrino mass matrix and the charged lepton Yukawa coupling matrix are diagonal  with three real eigenvalues each (seesaw flavour basis~\cite{Molinaro:2010kxa}). For the latter case those are the charged lepton Yukawa couplings $h_e$, $h_\mu$ and $h_\tau$ (at energies below the breaking of the electroweak symmetric phase they provide the masses of the charged leptons, e.~g.~$m_e \approx h_e v$). According to this choice of the basis, the Yukawa matrix $F_{fI}$ is a complex matrix with 18 parameters from which three phases can be removed by field redefinitions of $L_f$, leaving 9 moduli and 6 phases as physical parameters. Hence there are in total 21 real parameters in the lepton sector.

Different flavours may be distinguished during leptogenesis if the $h_f$-mediated interactions are
fast compared to those of leptogenesis and to the universe expansion rate. The authors in~\cite{Campbell:1992jd, Cline:1993bd} showed that the interaction rate for the charged Yukawa couplings can be estimated as 
\be
\Gamma_f \simeq 5 \times 10^{-3} h_f^2 \, T \, ,
\ee
and by requiring the rate $\Gamma_f$ to be larger than the universe expansion rate $H$ (see eq.~\eqref{hotbb_15}), one can extract the temperatures for which different flavours are resolved. It is found that at $T \approx 10^{12}$ GeV, the interaction rate involving the $\tau$-doublet is faster than the universe expansion rate. 
Hence the $\tau$-flavour is resolved by the thermal bath, while the $e$- and $\mu$-flavours remain still unresolved. 
At temperatures of about $10^9$ GeV also the interaction rates involving the $\mu$-doublet enter in equilibrium, so that three flavours are resolved and measured by the heat bath.
The importance of flavour effects in leptogenesis has been investigated in the literature in many different directions e.~g.~\cite{Blanchet:2006be, DeSimone:2006nrs, Antusch:2006cw}. 

In the case different flavour states are resolved during leptogenesis, the CP asymmetries have to be recast in a way that makes transparent how the matter-antimatter asymmetry is stored into different flavour components. In order to embed flavour effects in our approach, we start with the definition of the CP asymmetry, $\epsilon_{fI}$, 
generated by the $I$-th heavy neutrino decaying into leptons and antileptons of flavour $f$, it reads: 
\begin{equation}
\epsilon_{fI}=
\frac{ \Gamma(\nu_{R,I} \to \ell_{f} + X)-\Gamma(\nu_{R,I} \to \bar{\ell}_{f}+ X )  }
{\sum_f \Gamma(\nu_{R,I} \to \ell_{f} + X ) + \Gamma(\nu_{R,I} \to \bar{\ell}_{f}+ X)} \, .
\label{eq:adef_fla}
\end{equation}
The difference with respect to  eq.~\eqref{CP_CPdege} is that we do not sum over the flavour index $f$ in the numerator. The quantity in eq.~(\ref{eq:adef_fla}) would then be the right one to be inserted in the Boltzmann equations in the flavoured regime \cite{Nardi:2006fx,Abada:2006ea}.   

In order to show how the different CP flavour components are relevant for leptogenesis, let us assume a hierarchical spectrum of heavy neutrino masses, $M_1 \ll M_i$, so that the time scale of leptogenesis is set to $T \sim M_1$ (see section~\ref{sec_lepto2}). For $T \sim M_1 >  10^{12}$ GeV the lepton flavours are indistinguishable
and the one-flavour approximation is valid. The relevant CP asymmetry in this case is $\epsilon_1 = \epsilon_{e1}+\epsilon_{\mu 1}+\epsilon_{\tau 1}$ and this is equivalent to (\ref{CP_CPdege}). For $10^9 \, \text{GeV} < T \sim M_1 < 10^{12} \, \text{GeV}$ the $\tau$-Yukawa interactions are in equilibrium and the time evolution of the lepton charge $L_\tau$, proportional to $\epsilon_{\tau 1}$, is different from the evolution of the $(e + \mu)$ lepton charge, $L_{e + \mu}$, in turn proportional to $\epsilon_{e 1} + \epsilon_{\mu 1}$. Therefore two different flavoured CP asymmetries have to be considered if one aims at studying quantitatively leptogenesis for temperatures below $10^{12}$ GeV.

\section{Flavoured CP asymmetries for nearly degenerate neutrino masses}
\label{sec_fla2}
In this section the derivation of the CP asymmetries for two heavy neutrinos nearly degenerate in mass is provided. 
Following the same order adopted for the unflavoured case, we will, 
first, compute the flavoured direct and indirect CP asymmetries at $T=0$, and then the CP asymmetries at finite temperature.
\subsection{CP asymmetries at T=0}
It is straightforward to extend the derivation of section~\ref{sec:zeroT} for the direct CP asymmetry at $T=0$ in the unflavoured case 
to the CP asymmetry in the flavoured case.
In the latter case one has simply to omit the sum over the flavour index $f$ in \eqref{CPdege_b3} and \eqref{CPdege_b6},  
obtaining for the CP asymmetry in the neutrino of type~1 decays
\begin{eqnarray}
&& \epsilon_{f1} =   
\nonumber\\
&& \sum_J \resizebox{.9 \textwidth}{!} { $  \frac{\left( {\rm{Re}}(B)-{\rm{Re}}(C) \right) {\rm{Re}} \left[ (F_{1}^{*}F_{J})(F_{f1}^*F_{fJ})\right] 
- \left( {\rm{Im}}(B)+{\rm{Im}}(C) \right) {\rm{Im}} \left[ (F_{1}^{*}F_{J})(F_{f1}^*F_{fJ}) \right] }{|F_{1}|^2} .
$ }\nonumber\\
\label{cpzero_fla}
\end{eqnarray}
The calculation of the diagrams in figure~\ref{fig:fig3_CPdege} leads to the same results for the functions $A$, $B$ and $C$: 
the loop calculation is unaffected by the different treatment of the flavour. 
Note that additional two-loop diagrams, similar to 2) and 3) of figure~\ref{fig:fig3_CPdege} but involving only lepton (or antilepton) internal lines,  
are not allowed by the Feynman rules of \eqref{lepto_8}.
Therefore the direct CP asymmetry at $T=0$ for the neutrino of type~1 decay into leptons of flavour $f$ 
reads up to order $\Delta /M$ 
\begin{equation}
\epsilon_{f1,{\rm direct}}^{T=0} = 
\left[   (1 -2 \ln 2) + (3 - 4 \ln 2)\frac{\Delta}{M}   \right] \frac{{\rm{Im}}\left[ (F_{1}^{*}F_{2})(F_{f1}^*F_{f2}) \right]}{8 \pi |F_{1}|^2} .
\label{CPdir_1fla}
\end{equation}
The result for $\epsilon_{f2,{\rm direct}}^{T=0}$ can be obtained from the above formula by changing $F_1 \leftrightarrow F_2$ and $\Delta \to -\Delta$.
The results agree in the nearly degenerate limit with the flavoured CP asymmetry obtained in~\cite{Fong:2013wr}. 

\begin{figure}[t!]
\centering
\includegraphics[scale=0.57]{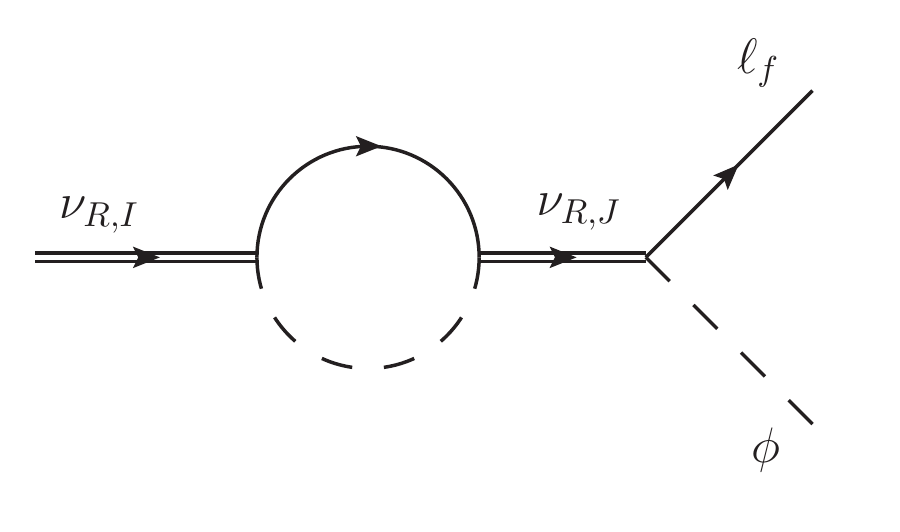}
\caption{One-loop self-energy diagram responsible for an additional contribution to the indirect CP asymmetry in the flavoured case. 
Note that only heavy-neutrino propagators with forward arrow appear, namely $\langle 0| T(\psi \bar{\psi}) |0 \rangle$. }
\label{fig:flavor_1} 
\end{figure}

\begin{figure}[t!]
\centering
\includegraphics[scale=0.53]{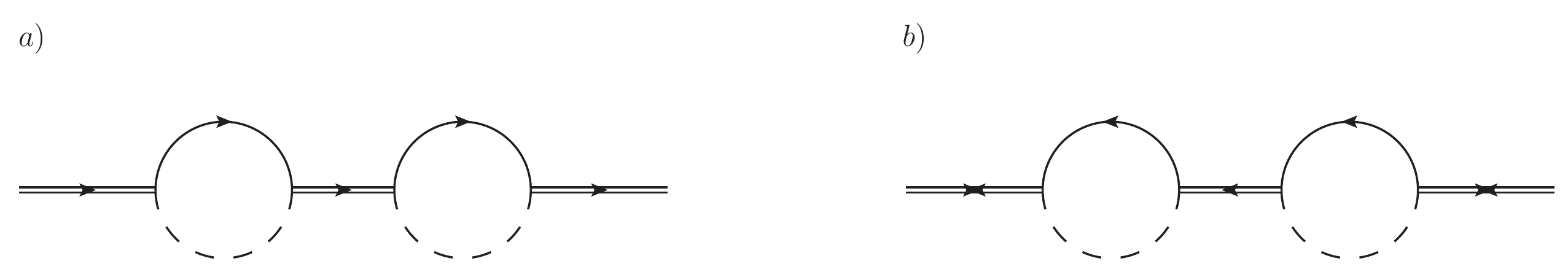}
\caption{Two-loop self-energy diagrams in the fundamental theory contributing to the indirect CP asymmetry at $T=0$ in the flavoured case only. 
Diagram $a)$ admits two cuts through lepton lines, whereas diagram $b)$ admits two cuts through antilepton lines.}
\label{fig:flavor_2} 
\end{figure}

We can compute the flavoured indirect CP asymmetry at $T=0$ either in the fundamental or in the effective theory.
In the fundamental theory, besides the diagrams that appear in the unflavoured case,  
one has to consider also the interference between the tree-level diagram of figure~\ref{fig:dirind_degeCP} 
with the additional one-loop diagram shown in figure~\ref{fig:flavor_1}. 
This contribution is equivalent to cutting through lepton or antilepton lines respectively the two-loop diagrams $a)$ and $b)$ shown in figure~\ref{fig:flavor_2}. 
The additional diagrams give a contribution to the CP asymmetry that is proportional to ${\rm Im}\left[(F_1 F^*_2)  (F^*_{f1} F_{f2})\right]$.
Clearly this contribution vanishes if summed over all flavours $f$. 
For this reason it has not been considered in the unflavoured case in chapter~\ref{chap:CPdege}.

As argued in section~\ref{sec:indirect}, it is particularly convenient to compute the indirect CP asymmetry in the EFT. 
In fact, the relevant diagrams are the same computed in the unflavoured case, i.~e.~, those shown in figure~\ref{fig:indirectEFT}. 
They already comprise the two additional diagrams of figure~\ref{fig:flavor_2}, 
the only difference being that now the cut through the mixing vertex selects the decay into a specific leptonic (or antileptonic) flavour family. 
More specifically the cut stands for $M(F_{fI}^*F_{fJ})/(16\pi)$ (or $M(F_{fJ}^*F_{fI})/(16\pi)$), 
where $I$ is the type of the outgoing and $J$ the type of the incoming neutrino.
Hence the result for the leptonic width of a neutrino of type 1 decaying into a lepton of flavour $f$
can be read off \eqref{gammalt0indirect} by refraining of summing over the flavours in the leptonic cuts
\begin{equation}
\Gamma^{\ell,T=0}_{f11,{\rm indirect}} = \frac{M}{16\pi} F_{f1}^*F_{f2}\frac{i}{-\Delta+i(\Gamma_{22}^{T=0}-\Gamma_{11}^{T=0})/2}
\left(-\frac{M}{16\pi}\right)\frac{F_1^*F_2+F_2^*F_1}{2} + {\rm c.c.}\;.
\label{gammalt0indirect_fla}
\end{equation}
For antileptonic cuts the diagrams in figure~\ref{fig:indirectEFT} give the antileptonic width, $\Gamma^{\bar{\ell},T=0}_{f11,{\rm indirect}}$, 
which is the same as \eqref{gammalt0indirect_fla} but with the change $F_{f1}^*F_{f2} \leftrightarrow F_{f2}^*F_{f1}$ in the mixing vertices.
The flavoured indirect CP asymmetry at $T=0$ for a Majorana neutrino of type~1 then is\footnote{
A more compact expression follows from 
${\rm{Im}}\left[(F^*_1 F_2)  (F^*_{f1} F_{f2})\right] + {\rm{Im}}\left[(F_1 F^*_2)  (F^*_{f1} F_{f2})\right] = 
2 \,{\rm{Re}}\left[(F^*_1 F_2) \right] \, {\rm{Im}}\left[(F^*_{f1} F_{f2})\right]$.}
\begin{eqnarray}
\epsilon_{f1,{\rm indirect}}^{T=0}=
&-& \frac{{\rm{Im}}\left[(F^*_1 F_2)  (F^*_{f1} F_{f2})\right] }{16 \pi |F_{1}|^2}  \frac{M \, \Delta}{\Delta^2 + (\Gamma_{22}^{T=0}-\Gamma_{11}^{T=0})^2/4}
\nonumber \\ 
&-& \frac{{\rm{Im}}\left[(F_1 F^*_2)  (F^*_{f1} F_{f2})\right] }{16 \pi |F_{1}|^2}  \frac{M \, \Delta}{\Delta^2 + (\Gamma_{22}^{T=0}-\Gamma_{11}^{T=0})^2/4}.
\label{indirect1T0_fla}
\end{eqnarray}
The first line, if summed over all flavours, gives back \eqref{indirect1T0}.
The second line is specific of the flavoured CP asymmetry and would vanish if summed over all flavours, 
indeed, $\sum_f {\rm{Im}}\left[ (F_1 F^*_2) (F^*_{f1} F_{f2})\right] = {\rm{Im}}\left[ |(F_1 F^*_2)|^2 \right]=0$. 
A similar calculation leads to the expression for the flavoured indirect CP asymmetry at $T=0$ for a Majorana neutrino of type~2, 
which follows from \eqref{indirect1T0_fla} after the changes $F_1 \leftrightarrow F_2$ and $\Delta \to -\Delta$:
\begin{eqnarray}
\epsilon_{f2,{\rm indirect}}^{T=0}=
&-& \frac{{\rm{Im}}\left[(F^*_1 F_2)  (F^*_{f1} F_{f2})\right] }{16 \pi |F_{2}|^2}  \frac{M \, \Delta}{\Delta^2 + (\Gamma_{22}^{T=0}-\Gamma_{11}^{T=0})^2/4}
\nonumber \\ 
&-& \frac{{\rm{Im}}\left[(F_1 F^*_2)  (F^*_{f1} F_{f2})\right] }{16 \pi |F_{2}|^2}  \frac{M \, \Delta}{\Delta^2 + (\Gamma_{22}^{T=0}-\Gamma_{11}^{T=0})^2/4}.
\label{indirect2T0_fla}
\end{eqnarray}
The expressions for $\epsilon_{f1,{\rm indirect}}^{T=0}$ and $\epsilon_{f2,{\rm indirect}}^{T=0}$ agree with those that can be found in~\cite{Fong:2013wr} 
when taking the nearly degenerate limit and resumming the widths of both types of neutrino in the heavy-neutrino propagators. 
\subsection{CP asymmetries at finite temperature}
We conclude by computing the flavoured CP asymmetries at finite temperature. 
Concerning the direct asymmetry, we may identify two type of contributions. 
First, there are contributions coming from the same diagrams considered for the unflavoured case. 
These diagrams contribute also to the flavoured CP asymmetry if the final lepton (or antilepton) flavour is resolved.
This amounts at replacing 
\begin{equation}
{\rm{Im}}\left[ (F_1^* F_2)^2\right] \to {\rm{Im}}\left[ (F^{*}_{1}F_{2})(F^{*}_{f1}F_{f2})\right]   \, ,
\label{replacement}
\end{equation}
in the expressions of the Feynman diagrams given in sections~\ref{appHiggs} and~\ref{appgauge} of appendix~\ref{appC:CPdegematch}.

A second type of contributions comes from diagrams involving only lepton (or antilepton) lines.
They would potentially give rise to a CP asymmetry that is proportional to ${\rm{Im}}\left[(F_1F_2^{*})(F^*_{f1}F_{f2})\right]$ 
and that would vanish in the unflavoured case. 
We have examined these diagrams in appendix~\ref{appC:CPdegematch}, section~\ref{appflaCPdege}, and found that they do not contribute.
Hence, the complete contribution to the matching coefficients ${\rm{Im}} \, a^{\ell}_{II}$ and ${\rm{Im}} \, a^{\bar{\ell}}_{II}$ 
from cuts selecting a lepton or an antilepton of flavour $f$ comes only from the diagrams discussed in the previous paragraph 
and can be read off equations \eqref{match1} and \eqref{match2} by simply performing the replacement~\eqref{replacement}. 

As discussed in section~\ref{sec:direct2}, the Majorana neutrino of type $2$, if heavier than the Majorana neutrino 
of type $1$, has an additional source of CP asymmetry whose ultimate origin is the kinematically allowed transition 
$\nu_{R,2} \to \nu_{R,1} + $ Higgs boson. This asymmetry is described in the EFT by the diagrams shown in figure~\ref{fig:DeltaEFT}.
The only difference with the unflavoured case is that we now require for the cut to select a lepton (or antilepton) with a specific flavour $f$. 
Hence the cut stands for $-3(F_{fJ} F^*_{fI})\lambda/(8\pi M)$ (or $-3(F_{fI} F^{*}_{fJ})\lambda/(8\pi M)$ in the antileptonic case), 
where $I$ is the type of outgoing and $J$ the type of incoming neutrino.
Going through the same derivation as in section~\ref{sec:direct2}, we find
\begin{equation}
\Delta\Gamma_{f2,{\rm direct}}^{\rm mixing} = 
\frac{{\rm{Im}}\left[ (F^{*}_1 F_2) (F^{*}_{f1} F_{f2})\right] + {\rm{Im}}\left[ (F_1 F^{*}_2) (F^{*}_{f1} F_{f2})\right]}{16 \pi^2} \lambda \frac{T^2\Delta}{M^2}.
\label{gammaDelta_flavour}
\end{equation}
The quantity $\Delta\Gamma_{f2,{\rm direct}}^{\rm mixing}$ is the equivalent of $\Delta\Gamma_{2,{\rm direct}}^{\rm mixing}$ in the flavoured case. 
It reduces to $\Delta\Gamma_{2,{\rm direct}}^{\rm mixing}$, given in \eqref{gammaDelta}, when summed over the flavours $f$.

Rewriting the thermal contributions to the direct CP asymmetry given in \eqref{CPnu1} and \eqref{CPnu2} 
for the flavoured case through \eqref{replacement} and adding to the CP asymmetry of the Majorana neutrino of type 2 
the contribution in \eqref{gammaDelta_flavour} proportional to ${\rm{Im}}\left[ (F_1 F^{*}_2) (F^{*}_{f1} F_{f2})\right]$  
gives at order $T^2/M^2$ and at order $\Delta/M$ 
\begin{eqnarray}
\epsilon^{T}_{f1,{\rm direct}} &=& \frac{{\rm{Im}}\left[  (F^{*}_{1}F_{2})(F^{*}_{f1}F_{f2})\right] }{8 \pi |F_{1}|^2}  \left(  \frac{T}{M} \right)^2 
\left\lbrace   \lambda \left[ 2-\ln 2+\left( 1-3\ln 2 \right) \frac{\Delta}{M}\right]  \right.
\nonumber \\
&& 
\left.
- \frac{g^2}{16}\left[ 2- \ln2 +\left( 3 - 5 \ln 2\right) \frac{\Delta}{M}  \right]  
- \frac{g'^2}{48}\left[ 4- \ln2 +\left( 1 - 5 \ln 2\right) \frac{\Delta}{M}  \right]  \right\rbrace,
\label{CPnu1_flavour}
\nonumber 
\\
\end{eqnarray}
and 
\begin{eqnarray}
\epsilon^{T}_{f2,{\rm direct}} &=& -\frac{{\rm{Im}}\left[ (F^{*}_{1}F_{2})(F^{*}_{f1}F_{f2}) \right] }{8 \pi |F_{2}|^2}  \left(  \frac{T}{M} \right)^2 
\left\lbrace   \lambda \left[ 2-\ln 2-\left( 9 - 5\ln 2 \right) \frac{\Delta}{M}\right]  \right.
\nonumber \\
&& 
\left.
- \frac{g^2}{16}\left[ 2- \ln2 - 7 \left( 1 - \ln 2\right) \frac{\Delta}{M}  \right]  
- \frac{g'^2}{48}\left[ 4- \ln2 -\left( 9 - 7 \ln 2\right) \frac{\Delta}{M}  \right]  \right\rbrace
\nonumber\\
&& + \frac{{\rm{Im}}\left[ (F_{1}F^{*}_{2})(F^{*}_{f1}F_{f2}) \right] }{2 \pi |F_{2}|^2}  \left(  \frac{T}{M} \right)^2\lambda\frac{\Delta}{M} \,.
\label{CPnu2_flavour}
\nonumber 
\\
\end{eqnarray}

Finally, the thermal corrections to the indirect CP asymmetry are easily computed in the EFT. 
The analysis carried out in section~\ref{sec:indirect} is valid also in the flavoured regime. 
The thermal corrections to the indirect CP asymmetry have the same form as \eqref{indirect1T} and \eqref{indirect2T}, 
namely for the two neutrino species
\begin{equation}
\epsilon_{f1,{\rm indirect}}^{T}= -\frac{\epsilon_{f1,{\rm indirect}}^{T=0}}{3} \,\left(|F_2|^2-|F_1|^2\right)\,
\frac{M\Delta}{\Delta^2 + (\Gamma_{22}^{T=0}-\Gamma_{11}^{T=0})^2/4}\,\frac{T^2}{M^2}\,,
\label{indirect1T_fla}
\end{equation}
and
\begin{equation}
\epsilon_{f2,{\rm indirect}}^{T}= -\frac{\epsilon_{f2,{\rm indirect}}^{T=0}}{3} \,\left(|F_2|^2-|F_1|^2\right)\,
\frac{M\Delta}{\Delta^2 + (\Gamma_{22}^{T=0}-\Gamma_{11}^{T=0})^2/4}\,\frac{T^2}{M^2}\,.
\label{indirect2T_fla}
\end{equation}
Note that the first factor in the right-hand side of each asymmetry is the flavoured indirect CP asymmetry at $T=0$
computed in \eqref{indirect1T0_fla} and \eqref{indirect2T0_fla}.

\section{Flavoured CP asymmetry for $ M_1 \ll M_i$}
\label{sec_fla3}
In this section we address the generalization of the CP asymmetries in eqs.~(\ref{CPhiera}) and (\ref{finalRes}) to the flavoured regime. We divide the discussion in two parts: first we study the impact of flavour on the EFT$_1$ introducing dimension-six operators to the Lagrangian (\ref{Lag2_hiera}), second we re-derive the expression for the CP asymmetry at finite temperature in the EFT$_2$. 

\begin{figure}[h!]
\centering
\includegraphics[scale=0.585]{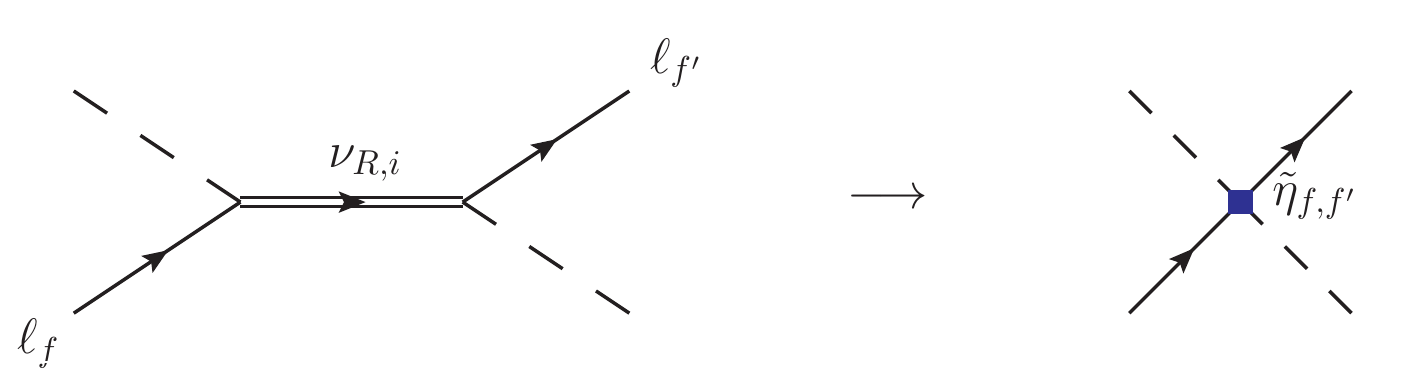}
\caption{\label{fig:eft_ll_hiera} The figure shows the tree-level matching between the fundamental theory and the EFT$_1$. This process contributes only in the flavoured regime. The diagram on the left hand side is the sub-diagram appearing in the self-energy two-loop topology in figure~\ref{fig:flavor_2}. On the right-hand side the four-particle diagram stands for the effective interaction in the EFT$_1$ with one incoming and one outgoing lepton.}
\end{figure}
\subsection{EFT$_1$ and dimension-six operators}
At leading order in $M_1/M_i$ and at zero temperature, the CP asymmetry in the flavoured case can be easily inferred by substituting the Yukawa couplings combination ${\rm{Im}}[(F^*_1F_i)^2]$ with ${\rm{Im}}[(F_{1}^{*}F_{i})(F_{f1}^*F_{fi})]$ in (\ref{CPhiera}). This is in complete analogy with the discussion carried out in the nearly degenerate case in section~\ref{sec_fla2}. Without summing over the final lepton (antilepton) as a product of the heavy neutrino decays, we become sensitive to the flavour $f$ in the CP asymmetry. The calculation is the same as that done in section~\ref{CPhiera_sec2} but the flavour sum. 
\begin{figure}[t!]
\centering
\includegraphics[scale=0.585]{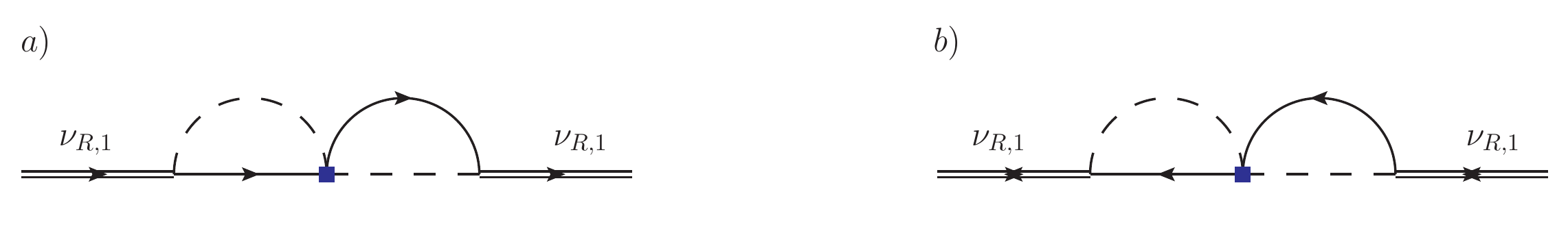}
\caption{\label{fig:twoloop_ll_hiera} Two-loop self-energy diagrams for the lightest neutrino $\nu_{R,1}$ in the EFT$_{1}$. The internal propagator corresponding to the heavier neutrino states is shrunk into a point, accounting for the effective vertices described in figure~\ref{fig:eft_ll_hiera}.}
\end{figure}
We study now to the additional diagram for heavy neutrino decays at one-loop shown in figure~\ref{fig:flavor_1} and responsible for the two-loop self energies in figure~\ref{fig:flavor_2}. They induce the Yukawa couplings combination ${\rm{Im}}[(F_{1}F_{i}^{*})(F_{f1}^*F_{fi})]$, that vanishes in the unflavoured case. This is the reason why we did not considered the corresponding Higgs-lepton scatterings in matching the fundamental theory in (\ref{lepto_8}) with the EFT$_1$ in (\ref{Lag2_hiera}).  Here we extend the matching by including those contributions. The tree-level matching is shown in figure~\ref{fig:eft_ll_hiera}, where one may understand the diagram on the left-hand side of the matching equation to be s-channel Higgs-lepton scattering (subdiagram of the two-loop diagrams in figure~\ref{fig:flavor_2}). They involve lepton-number conserving scatterings, $\ell\phi \to \ell \phi$ and its complex conjugate process $\bar{\ell}\phi^\dagger \to \bar{\ell} \phi^\dagger$.  For details on the matching we refer to appendix \ref{appD:CPhieramatch}. The main difference with the scatterings shown in figure~\ref{fig:eft1_hiera} lies on the combination of the chiral projectors  that select the internal heavy neutrino momentum instead of its mass
\be
P_R \frac{i(\slashed{p}+M_i)}{p^2-M_i^2+i\eta} P_L \to P_R \frac{i \slashed{p}}{p^2-M_i^2+i\eta} \, .
\ee
Then, by expanding the heavy neutrino propagator in $p \ll M_i$, where $p$ is the sum of the incoming lepton and Higgs momenta, we obtain dimension-six operators to be added to the EFT$_1$ Lagrangian. At order $1/M_i^2$ it reads
\bea 
\mathcal{L}_{{\rm{EFT}_1}}=\mathcal{L}_{\hbox{\tiny SM}} 
&+& \frac{1}{2} \,\bar{\psi}_{1} \,i \slashed{\partial}  \, \psi_{1}  - \frac{M_1}{2} \,\bar{\psi}_{1}\psi_{1}  - F_{f1}\,\bar{L}_{f} \tilde{\phi} P_{R}\psi_{1}  - F^{*}_{f1}\,\bar{\psi}_{1}P_{L} \tilde{\phi}^{\dagger}  L_{f}
\nonumber 
\\
&+& \left( \frac{\eta_{ff'}^i}{M_i} \bar{L}_f \tilde{\phi} \, C P_R \,  \tilde{\phi}^T  \bar{L}^T_{f'} + h. c.\right) + \frac{\tilde{\eta}^i_{ff'}}{M^2_i} \bar{L}_f  \tilde{\phi}  P_R \, i \slashed{\partial} (\tilde{\phi}^\dagger L_{f'})+  \cdots \, ,
\nn
\\
\label{Lag2_fla}
\eea
where 
\be 
\tilde{\eta}^i_{f,f'} =  F_{f,i}F^*_{f',i} \,  ,
\label{new_math_dim_six}
\ee
and the dots stand for higher order operators further suppressed in the large scale $M_i$. The additional vertex, induced by the dimension-six operators in (\ref{Lag2_fla}), leads to the two-loop self-energy diagrams shown in figure~\ref{fig:twoloop_ll_hiera}. They are equivalent to the diagrams in figure~\ref{fig:flavor_2} when the heavy neutrino states with masses $M_i \gg M_1$ are integrated out from the theory. Following the derivation carried out in section~\ref{CPhiera_sec2} one obtains for the in-vacuum CP asymmetry in the flavoured regime
\begin{equation}
\epsilon^{T=0}_{f1}= -\frac{3}{16 \pi} \frac{M_1}{M_i} \frac{{\rm{Im}} \left[ (F_1^* F_i)(F_{f1}^* F_{fi}) \right] }{|F_1|^2} - \frac{1}{8 \pi} \left( \frac{M_1}{M_i}\right)^2 \frac{{\rm{Im}} \left[ (F_1 F_i^*)(F_{f1}^* F_{fi}) \right] }{|F_1|^2}   \, ,
\label{CPhiera_fla}
\end{equation}
that agrees with \cite{Fong:2013wr} when expanding in powers of $M_1/M_i$.
The effect of the two-loop self energy diagrams in~\ref{fig:twoloop_ll_hiera} is suppressed by one additional power of $M_1/M_i$ with respect to those given in figure~\ref{fig:eft1_hiera}. This totally reflects the counting in the EFT$_1$: the vertices relevant only in the flavoured case are induced by dimension-six operators and then suppressed by one power more in the large scale $M_i$. 

\subsection{EFT$_2$ and flavoured CP asymmetries}
As regards the CP asymmetry at finite temperature we shall keep only the vertices induced by the dimension-five operators in (\ref{Lag2_fla}), neglecting the effects at order $(M_1/M_i)^2$ and proportional to ${\rm{Im}} [ (F_1 F_i^*)(F_{f1}^* F_{fi}) ]$. Then the derivation of the CP asymmetry at finite temperature is straightforward in the hierarchical case. We only have to select a lepton flavour $f$ when cutting through lepton and antilepton lines in the derivation of the matching coefficients in appendix~\ref{appD:CPhieramatch}. In particular we can perform the substitution  
\be 
{\rm{Im}}[(F^*_1F_i)^2] \to {\rm{Im}}[(F_{1}^{*}F_{i})(F_{f1}^*F_{fi})] 
\label{sub_hiera_fla}
\ee
in eqs.~(\ref{match_a})-(\ref{match_cLbar}), and hence we obtain for the difference between the leptonic and antileptonic thermal widths the following expression
\begin{eqnarray}
&&\Gamma^{\ell,T}-\Gamma^{\bar{\ell},T}
\nn
\\
&&\hspace{0.5 cm}=\frac{1}{64 \pi^2} \frac{M^2_1}{M_i}  {\rm{Im}}\left[ \left( F^{*}_1 F_i\right)(F^{*}_{f1} F_{fi})\right] \left[   \left( 4 \lambda - \frac{2g^2+g'^2}{8} \right)  \frac{T^2}{M^2_1} - |\lambda_t|^2 \frac{7 \pi^2}{ 20} \left( \frac{T}{M_1}\right)^4   \right]  \, .
\label{difftherm_hiera_fla}
\nn
\\
\end{eqnarray}
Then the $T=0$ difference between the leptonic and antileptonic thermal widths in the flavoured case reads, 
\begin{equation}
\epsilon^{T=0}_1=\frac{\Gamma^{\ell,T=0}-\Gamma^{\bar{\ell},T=0}}{\Gamma^{\ell,T=0}+\Gamma^{\bar{\ell},T=0}}= -\frac{3}{16 \pi} \frac{M_1}{M_i} \frac{{\rm{Im}} \left[ (F_1^* F_i)(F^{*}_{f1} F_{fi}) \right] }{|F_1|^2} \, .
\label{CPhiera_fla}
\end{equation}
and combining (\ref{difftherm_hiera_fla}), (\ref{CPhiera_fla}) and the total width in~(\ref{widthfour_hiera}), we obtain for the flavoured CP asymmetry 
\begin{equation}
\epsilon_1^{T}=-\frac{3}{16 \pi} \frac{ {\rm{Im}}\left[(F_1^* F_i)(F^{*}_{f1} F_{fi}) \right]}{|F_1|^2} \frac{M_1}{M_i} \left[   \left(  -\frac{5}{3} \lambda + \frac{2g^2+g'^2}{12} \right) \left( \frac{T}{M_1}\right)^2  +\frac{7\pi^2}{20} |\lambda_t|^2 \left( \frac{T}{M_1}\right)^4   \right] \, .
\label{finalRes_fla}
\end{equation}
Finally one can derive the flavoured version of eq.~(\ref{finRes2_hiera}) by following the above mentioned procedure. The result for the momentum dependent CP asymmetry reads
\begin{equation}
\epsilon_{1,{\hbox{\tiny mom.\,dep.}}}^{T}=-\frac{3}{16 \pi} \frac{ {\rm{Im}}\left[ (F_1^* F_i)(F^{*}_{f1} F_{fi}) \right]}{|F_1|^2} \frac{M_1}{M_i} \left[   \left(  \frac{5}{6} \lambda - \frac{2g^2+g'^2}{24} \right) \frac{\bm{k}^2 \,T^2 }{M^4_1} \right] \, .
\label{finRes2_hiera_fla}
\end{equation}

%% file: conclusions.tex
In this thesis we have discussed the construction of an EFT 
for non-relativistic Majorana fermions and we have shown 
how to use it to calculate observables in a thermal medium. 
The EFT presented here is similar to HQET but keeps track 
of the Majorana nature of the fermion by describing both the 
particle and the antiparticle with the same field.

Although the approach is quite general, we apply it to a particle physics model that comprises some species of right-handed neutrinos coupled to the SM Higgs boson and lepton doublets via Yukawa interactions (see the Lagrangian \eqref{lepto_8}). Such model provides the fundamental ingredients to achieve a successful baryogenesis via leptogenesis in the early universe. Here heavy neutrinos with a large Majorana masses are at the origin of the matter-antimatter asymmetry. Interactions between heavy neutrinos and SM particles occur in a thermal medium. We assume that the right-handed neutrino mass and the temperature of the plasma satisfy the condition $M \gg T$, where the temperature is still larger than the electroweak scale. In this regime the heavy neutrinos are non-relativistic objects and it is conceivable that the lepton asymmetry is effectively generated when the temperature drops below the heavy neutrino mass. We addressed the calculation of observables related to leptogenesis: the right-handed neutrino production rate and the CP asymmetries generated in heavy Majorana neutrino decays  in a heat bath.

As regards the former observable, our result given in (\ref{eq20}) agrees with earlier findings~\cite{Salvio:2011sf,Laine:2011pq}; 
the derivation however appears simpler.  
At our accuracy, i.e. first order in the SM couplings and order $T^4/M^3$, 
the two-loop thermal field theory computation necessary to describe the process 
in the full theory splits into two one-loop computations in the EFT.
The first one-loop computation is required to match the full theory with the EFT.
This can be done setting the temperature to zero, so it amounts at the calculation 
of typical in-vacuum matrix elements. The second one-loop computation 
is required to calculate the thermal corrections in the EFT.
At the accuracy of this work, only tadpole diagrams are involved.
These may be easily computed with the real-time formalism or with other methods. 
The use of the real-time formalism is particularly convenient with heavy particles: 
since they do not thermalize, heavy particles and particles coupled to them 
are not affected by the doubling of degrees of freedom typical of the formalism. 
The situation is again analogous to the one faced when studying heavy quarks in a thermal bath~\cite{Brambilla:2008cx}.

The total width of the Majorana neutrino, $\Gamma = \Gamma^{T=0} + \Gamma^T$, 
is organized as a double expansion in the SM couplings and in $T/M$.
At the present accuracy, the double expansion reflects the hierarchy of energy scales $M \gg T$
and corresponds in the EFT to the two steps of the computation: matching and thermal loops. 
The SM couplings entering in the Wilson coefficients of the EFT 
are computed at the heavy neutrino mass scale, $M$, and one can evolve them down to the scale $T$ to make the theory homogeneous.
Whether terms in one expansion are more relevant than terms in the other 
depends on the considered temperature regime. A temperature close to the 
Majorana neutrino mass makes terms in the $T/M$ expansion more relevant, 
although a temperature too close to it may spoil the convergence and 
signal a breakdown of the non-relativistic treatment.

Besides simplifying existing results the EFT approach provides a useful framework to address even more involved observables. In particular, we have taken a step forward a systematic improvement of the CP asymmetry at NLO in heavy neutrino decays into leptons and antileptons. This is one of the key ingredients entering the rate equations for leptogenesis. To the best of our knowledge thermal corrections to the CP asymmetry at first order in the SM couplings are unknown. We believe that having any information about them, even just for the case $T \ll M$, should be seen as an advancement, in the same way as it has been for the thermal corrections to the production rate
computed in~\cite{Salvio:2011sf,Laine:2011pq}.

The EFT allows to address different configurations of the heavy-neutrino mass patterns. We have computed the leading thermal corrections to the direct and indirect CP asymmetries 
in an extension of the SM that includes two 
generations of heavy Majorana neutrinos with nearly degenerate masses $M$ and $M+\Delta$. In order to describe a condition that occurred in the early universe, 
we have assumed the SM particles to form a plasma whose temperature $T$ is larger 
than the electroweak scale but smaller than $M$. The main original results are eqs.~\eqref{CPnu1} and \eqref{CPnu2} for the thermal corrections to the direct CP asymmetry, 
and eqs.~\eqref{indirect1T} and \eqref{indirect2T} for the thermal corrections to the indirect CP asymmetry. 
Thermal corrections to the CP asymmetry arise at order $F^4$ in the Yukawa couplings.
As regards the direct CP asymmetry corrections are further suppressed by one SM coupling.
Hence the calculation of the thermal effects to the direct CP asymmetry is a three-loop 
calculation in the fundamental theory \eqref{lepto_8}. We have performed the calculation in 
the EFT framework introduced in~\cite{Biondini:2013xua}, which is valid for $T\ll M$. 
The three-loop thermal calculation of the original theory splits into the calculation of the 
imaginary parts of two-loop diagrams that match the Wilson coefficients of the EFT \eqref{eq:efflag_CPdege}, 
a calculation that can be performed in vacuum, and the calculation of a thermal one-loop diagram in the EFT (see figure~\ref{fig:tadpoles_CPdege}).
Therefore, in its range of applicability, the EFT framework provides a significantly simpler method of calculation.

The same formalism may prove to be a useful tool to calculate the CP asymmetry 
also in other arrangements of the heavy-neutrino masses, such as a hierarchically ordered 
neutrino mass spectrum, where the direct and  the indirect CP asymmetries are of comparable size (see chapter~\ref{chap:CPhiera}).
In this case one heavy neutrino is much lighter than the other neutrino species. The strategy to obtain thermal corrections for the CP asymmetry follows closely the one carried out for the nearly degenerate case. The hierarchy of scales, $M_i \gg M_1 \gg T \gg M_W$, actually allows for constructing an EFT where the heavier neutrino states are integrated out from the theory (EFT$_1$). This is the starting point to build the subsequent EFT where only non-relativistic excitations of the lightest heavy neutrino are dynamical (EFT$_2$). We obtained the thermal corrections at leading order in the expansion $1/M_i$ and fully at order $(T/M_1)^2$. 

At relative order $(T/M)^2$ only the Higgs self-coupling, $\lambda$, 
and the SU(2)$_L\times$U(1)$_Y$ gauge couplings, $g$ and $g'$, enter the expression of the CP asymmetry. 
Higher-order operators in the $1/M$ expansion have not been considered in the nearly degenerate case. 
However, higher-order operators, most importantly the dimension-seven operators described in chapter~\ref{chap:part_prod}, 
may contribute to the CP asymmetry as well. The power counting of the EFT
shows that they can induce thermal corrections that scale like $g_{\hbox{\tiny SM}}(T/M)^4$, 
where $g_{\hbox{\tiny SM}}$ is understood as either $\lambda$, $(3g^2+g'^2)$ or the top Yukawa coupling $|\lambda_t|^2$. Even though these corrections are further suppressed in the expansion in $T/M$, 
the particular values of the SM couplings at high energies can make 
$g_{\hbox{\tiny SM}}(T/M)^4$ corrections numerically comparable with or larger 
than those calculated at order $(T/M)^2$. 
As a reference, at a scale of $10^4$~TeV the Higgs self coupling is about $\lambda\approx 0.02$, 
the top-Yukawa coupling is about $|\lambda_t|^2\approx 0.4$ and $(3g^2+g'^2) \approx 1.2$,
whereas  at a scale of $1$~TeV $\lambda\approx 0.1$, $|\lambda_t|^2\approx 0.7$ and $(3g^2+g'^2) \approx 1.6$~\cite{Rose15,Buttazzo:2013uya}. 
To shape better this issue the effect of, at least, some higher-order operators should be calculated. Indeed we studied the thermal corrections comprising the top-Yukawa coupling in the hierarchical case. The complete set of corrections at order $(T/M)^4$ is in the reach of the proposed EFT approach and possibly subject of future investigations.    

A quantitative study of leptogenesis requires flavour to be included in the formalism. Indeed the unflavoured approximation is valid only at very high temperatures, typically $T > 10^{12}$ GeV. Indeed in the flavoured regime the CP asymmetry stored in a single flavour component is found to be relevant one for solving the Boltzmann equations for leptogenesis. 
The impact of flavour on our approach was discussed, both for the $T=0$ and finite temperature CP asymmetries, and for the two different heavy neutrino mass patterns. The results are collected in chapter~\ref{chap:CPfla}.    

The expansion $T/M$ is adopted in the derivation of the results presented in this thesis. We discussed in detail such topic in the case of the right-handed neutrino production rate. The issue on the convergence of such expansion for not too small values of $T/M$ could be also for the CP asymmetry. However, to the best of our knowledge, the expression valid for $T \sim M$ at leading order in the SM couplings does not exist for the CP asymmetry at variance with the neutrino production rate.  The proposal presented in section~\ref{expansion_partprod}, for the neutrino production rate, may be applied for the CP asymmetry as well:  include exponentially suppressed terms, $e^{-M/T}$, in the zeroth order term in the SM couplings. Such result actually exists in the literature and it has been derived in the framework of the Kadanoff-Baym equations \cite{Garny:2009qn,Garny:2010nj,Garny:2009rv}.

Another question is how the corrections in $T/M$ compare with the yet unknown radiative corrections to the CP asymmetry at zero temperature. 
First, we note that for the indirect CP asymmetry, which is the dominant part of the asymmetry in particular 
for the resonant case or close to it, the computed $(T/M)^2$ corrections are not suppressed by the SM couplings. 
Hence they are likely to be larger than or of the same size as radiative corrections for a wide range of temperatures.
Second, we observe that thermal corrections to the direct CP asymmetry, which are suppressed in the SM couplings,  
are indeed of relative size $\lambda (T/M)^2$ and $(3g^2+g'^2)(T/M)^2$ 
(cf.~with \eqref{CPnu1} and \eqref{CPnu2}). 
These should be compared with radiative corrections of possible relative size $\lambda/\pi^2$, $|\lambda_t|^2/\pi^2$ or $(3g^2+g'^2)/\pi^2$
(cf.~with the radiative corrections to the production rate in~\cite{Salvio:2011sf}).
The factor $1/\pi^2$ is typical of radiative corrections, but absent in thermal corrections.
The two are of comparable size for $T/M \sim 1/\pi$, which is inside the range of convergence of the expansion in $T/M$.
Clearly radiative corrections are a missing ingredient for a complete quantitative evaluation of the CP asymmetry.
Following the above discussion, their evaluation seems most needed when the CP asymmetry 
is dominated by direct contributions and at lower temperatures.  

The EFTs \eqref{eq:efflag_CPdege} and \eqref{Lag3_hiera} are also the natural starting point to establish the rate equations 
for the time evolution of the particle densities in the regime where the Majorana neutrinos are non-relativistic. The way to proceed would be similar to that developed recently using CTP formalism~\cite{Garny:2010nj,Anisimov:2010dk,Kiessig:2011fw,Garbrecht:2011aw}: one can derive the evolution equations for the heavy-neutrino and lepton-number expectation values at finite temperature from Green's functions. Over exploiting the EFT approach, one can start from the beginning with the suitable degrees of freedom which are dynamical at the temperature scale, whereas effects of larger scales are already encoded in the Wilson coefficients of the EFT.    
A first study of the non-relativistic approximation for the rate equations can be found in~\cite{Bodeker:2013qaa}, where our result for the CP asymmetry in the hierarchical case may be included in a rather straightforward way in the numerical calculations. 

Finally the effective field theory presented here is suitable to be used 
for a variety of different models involving non-relativistic Majorana fermions, such as possible applications to dark matter production at finite temperature in the early universe.

%% file: matchwidth.tex
In this appendix, we compute the Wilson coefficients \eqref{coa}-\eqref{cod34}.
They are obtained by matching matrix elements calculated in the fundamental theory 
(\ref{eq3_partprod}) with matrix elements calculated in the EFT \eqref{eq10}.
The fundamental theory contains the SM with unbroken gauge symmetries, whose Lagrangian reads
\begin{eqnarray}
\mathcal{L}_{\hbox{\tiny {SM}}} &=&
\bar{L}_{f} P_R\, i \slashed{D} \, L_{f} + \bar{Q}P_R\, i\slashed{D} \, Q  +\bar{t}P_L\, i\slashed{D} \, t -\frac{1}{4}W_{\mu\nu}^aW^{a\,\mu\nu} -\frac{1}{4}F_{\mu\nu}F^{\mu\nu}
\nonumber\\
&& + \left( D_{\mu} \phi \right)^{\dagger}\left( D^{\mu} \phi \right)  - \lambda \left( \phi^{\dagger}\phi \right)^2 
- \lambda_{t}    \, \bar{Q} \, \tilde{\phi} \, P_{R} t
- \lambda^{*}_{t} \, \bar{t} P_{L} \, \tilde{\phi}^{\dagger} \, Q  + \dots \,.
\label{SMlag}
\end{eqnarray}
The dots stand for terms that are irrelevant for our calculation, e.g. those involving light quarks or right-handed leptons. 
The covariant derivative is given by 
\begin{equation}
D_{\mu}=\partial_{\mu} -igA^{a}_{\mu} \tau^{a} -ig'YB_{\mu} \, ,
\label{SMCov}
\end{equation}
where $\tau^{a}$ are the SU(2) generators and $Y$ is the hypercharge ($Y=1/2$ for the Higgs, 
$Y=-1/2$ for left-handed leptons). The fields $L_{f}$ are the SU(2) lepton doublets with flavor $f$, 
$Q^T=(t,b)$ is the heavy-quark SU(2) doublet, $t$ is the SU(2)-singlet top quark field for which there is no coupling with the SU(2) gauge boson in eq.~\eqref{SMCov}, $\phi$ the Higgs doublet, 
$A^{a}_{\mu}$ are the SU(2) gauge fields, $B_{\mu}$ the U(1) gauge fields and 
$W^{a\,\mu\nu}$, $F_{\mu\nu}$ the corresponding field strength tensors.  
The couplings $g$, $g'$, $\lambda$ and $\lambda_t$ are the SU(2) and U(1) gauge 
couplings, the four-Higgs coupling and the top Yukawa coupling respectively.

Many one-loop diagrams are needed for the matching. We adopt in all the calculations dimensional regularization.
Therefore loop diagrams in the EFT vanish in dimensional 
regularization because scaleless.
The Wilson coefficients that we need to compute are those appearing in \eqref{Wa} and \eqref{Wb}.
We compute them by matching four-field matrix elements involving 
two Majorana fields and either two Higgs, two lepton, two quark or two gauge 
fields. We will discuss the matching of these matrix elements one by one in the 
rest of the appendix. Before, we add few general considerations.

We perform the matching in the reference frame $v^{\mu}=(1,\bm{0}\,)$, 
where we assume the plasma to be at rest.
The leading momentum dependent operator \eqref{Wmomdep} is fixed by symmetry and does not need to be calculated. 
Since we are interested in the imaginary parts of the Wilson coefficients,
we evaluate the imaginary parts of $-i {\mathcal{D}}$, 
where ${\mathcal{D}}$ are generic Feynman diagrams,  
by taking the Majorana neutrino mass at $M+i\eta$.
We choose the incoming and outgoing SM particles to carry the same momentum $q^\mu$.
Because $q^{\mu}$ is much smaller than $M$, diagrams in the fundamental theory are expanded 
in powers of $q^{\mu}$ that eventually matches the operator expansion in the EFT.

The fundamental theory \eqref{eq3_partprod} is SU(2)$\times$U(1) gauge invariant, 
so are all operators in the EFT. Hence, the Wilson coefficients are gauge independent.
As a practical choice, however, we will present results for single diagrams in Landau gauge.
This is a convenient gauge in the presence of momentum dependent vertices like those between the Higgs and 
the gauge bosons. We have explicitly checked gauge invariance by computing the Wilson coefficients 
also in Feynman gauge. 

\begin{figure}[htb]
\centering
\includegraphics[scale=0.55]{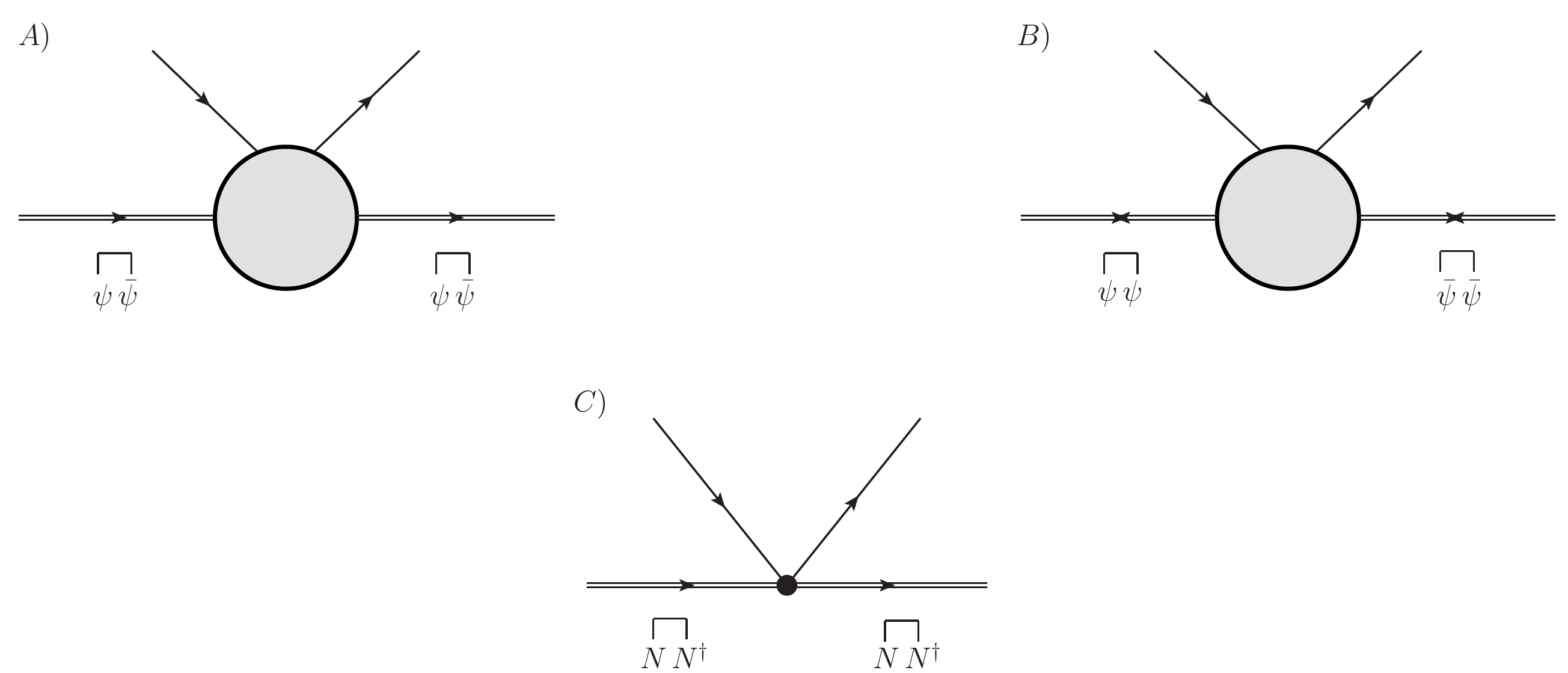}
\caption{\label{Fig0A} The diagrams represent matrix elements with two 
Majorana neutrino fields and two SM fields in the fundamental theory 
(diagrams $A$ and $B$) and in the EFT (diagram $C$). The
bubbles in $A$ and $B$ denote generic loops. 
The diagrams $A$ and $B$ in the relativistic theory
allow for two possible contractions of the neutrino fields, 
while the diagram $C$ in the non-relativistic EFT allows just for one.}
\end{figure}

When computing matrix elements involving Majorana fermions, one has to keep in 
mind that the relativistic Majorana field $\psi$ may be contracted 
in two possible ways, \eqref{eq5_partprod} and \eqref{eq6_partprod},
as a consequence of the indistinguishability of the particle from the antiparticle. 
A similar observation holds for the field $\bar{\psi}$.
For our calculation, involving matrix elements with two external Majorana neutrinos, 
this implies that in the fundamental theory we have to consider for each diagram 
two possible configurations: each one corresponding to the two possible 
way to contract the Majorana fields $\psi$ and $\bar{\psi}$. See diagrams $A$ and $B$ in figure~\ref{Fig0A}. 
In the non-relativistic EFT, we have only one possible way to contract the Majorana field $N$, 
which is \eqref{effpropagator}. See diagram $C$ in figure~\ref{Fig0A}. 
One has to properly account for this when matching the relativistic matrix elements  
with the ones in the EFT. In our calculation, with the exception of diagrams with external leptons,  
the two possible configurations give the same result as a consequence of 
\begin{equation}
C \gamma^{\mu_1\,T}...\gamma^{\mu_{2n+1}\,T}C=\gamma^{\mu_1}...\gamma^{\mu_{2n+1}} \, ,
\end{equation}  
and because of the insensitivity of the result to the direction of the momentum carried by the Majorana neutrino.

\section{Higgs}
\begin{figure}[htb]
\centering
\includegraphics[scale=0.435]{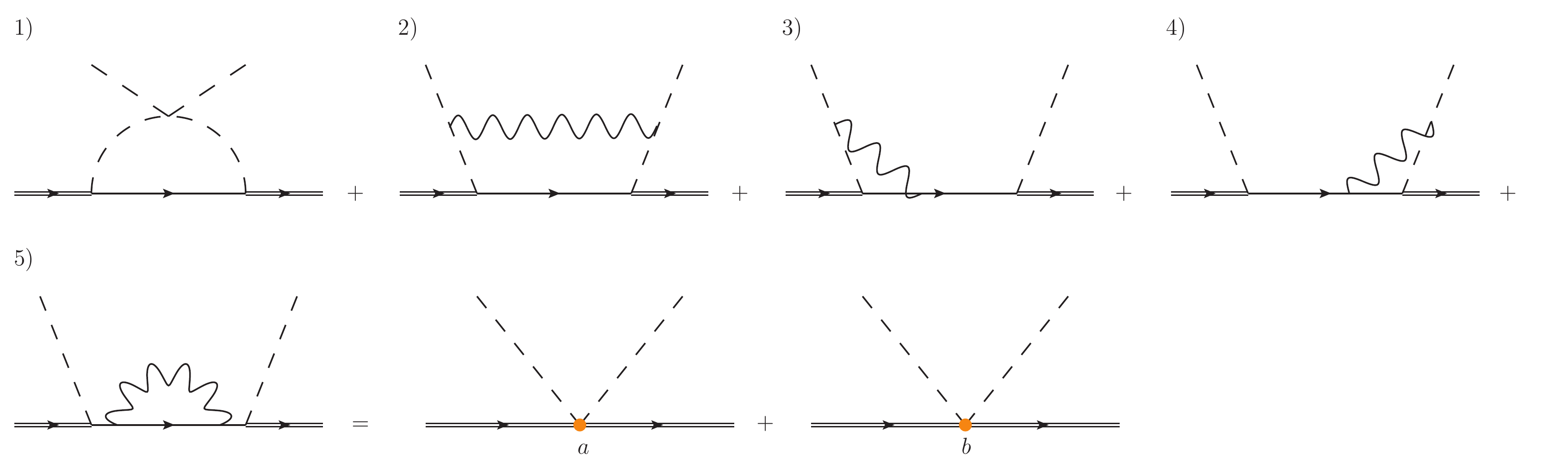}
\caption{\label{Fig1A} Diagrams in the full theory (left-hand side of the equality)
contributing to the Majorana neutrino-Higgs four-field operators in the EFT (right-hand side). 
The solid double lines stand for heavy Majorana neutrinos, the solid single lines 
for leptons, the dashed lines for Higgs particles and the wiggled lines for gauge bosons.}
\end{figure}

In order to determine the Wilson coefficients $a$ and $b$, we compute in the fundamental theory the matrix element
\begin{equation}
-i \left.\int d^{4}x\,e^{i p \cdot x} \int d^{4}y \int d^{4}z\,e^{i q \cdot (y-z)}\, 
\langle \Omega | T(\psi^{\mu}(x) \bar{\psi}^{\nu }(0) \phi_{m}(y) \phi_{n}^{\dagger}(z) )| \Omega \rangle
\right|_{p^\mu =(M + i\eta,\bm{0}\,)},
\label{A1}
\end{equation} 
where $\mu$ and $\nu$ are Lorentz indices, $m$ and $n$ are SU(2) indices and $|\Omega \rangle$ is the ground state 
of the fundamental theory. The matrix element \eqref{A1} describes a $2 \rightarrow 2$ scattering between 
a heavy Majorana neutrino at rest and a Higgs boson carrying momentum $q^\mu$.
In figure~\ref{Fig1A}, we show on the left-hand side of the equality all diagrams 
that in the fundamental theory contribute to the effective vertices shown on its right-hand side. 

In order to compute the imaginary parts of the Wilson coefficients $a$ and $b$, 
we need to consider only the imaginary parts of the diagrams shown in figure~\ref{Fig1A}.
In Landau gauge, the diagrams in the fundamental theory read\footnote{ 
To keep the notation simple, we drop, from now and in the rest of the appendix, 
propagators on external legs, and we label the so-obtained amputated Green's functions 
with the same indices used for the unamputated ones. 
}
\begin{eqnarray}
&& {\rm{Im}}\, (-i \mathcal{D}_{1}) = 
- \frac{3}{8 \pi} \frac{\lambda|F|^{2}}{M}\delta_{mn} \delta^{\mu \nu} + \dots \, ,   
\\
&& {\rm{Im}}\, (-i \mathcal{D}_{2}) =  
- \frac{1}{96 \pi} \frac{(3g^{2}+g'^{\,2})|F|^{2}}{M^{3}}\delta_{mn}\delta^{\mu \nu}    (q_{0})^{2}  + \dots \, ,  
\\
&& {\rm{Im}}\,(-i \mathcal{D}_{3}) + {\rm{Im}}\,(-i \mathcal{D}_{4}) = 
- \frac{7}{48 \pi} \frac{(3g^{2}+g'^{\,2})|F|^{2}}{M^ {3}}\delta_{mn}\delta^{\mu \nu}   (q_{0})^{2} + \dots \, ,   
\\
&& {\rm{Im}}\,(-i \mathcal{D}_5) = 0\, ,    
\label{zero_special_case}
\end{eqnarray}
where the subscripts refer to the diagrams as listed in figure~\ref{Fig1A}.\footnote{
The vanishing of diagram 5 is specific of the Landau gauge.}
The dots stand for terms that are either proportional to $q^\mu/M^2$, or to 
$q_0q_i/M^3$ ($i=1,2,3$) or to $q^2/M^3$; we have not displayed terms that are of order $1/M^4$ or smaller.
Such terms do not contribute to the matching of the operators in \eqref{Wa} and \eqref{Wb}. 
Summing up all contributions we get 
\begin{equation}
- \frac{3}{8 \pi} \frac{\lambda|F|^{2}}{M}\delta_{mn}\delta^{\mu \nu}  
- \frac{5}{32 \pi} \frac{(3g^{2}+g'^{\,2})|F|^{2}}{M^ {3}}\delta_{mn} \delta^{\mu \nu}  (q_{0})^{2}  + \dots \, .
\label{A10}
\end{equation}  

The symmetries of the EFT enforce that the matrix element \eqref{A1} is reproduced by the following expression 
\begin{equation}
\frac{a}{M} \delta_{mn}\delta^{\mu \nu}  +   \frac{b}{M^{3}} \delta_{mn}\delta^{\mu \nu} (q_{0})^{2} + \dots \,,
\label{A3}
\end{equation}
where the dots stand for contributions coming from operators that are not listed in \eqref{Wa} and \eqref{Wb}. 

Matching the imaginary part of (\ref{A3}) with (\ref{A10}) fixes the imaginary parts of $a$ and $b$:
\begin{equation}
{\rm Im}\,a = -\frac{3}{8\pi}|F|^{2}\lambda \, , 
\qquad 
{\rm Im}\,b = - \frac{5}{32 \pi} (3g^{2}+g'^{\,2})|F|^{2} \, .
\end{equation} 
Note that only the first diagram of figure~\ref{Fig1A}
contributes to the effective operator \eqref{Wa}, which provides the leading contribution 
to the Majorana neutrino thermal width. The remaining diagrams contribute to the subleading 
operator $b \; N^{\dagger}N \, (D_0 \phi^{\dagger}) \, (D_0 \phi)/M^3$.

\begin{figure}[htb]
\includegraphics[scale=0.384]{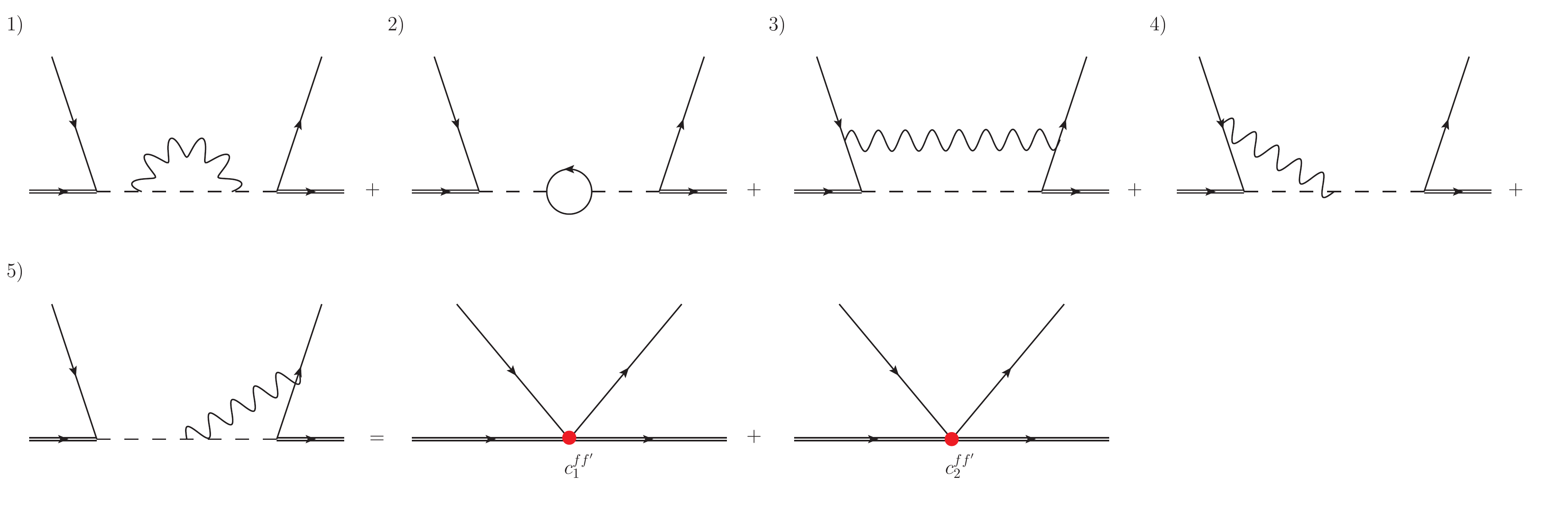}
\caption{\label{Fig2A} 
Diagrams in the full theory (left-hand side of the equality)
contributing to the Majorana neutrino-lepton four-fermion operators in the EFT (right-hand side).
The lines stand for the same particle propagators as in figure~\ref{Fig1A}.
}
\end{figure}

\section{Leptons}
In the fundamental theory, the matrix element 
\begin{equation}
-i \left.\int d^{4}x\,e^{i p \cdot x} \int d^{4}y \int d^{4}z\,e^{i q \cdot (y-z)}\, 
\langle \Omega | T(\psi^{\mu}(x) \bar{L}^{\beta}_{f,m}(z) L^{\alpha}_{f',n}(y) \bar{\psi}^{\nu }(0))  | \Omega \rangle 
\right|_{p^\mu =(M + i\eta,\bm{0}\,)},
\label{B1}
\end{equation}
where $f$ and $f'$ are flavor indices, $\alpha$, $\beta$, $\mu$ and $\nu$ Lorentz indices, and $m$ and $n$ SU(2) indices, 
describes a $2 \rightarrow 2$ scattering between a heavy Majorana neutrino at rest and a lepton carrying momentum $q^\mu$.
The diagrams contributing to the matrix element in the fundamental theory are shown on the left-hand side of the equality 
of figure~\ref{Fig2A}. 
Their imaginary part in Landau gauge gives 
\begin{eqnarray}
{\rm{Im}}\,(-i\mathcal{D}_{1}) &=& 
- \delta_{mn} F_{f'} F^{*}_{f} \left( \frac{3(3g^{2}+g'^{\,2})}{32 \pi M^{3}}\right) 
\left[(P_{L})^{\mu \beta}(P_{R})^{ \alpha \nu} \right.
\nonumber\\
&& \hspace{4.0cm} \left.
+ (C\,P_{R})^{\mu \alpha}(P_{L}\,C)^{\beta \nu}\right] q_{0} 
+\dots\,,
\\
{\rm{Im}}\,(-i\mathcal{D}_{2}) &=&  
\delta_{mn} F_{f'} F^{*}_{f} \left( \frac{3|\lambda_{t}|^2}{8 \pi M^{3}}\right)     
\left[(P_{L})^{\mu \beta}(P_{R})^{ \alpha \nu} \right.
\nonumber\\
&& \hspace{4.0cm} \left.
+ (C\,P_{R})^{\mu \alpha}(P_{L}\,C)^{\beta \nu}\right] q_{0}
+\dots\,,
\\
{\rm{Im}}\,(-i\mathcal{D}_{3}) &=& 
- \delta_{mn} F_{f'} F^{*}_{f} \left( \frac{(3g^{2}+g'^{\,2})}{32 \pi M^{3}} \right) 
\left[(P_{L})^{\mu \beta}(P_{R})^{ \alpha \nu} + (C\,P_{R})^{\mu \alpha}(P_{L}\,C)^{\beta \nu}\right] q_{0}
\nonumber \\
&&+ \delta_{mn} F_{f'} F^{*}_{f}  \left( \frac{(3g^{2}+g'^{\,2})}{384 \pi M^{3}} \right)    
\left[ (P_{L}\,\gamma_{\lambda}\gamma_{\sigma})^{\mu \beta}(\gamma^{\sigma}\gamma^{\lambda}\, P_{R})^{ \alpha \nu} \right.
\nonumber\\
&& \hspace{4.0cm} \left.
+ (C\, P_{R}\,\gamma_{\lambda}\gamma_{\sigma})^{\mu \alpha}(\gamma^{\sigma}\gamma^{\lambda} \,P_{L}\,C)^{\beta\nu} \right]q_{0}  
+\dots\,,
\nn
\\
\phantom{x}
\\
{\rm{Im}}\,(-i\mathcal{D}_{4}) &+& {\rm{Im}}\,(-i\mathcal{D}_{5}) = 
- \delta_{mn} F_{f'} F^{*}_{f} \left( \frac{(3g^{2}+g'^{\,2})}{16 \pi M^{3}}\right) 
\left[(P_{L})^{\mu \beta}(P_{R})^{ \alpha \nu} \right.
\nonumber\\
&& \hspace{4.0cm} \left.
+ (C\,P_{R})^{\mu \alpha}(P_{L}\,C)^{\beta \nu}\right] q_{0} 
+\dots\,,
\label{B2}
\end{eqnarray}
where the subscripts refer to the diagrams as listed in figure~\ref{Fig2A} and the dots stand either for higher-order terms 
in the $1/M$ expansion or for terms of order $1/M^2$ but that do not depend on the momentum $q^\mu$.
Summing up all contributions and comparing with the corresponding expression in the EFT, which is 
\begin{eqnarray}
&& \hspace{-1cm}
\frac{c^{ff'}_{1}}{M^{3}} \delta_{mn} \left[(P_{L})^{\mu \beta}(P_{R})^{ \alpha \nu} + (C\,P_{R})^{\mu \alpha}(P_{L}\,C)^{\beta \nu}\right] q_{0} 
\nonumber\\
&& \hspace{-1cm}
+ \frac{c^{ff'}_{2}}{M^{3}} \delta_{mn} 
\left[ (P_{L}\,\gamma_{\lambda}\gamma_{\sigma})^{\mu \beta}(\gamma^{\sigma}\gamma^{\lambda} \,P_{R})^{ \alpha \nu} 
+ (C\, P_{R}\,\gamma_{\lambda}\gamma_{\sigma})^{\mu \alpha}(\gamma^{\sigma}\gamma^{\lambda} \,P_{L}\,C)^{\beta\nu} \right] q_{0} 
+\dots\,,
\nn
\\
\phantom{x}
\label{B4}
\end{eqnarray}
we obtain
\begin{eqnarray}
&&{\rm Im}\,c^{ff'}_{1} = \frac{3}{8\pi}|\lambda_{t}|^{2} {\rm{Re}}(F_{f'} F^{*}_{f})  - \frac{3}{16 \pi}(3g^{2}+g'^{\,2}) {\rm{Re}}(F_{f'} F^{*}_{f}) \,, 
\\
&&{\rm Im}\,c^{ff'}_{2} =  \frac{1}{384 \pi} (3g^{2}+g'^{\,2}) {\rm{Re}}(F_{f'} F^{*}_{f}) \, .
\label{B5}
\end{eqnarray} 
The dots in \eqref{B4} stand for contributions coming from operators that are not listed in~\eqref{Wb}.

\begin{figure}[htb]
\centering
\includegraphics[scale=0.4]{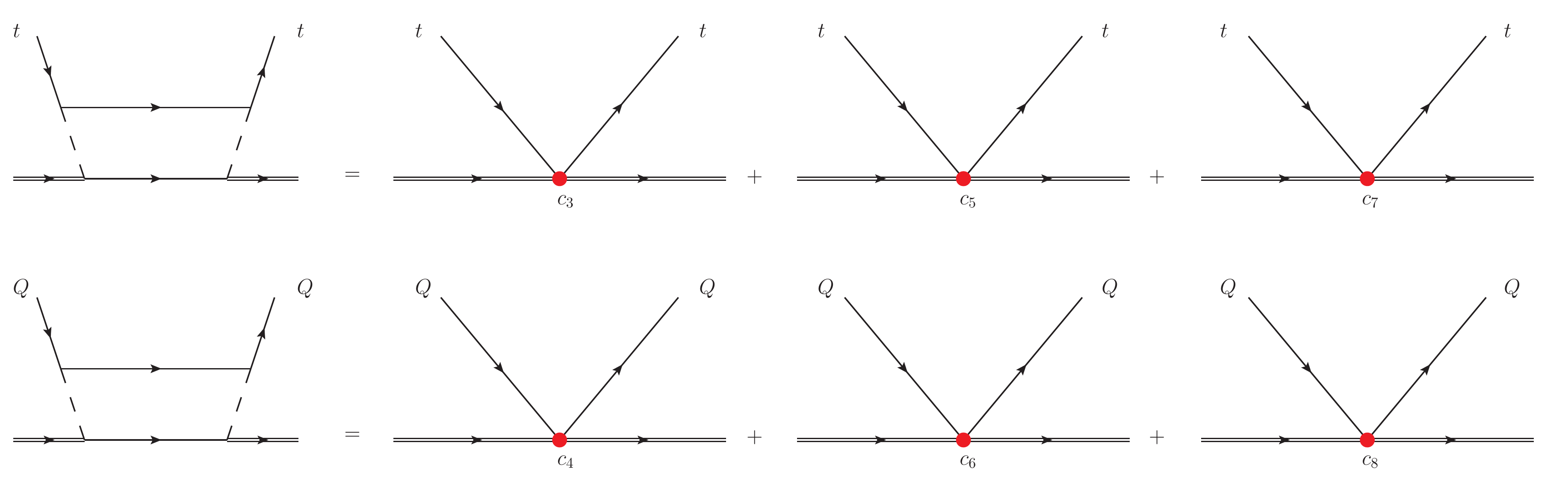}
\caption{\label{Fig3A} In the top panel, the diagram in the full theory (left-hand side)
contributing to the Majorana neutrino-top-quark singlet four-fermion operators in the EFT (right-hand side).
In the bottom panel, the diagram in the full theory (left-hand side)
contributing to the Majorana neutrino-heavy-quark doublet four-fermion operators in the EFT (right-hand side).
The solid single lines marked $t$ stand for top singlets, the solid single lines marked $Q$ 
for heavy-quark doublets, unmarked  solid lines connecting top lines and heavy-quark doublets stand for 
heavy-quark doublets and top singlets respectively. All other lines stand for the same particle propagators as in figure~\ref{Fig1A}.}
\end{figure}

\section{Quarks}
We consider only couplings with top quarks, for $\lambda_{t} \sim 1$ while 
all other Yukawa couplings are negligible.
In the fundamental theory, we compute the two matrix elements
\begin{eqnarray}
&&\hspace{-10mm}
-i \left.\int d^{4}x\,e^{i p \cdot x} \int d^{4}y \int d^{4}z\,e^{i q \cdot (y-z)}\, 
\langle \Omega | T(\psi^{\mu}(x) \bar{\psi}^{\nu }(0) \, t^{\alpha}(y) \bar{t}^{\beta}(z)) | \Omega \rangle 
\right|_{p^\mu =(M + i\eta,\bm{0}\,)}\!,
\nn
\\
\label{C11_zero}\\
&&\hspace{-10mm}
-i \left.\int d^{4}x\,e^{i p \cdot x} \int d^{4}y \int d^{4}z\,e^{i q \cdot (y-z)}\, 
\langle \Omega | T(\psi^{\mu}(x) \bar{\psi}^{\nu }(0) \, Q_{m}^{\alpha}(y) \bar{Q}_{n}^{\beta}(z)) | \Omega \rangle 
\right|_{p^\mu =(M + i\eta,\bm{0}\,)}\!,
\nn
\\
\label{C1_zero}
\end{eqnarray}
describing respectively a $2 \rightarrow 2$ scattering between a heavy Majorana neutrino at rest and a right-handed top quark carrying momentum $q^\mu$, 
and a $2 \rightarrow 2$ scattering between a heavy Majorana neutrino at rest and a left-handed heavy quark carrying momentum $q^\mu$. 
The indices $\alpha$, $\beta$, $\mu$ and $\nu$ are Lorentz indices, whereas $m$ and $n$ are the SU(2) indices 
of the heavy-quark doublet.
The diagrams contributing to the matrix elements in the fundamental theory are shown in figure~\ref{Fig3A}. 
We call $\mathcal{D}_{t}$ the diagram with external top lines and  $\mathcal{D}_{Q}$ the diagram with external 
heavy-quark lines. The imaginary parts of $-i\mathcal{D}_{t}$ and $-i\mathcal{D}_{Q}$ read 
\begin{eqnarray}
&&\hspace{-10mm}
{\rm{Im}}\,(-i\mathcal{D}_{t}) = 
\frac{|F|^{2}|\lambda_{t}|^2}{24 \pi M^{3}}   \delta^{\mu \nu}   \left( P_{L} \gamma^{0} \right)^{\alpha \beta} q_{0} 
\nonumber\\
&& \hspace{10mm}
+ \frac{|F|^{2}|\lambda_{t}|^2}{48 \pi M^{3}}\left[   
\left(\gamma^5\gamma^i\right)^{\mu \nu}   \left( P_{L} \gamma^{0} \right)^{\alpha \beta} q_{i}
+\left(\gamma^5\gamma^i\right)^{\mu \nu}  \left( P_{L} \gamma_{i} \right)^{\alpha \beta} q_{0}
\right] + \dots\,,
\label{C2_zero} \\
&&\hspace{-10mm}
{\rm{Im}}\,(-i\mathcal{D}_{Q}) = 
\frac{|F|^{2}|\lambda_{t}|^2}{48 \pi M^{3}}  \delta_{mn} \delta^{\mu \nu} \left( P_{R} \gamma^{0} \right)^{\alpha \beta}   q_{0} 
\nonumber\\
&& \hspace{10mm}
+ \frac{|F|^{2}|\lambda_{t}|^2}{96 \pi M^{3}} \delta_{mn} \left[   
\left(\gamma^5\gamma^i\right)^{\mu \nu}   \left( P_{R} \gamma^{0} \right)^{\alpha \beta} q_{i}
+\left(\gamma^5\gamma^i\right)^{\mu \nu}  \left( P_{R} \gamma_{i} \right)^{\alpha \beta} q_{0}
\right] + \dots\,,
\label{C3_zero}
\nn
\\
\end{eqnarray}
where the dots stand for higher-order terms in the $1/M$ expansion or terms that are of order $1/M^2$ 
but do not depend on the momentum $q^\mu$. 

The matrix element \eqref{C11_zero} is matched in the EFT by 
\begin{equation}
\frac{c_{3}}{M^{3}}  \delta^{\mu \nu}   \left( P_{L} \gamma^{0} \right)^{\alpha \beta}  q_{0}   
+ \frac{c_{5}}{M^{3}}  \left(\gamma^5\gamma^i\right)^{\mu \nu}  \left( P_{L} \gamma^{0} \right)^{\alpha \beta} q_{i}
+ \frac{c_{7}}{M^{3}}  \left(\gamma^5\gamma^i\right)^{\mu \nu}  \left( P_{L} \gamma_{i} \right)^{\alpha \beta} q_{0}
+ \dots,
\label{C6_zero}
\end{equation}
and the matrix element \eqref{C1_zero} by
\begin{eqnarray}
&&\frac{c_{4}}{M^{3}}  \delta_{mn} \delta^{\mu \nu}  \left( P_{R} \gamma^{0} \right)^{\alpha \beta}      q_{0} 
+ \frac{c_{6}}{M^{3}} \delta_{mn}\left(\gamma^5\gamma^i\right)^{\mu \nu}  \left( P_{R} \gamma^{0} \right)^{\alpha \beta} q_{i}
\nn
\\
&&+ \frac{c_{8}}{M^{3}} \delta_{mn}\left(\gamma^5\gamma^i\right)^{\mu \nu}  \left( P_{R} \gamma_{i} \right)^{\alpha \beta} q_{0}
+ \dots,
\label{C7_zero}
\end{eqnarray}
where the dots in \eqref{C6_zero} and \eqref{C7_zero} stand for contributions coming from operators not listed in~\eqref{Wb}.
Comparing \eqref{C2_zero} and \eqref{C3_zero} with the imaginary parts of \eqref{C6_zero} and \eqref{C7_zero} respectively, we obtain 
\begin{eqnarray}
{\rm Im}\,c_{3}= \frac{1}{24\pi}|\lambda_{t}|^{2}|F|^{2} \,, && \qquad {\rm Im}\,c_{4}= \frac{1}{48\pi}|\lambda_{t}|^{2}|F|^{2} \,,
\label{C81}\\
{\rm Im}\,c_{5}= \frac{1}{48\pi}|\lambda_{t}|^{2}|F|^{2} \,, && \qquad {\rm Im}\,c_{6}= \frac{1}{96\pi}|\lambda_{t}|^{2}|F|^{2} \,,
\label{C82}\\
{\rm Im}\,c_{7}= \frac{1}{48\pi}|\lambda_{t}|^{2}|F|^{2} \,, && \qquad {\rm Im}\,c_{8}= \frac{1}{96\pi}|\lambda_{t}|^{2}|F|^{2} \,.
\label{C83}
\end{eqnarray}

\begin{figure}[htb]
\centering
\includegraphics[scale=0.46]{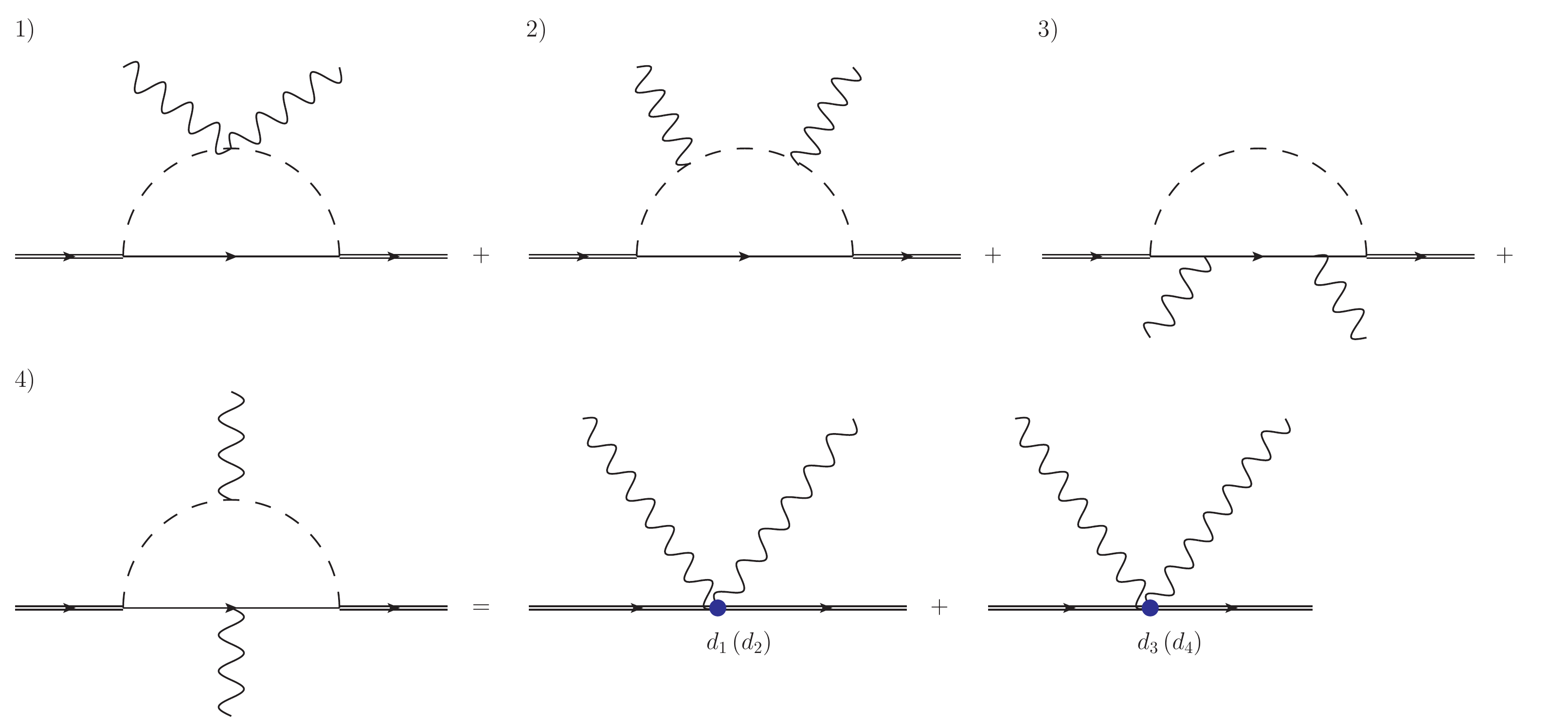}
\caption{\label{Fig4A} 
Diagrams in the full theory (left-hand side of the equality)
contributing to the Majorana neutrino-gauge boson four-field operators in the EFT (right-hand side).
Diagrams with crossed gauge bosons have not been explicitly displayed.
External gauge fi\-elds are ei\-ther SU(2) or U(1) gauge fields. 
In one case they contribute to the operators $d_{1} \, N^{\dagger}N \, W^{a}_{i0} W^a_{i0}/M^3$ 
and $d_{3} \, N^\dagger N \, W^{a}_{\mu\nu} W^{a\,\mu\nu}\!/M^3$, 
in the other case to the operators $d_{2} \, N^{\dagger}N \, F_{i0} F_{i0}/M^3$ 
and $d_{4} \, N^\dagger N $ $\times F_{\mu\nu} F^{\mu\nu}\!/M^3$ in the EFT.
The lines stand for the same particle propagators as in figure~\ref{Fig1A}.
}
\end{figure}

\section{Gauge bosons}
The couplings $d_{i}$ of the Majorana neutrino with the gauge bosons 
are conveniently computed by considering in the fundamental theory the following two matrix elements 
\begin{equation}
-i \left.\int d^{4}x\,e^{i p \cdot x} \int d^{4}y \int d^{4}z\,e^{i q \cdot (y-z)}\, 
\langle \Omega | T(\psi^{\mu}(x) \bar{\psi}^{\nu}(0)  \, A^a_i(y) \, A^b_j(z) )| \Omega\rangle 
\right|_{p^\mu =(M + i\eta,\bm{0}\,)},
\label{D10}
\end{equation}
and 
\begin{equation}
-i \left.\int d^{4}x\,e^{i p \cdot x} \int d^{4}y \int d^{4}z\,e^{i q \cdot (y-z)}\, 
\langle \Omega | T(\psi^{\mu}(x) \bar{\psi}^{\nu}(0)  \, B_i(y) \, B_j(z) )| \Omega\rangle 
\right|_{p^\mu =(M + i\eta,\bm{0}\,)},
\label{D1}
\end{equation}
where $a$ and $b$ are indices labeling fields in the adjoint representation of SU(2), and $i$ and $j$ are spatial Lorentz indices.
The matrix elements \eqref{D10} and \eqref{D1} describe $2 \rightarrow 2$ scatterings between heavy Majorana neutrinos 
at rest and gauge bosons carrying momentum $q^\mu$. 
Each diagram in the full theory, labeled according to figure~\ref{Fig4A},  
contributes with an imaginary part that reads for the \eqref{D10} matrix element 
\begin{eqnarray}
{\rm{Im}}\,(-i\mathcal{D}_{1}) &=& 
- \frac{g^{2} |F|^{2}}{16 \pi M} \,  \delta^{\mu \nu} \delta^{ab}  \delta_{ij} 
+\dots \,,   
\label{D41}\\
{\rm{Im}}\,(-i\mathcal{D}_{2}) &=& 
\frac{g^{2} |F|^{2}}{16 \pi M}  \, \delta^{\mu \nu}  \delta^{ab} 
\left( \delta_{ij} + \delta_{ij} \frac{(q_{0})^{2}}{3 M^{2}}  + \frac{q_iq_j}{6 M^{2}} \right)   
+\dots \,,   
\label{D42}\\
{\rm{Im}}\,(-i\mathcal{D}_{3}) &=& 
- \frac{g^{2} |F|^{2}}{24 \pi M^{3}}  \, \delta^{\mu \nu} \delta^{ab}  \, 
\left( \delta_{ij} (q_{0})^{2} - \frac{q_iq_j}{2} \right) 
+\dots \,,   
\label{D43}\\
{\rm{Im}}\,(-i\mathcal{D}_{4}) &=& 
- \frac{g^{2} |F|^{2}}{48 \pi M^{3}}  \, \delta^{\mu \nu} \delta^{ab}  \, q_iq_j 
+\dots \,.
\label{D44}
\end{eqnarray}
For the matrix element \eqref{D1} the result is the same after the replacement 
$g^2\delta^{ab} \to g'^2$. The dots stand for $1/M^3$ terms that are proportional to $q^2$ or $q_0q_i$ 
or for terms of order $1/M^4$ or smaller.

The matrix element \eqref{D10} is matched in the EFT by 
\begin{equation}
\frac{2 d_{1} }{M^{3}} \, \delta^{\mu \nu}  \delta^{ab} \delta_{ij} \, (q_{0})^{2}  
- \frac{4 d_{3} }{M^{3}} \, \delta^{\mu \nu}  \delta^{ab} \, q_iq_j  
+\dots\,,
\label{D2}
\end{equation}
and the matrix element \eqref{D1} by 
\begin{equation}
\frac{2 d_{2} }{M^{3}} \, \delta^{\mu \nu}  \delta_{ij} \, (q_{0})^{2}  
- \frac{4 d_{4} }{M^{3}} \, \delta^{\mu \nu}  \, q_iq_j  
+\dots\,,
\label{D3}
\end{equation}
where the dots stand for contributions coming from operators not listed in~\eqref{Wb}.
Summing up all contributions \eqref{D41}-\eqref{D44} for each of the two matrix elements and comparing 
with the imaginary parts of \eqref{D2} and \eqref{D3}, we finally find
\begin{eqnarray}
{\rm{Im}}\,d_{1}= -\frac{g^{2}|F|^{2}}{96 \pi} \,, && \qquad {\rm{Im}}\,d_{2}=-\frac{g'^{\,2}|F|^{2}}{96 \pi} \,,
\label{D5}\\
{\rm{Im}}\,d_{3}= -\frac{g^{2}|F|^{2}}{384 \pi} \,, && \qquad {\rm{Im}}\,d_{4}=-\frac{g'^{\,2}|F|^{2}}{384 \pi} \,.
\label{D51}
\end{eqnarray}
The same Wilson coefficients satisfy the matching conditions for matrix elements with temporal gauge bosons.

%% file: CPdegematch.tex
In this appendix we discuss in detail the diagrams relevant for the derivation of the matching coeffcients in eqs.~\eqref{match1} and \eqref{match2}. The Wilson coefficients are split into a leptonic and antileptonic contribution, in turn related to the cuts performed in the two-loop amplitudes considered in the matching. Therefore the cutting rules are discussed in section~\ref{appcutting} being a fundamental tool for the derivation of the matching coefficients. Then in section~\ref{appHiggs} the diagrams involving the Higgs four-coupling are presented whereas in section~\ref{appgauge} we discuss those comprising gauge interactions.  Finally diagrams and corresponding matching calculations relevant for the flavoured case are shown in section~\ref{appflaCPdege}. Many of the loop diagrams computed here analytically were also crosschecked with tools for automated loop calculations~\cite{Shtabovenko:2016sxi}.

\section{Cutting rules}
\label{appcutting}
A way of computing the imaginary part of $-i {\mathcal{D}}$, where ${\mathcal{D}}$ is a Feynman diagram, is by means of cutting rules.
Here we describe briefly the cutting rules at zero temperature and the notation that we will use; we also illustrate them with an example. 
We refer to~\cite{Cutkosky:1960sp,Remiddi:1981hn,LeBellac} for some classical presentations and to~\cite{Denner:2014zga} for a more recent one 
suited to include complex masses and couplings.

\begin{figure}[ht]
\centering
\includegraphics[scale=0.55]{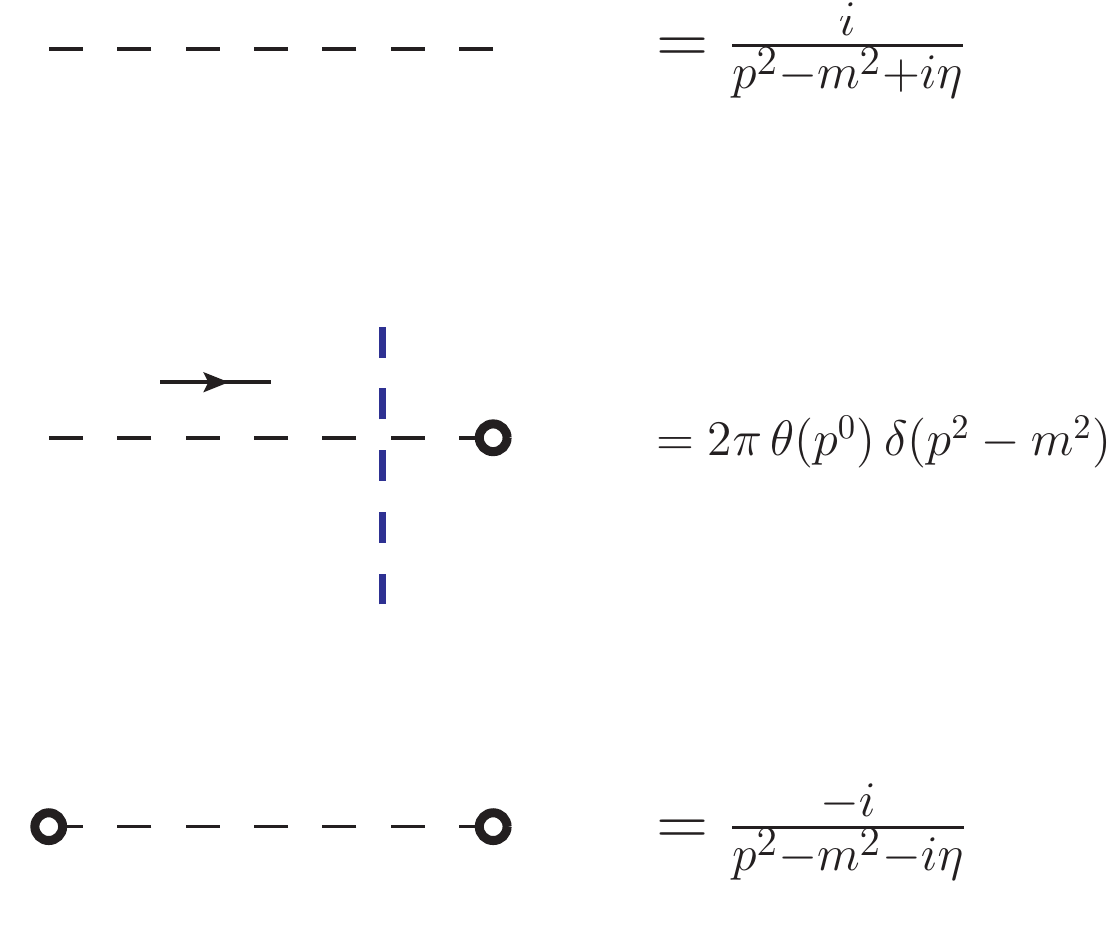}
\caption{The relevant cutting rules for a scalar propagator at zero temperature 
in the convention of~\cite{LeBellac}. The momentum direction is represented by the arrow. 
The blue thick dashed line stands for the cut. Vertices on the right of the cut are circled. 
Circled vertices have opposite sign than non-circled vertices.}
\label{fig:cutting} 
\end{figure}

At the core of the method is the cutting equation, which relates ${\rm{Im}}(-i {\mathcal{D}})$ with cut diagrams of ${\mathcal{D}}$.
It reads 
\begin{equation}
{\rm{Im}}(-i \mathcal{D})= - {\rm{Re}}(\mathcal{D})=\frac{1}{2} \sum_{\rm cuts} \mathcal{D} \, .
\label{cuttingEquation}
\end{equation} 
A cut diagram consists in separating the Feynman diagram into two disconnected diagrams by putting on shell some of its internal propagators. 
The cut is typically represented by a line ``cutting'' through these propagators: in our case it is a blue thick dashed line.
Vertices on the right of the cut are circled.  Circled vertices have opposite sign than uncircled vertices. 
We can have three types of propagators. Propagators between two circled vertices, propagators between uncircled vertices 
and propagators between one circled and one uncircled vertex. This last situation occurs when the cut goes through the propagator.
The expressions for these three propagators are shown in the case of a scalar particle in figure~\ref{fig:cutting};
the extension to fermions and gauge bosons is straightforward.
Note that when the cut goes through the propagator the particle is put on shell.
The sum in \eqref{cuttingEquation} extends over all possible cuts of the diagram ${\mathcal{D}}$. 

As an example, we show how to obtain the imaginary part of the Wilson coefficient of the operator~\eqref{Wa} 
in the case of just one neutrino generation. We call this single Wilson coefficient $a$.
It was first derived in~\cite{Biondini:2013xua} without using cutting rules.
Cutting rules have the advantage that they allow to disentangle the contribution coming from the decay into a lepton,
which we call ${\rm Im}\,a^\ell$, from the contribution coming from the decay into an antilepton, which we call ${\rm Im}\,a^{\bar{\ell}}$.
The coefficient ${\rm Im}\,a$ is at leading order the sum of these two contributions: ${\rm Im}\,a= {\rm Im}\,a^\ell + {\rm Im}\,a^{\bar{\ell}}$.
It can be obtained by matching the following matrix element of the fundamental theory
\begin{equation}
-i \left.\int d^{4}x\,e^{i p \cdot x} \int d^{4}y \int d^{4}z\,e^{i q \cdot (y-z)}\, 
\langle \Omega | T(\psi^{\mu}(x) \bar{\psi}^{\nu }(0) \phi_{m}(y) \phi_{n}^{\dagger}(z) )| \Omega \rangle
\right|_{p^\alpha =(M + i\eta,\bm{0}\,)},
\label{matrix}
\end{equation} 
with the corresponding matrix element of the EFT. The field $\psi$ identifies the only Majorana 
neutrino field available in this case, $\mu$ and $\nu$ are Lorentz indices and $m$ and $n$ SU(2)$_L$ indices.

\begin{figure}[ht]
\centering
\includegraphics[scale=0.5]{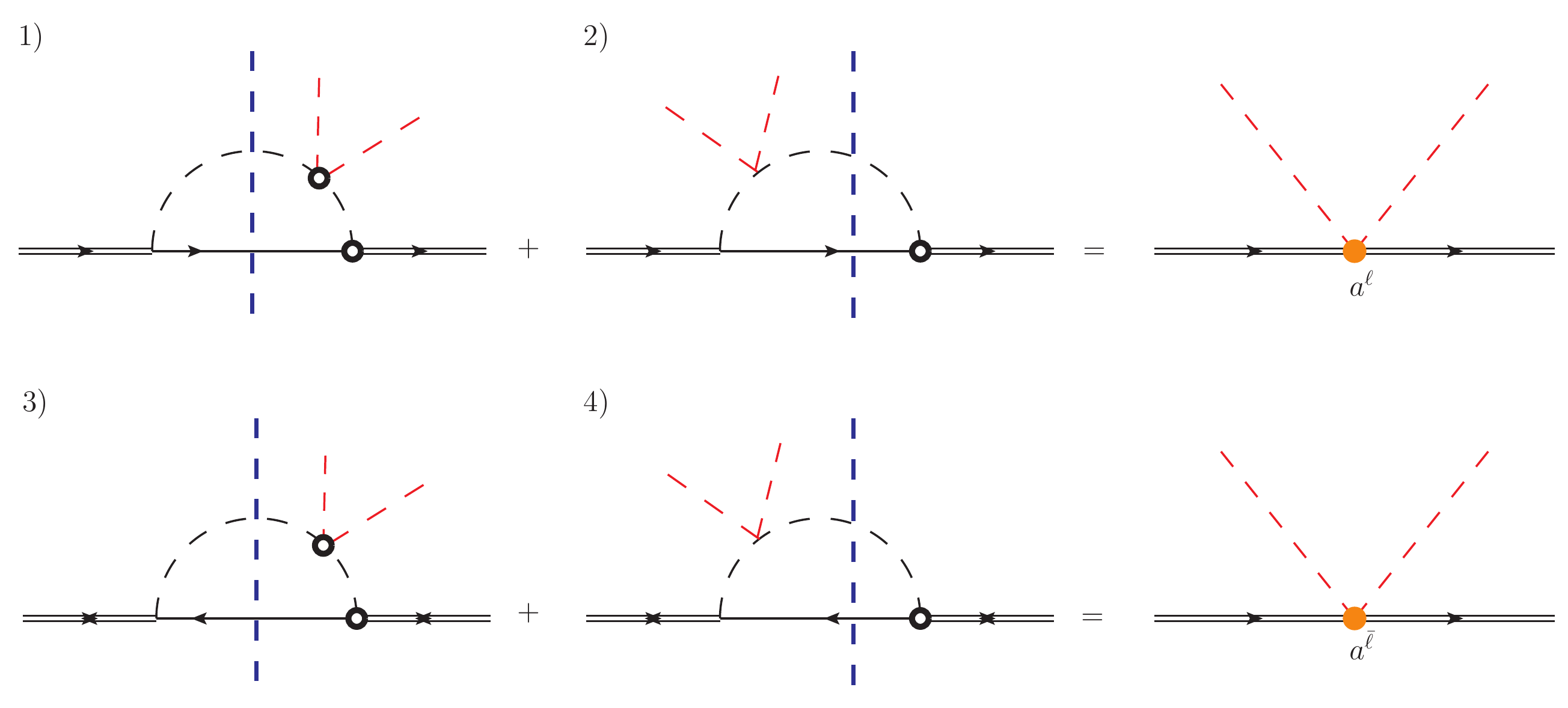}
\caption{Diagrams in the full theory contributing to the Majorana neutrino-Higgs boson dimension-five operator. 
On the left-hand side are the diagrams in the full theory, whereas on the right-hand side are the diagrams in the EFT.
As in figure~\ref{fig:treeMatch} and in the rest of the paper, red dashed lines indicate external Higgs bosons  
with a soft momentum much smaller than the mass of the Majorana neutrino. The cuts on the diagrams 
in the fundamental theory are explicitly shown.}
\label{fig8_CPdege} 
\end{figure}

When computing matrix elements involving Majorana fermions, one has to consider 
that the relativistic Majorana field may be contracted in more ways than if it was a Dirac field, 
this reflecting the indistinguishability of the Majorana particle and anti-particle.
The different contractions give rise to the different propagators listed in \eqref{eq5_partprod}-\eqref{eq7_partprod}.
When contracting the Majorana fields in \eqref{matrix} according to \eqref{eq5_partprod}, one obtains at leading order 
\begin{eqnarray}
\left[\hat{P}\left( -i \mathcal{D}\right)\hat{P}\right]^{\mu\nu}
= 6 |F|^2 \lambda \, \delta_{mn} \,  \int \frac{d^{4} \ell }{(2\pi)^{4}}  
\left( \hat{P} \, P_L \slashed{\ell} \, \hat{P} \right)^{\mu \nu} \frac{i}{\ell^{2}+i \eta} 
\left( \frac{i}{(M v - \ell )^2+i \eta} \right)^{2} ,
\label{3a}
\end{eqnarray}
where we have dropped all external propagators and $\mathcal{D}$ is the amputated (uncut) diagram shown in the upper raw and left-hand side of figure~\ref{fig8_CPdege}.
The external heavy-neutrino propagators reduce in the non-relativistic limit and in the rest frame 
to a matrix proportional to $\hat{P}=(1+\gamma^{0})/2$ (see \eqref{effpropagator}).
We have kept the matrix $\hat{P}$ on the left- and right-hand side of \eqref{3a}, because it helps projecting 
out the contributions relevant in the heavy-neutrino mass limit, e.g., $ \hat{P} \, P_L \, \hat{P} = \hat{P}/2$.
After projection, also the matrix $\hat{P}$ may be eventually dropped from the left- and right-hand side of the matching equation.
The internal loop momentum is $\ell^{\mu}$, $Mv^{\mu}=(M,\bm{0})$ is the neutrino momentum in the rest frame and $|F|^2=\sum_{f} F^{*}_{f} F_{f}$. 

The diagram $\mathcal{D}$ admits two cuts labelled $1)$ and $2)$ and shown in the upper raw and left-hand side of figure~\ref{fig8_CPdege}.
Both cuts select a final state made of a lepton and, therefore, contribute to~$a^\ell$.
Using \eqref{cuttingEquation} and the cutting rules we obtain for the two cuts:
\begin{eqnarray}
\left[\hat{P}\, {\rm{Im}}(-i\mathcal{D}^{\ell}_{1,\hbox{\tiny fig.\ref{fig8_CPdege}}})\hat{P}\right]^{\mu\nu}
=3 |F|^2 \lambda \,(-1)^2 \, \delta_{mn} \, \int \frac{d^{4} \ell}{(2\pi)^{4}}    
\left( \hat{P} \, P_L \slashed{\ell} \, \hat{P} \right)^{\mu \nu}\,2\pi \theta(\ell^{0}) \delta(\ell^2) 
\nonumber \\ 
\times 2\pi\theta(M-\ell^{0})\delta((Mv-\ell)^2) \frac{-i}{(M v - \ell)^2  -i \eta} ,
\label{cut1}
\end{eqnarray}
\begin{eqnarray}
\left[\hat{P}\, {\rm{Im}}(-i\mathcal{D}^{\ell}_{2,\hbox{\tiny fig.\ref{fig8_CPdege}}})\hat{P}\right]^{\mu\nu}
=3 |F|^2 \lambda \, (-1)  \, \delta_{mn} \,  
\int \frac{d^{4}\ell}{(2\pi)^{4}} \left( \hat{P} \, P_L \slashed{\ell} \, \hat{P} \right)^{\mu \nu}\,2\pi  \theta(\ell^{0}) \delta(\ell^2)  
\nonumber \\
\times 2\pi\theta(M-\ell^{0})\delta((Mv-\ell)^2) \frac{i}{(M v - \ell )^2  +i \eta} .
\label{cut2}
\end{eqnarray}
Both ${\rm{Im}}(-i\mathcal{D}^{\ell}_{1,\hbox{\tiny fig.\ref{fig8_CPdege}}})$ and ${\rm{Im}}(-i\mathcal{D}^{\ell}_{2,\hbox{\tiny fig.\ref{fig8_CPdege}}})$ have a pinch singularity 
whose origin is the soft limit of the Higgs momentum pair. A way to regularize the singularity 
is to give a small finite momentum to the Higgs pair and set it to zero after cancellation of the singularity. 
The singularity cancels in the sum of the two cuts, which reads
\begin{equation}
{\rm{Im}}(-i\mathcal{D}^{\ell}_{1,\hbox{\tiny fig.\ref{fig8_CPdege}}}) + {\rm{Im}}(-i\mathcal{D}^{\ell}_{2,\hbox{\tiny fig.\ref{fig8_CPdege}}})=
-\frac{3}{16 \pi M} |F|^2 \lambda \, \delta^{\mu \nu} \delta_{mn}, 
\label{E}
\end{equation} 
where we have used for the amputated Green function the same indices used for the unamputated one, a convention that we will keep in the following. 

When contracting the Majorana fields in~\eqref{matrix} according to~\eqref{eq6_partprod} and~\eqref{eq7_partprod} 
one obtains at leading order a contribution encoded in the diagram shown in the lower raw and left-hand side of figure~\ref{fig8_CPdege}.
The expression for this diagram is the same as the one in~\eqref{3a} up to an irrelevant change $P_L \to P_R$
(the expression is also unsensitive to the change $F_f  \leftrightarrow F_f^{*}$).
The diagram admits two cuts labeled $3)$ and $4)$ and shown in the lower raw and left-hand side of figure~\ref{fig8_CPdege}.
Both cuts select a final state made of an antilepton and, therefore, contribute to  $a^{\bar{\ell}}$.
The contributions from these two cuts are the same as the ones in \eqref{cut1} and \eqref{cut2} and give eventually the same result for the sum
\begin{equation}
{\rm{Im}}(-i\mathcal{D}^{\bar{\ell}}_{3,\hbox{\tiny fig.\ref{fig8_CPdege}}}) + {\rm{Im}}(-i\mathcal{D}^{\bar{\ell}}_{4,\hbox{\tiny fig.\ref{fig8_CPdege}}})=
-\frac{3}{16 \pi M} |F|^2 \lambda \, \delta^{\mu \nu} \delta_{mn}.
\label{Ebis}
\end{equation} 

Comparing \eqref{E} and \eqref{Ebis} with the corresponding expressions in the EFT, which are 
$({\rm{Im}} \,a^{\ell}/M) \, \delta^{\mu \nu} \delta_{mn}$ and $({\rm{Im}} \,a^{\bar{\ell}}/M) \, \delta^{\mu \nu} \delta_{mn}$ respectively, one obtains
\begin{eqnarray}
&& {\rm{Im}} \, a^{\ell} = {\rm{Im}} \, a^{\bar{\ell}} = -\frac{3}{16\pi}|F|^{2}\lambda ,
\label{alalbar}\\
&& {\rm{Im}} \, a={\rm{Im}} \, a^{\ell}+{\rm{Im}} \, a^{\bar{\ell}}= -\frac{3}{8\pi}|F|^{2}\lambda .
\label{atot}
\end{eqnarray} 
Equation \eqref{atot} agrees with the result found in eq.~\eqref{match_final_a}, the latter calculated without the cutting rules.

\section{Matching diagrams with four-Higgs interaction}
\label{appHiggs}

We compute in the fundamental theory the matrix element 
\begin{equation}
-i \left.\int d^{4}x\,e^{i p \cdot x} \int d^{4}y \int d^{4}z\,e^{i q \cdot (y-z)}\, 
\langle \Omega | T(\psi^{\mu}_{1}(x) \bar{\psi}^{\nu }_{1}(0) \phi_{m}(y) \phi_{n}^{\dagger}(z) )| \Omega \rangle
\right|_{p^\alpha =(M + i\eta,\bm{0}\,)}.
\label{B1_CPdege}
\end{equation} 
The matrix element is similar to \eqref{matrix}, but now in a theory with two types of heavy Majorana neutrinos.
External neutrinos are of type 1, whereas neutrinos of type 2 appear only as intermediate states. 
The result can be extended straightforwardly to the case of external neutrinos of type 2, leading to \eqref{match2}. 
The matrix element describes a $2 \rightarrow 2$ scattering between a heavy Majorana neutrino 
of type 1 at rest and a Higgs boson carrying momentum $q^\mu$. 
Since the momentum $q^\mu$ is much smaller than the neutrino mass 
and we are not matching derivative operators, $q^\mu$ can be set to zero in the matching.
Here, we compute the diagrams contributing to \eqref{B1_CPdege} that enter the matching of $a^{\ell}_{11}$ 
(and $a^{\bar{\ell}}_{11}$) up to first order in $\lambda$ and are relevant for the direct CP asymmetry;
in the next section, we will compute the diagrams of order $g^2$ and $g'^2$.
It may be useful to cast the diagrams into three different typologies as we will do in the following.
All diagrams are understood as amputated of their external legs when writing the corresponding amplitudes.

\begin{figure}[ht]
\centering
\includegraphics[scale=0.55]{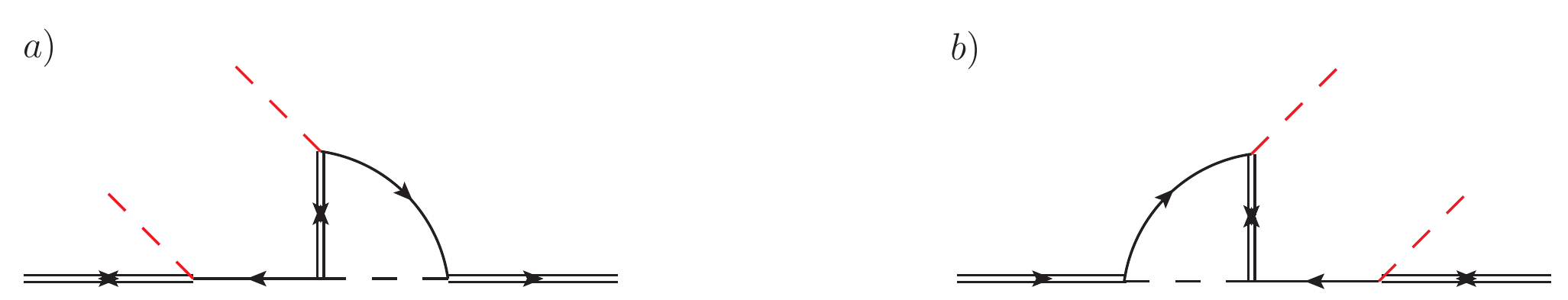}
\caption{Diagrams contributing to $a^\ell_{II}$ at order $F^4$. One diagram is the complex conjugate of the other.}
\label{fig:F4_CPdege}
\end{figure}

A first class of diagrams is obtained by opening-up a Higgs line in the two-loop diagrams of figure~\ref{fig:fig3_CPdege}. 
These diagrams are of order $F^4$. The subset contributing to $a^\ell_{II}$ is shown in figure~\ref{fig:F4_CPdege}.
Diagrams $a)$ and $b)$ are one the complex conjugate of the other; their sum is real.
By cutting the loops so to bring one lepton on shell and summing both diagrams the result is proportional 
to the Yukawa coupling combination ${\rm{Re}}\left[ (F_1^{*}F_J)^2\right]$ only.
The reason is that, after the cuts, the diagrams do not contain loops anymore and cannot develop any additional complex phase. 
If we consider the subset of diagrams contributing to $a^{\bar{\ell}}_{II}$, which are diagrams where the antilepton can be put on shell, 
we obtain through a similar argument that the sum of diagrams is proportional again to the Yukawa coupling combination ${\rm{Re}}\left[(F_1^{*}F_J)^2\right]$. 
It follows that the matching coefficients obtained for leptons and antileptons and the corresponding leptonic and antileptonic widths 
cancel in the difference. One-loop diagrams of order $F^4$ with two external Higgs bosons do not contribute to the direct CP asymmetry.  

\begin{figure}[ht]
\centering
\includegraphics[scale=0.5]{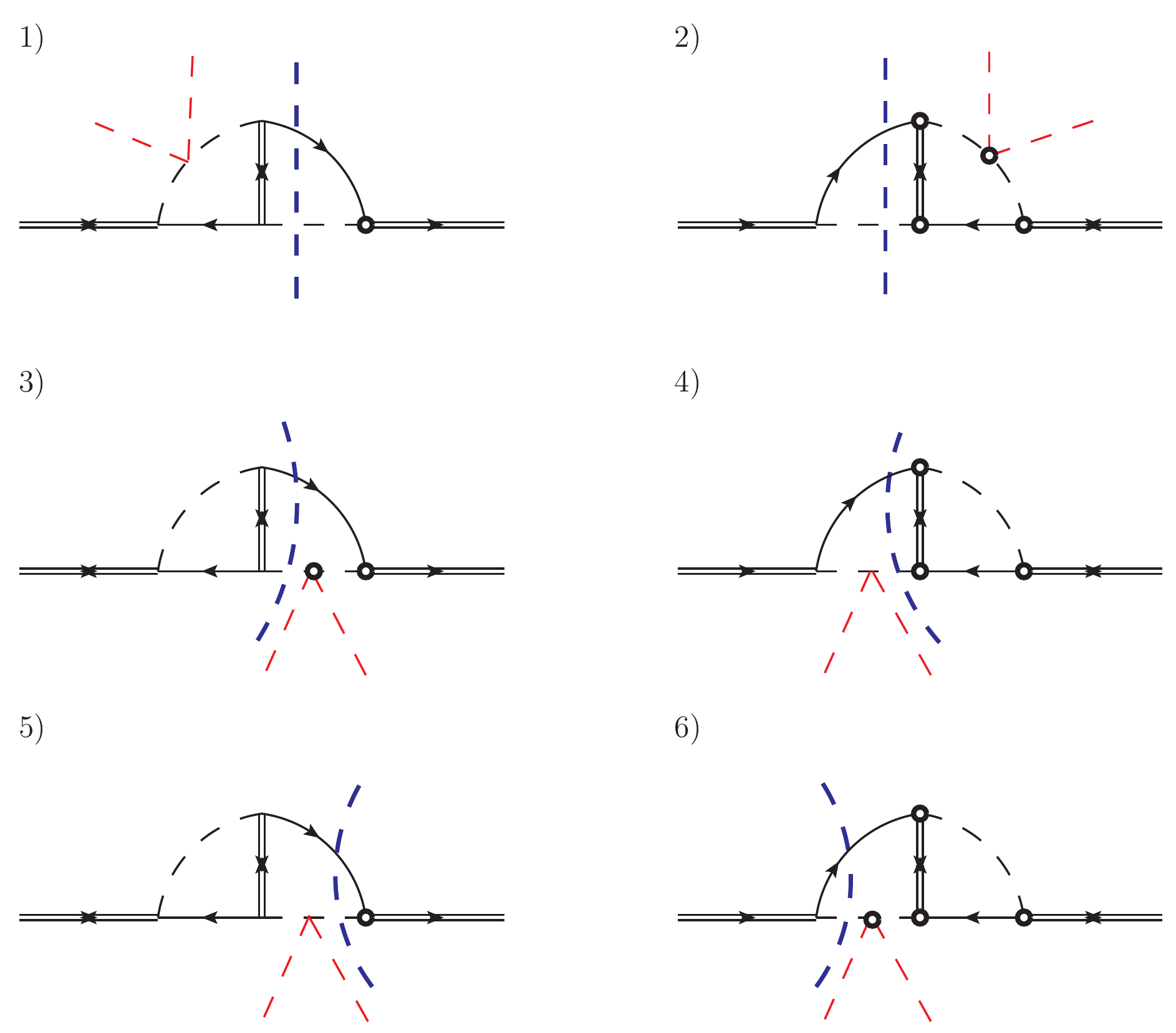}
\caption{Diagrams contributing to $a^\ell_{II}$ and $a^{\bar{\ell}}_{II}$ at order $F^4\lambda$.
The cuts through leptons are explicitly shown and implemented according to the rules of figure~\ref{fig:cutting}.}
\label{fig:fig10_CPdege} 
\end{figure}

A second class of diagrams is obtained by attaching a four-Higgs vertex to an existing Higgs line in the two-loop diagrams of figure~\ref{fig:fig3_CPdege}.
These diagrams are of order $F^4\lambda$ and are shown with the relevant cuts in figure~\ref{fig:fig10_CPdege}. 
In each raw we show a diagram and its complex conjugate and we draw explicitly the cuts that put a lepton on shell.
This amounts at selecting in all the diagrams in figure~\ref{fig:fig10_CPdege}
the decay of a heavy Majorana neutrino into a lepton. 
The decay width into an antilepton can be computed by cutting antilepton lines. 
In general, the sum of each couple of diagrams in figure~\ref{fig:fig10_CPdege}
is a linear combination of the real and the imaginary parts of $(F_{1}^{*}F_{J})^2$. 
The appearance of a term proportional to ${\rm{Im}}\left[(F_1^{*}F_2)^2\right]$
in addition to ${\rm{Re}}\left[(F_1^{*}F_J)^2\right]$ reflects the fact that after the cut we are 
left with a loop that also develops an imaginary part.
For each couple of diagrams, contributions coming from the lepton and the antilepton cuts give 
the same terms proportional to  ${\rm{Re}}\left[(F_1^{*}F_J)^2\right]$  but terms 
proportional to  ${\rm{Im}}\left[(F_{1}^{*}F_{2})^2\right]$ with opposite signs, 
since ${\rm{Re}}\left[(F_1^{*}F_J)^2\right] = {\rm{Re}}\left[(F_1F_J^{*})^2\right]$
while ${\rm{Im}}\left[(F_{1}^{*}F_{2})^2\right] = - {\rm{Im}}\left[(F_{1}F_{2}^{*})^2\right]$.
So that, when calculating the CP asymmetry, terms proportional to ${\rm{Re}}\left[(F_1^{*}F_J)^2\right]$ cancel, 
and only those proportional to ${\rm{Im}}\left[(F_{1}^{*}F_{2})^2\right]$ remain.
Hence for each diagram we only need to calculate the terms proportional to ${\rm{Im}}\left[(F_{1}^{*}F_{2})^2\right]$, 
consistently with the discussion in section~\ref{sec:zeroT}.
Up to relative order $\Delta/M$ they are:
\begin{eqnarray}
&&\!\!\!\!\! \!\!\!\!\! 
{\rm{Im}}\, (-i \mathcal{D}^{\ell}_{1,\hbox{\tiny fig.\ref{fig:fig10_CPdege}}}) + {\rm{Im}}\, (-i \mathcal{D}^{\ell}_{2,\hbox{\tiny fig.\ref{fig:fig10_CPdege}}}) = 
\nonumber \\
&&\hspace{3.7cm}  
\frac{3\,{\rm{Im}}\left[ (F_{1}^{*}F_{2})^2\right]}{(16 \pi)^2M} \lambda \left[ \ln 2 -(1-\ln 2) \frac{\Delta}{M}\right] 
\delta^{\mu \nu} \delta_{mn} + \dots \, ,  
\label{Higgs1} 
\nn
\\
\phantom{x}
\\
&&\!\!\!\!\! \!\!\!\!\! 
{\rm{Im}}\,(-i \mathcal{D}^{\ell}_{3,\hbox{\tiny fig.\ref{fig:fig10_CPdege}}}) + {\rm{Im}}\,(-i \mathcal{D}^{\ell}_{4,\hbox{\tiny fig.\ref{fig:fig10_CPdege}}}) 
+ {\rm{Im}}\,(-i \mathcal{D}^{\ell}_{5,\hbox{\tiny fig.\ref{fig:fig10_CPdege}}}) + {\rm{Im}}\,(-i \mathcal{D}^{\ell}_{6,\hbox{\tiny fig.\ref{fig:fig10_CPdege}}}) = 
\nonumber \\
&&\hspace{3.7cm}  
\frac{ 3\,{\rm{Im}}\left[ (F_{1}^{*}F_{2})^2\right] }{(16 \pi)^2M} \lambda\left[ \ln 2 -(1-\ln 2) \frac{\Delta}{M}\right]
\delta^{\mu \nu} \delta_{mn} + \dots \, .
\nn 
\\
\label{Higgs2}
\end{eqnarray}   
The dots stand for terms proportional to the Yukawa coupling combination ${\rm{Re}}\left[(F_{1}^{*}F_{J})^2\right]$ 
and higher-order terms in the expansion in $\Delta/M$.  
The superscript $\ell$ reminds that we have cut through leptons only; 
as we argued above, the contribution of antileptons has opposite sign.
We give the result in \eqref{Higgs2} as the sum of four diagrams to cancel 
a pinch singularity that arises in the soft momentum limit of the Higgs boson. 
This is analogous to the calculation carried out in section~\ref{appcutting}.

\begin{figure}[ht]
\centering
\includegraphics[scale=0.5]{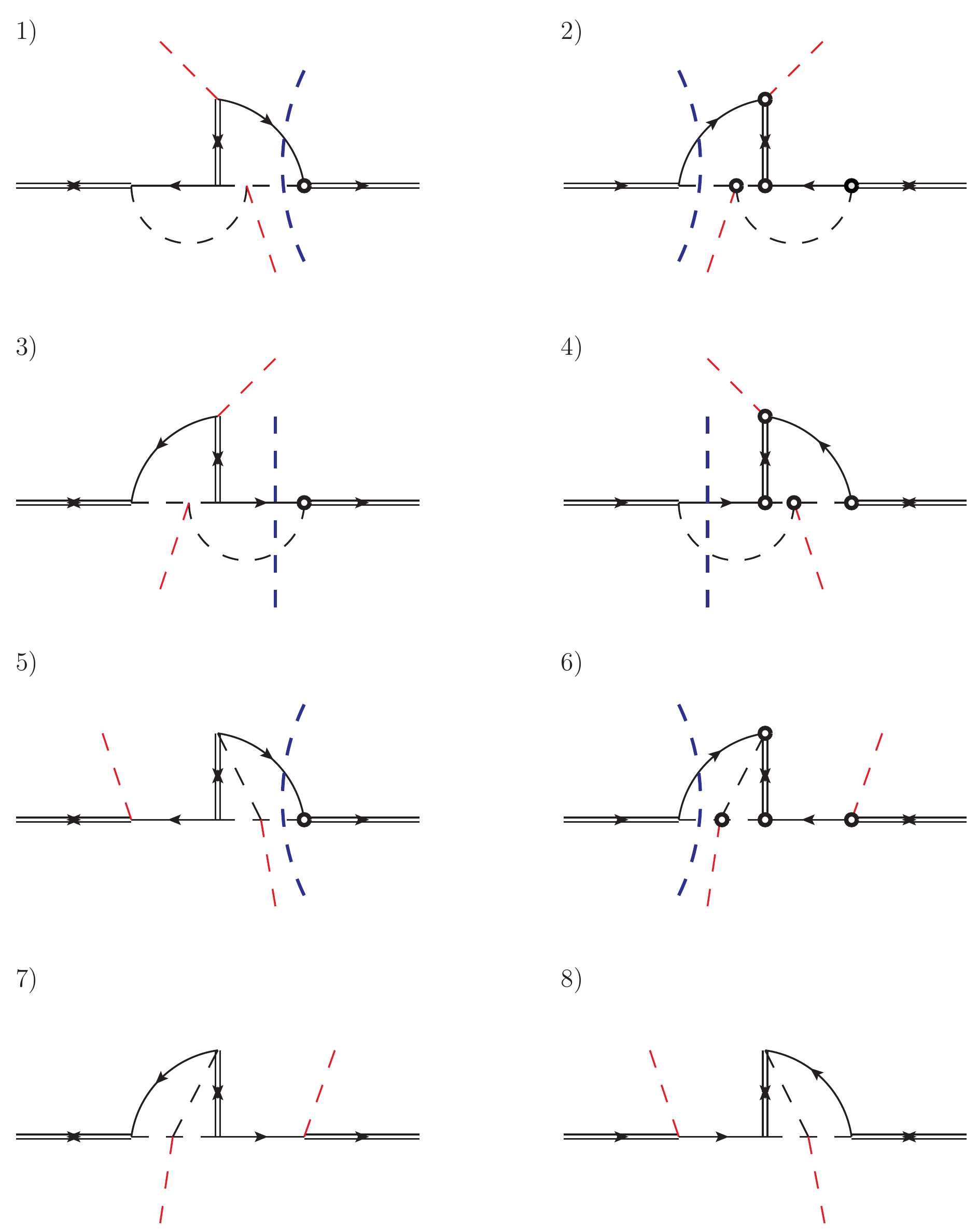}
\caption{Diagrams contributing to $a^\ell_{II}$ and $a^{\bar{\ell}}_{II}$ at order $F^4\lambda$. The cuts through leptons are explicitly shown.}
\label{fig:new_Higgs_direct} 
\end{figure}

Once the four-Higgs vertices are removed, the diagrams of figure~\ref{fig:fig10_CPdege} preserve the topology of the $T=0$ two-loop diagrams of figure~\ref{fig:fig3_CPdege}.
There is, finally, a third class of diagrams where this topology is not preserved. 
A way to construct them is from the diagrams of figure~\ref{fig:F4_CPdege} (and the corresponding ones with an antilepton in the loop)
by adding a four-Higgs vertex to the internal Higgs line; we show the diagrams with the relevant cuts in figure~\ref{fig:new_Higgs_direct}. 
The results for the cuts through leptons read
\begin{eqnarray}
&&\!\!\!\!\! \!\!\!\!\! 
{\rm{Im}}\, (-i \mathcal{D}^{\ell}_{1,\hbox{\tiny fig.\ref{fig:new_Higgs_direct}}}) + {\rm{Im}}\, (-i \mathcal{D}^{\ell}_{2,\hbox{\tiny fig.\ref{fig:new_Higgs_direct}}}) = 
\nonumber \\
&&\hspace{3.7cm}  
\frac{3\,{\rm{Im}}\left[ (F_{1}^{*}F_{2})^2\right]}{(16 \pi)^2M} \lambda  \left(1-\frac{\Delta}{M} \right)
\delta^{\mu \nu} \delta_{mn} + \dots \, ,  
\label{Higgs3} 
\end{eqnarray}
\begin{eqnarray}
&&{\rm{Im}}\,(-i \mathcal{D}^{\ell}_{3,\hbox{\tiny fig.\ref{fig:new_Higgs_direct}}}) + {\rm{Im}}\,(-i \mathcal{D}^{\ell}_{4,\hbox{\tiny fig.\ref{fig:new_Higgs_direct}}}) =
\nonumber \\
&&\hspace{3.7cm}  
\frac{3\,{\rm{Im}}\left[ (F_{1}^{*}F_{2})^2\right]}{(16 \pi)^2M} \lambda  \left(1-\frac{\Delta}{M} \right)
\delta^{\mu \nu} \delta_{mn} + \dots \, ,  
\label{Higgs4}
\\
&&{\rm{Im}}\,(-i \mathcal{D}^{\ell}_{5,\hbox{\tiny fig.\ref{fig:new_Higgs_direct}}}) + {\rm{Im}}\,(-i \mathcal{D}^{\ell}_{6,\hbox{\tiny fig.\ref{fig:new_Higgs_direct}}}) =0\,.
\label{Higgs5}
\end{eqnarray}    

Some remarks, which will be of use also in the following to simplify the calculation, are in order.
First, in the Feynman diagrams, integrals over momentum regions where the intermediate neutrino is on shell do no contribute to the matching. 
Such momentum regions are either kinematically forbidden, if the intermediate neutrino is heavier 
than the initial one, or they are reproduced in the EFT, if the intermediate neutrino is lighter than the initial one
(see diagrams in figure~\ref{fig:DeltaEFT} and the related discussion in section~\ref{sec:direct2}).
In the last case, the momentum is necessarily of order $\Delta$.
Modes with energy or momentum of order $\Delta \ll M$ are still dynamical in the effective theory 
and should not be integrated out with the mass scale (if they are, then they would need to be subtracted 
by computing suitable loops in the effective theory). Second, also momentum regions where three massless particles happen 
to be on-shell and enter the same vertex do not contribute to the matching, because the available phase space vanishes in dimensional regularization.
These general remarks apply in the present case to the diagrams 5) and 6) of figure~\ref{fig:new_Higgs_direct}.
After the cuts through the lepton propagators shown in the diagrams have been implemented, the remaining one-loop diagrams may 
develop an imaginary part only if two of the particles in the loop can be put on shell. If the neutrino 
is put on shell, then the one-loop integral is either over a kinematically forbidden momentum region or 
over a momentum region which is much smaller than $M$, according to the first remark above. 
If the light particles are put on shell, then, for we can neglect the 
momentum of the external Higgs boson, we have a situation equivalent to a vertex with three on-shell massless 
particles and the second remark above applies. The result is that diagrams 5) and 6) of figure~\ref{fig:new_Higgs_direct}
do not contribute to the CP asymmetry at the scale $M$, which is the result~\eqref{Higgs5}.

\section{Matching diagrams with gauge interactions}
\label{appgauge}
At order $F^4$ and at first order in the SM couplings, besides the Feynman diagrams with four-Higgs vertices 
computed in the previous section, also diagrams with a gauge boson can contribute.
We will compute them here.

\begin{figure}[ht]
\centering
\includegraphics[scale=0.45]{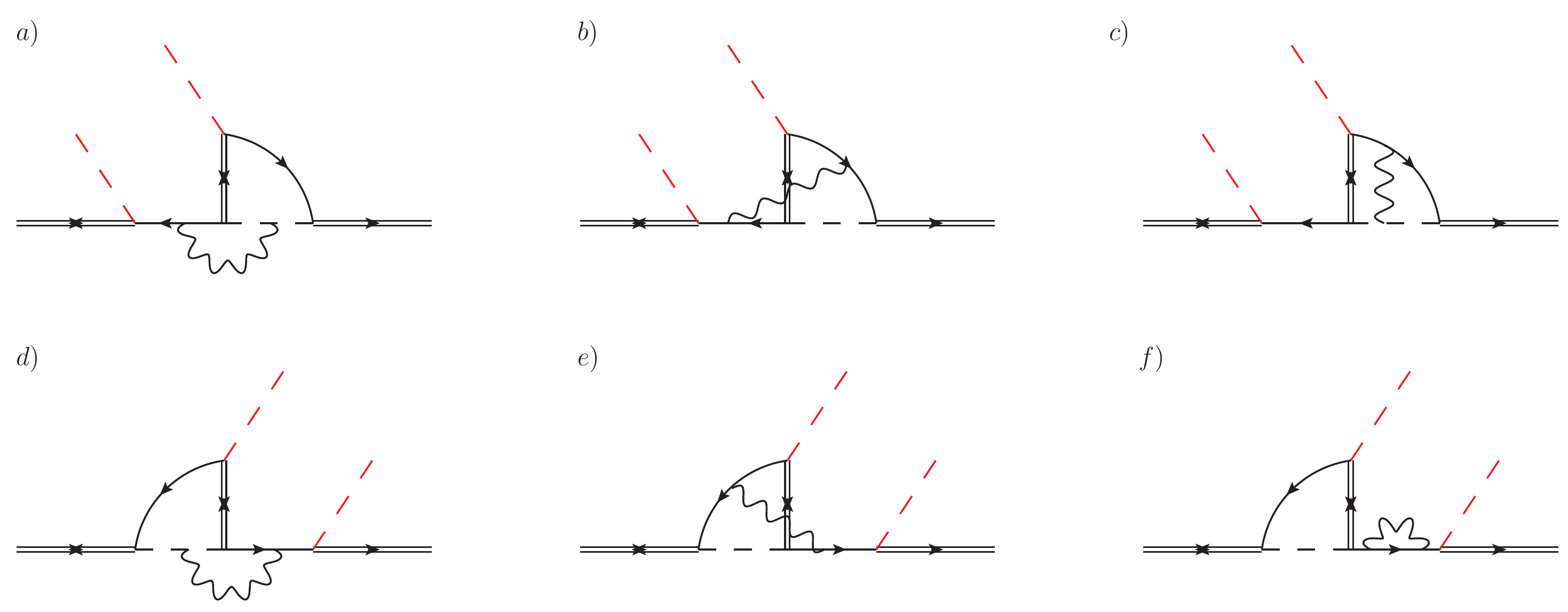}
\caption{If the incoming and outgoing Majorana neutrinos are conventionally chosen to be of type 1, 
then the displayed diagrams contribute to $a^\ell_{11}$ at order $F^4$ and at first order in the gauge couplings. 
The diagrams contribute also to $a^{\bar{\ell}}_{11}$ if cut through the antilepton.
Only diagrams proportional to $(F_{1}^{*} F_{2})^2$ are displayed.}
\label{fig:diagrams1_CPdege}
\end{figure}

By cutting this kind of diagrams we distinguish two different type of processes:
processes with a gauge boson in the final state or processes without a gauge boson in the final state. 
These being two distinct physical processes, we can compute them in different gauges. 
It is advantageous to adopt the Coulomb gauge in the first type of processes and the Landau gauge in the second one. The advantages are twofold.
First, with this choice of gauge we can neglect, for the purpose of matching the dimension-five operators
in the EFT in (\ref{eq:efflag_CPdege}), all diagrams with a gauge boson attached to an external Higgs boson leg.
The reason is that the coupling of the gauge boson with the Higgs boson is proportional to the momentum of the latter (see~\eqref{SMlag} and \eqref{SMCov}). 
If it depends on the external momentum, then the diagram will contribute to the matching of a higher-dimensional operator in the EFT, 
for the dimension-five operators do not contain derivatives. If it depends on the internal momentum then its contraction 
with the gauge boson propagator vanishes both in Landau gauge, if the gauge boson is uncut, and in Coulomb gauge, if the gauge boson is cut.
In the latter case, only transverse gauge bosons can be cut.
Second, the physical Coulomb gauge does not generate spurious singularities when the gauge boson is cut.

\begin{figure}[ht]
\centering
\includegraphics[scale=0.45]{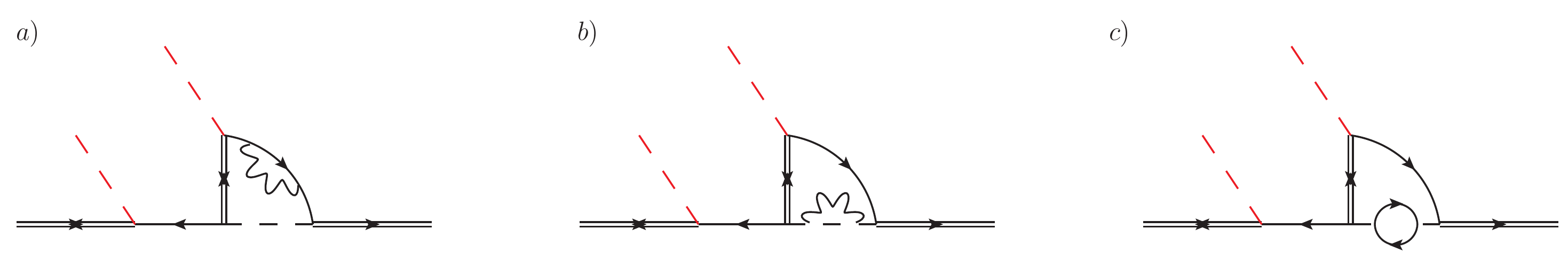}
\caption{Diagrams as in figure~\ref{fig:diagrams1_CPdege}. 
In diagram $c)$, the particles in the small loop coupled to a Higgs boson are a top quark and a heavy-quark doublet.}
\label{fig:diagrams2_CPdege} 
\end{figure}

With the above choice of gauges, it is convenient to divide the remaining diagrams contributing to the matching of the dimension-five operators  
into the four sets shown in figures~\ref{fig:diagrams1_CPdege}, \ref{fig:diagrams2_CPdege}, \ref{fig:diagram_single_CPdege} and~\ref{fig:fig_ind_new_CPdege} 
for the leptonic contribution. After closer inspection, diagram~$c)$ in figure~\ref{fig:diagrams1_CPdege} turns out not to contribute to the CP asymmetry.
The diagram may be cut through the lepton propagator in two ways leaving in each case an uncut one-loop subdiagram.
The only cuts for these subdiagrams that are relevant for the matching (see discussion at the end of section~\ref{appHiggs}) 
give rise to two identical but opposite contributions (they differ only in the number of circled vertices), which cancel. 
We have checked the cancellation also by explicit calculation. 

\begin{figure}[ht]
\centering
\includegraphics[scale=0.385]{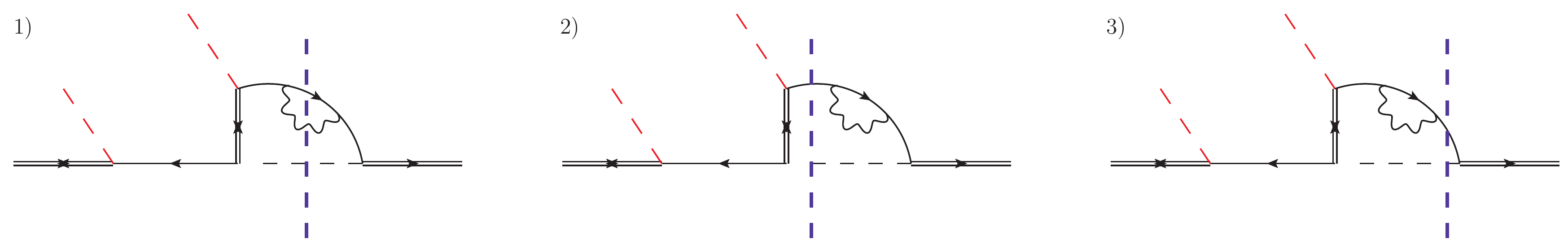}
\caption{Cuts on diagram $a)$ of figure~\ref{fig:diagrams2_CPdege}. 
The first cut does not contain any loop. The other two cut diagrams do contain a
remaining loop that however does not develop an imaginary part.}
\label{fig:cutexample} 
\end{figure}

We consider now the three diagrams in figure~\ref{fig:diagrams2_CPdege}. 
It turns out that these diagrams cannot introduce an additional complex phase, i.e., they do not develop 
an imaginary part of the loop amplitude, the quantity that we called ${\rm{Im}}(B)$ in section \ref{sec:zeroT}. 
In order to prove this statement, let us pick up diagram $a)$ in figure~\ref{fig:diagrams2_CPdege} 
and consider all possible cuts that put a lepton on shell. These are shown in figure~\ref{fig:cutexample}. 
The first cut does not contain any loop, hence it does not generate any additional complex phase besides the Yukawa couplings. 
In the second and third cut, in order to generate a complex phase, the remaining loop diagrams would need to develop an imaginary part. 
However, this is not the case since the (on-shell) incoming and outgoing particles in the loop and the particles in the loop itself are massless, 
a situation already discussed at the end of section~\ref{appHiggs}.
Therefore, also in this case, the diagram and its complex conjugate contribute with a term proportional to ${\rm{Re}}\left[(F_{1}^{*}F_{2})^2\right]$, 
which cancels eventually against the antileptonic width in the CP asymmetry. The same argument applies to both diagrams $b)$ and $c)$ in figure~\ref{fig:diagrams2_CPdege}
(as well as to diagrams with loops inserted in the external Higgs legs that we have not displayed).  
As an important consequence, there are not thermal corrections to the CP asymmetry of order~$T^2/M^2$ that are proportional  
to the top-Yukawa coupling, $\lambda_t$.

\begin{figure}[ht]
\centering
\includegraphics[scale=0.47]{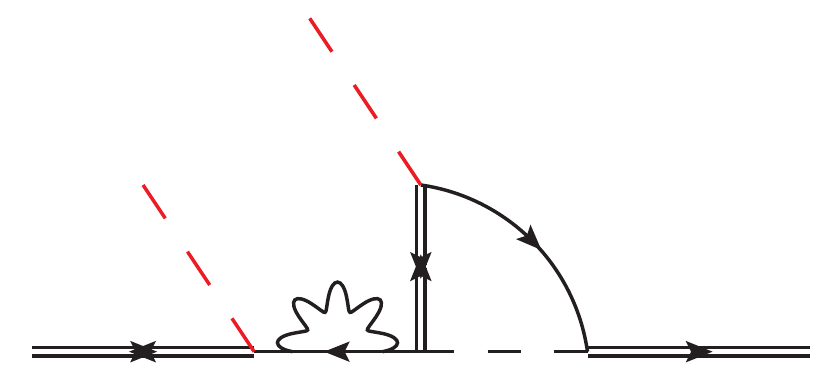}
\caption{Diagram as in figure~\ref{fig:diagrams1_CPdege}.}
\label{fig:diagram_single_CPdege} 
\end{figure}

The diagram in figure~\ref{fig:diagram_single_CPdege} does not contribute as well to the CP asymmetry. 
Indeed, once it has been cut in a way that the lepton and Higgs boson are on shell, 
what is left is a subdiagram with a vanishing imaginary part in Landau gauge. 
This has been shown by direct computation in~\cite{Biondini:2013xua}\footnote{
See figure~4, diagram 5), and eq.~(A.8) there.} or see~\ref{zero_special_case}.

\begin{figure}[ht]
\centering
\includegraphics[scale=0.5]{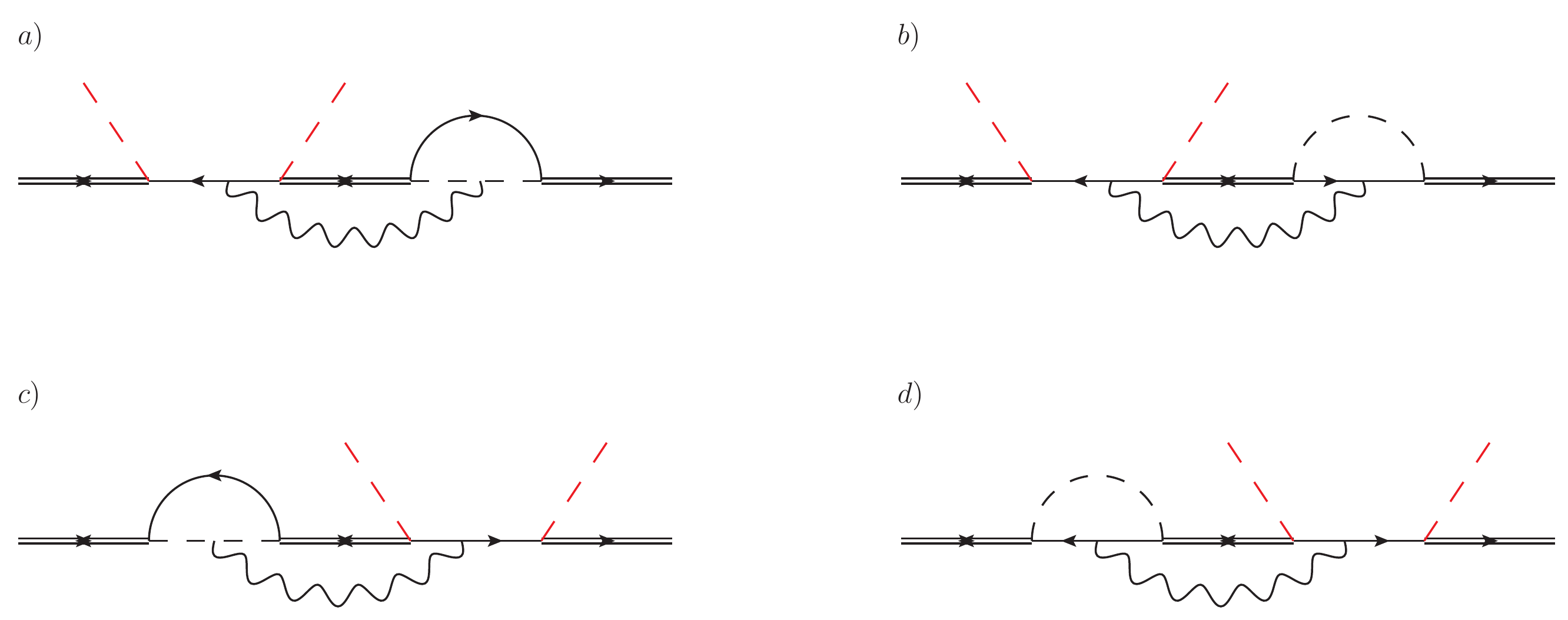}
\caption{Four diagrams that would be resonant without the gauge boson.
Only diagrams proportional to $(F_{1}^{*} F_{2})^2$ are displayed.}
\label{fig:fig_ind_new_CPdege} 
\end{figure}

\begin{figure}[ht]
\centering
\includegraphics[scale=0.5]{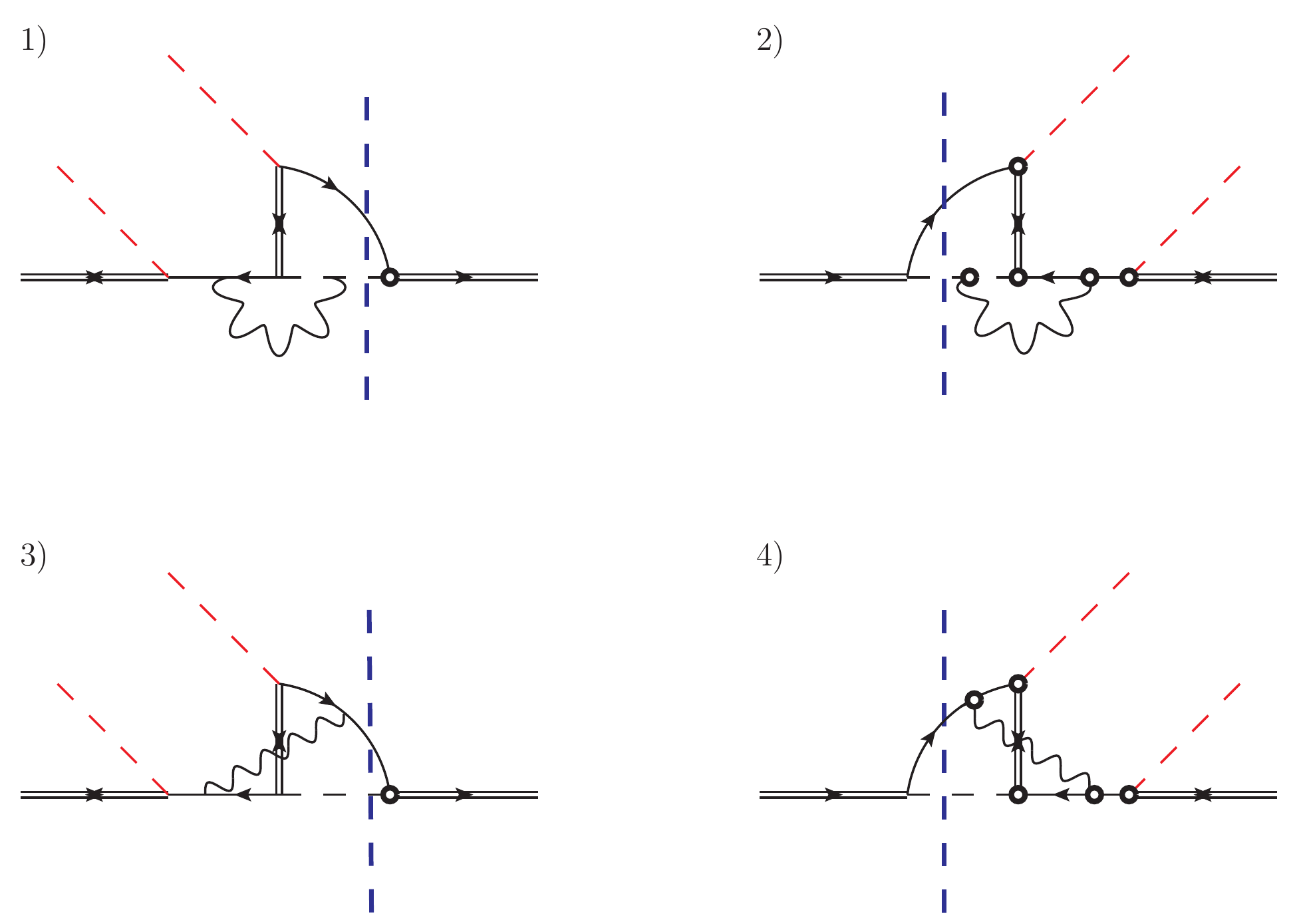}
\caption{On each raw we show the diagrams $a)$ and $b)$ of figure~\ref{fig:diagrams1_CPdege} 
together with their complex conjugates. Higgs bosons and leptons are cut.}
\label{appendix3_CPdege}
\end{figure}

\begin{figure}[ht]
\centering
\includegraphics[scale=0.5]{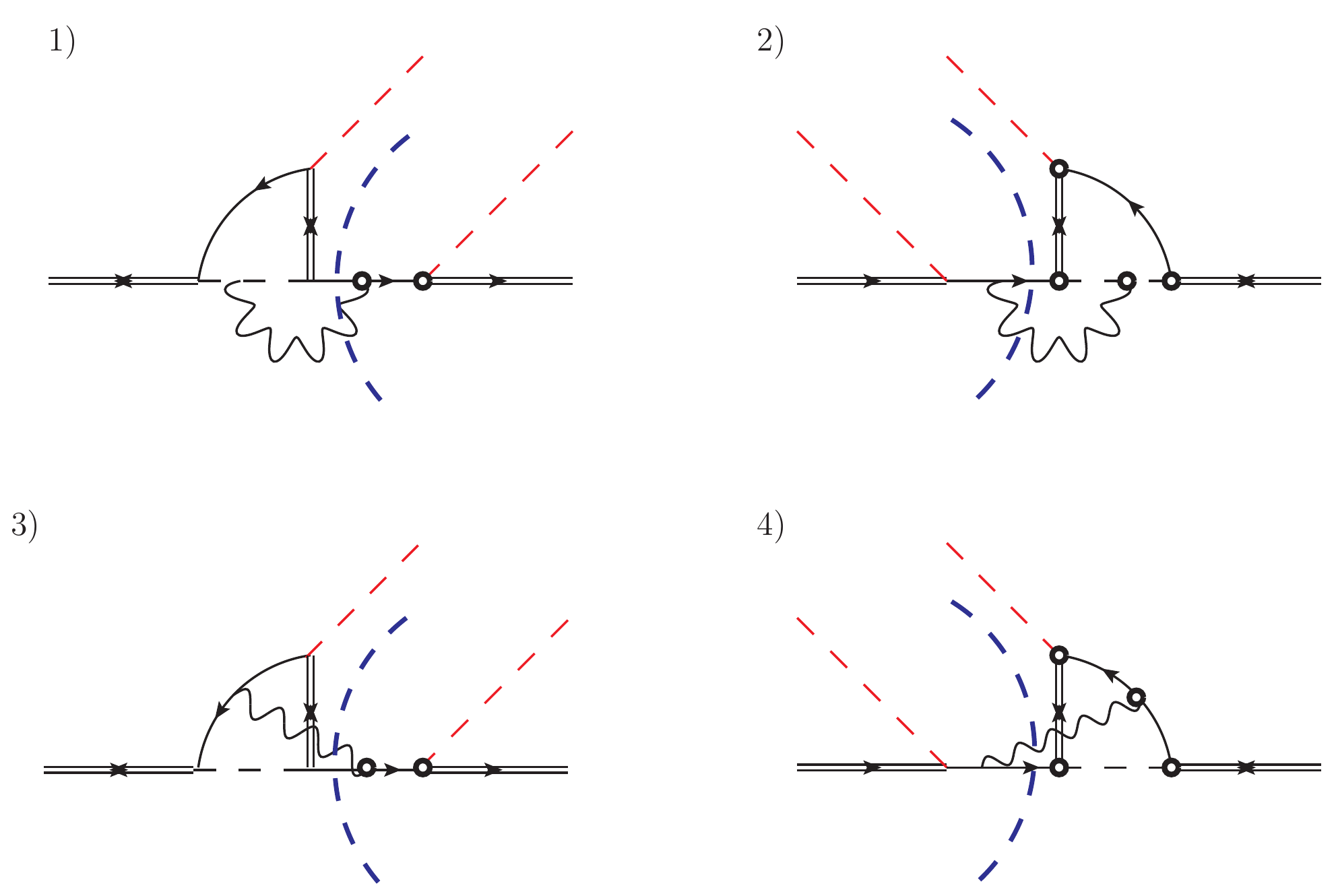}
\caption{On each raw we show the diagrams $d)$ and $e)$ of figure~\ref{fig:diagrams1_CPdege} 
together with their complex conjugates. Gauge bosons and leptons are cut.}
\label{appendix5_CPdege} 
\end{figure}

\begin{figure}[ht]
\centering
\includegraphics[scale=0.5]{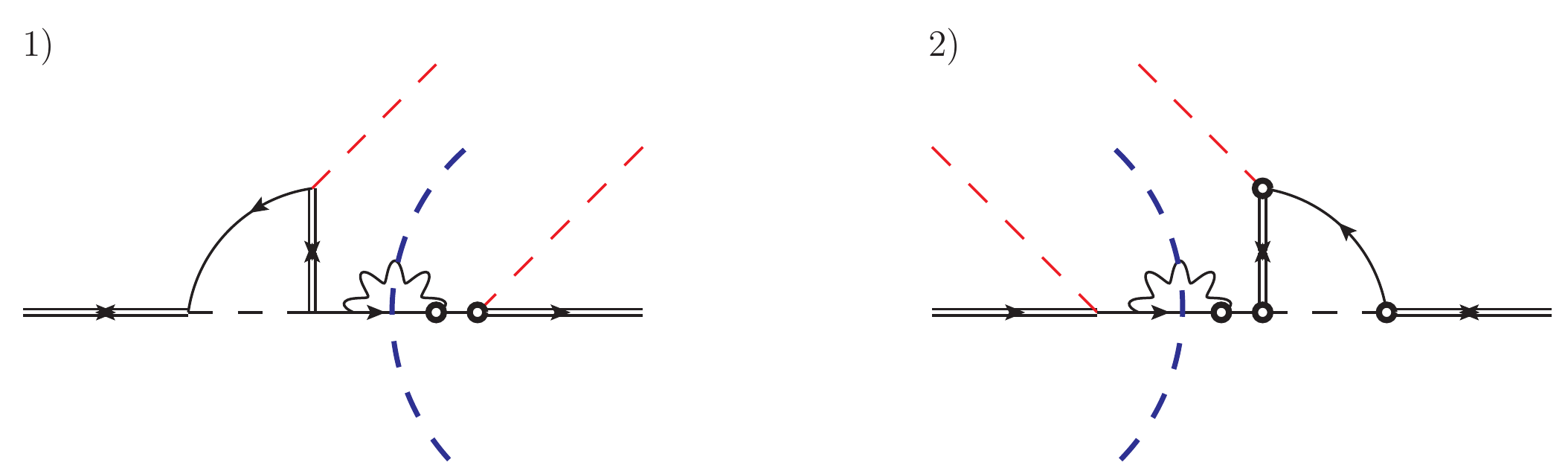}
\caption{Diagram $f)$ of figure~\ref{fig:diagrams1_CPdege} together with its complex conjugate. Gauge bosons and leptons are cut.}
\label{appendix6_CPdege} 
\end{figure}

We compute now the part of $a^\ell_{11}$ relevant for the CP asymmetry coming from the diagrams 
of figure~\ref{fig:diagrams1_CPdege} that have not been already excluded on the basis of the previous arguments.
We organize the calculation as follows: first, we compute the cuts that go through the lepton 
but not the gauge boson, i.e., the gauge boson contributes only as a virtual particle in the loop, 
then we compute the cuts that go through both the lepton and the gauge boson.  
In figure~\ref{appendix3_CPdege}, we show the cuts in the first case, 
whereas in figure~\ref{appendix5_CPdege} and \ref{appendix6_CPdege} we show them in the second one. 
On each raw we draw a diagram and its complex conjugate. 
As argued before, cuts that do not leave a loop uncut do not generate any additional complex phase 
and therefore do not contribute to the CP asymmetry. These cuts are not displayed.

We start with computing the cuts shown in figure~\ref{appendix3_CPdege}. 
In Landau gauge, the result is 
\begin{eqnarray}
&&\!\!\!\!\!\!\!\! 
{\rm{Im}}\, (-i \mathcal{D}^{\ell}_{1,\hbox{\tiny fig.\ref{appendix3_CPdege}}}) + {\rm{Im}}\, (-i \mathcal{D}^{\ell}_{2,\hbox{\tiny fig.\ref{appendix3_CPdege}}}) = 0 \, , 
\label{boson1} \\
&&\!\!\!\!\!\!\!\! 
{\rm{Im}}\, (-i \mathcal{D}^{\ell}_{3,\hbox{\tiny fig.\ref{appendix3_CPdege}}}) + {\rm{Im}}\, (-i \mathcal{D}^{\ell}_{4,\hbox{\tiny fig.\ref{appendix3_CPdege}}}) = 
\nonumber\\
&&\hspace{2cm}
-  \frac{{\rm{Im}}\left[ (F^{*}_{1}F_{2})^2\right]}{(16 \pi)^2M}  \frac{3g^2+g'^{2}}{8} 
\left[ \ln 2 - \left( 1-\ln 2 \right) \frac{\Delta}{M}\right] \,\delta^{\mu \nu}  \delta_{mn} + \dots  \, ,  
\label{boson2}
\nn
\\
\phantom{x}
\end{eqnarray}
where the superscript $\ell$ refers to having cut a lepton line. 
The dots stand for higher-order terms in the $\Delta/M$ expansion and for terms that do not contribute to the CP asymmetry.

We compute now cuts through gauge bosons. As argued at the beginning of this section, we can use for this kind of cuts a different gauge, 
namely the Coulomb gauge. The result for the cuts shown in figure~\ref{appendix5_CPdege} reads
\begin{eqnarray}
&&\!\!\!\! {\rm{Im}}\, (-i \mathcal{D}^{\ell}_{1,\hbox{\tiny fig.\ref{appendix5_CPdege}}}) + {\rm{Im}}\, (-i \mathcal{D}^{\ell}_{2,\hbox{\tiny fig.\ref{appendix5_CPdege}}}) =
\nonumber \\
&& \hspace{1.2cm}
- \frac{{\rm{Im}}\left[ (F^{*}_{1}F_{2})^2\right]}{(16 \pi)^2M} \frac{3g^2+g'^2}{8}  \left(-1+\frac{\Delta}{M} \right) \delta^{\mu \nu} \delta_{mn}  + \dots  \, , 
\label{boson5}
\\
&&\!\!\!\! {\rm{Im}}\, (-i \mathcal{D}^{\ell}_{3,\hbox{\tiny fig.\ref{appendix5_CPdege}}}) + {\rm{Im}}\, (-i \mathcal{D}^{\ell}_{4,\hbox{\tiny fig.\ref{appendix5_CPdege}}}) = 
\nonumber \\
&& \hspace{1.2cm}
- \frac{{\rm{Im}}\left[ (F^{*}_{1}F_{2})^2\right]}{(16 \pi)^2M} \frac{3g^2+g'^2}{4}  
\left[  \left(1-\ln 2 \right) +\left( 2-3 \ln 2\right) \frac{\Delta}{M} \right]   \,  \delta^{\mu \nu} \delta_{mn} + \dots \, .
\label{boson6}
\nn
\\
\phantom{x}
\end{eqnarray}

Two more diagrams that contribute to the part of $a^\ell_{11}$ that matters for the CP asymmetry with the relevant cuts are shown in figure~\ref{appendix6_CPdege}. 
They give
\begin{equation}
{\rm{Im}}\, (-i \mathcal{D}^{\ell}_{1,\hbox{\tiny fig.\ref{appendix6_CPdege}}}) + {\rm{Im}}\, (-i \mathcal{D}^{\ell}_{2,\hbox{\tiny fig.\ref{appendix6_CPdege}}}) =
- \frac{{\rm{Im}}\left[ (F^{*}_{1}F_{2})^2\right]}{(16 \pi)^2M} \frac{3g^2+g'^{2}}{8} \left(1-\frac{\Delta}{M} \right) \delta^{\mu \nu} \delta_{mn}  + \dots\,.
\label{boson7}
\end{equation}

Finally, we consider the diagrams shown in figure~\ref{fig:fig_ind_new_CPdege}. Removing the gauge boson, these diagrams could become resonant and contribute 
to the indirect CP asymmetry discussed in section~\ref{sec:indirect}. Indeed their contribution is accounted for by the diagrams in 
the EFT shown in figure~\ref{fig:indirectEFT}. With the gauge bosons included these diagrams cannot become resonant when the gauge boson carries 
away an energy of order $M$ and, according to the definition adopted in this paper, they contribute to the direct CP asymmetry. 
Clearly they do contribute to the Wilson coefficients ${\rm{Im}} \,a_{II}^\ell$ and ${\rm{Im}} \,a_{II}^{\bar{\ell}}$.

\begin{figure}[ht]
\centering
\includegraphics[scale=0.5]{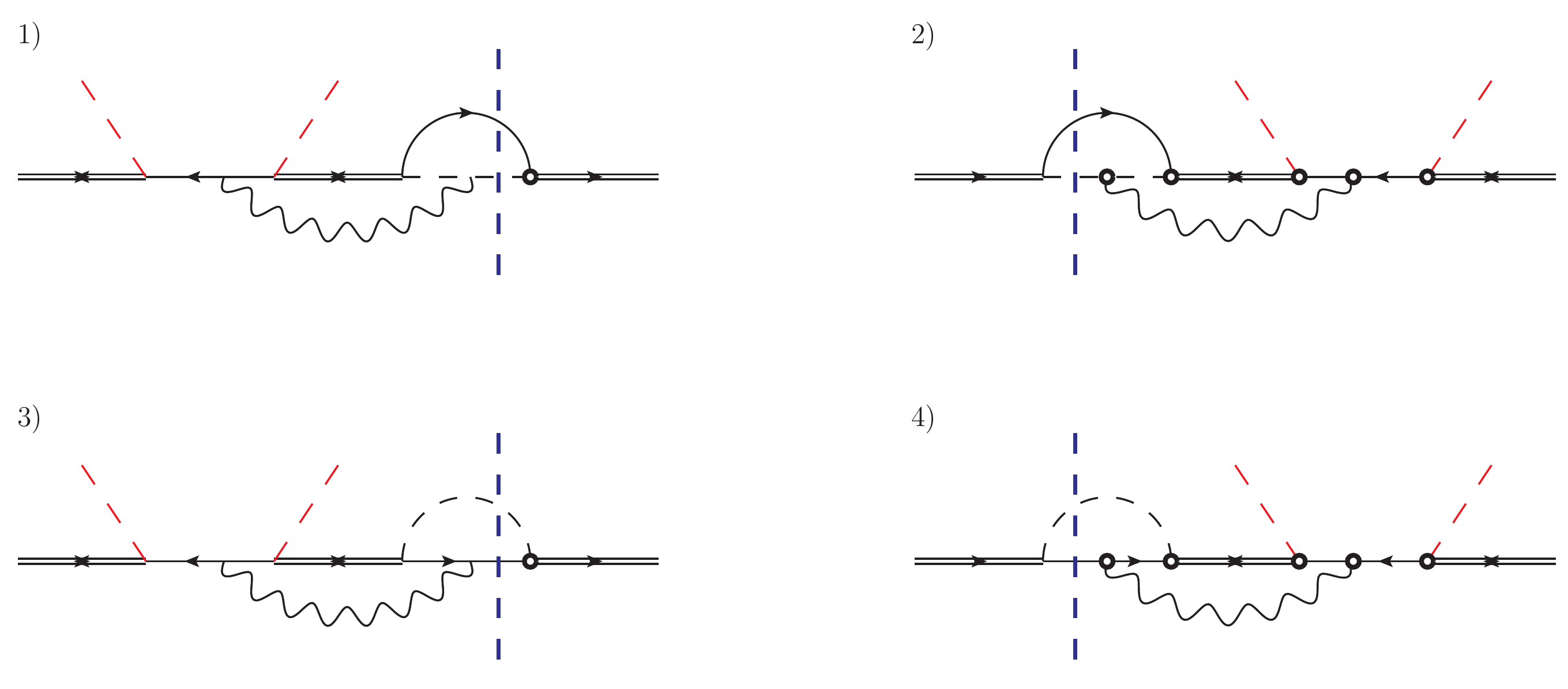}
\caption{On each raw we show the diagrams $a)$ and $b)$ of figure~\ref{fig:fig_ind_new_CPdege} together with their complex conjugates.
Higgs bosons and leptons are cut.}
\label{appendix7_CPdege}  
\end{figure}

As before, we start considering cuts through leptons and Higgs bosons. Only diagrams $a)$ and $b)$ of figure~\ref{fig:fig_ind_new_CPdege}
may be cut in this way and contribute to the CP asymmetry. The diagrams and the relevant cuts are shown in figure~\ref{appendix7_CPdege}. 
The result in Landau gauge reads
\begin{eqnarray}
&&\!\!\!\! {\rm{Im}}\, (-i \mathcal{D}^{\ell}_{1,\hbox{\tiny fig.\ref{appendix7_CPdege}}}) + {\rm{Im}}\, (-i \mathcal{D}^{\ell}_{2,\hbox{\tiny fig.\ref{appendix7_CPdege}}}) =0     \, , 
\label{boson8} \\
&&\!\!\!\! {\rm{Im}}\, (-i \mathcal{D}^{\ell}_{3,\hbox{\tiny fig.\ref{appendix7_CPdege}}}) + {\rm{Im}}\, (-i \mathcal{D}^{\ell}_{4,\hbox{\tiny fig.\ref{appendix7_CPdege}}}) = 
\nonumber \\
&& \hspace{1.2cm}
- \frac{{\rm{Im}}\left[ (F^{*}_{1}F_{2})^2\right]}{(16 \pi)^2M} \frac{g'^2}{4} \left(1-\frac{\Delta}{M} \right) \delta^{\mu \nu} \delta_{mn}  + \dots \, .
\label{boson9}
\end{eqnarray}

\begin{figure}[h!]
\centering
\includegraphics[scale=0.5]{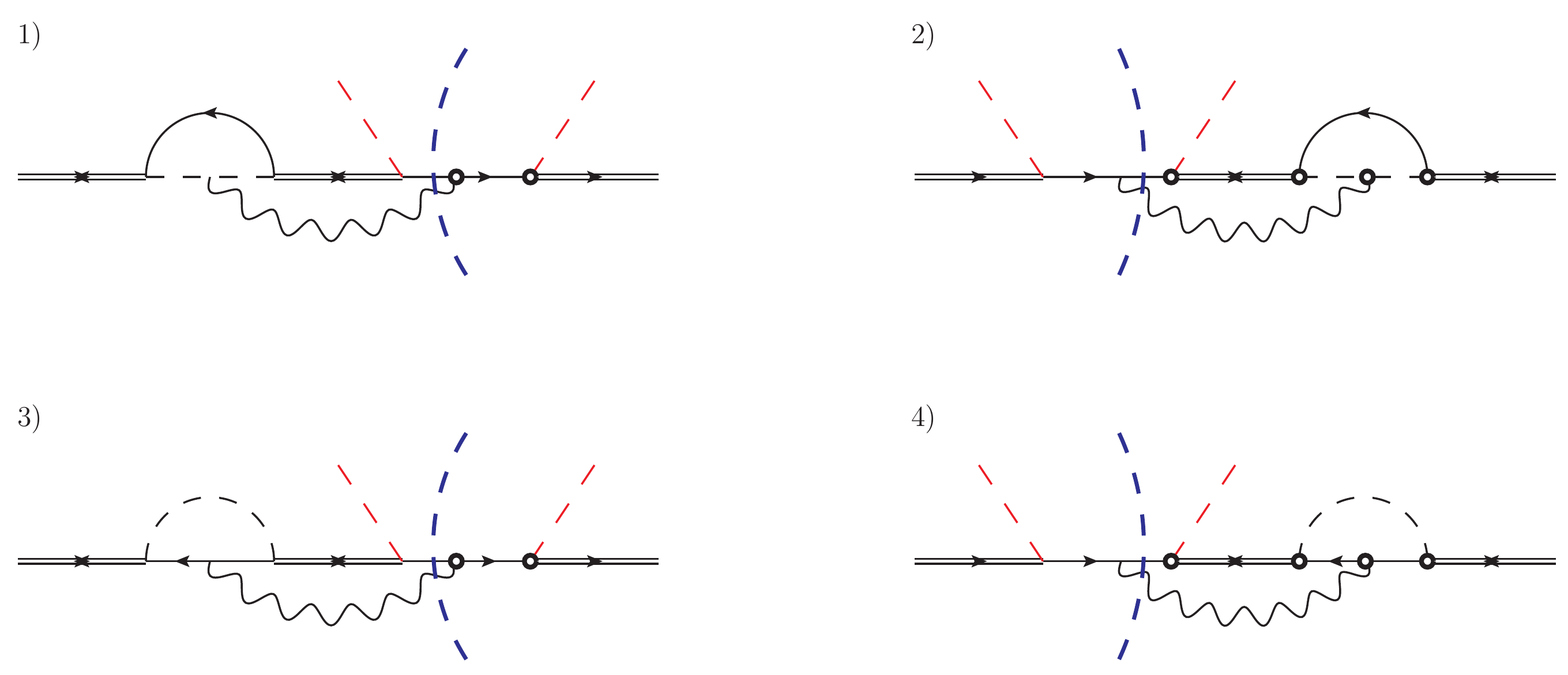}
\caption{On each raw we show the diagrams $c)$ and $d)$ of figure~\ref{fig:fig_ind_new_CPdege} together with their complex conjugates.
Gauge bosons and leptons are cut.}
\label{appendix8_CPdege} 
\end{figure}

On the other hand, only diagrams $c)$ and $d)$ of figure~\ref{fig:fig_ind_new_CPdege}
may be cut through a lepton and a gauge boson. The diagrams and the relevant cuts are shown in figure~\ref{appendix8_CPdege}. 
The result in Coulomb gauge reads
\begin{eqnarray}
&&\!\!\!\! {\rm{Im}}\, (-i \mathcal{D}^{\ell}_{1,\hbox{\tiny fig.\ref{appendix8_CPdege}}}) + {\rm{Im}}\, (-i \mathcal{D}^{\ell}_{2,\hbox{\tiny fig.\ref{appendix8_CPdege}}}) =
\nonumber \\
&& \hspace{1.2cm}
- \frac{{\rm{Im}}\left[ (F^{*}_{1}F_{2})^2\right]}{(16 \pi)^2M} \frac{g'^2}{4}  \left(-1+\frac{\Delta}{M} \right) \delta^{\mu \nu} \delta_{mn}  + \dots  \, , 
\label{boson10} \\
&&\!\!\!\! {\rm{Im}}\, (-i \mathcal{D}^{\ell}_{3,\hbox{\tiny fig.\ref{appendix8_CPdege}}}) + {\rm{Im}}\, (-i \mathcal{D}^{\ell}_{4,\hbox{\tiny fig.\ref{appendix8_CPdege}}}) = 
\nonumber \\
&& \hspace{1.2cm}
- \frac{{\rm{Im}}\left[ (F^{*}_{1}F_{2})^2\right]}{(16 \pi)^2M} \frac{g'^2}{4}  
\left(1-\frac{\Delta}{M} \right) \delta^{\mu \nu} \delta_{mn}  + \dots \, .
\label{boson11}
\end{eqnarray}

Summing up all diagrams \eqref{Higgs1}-\eqref{boson11}, and comparing with the expression of 
the matrix element \eqref{B1_CPdege} in the EFT, which is $({\rm{Im}} \,a_{11}^\ell/M) \delta^{\mu \nu} \delta_{mn}$ for the leptonic contribution 
and $({\rm{Im}} \,a_{11}^{\bar{\ell}}/M) \delta^{\mu \nu} \delta_{mn} $ for the antileptonic one, we obtain~\eqref{match1}.
The expression for the Wilson coefficient involving the Majorana neutrino of type 2 can be inferred from the above 
results after the substitutions $F_1 \leftrightarrow F_2$, $M \to M_2$ and $\Delta \to -\Delta$ in~\eqref{Higgs1}-\eqref{boson11} or just in~\eqref{match1}. 
The result, in terms of the lightest neutrino mass, $M$, has been written in~\eqref{match2}.
That the above substitutions work follows from the fact that the real transition from a heavier neutrino of type 2 to a lighter neutrino of type 1, 
which is a decay channel absent in the case of neutrinos of type 1, is a process accounted for by the EFT (see section~\ref{sec:direct2}), 
and, therefore, it does not contribute to the matching. 
In fact, the energy emitted in such a transition is of order~$\Delta$; 
this is, in the nearly degenerate case considered in this work, much smaller than~$M$. 

\section{Matching in the flavoured case}
\label{appflaCPdege}
There are diagrams contributing to the matching coefficients ${\rm{Im}} \,a_{II}^\ell$ and ${\rm{Im}} \,a_{II}^{\bar{\ell}}$ 
that are relevant only for the flavoured CP asymmetry.
These are diagrams involving only lepton (or antilepton) propagators. 
They could contribute to the CP asymmetry with terms proportional to ${\rm{Im}}\left[(F_1F_2^{*})(F^*_{f1}F_{f2})\right]$.
Clearly such terms vanish in the unflavoured case. 
Here we examine these diagrams and find that they do not contribute.

\begin{figure}[ht]
\centering
\includegraphics[scale=0.53]{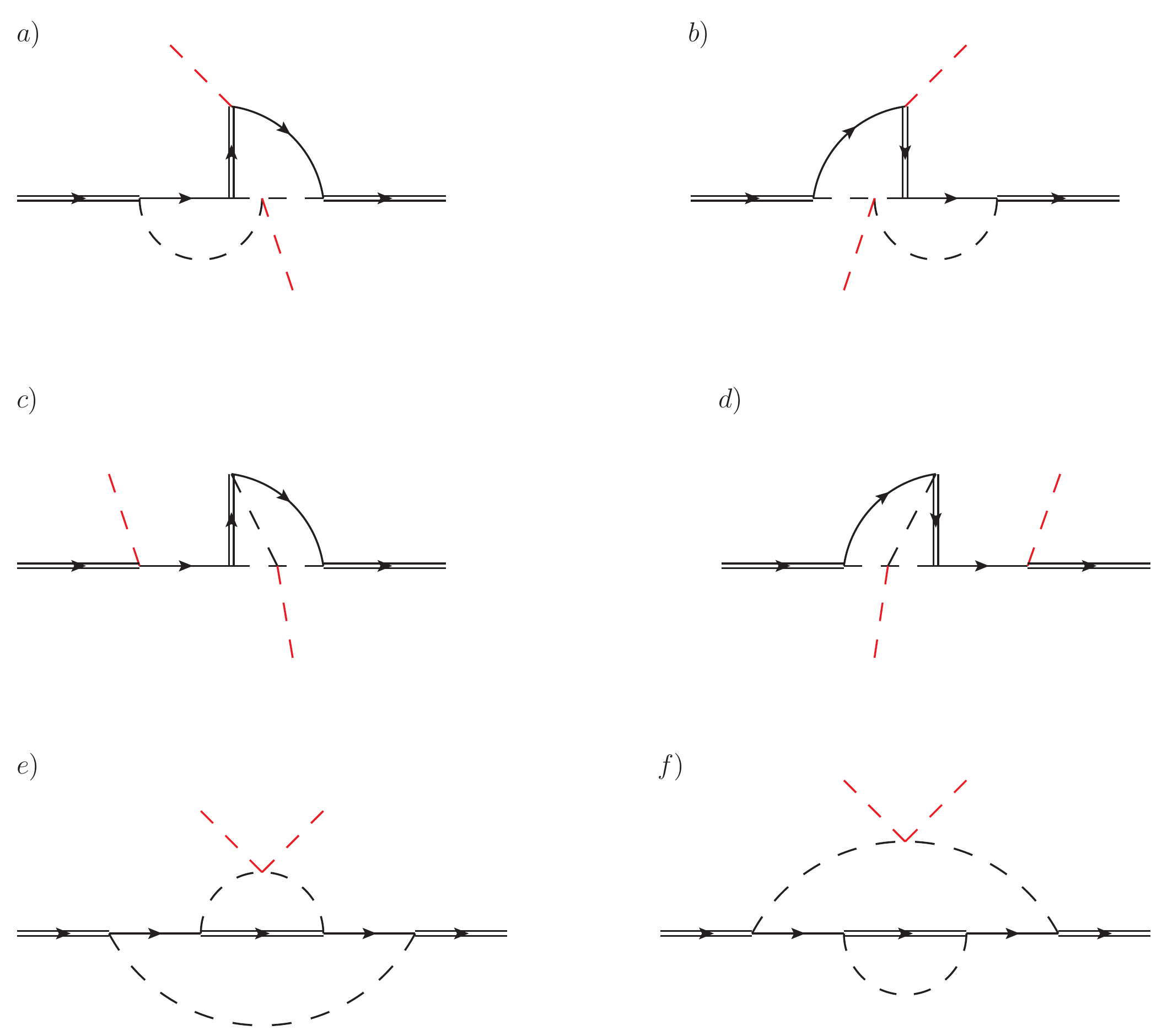}
\caption{Diagrams contributing to the matching coefficients \eqref{match1} and \eqref{match2} involving the four-Higgs coupling.
Diagrams $a)$-$d)$ may be inferred from the diagrams of figure~\ref{fig:new_Higgs_direct} by changing an antilepton line in a lepton line.
 The topologies of diagrams $e)$ and $f)$ are relevant only for the flavoured case. 
 We display only diagrams that admit leptonic cuts.}
\label{fig:flavor_Higgs} 
\end{figure} 

\begin{figure}[ht]
\centering
\includegraphics[scale=0.53]{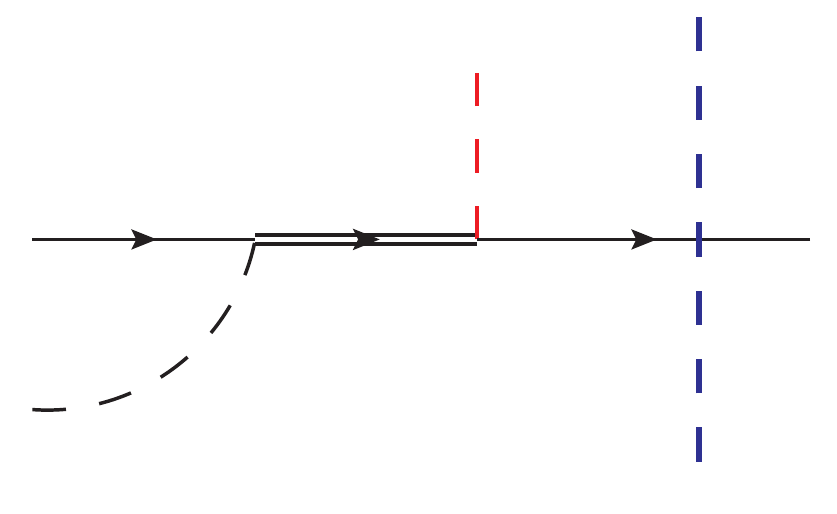}
\caption{The blue dashed line on the right is the cut, the red central dashed line is an external Higgs boson whose momentum can be set to zero and
the black dashed line on the left may identify a Higgs boson in a loop or an external one.}
\label{fig:flavour_cut} 
\end{figure} 

\begin{figure}[ht]
\centering
\includegraphics[scale=0.53]{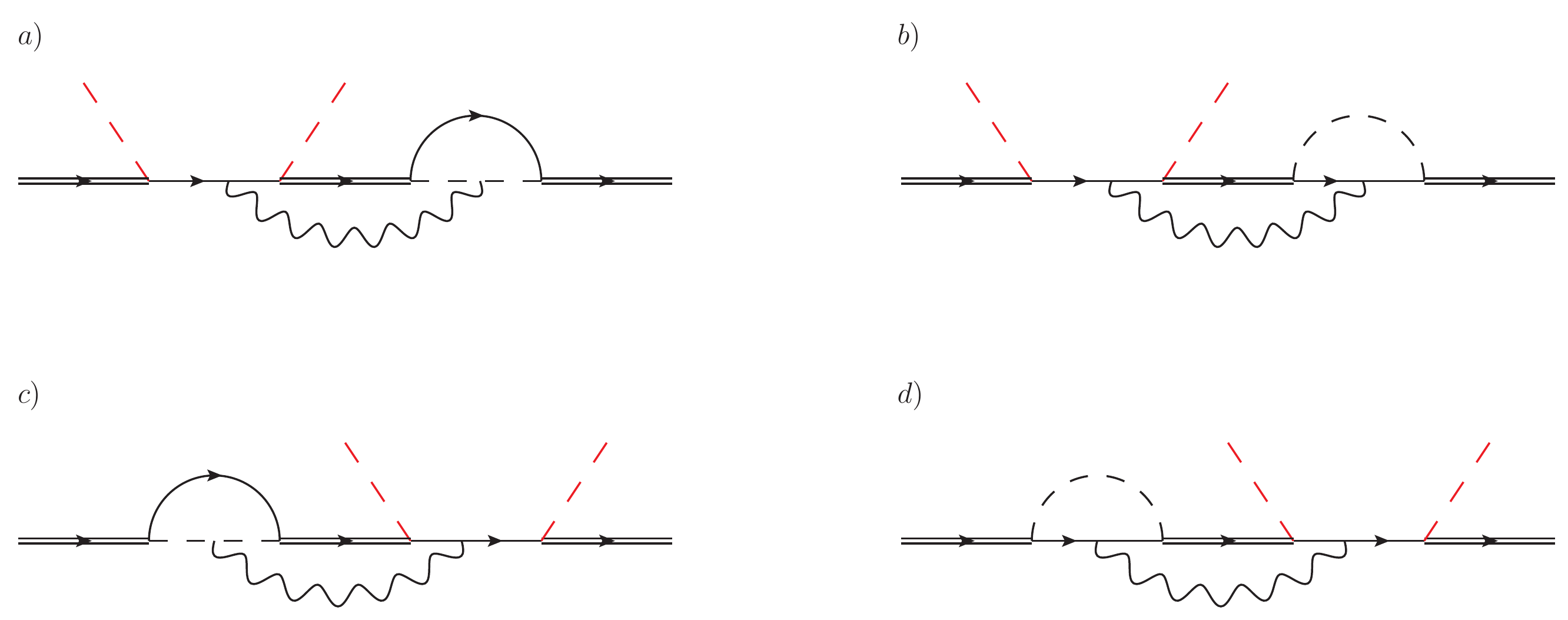}
\caption{Diagrams contributing to the matching coefficients \eqref{match1} and \eqref{match2} involving gauge couplings.
The diagrams may be inferred from the diagrams of figure~\ref{fig:fig_ind_new_CPdege} by changing an antilepton line in a lepton line.
We display only diagrams that admit leptonic cuts.}
\label{fig:flavor_gauge} 
\end{figure}

\begin{figure}[ht]
\centering
\includegraphics[scale=0.53]{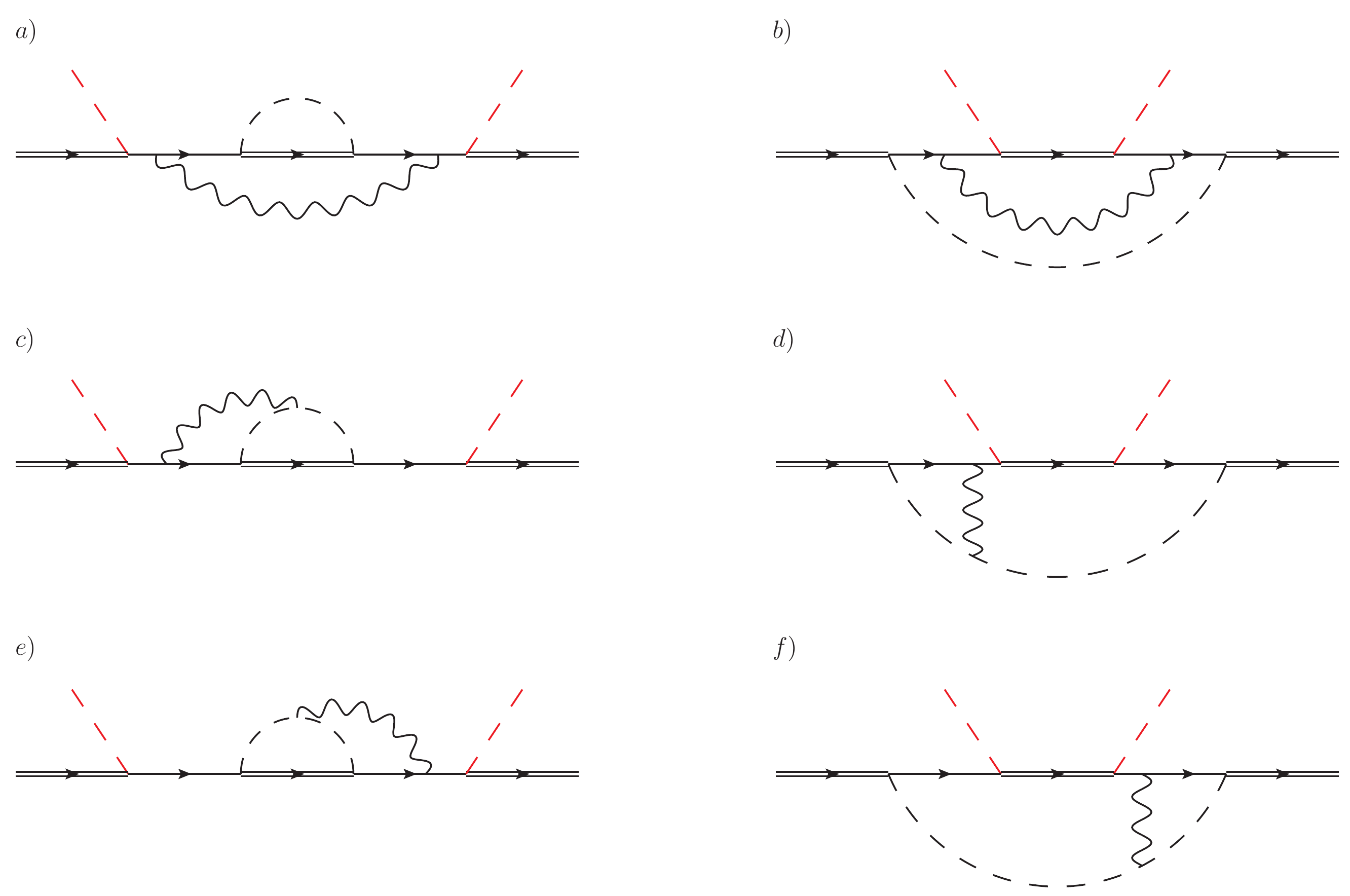}
\caption{Diagrams contributing to the matching coefficients \eqref{match1} and \eqref{match2} involving gauge couplings.
The topologies of these diagrams are relevant only for the flavoured case. 
We display only diagrams that admit leptonic cuts.}
\label{fig:flavor_gauge_rainbow} 
\end{figure}

We may divide these diagrams into two classes: diagrams that involve the four-Higgs coupling, shown in figure~\ref{fig:flavor_Higgs}, 
and diagrams involving gauge couplings, shown in figures~\ref{fig:flavor_gauge} and~\ref{fig:flavor_gauge_rainbow}.
Let us consider diagram $a)$ of figure~\ref{fig:flavor_Higgs}. If we cut the lepton in the loop on the right, then the cut gives rise to 
the Feynman subdiagram shown in figure~\ref{fig:flavour_cut}. This is proportional to ($\ell^\mu$ is the momentum of the lepton) 
\begin{equation}
\delta(\ell^2)\slashed{\ell} \, P_R \, \frac{\slashed{\ell}+M_J}{\ell^2-M_J^2+i\eta}\, P_L = P_L \, \delta(\ell^2) \ell^2 \frac{1}{\ell^2-M_J^2+i\eta} = 0,
\label{flavour_identity}
\end{equation} 
and therefore vanishes.\footnote{
The corresponding Feynman subdiagram of 1) in figure~\ref{fig:new_Higgs_direct} involves a neutrino propagator of the type \eqref{eq6_partprod}
and an antilepton on the left. Hence it is proportional to 
$$
\delta(\ell^2)\slashed{\ell} \, P_R \, \frac{\slashed{\ell}+M_J}{\ell^2-M_J^2+i\eta}\, P_R = 
P_L \, \delta(\ell^2) \slashed{\ell} M_J \frac{1}{\ell^2-M_J^2+i\eta} \neq 0.
$$
} 
If we cut the lepton in the loop on the left, then we need the imaginary part of the remaining (uncut) loop on the right.
The imaginary part of the loop on the right may be computed by considering all its possible cuts. Those include cuts through the lepton, which 
vanish according to the above argument, cuts through the Higgs-boson propagator, which vanish because they involve three massless on-shell particles 
entering the same vertex, and cuts through the Majorana-neutrino propagator, which are either kinematically forbidden or involve momenta 
of order $\Delta$ that are accounted for by the EFT (for more details see the discussion at the end of section~\ref{appHiggs}).

The same arguments may be applied to all remaining diagrams shown in figures~\ref{fig:flavor_Higgs}, \ref{fig:flavor_gauge} and \ref{fig:flavor_gauge_rainbow}.
In particular, for many of them the argument based on the identity~\eqref{flavour_identity} is crucial.
The identity~\eqref{flavour_identity} is relevant only for the flavoured case.

%% file: CPhieramatch.tex
In this appendix we present the diagrams necessary to obtain the matching coefficients of the EFT$_2$ in (\ref{match_a})-(\ref{match_cLbar}). The Wilson coefficients are obtained by matching four-point Green's functions calculated in the EFT$_1$ with four-point Green's functions in the EFT$_2$. Since we are going to consider the effects induced by  the particle of the thermal bath, we need to specify the SM Lagrangian that reads off~\eqref{SMlag}.
We can set the temperature to zero so that all loop diagrams in the EFT$_2$ are scaleless in dimensional regularization. This comes from the fact that we integrate out high energy modes, of order $M_1 \gg T$,  and any other low energy scale can be put to zero. Dimensional regularization is used for all the loop diagrams that we discuss in the following. The operators that we need to consider are the dimension-five heavy neutrino-Higgs operator and the dimension-seven heavy neutrino-top (heavy-quark doublet), heavy neutrino-lepton doublet operators. Therefore we consider matrix elements with external heavy neutrinos and Higgs bosons in section~\ref{five_hiera_match}, whereas top quarks (heavy-quark doublets) and lepton doublets are external legs together with heavy neutrinos in section~\ref{seven_hiera_match}.

\section{EFT$_1$: matching dimension-five and dimension-six operators}
In this appendix we give some details on the derivation of the CP asymmetry presented in section~\ref{CPhiera_sec1} in the limit $M_1 \ll M_i$, and the tree-level matching to define the parameters of the EFT$_1$. To keep the notation simple, we drop
propagators on external legs, and we label the so-obtained amputated Green's functions 
with the same indices used for the unamputated ones. 

Let us start with the calculation of the matching coefficient $\eta_{f,f'}$ of the effective dimension-five operator in (\ref{Lag2_hiera}). In order to carry out the tree level matching we consider the following matrix element in the full theory in (\ref{lepto_8}) and in the EFT$_1$ in (\ref{Lag2_hiera}):
\begin{eqnarray}
&&-i \int d^{4}x\,e^{i p_1 \cdot x} \int d^{4}ye^{i k_1 \cdot y} \int d^{4}z\,e^{i k_2 \cdot z}\, 
\langle \Omega |T( L^{\mu}_{h,m}(x) L^{\nu}_{h',n}(0) \phi_{r}(y) \phi_{s}(z) ) | \Omega \rangle \, ,
\label{matrix4part}
\nn
\\
\end{eqnarray}
where $h,h'$ are flavour indices.
The result when evaluating the the matrix element in (\ref{matrix4part}) in the fundamental theory reads
\begin{eqnarray}
&&-i \int d^{4}x\,e^{i p_1 \cdot x} \int d^{4}ye^{i k_1 \cdot y} \int d^{4}z\,e^{i k_2 \cdot z}\, 
\langle \Omega |T( L^{\mu}_{h,m}(x) L^{\nu}_{h',n}(0) \phi_{r}(y) \phi_{s}(z) ) | \Omega \rangle \, = \nonumber \\
&& \hspace{4.5 cm} \frac{F_{h,i} F_{h',i}}{M_i} (P_R C)^{\mu \nu} (\sigma^2_{mr} \sigma^2_{ns}+\sigma^2_{ms} \sigma^2_{nr} ) \, ,
\label{matrix4partbis}
\end{eqnarray}
whereas the result in the EFT$_1$ is
\begin{eqnarray}
&&-i \int d^{4}x\,e^{i p_1 \cdot x} \int d^{4}ye^{i k_1 \cdot y} \int d^{4}z\,e^{i k_2 \cdot z}\, 
\langle \Omega |T( L^{\mu}_{h,m}(x) L^{\nu}_{h',n}(0) \phi_{r}(y) \phi_{s}(z) ) | \Omega \rangle \, = \nonumber \\
&& \hspace{4.5 cm} +\frac{2 \eta^i_{f,f'}}{M_i}  \delta_{f,h} \delta_{f',h'} (P_R C)^{\mu \nu} (\sigma^2_{mr} \sigma^2_{ns}+\sigma^2_{ms} \sigma^2_{nr} ) \, .
\label{matrix4parttris}
\end{eqnarray}
Then comparing (\ref{matrix4partbis}) and (\ref{matrix4parttris}), we find the matching coefficient to be $\eta^i_{f,f'}$ in (\ref{eq8}). The Wilson coefficients of the dimension-six operators in (\ref{Lag2_fla}) can be obtained the same way. For that describing the $\ell \phi \to \ell \phi$ scattering we find
\begin{eqnarray}
&&-i \int d^{4}x\,e^{i p_1 \cdot x} \int d^{4}ye^{i k_1 \cdot y} \int d^{4}z\,e^{-i k_2 \cdot z}\, 
\langle \Omega |T( L^{\mu}_{h,m}(x) \bar{L}^{\nu}_{h',n}(0) \phi_{r}(y) \phi_{s}^\dagger(z) ) | \Omega \rangle \,  =
\nonumber \\
&& \hspace{4.5 cm} \frac{F_{h,i} F_{h',i}}{M^2_i} P_R^{\mu \nu} (\slashed{p}_1+\slashed{k}_1) \sigma^2_{mr} \sigma^2_{sn} \, ,
\label{matrix4ll}
\end{eqnarray}
and from the EFT$_1$ side one finds correspondingly
\begin{eqnarray}
&&-i \int d^{4}x\,e^{i p_1 \cdot x} \int d^{4}ye^{i k_1 \cdot y} \int d^{4}z\,e^{-i k_2 \cdot z}\, 
\langle \Omega |T( L^{\mu}_{h,m}(x) \bar{L}^{\nu}_{h',n}(0) \phi_{r}(y) \phi_{s}^\dagger(z) ) | \Omega \rangle \,  =
\nonumber \\
&& \hspace{4.5 cm} \frac{\tilde{\eta}^i_{f,f'}  \delta_{f,h} \delta_{f',h'} }{M_i} P_R^{\mu \nu} (\slashed{p}_1+\slashed{k}_1) \sigma^2_{mr} \sigma^2_{sn} \, .
\label{matrix4llEFT}
\end{eqnarray}
The energy-momentum conservation in the s-channel allows for interchanging the sum $p_1+k_1$ with $p_2+k_2$ and comparing \eqref{matrix4ll} with \eqref{matrix4llEFT} one finds $\tilde{\eta}^i_{f,f'}$ in (\ref{new_math_dim_six}).

\section{Matching the dimension-five operator in EFT$_2$}
\label{five_hiera_match}
In order to determine the Wilson coefficient of the dimension-five operator we consider the following matrix element in the heavy Majorana neutrino rest frame 
\begin{equation}
-i \left.\int d^{4}x\,e^{i p \cdot x} \int d^{4}y \int d^{4}z\,e^{i q \cdot (y-z)}\, 
\langle \Omega | T(\psi^{\mu}_{1}(x) \bar{\psi}^{\nu }_{1}(0) \phi_{m}(y) \phi_{n}^{\dagger}(z) )| \Omega \rangle
\right|_{p^\alpha =(M_1 + i\eta,\bm{0}\,)},
\label{B1_hiera}
\end{equation} 
where $\mu$ and $\nu$ are Lorentz indices, $m$ and $n$ are SU(2) indices.  The matrix element \eqref{B1_hiera} can be understood as a $2 \rightarrow 2$ scattering between a heavy Majorana neutrino at rest and a Higgs boson carrying momentum $q^\mu$ much smaller than $M_1$.  We divide the analysis of the diagrams as follows. First we discuss diagrams involving the Higgs self-coupling, $\lambda$, and then we address the case of diagrams with gauge bosons.

\begin{figure}[h]
\centering
\includegraphics[scale=0.55]{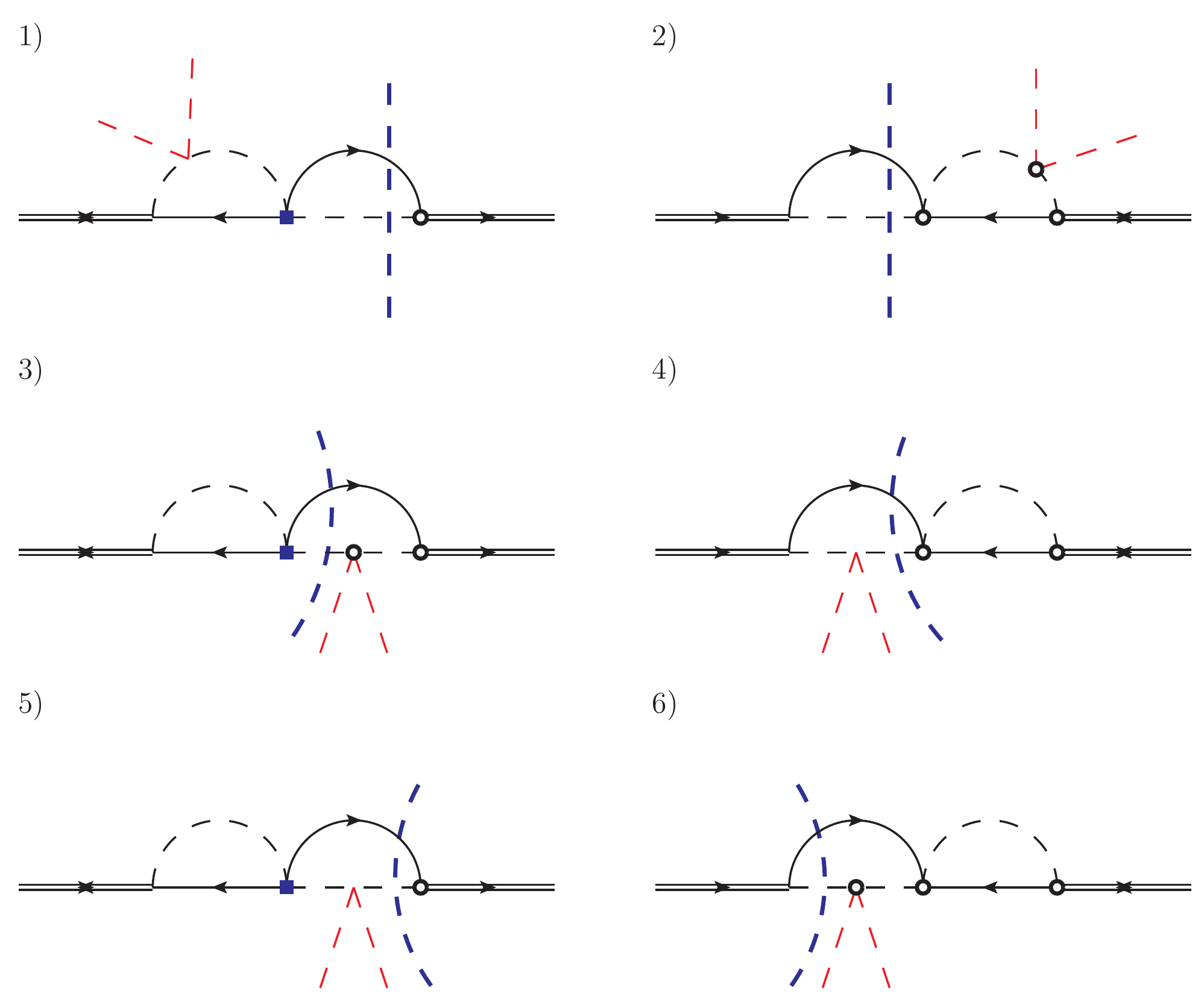}
\caption{\label{fig:Higgsmatch_hiera} First set of diagrams involving the Higgs self-coupling $\lambda$ and the corresponding cuts that put leptons on shell. The dashed blue line stands for the cut and the circled vertices are explicitly shown.  }
\end{figure}

\begin{figure}[h]
\centering
\includegraphics[scale=0.55]{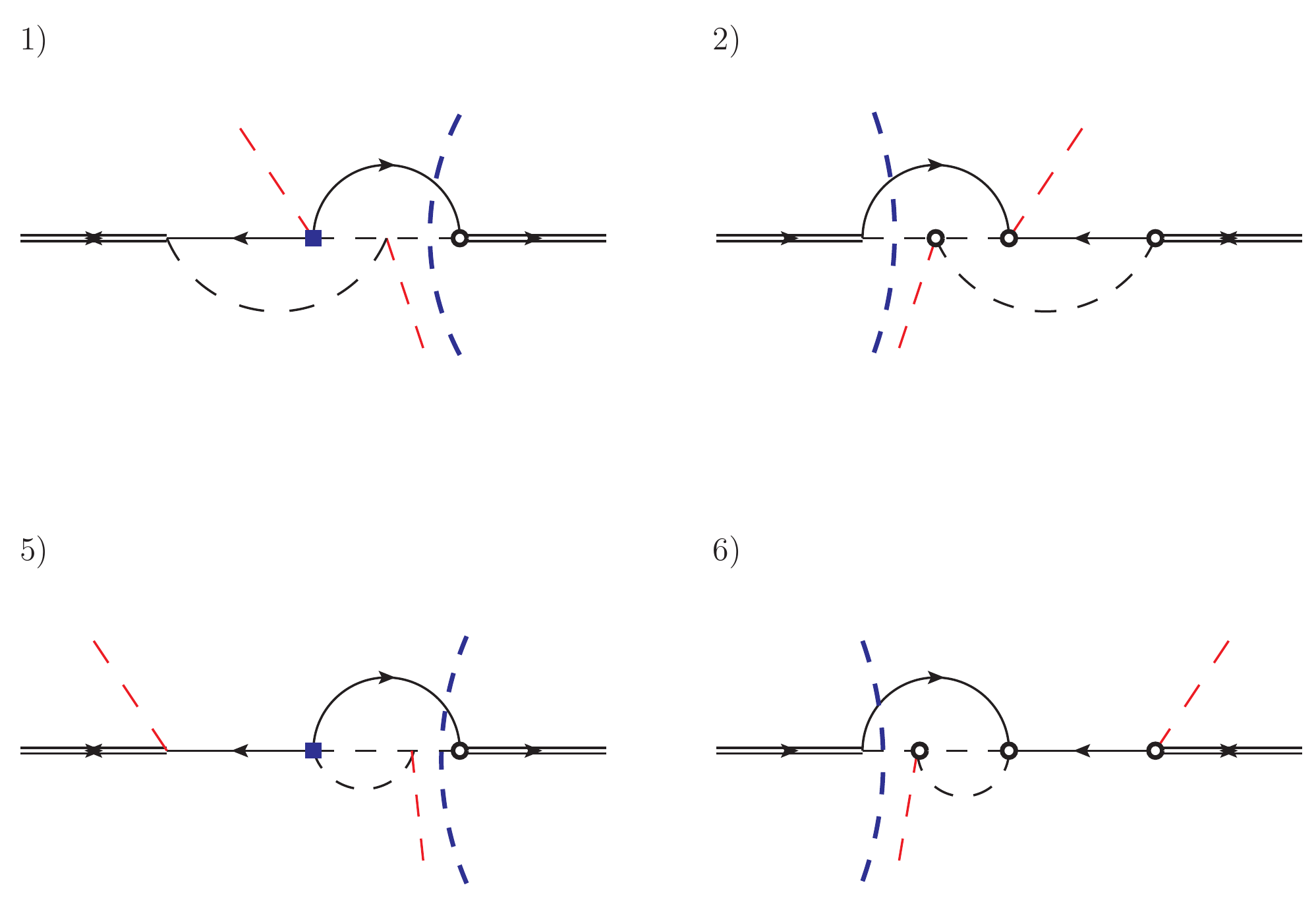}
\caption{\label{fig:Higgsmatchbis_hiera} Second set of diagrams involving the Higgs self-coupling $\lambda$ and the corresponding cuts that put leptons on shell.  }
\end{figure}

In figure \ref{fig:Higgsmatch_hiera} and \ref{fig:Higgsmatchbis_hiera} we show the diagrams contributing to the  Wilson coefficient of the dimension-five operator that involve the Higgs self-coupling. In each raw we show a diagram and its complex conjugate and we draw explicitly the cut that put a lepton on shell (dashed blue line). 
The first set of diagrams in figure \ref{fig:Higgsmatch_hiera} is obtained by adding a four-Higgs vertex to the diagrams in figure \ref{fig:eft1cp}.  
On the other hand, one can also open up one of the Higgs propagator in the two-loop diagrams in figure \ref{fig:eft1cp}. In this way one reduces to one loop diagrams. However, we can add a four-Higgs vertex to the remaining internal Higgs line and a two-loop diagram can be again obtained by connecting one of the pre-existing external legs with one of those induced by the four-Higgs vertex. These diagrams are shown in figure \ref{fig:Higgsmatchbis_hiera}. 
Now we list the result of the diagrams and we start with those in figure \ref{fig:Higgsmatch_hiera}. We show the case with cuts on the lepton and we obtain 
\begin{eqnarray}
&& {\rm{Im}}\, (-i \mathcal{D}^{\ell}_{1,\hbox{\tiny fig.\ref{fig:Higgsmatch_hiera}}})+{\rm{Im}}\, (-i \mathcal{D}^{\ell}_{2,\hbox{\tiny fig.\ref{fig:Higgsmatch_hiera}}}) = \frac{\lambda}{M_i}\frac{9 }{(16 \pi)^2}  {\rm{Im}}\left[ \left( F^{*}_1 F_i\right)^2\right] \delta^{\mu \nu}  \delta_{mn} 
 + \dots \, , 
\nn
\\
\label{B2_hiera}  
\\
&&{\rm{Im}}\, (-i \mathcal{D}^{\ell}_{3,\hbox{\tiny fig.\ref{fig:Higgsmatch_hiera}}})+ {\rm{Im}}\, (-i \mathcal{D}^{\ell}_{4,\hbox{\tiny fig.\ref{fig:Higgsmatch_hiera}}}) +{\rm{Im}}\, (-i \mathcal{D}^{\ell}_{5,\hbox{\tiny fig.\ref{fig:Higgsmatch_hiera}}})+ {\rm{Im}}\, (-i \mathcal{D}^{\ell}_{6,\hbox{\tiny fig.\ref{fig:Higgsmatch_hiera}}})  = \nonumber \\
&&\hspace{2.6 cm}\frac{\lambda}{M_i}\frac{9 }{(16 \pi)^2}  {\rm{Im}}\left[ \left( F^{*}_1 F_i\right)^2\right] \delta^{\mu \nu}  \delta_{mn}   
+ \dots \, 
\label{B3_hiera}
\end{eqnarray}
where the subscripts refer to the diagrams as listed in figure~\ref{fig:Higgsmatch_hiera} and the superscript $\ell$ stands for a lepton put on shell in the cuts.  %In the diagrams we have suppressed the symbol for the effective vertex that might be confused with the circled vertices due to the cutting rules.
The dots in eqs.~(\ref{B2_hiera}) and (\ref{B3_hiera}) stand for terms that are of higher order in the neutrino mass expansion and for terms proportional to the real part of the Yukawa $(F_1F_i^*)^2$, irrelevant for the calculation of the CP asymmetry. The result for the antileptons differs for an overall minus sign, according to the substitution $F_1 \leftrightarrow F_i$. 
We move to the diagrams shown in figure ~\ref{fig:Higgsmatchbis_hiera}, and we obtain 
\begin{eqnarray}
&&{\rm{Im}}\, (-i \mathcal{D}^{\ell}_{1,\hbox{\tiny fig.\ref{fig:Higgsmatchbis_hiera}}})+ {\rm{Im}}\, (-i \mathcal{D}^{\ell}_{2,\hbox{\tiny fig.\ref{fig:Higgsmatchbis_hiera}}})  = \frac{\lambda}{M_i}\frac{6 }{(16 \pi)^2}  {\rm{Im}}\left[ \left( F^{*}_1 F_i\right)^2\right] \delta^{\mu \nu}  \delta_{mn}   
+ \dots \, 
\nn
\\
\label{B4_hiera}
\\
&&{\rm{Im}}\, (-i \mathcal{D}^{\ell}_{3,\hbox{\tiny fig.\ref{fig:Higgsmatchbis_hiera}}})+ {\rm{Im}}\, (-i \mathcal{D}^{\ell}_{4,\hbox{\tiny fig.\ref{fig:Higgsmatchbis_hiera}}})  = 0 \, .
\label{B6_hiera}
\end{eqnarray}
We can understand the result in (\ref{B6_hiera}) as follows. After the cut on the lepton line the remaining loop amplitude gives a vanishing imaginary part. Indeed, as we notice in \cite{Biondini:2015gyw}, the momentum of the external Higgs boson can be put to zero and hence we have three massless particle entering the same vertex. In this case the corresponding phase space in dimensional regularization vanishes (we find the same situation in section~\ref{appHiggs}).
  
We move now to the diagrams that involve gauge bosons in the matching calculation. They contribute to the Wilson coefficient of the dimension-five operator, and induce a dependence on the couplings of the unbroken SU(2)$_{L}\times$U(1)$_{Y}$ gauge group, $g$ and $g'$ respectively. Differently to what happens at order $|F_1|^2$ for the thermal width, where the gauge interactions appear in the matching of dimension-seven operators \cite{Biondini:2013xua}, here, at order $(F_1^*F_i)^2$, they contribute already in the matching of the dimension-five operator in (\ref{Ope_Higgs}). We have discussed rather extensively how to address the calculation of diagrams involving the gauge bosons in appendix~\ref{appC:CPdegematch}, therefore we remind the main points in short. 

\begin{figure}[h]
\centering
\includegraphics[scale=0.48]{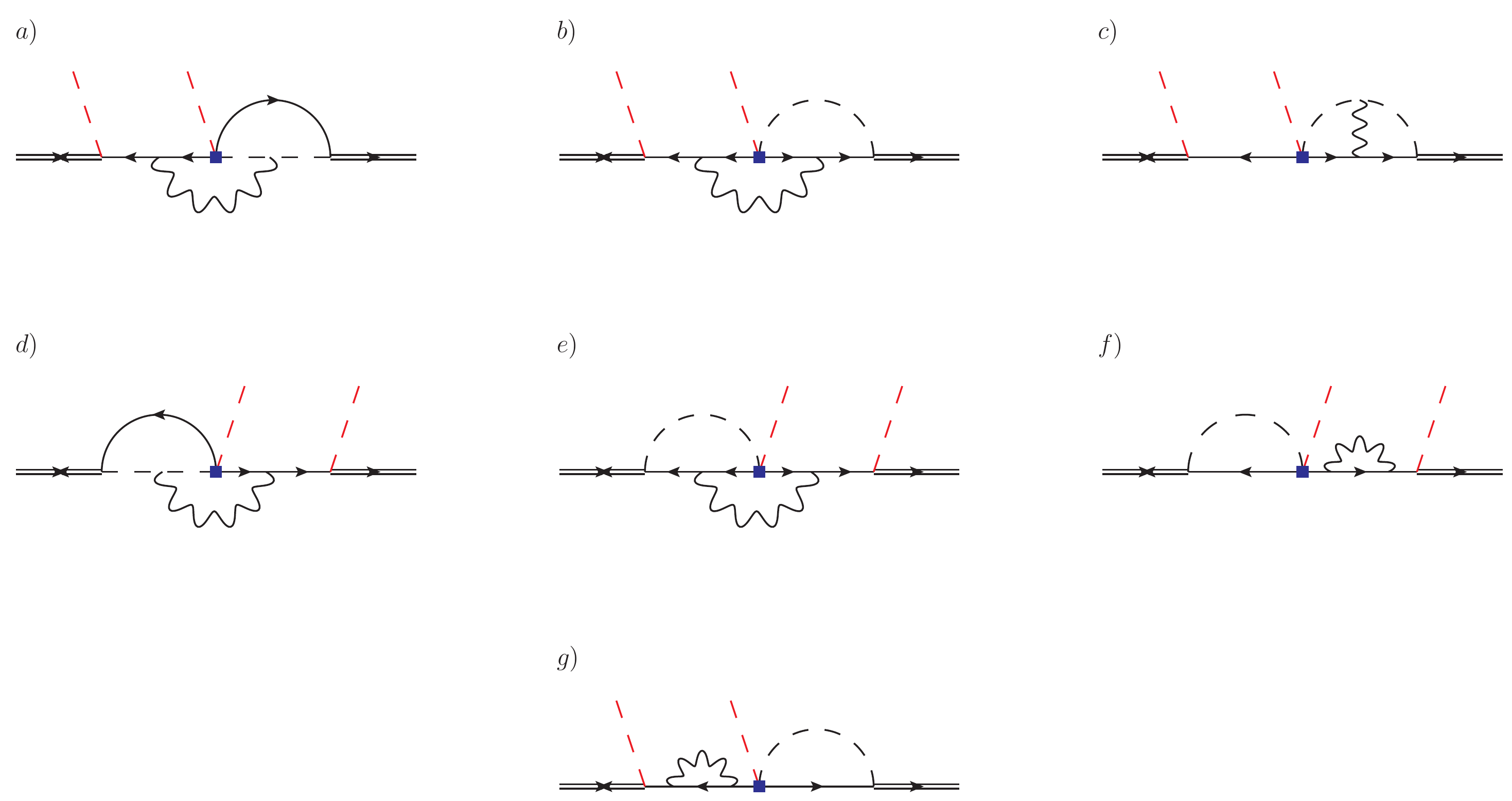}
\caption{\label{fig:fig_gauge_set1_hiera} We display the diagrams at order $(F_1^*F_i)^2$ and at leading order in the gauge couplings relevant for the matching calculation. According to the cut performed, either on lepton or antilepton lines, the diagrams contribute to $a^{\ell}$ or $a^{\bar{\ell}}$. }
\end{figure}

\begin{figure}[h]
\centering
\includegraphics[scale=0.45]{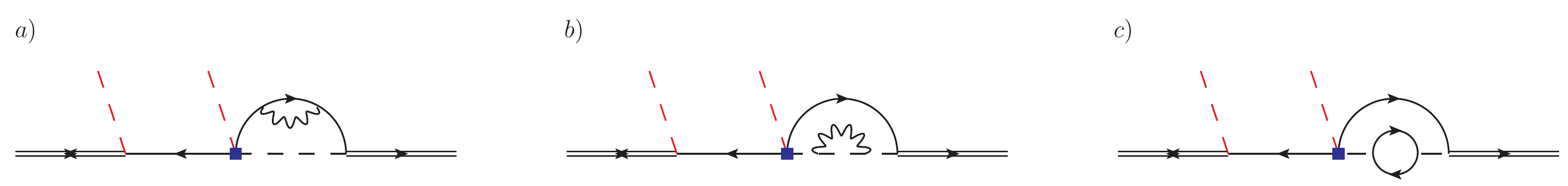}
\caption{\label{fig:noCP_hiera} We show three diagrams proportional to $(F_1^{*}F_i)^2$ that do not contribute to the CP asymmetry. The fermion loop in the diagram $c)$ is a top quark loop.}
\end{figure}

The topologies of the diagrams that potentially contribute to the matching coefficient are shown in figures \ref{fig:fig_gauge_set1_hiera} and \ref{fig:noCP_hiera}. 
 
One needs to fix a gauge for the actual calculation. We observe that we can distinguish two different situations by cutting a lepton line and leaving one loop in the diagrams in figure \ref{fig:fig_gauge_set1_hiera}: first, a Higgs boson and a lepton put on shell, second, a gauge boson and a lepton simultaneously on shell. Therefore we obtain processes with one or any gauge boson in the final state.  Being different physical processes, one can treat them within different gauges.

We adopt the Landau gauge for the diagrams in which the Higgs boson is cut together with a lepton (the gauge boson uncut). On the other hand, the Coulomb gauge is used when a gauge boson is cut. According to such choice, we can neglect all the diagrams with a gauge boson attached to an external Higgs boson leg. Indeed, the vertex interaction between a gauge and a Higgs boson is proportional to the momentum of the latter both in Landau and Coulomb gauge (see (\ref{SMlag}) and (\ref{SMCov})). If it depends on the external momentum, the diagrams develop a derivative that cannot be matched in the dimension-five operator (it will go in the matching of higher order operators containing derivatives). On the other hand, if it depends on the internal momentum then its contraction with the propagator vanishes both in Coulomb gauge, if the gauge boson is cut, and in Landau gauge if the gauge boson is uncut. Moreover the Coulomb gauge avoids singularities when a gauge boson is cut. 

The diagram $c)$ in figure \ref{fig:fig_gauge_set1_hiera} is similar to that studied in the case of nearly degenerate neutrino masses. The diagram may be cut in two different ways in order to put on shell a lepton together with a Higgs boson. It turns out that the only difference between the cuts lies in the number of circled vertices that brings to two opposite sign contributions eventually cancelling each other. The diagram $g)$ contains a subdiagram that vanishes in Landau gauge after the cut on the Higgs and lepton is performed (see figure~\ref{Fig1A} and eq.~\ref{zero_special_case}). 

We now discuss the three diagrams in figure \ref{fig:noCP_hiera}. These diagrams do not develop an imaginary part for the reaming loop amplitude after the cut on the lepton line. This has been discussed in the case of the corresponding diagrams for nearly degenerate neutrino masses. The different heavy neutrino mass arrangement do not change the argument. In order to remind our point, let us consider the diagram $a)$ in figure \ref{fig:noCP_hiera}, and let us cut it in all possible ways that put a lepton on shell. A first cut separates the diagram into tree-level subdiagrams. Since there is no loop uncut, we cannot have any additional phase. A second and a third cut are such to leave a one loop diagram after cutting through the lepton line. However no additional phase is generated by these diagrams. The incoming and outgoing particles are on shell and massless, and the particles in the loop are massless as well. The imaginary part of these diagrams corresponds to processes in which three massless particle enter the same vertex, and the available phase space vanishes in dimensional regularization. Therefore  the diagrams in figure \ref{fig:noCP_hiera} can develop only terms proportional to ${\rm{Re}}[\left( F^{*}_1 F_i\right)^2]$ that eventually cancel in the CP asymmetry. 
\begin{figure}[t]
\centering
\includegraphics[scale=0.55]{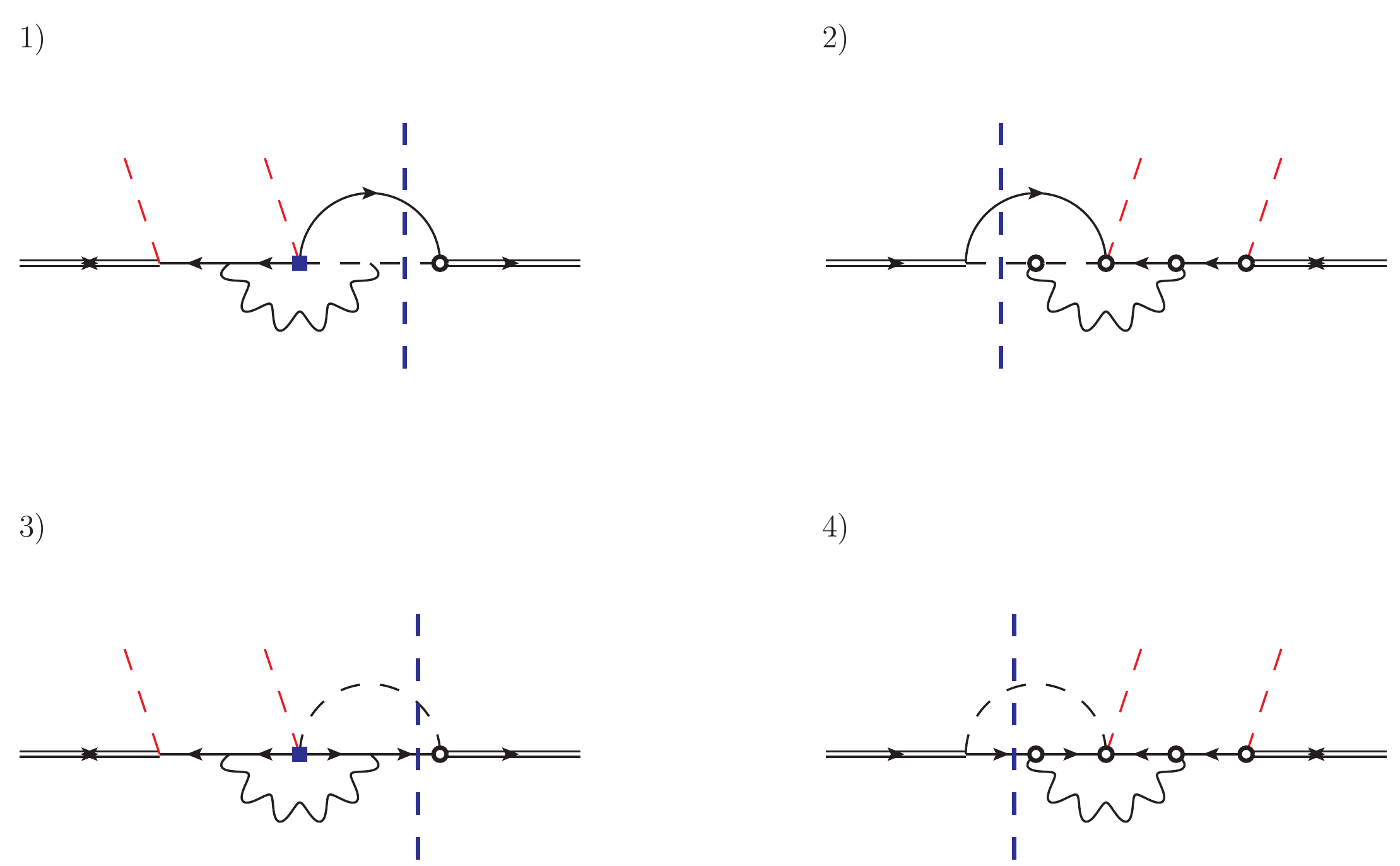}
\caption{\label{fig:leptohiggs} In each raw we draw a diagram and its complex conjugate where a lepton line is cut together with a Higgs boson. This set of diagrams are computed in Landau gauge.}
\end{figure}
%\begin{figure}[tbp]
%\centering
%\includegraphics[scale=0.42]{noCPexample}
%\caption{\label{fig:noCPexample} We show the three cuts that can be implemented in the diagram $a)$ in figure \ref{fig:noCP}. }
%\end{figure}

\begin{figure}[tbp]
\centering
\includegraphics[scale=0.55]{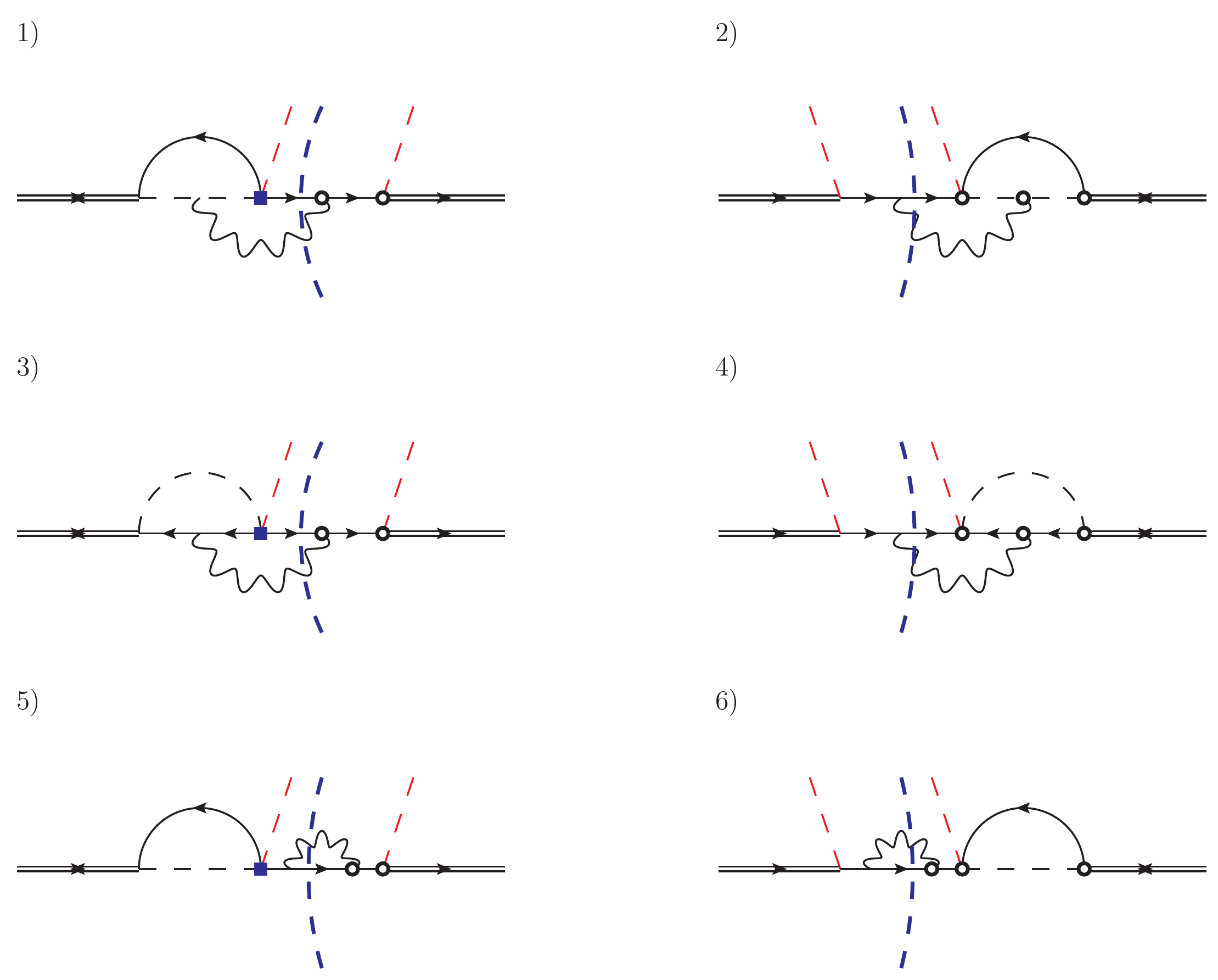}
\caption{\label{fig:leptogauge}In each raw we draw a diagram and its complex conjugate where a lepton line is cut together with a gauge boson. This set of diagrams are computed in Coulomb gauge.}
\end{figure}

We now discuss the diagrams that are not excluded by the above arguments. They are shown in figure \ref{fig:leptohiggs} and \ref{fig:leptogauge}, where the lepton line is cut together with a Higgs boson or a gauge boson respectively. In each raw a diagram and its complex conjugate are shown. We start with the diagrams in figure \ref{fig:leptohiggs} and we recall that the Landau gauge is adopted for these diagrams.  The result reads
\begin{eqnarray}
&& {\rm{Im}}\, (-i \mathcal{D}^{\ell}_{1,\hbox{\tiny fig.\ref{fig:leptohiggs}}})+{\rm{Im}}\, (-i \mathcal{D}^{\ell}_{2,\hbox{\tiny fig.\ref{fig:leptohiggs}}}) = 0 \,  , 
 \label{B7_hiera}  
\\
&& {\rm{Im}}\,  (-i \mathcal{D}^{\ell}_{3,\hbox{\tiny fig.\ref{fig:leptohiggs}}})+{\rm{Im}}\, (-i \mathcal{D}^{\ell}_{4,\hbox{\tiny fig.\ref{fig:leptohiggs}}})= -\frac{3\left( g^2+g'^2\right)}{8(16 \pi)^2 M_i}  {\rm{Im}}\left[ \left( F^{*}_1 F_i\right)^2\right]  \delta^{\mu \nu}  \delta_{mn}
  + \dots \, ,  \nonumber \\
  \phantom{x} 
\label{B8_hiera}
\end{eqnarray}
where the superscript $\ell$ stands for cutting a lepton in the diagram and the subscript refers to the diagram labels as in listed in figure \ref{fig:leptohiggs}. The dots stand for higher order terms in the heavy neutrino mass expansion and for the Yukawa coupling combination ${\rm{Re}}\left[ (F_1^{*}F_i)^2\right] $ not relevant for the CP asymmetry.  

Let us show the result for the diagrams shown in figure \ref{fig:leptogauge}. We use the Coulomb gauge to evaluate these diagrams where a gauge boson appears in the final state. The result reads as follows
\begin{eqnarray}
\footnotesize
&& {\rm{Im}}\, (-i \mathcal{D}^{\ell}_{1,\hbox{\tiny fig.\ref{fig:leptogauge}}})+{\rm{Im}}\, (-i \mathcal{D}^{\ell}_{2,\hbox{\tiny fig.\ref{fig:leptogauge}}}) = \frac{3\left( g^2+g'^2\right)}{8(16 \pi)^2 M_i}  {\rm{Im}}\left[ \left( F^{*}_1 F_i\right)^2\right]  \delta^{\mu \nu}  \delta_{mn} + \cdots \,  , 
\nonumber \\
\phantom{x}
 \label{B9_hiera}  
\end{eqnarray}
\begin{eqnarray}
&& {\rm{Im}}\,  (-i \mathcal{D}^{\ell}_{3,\hbox{\tiny fig.\ref{fig:leptogauge}}})+{\rm{Im}}\, (-i \mathcal{D}^{\ell}_{4,\hbox{\tiny fig.\ref{fig:leptogauge}}})= -\frac{3\left( g^2+g'^2\right)}{8(16 \pi)^2 M_i}   {\rm{Im}}\left[ \left( F^{*}_1 F_i\right)^2\right]  \delta^{\mu \nu}  \delta_{mn}
  + \dots \, ,  
  \nonumber \\
\phantom{x}
\label{B10_hiera} 
\end{eqnarray}
\begin{eqnarray}
&& {\rm{Im}}\,  (-i \mathcal{D}^{\ell}_{5,\hbox{\tiny fig.\ref{fig:leptogauge}}})+{\rm{Im}}\, (-i \mathcal{D}^{\ell}_{6,\hbox{\tiny fig.\ref{fig:leptogauge}}})=-\frac{3\left( 3g^2+g'^2\right)}{8(16 \pi)^2 M_i}  {\rm{Im}}\left[ \left( F^{*}_1 F_i\right)^2\right]  \delta^{\mu \nu}  \delta_{mn}
  + \dots \, ,
  \nonumber \\
\phantom{x} 
\label{B11_hiera}
\end{eqnarray}
where the superscript $\ell$ stands for cutting a lepton in the diagram and the subscript refers to the diagram labels as in listed in figure \ref{fig:leptogauge}. The matching coefficient of the dimension five-operator can be now fixed. The matrix element in the EFT$_2$ reads for the lepton case
\begin{equation}
\frac{{\rm{Im}} \, a^{\ell}}{M_1}\delta^{\mu \nu}  \delta_{mn} \, .
\label{B12_hiera}
\end{equation}
An analogous expression holds for the antilepton counterpart.
Therefore summing up the results (\ref{B2_hiera})-(\ref{B11_hiera}) and comparing with the expression in (\ref{B12_hiera}), we obtain for the imaginary part of Wilson coefficient contributing to the decay of $\nu_{R,1}$ into leptons (antileptons) the result in (\ref{match_a}).
%\begin{equation}
%{\rm{Im}} \, a^{\ell}=-{\rm{Im}} \, a^{\bar{\ell}}=\frac{3}{(16 \pi)^2} \frac{M}{M_i} \left[  10 \lambda - \frac{\left( 3g^2+g'^2\right)}{4} \right]   {\rm{Im}}\left[ \left( F^{*}_1 F_i\right)^2\right]
%\label{B10}
%\end{equation} 

\section{Matching dimension-seven operators in EFT$_2$}
\label{seven_hiera_match}
Now we want to address the matching of the dimension-seven operators that induce a thermal correction of order $|\lambda_t|(T/M_1)^4$ to the CP asymmetry in heavy Majorana neutrino decays. To this aim one has to consider  the dimension-seven operators describing the effective interaction between the non-relativistic heavy neutrinos and top-quark SU(2) singlet, the SU(2) heavy-quark doublet and the lepton doublet. As mentioned in chapter~\ref{chap:CPhiera}, these operators induce only part of the full set of corrections at order $(T/M_1)^4$ because other dimension-seven operators appear in the effective Lagrangian at order $1/M_1^3$.

\begin{figure}[h]
\centering
\includegraphics[scale=0.52]{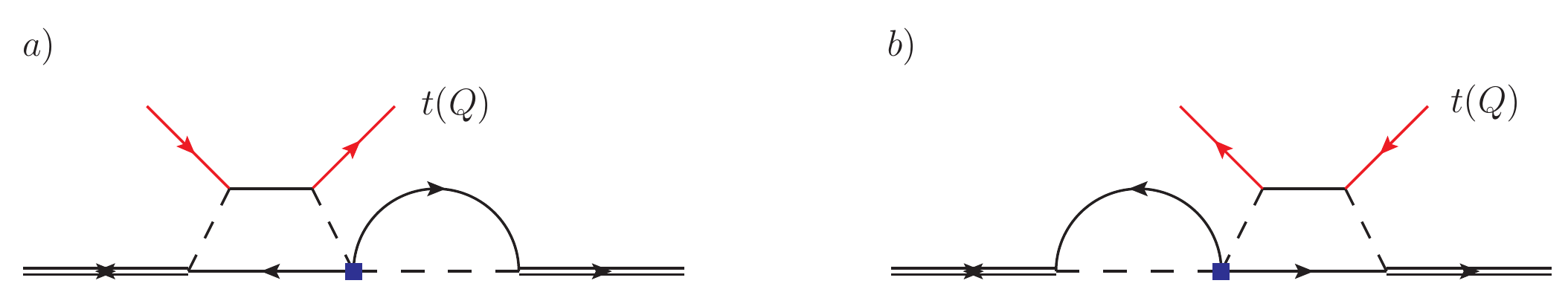}
\caption{\label{fig:matchtop} The two diagrams contributing to the heavy neutrino-top quark (heavy quark doublet) operator  are shown. Diagrams proportional to the Yukawa coupling combination $(F^*_1F_i)^2$ are displayed here.  Top (heavy-quark doublet) external legs are in solid red lines. We drop the arrow for internal top quark (heavy-quark doublet) in order to avoid confusion with lepton lines (arrows kept).}
\end{figure}

A quite limited number of diagrams allows to completely specify the matching coefficient of the heavy neutrino-top quark (heavy-quark doublet) operator and we show them in figure \ref{fig:matchtop}. The external fermion legs have to be understood as a top quark singlet or a heavy-quark doublet, as explicitly indicated.  As usual we show the diagrams proportional to the Yukawa coupling combination $(F^*_1F_i)^2$. 
We consider the following matrix elements in the fundamental theory
\begin{eqnarray}
&&\hspace{-10mm}
-i \left.\int d^{4}x\,e^{i p \cdot x} \int d^{4}y \int d^{4}z\,e^{i q \cdot (y-z)}\, 
\langle \Omega | T(\psi^{\mu}(x) \bar{\psi}^{\nu }(0) \, t^{\sigma}(y) \bar{t}^{\lambda}(z)) | \Omega \rangle 
\right|_{p^\alpha =(M_1 + i\eta,\bm{0}\,)}\!,
\nn
\\
\label{C0}\\
&&\hspace{-10mm}
-i \left.\int d^{4}x\,e^{i p \cdot x} \int d^{4}y \int d^{4}z\,e^{i q \cdot (y-z)}\, 
\langle \Omega | T(\psi^{\mu}(x) \bar{\psi}^{\nu }(0) \, Q_{m}^{\sigma}(y) \bar{Q}_{n}^{\lambda}(z)) | \Omega \rangle 
\right|_{p^\alpha =(M_1 + i\eta,\bm{0}\,)}\!,
\nn
\\
\label{C1}
\end{eqnarray}
describing respectively a $2 \rightarrow 2$ scattering between a heavy Majorana neutrino at rest and a right-handed top quark carrying momentum $q^\mu$, 
and a $2 \rightarrow 2$ scattering between a heavy Majorana neutrino at rest and a left-handed heavy quark doublet carrying momentum $q^\mu$. The indices  $\mu$, $\nu$, $\sigma$ and $\lambda$, are Lorentz indices, and $m$ and $n$ are the SU(2) indices 
of the heavy-quark doublet.

\begin{figure}[t]
\centering
\includegraphics[scale=0.52]{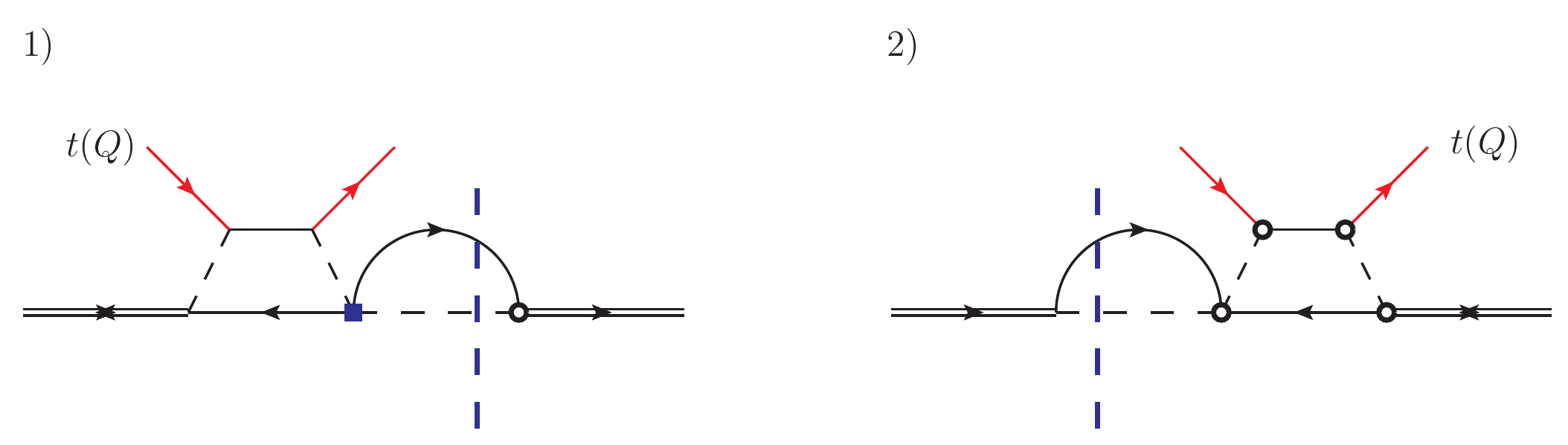}
\caption{\label{fig:singletop} We show the diagram a) of figure \ref{fig:matchtop} and its complex conjugate where a lepton line is cut together with a Higgs boson. Top (heavy-quark doublet) external legs are in solid red lines.}
\end{figure}

In order to distinguish the process with a top or a heavy-quark doublet as external fields and their contribution to the matching coefficients, we label the corresponding diagrams as $-i\mathcal{D}_t$ and $-i\mathcal{D}_Q$ respectively. We start with the diagram $a)$ in figure \ref{fig:matchtop}. In this case we can perform only one cut thorough the lepton line as shown in figure \ref{fig:singletop}. The result reads 
\begin{eqnarray}
&& {\rm{Im}}\, (-i \mathcal{D}^{\ell}_{t,1,\hbox{\tiny fig.\ref{fig:singletop}}})+{\rm{Im}}\, (-i \mathcal{D}^{\ell}_{t,2,\hbox{\tiny fig.\ref{fig:singletop}}}) 
\nn
\\
&&\hspace{4.0 cm}= -\frac{|\lambda_t|^2}{M_iM_1^2} \frac{{\rm{Im}}\left[ \left( F^{*}_1 F_i\right)^2\right] }{(16 \pi)^2}   \delta^{\mu \nu}  \delta_{mn} \left( P_{L} \gamma^{0} \right)^{\sigma \lambda} q_{0} + \cdots \, , 
\label{top1}
\nn
\\
\end{eqnarray}
\begin{eqnarray}
&& {\rm{Im}}\, (-i \mathcal{D}^{\ell}_{Q,1,\hbox{\tiny fig.\ref{fig:singletop}}})+{\rm{Im}}\, (-i \mathcal{D}^{\ell}_{Q,2,\hbox{\tiny fig.\ref{fig:singletop}}})
\nn
\\
&& \hspace{4.0 cm} = -\frac{|\lambda_t|^2}{M_iM_1^2} \frac{{\rm{Im}} \left[ \left( F^{*}_1 F_i\right)^2\right] }{2(16 \pi)^2}   \delta^{\mu \nu} \delta_{mn} \left( P_{R} \gamma^{0} \right)^{\sigma \lambda} q_{0} + \cdots \, .
\label{top2}
\nn
\\
%\nonumber \\
%&&+\frac{\kappa}{M_iM^2} \frac{|\lambda_t|^2}{6(16 \pi)^2}  {\rm{Im}}\left[ \left( F^{*}_1 F_i\right)^2\right] \left[   
%\left(\gamma^5\gamma^i\right)^{\mu \nu}   \left( P_{L} \gamma^{0} \right)^{\alpha \beta} q_{i}
%+\left(\gamma^5\gamma^i\right)^{\mu \nu}  \left( P_{L} \gamma_{i} \right)^{\alpha \beta} q_{0}
%\right] + \dots ,    \nonumber \\
%\phantom{x} 
\end{eqnarray} 
\begin{figure}[h]
\centering
\includegraphics[scale=0.55]{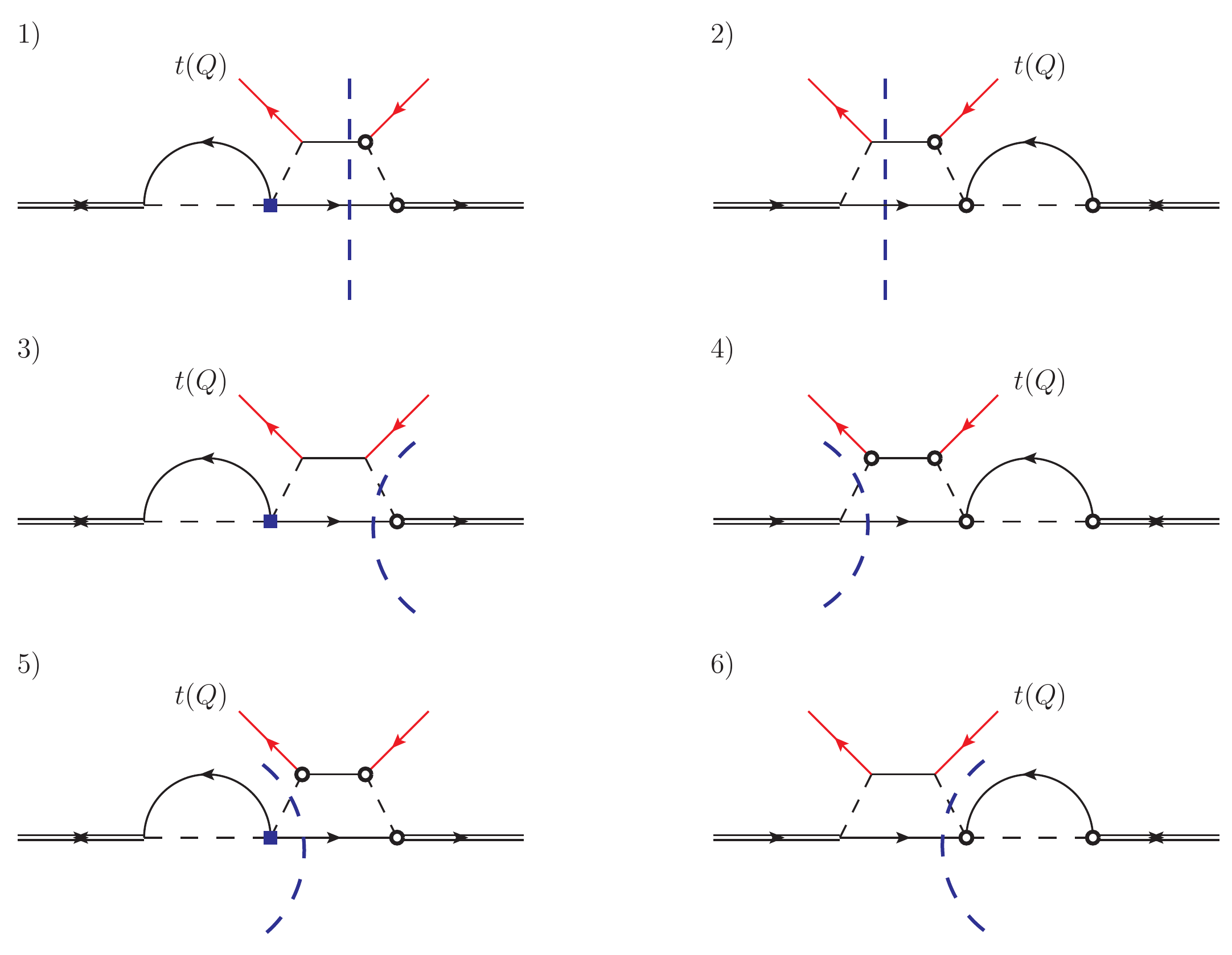}
\caption{\label{fig:tripletop} We show the diagram b) of figure \ref{fig:matchtop} and its complex conjugate where a lepton line is cut together with a Higgs boson or a top-quark line. Top (heavy-quark doublet) external legs are in solid red lines.}
\end{figure}
In (\ref{top1}) and (\ref{top2}) the dots stand for the real part of the Yukawa couplings combination $(F_1^*F_i)^2$ not relevant for the CP asymmetry, for higher order terms in the neutrino mass expansion and for terms that contain the coupling between the heavy Majorana neutrino spin and the medium. 

We then consider diagram $b)$ in figure \ref{fig:matchtop}. In this case the lepton line can be cut in three different ways as displayed in figure \ref{fig:tripletop}. This cuts were also discussed in section~\ref{sec_hiera_top_close} and it was highlighted that the whole amplitude is finite, whereas each single cut shows an IR divergence due to a massless Higgs boson. However, such divergence does not enter the term in the momentum expansion that we need for matching the dimension-seven operators in eqs.~(\ref{Ope_t}) and (\ref{Ope_Q}) and eventually responsible for a thermal correction to the neutrino decay widths. Also when calculating the top-quark (heavy-quark doublet) thermal condensates in the EFT$_2$ such terms would vanish (even number of momentum powers in the fermion tadpole). If we were interested in the CP asymmetries in  heavy neutrino-top scatterings then we would be sensitive to such IR divergence. The cuts in the first raw of figure~\ref{fig:tripletop} gives a contribution proportional to $q^0/M_1$, whereas the cuts on the second and third raw do not. The result reads
\begin{eqnarray}
&& \sum_{n=1}^{6} {\rm{Im}}\, (-i \mathcal{D}^{\ell}_{t,n,\hbox{\tiny fig.\ref{fig:tripletop}}}) = -\frac{3}{2}\frac{|\lambda_t|^2}{M_iM_1^2} \frac{{\rm{Im}}\left[ \left( F^{*}_1 F_i\right)^2\right] }{(16 \pi)^2}   \delta^{\mu \nu}  \delta_{mn} \left( P_{L} \gamma^{0} \right)^{\sigma \lambda} q_{0} + \cdots \, , 
\label{top1_3cut}
\nonumber
\\
\end{eqnarray}
\begin{eqnarray}
&& \sum_{n=1}^{6} {\rm{Im}}\, (-i \mathcal{D}^{\ell}_{Q,n,\hbox{\tiny fig.\ref{fig:tripletop}}}) = -\frac{3}{4}\frac{|\lambda_t|^2}{M_iM_1^2} \frac{{\rm{Im}}\left[ \left( F^{*}_1 F_i\right)^2\right] }{(16 \pi)^2}   \delta^{\mu \nu}  \delta_{mn} \left( P_{R} \gamma^{0} \right)^{\sigma \lambda} q_{0} + \cdots \, , 
\label{top2_3cut}
\nonumber
\\
\end{eqnarray}
where the dots stand for terms irrelevant for the CP asymmetry and powers of $q^0/M_1$ not contributing to the matching of the dimension-seven operators (\ref{Ope_t}) and (\ref{Ope_Q}). The sum of the cuts is IR finite.
The matrix element  is matched in the EFT$_2$ by assuming an isotropic medium 
\begin{equation}
\frac{{\rm{Im}} \, c^{\ell}_3}{M_1^3} \delta^{\mu \nu}  \left( P_{L} \gamma^{0} \right)^{\sigma \lambda} q_{0} \, ,
\label{eft2top}
\end{equation}
for the top-quark field, and 
\begin{equation}
\frac{{\rm{Im}} \, c^{\ell}_4}{M_1^3} \delta^{\mu \nu}  \delta_{mn} \left( P_{R} \gamma^{0} \right)^{\sigma \lambda} q_{0} \, .
\label{eft2doublet}
\end{equation}
for the heavy-quark doublet. Therefore we compare the sum of (\ref{top1}) and (\ref{top1_3cut}) with (\ref{eft2top}), and the sum of (\ref{top2}) and (\ref{top2_3cut}) with (\ref{eft2doublet}) respectively. In so-doing we obtain the result in (\ref{match_ct}) and (\ref{match_cQ}). 
%\begin{equation}
%c^{\ell}_3 = - \frac{|\lambda_t|^2}{(16 \pi)^2} \frac{M_1}{M_i} {\rm{Im}} \left[ \left( F^{*}_1 F_i\right)^2\right] \, , \quad c^{\ell}_4 = - \frac{|\lambda_t|^2}{2(16 \pi)^2} \frac{M_1}{M_i} {\rm{Im}} \left[ \left( F^{*}_1 F_i\right)^2\right] \, ,
%\end{equation}
Also in this case, the result for the antileptonic decays may be obtained by the substitution $F_1 \leftrightarrow F_i$.

\begin{figure}[tpb]
\centering
\includegraphics[scale=0.55]{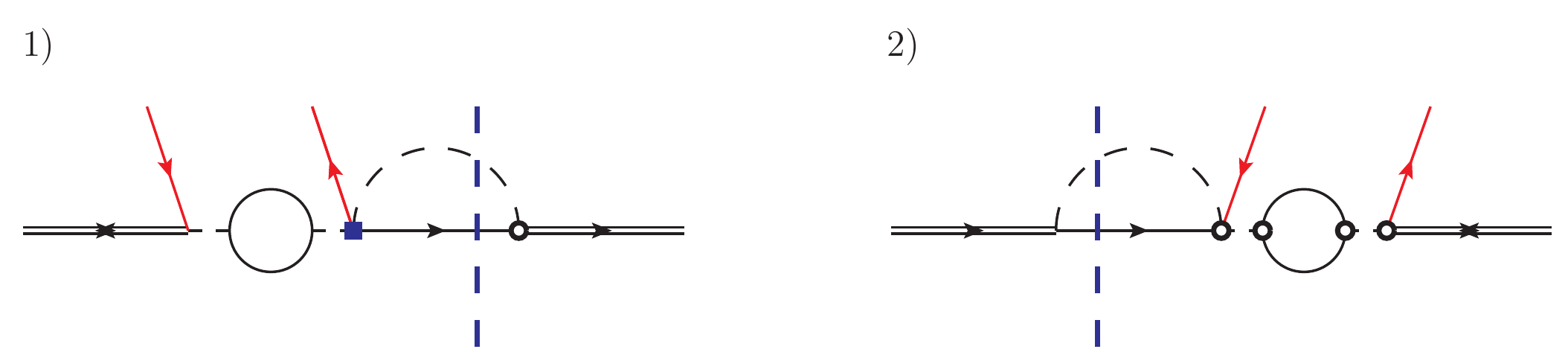}
\caption{\label{fig:lepto_cut} In each raw we draw a diagram and its complex conjugate where a lepton line is cut together with a Higgs boson. Lepton doublets as external legs are in solid red lines.}
\end{figure}

We now discuss the two diagrams that involve the lepton doublet $L_f$ as an external particle together with the heavy Majorana neutrinos. In this respect we consider the following matrix element where the heavy neutrino is at rest
\begin{equation}
-i \left.\int d^{4}x\,e^{i p \cdot x} \int d^{4}y \int d^{4}z\,e^{i q \cdot (y-z)}\, 
\langle \Omega | T(\psi^{\mu}(x) \bar{L}^{\lambda}_{f,m}(z) L^{\sigma}_{f',n}(y) \bar{\psi}^{\nu }(0))  | \Omega \rangle 
\right|_{p^\alpha =(M_1 + i\eta,\bm{0}\,)}, 
\label{C2}
\end{equation}
where $f$ and $f'$ are flavour indices, $\mu$, $\nu$, $\sigma$ and $\lambda$ are Lorentz indices, and $m$ and $n$ SU(2) indices. The matrix element in (\ref{C2}) 
describes a $2 \rightarrow 2$ scattering between a heavy Majorana neutrino at rest and a lepton doublet carrying momentum $q^\mu$. We consider only the diagrams involving the top-quark Yukawa coupling. In principle many other diagrams with gauge boson may contribute to the matrix element as well. Differently from the diagrams discussed so far, we have to treat separately the diagrams that admit a cut on a lepton line from those that allow for a cut on antilepton line. In the end, we see that leptonic cuts contributes to  $\left( \bar{N} P_{R} \, i v\cdot D L^{c}_{f'} \right) \left( \bar{L}^{c}_{f} P_{L}  N \right)$, whereas the cuts on antileptons contribute to   $\left(\bar{N}P_{L} \, iv\cdot D L_{f}\right) \left(\bar{L}_{f'} P_{R} N \right)$. We start from the diagrams in figure \ref{fig:lepto_cut}, where we can select a lepton in the final state and the result reads
\begin{eqnarray}
&&{\rm{Im}}\, (-i \mathcal{D}^{\ell}_{1,\hbox{\tiny fig.\ref{fig:lepto_cut}}}) +{\rm{Im}}\, (-i \mathcal{D}^{\ell}_{2,\hbox{\tiny fig.\ref{fig:lepto_cut}}})  \nonumber \\
&&=-\frac{9|\lambda_t|^2}{(16 \pi)^2} \, {\rm{Im}} \left[ (F_1^*F_i) (F^*_{f1}F_{f'i}) - (F_1F^*_i) (F_{f'1}F^*_{fi}) \right]  \frac{q_0}{M_1^2 M_i} (C\,P_{R})^{\mu \sigma}(P_{L}\,C)^{\lambda \nu} \, \delta_{mn} \, ,   \nonumber \\
\phantom{x}
\label{lepto_1}
\end{eqnarray}

\begin{figure}[tpb]
\centering
\includegraphics[scale=0.55]{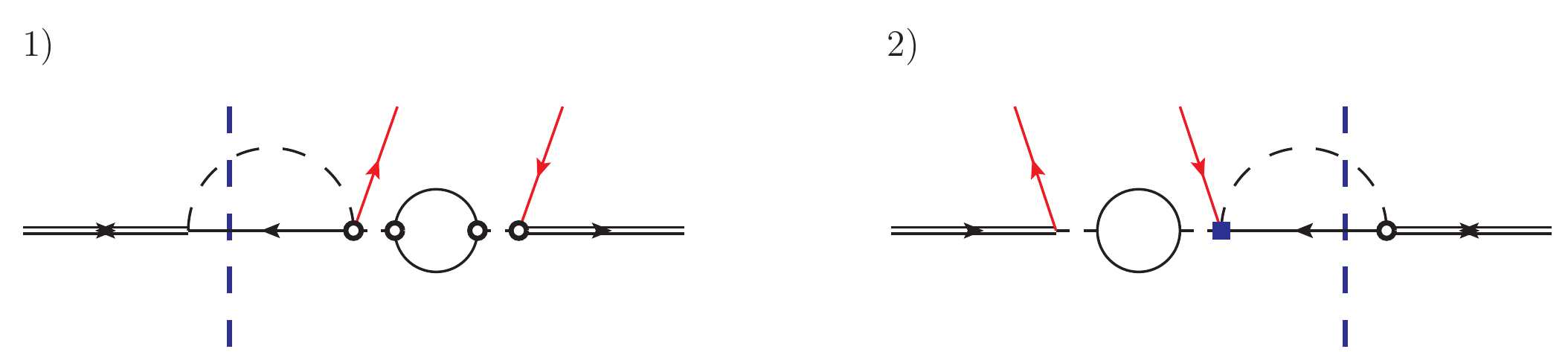}
\caption{\label{fig:lepto_cut2} In each raw we draw a diagram and its complex conjugate where an antilepton line is cut together with a Higgs boson. Lepton doublets as external legs are in solid red lines.}
\end{figure}

In this case the combination of the Yukawa coupling does not allow to combine them yet in the structure ${\rm{Im}}[(F_1^*F_i)^2]$, that is recovered when the tadpole in the EFT$_2$ is considered. On the EFT$_2$ side, the result in is matched with the following expression
\begin{equation}
\frac{{\rm{Im}}(c_{1,c}^{\ell, ff'})}{M_1^3} \, q_0 \, (C\,P_{R})^{\mu \sigma}(P_{L}\,C)^{\lambda \nu} \, \delta_{mn} \, ,
\label{leptoEFT2}
\end{equation}
and one obtains the result given in (\ref{match_cL}).
%\begin{equation}
%{\rm{Im}}(c_1^{\ell, ff'}) = -\frac{6|\lambda_t|^2}{(16 \pi)^2} \frac{M_1}{M_i} {\rm{Im}} \left[ (F_1^*F_i) (F^*_{f1}F_{f'i}) - (F_1F^*_i) (F_{f'1}F^*_{fi}) \right] \, .
%\label{matchlepto}
%\end{equation}
The result for cut on antileptons, namely the contribution to ${\rm{Im}}(c_1^{\bar{\ell}, ff'})$ has a very similar structure and it involves the second operator on the right-hand side in (\ref{Ope_L}) without the charge conjugation matrix. From the diagrams in figure \ref{fig:lepto_cut2} we find the following result 
\begin{eqnarray}
&&{\rm{Im}}\, (-i \mathcal{D}^{\ell}_{1,\hbox{\tiny fig.\ref{fig:lepto_cut2}}}) +{\rm{Im}}\, (-i \mathcal{D}^{\ell}_{2,\hbox{\tiny fig.\ref{fig:lepto_cut2}}})  \nonumber \\
&&=-\frac{9|\lambda_t|^2}{(16 \pi)^2} \, {\rm{Im}} \left[ (F_1F_i^*) (F_{f1}F^*_{f'i}) - (F_1^*F_i) (F^*_{f'1}F_{fi}) \right]  \frac{q_0}{M_1^2 M_i} (P_{L})^{\mu \lambda}(P_{R})^{\sigma \nu} \, \delta_{mn} \, , \nonumber 
\\
\phantom{x}
\label{lepto_2} 
\end{eqnarray}
and in the EFT$_2$ the matrix element in (\ref{C2}) reads
\begin{equation}
\frac{{\rm{Im}}(c_1^{\bar{\ell}, ff'})}{M_1^3} \, q_0 \, (P_{L})^{\mu \lambda}(P_{R})^{\sigma \nu} \, \delta_{mn}\, ,
\label{leptoEFT2_2}
\end{equation}
and we obtain the matching coefficient in (\ref{match_cLbar}). 
%\begin{equation}
%{\rm{Im}}(c_1^{\bar{\ell}, ff'}) = -\frac{6|\lambda_t|^2}{(16 \pi)^2} \frac{M_1}{M_i} {\rm{Im}} \left[ (F_1F_i^*) (F_{f1}F^*_{f'i}) - (F_1^*F_i) (F^*_{f'1}F_{fi}) \right] \, .
%\label{matchlepto2}
%\end{equation}
We notice that the first operator on the right-hand side in (\ref{Ope_L}) receives contribution only from the diagrams that admit a cut on a lepton line. Conversely the second term on the right-hand side in (\ref{Ope_L}) contains the contributions coming from those diagrams in which an antilepton is cut.